\documentclass{report}
\pdfoutput=1

\usepackage{amsmath}
\usepackage{amssymb}
\usepackage{amsthm}
\usepackage{graphicx}
\usepackage{verbatim}
\usepackage{makeidx}
\usepackage{ifthen}
\usepackage[refpage]{nomencl}
\usepackage{pict2e}
\usepackage{tikz-cd}
\usepackage{mathpartir}
\usepackage{hyperref}
\usepackage{enumitem}

\makeindex
\makenomenclature 

\RequirePackage{ifthen}
\renewcommand{\nomgroup}[1]{%
  \ifthenelse{\equal{#1}{B}}{\item[\textbf{Meta-Identifier Naming Conventions}]}{%
	\ifthenelse{\equal{#1}{C}}{\item[\textbf{Syntax of \dcalc}]}{%
	\ifthenelse{\equal{#1}{D}}{\item[\textbf{Syntactic Sugaring}]}{%
	\ifthenelse{\equal{#1}{F}}{\item[\textbf{Basic Sets and Functions of \dcalc}]}{%
	\ifthenelse{\equal{#1}{G}}{\item[\textbf{Basic Relations of \dcalc}]}{%
	\ifthenelse{\equal{#1}{E}}{\item[\textbf{Auxiliary Notations and Functions}]}{%
	\ifthenelse{\equal{#1}{J}}{\item[\textbf{Notations for Norms}]}{%
	\ifthenelse{\equal{#1}{K}}{\item[\textbf{Appendix B: Semantic Mapping}]}{
	\ifthenelse{\equal{#1}{L}}{\item[\textbf{Appendix D: Bounded Polymorphism, \dcalcb}]}{
}}}}}}}}}}

\chardef\myat=`\@ 

\def\eg{\emph{e.g.}}
\def\wrt{w.r.t.}

\newcommand{\dcalc}{$\mathtt{d}$}
\newcommand{\dcalct}{\tt{d}}
\newcommand{\dcalcb}{$\mathtt{d}_{\leq}$}
\newcommand{\dcalcbt}{\tt{d}} 

\title{An extended type system with lambda-typed lambda-expressions (extended version)\\
{\bf\small December 2024}}
\author{Matthias Weber}
\date{\normalsize TU Berlin, Technical Report,\\
mattnweber\myat $\!\!$ gmail.com}

\newenvironment{meditemize}{\begin{description}[leftmargin=3.5em,style=nextline,font=\normalfont,align=right]}{\end{description}}
\newenvironment{largeitemize}{\begin{description}[leftmargin=5em,style=nextline,font=\normalfont,align=right]}{\end{description}}
\newenvironment{hugeitemize}{\begin{description}[leftmargin=6.5em,style=nextline,font=\normalfont,align=right]}{\end{description}}
\newenvironment{giantitemize}{\begin{description}[leftmargin=8em,style=nextline,font=\normalfont,align=right]}{\end{description}}

\newcommand{\prim}{\tau}
\newcommand{\sdef}{!}
\newcommand{\prdef}[4]{[#1\!\doteq\!#2,#3:#4]}
\newcommand{\mydef}{\!:=\!}
\newcommand{\smydef}{:=}
\newcommand{\fun}{\!\Rightarrow\!}
\newcommand{\sfun}{\Rightarrow}
\newcommand{\injl}[2]{[#1,:\!#2]}
\newcommand{\injr}[2]{[:\!#1,#2]}
\newcommand{\pleft}[1]{#1.1}
\newcommand{\pright}[1]{#1.2}
\newcommand{\case}[2]{[#1\,?\,#2]}
\newcommand{\myneg}{\neg}

\newcommand{\any}[1]{\prim_{#1}}

\newcommand{\rtyp}{:^{\circ}}
\newcommand{\rsinc}{<^{\circ}}
\newcommand{\rleq}{\leq^{\circ}}
\newcommand{\rsgv}{\vdash^{\circ}}
\newcommand{\rincl}{\leq_{\lambda}^{\circ}}
\newcommand{\mint}[2]{M_{#1}(#2)}

\newcommand{\binop}[2]{\oplus(#1,#2)}
\newcommand{\prsumop}[2]{[#1\oplus#2]}
\newcommand{\prsuminjop}[2]{[#1\otimes#2]}
\newcommand{\prsumopd}[2]{[#1\oplus'#2]}
\newcommand{\prsumopdd}[2]{[#1\oplus''#2]}
\newcommand{\binbop}[3]{\ensuremath{\oplus_{#1}(#2,#3)}}
\newcommand{\binbopd}[3]{\ensuremath{\oplus'_{#1}(#2,#3)}}

\newcommand{\gv}{\vdash\,}
\newcommand{\sgv}{\!\vdash\!}
\newcommand{\ngv}{\models}
\newcommand{\sngv}{\!\models\!}
\newcommand{\rd}{\to^*}
\newcommand{\srd}{\to}
\newcommand{\rdn}[1]{\to_{#1}}
\newcommand{\rdr}{\nabla}
\newcommand{\eqv}{=_{\lambda}}
\newcommand{\eqAlph}{=_{\alpha}}
\newcommand{\neqv}{\neq_{\lambda}}

\newcommand{\gincl}{\leq_{\lambda^*}}
\newcommand{\ginc}{{{}\cdot\!{\!}\!\leq{}}}
\newcommand{\rginc}{{{}\cdot\!{\!}\!\leq{}^{\!\!{\circ}}\;}}

\newcommand{\finc}{{{}\ast\!{\!}\!\leq{}}}

\newcommand{\mrd}{\to_{:=}^*}
\newcommand{\smrd}{\to_{:=}}
\newcommand{\mrdn}[1]{\to_{:=}^{#1}}
\newcommand{\mrdr}{\nabla_{:=}}

\newcommand{\snur}{\to_{\myneg}}
\newcommand{\nur}{\to_{\myneg}^*}
\newcommand{\nurn}[1]{\to_{\myneg}^{#1}}
\newcommand{\dmrd}{\to_{=}}
\newcommand{\nurp}{\to_{\myneg}^+}
\newcommand{\nuro}{\to_{\myneg}^{01}}

\newcommand{\incl}{\leq_{\lambda}}

\newcommand{\ax}{\emph{ax}}
\newcommand{\axm}{\it{(ax)}}
\newcommand{\mystart}{\emph{start}}

\newcommand{\mystartm}{\it{(start)}}

\newcommand{\weak}{\emph{weak}}
\newcommand{\weakm}{\it{(weak)}}
\newcommand{\conv}{\emph{conv}}
\newcommand{\convm}{\it{(conv)}}
\newcommand{\sinc}{\emph{inc}}
\newcommand{\sincm}{\it{(inc)}}

\newcommand{\absu}{\emph{abs$_U$}}
\newcommand{\absum}{\it{(abs_U)}}

\newcommand{\appl}{\emph{appl}}
\newcommand{\applm}{\it{(appl)}}
\newcommand{\abse}{\emph{abs$_E$}}
\newcommand{\absem}{\it{(abs_E)}}
\newcommand{\chin}{\emph{ch$_I$}}
\newcommand{\chinm}{\it{(ch_I)}}
\newcommand{\chba}{\emph{ch$_B$}}
\newcommand{\chbam}{\it{(ch_B)}}
\newcommand{\pdef}{\emph{def}}
\newcommand{\pdefm}{\it{(def)}}

\newcommand{\bprod}{\emph{prd}}
\newcommand{\bprodm}{\it{(prd)}}

\newcommand{\prl}{\emph{pr$_L$}}
\newcommand{\prlm}{\it{(pr_L)}}
\newcommand{\prr}{\emph{pr$_R$}}
\newcommand{\prrm}{\it{(pr_R)}}
\newcommand{\bsum}{\emph{sum}}
\newcommand{\bsumm}{\it{(sum)}}
\newcommand{\injll}{\emph{inj$_L$}}
\newcommand{\injllm}{\it{(inj_L)}}
\newcommand{\finjll}{\emph{inj$_{L,\ast\leq}$}}
\newcommand{\finjllm}{\it{(inj_{L,\ast\leq})}}
\newcommand{\injlr}{\emph{inj$_R$}}
\newcommand{\injlrm}{\it{(inj_R)}}

\newcommand{\finjlrm}{\it{(inj_{R,\ast\leq})}}
\newcommand{\cased}{\emph{case}}
\newcommand{\casedm}{\it{(case)}}
\newcommand{\casedr}{\emph{case}$^{\circ}$}
\newcommand{\casedrm}{\it{(case^{\circ})}}
\newcommand{\negate}{\emph{neg}}
\newcommand{\negatem}{\it{(neg)}}

\newcommand{\srefl}{\emph{refl$_{\leq}$}}
\newcommand{\sreflm}{\it{(refl_{\leq})}}

\newcommand{\sstart}{\emph{ord$_<$}}
\newcommand{\sstartm}{\it{(ord_<)}}

\newcommand{\sembed}{\emph{emb$_{\leq}$}}
\newcommand{\sembedm}{\it{(emb_{\leq})}}
\newcommand{\sfembed}{\emph{emb$_{\ast\leq}$}}
\newcommand{\sfembedm}{\it{(emb_{\ast\leq})}}

\newcommand{\sstyp}{\emph{typ$_<$}}
\newcommand{\sstypm}{\it{(typ_<)}}

\newcommand{\ssabsu}{\emph{abs$_<$}}
\newcommand{\ssabsum}{\it{(abs_<)}}
\newcommand{\ssinjl}{\emph{inj$_{L,<}$}}
\newcommand{\ssinjlm}{\it{(inj_{L,<})}}
\newcommand{\ssinjr}{\emph{inj$_{R,<}$}}
\newcommand{\ssinjrm}{\it{(inj_{R,<})}}
\newcommand{\sspdef}{\emph{def$_<$}}
\newcommand{\sspdefm}{\it{(def_<)}}

\newcommand{\sabs}{\emph{abs}$_{\leq}$}
\newcommand{\sabsm}{(\it{abs}_{\leq})}
\newcommand{\sfabs}{\emph{abs}$_{\ast\leq}$}
\newcommand{\sfabsm}{(\it{abs}_{\ast\leq})}
\newcommand{\sfabse}{\emph{abs}$_{E,\ast\leq}$}
\newcommand{\sfabsem}{(\it{abs}_{E,\ast\leq})}

\newcommand{\ssbprod}{\emph{prsu$_<$}}
\newcommand{\ssbprodm}{\it{(prsu_<)}}
\newcommand{\sbprod}{\emph{prsu$_{\leq}$}}
\newcommand{\sbprodm}{\it{(prsu_{\leq})}}
\newcommand{\sfbprod}{\emph{prsu$_{\ast\leq}$}}
\newcommand{\sfbprodm}{\it{(prsu_{\ast\leq})}}

\newcommand{\sfpdef}{\emph{pdef$_{\ast\leq}$}}
\newcommand{\sfpdefm}{(\it{pdef_{\ast\leq}})}

\newcommand{\ostart}{\emph{start$_{\nord}$}}
\newcommand{\ostartm}{(\it{start}_{\nord})}
\newcommand{\ovar}{\emph{var$_{\nord}$}}
\newcommand{\ovarm}{(\it{var}_{\nord})}
\newcommand{\oabsl}{\emph{abs$_{\nord,1}$}}

\newcommand{\oabsr}{\emph{abs$_{\nord,2}$}}

\newcommand{\oabs}{\emph{abs$_{\nord,3}$}}

\newcommand{\oabsp}{\emph{abs$_{\nord,4}$}}

\newcommand{\dexp}{\mathcal{E}}
\newcommand{\dnrm}{\mathcal{\bar{E}}}

\newcommand{\dexps}{\mathcal{\dot{E}}}
\newcommand{\dvar}{\mathcal{V}}
\newcommand{\dsuper}{\mathcal{C}}
\newcommand{\nf}{\mathcal{N}}
\newcommand{\de}{\mathcal{D}}
\newcommand{\sn}[1]{{\mathcal{S}_{#1}}}
\newcommand{\ce}{\mathit{C}}
\newcommand{\dnf}[1]{\mathcal{N}^=_{#1}}
\newcommand{\dnfe}{\mathcal{N}^=}

\newcommand{\free}{\it{FV}}
\newcommand{\dom}{\it{dom}}
\newcommand{\domdec}{\it{dom}_:}
\newcommand{\domdef}{\it{dom}_{:=}}
\newcommand{\ran}{\it{ran}}

\newcommand{\nrm}[2]{\,\parallel\!\!#2\!\!\parallel_{#1}}
\newcommand{\snrm}[2]{\parallel#2\parallel_{#1}}

\newcommand{\asize}[3]{\sz_{#1}^{(#3)}(#2)}

\newcommand{\nord}{\ll}
\newcommand{\wnord}{\lesssim}
\newcommand{\rnord}{\gg}

\newcommand{\gsub}[2]{\lbrack#2/#1\rbrack}
\newcommand{\sth}{\mid}
\newcommand{\union}{\cup}
\newcommand{\sz}{{\mathit{S}}}
\newcommand{\negwt}{\mathit{W_{\myneg}}}

\newcommand{\inst}[1]{\mathbb{I}_{#1}}

\newcommand{\wff}{\mathcal{F}}
\newcommand{\vwff}{\mathcal{F}_{val}}

\newcommand{\ff}{\mathbf{ff}}
\newcommand{\tr}{\mathbf{tt}}

\newcommand{\noarg}{(.)}
\newcommand{\nats}{N}
\newcommand{\pnats}{N^{>0}}
\newcommand{\minimal}{\textit{Min}}

\newcommand{\negax}{\myneg}
\newcommand{\negaxx}{\myneg'}
\newcommand{\cast}{()}
\newcommand{\castin}[1]{()_{#1}^+}
\newcommand{\castout}[1]{()_{#1}^-}
\newcommand{\dcastin}[1]{()_{#1}^{++}}
\newcommand{\dcastout}[1]{()_{#1}^{--}}

\newcommand{\eq}{=_{\beta}}
\newcommand{\brd}{\to_{\beta}}
\newcommand{\snf}{\mathcal{\dot{N}}}
\newcommand{\sde}{\mathcal{\dot{D}}}
\newcommand{\ssem}{\mathbb{\dot{M}}}
\newcommand{\sdgv}{\:\dot{\vdash}\:}
\newcommand{\sem}{\mathbb{M}}
\newcommand{\primfun}{\prim}

\newcommand{\PTStoD}{\delta}
\newcommand{\Type}{\mathsf{type}}
\newcommand{\Prop}{\mathsf{Prop}}

\theoremstyle{definition} \newtheorem*{property}{Property}
\theoremstyle{definition} \newtheorem{law}{Law}
\theoremstyle{definition} \newtheorem{definition}{Definition}
\theoremstyle{remark} \newtheorem*{remark}{Remark}
\begin{document}
\rm
\setcounter{page}{1}

\hypersetup{pageanchor=false}
\begin{titlepage}
\maketitle
\begin{abstract}
We present the system \dcalc, an extended type system with lambda-typed lambda-expressions.
It is related to type systems originating from the Automath project.
\dcalc\ extends existing lambda-typed systems by an existential abstraction operator as well as propositional operators.
$\beta$-reduction is extended to also normalize negated expressions using a subset of the laws of classical negation,
hence \dcalc\ is normalizing both proofs and formulas which are handled uniformly as functional expressions.
\dcalc\ is using a reflexive type axiom for a constant $\prim$ to which no function can be typed.
Some properties are shown including confluence, subject reduction, uniqueness of types, strong normalization, and consistency.
We illustrate how, when using \dcalc, due to its limited logical strength, additional axioms must be added both for negation and for the mathematical structures whose deductions are to be formalized.
A condensed version of the main part can be found in \cite{Web20}.
Several appendices deal with extensions and variations of the proposed system.
\end{abstract}

\end{titlepage}
\hypersetup{pageanchor=true}
\pagenumbering{arabic}
\tableofcontents
\chapter{Introduction}

\label{intro}
\section{Overview of this document}%
The following section provides a brief overview of \dcalc\ including some simple examples.
The subsequent sections of this chapter contain a more detailed motivation of the concepts of \dcalc.
Chapter~\ref{definition} contains a formal definition of \dcalc.
Chapter~\ref{examples} presents some examples of the use of \dcalc.
Chapter~\ref{properties} contains proofs of the main properties of \dcalc.
Chapter~\ref{related} discusses possible variations of \dcalc\ and its relation to other systems.
Appendix~\ref{semantics} defines and briefly analyzes two mappings from \dcalc\ to untyped $\lambda$-calculus.
Appendix~\ref{formalization} illustrates the use of \dcalc\ as a logical framework.
\section{Overview of \dcalct}%
\label{overview}
In this paper, we will present an extended type system with lambda-typed lambda-expressions. Since such systems have received little attention we begin with some general remarks to provide some context.

Most type systems contain subsystems that can be classified as instances of \emph{pure type systems} (PTS) (\eg~\cite{Bar:93}).
As one of their properties, these systems use distinct operators to form dependent products and functional (i.e.~$\lambda$)-abstractions.
This reflects their underlying semantic distinction between the domains of functions and types.
In contrast, in the semantics of lambda-typed systems all entities are (partial) functions and typing is a binary relation between total functions.
As a consequence a function participating in the type relation may play the role of an element or of a type.
Moreover, since in general the domain and range of the type relation are not disjoint a function can (and will usually) have a double role both as element and as type. 
Therefore, in lambda-typed calculi one has to separate three aspects of an operator: its functional interpretation, i.e.~the equivalence class of entities it represents, its role as a type, i.e.~the entities on the element side related to its entities in the type relation, and
its role as an element, i.e.~the entities on the type side related to its entities in the type relation.
From this point of view, the semantic distinction between dependent products and $\lambda$-abstractions can be reduced to the distinction between the type-role or the element-role of a single underlying function and therefore, when defining a calculus, a single operator is sufficient.

Type systems outside of PTS using a single operator for both dependent products and $\lambda$-abstractions (i.e.~using \emph{$\lambda$-structured types}) have been investigated in early type systems such as $\Lambda$~\cite{deBruijn94,Nederpelt73}, as well as more recent approaches~\cite{Guidi09,guidi2021formal,JFP05}. In particular~\cite{JFP05} introduces the single-binder based $\flat$-cube, a variant of the $\beta$-cube which does not keep uniqueness of types and studies $\lambda$-structured type variants of well-known systems within this framework.

Before we present the basic elements of \dcalc, we would like to point out its additional underlying semantic assumptions:
\begin{itemize}
\item The range of the type relation is a subset of its domain, or in other words, every type has a type.
This allows for modelling type hierarchies of arbitrary size.
\item The type relation includes entities typing to itself, which allows for generating type hierarchies of arbitrary size from a finite set of distinct base entities. We discuss below why this assumption does not lead to well-known paradoxes.
\item There should be no inequivalent types of an element, or in other words, the type relation is restricted to be a function.
This means that \dcalc\ needs to satisfy the type uniqueness property (see Section~\ref{typing}).
We will review this assumption in Section~\ref{related.unique}.
\end{itemize}
It is well known that types can be interpreted as propositions and elements as proofs~\cite{Howard69}.
In our semantic setting this analogy is valid and we will make use of it throughout the paper, for example when motivating and describing the roles of operators we will frequently use the viewpoint of propositions and proofs.

Note that in our setting the interest is to formalize structured mathematical reasoning, hence there is no interest in the computational content of a proof.

Finally we would like to make a notational remark related to \dcalc\, but also to lambda-typed calculi in general.
Since an operator in \dcalc\ can be interpreted both as element and as type (or proposition), there is, in our view, a notational and naming dilemma.
For example, one and the same entity would be written appropriately as lambda-abstraction $\lambda x\!:\!a.b$ in its role as an element and universal quantification $\forall x\!:\!a.b$ in its role as a proposition
and neither notation would adequately cover both roles.
Therefore, in this case we use a more neutral notation $[x:a]b$ called \emph{universal abstraction}.
As will be seen this notational issue also applies to other operators of \dcalc.

After these remarks we can now turn to the core of \dcalc\, which is the system $\lambda^{\lambda}$~\cite{PdG93}, a reconstruction of a variation~\cite{Nederpelt73} of $\Lambda$, modified with a reflexive type axiom, see Table~\ref{lamlammod} for its type rules.
\begin{table}[!htb]
\fbox{
\begin{minipage}{0.96\textwidth}
\begin{align*}
\\[-8mm]
\it{(ax)}\;&\frac{}{\gv\prim:\prim}&
\it{(start)}\;\frac{\Gamma\gv a:b}{\Gamma,x:a\gv x:a}\qquad\qquad\qquad\qquad\\[1mm]
\it{(weak)}\;&\frac{\Gamma\gv a:b\qquad\Gamma\gv c:d}{\Gamma,x:c\gv a:b}&
\it{(conv)}\;\frac{\Gamma\gv a:b\qquad b\eqv c\qquad\Gamma\gv c:d}{\Gamma\gv a:c}\\[1mm]
\absum\;&\frac{\Gamma,x:a\gv b:c}{\Gamma\gv[x:a]b:[x:a]c}&
\applm\;\frac{\Gamma\gv a:[x:b]c\qquad\Gamma\gv d:b}{\Gamma\gv (a\,d):c\gsub{x}{d}}\qquad\;\\[-5mm]
\end{align*}
\end{minipage}
}
\caption{The kernel of~\dcalc: The system $\lambda^{\lambda}$~\cite{PdG93} modified with $\prim:\prim$\label{lamlammod}}
\end{table} 
As usual, we use contexts $\Gamma=(x_1:a_1,\ldots,x_n:a_n)$ declaring types of distinct variables and a $\beta$-conversion induced congruence $\eqv$ on expressions.
$\Gamma,x:a$ denotes the extension of $\Gamma$ on the right by a binding $x:a$ where $x$ is a variable not yet declared in $\Gamma$.
We use the notation $a\gsub{x}{b}$ to denote the substitution of free occurrences of $x$ in $a$ by $b$.

Element-roles and type-roles can now be illustrated by two simple examples:
Let $C=[x:\prim]\prim$ be the constant function delivering $\prim$, for any given argument of type $\prim$.
In its role as a type $C$ corresponds to an implication (``$\prim$ implies $\prim$'') and its
proofs include the identity $I=[x:\prim]x$ over $\prim$, i.e.~$\gv I:C$ using essentially the rules \mystart~and \absu.
In its role as an element $C$ corresponds to a constant function and it will type to itself, i.e.~$\gv C:C$ using essentially the rules \ax~and \absu.
$I$, which as element is obviously the identity function, in its role as a type corresponds to the proposition ``everything of type $\prim$ is true'' which should not have any elements (consistency is shown in Section~\ref{regular}).

More generally, in the rule \appl~the type-role of a universal abstraction ($[x:b]c$) can be intuitively understood as an infinite conjunction of instances ($c\gsub{x}{b_1}\wedge c\gsub{x}{b_2}\ldots$ where $b_i:b$ for all $i$).
Individual instances ($c\gsub{x}{b_i}$) can be projected by means of the application operator.
Intuitively, a proof of a universal abstraction ($[x:a]c$) must provide an infinite list of instances ($b\gsub{x}{a_1},b\gsub{x}{a_2},\ldots$ where $b\gsub{x}{a_i}:c\gsub{x}{a_i}$ for all $i$).
Consequently, in the rule \absu~a universal abstraction ($[x:a]b$ where $x:a\sgv b:c$) has the element-role of being a proof of another universal abstraction ($[x:a]c$).
 
As another example, the introduction and elimination rules for universal quantification can be derived almost trivially.
Note that in this and the following examples we write $[a\fun b]$ to denote $[x:a]b$ if $x$ is not free in $b$:
\begin{eqnarray*}
P:[a\fun b]\;\;\gv\;\;[x:[y:a](P\,y)]x&:&[[y:a](P\,y)\fun[y:a](P\,y)]\\{}
P:[a\fun b]\;\;\gv\;\;[x:a][z:[y:a](P\,y)](z\,x)&:&[x:a][[y:a](P\,y)\fun(P\,x)]
\end{eqnarray*}
When substituting $P$ by a constant function $[a\fun c]$ (where $\sgv c:b$) the deductions simplify to two variants of the modus ponens rule:
\begin{eqnarray*}
\gv\;\;[x:[a\fun c]]x&:&[[a\fun c]\fun[a\fun c]]\\{}
\gv\;\;[x:a][z:[a\fun c]](z\,x)&:&[a\fun[[a\fun c]\fun c]]
\end{eqnarray*}
In contrast to other type systems with $\lambda$-structured types, \dcalc\ is using a reflexive axiom (\ax).
This might seem very strange as the use of a reflexive axiom in combination with basic rules of PTS leads to paradoxes, \eg~\cite{Coquand86,Hurkens1995}, see also~\cite[Section 5.5]{Bar:93}.
However, as will be seen, in our setting of $\lambda$-structured types, the axiom \ax\ leads to a consistent system.
This is actually not very surprising since $\lambda^{\lambda}$ does not have an equivalent to the product rule used in PTS:\@
\begin{align*}
\it{(product)}\;\;&\frac{\Gamma\gv a:s_1\quad\Gamma,x:a\gv b:s_2}{\Gamma\gv(\Pi x:a.b):s_3}\quad\text{where}\;s_i\in S,\;\text{$S$ is a set of sorts}
\end{align*}
Adding such a rule, appropriately adapted, to the kernel of \dcalc\ would violate uniqueness of types~\cite{JFP05} and allow for reconstructing well-known paradoxes (see Sections~\ref{related.unique} and~\ref{paradox.subtyping}).

As a consequence of this restriction, functions from $\prim$ such as $I$ do not accept functional arguments such as $C$\footnote{In fact, $(I\;C)$ cannot be typed since $()\sgv I:[x:\prim]\prim$, hence $I$ is expecting arguments of type $\prim$, but $()\sgv C:C$ and $\prim\neq_{\lambda}C$}.
The lack of functions of type $\prim$ is the reason for achieving consistency with axiom $\prim:\prim$.

Unlike instances of PTS~(\eg~\cite{luo1989extended}), systems with $\lambda$-structured types have never been extended by existential or classical propositional operators.
While the kernel of \dcalc\ is sufficiently expressive to axiomatize basic mathematical structures (Sections~\ref{equality},~\ref{examples.nats}) 
the expressive and structuring  means of deductions can be enhanced by additional operators.
We begin with an operator that effectively provides for a \emph{deduction interface}, i.e.~a mechanism to hide details of interdependent deductions.
For this purpose, in analogy to a universal abstraction ($[x:a]b$), \dcalc\ introduces an \emph{existential abstraction} ($[x!a]b$) (see Table~\ref{existential} for its type rules).
\begin{table}[!htb]
\fbox{
\begin{minipage}{0.96\textwidth}
\begin{align*}
\\[-8mm]
\pdefm\;\;&\frac{\Gamma\sgv a: b\quad\Gamma\sgv c:d\gsub{x}{a}\quad\Gamma,x:b\sgv d:e}{\Gamma\sgv\prdef{x}{a}{c}{d}:[x!b]d}&
\absem\;\;&\frac{\Gamma,x:a\sgv b:c}{\Gamma\sgv[x!a]b:[x:a]c}\\[-5mm]
\end{align*}
\begin{align*}
\chinm\;\;&\frac{\Gamma\sgv a:[x!b]c}{\Gamma\sgv\pleft{a}:b}&
\chbam\;\;&\frac{\Gamma\sgv a:[x!b]c}{\Gamma\sgv\pright{a}:c\gsub{x}{\pleft{a}}}\\[-5mm]
\end{align*}
\end{minipage}
}
\caption{Type rules for existential abstractions\label{existential}}
\end{table} 
The notation is intended to maximise coherence with universal abstraction.
The type-role of an existential abstraction ($[x!b]d$) can be intuitively understood as an infinite disjunction ($d\gsub{x}{b_1}\vee d\gsub{x}{b_2}\ldots$ where $b_i:b$ for all $i$).
A proof of an existential abstraction ($[x!b]d$) must prove one of the instances (say $d\gsub{x}{b_j}$), i.e.~it must provide an instance of the quantification domain ($b_j:b$) and an element proving the instantiated formula ($c:d\gsub{x}{b_j}$).
This is formalized in rule \pdef\ with a new operator called \emph{protected definition} combining the two elements and a tag for the abstraction type ($\prdef{x}{a}{c}{d}$)\footnote{The type tag $d$ in $\prdef{x}{a}{c}{d}$ is necessary to ensure uniqueness of types}.
This is formalized in rule \pdef\ with a new operator called \emph{protected definition} combining the two elements and a tag for the abstraction type ($\prdef{x}{a}{c}{d}$)\footnote{The type tag $d$ in $\prdef{x}{a}{c}{d}$ is necessary to ensure uniqueness of types}.
This means that two deductions ($a$ and $c$), where one ($c$) is using the other one ($a$) in its type, are simultaneously abstracted into an existential abstraction type
\footnote{In Section~\ref{related.definition}, we discuss the restriction that $x$ may not appear in $c$ and how it could be relaxed.}.  
The element-role of an existential abstraction ($[x!a]b$) is not the one of a logical operator but of an entity providing an infinite list of instances ($b\gsub{x}{a_1}$, $b\gsub{x}{a_2}$, $\ldots$ where $a_i:a$ for all $i$).
But, as we just discussed above with respect to typing of universal abstractions (\absu), this is sufficient to type it to a universal abstraction (\abse).
Hence the element-roles of universal and existential abstraction are equivalent.
Pragmatically the element-role of existential abstraction is less frequently used and of less importance.
This is the reason why the notation for existential abstractions is more ``type-oriented'' than that of universal abstraction.

In analogy to the type-elimination of universal abstraction, \emph{projections} ($\pleft{a}$ and $\pright{a}$
\footnote{A postfix notation has been chosen since it seems more intuitive for projection sequences.}) 
are introduced as type-eliminators for existential abstraction with obvious equivalence laws.
\[
\pleft{\prdef{x}{a}{b}{c}}\eqv a\qquad
\pright{\prdef{x}{a}{b}{c}}\eqv b
\]
The type rules \chin~and \chba~for projections are similar to common rules for $\Sigma$ types, \eg~\cite{luo1989extended}.

As an example, the introduction and elimination rules for existential quantification can be derived as follows (with $\Gamma=(P,Q:[a\fun b])$):
\begin{eqnarray*}
\Gamma\;\gv\;[x\!:\!a][z\!:\!(P\,x)]\prdef{y}{x}{z}{(P\,y)}&:&[x:a][(P\,x)\fun[y!a](P\,y)]\\[1mm]
\Gamma\;\gv\;[x\!:\![y_1!a](P y_1)][z\!:\![y_2\!:\!a][(P y_2)\fun(Q y_2)]]\\
\prdef{y_3}{\pleft{x}}{((z\:\pleft{x})\pright{x})}{(Q y_3)}&:&[[y_1!a](P y_1)\\
&&\;\;\fun[[y_2\!:\!a][(P y_2)\fun(Q y_2)]\\
&&\quad\fun[y_3!a](Q y_3)]]
\end{eqnarray*}
A more detailed explanation of these deductions is given in Sections~\ref{reduction} and~\ref{typing}.
Note how the second example illustrates in a nutshell the role of existential abstraction as a deduction interface.
It can be understood as the transformation of a deduction on the basis of a reference ($x$) to its interface ($[y_1!a](P y_1)$) followed by the creation of a new interface ($[y_3!a](Q y_3)$) hiding the transformation details (application of $z$ to the extracted interdependent elements $\pleft{x}$ and $\pright{x}$).
More applications will be shown in Section~\ref{examples}, in particular in Sections~\ref{example.partial}(Partial functions),~\ref{example.functions}(Defining functions from deductions), and~\ref{example.groups}(Proof structuring).

The rule \abse~has the consequence that existential abstractions can now be instantiated as elements in the elimination rule for universal abstraction, i.e.~the rule \appl~can be instantiated as follows:
\[
(\it{appl})\gsub{a}{[x!a_1]a_2}\quad \frac{\Gamma\sgv [x!a_1]a_2:[x:b]c\quad \Gamma\sgv d:b}{\Gamma\sgv([x!a_1]a_2\,d):c\gsub{x}{d}}
\]
This motivates the extension of $\beta$-equality to existential abstractions, i.e.~we have
\[
([x:a]b\;c)\;\eqv\;b\gsub{x}{c}\;\eqv\;([x!a]b\;c)
\]
Note that these properties merely state that both abstractions have equivalent functional interpretations.
Their semantic distinction is represented by their role in the type relation.
For example, note that from $x:[y!a]b$ and $z:a$ one cannot conclude $(x\,z):([y!a]b\,z)$.
Note also that the extension of $\beta$-equality to existential abstraction precludes $\eta$-equality, i.e.~to uniquely determine a function by its value at each point, as this would directly lead to an inconsistency: When assuming $[x:a](b\, x)\eqv b$ for arbitrary $a$ and $b$ where $x$ not free in $b$ then
\[
[x:a](b\, x)\;\eqv\;[x:a]([x!a](b\, x)\, x)\;\eqv\;[x!a](b\, x)
\]
Finally, one might wonder why not type an existential abstraction $[x!a]b$ to an existential abstraction $[x!a]c$ (assuming  $x:a\sgv b:c$)?
According to the intuitive understanding of the type role of an existential abstraction as an infinite disjunction this would be logically 
invalid and indeed it is quite easy to see such a rule would lead to inconsistency\footnote{
First, with such a rule the type $P=[x:\prim][y!x]\prim$ would have itself as an element, i.e.~$\sgv P : P$.
However from a declaration of this type one could then extract a proof of $[z:\prim]z$ as follows
$y:P\sgv[z:\prim]\pleft{(y\,z)}:[z:\prim]z$.}.

The remaining part of \dcalc\ consists of some propositional operators (see Table~\ref{propositional} for their type rules).
\begin{table}[!htb]
\fbox{
\begin{minipage}{0.96\textwidth}
\begin{align*}
\\[-8mm]
\bprodm\;\;&\frac{\Gamma\sgv a:c\quad \Gamma\sgv b:d}{\Gamma\sgv[a,b]:[c,d]}\quad\;\;\;&
\bsumm\;\;&\frac{\Gamma\sgv a:c\quad \Gamma\sgv b:d}{\Gamma\sgv[a+b]:[c,d]}\\[0mm]
\prlm\;\;&\frac{\Gamma\sgv a:[b,c]}{\Gamma\sgv\pleft{a}:b}&
\prrm\;\;&\frac{\Gamma\sgv a:[b,c]}{\Gamma\sgv\pright{a}:c}\\[0mm]
\injllm\;\;&\frac{\Gamma\sgv a:b\quad\Gamma\sgv c:d}{\Gamma\sgv\injl{a}{c}:[b+c]}&
\injlrm\;\;&\frac{\Gamma\sgv a:b\quad\Gamma\sgv c:d}{\Gamma\sgv\injr{c}{a}:[c+b]}\\[-6mm]
\end{align*}
\begin{align*}
\casedm\;\;&\frac{\Gamma\sgv a:[x:c_1]d\quad\Gamma\sgv b:[y:c_2]d\quad\Gamma\sgv d:e}{\Gamma\sgv\case{a}{b}:[z:[c_1+c_2]]d}\qquad\qquad\qquad\qquad\quad\\[-7mm]
\end{align*}
\begin{align*}
\negatem\;\;&\frac{\Gamma\sgv a:b}{\Gamma\sgv\myneg a:b}&&\qquad\qquad\qquad\qquad\qquad\qquad\qquad\qquad\qquad\;\;\\[-5mm]
\end{align*}
\end{minipage}
}
\caption{Type rules for propositional operators\label{propositional}}
\end{table} 
\dcalc\ adds a \emph{product} ($[a,b]$) as a binary variation of universal abstraction ($[x:a]b$), i.e.~with an element-role as a binary pair and an type-role as a binary conjunction.
Due to the same notation dilemma as for universal abstraction, i.e.~neither the notation for pairs $\langle a,b\rangle$ nor for conjunctions $a \wedge b$ would be satisfactory, we use the neutral notation $[a,b]$.
The type rules \bprod, \prr, and \prl~for products directly encode the introduction and elimination rules for conjunctions.
The equivalence laws for projection are extended in an obvious way.
\[
\pleft{[a,b]}\eqv a\qquad
\pright{[a,b]}\eqv b
\]
As a example we show below that existential abstractions without dependencies (i.e.~$[x!a]b$, where $x$ is not free in $b$) are logically equivalent to products:
\begin{eqnarray*}
\gv\;\;[y:[x!a]b][\pleft{y},\pright{y}&:&[[x!a]b\fun[a,b]]\\{}
\gv\;\;[y:[a,b]]\prdef{x}{\pleft{y}}{\pright{y}}{b}&:&[[a,b]\fun[x!a]b]
\end{eqnarray*}
However, note that in comparison to existential abstractions, products have an intuitive symmetric type rule for this case (cf.~\abse\ vs.~\bprod).

Similarly \dcalc\ adds a sum ($[b_1+b_2]$) as a binary variation of the existential abstraction ($[x!a]b$), i.e.~with an element-role as a binary pair and an type-role as a binary disjunction.
Consequently, there is a type sequence analogous to existential abstraction: Expressions ($a_1:b_1$, $a_2:b_2$) may be used within \emph{injections} ($\injl{a_1}{b_2}$, $\injr{b_1}{a_2}$) (injections are the analogue to protected definitions), which type to a sum ($[b_1+b_2]$) which types to a product.
The bracket notation for products and abstractions is extended to injections and sums for notational coherence.
Similarly to existential abstraction we use a more type-oriented notation for sums.
The injection rules directly encode the introduction rules for disjunctions and rule \cased~introduces a case distinction operator.
Two equivalence laws describe its functional interpretation.
\[
(\case{a}{b}\,\injl{c}{d})\eqv(a\,c)\qquad
(\case{a}{b}\,\injr{c}{d})\eqv(b\,d)
\]
As an example, the introduction and elimination rules for disjunction can be derived as follows:
\begin{eqnarray*}
\gv\;\;[[x:a]\injl{x}{b}],[x:a]\injr{b}{x}]&:&[[a\fun[a+b]],[a\fun[b+a]]]\\{}
\gv\;\;[x:[a+b]][y:[a\fun c]]\\{}
[z:[b\fun c]](\case{y}{z}\,x)&:&[[a +b]\fun[[a\fun c]\fun [[b\fun c]\fun c]]]
\end{eqnarray*}
In the second example, in analogy to existential abstractions, a reference ($x$) to an interface ($[a+b]$) is hiding the information on which particular proposition ($a$ or $b$) is proven.

\noindent
In analogy to the $\beta$-equality for existential abstraction, projection is extended to sums.
\[
\pleft{[a+b]}\eqv a\qquad\pright{[a+b]}\eqv a
\]
Finally, to support common mathematical reasoning practices,
\dcalc\ introduces a classical negation operator $\myneg a$ which has a neutral type rule \negate~and which defines an equivalence class of propositions \wrt\ classical negation.
The central logical properties are the following: 
\[ 
a\eqv\myneg\myneg a 
\qquad
[a,b]\eqv\myneg[\myneg a +\myneg b] 
\qquad
[x\!:\!a]b\eqv\myneg[x!a]\myneg b 
\]
Furthermore, negation has no effect on terms to which no elimination operator can be applied when they are used as type.
\[
\myneg\prim\eqv\prim
\qquad\myneg\prdef{x}{a}{b}{c}\eqv\prdef{x}{a}{b}{c}
\]
\[
\myneg\injl{a}{b}\eqv\injl{a}{b}
\qquad\myneg\injr{a}{b}\eqv\injr{a}{b}
\qquad\myneg\case{a}{b}\eqv\case{a}{b}
\]
The negation laws define many negated formulas as equivalent which helps to eliminate many routine applications of logical equivalences in deductions.
For example, the following laws can be derived for arbitrary well-typed expressions $a,b$.
\begin{eqnarray*}
\gv\;\;\;[x:a]x&:&[a\fun a]\;\eqv\;[\myneg\myneg a\fun a]\;\eqv\;[a\fun\myneg\myneg a]\\
\gv\;\;\;[x:\myneg[a,b]]x&:&[\myneg[a,b]\fun[\myneg a\!+\!\myneg b]]\;\eqv\;[[\myneg a\!+\!\myneg b]\fun\myneg[a,b]]\\
\gv\;\;\;[x:\myneg[x!a]b]x&:&[\myneg[x!a]b\,\fun\,[x\!:\!a]\myneg b]\;\eqv\;[[x\!:\!a]\myneg b\,\fun\,\myneg[x!a]b]
\end{eqnarray*}
Truth and falsehood can now be defined as follows:
\[
\ff := [x:\prim]x \qquad \tr := \myneg\ff
\]
Note that falsehood is just a new convenient notation for the type role of the identity ($\ff=I$).
The expected properties follow almost directly:
\begin{eqnarray*}
\gv\;\;[x:\prim][y:\ff](y\,x)&:&[x:\prim][\ff\fun x]\\
\gv\;\;\prdef{x}{\prim}{\prim}{\myneg x}&:&\tr
\end{eqnarray*}
In the proof of $\tr$ note that $\tr\eqv[x!\prim]\myneg x$ and $\prim :\prim\eqv\myneg\prim\eqv(\myneg x)\gsub{x}{\prim}$.

Equivalence rules for negation obviously do not yield all logical properties of negation, \eg~they are not sufficient to prove 
$[a+\myneg a]$. Therefore one has to assume additional axioms.
Similarly, due to the limited strength of \dcalc, additional axioms must also be added for the mathematical structures whose deductions are to be formalized (for both see Section~\ref{examples}).

As this completes the overview of \dcalc, one may ask for its general advantages \wrt\ PTS\@.
While essentially equivalent expressive means could probably also be defined in a semantic setting using different domains for functions and types, the purely functional setting of \dcalc\ can be considered as conceptually more simple.
Independently from the semantic setting the use of common logical quantifiers and propositional connectors including a classical negation with rich equivalence laws seems more suitable for describing mathematical deductions than encoded operators or operators with constructive interpretation.

The remainder of this paper is structured as follows: A formal definition of \dcalc\ is presented in Section~\ref{definition} and several application examples are shown in Section~\ref{examples}. Readers with less focus on the theoretical results can well read Section~\ref{examples} before Section~\ref{definition}. 
The main part of the paper is Section~\ref{properties} containing proofs of confluence, subject reduction, uniqueness of types, strong normalization, and consistency.
\section{Core concepts}
\label{concepts}
In this section, we motivate and present the core concepts of \dcalc\ in a semi-formal property-oriented style.
To understand the starting point of~\dcalc, consider the rules of \emph{pure type systems} (PTS) summarized in Table~\ref{pts} which are the basis for a large class of typed systems which can be used to formalize deductions (see \eg\cite{Bar:93}).
As usual we use the notation $\Gamma\sgv A:B$ where $\Gamma=(x_1:A_1,\cdots,x_n:A_n)$ to formalize type assumptions.
As usual, $A\eqv B$ denotes equality modulo $\beta$-reduction.
The substitution of free occurrences of $x$ in $A$ by $B$ is denoted by $A\gsub{x}{B}$.
In the rules of Table~\ref{pts}, $A$, the set of \emph{axioms}, is a set of pairs $(s_1,s_2)$, where $s_i$ are from a set of sort $S$; and $R$, the set of \emph{rules}, is a set of pairs $(s_1,s_2,s_3)$ where $s_i\in S$.
\begin{table}[!htb]
\fbox{
\begin{minipage}{0.96\textwidth}
\begin{align*}
\\[-7mm]
\it{(axioms)}\;\;&\frac{}{()\gv s_1:s_2}\quad\text{where}\;(s_1,s_2)\in A \\
\it{(start)}\;\;&\frac{\Gamma\gv A:s}{\Gamma,x:A\gv x:A}\\
\it{(weakening)}\;\;&\frac{\Gamma\gv A:B\quad\Gamma\gv C:s}{\Gamma,x:C\gv A:B}\\
\it{(product)}\;\;&\frac{\Gamma\gv A:s_1\quad\Gamma,x:A\gv B:s_2}{\Gamma\gv(\Pi x:A.B):s_3}\quad\text{where}\;(s_1,s_2,s_3)\in R\\
\it{(application)}\;\;&\frac{\Gamma\gv F:(\Pi x:A.B)\quad\Gamma\gv a:A}{\Gamma\gv(F\,a):B\gsub{x}{a}}\\
\it{(abstraction)}\;\;&\frac{\Gamma,x:A\gv b:B\quad\Gamma\gv(\Pi x:A.B):s}{\Gamma\gv(\lambda x:A.b):(\Pi x:A.B)}\\
\it{(conversion)}\;\;&\frac{\Gamma\gv A:B\quad\Gamma\gv B':s\quad B\eqv B'}{\Gamma\gv A:B'}
\end{align*}
\end{minipage}
}
\caption{Rules of pure type systems\label{pts}}
\end{table}
Many instances of these typing systems allow for representing deductions as well-sorted $\lambda$-expressions which reduce to a unique normal form with respect to $\beta$-reduction, \eg\ \cite{Bar:93}. 
Depending on their axioms and rules, these systems allow to represent logical propositions as types and to identify sorting of terms and typing of formulae.

A well-known example of these systems is the calculus of constructions (CoC)~\cite{COQUAND198895} which can essentially be seen as a PTS with the following configuration (as usual, for $R$, we write $(s_1,s_2)$ for $(s_1,s_2,s_2)$):
\begin{eqnarray*}
S&=&\{\Prop,\Type\}\\
A&=&\{(\Prop,\Type)\}\\
R&=&\{(\Prop,\Prop),(\Prop,\Type),(\Type,\Prop),(\Type,\Type)\}
\end{eqnarray*}
We will now discuss the differences of the system $\lambda^{\lambda}$~\cite{PdG93}\footnote{which is itself a reconstruction of~\cite{Nederpelt73} which is a variation of Automath system $\Lambda$~\cite{deBruijn80}} to PTS.

The first deviation from PTS that is characteristic 
for the system $\lambda^{\lambda}$ is to use a single abstraction mechanism for propositions and expressions on all levels which we will denote as \emph{universal abstraction} and write as $[x:A]B$.
Some of the resulting types of systems have been studied as $\flat$-cubes in~\cite{JFP05}.
Semantically, a universal abstractions can be seen as a function specified by a $\lambda$-expression\footnote{Nevertheless, we use the term \emph{universal abstraction} to emphasize their  logical roles as functional \emph{abstraction} and \emph{universal} quantification}.
As a consequence, typing does not correspond to set inclusion anymore but becomes a relation between functions and types become roles that functions can play in this relation.
Consequently, $\beta$-equality can now be applied also to expressions playing the role of types.
Merging the abstraction mechanisms would lead to the following adapted typing rules, replacing the rules \emph{product}, \emph{application}, and \emph{abstraction}.
\begin{align*}
\it{(product')}\;\;&\frac{\Gamma\gv a:s_1\quad\Gamma,x:a\gv b:s_2}{\Gamma\gv[x:a]b:s_3}\quad\text{where}\;(s_1,s_2,s_3)\in R\\
\it{(application')}\;\;&\frac{\Gamma\gv a:[x:b]c\qquad\Gamma\gv d:b}{\Gamma\gv(a\,d):c\gsub{x}{d}}\\
\it{(abstraction')}\;\;&\frac{\Gamma,x:a\sgv b:c\quad\Gamma\sgv[x:a]c:s}{\Gamma\gv[x:a]b:[x:a]c}
\end{align*}
Note that to emphasize the difference to PTS we use $a$, $b$, $c$, etc.~to denote expressions using a single binding mechanism.
An important reason that this unification works smoothly is that abstraction works as introduction operator for itself.
We will see later how this becomes more complex when existential operators are introduced.

Note also that $\it{product'}$  and $\it{abstraction'}$ together would violate \emph{uniqueness of types}~\cite{JFP05}:
\[
\frac{\Gamma\gv a:b\quad \Gamma\gv a:c}{b\eqv c}
\]  
For example if $\ast,\square\in S$, $(\ast,\square)\in A$, and $(\ast,\square)\in R$ we would have
\[
\gv[x:\ast]x:\square\quad\text{and}\quad\gv[x:\ast]x:[x:\ast]\square
\]
This motivates the second difference of $\lambda^{\lambda}$ \wrt\ PTS which is to reject the rule $\it{product'}$ in order to ensure uniqueness of types.

The third and final difference to PTS that is characteristic for $\lambda^{\lambda}$  is to strengthen the rule $\it{abstraction'}$ to allow using any typable universal abstraction.
The rules \emph{weakening}, \emph{conversion}, and \emph{start} are adapted accordingly.
\begin{align*}
\it{(abstraction'')}\;\;&\frac{\Gamma,x:a\gv b:c}{\Gamma\gv[x:a]b:[x:a]c}\\ 
\it{(weakening')}\;\;&\frac{\Gamma\sgv a:b\quad\Gamma\sgv c:d}{\Gamma,x:c\sgv a:b}\\
\it{(conversion')}\;\;&\frac{\Gamma\sgv a:b\quad\Gamma\sgv c:d\quad b\eqv c}{\Gamma\sgv a:c}\\
\it{(start')}\;\;&\frac{\Gamma\sgv a:b}{\Gamma,x:a\sgv x:a}
\end{align*}
Similar kind of choices are included in the system $\lambda\delta$~\cite{Guidi09}.

\dcalc\ deviates from $\lambda^{\lambda}$ by setting $S=\{\prim\}$ and $A=\{(\prim,\prim)\}$ instead of $S=\{\prim,\kappa\}$ and $A=\{(\prim,\kappa)\}$..
It is well know that systems with $\prim:\prim$ can lead to paradoxes, \eg\ \cite{Coquand86}\cite{Hurkens1995}, see also~\cite{Bar:93}(Section 5.5). It will turn out that rejection of the rule $\it{product'}$ is the key reason for avoiding these paradoxes\footnote{Note that the system does do have formation rules that yield \eg\ $([x:\prim]\prim):\prim$}.

These adaptations result in a system that correponds to the system $\lambda^{\lambda}$, modified by $\prim:\prim$, in inference rule notation and which will be used as backbone of \dcalc.
The basic properties of this system are summarized in Table~\ref{dcalculus.core}.

As usual we use the notation $a\rd b$ for \emph{reduction}, the reflexive and transitive closure of \emph{single-step reduction} $a\srd b$.
%
\begin{table}[!htb]
\fbox{
\begin{minipage}{0.96\textwidth}
\begin{flushleft}
\mbox{\emph{Syntax}:}
\end{flushleft}
\begin{align*}
\\[-16mm]
\qquad\prim\qquad&\mbox{\emph{primitive constant }}\\[-1mm]
x, y,z,\ldots\in\dvar\qquad&\mbox{\emph{variables}}\\[-1mm]
{}[x:a]b\qquad&\mbox{\emph{universal abstraction}}\\[-1mm]
(a\,b)\qquad&\mbox{\emph{application}}\\[-6mm]
\end{align*}
\begin{flushleft}
\mbox{\emph{Reduction}:}
\end{flushleft}
\begin{align*}
\\[-11mm]
&\qquad\text{reflexive, transitive relation}\;\rd\\[2mm]
&\qquad(\beta_1)\quad([x:a]b\;c)\;\srd\;b\gsub{x}{c}\\[2mm]
&\mathit{([\_:\_]\_)}\;\frac{a\rd c\quad b\rd d}{[x:a]b\rd [x:c]d}\qquad
\mathit{(\_\,\_)}\;\frac{a\rd c\quad b\rd d}{(a\,b)\rd(c\,d)}
\end{align*}
\begin{flushleft}
\mbox{\emph{Typing}:}
\end{flushleft}
\begin{align*}
\\[-11mm]
&\qquad\text{ context  }\Gamma=(x_1:a_1,\cdots, x_n:a_n), x_i\neq x_j\\[-4mm]
\end{align*}
\begin{align*}
\\[-10mm]
&\axm\;\frac{}{\sgv\prim:\prim}\qquad
\mystartm\;\frac{\Gamma\sgv a:b}{\Gamma,x:a\sgv x:a}\\[1mm]
&\weakm\;\frac{\Gamma\sgv a:b\quad\Gamma\sgv c:d}{\Gamma,x:c\sgv a:b}\quad
\convm\;\frac{\Gamma\sgv a:b\quad b\eqv c\quad\Gamma\sgv c:d}{\Gamma\sgv a:c}\\[-4mm]
\end{align*}
\begin{align*}
\\[-10mm]
&\absum\;\frac{\Gamma,x:a\sgv b:c}{\Gamma\sgv[x:a]b:[x:a]c}\quad
\applm\;\frac{\Gamma\sgv a:[x:b]c\quad \Gamma\sgv d:b}{\Gamma\sgv(a\,d):c\gsub{x}{d}}
\end{align*}
\end{minipage}
}
\caption{Properties of the core of~\dcalc\label{dcalculus.core}}
\end{table}
We require two more properties for its meaningful use as a typing systems.
First, the typing relation should be decidable in the sense that there is an algorithm which takes an expression $a$ and either fails or delivers an expression $b$ with $()\gv a:b$.
Second, the calculus should be consistent in the sense that there is an expression which is not the type of any other expression. In~\dcalc, $[x:\prim]x$ is such an expression:
\[
\text{there is no $a$ with }()\gv a:[x:\prim]x
\] 

Due to the properties of confluence and strong normalisation we can define the unique \emph{normal forms} of a typable expression.
In the core of \dcalc, we have the following such normal forms.
\begin{eqnarray*}
\nf&=&\{\prim\}\;\union\;\{[x:a]b\sth a,b\in\nf\}\;\union\;\de\\
\de&=&\{x\sth x\in\dvar\}\;\union\;\{(a\,b)\sth a\in\de,b\in\nf\}
\end{eqnarray*}
Here the set $\de$ denotes the set of \emph{dead ends} (of reduction).
\section{Why additional operators?}
\label{opencoding}
In $\lambda^{\lambda}$, common encodings of logical operators can be used (where $[a\fun b]$ abbreviates $[x:a]b$ if $x$ is not free in $b$):
\begin{eqnarray*}
\mathrm{false}&:=&[x:\prim]x \\
\mathrm{true}&:=&[x:\prim][y:x]y\\
\mathrm{implies}&:=&[x:\prim][y:\prim][x\fun y]\\
\mathrm{not}&:=&[x:\prim][x\fun\mathrm{false}]\\
\mathrm{and}&:=&[x:\prim][y:\prim][z:\prim][[x\fun[y\fun z]]\fun z]\\
\mathrm{or}&:=&[x:\prim][y:\prim][z:\prim][[x\fun z]\fun[[y\fun z]\fun z]]\\
\mathrm{forall}&:=&[x:\prim][y:[x\fun\prim]][z:x](y\,z)\\
\mathrm{exists}&:=&[x:\prim][y:[x\fun\prim][[z:x][(y\,z)\fun x]\fun x]  
\end{eqnarray*}
Given these definitions, one can derive further logical properties, for example the following one (where $(a_1\,a_2\ldots a_n)$ abbreviates $(\ldots(a_1\,a_2)\ldots a_n))$:
\[
x_1,x_2:\prim \gv [y:(\mathrm{and}\:x_1\:x_2)](y\,x_1\,[z_1:x_1][z_2:x_2]z_1):[(\mathrm{and}\:x_1\:x_2)\fun x_1]
\]
Hence one could argue that no further logical operators (apart from the law of the excluded middle) seem necessary.
We do not follow this argument because as we have already argued in Section~\ref{overview}, due to the typing rules of the core of \dcalc, declarations such as $x:\prim$ cannnot be instantiated to functions $[x:a]b$.
This limitation of expressive power is a drawback of such encodings.

To overcome this limitation, we could introduce abbreviations for expression schemas, \eg
\begin{eqnarray*}
\mathrm{and}(a,b)&:=&[z:\prim][[a\fun[b\fun z]]\fun z]
\end{eqnarray*}
However, regardless of the use of schemas we do not adopt the encoding approach in \dcalc\ because of properties such as 
\[
\frac{c:a\quad d:b}{[z:P][w:[x:a][y:b]z]((w\,c)\,d):\mathrm{and}_P(a,b)}
\]
which we consider less intuitive for deductions involving conjunction as compared to the approach in \dcalc\ 
which is to introduce additional logical operators with their specific typing and congruence laws.
This will lead to the law:
\[
\frac{c:a\quad d:b}{[c,d]:[a,b]}
\]
\section{Existential abstraction}
\label{existential2}
The restricted structuring means of the core of \dcalc\ become apparent when modeling the assumption of some variable $x$ of type $a$ and with a constraining property $P_x$ depending on $x$. In the core of \dcalc\ one is forced to use two consecutive declarations as assumption and consequently a nested application as resolution.
\[
\frac{\Gamma\gv b:[x_1:a][x_2:P_{x_1}] Q_{x_1,x_2} \quad \Gamma\gv c_1:a \quad \Gamma\gv c_2:P_{c_1}}
{\Gamma\gv ((b\,c_1)\,c_2): Q_{c_1,c_2}}
\]
A key mechanism of \dcalc\ is to extend this system by \emph{existential abstraction}, written as $[x!a]b$, and \emph{left and right projection}, written as $\pleft{a}$ and $\pright{a}$. Using these operators, the two assumptions and the two applications can be merged.
\[
\frac{\Gamma\sgv b:[x:[y!a] P_y]Q_{\pleft{x},\pright{x}}\quad \Gamma\sgv c:[y!a]P_y}
{\Gamma\sgv (b\,c):Q_{\pleft{c},\pright{c}}}
\]
We now turn to the intended additional typing laws. The elimination laws are relatively straightforward and similar to common laws for $\Sigma$ types, \eg\ \cite{luo1989extended}.
\[
\chinm\;\frac{\Gamma\sgv a:[x!b]c}{\Gamma\sgv\pleft{a}:b}\qquad 
\chbam\;\frac{\Gamma\sgv a:[x!b]c}{\Gamma\sgv\pright{a}:c\gsub{x}{\pleft{a}}}\qquad
\]
Note that the type of $\pright{a}$ contains a projection $\pleft{a}$, i.e.~$\pleft{a}$ is used on both sides of the elimination rules of $[x!b]c$\footnote{This is related to the remark about existential quantification in \S12 of~\cite{Howard69}.}.
Note that logically left projection has similarities to a skolem function and right projection can be seen as a skolemisation operator\footnote{Left projection has also similarities with Hilbert's $\epsilon$-operator, see Section~\ref{related.hol}}.

Next, we define the introduction law for existential abstraction, which turns out to be somewhat more complicated. As for universal abstractions, some sort of unique introduction operation for existential abstraction seems necessary\footnote{Otherwise expressions would have types of very different structures at the same time. This is not investigated further in this context.}. A first approximation of the corresponding typing rule would be to use a tuple to introduce existential abstractions:
\[
\frac{\Gamma\sgv a:b\quad\Gamma\sgv c:d\gsub{x}{a}}{\Gamma\sgv[a,c]:[x!b]d}
\]
However, this rule is inappropriate in our setting as it allows for many incompatible types of the same expression and thus violates uniqueness of types, for example:
\[
\frac{\Gamma\sgv a:b\quad c:[d\fun d]}{\Gamma\sgv [a,c]:[x!b][x\fun d]}
\qquad
\frac{\Gamma\sgv a:b\quad c:[d\fun d]}{\Gamma\sgv [a,c]:[x!b][d\fun x]}
\]
A solution to this issue is to specify the intended type explicitly in the protected definition. Instead of using a notation such as $[a,c]_{[x!b]d}$ we can omit the expression $b$ by introducing an operator called \emph{protected definition}, written as $\prdef{x}{a}{c}{d}$ where we allow $d$, but not $c$ to use the variable $x$. The second attempt for the typing rule then becomes
\[
\frac{\Gamma\sgv a:b\quad\Gamma\sgv c:d\gsub{x}{a}}{\Gamma\sgv\prdef{x}{a}{c}{d}:[x!b]d}
\]
There is a logical issue with this naive characterization of existential quantification:
In the expression $d\gsub{x}{a}$ in the rule-antecedent it would be possible to unfold $x$ to $a$ at arbitrary places in $d$. This may lead to problematic instantiation of the rules which leads to expressions that cannot be typed. For example, we can instantiate the proposed rule to obtain (with $\dot{\prim}$ abbreviating $[\prim\fun\prim]$):
\[
\frac{\Gamma\sgv\dot{\prim}:\dot{\prim}\qquad
\Gamma\sgv[x:\dot{\prim}][(x\,\prim)\fun\prim]\;:\;([x:y][(x\,\prim)\fun\prim])\gsub{y}{\dot{\prim}}}
{\prdef{y}{\dot{\prim}}{[x:\dot{\prim}][(x\,\prim)\fun\prim]}{[x:y][(x\,\prim)\fun\prim]}\;:\;[y!\dot{\prim}][x:y][(x\,\prim)\fun\prim]}
\]
and therefore, since both antecedents of the rule are true, i.e.
\[
\Gamma\sgv\dot{\prim}:\dot{\prim}\qquad[x:\dot{\prim}][(x\,\prim)\fun\prim]:[x:\dot{\prim}][(x\,\prim)\fun\prim]
\] 
the consequence of this rule would be true as well. Note however, that in the type expression in the rule-consequence the expression $(x\,\prim)$ is not typable.
The cause of this issue is that the unfolding of the definition of $x$, which is necessary to ensure that $(x\,\prim)$ is typable, has been lost in the typing step.
We therefore need to add a condition to this rule requiring typability of the intended type. This leads to the final form
\[
\pdefm\;\frac{\Gamma\sgv a:b\quad\Gamma\sgv c:d\gsub{x}{a}\quad\Gamma,x:b\sgv d:e}{\Gamma\sgv\prdef{x}{a}{c}{d}:[x!b]d}
\]
Furthermore, protected definitions are subject to the following reduction axioms:
\[
\mathit{(\pi_1)}\;\;\pleft{\prdef{x}{a}{b}{c}}\srd a\qquad
\mathit{(\pi_2)}\;\;\pright{\prdef{x}{a}{b}{c}}\srd b
\] 

Finally, we need to define a typing law for existential abstraction itself, in order to define well-defined existential abtractions.
With respect to type elimination, an expression of type $[x!a]b$ embodies a proof of $b$ for some concrete expression $a_0$.
However, with respect to type introduction, i.e.~when playing the role of a proof, an expression $[x!a]b$ has the logical strength identical to a universal abstraction: it is a function which given some $x:a$ it produces a $b:c$, hence we use the rule: 
\[ 
\absem\;\frac{\Gamma\sgv a:d\quad \Gamma,x:a\sgv b:c}{\Gamma\sgv[x!a]b:[x:a]c}
\]
It is important to note that this law has the effect that existential abstractions can now be instantiated as elements in the elimination rule for universal abstraction, i.e.
\[
\frac{\Gamma\sgv [x!a]b:[x:a]c\quad \Gamma\sgv d:a}{\Gamma\sgv([x!a]b\:d):c\gsub{x}{d}}
\]
Hence, we extend the law of $\beta$-equality to existential abstractions as well
\[
(\beta_2)\;\;([x!a]b\:c)\srd b\gsub{x}{c}
\]
Note that this does not lead to logical inconsistency because from $x:[y!a]b$ and $z:a$ one cannot conclude $(x\,z):([y!a]b\:z)$.

One might ask why not type an existential abstraction $[x!a]b$ to an existential abstraction $[x!a]c$ (assuming  $x:a\sgv b:c$).
This would logically be invalid since we could prove existential statements by magic, i.e.~without providing an instance expressions.
Actually it is quite easy to see such a rule would lead to inconsistency.
First, the type $[x:\prim][y!x]\prim$ has an element, namely itself! 
\[
\gv [x:\prim][y!x]\prim : [x:\prim][y!x]\prim
\]
However from a declaration of this type one can extract a proof of $[z:\prim]z$ as follows
\[
y:[x_1:\prim][x_2!x_1]\prim \gv [z:\prim]\pleft{(y\,z)}:[z:\prim]z
\]
The discussion about the typing rules for abstractions can be summarized as follows:
\begin{itemize}
\item From the point of view of reduction, universal abstraction and existential abstraction both share the properties of $\lambda$-abstraction.
\item From the point of view of typing, when playing the role of a type, universal abstraction is its own \emph{introducer} and existential abstraction requires a proof of a concrete instance in its introduction rule, however when playing the role of an element, universal abstraction and existential abstraction have the same contraction and typing rules.
\end{itemize}
\noindent
The set of normal forms $\nf$ is extended as follows::
\begin{eqnarray*}
\nf&=&\{\prim\}\;\union\;\{[x:a]b,[x!a]b,\prdef{x}{a}{b}{c}\sth a,b,c\in\nf\}\;\union\;\de\\
\de&=&\{x\sth x\in\dvar\}\;\union\;\{(a\,b),\pright a,\pleft a\sth a\in\de,b\in\nf\}
\end{eqnarray*}

\noindent
The basic properties of existential abstraction are summarized in Table~\ref{dcalculus.existential}.
\begin{table}[!htb]
\fbox{
\begin{minipage}{0.96\textwidth}
\begin{flushleft}
\mbox{\emph{Syntax}:}
\end{flushleft}
\begin{align*}
\\[-16mm]
[x!a]c\qquad&\mbox{\emph{existential abstraction}}\\[-1mm]
\prdef{x}{a}{b}{c}\qquad&\mbox{\emph{protected definition ($x$ may be used in $c$)}}\\[-1mm]
\pleft{a}, \pright{a}\qquad&\mbox{\emph{left projection}, \emph{right projection}}\\[-6mm]
\end{align*}
\begin{flushleft}
\mbox{\emph{Reduction}:}
\end{flushleft}
\begin{align*}
\\[-8mm]
&\mathit{(\pi_1)}\;\;\pleft{\prdef{x}{a}{b}{c}}\srd a\qquad
\mathit{(\pi_2)}\;\;\pright{\prdef{x}{a}{b}{c}}\srd b\\
&(\beta_2)\;\;([x!a]b\:c)\srd b\gsub{x}{c}\\[1mm]
&(\pleft{\_})\;\frac{a\rd b}{\pleft{a}\rd\pleft{b}}\qquad
(\pright{\_})\;\frac{a\rd b}{\pright{a}\rd\pright{b}}\\
&([x!\_]\_)\;\frac{a\rd c\quad b\rd d}{[x!a]b\rd [x!c]d}\qquad
(\prdef{x}{\_}{\_}{\_})\;\frac{a_1\rd a_2\quad b_1\rd b_2\quad c_1\rd c_2}{\prdef{x}{a_1}{b_1}{c_1}\rd\prdef{x}{a_2}{b_2}{c_2}}
\\[-4mm]
\end{align*}
\begin{flushleft}
\mbox{\emph{Typing}:}
\end{flushleft}
\begin{align*}
\\[-12mm]
&\chinm\;\frac{\Gamma\sgv a:[x!b]c}{\Gamma\sgv\pleft{a}:b}\quad 
\chbam\;\frac{\Gamma\sgv a:[x!b]c}{\Gamma\sgv\pright{a}:c\gsub{x}{\pleft{a}}}\\[1mm]
&\absem\;\frac{\Gamma\!,x:a\sgv b:c}{\Gamma\sgv[x!a]b:[x:a]c}\\[1mm]
&\pdefm\;\frac{\Gamma\sgv a:b\quad\Gamma\sgv c:d\gsub{x}{a}\quad\Gamma\!,x:b\sgv d:e}{\Gamma\sgv\prdef{x}{a}{c}{d}:[x!b]d}
\end{align*}
\end{minipage}
}
\caption{Properties of existential abstraction.\label{dcalculus.existential}}
\end{table}
\section{Products and sums}
\label{prodsum}
While quantifiers now have a direct intuitive correspondence as the type roles of operators of \dcalc.
\begin{eqnarray*}
[x:a]b&\sim&\forall x:a.b\\
{}[x!a]b&\sim&\exists x:a.b
\end{eqnarray*}
this is not yet the case for propositional operators.
One could introduce the notational convention 
\begin{eqnarray*}
\mathrm{and}(a,b)&:=&[x!a]b
\end{eqnarray*}
where $x$ may not occur free in $b$.
However we would then obtain the property
\[
\frac{\Gamma\gv a:b\quad \Gamma\gv c:d}{\Gamma\gv and(a,c):[x:a]d}
\]
which we consider as unintuitive (\emph{cf.} discussion in Section~\ref{opencoding}).

\noindent
In this section, we therefore introduce explicit propositional operators. First, we introduce binary products $[a,b]$ as a finite version of universal abstractions.
\[ 
\bprodm\;\frac{\Gamma\sgv a:c\quad\Gamma\sgv b:d}{\Gamma\sgv[a,b]:[c,d]} 
\]
Furthermore, the left and right projection operators $\pleft{}$ and $\pright{}$ are extended to products:
\[
\prlm\;\frac{\Gamma\sgv a:[b,c]}{\Gamma\sgv\pleft{a}:b}\qquad\prrm\;\frac{\Gamma\sgv a:[b,c]}{\Gamma\sgv\pright{a}:c}
\]
\[
\mathit{(\pi_3)}\;\;\pleft{[a,b]}\srd a\qquad\mathit{(\pi_4)}\;\;\pright{[a,b]}\srd b
\]
Analogously to binary products, we introduce binary sums $[a+b]$ as a finite version of existential abstractions. Together with binary sums, we introduce left- and right-injections $\injl{a}{b}$, $\injr{a}{b}$ of expressions  into sums and a case distinction operator $\case{a}{b}$.
The notations with sharp brackets are used to emphasize the fact these constructs are related to 
universal and existential abstraction in the sense that they represent finite variations of these constructs.
These constructs have the following typing and rules:
\[
\injllm\;\frac{\Gamma\sgv a:b\quad\Gamma\sgv c:d}{\Gamma\sgv\injl{a}{c}:[b+c]}\qquad
\injlrm\;\frac{\Gamma\sgv a:c\quad\Gamma\sgv b:d}{\Gamma\sgv\injr{b}{a}:[b+c]}
\]
\[
\casedm\;\frac{\Gamma\sgv a:[x:c_1]d\quad\Gamma\sgv b:[y:c_2]d\quad\Gamma\sgv d:e}{\Gamma\sgv\case{a}{b}:[z:[c_1+c_2]]d}
\]
Note that left-injection $\injl{a}{c}$ carries a type tag $c$, similar to the tag in $\prdef{x}{a}{b}{c}$. Similar for right-injections.
Note that the third condition in the typing rule for case distinction implies that $x$ does not appear free in $d$.

Analogously to the extension of projection to products, application is extended to case distinctions:
\[
\mathit{(\beta_3)}\;\;(\case{a}{b}\,\injl{c}{d})\srd (a\,c)\qquad\mathit{(\beta_4)}\;\;(\case{a}{b}\,\injr{c}{d})\srd (b\,d)
\]
Concerning typing and reduction of sums, since a sum can be seen as a finite variant of an existential abstraction we use \emph{finite} variants of the
rules motivated in Section~\ref{existential2}: 
The typing of sums is analogous to that of existential abstraction:
\[
\bsumm\;\frac{\Gamma\sgv a:c\quad\Gamma\sgv b:d}{\Gamma\sgv[a+b]:[c,d]}
\]
Obviously, there should be similar treatment for the finite variants $\pleft{[a+b]}$ and $\pright{[a+b]}$ of the expression $([x!a]b\:d)$.
Hence we also include the following equivalences:
\[
\mathit{(\pi_5)}\;\;\pleft{[a+b]}\srd a\qquad
\mathit{(\pi_6)}\;\;\pright{[a+b]}\srd b
\]
As a simple example of the use of these rules consider the following expression ($C_{a,b}$) proving commutativity of sums:
\[
[x:[a+b]](\case{[y:a]\injr{b}{y}}{[y:b]\injl{y}{a}}\,x):[[a+b]\fun[b+a]]\qquad(C_{a,b})
\]
The set of normal  forms $\nf$ now looks as follows: 
\begin{eqnarray*}
\nf&=&\{\prim\}\;\union\;\{[x:a]b,[x!a]b,\prdef{x}{a}{b}{c}\sth a,b,c\in\nf\}\\
&&\;\union\;\{[a,b],[a+b],\injl{a}{b},\injr{a}{b},\case{a}{b}\sth a,b\in\nf\}\;\union\;\de\\
\de&=&\{x\sth x\in\dvar\}\;\union\;\{(a\,b),\pright a,\pleft a,(\case{b}{c}\,a)\sth a\in\de,b,c\in\nf\}
\end{eqnarray*}

\noindent
The basic properties of products and sum are summarized in Table~\ref{dcalculus.finite}.
\begin{table}[!htb]
\fbox{
\begin{minipage}{0.96\textwidth}
\begin{flushleft}
\mbox{\emph{Syntax}:}
\end{flushleft}
\begin{align*}
\\[-16mm]
[a,b]\qquad&\mbox{\emph{product }}\\[-1mm]
[a+b]\qquad&\mbox{\emph{sum }}\\[-1mm]
\injl{a}{b},\injr{a}{b}\qquad&\mbox{\emph{left injection}, \emph{right injection}}\\[-1mm]
\case{a}{b}\qquad&\mbox{\emph{case distinction}}\\[-7mm]
\end{align*}
\begin{flushleft}
\mbox{\emph{Reduction}:}
\end{flushleft}
\begin{align*}
\\[-10mm]
&\mathit{(\pi_3)}\;\;\pleft{[a,b]}\srd a\qquad
\mathit{(\pi_4)}\;\;\pright{[a,b]}\srd b\qquad\\
&\mathit{(\pi_5)}\;\;\pleft{[a+b]}\srd a\qquad
\mathit{(\pi_6)}\;\;\pright{[a+b]}\srd b\\
&\mathit{(\beta_3)}\;\;(\case{a}{b}\,\injl{c}{d})\srd (a\,c)\qquad
\mathit{(\beta_4)}\;\;(\case{a}{b}\,\injr{c}{d})\srd (b\,d)
\\[1mm]
&([\_,\_])\;\frac{a\rd c\quad b\eqv d}{[a,b]\rd[c,d]}\qquad
([\_+\_])\;\frac{a\rd c\quad b\rd d}{[a+b]\rd[c+d]}\\[1mm]
&(\injl{\_}{\_})\;\frac{a\rd c\quad b\rd d}{\injl{a}{b}\rd\injl{c}{d}}\qquad
(\injr{\_}{\_})\;\frac{a\rd c\quad b\rd d}{\injr{a}{b}\rd\injr{c}{d}}\\[1mm]
&(\case{\_}{\_})\;\frac{a\rd c\quad b\rd d}{\case{a}{b}\rd\case{c}{d}}
\\[-4mm]
\end{align*} 
\begin{flushleft}
\mbox{\emph{Typing}:}
\end{flushleft}
\begin{align*}
\\[-11mm]
&\bprodm\;\frac{\Gamma\sgv a:c\quad\Gamma\sgv b:d}{\Gamma\sgv[a,b]:[c,d]}\qquad
\bsumm\;\frac{\Gamma\sgv a:c\quad\Gamma\sgv b:d}{\Gamma\sgv[a+b]:[c,d]}\\[1mm]
&\prlm\;\frac{\Gamma\sgv a:[b,c]}{\Gamma\sgv\pleft{a}:b}\qquad
\prrm\;\frac{\Gamma\sgv a:[b,c]}{\Gamma\sgv\pright{a}:c}\\[1mm]
&\injllm\;\frac{\Gamma\sgv a:b\quad\Gamma\sgv c:d}{\Gamma\sgv\injl{a}{c}:[b+c]}\qquad
\injlrm\;\frac{\Gamma\sgv a:c\quad\Gamma\sgv b:d}{\Gamma\sgv\injr{b}{a}:[b+c]}\\[1mm]
&\casedm\;\frac{\Gamma\sgv a:[x:c_1]d\quad\Gamma\sgv b:[y:c_2]d\quad\Gamma\sgv d:e}{\Gamma\sgv\case{a}{b}:[z:[c_1+c_2]]d}\\[-3mm]
\end{align*}
\end{minipage}
}
\caption{Properties of product and sum.\label{dcalculus.finite}}
\end{table}
\section{Negation}
\label{negation}
The rules given sofar allow, within the paradigm of natural deduction, for defining negation as described in Section~\ref{opencoding}:
\[
\text{not}(a):=[a\fun\ff]
\]
where $\ff$ is shorthand for $[x:\prim]x$,
and then to derive the laws of intuitionistic logic, \eg\ by the following typing 
\[
\sgv [y:a][z:\text{not}(a)](z\,y):[a\fun\text{not}(\text{not}(a))]
\]
Similarly it is for example easy to define expressions $b_{a,b}$, and $c_{a,b}$ where
\[
\sgv b_{a,b}:[[\text{not}(a),\text{not}(b)]\fun\text{not}([a+b])]
\]
\[
\sgv c_{a,b}:[[\text{not}(a)+\text{not}(b)]\fun\text{not}([a,b])]
\]
The remaining laws of classical logic could then be obtained assuming the law of the excluded middle
\[
\it{tnd}_a:[a+\text{not}(a)]
\]
\dcalc\ is following a different approach and introduces negation as an explicit operator $\myneg a$ that is neutral with respect to typing 
\[
\negatem\;\frac{\Gamma\sgv a:b}{\Gamma\sgv\myneg a:b}
\] 
In order to simplify deductions involving negated formulas, \dcalc\ is encoding  into its contraction relation some logical equivalence laws involving negation, thus defining a \emph{negation normal form}.
Semantically this is possible since in \dcalc\ proofs and formulas are treated uniformly as functional expressions.

In particular, we postulate that negation reduces according to the laws of double negation and the De Morgan laws:
\[ 
\mathit{(\nu_1)}\;\;\myneg\myneg a\srd a 
\qquad
\mathit{(\nu_2)}\;\;\myneg[a,b] \srd [\myneg a +\myneg b] 
\qquad 
\mathit{(\nu_3)}\;\;\myneg[a+b] \srd [\myneg a,\myneg b]
\]
These laws are extended to infinite sums and products as follows:
\[
\mathit{(\nu_4)}\;\;\myneg[x:a]b \srd [x!a]\myneg b 
\qquad
\mathit{(\nu_5)}\;\;\myneg[x!a]b \srd [x:a]\myneg b 
\]
This defines a procedure of normalization with respect to negation.

These normalization rules define many negated formulas as equivalent which helps to eliminate many routine applications of logical equivalences in deductions.
For example it suffices to prove an instance of a negated proposition in order to prove a negated universal abstraction:
\[
\frac{\Gamma\gv a:b\quad \Gamma\gv c:\myneg d\gsub{x}{a}}{\Gamma\gv\prdef{x}{a}{c}{d}:\myneg[x:b]d\;(\eqv\;[x!b]\myneg d)
}
\]
Conversely, it is sufficient to prove a negated proposition assuming an arbitrary variable to prove a negated existential abstraction over that variable:
\[
\frac{\Gamma,x:a \gv b:\myneg c}{\Gamma\gv [x:a]b:\myneg[x!a]c\;(\eqv\;[x:a]\myneg c)}
\]

In order to avoid many uninteresting variations of normal forms we also add the properties that the application of negation to any constructor other than the above one does not yield any effect.
These properties $(\nu_6)$ to $(\nu_{10})$ are non-logical in the sense that they cannot be derived as equivalences from the negation mechanism for sequent calculus.
\[
\mathit{(\nu_6)}\;\;\myneg\prim\srd\prim 
\qquad
\mathit{(\nu_7)}\;\;\myneg\prdef{x}{a}{b}{c}\srd\prdef{x}{a}{b}{c}
\]
\[
\mathit{(\nu_8)}\;\;\myneg\injl{a}{b}\srd\injl{a}{b}
\qquad 
\mathit{(\nu_9)}\;\;\myneg\injr{a}{b}\srd\injr{a}{b}
\qquad 
\mathit{(\nu_{10})}\;\;\myneg\case{a}{b}\srd\case{a}{b}
\]
These reduction rules have the consequence that the direct encoding of the negation of $a$ as $[a\fun\ff]$, where $\ff$ abbreviates $[x:\prim]x$, becomes problematic, since, for example, the axiom $\nu_1$ would then imply that
\[
[[a\fun\ff]\fun\ff]\rd a
\]
and hence we could derive the following two reductions obviously violating the confluence property of reduction.
\[
([x:[y:a]\ff]\ff\;b)\srd\ff\gsub{x}{b}=\ff\qquad([x:[y:a]\ff]\ff\;b)\srd(a\;b)
\]
The negation axioms do not yield all desired properties of negation, \eg\ it is not possible to produce an expression of type $[a\fun[\myneg a\fun b]]$.
Therefore one has to assume additional axioms
\footnote{In Section~\ref{related.negation} we discuss the problems of internalizing the complete properties of negation into \dcalc.}.
In principle many different axioms schemes are possible.
For example, one could be inspired by the rules for negation in sequent calculus ($\underline{A}$ stands for a sequence of formulas $A_1$, $\ldots$, $A_n$):
\[
\frac{\underline{A}\sgv B,\underline{C}}{\underline{A},\myneg B\sgv\underline{C}}\;(L\myneg)\qquad
\frac{\underline{A},B\sgv\underline{C}}{\underline{A}\sgv \myneg B,\underline{C}}\;(R\myneg)
\]
This can be reformulated as the following axiom schemes:
\[
\negax^+_{a,b}:[[a+b]\fun[\myneg a\fun b]]\qquad\negax^-_{a,b}:[\myneg a\fun b]\fun[a+b]]
\]
Formally such axiom schemes can be defined as an appropriate context $\Gamma'$ prefixing a context $\Gamma$ to yield a typing $\Gamma',\Gamma\sgv a:b$ (see Appendix~\ref{axioms}).
The basic properties of negation by negation normalization are summarized in Table~\ref{dcalculus.negation}.
\begin{table}[!htb]
\fbox{
\begin{minipage}{0.96\textwidth}
\begin{flushleft}
\mbox{\emph{Syntax}:}
\end{flushleft}
\begin{align*}
\\[-16mm]
\myneg a\qquad&\mbox{\emph{negation }}\\[-7mm]
\end{align*}
\begin{flushleft}
\mbox{\emph{Reduction}:}
\end{flushleft}
\begin{align*}
\\[-10mm]
&\mathit{(\nu_1)}\;\;\myneg\myneg a\srd a\qquad
\mathit{(\nu_2)}\;\;\myneg[a,b] \srd [\myneg a +\myneg b]\\[1mm]
&\mathit{(\nu_3)}\;\;\myneg[a+b] \srd [\myneg a,\myneg b]\\[1mm]
&\mathit{(\nu_4)}\;\;\myneg[x:a]b \srd [x!a]\myneg b\qquad
\mathit{(\nu_5)}\;\;\myneg[x!a]b \srd [x:a]\myneg b\\[3mm]
&\mathit{(\nu_6)}\;\;\myneg\prim\srd\prim\qquad 
\mathit{(\nu_7)}\;\;\myneg\prdef{x}{a}{b}{c}\srd\prdef{x}{a}{b}{c}\\[3mm]
&\mathit{(\nu_8)}\;\;\myneg\injl{a}{b}\srd\injl{a}{b}\qquad 
\mathit{(\nu_9)}\;\;\myneg\injr{a}{b}\srd\injr{a}{b}\\[1mm]
&\mathit{(\nu_{10})}\;\;\myneg\case{a}{b}\srd\case{a}{b}\\[3mm]
&\mathit{(\myneg\_)}\;\frac{a\rd b}{\myneg a\rd\myneg b}\\[-4mm]
\end{align*} 
\begin{flushleft}
\mbox{\emph{Typing}:}
\end{flushleft}
\begin{align*}
\\[-11mm]
&\negatem\;\frac{\Gamma\sgv a:b}{\Gamma\sgv\myneg a:b}\\[-4mm]
\end{align*}
\end{minipage}
}
\caption{Properties of negation.\label{dcalculus.negation}}
\end{table}

\noindent
Note that the above axioms scheme allows for deriving mutual implication of the left-hand sides and right-hand sides of axioms $\nu_1$ to $\nu_5$, without of course using these axioms.
For example  to prove $[\myneg\myneg a\fun a]$ we can use the following expression (see Section~\ref{prodsum} for $C_{a,b}$):
\[
(\negax^+_{\myneg a,a}\,(C_{a,\myneg a}\,(\negax^-_{a,\myneg a}\,[x:\myneg a]x))):[\myneg\myneg a\fun a]
\] 
Note however, that, due to the use of $\lambda$-structured types, this does not mean the axioms $\nu_1$ to $\nu_5$ are logically redundant if we assume additional axioms.
For example when assuming $\Gamma=(x:[\prim\fun\prim],y:\prim)$, it is only with the use of $\nu_1$ that we can establish the following typing:
\[
[z:(x\,\myneg\myneg y)]z:[(x\,\myneg\myneg y)\fun(x\,y)]
\] 
Through negation operators, the set of normal forms $\nf$ is extended as follows:
\begin{eqnarray*}
\nf&=&\{\prim\}\;\union\;\{[x:a]b,[x!a]b,\prdef{x}{a}{b}{c}\sth a,b,c\in\nf\}\\
&&\;\union\;\{[a,b],[a+b],\injl{a}{b},\injr{a}{b},\case{a}{b}\sth a,b\in\nf\}\;\union\;\de\\
\de&=&\{x\sth x\in\dvar\}\;\union\;\{(a\,b),\pleft a, \pright a,(\case{b}{c}\,a)\sth a\in\de,b,c\in\nf\}\\
&&\;\union\;\{\myneg a\sth a\in\de,a\;\text{is not a negation}\}
\end{eqnarray*}
\noindent
\section{Casting types to the primitive constant}
\label{typecasting}
A restriction of $\lambda^{\lambda}$ is related to its type formation rules. In particular if $(x:b)\sgv a:\prim$ we may not conclude $\gv[x:b]a\,:\,\prim$ but only $\gv[x:b]a:[x: b]\prim$. Therefore in an abstraction $[x:\prim]b$, $x$ is restricted in the sense that, i.e.~it does not abstract over expressions of type $[y:\prim]\prim$ or even more complicated ones.

One possibility to resolve this restriction would be to introduce an inclusion relation $\leqslant$. The idea would then be to extend the type extension property to subsumptive subtyping (\eg\ \cite{LuoSX13}):
\[
\frac{\Gamma\gv a:b\qquad b\leqslant c\qquad \Gamma\gv c:d}{\Gamma\gv a:c}
\]
and to use inclusion laws including the axiom
\[
\frac{}{a\leqslant\prim}
\]
However it turns out (see Section~\ref{paradox.subtyping}) this would lead to a calculus essentially satisfying the laws of the system $\lambda\ast$ (\cite{Bar:93}, Section 5.5) which is known to be inconsistent in the sense that every type has at least on expression typing to it.
Moreover $\lambda\ast$ does not have the strong normalization property.

An alternative approach is to use a very specific form of explicit type casting.

This can be done by an axiom scheme  $\cast_a$ to cast an expression $a$ of any type to the type $\prim$.
\[
\cast_a:[a\fun\prim]
\]
Furthermore, similarly to negation, there are axioms schemes  $\castin{a}$ and $\castout{a}$ to add and remove $\prim$-casts:
\[
\castin{a}:[x:a][x\fun(\cast_a\,x)]
\qquad
\castout{a}:[x:a][(\cast_a\,x)\fun x]
\]
As for negation, such axiom schemes can be defined as an appropriate context $\Gamma'$ prefixing a context $\Gamma$ to yield a typing $\Gamma',\Gamma\sgv a:b$ (see Appendix~\ref{axioms}).
As an example of the use of these axioms, consider a function $z$ parametrized over a type  $x$ of type $\prim$.
\[
\Gamma\;\;=\;\;(z:[x:\prim][y:x]x)
\]
Using the casting axioms, it can now be instantiated with arguments of a more complex type structure (with $\dot{\prim}$ abbreviating $[\prim\fun\prim]$).
\begin{eqnarray*}
\Gamma,x:\dot{\prim}&\gv&(\castout{\dot{\prim}}\,\dot{\prim}\,(z\,(\cast_{\dot{\prim}}\,\dot{\prim})\,(\castin{\dot{\prim}}\,\dot{\prim}\,x))):\dot{\prim}
\end{eqnarray*}
When instantiating $z$ by the identity function $[x:\prim][y:x]y$ in the above example we obtain
\begin{eqnarray*}
x:\dot{\prim}&\gv&(\castout{\dot{\prim}}\,\dot{\prim}\,(z\,(\cast_{\dot{\prim}}\,\dot{\prim})\,(\castin{\dot{\prim}}\,\dot{\prim}\,x)))\\
&\rd&(\castout{\dot{\prim}}\,\dot{\prim}\,(\castin{\dot{\prim}}\,\dot{\prim}\,z))
\end{eqnarray*}
Using the casting axioms, the gist of the rules \emph{product}, \emph{application}, and \emph{abstraction} from PTS can be reconstructed as follows.  
\begin{align*}
\it{(product^*)}\;\;&\frac{\Gamma\sgv a:\prim\quad\Gamma,x:a\sgv b:c}{\Gamma\sgv(\cast_{[x:a]c}\,[x:a]b):\prim}\\
\it{(application^*)}\;\;&\frac{\Gamma\sgv c:\cast_{[x:a]b'}([x\!:\!a]b)\quad\Gamma\sgv d:a}{\Gamma\sgv(\castout{[x:a]b'}\,[x\!:\!a]b\,c\,d):b\gsub{x}{d}}\;\;\text{where}\;\Gamma,x:a\sgv b:b'\\
\it{(abstraction^*)}\;\;&\frac{\Gamma,x:a\sgv b:c\quad\Gamma\sgv(\cast_{[x:a]c'}\,[x:a]c):\prim}{\Gamma\sgv(\castin{[x:a]c'}\,[x\!:\!a]c\,[x\!:\!a]b):(\cast_{[x:a]c'}\,[x\!:\!a]c)}\;\;\text{where}\;\Gamma,x:a\sgv c:c'
\end{align*}
This motivates a mapping $\PTStoD_{\Gamma}$ from PTS-expressions into \dcalc:
\begin{eqnarray*}
\PTStoD_{\Gamma}(*)&=&\prim\\
\PTStoD_{\Gamma}(x)&=&x\\
\PTStoD_{\Gamma}(\Pi x:A.B)&=&(\cast_{[x:\PTStoD_{\Gamma}(A)]c}\,[x:\PTStoD_{\Gamma}(A)]\PTStoD_{\Gamma,x:A}(B))\\
&&\;\;\text{where}\;\;\PTStoD(\Gamma,x:A) \sgv \PTStoD_{\Gamma,x:A}(B):c\\
\PTStoD_{\Gamma}(\lambda x:A.B)&=&(\castin{[x:\PTStoD_{\Gamma}(A)]d}\:[x:\PTStoD_{\Gamma}(A)]c\:[x:\PTStoD_{\Gamma}(A)]\PTStoD_{\Gamma,x:A}(B))\\
&&\;\;\text{where}\;\;\PTStoD(\Gamma,x:A)\sgv\PTStoD_{\Gamma,x:A}(B):c,\;\;\PTStoD(\Gamma,x:A)\sgv c:d\\
\PTStoD_{\Gamma}((A\,B))&=&(\castout{d}c\:\PTStoD_{\Gamma}(A)\:\PTStoD_{\Gamma}(B))\\
&&\;\;\text{where}\;\;\PTStoD(\Gamma)\sgv \PTStoD_{\Gamma}(A):(\cast_d\,c),\;\;\PTStoD(\Gamma)\sgv c:d
\end{eqnarray*}
Here, the mapping $\PTStoD(\Gamma)$ from PTS-contexts to contexts in \dcalc\ is recursively defined as follows:
\begin{eqnarray*}
\PTStoD(\Gamma)&=&\PTStoD_{()}(\Gamma)\\
\PTStoD_{\Gamma}(())&=&()\\
\PTStoD_{\Gamma}(x:A,\Gamma')&=&(x:\PTStoD_{\Gamma}(A),\PTStoD_{\Gamma,x:A}(\Gamma'))
\end{eqnarray*}
This mapping illustrates that the operators of \dcalc\ together with the axioms for casting are, informally speaking, on a more elementary level than the operators of PTS, in the sense that
PTS operators can be reconstructed using basic \dcalc-operators and casting axioms.

However, the casting axioms are obviously not able to reconstruct $\beta$-equality in PTS.
A simple analysis of the above mapping shows that in order to reconstruct $\beta$-equality in PTS  
the successive application of $\castin{a}$ and $\castout{a}$ must cancel each other out.
This effect can be partially achieved by the following scheme of substitution axioms:
\begin{eqnarray*}
\dcastin{a,b}:[x:a;y:[x\fun b];z:x][(y\,z)\fun (y\,(\castout{a}\,x\,(\castin{a}\,x\,z)))]\\
\dcastout{a,b}:[x:a;y:[x\fun b];z:x][(y\,(\castout{a}\,x\,(\castin{a}\,x\,z)))\fun(y\,z)]
\end{eqnarray*}
\noindent
One might as why not add $\cast$, $\castin{}$, and $\castout{}$ as direct operators do \dcalc\
and add the reduction axiom
\[
(\castout{}\,(\castin{}\,a))\srd a
\]
It turns out (see Section~\ref{paradox.casting}) this would also lead to a calculus essentially satisfying the laws of the system $\lambda\ast$ (\cite{Bar:93}, Section 5.5).
\section{Summary}
The various operations can be conceptually organized by means of two typing sequences:
\begin{itemize}
\item
There is the typing sequence of finite aggregations: Expressions may be injected into sums, sums type to products, products type to products, and so on.
\[
\frac{\Gamma\gv a_1:b_1:c_1:d_1:\ldots \quad \Gamma\gv a_2:b_2:c_2:d_2:\ldots}{\Gamma\gv\injl{a_1}{b_2},\injr{b_1}{a_2}:[b_1+b_2]: [c_1,c_2]: [d_1,d_2]:\ldots}
\]
Elimination goes via projection from a product, or case distinction from a sum.
\item
There is the typing sequence of infinite aggregations: Index and body may be injected into existential abstractions, existential abstractions type to universal abstractions, universal abstractions type to universal abstractions, and so on.
\[
\frac{\Gamma\gv a_i:a\qquad\Gamma\gv b_i:b\gsub{x}{a_i}\qquad\Gamma\gv b:c:d:\ldots}{\Gamma\gv \prdef{x}{a_i}{b_i}{b}:[x!a]b:[x:a]c:[x:a]d:\ldots}
\]
Operations are projections from an existential abstractions $\pright{a}$, $\pleft{a}$ and projection to a member of a universal abstractions using $(a\,b)$.
\item
Negation normalizes expressions on each level and, when assuming additional axioms, embeds the system into classical logic.
\end{itemize}
This completes an informal motivation of \dcalc. The following sections contain a more rigorous presentation.
\chapter{Definition of \dcalct}%
\label{definition}
In this chapter, we will propose a formal definition of \dcalc:
First we summarize the syntax (which has been motivated in Section~\ref{overview}) and define some basic notions such as free occurrences of variables.
We then define the congruence relation $a\eqv b$ as transitive and symmetric closure of a reduction relation $a\rd b$.
Finally we define contexts $\Gamma$ and the type relation $\Gamma\sgv a:b$.
\section{Basic definitions}	%
\label{basic.definitions}%
\begin{definition}[Expression]%
\label{expression}
\nomenclature[fBasic01]{$\dvar$}{set of variables}
\index{variable}
\nomenclature[fBasic02]{$\dexp$}{set of expressions}
\index{expression}
\nomenclature[cCalc02]{$\prim$}{primitive constant}
\index{constant!primitive}
\nomenclature[bSets1]{$x,y,z,\cdots$}{variables}
\index{expression!variable}
\nomenclature[bSets2]{$a,b,c,\cdots$}{expressions}
\index{expression}
\nomenclature[cCalc04]{$[x:a]b$}{universal abstraction}
\index{abstraction!universal}
\nomenclature[cCalc05]{$(a\,b)$}{application}
\index{application}
\nomenclature[cCalc06]{$[x\sdef a]b$}{existential abstraction}
\index{abstraction!existential}
\nomenclature[cCalc07]{$\prdef{x}{a}{b}{c}$}{protected definition}
\index{definition!protected}
\nomenclature[cCalc13]{$[a,b]$}{product}
\index{product}
\nomenclature[cCalc14]{$\pleft{a}$}{left-projection}
\index{projection!left}
\nomenclature[cCalc15]{$\pright{a}$}{right-projection}
\index{projection!right}
\nomenclature[cCalc16]{$[a+b]$}{sum}
\index{sum}
\nomenclature[cCalc17]{$\injl{a}{b}$}{left-injection}
\index{injection!left}
\nomenclature[cCalc18]{$\injr{a}{b}$}{right-injection}
\index{injection!right}
\nomenclature[cCalc19]{$\case{a}{b}$}{case distinction}
\index{case distinction}
\nomenclature[cCalc20]{$\myneg a$}{negation}
\index{negation}
\nomenclature[eCalc001]{$\prsumop{a}{b}$}{product/sum operator}
\index{operator!product/sum}
\nomenclature[eCalc002]{$\binop{a_1}{\ldots a_n}$}{non-binding operator}
\index{operator!binary}
\nomenclature[eCalc0051]{$\binbop{x}{a_1}{\ldots,a_n}$}{binding operator}
\index{operator!general binding}
Let $\dvar=\{x,y,z,\cdots\}$ be an infinite set of variables.
The set of expressions $\dexp$ is generated by the following rules
\begin{eqnarray*}
\dexp&\!::=\!&\{\prim\}\,\mid\,\dvar\\
&&\,\mid\,[\dvar:\dexp]\dexp\,\mid\,(\dexp\,\dexp)\\
&&\,\mid\,[\dvar!\dexp]\dexp\,\mid\,\prdef{\dvar}{\dexp}{\dexp}{\dexp}\,\mid\,\pleft{\dexp}\,\mid\,\pright{\dexp}\\
&&\,\mid\,[\dexp,\dexp]\,\mid\,[\dexp+\dexp]\,\mid\,\injl{\dexp}{\dexp}\,\mid\,\injr{\dexp}{\dexp}\,\mid\,\case{\dexp}{\dexp}\,\mid\,\myneg\dexp
\end{eqnarray*}
Expressions will be denoted by $a,b,c,d,\cdots$, 
\begin{itemize}
\item 
$\prim$ is the \emph{primitive constant}, $x,y,z,\ldots\in\dvar$ are \emph{variables}, 
\item 
$[x:a]b$ is a \emph{universal abstraction}, $(a\,b)$ is an \emph{application},
\item
$[x!a]b$ is an \emph{existential abstraction}, $\prdef{x}{a}{b}{c}$ is a \emph{protected definition}, $\pleft{a}$ and $\pright{a}$ are \emph{left-} and \emph{right-projection},
\item 
$[a,b]$ is a \emph{product}, $[a+b]$ is a \emph{sum}, $\injl{a}{b}$ and $\injr{a}{b}$ are \emph{left-} and \emph{right-injection}, $\case{a}{b}$ is a \emph{case distinction}, $\myneg a$ is \emph{negation}, and
\end{itemize}
We use additional brackets to disambiguate expressions, \eg\ $([x:\prim]x)(\prim)$.

For the sake of succinctness and homogenity in the following definitions we are using some additional notations for groups of operations:
\begin{eqnarray*}
\prsumop{a}{b}&\text{stands for}&[a,b]\;\text{or}\;[a+b]\\
\binop{a_1}{\ldots,a_n}&\text{stands for}&
\begin{cases}
\pleft{a_1},\pright{a_1},\text{or}\;\myneg a_1&\text{if}\;n=1\\
(a_1\,a_2),\prsumop{a_1}{a_2},\\
\injl{a_1}{a_2},\injr{a_1}{a_2},\;\text{or}\;\case{a_1}{a_2}
&\text{if}\;n=2
\end{cases}\\
\binbop{x}{a_1}{\ldots,a_n}&\text{stands for}&
\begin{cases}
[x:a_1]a_2\;\text{or}\;[x!a_1]a_2&\text{if}\;n=2\\
\prdef{x}{a_1}{a_2}{a_3}&\text{if}\;n=3
\end{cases}
\end{eqnarray*}
If one of these notations is used more than once in an equation or inference rule it always denotes the same concrete notation of \dcalc.
\end{definition}
\noindent
\begin{definition}[Free variables]%
\label{free}
\nomenclature[fBasic03]{$\free(a)$}{free variables}
\index{variable!free}
Variables occurring in a expression which do not occur in the range of a binding \emph{occur free} in the expression.
$\free(a)$ which denotes the set of \emph{free variables} of an expression $a$ is defined as follows:
\begin{eqnarray*}
\free(\prim)&=&\{\}\\
\free(x)&=&\{x\}\\
\free(\binop{a_1}{\ldots,a_n})&=&\free(a_1)\union\ldots\union\free(a_n)\\
\free(\binbop{x}{a_1}{\ldots,a_n}&=&\free(a_1)\union\ldots\union\free(a_{n-1})\union(\free(a_n)\!\setminus\!\{x\})
\end{eqnarray*}
Note that in a protected definition $\prdef{x}{a}{b}{c}$ the binding of $x$ is for $c$ only.
\end{definition}
\begin{definition}[Substitution of free variables]%
\label{substitute}
The \emph{substitution} $a\gsub{x}{b}$ of all free occurrences of variable $x$ in expression $a$ by expression $b$ is defined as follows:
\nomenclature[eCalc01]{$a\gsub{x}{b}$}{substitution of variables in expression}
\index{substitution!free variables in expression}
\begin{eqnarray*}
\prim\gsub{x}{b}&=&\prim\\
y\gsub{x}{b}&=&\begin{cases}b&\text{if}\;x=y\\y&\text{otherwise}\end{cases}\\
\binop{a_1}{\ldots,a_n}\gsub{x}{b}&=&\binop{a_1\gsub{x}{b}}{\ldots,a_n\gsub{x}{b}}\\
\binbop{y}{a_1}{\ldots,a_n}\gsub{x}{b}&=&\begin{cases}\binbop{y}{a_1\gsub{x}{b}}{\ldots,a_{n-1}\gsub{x}{b},a_n}\\
           \qquad\text{if}\;x=y\\
           \binbop{y}{a_1\gsub{x}{b}}{\ldots,a_n\gsub{x}{b}}\\
           \qquad\text{otherwise}\end{cases}
\end{eqnarray*}
\end{definition}
\noindent
\noindent
A substitution $a\gsub{x}{b}$ may lead to name clashes in case a variable $y$ occurring free in the inserted expression $b$ comes into the range of a binding of $y$ in the original expression. These name clashes can be avoided by renaming of variables.
\begin{definition}[$\alpha$-conversion, renaming of bound variables]%
\label{alpha}
\label{rename}
\nomenclature[gRel00]{$a\eqAlph b$}{renaming of bound variables}
\index{variable!renaming of bound}
The renaming relation on bound variables is usually called $\alpha$-conversion and induced by the following axiom (using the notation $=_{\alpha}$).
\[
\frac{y\notin\free(a_n)}{\binbop{x}{a_1}{\ldots,a_{n-1},a_n} \;=_{\alpha}\;\binbop{y}{a_1}{\ldots,a_{n-1},a_n\gsub{x}{y}}}
\]
\end{definition}

\begin{remark}[Implicit renamings, name-independent representations]%
\label{nameclashes}
In order not to clutter the presentation, we will write variables as strings but always assume appropriate renaming of bound variables in order to avoid name clashes. This assumption is justified because one could also use a less-readable but name-independent presentation of expressions using \eg\ de Bruijn indexes~\cite{Bruijn72lambdacalculus} which would avoid the necessity of renaming all together.
\end{remark}
\section{Reduction and congruence}%
\label{reduction}
The most basic semantic concept defines the reduction of a expression into a more basic expression. We will later show that reduction, if it terminates, always leads to a unique result.
\begin{definition}[Single-step reduction]%
\label{sred}
\nomenclature[gRel00]{$a\srd b$}{single-step reduction}
\index{reduction!single step}
\emph{Single-step reduction} $a\srd b$ is the smallest relation satisfying the axioms and inference rules of Table~\ref{red.rules}.
\begin{table}[!htb]
\fbox{
\begin{minipage}{0.96\textwidth}
\begin{tabular}{@{$\;$}l@{$\;\;$}r@{$\;\;$}c@{$\;\;$}ll@{$\;\;$}r@{$\;\;$}c@{$\;\;$}l}
$\;$\\[-3mm]
$\mathit{(\beta_1)}$&$([x:a]b\:c)$&$\srd$&$b\gsub{x}{c}$&$\mathit{(\beta_2)}$&$([x!a]b\:c)$&$\srd$&$b\gsub{x}{c}$\\
$\mathit{(\beta_3)}$&$(\case{a}{b}\,\injl{c}{d})$&$\srd$&$(a\,c)$&$\mathit{(\beta_4)}$&$(\case{a}{b}\,\injr{c}{d})$&$\srd$&$(b\,d)$\\[2mm]
$\mathit{(\pi_1)}$&$\pleft{\prdef{x}{a}{b}{c}}$&$\srd$&$a$&$\mathit{(\pi_2)}$&$\pright{\prdef{x}{a}{b}{c}}$&$\srd$&$b$\\
$\mathit{(\pi_3)}$&$\pleft{[a,b]}$&$\srd$&$a$&$\mathit{(\pi_4)}$&$\pright{[a,b]}$&$\srd$&$b$\\
$\mathit{(\pi_5)}$&$\pleft{[a+b]}$&$\srd$&$a$&$\mathit{(\pi_6)}$&$\pright{[a+b]}$&$\srd$&$b$\\[2mm]
$\mathit{(\nu_1)}$&$\myneg\myneg a$&$\srd$&$a$\\
$\mathit{(\nu_2)}$&$\myneg[a,b]$&$\srd$&$[\myneg a+\myneg b]$&$\mathit{(\nu_3)}$&$\myneg[a+b]$&$\srd$&$[\myneg a,\myneg b]$\\
$\mathit{(\nu_4)}$&$\myneg[x:a]b$&$\srd$&$[x!a]\myneg b$&$\mathit{(\nu_5)}$&$\myneg[x!a]b$&$\srd$&$[x:a]\myneg b$\\
$\mathit{(\nu_6)}$&$\myneg\prim$&$\srd$&$\prim$&$\mathit{(\nu_7)}$&$\myneg\prdef{x}{a}{b}{c}$&$\srd$&$\prdef{x}{a}{b}{c}$\\
$\mathit{(\nu_8)}$&$\myneg\injl{a}{b}$&$\srd$&$\injl{a}{b}$&$\mathit{(\nu_9)}$&$\myneg\injr{a}{b}$&$\srd$&$\injr{a}{b}$\\
$\mathit{(\nu_{10})}$&$\myneg\case{a}{b}$&$\srd$&$\case{a}{b}$
\end{tabular}
\begin{align*}
\\[-6mm]
\mathit{(\oplus{\overbrace{(\_,\ldots,\_)}^{n}}_i)}\quad&\frac{a_i\srd b_i}{\binop{a_1,\ldots,a_i}{\ldots,a_n}\srd \binop{a_1,\ldots,b_i}{\ldots,a_n}}\\
\mathit{(\oplus_x{\overbrace{(\_,\ldots,\_)}^{n}}_i)}\quad&\frac{a_i\srd b_i}{\binbop{x}{a_1,\ldots,a_i}{\ldots,a_n}\srd \binbop{x}{a_1,\ldots,b_i}{\ldots,a_n}}\\[-4mm]
\end{align*}
\end{minipage}
}
\caption{Axioms and rules for single-step reduction.\label{red.rules}}
\end{table}
\end{definition}
\begin{definition}[Reduction]%
\nomenclature[gRel01]{$a\rd b$}{reduction}
\index{reduction!arbitrary number of steps}
\emph{Reduction} $a\rd b$ of an expression $a$ to $b$ is defined as the reflexive and transitive closure of single-step reduction $a\srd b$.
We use the notation $a\rd b\rd c\ldots$ to indicate reduction sequences.
To show argument sequences in arguments about reduction we use the notation $a_1=\cdots=a_n\;\rd\;b_1=\cdots=b_m\:\rd\: c_1=\cdots=c_k\cdots$ to indicate equality of expressions in reduction sequences.
This will also be used for sequences of $n$-step reductions $\rdn{n}$ and accordingly for sequences containing both notations.
\end{definition}
\begin{definition}[Congruence]%
\nomenclature[gRel02]{$a\eqv b$}{congruence}
\index{congruence}
\emph{Congruence} of expressions, denoted by $a\eqv b$, is defined as the symmetric and transitive closure of reduction.
The notations for reduction sequences are extended to contain congruences as well.
\end{definition}
\begin{definition}[Reduction to common expressions]%
\nomenclature[gRel03]{$a\rdr b$}{reduction to common expression}
\index{reduction!to common expression}
If two expressions reduce to a common expression we write $a\rdr b$.
\end{definition}
\noindent
The following simple examples illustrate reduction and congruence (where $x\notin\free(a)\cup\free(b)$)
\begin{eqnarray*}
(([y_2:a][y:(P\:y_2)](Q\:y_2)\:\pleft{x})\:\pright{x})
&\srd_{(\beta_1)}&([y:(P\:\pleft{x})](Q\:\pleft{x})\:\pright{x})\\
&\srd_{(\beta_1)}&(Q\:\pleft{x})\\{}
\pright{\pleft{[\prdef{x}{a}{b}{c},d]}}&\srd_{(\pi_3)}&\pright{\prdef{x}{a}{b}{c}}\\{}
&\srd_{(\pi_2)}&b\\{}
[x:\myneg[y:\prim]\prim]\myneg[a,b]&\srd_{(\nu_4)}&[x:[y!\prim]\myneg\prim]\myneg[a,b]\\{}
&\srd_{(\nu_6)}&[x:[y!\prim]\prim]\myneg[a,b]\\{}
&\srd_{(\nu_2)}&[x:[y!\prim]\prim][\myneg a+\myneg b]\\{}
[x:\myneg[a+b]][\myneg a,\myneg b]&\eqv&[x:[\myneg a,\myneg b]]\myneg[a+b]
\end{eqnarray*}
\section{Typing and Validity}%
\label{typing}
\begin{definition}[Context]
\nomenclature[bSets3]{$\Gamma, \Gamma_1, \Gamma_2,\cdots$}{contexts}
\index{context}
\nomenclature[eCalc03]{$\Gamma(x)$}{variable lookup}
\index{context!lookup of variable}
\nomenclature[eCalc04]{$\dom(\Gamma)$}{domain of context}
\index{context!domain}
\nomenclature[eCalc04]{$\ran(\Gamma)$}{range of context}
\index{context!range}
\nomenclature[eCalc06]{$\Gamma,x:a$}{context extension}
\index{context!extension}
\nomenclature[eCalc06]{$\Gamma_1,\Gamma_2$}{context concatenation}
\index{context!concatenation}
\nomenclature[eCalc05]{$()$}{empty context}
\index{context!empty}
\nomenclature[eCalc07]{$[\Gamma]a$}{context abstraction}
\index{abstraction!of context}
\emph{Contexts}, denoted by $\Gamma$, are finite sequences of declarations $(x_1 :a_1,\ldots,x_n :a_n)$, where $x_i$ are variables with $x_i\neq x_j$ and $a_i$ are  expressions.
The assumption about name-free representation of bound variables justifies the uniqueness assumption.
The lookup of an variable in a context $\Gamma(x)$ is a partial function defined by $\Gamma(x_i)= a_i$.
$\dom(\Gamma)=\{x_1,\ldots,x_n\}$ and $\ran(\Gamma)=\{a_1,\ldots,a_n\}$ denote  the domain and range of a context $\Gamma$.
$\Gamma,x:a$ denotes the extension of $\Gamma$ on the right by a binding $x:a$ where $x$ is an variable not yet declared in $\Gamma$.
$\Gamma_1,\Gamma_2$ denotes the concatenation of two contexts declaring disjoint variables.
The empty context is written as $()$ or all together omitted.
$[\Gamma]a= [x_1:a_1]\cdots[x_1:a_n]a$ denotes the abstraction of a context over an expression.
\end{definition}
\begin{definition}[Typing]%
\nomenclature[gRel04]{$\Gamma\sgv a:b$}{typing}%
\index{typing}
Typing $\Gamma\sgv a:b$ of $a$ to $b$, the \emph{type} of $a$, under a context $\Gamma$ is the smallest ternary relation on contexts and two expressions satisfying the inference rules of Table~\ref{typ.rules}.
\begin{table}[!htb]
\fbox{
\begin{minipage}{0.96\textwidth}
\begin{align*}
\\[-9mm]
\axm\;\;&\sgv\prim:\prim&
\mystartm\;\;&\frac{\Gamma\sgv a:b}{\Gamma,x:a\sgv x:a}\\[0mm]
\weakm\;\;&\frac{\Gamma\sgv a:b\quad\Gamma\sgv c:d}{\Gamma,x:c\sgv a:b}\qquad&
\convm\;\;&\frac{\Gamma\sgv a:b\quad b\eqv c\quad\Gamma\sgv c:d}{\Gamma\sgv a:c}\\[0mm]
\absum\;\;&\frac{\Gamma,x:a\sgv b:c}{\Gamma\sgv[x:a]b:[x:a]c}&
\absem\;\;&\frac{\Gamma,x:a\sgv b:c}{\Gamma\sgv[x!a]b:[x:a]c}\\[0mm]
\applm\;\;&\frac{\Gamma\sgv a:[x:c]d\quad\Gamma\sgv b:c}{\Gamma\sgv (a\,b):d\gsub{x}{b}}&\\[-11mm]
\end{align*}
\begin{align*}
\pdefm\;\;&\frac{\Gamma\sgv a: b\quad\Gamma\sgv c:d\gsub{x}{a}\quad\Gamma,x:b\sgv d:e}{\Gamma\sgv\prdef{x}{a}{c}{d}:[x!b]d}\qquad\qquad\qquad\qquad\\[-12mm]
\end{align*}
\begin{align*}
\chinm\;\;&\frac{\Gamma\sgv a:[x!b]c}{\Gamma\sgv\pleft{a}:b}&
\chbam\;\;&\frac{\Gamma\sgv a:[x!b]c}{\Gamma\sgv\pright{a}:c\gsub{x}{\pleft{a}}}\qquad\\[1mm]
\bprodm\;\;&\frac{\Gamma\sgv a:c\quad \Gamma\sgv b:d}{\Gamma\sgv[a,b]:[c,d]}\quad\;\;\;&
\bsumm\;\;&\frac{\Gamma\sgv a:c\quad \Gamma\sgv b:d}{\Gamma\sgv[a+b]:[c,d]}\\[0mm]
\prlm\;\;&\frac{\Gamma\sgv a:[b,c]}{\Gamma\sgv\pleft{a}:b}&
\prrm\;\;&\frac{\Gamma\sgv a:[b,c]}{\Gamma\sgv\pright{a}:c}\\[0mm]
\injllm\;\;&\frac{\Gamma\sgv a:b\quad\Gamma\sgv c:d}{\Gamma\sgv\injl{a}{c}:[b+c]}&
\injlrm\;\;&\frac{\Gamma\sgv a:b\quad\Gamma\sgv c:d}{\Gamma\sgv\injr{c}{a}:[c+b]}\\[-10mm]
\end{align*}
\begin{align*}
\casedm\;\;&\frac{\Gamma\sgv a:[x:c_1]d\quad\Gamma\sgv b:[y:c_2]d\quad\Gamma\sgv d:e}{\Gamma\sgv\case{a}{b}:[z:[c_1+c_2]]d}\qquad\qquad\qquad\qquad\\[-12mm]
\end{align*}
\begin{align*}
\negatem\;\;&\frac{\Gamma\sgv a: b}{\Gamma\sgv\myneg a:b}&&\qquad\qquad\qquad\qquad\qquad\qquad\qquad\qquad\qquad\;\;
\end{align*}
\end{minipage}
}
\caption{Axiom and rules for typing.\label{typ.rules}}
\end{table}
We use the notation $\Gamma\sgv a_1=\cdots=a_n\::\: b_1=\cdots=b_m$ to indicate arguments about equality of expressions in proofs about the typing relation.
Sometimes we also mix the use of $\eqv$ and $=$ in this notation.
Similarly we use the notation $\Gamma_1=\cdots=\Gamma_n\sgv a:b$ and combinations of both.
\end{definition}
\noindent
As a simple example consider the introduction and elimination laws for existential quantification (see Section~\ref{overview}).
The type determination of the first law can be seen as follows: With $\Gamma_1=(P:[y:a]b,x:a,z:(P\,x))$) where $y\notin\free(b)$ 
and since $(P\,y)\gsub{y}{x}=P(\,x)$ by the rules \mystart, \weak, and \appl\ we obtain
\[
\Gamma_1\gv x:a\qquad\Gamma_1\gv z:(P\,y)\gsub{y}{x}\qquad\Gamma_1\gv (P\,y):b
\]
which due to the rule \pdef\ implies that
\begin{eqnarray*}
\Gamma_1&\gv&\prdef{y}{x}{z}{(P\,y)}:[y!a](P\,y)
\end{eqnarray*}
The type relation in Section~\ref{overview} follows from rule \absu.
The type determination of the second law can be seen as follows, let 
\begin{eqnarray*}
\Gamma_2&=&(P:[y:a]b,Q:[y:a]b,x:[y_1!a](P\,y_1),z:[y_2\!:\!a][y:(P\,y_2)](Q\,y_2))
\end{eqnarray*}
By rules \mystart, \weak, and \appl\ we obtain:
\[
\Gamma_2\gv(z\,\pleft{x}):[u:(P\,\pleft{x})](Q\,\pleft{x})
\]
From this, by the same rules we obtain:
\[
\Gamma_2\gv((z\,\pleft{x})\:\pright{x}):(Q\,\pleft{x}) = (Q\,y_3)\gsub{y_3}{\pleft{x}}
\]
Similarly to the first law, by rule \pdef\ this implies 
\begin{eqnarray*}
\Gamma_2&\gv&\prdef{y_3}{\pleft{x}}{((z\:\pleft{x})\:\pright{x})}{(Q\,y_3)}\;\;:\;\;[y_3!a](Q\,y_3)
\end{eqnarray*}
\noindent
A classification of the typing rules is shown in Table~\ref{typ.class}.
\begin{table}[!htb]
\begin{center}
\begin{tabular}{|l|c|c|}
\hline
{\bf Basic Rules}&\multicolumn{2}{c|}{\em{(start)},\em{(weak)},\em{(conv)}}\\\hline
{\bf Neutral Rules}&\multicolumn{2}{c|}{\em{(neg)}}\\\hline
{\bf Operator Rules}&Introduction&Elimination\\
\hline
primitive constant&\em{(ax)}&\\
universal abstraction&$\it{(abs_U)},\it{(abs_E)}$&\em{(appl)}\\
existential abstraction&\em{(def)}&$\it{(ch_I)}$,$\it{(ch_B)}$\\
product&\em{(prd)},\em{(sum)}&$\it{(pr_L)}$,$\it{(pr_R)}$\\
sum&$\it{(inj_L)}$,$\it{(inj_R)}$&\em{(case)}
\\\hline
\end{tabular}
\end{center}
\caption{Classification of typing rules.\label{typ.class}}
\end{table}
One may differentiate between axioms, basic rules and operator-specific rules.
Note that while each operation of \dcalc\ has its own typing rule, only abstractions, products, and sums, can be introduced-to as well as eliminated-from type-level.
\begin{definition}[Validity]%
\nomenclature[gRel05]{$\Gamma\sgv a$}{validity}
\index{expression!valid}
Validity $\Gamma\sgv a$ of an expression $a$ under a context $\Gamma$ is defined as the existence of a type:
\begin{eqnarray*}
\Gamma\gv a&\equiv&\text{there is an expression $b$ such that}\;\;\Gamma\sgv a:b
\end{eqnarray*}
Similarly to typing we use the notation $\Gamma_1=\cdots=\Gamma_n\sgv a_1=\cdots=a_n$ to indicate arguments about equality of contexts and expressions in proofs about validity.
We also use the notation $\Gamma\sgv a_1,\cdots,a_n$ as an abbreviation for $\Gamma\sgv a_1$, $\cdots$, $\Gamma\sgv a_n$.
As for typing, we also omit writing the empty context.
\end{definition}
\chapter{Examples}%
\label{examples}
The purpose of this chapter is to illustrate the basic style of axiomatizing theories and describing deductions when using~\dcalc.
Note that the example are presented with fully explicit expressions of~\dcalc, i.e.~we do not omit any subexpressions which could be inferred from other other parts using pattern matching or proof tactics and we do not use a module concept for theories.
Such features should of course be part of deduction languages and support systems based on \dcalc.
\begin{remark}[Notational conventions]%
\nomenclature[dCalcAbbr01]{$[x_1:a_1\cdots x_n:a_n]a$}{repeated abstraction}
\nomenclature[dCalcAbbr02]{$[x_1,\cdots,x_n:a]b$}{repeated binding}
\nomenclature[dCalcAbbr04]{$[a\fun b]$}{implication}
\nomenclature[dCalcAbbr05]{$[a_1\cdots a_n\fun a]$}{repeated implication}
\nomenclature[dCalcAbbr06]{$a(a_1,\cdots a_n)$}{nested application}
For convenience, in the examples we write $[x_1:a_1]\cdots [x_n:a_n]a$ as $[x_1:a_1;\cdots;x_n:a_n]a$ and $[x_1:a]\cdots[x_n:a]b$ as $[x_1,\cdots,x_n:a]b$ and similar for existential abstractions.
We also use combinations of these abbreviations such that \eg\ $[x:a;y_1,y_2!b]c$ is shorthand for $[x:a][y_1!b][y_2!b]c$.
As in section \ref{concepts}, we write $[a\fun b]$ to denote $[x:a]b$ if $x\notin\free(b)$.
Similarly, we write $[a_1\fun\cdots[a_n\fun a]\cdots]$ as $[a_1;\cdots;a_n\fun a]$.
We write nested applications $(\cdots((a\,a_1)\,a_2)\cdots\,a_n)$, where $a$ is not an application, as $a(a_1,\cdots,a_n)$.
We also write $\Gamma\sgv a:b$ as $a:b$ if $\Gamma$ is empty or clear from the context. When writing $a:b\;(\it{name})$ we introduce $name$ as abbreviation for $a$ of type $b$.
\end{remark}
\section{Basic logical properties}%
\label{ex.logic}
The operators $[a\fun b]$, $[a,b]$, $[a+b]$, and $\myneg a$, show many properties of logical implication, conjunction, disjunction, and negation.
For example, the following laws can be derived for arbitrary well-typed expressions $a,b$.
We begin with laws whose deduction is trivial because they are directly built into the type introduction rules and/or the reduction relation:
\begin{eqnarray*}
\;[x:a]x&:&[a\fun a]\\
\;[x:a;y:b][x,y]&:&[a;b\fun[a,b]]\\
\;[[x:a]\injl{x}{b}],[x:a]\injr{b}{x}]&:&[[a\fun[a+b]],[a\fun[b+a]]]\\
\;[x:a]x&:&[\myneg\myneg a\fun a]\\
\;[x:a]x&:&[a\fun \myneg\myneg a]\\
\;[x:\myneg[a,b]]x&:&[\myneg[a,b]\fun[\myneg a+\myneg b]]\\
\;[x:\myneg[a,b]]x&:&[[\myneg a+\myneg b]\fun\myneg[a,b]]\\
\;[x:\myneg[a+b]]x&:&[\myneg[a+b]\fun[\myneg a,\myneg b]]\\
\;[x:\myneg[a+b]]x&:&[[\myneg a,\myneg b]\fun\myneg[a+b]]
\end{eqnarray*}
Note that $[x:a]x$ types to $[a\fun a]$, $[a\fun\myneg\myneg a]$, $[\myneg\myneg a\fun a]$ etc., as all these expressions are equivalent.
Next we present some laws which follow directly from using an elimination rule:
\begin{eqnarray*}
\;[x:a;y:[a\fun b]]y(x)&:&[a;[a\fun b]\fun b]\\
\;[[x:[a,b]]\pleft{x},[x:[a,b]]\pright{x}]&:&[[[a,b]\fun a],[[a,b]\fun b]]\\
\;[x:[a,b]][\pright{x},\pleft{x}]&:&[[a,b]\fun[b,a]]\\
\;[x:[a+b]]\case{[y:a]\injr{b}{y}}{[y:b]\injl{y}{b}]}(x)&:&[[a +b]\fun[b +a]]\;\;\qquad(\it{sym}_{a,b})
\end{eqnarray*}

In Section~\ref{overview} we explained the necessity of additional negation axioms.
Note that, without additional axioms, even constructive laws of negation are missing, as for example, the contrapositive laws is not valid.
In principle many different axioms schemes are possible.
For example, one could be inspired by the rules for negation in sequent calculus ($\underline{A}$ stands for a sequence of formulas $A_1$, $\cdots$, $A_n$):
\[
\frac{\underline{C}\sgv A,\underline{B}}{\underline{C},\myneg A\sgv\underline{B}}\;(L\myneg)\qquad
\frac{\underline{C},A\sgv\underline{B}}{\underline{C}\sgv \myneg A,\underline{B}}\;(R\myneg)
\]
Theses rule together with the negations axioms inspire the following axiom schemes indexed over expressions $a$ and $b$:
\[
\negax^+_{a,b}:[[a+b]\fun[\myneg a\fun b]]\qquad\negax^-_{a,b}:[[a\fun b]\fun[\myneg a+b]]\qquad(*)
\]
where $\negax^+_{a,b},\negax^-_{a,b}\in\dvar$ are from an infinite set of variables $\mathit{I_{Ax}}$.
Formally, typing an expression $c$ to an expression $d$ under the above axiom schemes could be defined so as to require that $\free(d)=\emptyset$ and that there is a context $\Gamma$ consisting of (a finite sequence of) declarations of variables from $\mathit{I_{Ax}}$ such that $\Gamma\sgv c:d$ (see Appendix~\ref{axioms}).

We can now show the following properties:
\begin{eqnarray*}
\;\negax^-_{\myneg a,\myneg a}([x:\myneg a]x)&:&[a+\myneg a]\qquad\qquad\qquad\quad(\it{tnd}_a)\\
\;\negax^+_{[\myneg a+a],b}(\injl{\it{tnd}_a}{b})&:&[[a,\myneg a]\fun b]\\{}
[x:[a\fun b]]\negax^+_{b,\myneg a}(\it{sym}_{\myneg a,b}(\negax^-_{\myneg a,b}(x)))&:&[[a\fun b]\fun[\myneg b\fun\myneg a]]\qquad\quad(\it{cp}_{a,b})
\end{eqnarray*}
The operators $[x:a]b$ and $[x!a]b$ show the properties of universal and existential quantification.
We begin with two properties which relate the existential abstraction operator with its propositional counterpart.
\begin{eqnarray*}
[y:[x!a]b][\pleft{y},\pright{y}]&:&[[x!a]b\fun[a,b]]\\[0mm]
[y:[a,b]]\it{cp}_{[a\sfun\myneg b],[\myneg a +\myneg b]}([z:[x:a]\myneg b]\injr{\myneg a}{z(\pleft{y})},y)&:&[[a,b]\fun[x!a]b]
\end{eqnarray*}
For the derivation of the second rule note that
\begin{eqnarray*}
\it{cp}_{[a\sfun\myneg b],[\myneg a+\myneg b]}&:&[[[x:a]\myneg b\fun[\myneg a+\myneg b]]\fun[[a,b]\fun[x!a]b]]
\end{eqnarray*}
The corresponding equivalence for universal abstractions is given by the axiom schemes $\negax^+_{a,b}$ and $\negax^-_{a,b}$.

However, one should note that with these additional axioms, we have to reject use of the reduction rules $\nu_6$ to $\nu_{10}$ as this would create logical
inconsistencies. As mentioned in Section \ref{negation} these rules have been introduced as a technical convenience to avoid cluttered definitions and proofs.
From a practical point of view they are not needed in the examples presented here.

We can define logical falsehood as follows:
\nomenclature[eCalc0052]{$\ff$}{false}
\begin{eqnarray*}
\ff&:=&[x:\prim]x
\end{eqnarray*}
Note that we cannot generalize in the definition of $\ff$ from $\prim$ to arbitrary $a$ as such some of the resulting formulas are not necessarily false.
This is discussed in a remark on page~\pageref{cons.negation}.
The expected property of $\ff$ can be deduced:
\begin{eqnarray*}
[x:\prim][y:\ff]y(x)&:&[x:\prim][\ff\fun x]
\end{eqnarray*}
We can define logical truth as follows:
\nomenclature[eCalc0053]{$\tr$}{true}
\begin{eqnarray*}
\tr&:=&\myneg\ff
\end{eqnarray*}
Note that $\tr\eqv[x!\prim]\myneg x$.


For universal quantification the general introduction and elimination rules can be almost trivially derived under the context $P:[a\fun b]$.
\begin{eqnarray*}
\;[x:[y:a]P(y)]x&:&[[y:a]P(y)\fun[y:a]P(y)]\\
\;[x:a;z:[y:a]P(y)]z(x)&:&[x:a][[y:a]P(y)\fun P(x)]
\end{eqnarray*}
For existential quantification the derivation of the introduction and elimination rules looks as follows (assuming $(P,Q:[a\fun b])$).
\begin{eqnarray*}
\;[x\!:\!a;z\!:\!P(x)]\prdef{y}{x}{z}{P(y)}&:&[x:a][P(x)\fun[y!a]P(y)]\\[2mm]
[x\!:\![y_1!a]P(y_1);z\!:\![y_2\!:\!a][P(y_2)\fun Q(y_2)]]\\
\prdef{y_3}{\pleft{x}}{z(\pleft{x},\pright{x})}{Q(y_3)}&:&[[y_1!a]P(y_1);[y_2\!:\!a][P(y_2)\fun Q(y_2)]\\
&&\fun[y_3!a]Q(y_3)]
\end{eqnarray*}
The derivation of the first rule can be seen as follows: We have
\begin{eqnarray*}
P:[a\fun b],x:a,z:P(x)&\gv&z\\
&:&P(x)\\
&=&P(y)\gsub{y}{x}
\end{eqnarray*}
which due to the typing rule for protected definitions implies that
\begin{eqnarray*}
P:[a\fun b],x:a,z:P(x)&\gv&\prdef{y}{x}{z}{P(y)}\\
&:&[y!a]P(y)
\end{eqnarray*}
which then due to the typing rule for universal abstraction implies that
\begin{eqnarray*}
P:[a\fun b]&\gv&[x:a;z:P(x)]\prdef{y}{x}{z}{P(y)}\\
&:&[x:a][P(x)\fun[y!a]P(y)]
\end{eqnarray*}
The derivation of the second rule can be seen as follows, let 
\begin{eqnarray*}
\Gamma&=&(P,Q:[a\fun b]),x:[y_1!a]P(y_1),z:[y_2:a;P(y_2)]Q(y_2)
\end{eqnarray*}
then the following typing statement is true:
\begin{eqnarray*}
\Gamma&\gv&z(\pleft{x},\pright{x})\\
&:& [y_2:a;P(y_2)]Q(y_2)](\pleft{x},\pright{x})\\
&\eqv&[P(\pleft{x})\fun Q(\pleft{x})](\pright{x})\\
&\eqv&Q(\pleft{x})\\
&=&Q(y_3)\gsub{y_3}{\pleft{x}}
\end{eqnarray*}
Therefore the following statements is true:
\begin{eqnarray*}
\Gamma&\gv&\prdef{y_3}{\pleft{x}}{z(\pleft{x},\pright{x})}{Q(y_3)}\\
&:& [y_3!a]Q(y_3)
\end{eqnarray*}
The initial proposition then follows by applying the typing rules for abstraction.
%
\section{Type casting}%
\label{ex.casting}
In many cases the of free variables to be of primitive type, i.e.~$x:\prim$, can be relaxed to arbitrary well typed expressions $a:b$ using $\prim$-casting.
For this reason we introduce an axiom scheme for a $\prim$-casting operator\footnote{See also Appendix~\ref{axioms}}:
\begin{eqnarray*}
\cast_a\noarg&:&[a\fun\prim]
\end{eqnarray*}
Note that, for better readability, we use the prefix notation $\cast_a b$ abbreviating the expression $\cast_a(b)$.
We also assume the following axioms schemes essentially stating equivalence between casted and uncasted types:
\begin{eqnarray*}
\castin{a}&:&[x:a][x\fun \cast_a x]\\
\castout{a}&:&[x:a][\cast_a x\fun x]
\end{eqnarray*}
As an application, we can generalize the property of $\ff$ in Section~\ref{examples} to arbitrary well-typed $a$:
\[
x:a\gv\castout{a}(x,[y:\ff]y(\cast_a x)):[\ff\fun x]
\]
This motivates an alternative axiom-scheme for negation
\[
\negaxx^+_{a}: [[a\fun\ff]\fun\myneg a]\qquad\negaxx^-_{a}:[\myneg a\fun[a\fun\ff]]
\]
It is an easy exercise to show the equivalence between the two axiom schemes.

\section{Minimal logic}%
\label{minimallogic}
In \dcalc, minimal logic can be axiomatized by the context $\minimal$.
\begin{eqnarray*}
\minimal&:=&(\\
F&:&\prim,\\
t,f&:&F,\\
I&:&[F;F\fun F],\\
i&:&[p,q:F][[p\fun q]\fun I(p,q)],\\
o&:&[p,q:F][I(p,q)\fun[p\fun q]]\\
) 
\end{eqnarray*}
Here are two deductions of logical properties under the context $(\minimal,(p,q:F))$.
\begin{eqnarray*}
i(p,I(q,p),[x:p]i(q,p,[y:q]x))&\!:\!&I(p,I(q,p))\\
i(p,I(I(p,q),q),[x\!:\!p]i(I(p,q),q,[f\!:\!I(p,q)]o(p,q,f,x)))&\!:\!&I(p,I(I(p,q),q))
\end{eqnarray*}
\section{Equality}%
\label{equality}
Basic axioms about an equality congruence relation on expressions of equal type can be formalized as context $Equality$:
\begin{eqnarray*}
Equality&:=&(\\
\noarg=_{\noarg}\noarg&:&[S:\prim][S;S\fun\prim],\\
E_1&:&[S:\prim;x:S]x=_{S}x,\\
E_2&:&[S:\prim;x,y:S][x=_{S}y\:\fun\:y=_{S}x],\\
E_3&:&[S:\prim;x,y,z:S][x=_{S}y;y=_{S}z\:\fun\:x=_{S}z],\\
E_4&:&[S_1,S_2:\prim;x,y:S_1;F:[S_1\fun S_2]][x=_{S_1}y\:\fun\:F(x)=_{S_2}F(y)]\\
)
\end{eqnarray*}
Here $S$ is an variable used to abstract over the type (or sort) of the expressions to be equal.
Note that, for better readability, we use the infix notation $x=_{S}y$, introduced by a declaration $\noarg=_{\noarg}\noarg:[S:\prim][S;S\fun S]$.
Equivalently we could have written $(\noarg=_{\noarg}\noarg)(S,x,y)$.
\section{Cartesian products}%
\label{cartesian}
Using equality one can axiomatize basic datatypes such as cartesian products.
\begin{eqnarray*}
CartesianProduct&:=&(\\
\noarg\!\!\times\!\!\noarg&\!\!:\!\!&[\prim;\prim\fun\prim]\\
<\!\!\noarg,\!\noarg\!\!>_{\noarg\times\noarg}&\!\!:\!\!&[S,T:\prim][S;T\fun S\times T]\\
L&\!\!:\!\!&[S,T:\prim][S\times T\fun S]\\
R&\!\!:\!\!&[S,T:\prim][S\times T\fun T]\\
Eq&\!\!:\!\!&[S,T\!:\!\prim;x,y\!:\!S;\!z,w\!:\!T]\\
&&[x=_{\!S} y;z=_{\!S} w\;\fun\;\; <\!\!x,\!z\!\!>_{S\times T}\;=_{\!S\times T}\;<\!\!y,\!w\!\!>_{S\times T}]\\
P_L&\!\!:\!\!&[S,T:\prim;x:S;y:T]L(S,T,<\!\!x,\!y\!\!>_{S\times T})=_{S}x\\
P_R&\!\!:\!\!&[S,T:\prim;x:S;y:T]R(S,T,<\!\!x,\!y\!\!>_{S\times T})=_{T}y\\
)
\end{eqnarray*}
Note that, for better readability, we use infix notations $S\!\!\times\!\!T$ and $<\!\!x,\!y\!\!>_{S\times T}$ abbreviating the expressions $(\noarg\!\!\times\!\!\noarg)(S,T)$ and $<\!\!\noarg,\!\noarg\!\!>_{\noarg\times\noarg}(S,T,x,y)$.
\section{Natural Numbers}%
\label{examples.nats}
Assuming the context \emph{Equality}, well-known axioms about naturals numbers including an induction principle can be formalized as context \emph{Naturals}:
\begin{eqnarray*}
\it{Naturals}:=\quad(\quad \nats&:&\prim\\
0&:&\nats\\
s&:&[\nats\fun\nats]\\
\noarg\!+\!\noarg,\noarg\!*\!\noarg&:&[\nats;\nats\fun\nats]\\
S_1&:&[n:\nats]\myneg((s\,n)\!=_{\nats}\!0)\\
S_2&:&[n,m:\nats][(s\,n)\!=_{\nats}\!(s\,m)\:\fun\:n\!=_{\nats}\!m]\\
A_1&:&[n:\nats] 0+n=_{\nats}n\\
A_2&:&[n,m:\nats](s\,n)+m=_{\nats}(s\,(n+m))\\
M_1&:&[n:\nats]0*n=_{\nats}0\\
M_2&:&[n,m:\nats](s\,n)*m=_{\nats}m+(n*m)\\
\it{ind}&:&[P:[\nats\fun\prim]][(P\,0);[n:\nats][(P\,n)\fun(P\,(s\,n))]\\
&&\qquad\qquad\qquad\qquad\qquad\qquad\quad\fun[n:\nats](P\,n)]\quad)
\end{eqnarray*}
Note that, for better readability, we use infix notations $n+m$ and $n*m$ abbreviating the expressions $(+\,n\,m)$ and $(*\,n\,m)$.
Note that the induction principle ranges over propositions of type $\prim$ only and is thus weaker than the original one (see also Section~\ref{ex.casting}).

As an example, a simple property can be (tediously) deduced under the context $(\it{Equality},\it{Naturals},n:\nats)$ where $1:= (s\,0)$:
\begin{center}
\begin{tabular}{l@{$\;\;$}c@{$\;\;$}l}
$(E_3\;\nats\;(1\!+\!n)\;(s\,(0+n))\;(s\,n)$\\
$\quad (A_2\,0\,n)\;(E_4\;\nats\;\nats\;(0\!+\!n)\;n\;[k:\nats](s\,k)\;(A_1\,n)))$&:&$1+n\!=_{\nats}\!(s\,n)$
\end{tabular}
\end{center}
We give two examples of predicates defined on natural numbers as follows (where $2:=(s\,1)$):
\begin{eqnarray*}
\noarg\geq\noarg&:=&[n,m:\nats;k!\nats]n=_{\nats}m+k\\
\it{even}&:=&[n:\nats;m!\nats]n=2*m
\end{eqnarray*}
\noindent
The property $[n:\nats][(\it{even}\:n)\fun(\it{even}\,(2+n))]$ about even numbers can be deduced on the basis of the following typing (with $\Gamma=(n\!:\!\nats,x\!:\!(\it{even}\:n))$):
\begin{eqnarray*}
\Gamma\gv\prdef{m}{1+\pleft{x}}{(\it{law}\,n\,\pleft{x}\,\pright{x})}{2+n=_{\nats}2*m}&\!\!:\!\!&(\it{even}\,(2+n))
\end{eqnarray*}
Here $\it{law}$ is an abbreviation for a proof of the following property 
\begin{eqnarray*}
\it{law}&:&[n,m:\nats][n\!=_{\nats}\!2\!*\!m\fun\;2\!+\!n\!=_{\nats}\!2\!*\!(1\!+\!m)]
\end{eqnarray*}
A definition of \emph{law} can be derived from the axioms in a style similar (and even more tedious) to the above deduction.
This deduction of the property about even numbers is correct since
\begin{eqnarray*}
(\it{law}\,n\,\pleft{x}\,\pright{x})&:&2+n=_{\nats}2*(1+\pleft{x})\\
&\eqv&(2+n=_{\nats}2*m)\gsub{m}{1\!+\!\pleft{x}}
\end{eqnarray*}
\section{Formalizing partial functions}%
\label{example.partial}
One can go on and prove properties about natural number based on the above axioms. One may use the application operator in \dcalc\ to model function application in mathematics. However, the above example was simple as it was dealing with total functions. As an example of a partial function consider the predecessor function on natural numbers. To introduce this function in~\dcalc, several approaches come to mind:
\begin{itemize}
\item
The predecessor function can be axiomatized as a total function over the type $\nats$.
\begin{eqnarray*}
p&:&[\nats\fun\nats]\\
P&:&[n:\nats]p(s(n))=_{\nats}n
\end{eqnarray*}
The (potential) problem of this approach is the interpretation of $p(0)$ which may lead to unintuitive or harmful consequences.
Furthermore, if additional axioms are to be avoided, the declaration of $p$ must eventually by instantiated by some (total) function which defines a value for $0$.
\item
The predecessor function can be defined with an additional argument formalizing the condition.
\begin{eqnarray*}
nonZero&:=&[i:\nats;j!\nats]i=_{\nats}s(j)\\
p&:=&[n:\nats;q:nonZero(n)]\pleft{q}
\end{eqnarray*}
While mathematically clean, this definition requires to always provide an additional argument $c$ when using the predecessor function.
\item
As a variant of the previous approach, the additional argument can be hidden into an adapted type of the predecessor function.
\begin{eqnarray*}
\pnats&:=&[i,j!\nats]i=_{\nats}s(j)\\
p&:=&[n:\pnats]\pleft{\pright{n}}
\end{eqnarray*}
While mathematically clean, this approach requires a more complex handling of the argument $n:\pnats$ when using the predecessor function.
For example, in $\pnats$ one can define the number {\bf 1} and apply the predecessor as follows:
\begin{eqnarray*}
{\bf 1}&:=&\prdef{i}{(s\,0)}{\prdef{j}{0}{(E_1\,N\,(s\,0))}{(s\,0)=_{\nats}(s\,j)}}{[j!\nats]i=_{\nats}(s\,j)} \\
&&:\pnats
\end{eqnarray*}
\end{itemize}
As a consequence $p({\bf 1})\eqv 0$. Which of these (or other) approaches is best to use seems to depend on the organization and the goals of the formalization at hand.
\section{Defining functions from deductions}%
\label{example.functions}
Note that while the predecessor function can be directly defined, more complex functions and their (algorithmic) properties can be derived from the proofs of properties.
As a sketch of an example consider the following well-known property
\begin{eqnarray*}
\it{GCD}&:=&[x,y:\nats;k!\nats]\it{gcd}(k,x,y)
\end{eqnarray*}
where $\it{gcd}(k,x,y)$ denotes the property that $k$ is the greatest common divisor of $x$ and $y$.
Given a (not necessarily constructive) deduction $P_{\it{GCD}}$ of type \emph{GCD}, one can then define the greatest common divisor $x\downarrow y$ and define deductions $d_1$ and $d_2$ proving the well-known algorithmic properties.
\begin{eqnarray*}
\noarg\downarrow\noarg&:=&[x,y:\nats]\pleft{(P_{\it{GCD}}(x,y))}\\
d_1&:&[x,y:\nats](x+y)\downarrow x=_{\nats}y\downarrow x\\
d_2&:&[x,y:\nats]x\downarrow(x+y)=_{\nats}x\downarrow y
\end{eqnarray*}
\section{Sets}%
\label{ex.sets}
When formalizing mathematical deductions, one needs formal systems for various basic theories of mathematics. For examples, sets can be axiomatized by the following context using a formalized set comprehension principle.
\begin{eqnarray*}
Sets&:=&(\\
\mathbb{P}&:&[\prim\fun\prim],\\
\noarg\in_{\noarg}\noarg&:&[S:\prim][S;\mathbb{P}(S)\fun\prim],\\
\{\noarg\}_{\noarg}&:&[S:\prim][[S\fun\prim]\fun\mathbb{P}(S)],\\
I&:&[S:\prim;x:S;P:[S\fun\prim]][P(x)\fun x\in_S\{[y:S]P(y)\}_S],\\
O&:&[S:\prim;x:S;P:[S\fun\prim]][x\in_{S}\{[y:S]P(y)\}_S\fun P(x)]\\
) 
\end{eqnarray*}
ote that, for better readability, we use the notation $x\in_{S}y$ and $\{P\}_S$ abbreviating the expressions $\noarg\!\in_{\noarg}\!\noarg(S,x,y)$ and
$\{\noarg\}_{\noarg}(S,P)$.

One can now define various sets using set comprehension. Note the use of the cast operator to ensure the set-defining properties are of type $\prim$.
\begin{eqnarray*}
\emptyset&:=& [S:\prim]\{[x:S\fun\cast_{[\prim\sfun\prim]}\ff]\}_S\\
&&:\quad[S:\prim]\mathbb{P}(S)\\
\noarg\!\cup_{\noarg}\!\noarg&:=& [S:\prim;A,B:\mathbb{P}(S)]\{[x:S]\cast_{[\prim,\prim]}[x\in_{S}A+x\in_{S}B]\}_S\\
&&:\quad[S:\prim][\mathbb{P}(S);\mathbb{P}(S)\fun \mathbb{P}(S)]\\
\it{Even}&:=&\{[x:\nats]\cast_{[\nats\fun\prim]} even(x)\}_{\nats}\\
&&:\quad\mathbb{P}(\nats)
\end{eqnarray*}
Properties of individual elements can be deduced on the basis of the axiom $O$, for example
we can extract the property $P:=[n!\nats](x=_{\nats}2*n)$ of a member $x$ of {\it Even} using the cast-removal axiom as follows (where $\it{Even}_P := [x:\nats]\cast_{[\nats\fun\prim]}\it{even}(x)$:
\begin{eqnarray*}
\Gamma&\sgv&\castout{[\nats\fun\prim]}(P,O(\nats,x,\it{Even}_P,\it{asm}))\;:\; P
\end{eqnarray*}
where $\Gamma\quad=\quad(x:\nats,asm:x\in_{\nats}Even)$.

Note that in this formalization of sets the axiom of choice can be immediately derived as follows:
We assume the usual assumptions 
\[
X,I:\prim,A:[I\fun\mathbb{P}(X)],u:[x:I;y!X]y\in_{X}A(x)
\] 
and deduce
\begin{eqnarray*}
AoC&:=&\prdef{F}{[i:I]\pleft{u(i)}}{[i:I]\pright{u(i)}}{[i:I]F(i)\in_{X}A(i)}\\
&:&[F![I\fun X];i:I]F(i)\in_{X}A(i)
\end{eqnarray*}
Note also that a alternative definition of naturals from integers can be given with sets as follows:
Assuming one already has axiomatized the more general theory of integers, \eg~with type {\em Int}, a constant $0_I$, and the relation $\geq$, 
one could instantiate the declaration of $N$ of the context {\em Naturals} using casting as follows:
\begin{eqnarray*}
(\nats\;:=\;(\cast_{(\mathbb{P}\,\it{Int})}\;\{[n:\it{Int}]\cast_{\prim}\,n\geq 0_I\}_{\it Int},\;\;0\;:=\ldots)
\end{eqnarray*}  
\section{Use of substitution axioms for casting}%
\label{ex.casting2}
Cartesian products have been axiomatized in Section~\ref{cartesian}.
Using the on-board means of \dcalc, there is a intuitive notion of cartesian product.
\begin{eqnarray*}
S\times T&\eqv&\cast[S,T]\\
<\!\!x,\!y\!\!>_{S\times T}&\eqv&\castin{[\prim,\prim]}([S,T],[x,y])\\
L(x)&\eqv&\pleft{x}\\
R(y)&\eqv&\pright{y}
\end{eqnarray*}
Casting is necessary since the equality notion used in Section~\ref{cartesian} works over expressions only whose type is of type $\prim$.
The type constructor, the pairing function, and the projection function declarations can be instantiated as follows:
\begin{eqnarray*}
\noarg\times\noarg&:=&[S;T:\prim\fun\cast_{[\prim,\prim]}([S,T])]\\
&:&[\prim;\prim\fun\prim]\\
P&:=&[S,T:\prim;x:S;y:T]\castin{[\prim,\prim]}([S,T],[x,y])\\
&:&[S,T:\prim][S;T\fun S\times T]\\
L&:=&[S,T:\prim;x:S\times T]\pleft{(\castout{[\prim,\prim]}([S,T],x))}\\
&:&[S,T:\prim][S\times T\fun S]\\
R&:=&[S,T:\prim;x:S\times T]\pright{(\castout{[\prim,\prim]}([S,T],x))}\\
&:&[S,T:\prim][S\times T\fun T]
\end{eqnarray*}
Based on these definitions, the equality law can be proven under the context $(S,T\!:\!\prim;x,y\!:\!S;z,w\!:\!T)$ as follows
\begin{eqnarray*}
&&[p:x=_S y;\,q:z=_t w]\\
&&E_4(T,S\!\times\!T,z,w,[u:T]\castin{[\prim,\prim]}([S,T],[x,u])=_{S\times T}\castin{[\prim,\prim]}([S,T],[y,u]),q,\\
&&\;\; E_4(S,S\!\times\!T,x,y,[u:S]\castin{[\prim,\prim]}([S,T],[u,z])=_{S\times T}\castin{[\prim,\prim]}([S,T],[u,z]),p,\\
&&\;\;\;\; E_1(\cast_{[\prim,\prim]}([S,T]),\castin{[\prim,\prim]}([S,T],[x,z]))))\\
&:&[x=_{S} y;z=_{S}w \fun\;\; <\!\!x,\!\!z>_{S\times T}\;=_{S\times T}\;<\!\!y,\!w\!\!>_{S\times T}]
\end{eqnarray*}
If we want to show the projection properties, \eg\ in case of left projection we need to show the following property
\begin{eqnarray*}
[S,T:\prim;x:S;y:T]\pleft{(\castout{[\prim,\prim]}([S,T],\castin{[\prim,\prim]}([S,T],[x,y])))}=_{S}x
\end{eqnarray*}
In order to show this law we need to use the substitution axioms for successive addition and removal of casts in types
\begin{eqnarray*}
\dcastin{a,b}:[x:a;y:[x\fun b];z:x][y(z)\fun y(\castout{a}(x,\castin{a}(x,z)))]\\
\dcastout{a,b}:[x:a;y:[x\fun b];z:x][y(\castout{a}(x,\castin{a}(x,z)))\fun y(z)]
\end{eqnarray*}
The first axiom can be instantiated in two steps as follows:
\begin{eqnarray*}
\dcastin{[\prim,\prim],\prim}&:&[x:[\prim,\prim];y:[x\fun\prim];z:x][y(z)\fun y(\castout{[\prim,\prim]}(x,\castin{[\prim,\prim]}(x,z)))]\\
\dcastin{[\prim,\prim],\prim}([S,T])&:&[y:[[S,T]\fun\prim];z:[S,T]]]\\
&&\quad[y(z)\fun y(\castout{[\prim,\prim]}([S,T],\castin{[\prim,\prim]}([S,T],z)))]
\end{eqnarray*}
Note that we cannot avoid the use of this kind of axiom since we cannot include a generic law into the reduction rules of \dcalc\ as this would lead to inconsistencies (Section~\ref{paradox.casting}).
The left projection law can now be shown as follows:
\begin{eqnarray*}
P_L&:=&[S,T:\prim;x:S;y:T]\\
&&\dcastin{[\prim,\prim],\prim}([S,T],[p:[S,T]]\pleft{p}=_S x,[x,y],E_1(S,\pleft{[x,y]}))\\
&:&[S,T:\prim;x:S;y:T]L(S,T,<\!\!x,\!y\!\!>_{S\times T})=_{S}x
\end{eqnarray*}
The proof of the right projection law runs analogously.
\section{Proof structuring}%
\label{example.groups}
To illustrate some proof structuring issues, we formalize the property of being a group as follows (writing $[a_1,a_2,\cdots]$ for $[a_1,[a_2,\cdots ]]$):
\begin{eqnarray*}
Group:=[S:\prim;\noarg\!*\noarg![S;S\fun S],e!S]&[&[x,y,z:S](x*y)*z =_S x*(y*z)\\
&,&[x:S]e*x =_S x\\
&,&[x:S,x'!S]x'*x =_S e\\
&]
\end{eqnarray*}
As an example, we can show that $+$, $-$, and $0$ form a group.
Obviously, we can construct a deduction $ded$ with 
\[
\it{Equality}, \it{Integers} \gv ded:P_g
\]
where  \emph{Integers} is the context \emph{Naturals} extended with a subtraction operator $a-b$ (to create elements inverse of addition) with corresponding axioms and where $P_g$ describes the group laws: 
\begin{eqnarray*}
P_g:=&[\;&[x,y,z:\nats]\,(x+y)+z=_{\nats}x+(y+z)\\
&,&[x:\nats]\,0+x=_{\nats}x\\
&,&[x:\nats,x'!\nats]\,x'+x=_{\nats}0\\
&]
\end{eqnarray*}
$pr$ can be turned into a proof of $Group(\nats)$ as follows:
\begin{eqnarray*}
isGroup&:=&\prdef{*}{+}{\prdef{e}{0}{ded}{P_g\gsub{0}{e}}}{P_g\gsub{0}{e}\gsub{+}{*}}\\
\it{Equality}, \it{Integers}&\gv& isGroup:Group(\nats)
\end{eqnarray*}
In group theory one can show that the left-neutral element is also right-neutral, this means, when assuming $g$ to be a group over $S$ there is a proof $p$ such that
\[
Equality,S:\prim,g:Group(S)\gv p:[x:S](\pleft{g})(x,\pleft{\pright{g}})=_S x
\]
Here $\pleft{g}$ is the $*$ function of $g$ and $\pleft{\pright{g}}$ is the neutral element of $g$.
Note that the use of existential declarations is supporting the proof structuring as it hides $*$ and $e$ in the assumptions inside the $g:Group(S)$ assumption.
On the other hand, one has to explicitly access the operators using $\pleft{}$ and $\pright{}$.

$p$ can be extended to a proof $p'$ of an implication about properties as follows:
\begin{eqnarray*}
p'&:=&[S:\prim;g:Group(S)]p\\
Equality&\gv& p':[S:\prim;Group(S);x:S]\;(\pleft{g})(x,\pleft{\pright{g}})=_S x
\end{eqnarray*}
Hence we can instantiate $p'$ to obtain the right-neutrality property of $0$ for integers.
\[
Equality,Integers\gv p'(\nats,isGroup):[x:\nats]\,x+0 =_{\nats}x
\]
\chapter{Properties of \dcalct}%
\label{properties}
\section{Overview}%
The following main properties will be shown.
\begin{property}[Church-Rosser property]%
$a\eqv b$ implies $a\rdr b$.
\end{property}
\begin{property}[Subject reduction]
$\Gamma\sgv a:c$ and $a\rd b$ imply $\Gamma\sgv b:c$.
\end{property}
\begin{property}[Uniqueness of types]%
$\Gamma\sgv a:b$ and $\Gamma\sgv a:c$ imply $b\eqv c$.
\end{property}
\begin{property}[Strong normalization]%
$\Gamma\sgv a$ implies that all one-step reduction sequences beginning with $a$ end in an irreducible expression $b$.
\end{property}
\noindent
Due to the Church-Rosser property, the irreducible \emph{normal form} of a valid expression is unique.
An immediate consequence of strong normalization and confluence is the decidability of the typing relation.
\begin{property}[Decidability of typing relation]
For expressions $a$ and $b$ and context $\Gamma$ the proposition $\Gamma\sgv a:b$ is decidable.
\end{property}
\noindent
Finally, we show logical consistency of \dcalc, in the sense that there is an empty type.
\begin{property}[Logical consistency]%
There is no expression $a$ such that $\sgv a$ and $a:[x:\prim]x$.
\end{property}
\noindent
%
\section{Notations and conventions used in proofs}%
Besides structural induction on the definition of $\dexp$, we will mainly use structural induction on expressions with context when showing properties of expressions.
\begin{definition}[Structural induction on expression with context]%
\index{induction!structural}
\index{induction!structural on expression with context}
A property $P(\Gamma,a)$ about an expression $a$  with context $\Gamma$ is shown by \emph{structural induction}, if we are using the following axioms and rules for arbitrary $x$, $y$, $a_1$,$\cdots$,$a_n$:
\[
\frac{}{P(\Gamma,\prim)}
\qquad 
\frac{x\in\dom(\Gamma)\;\;P(\Gamma,\Gamma(x))}{P(\Gamma,x)} 
\qquad
\frac{P(\Gamma,a_1)\;\ldots\;P(\Gamma,a_{n})}{P(\Gamma,\binop{a_1}{\ldots,a_n})}
\]
\[
\frac{P(\Gamma,a_1)\;\ldots\;P(\Gamma,a_{n-1})\;\;P((\Gamma,x:a_1),a_n)}{P(\Gamma,\binbop{y}{a_1}{\ldots,a_n})}
\]
\end{definition} 
\noindent
We frequently show properties about reduction relations by induction on the definition of single-step reduction.
\begin{definition}[Induction on the definition of single-step reduction]
\index{induction!on definition of single-step reduction}
Properties $P(a,b)$ which are proven for all $a$ and $b$ where $a\srd b$ can be shown by \emph{induction on the definition of single-step reduction} if the inductive base corresponds to the reduction axioms and the inductive step corresponds to the inference rules of reduction.
\end{definition} 
\begin{remark}[Renaming of variables in axioms or rules]%
Typically when we prove a property using some axiom, inference rule, or derived property, we just mention the identifier or this axiom, rule or, property and then use it with an instantiation \emph{renaming its variables so as to avoid name clashes with the proposition to be shown}.
In order not to clutter the presentation, these renamings are usually not explicitly indicated.
\end{remark}
\begin{remark}[Introduction of auxiliary identifiers]%
We usually introduce explicitly all auxiliary identifiers appearing in deduction steps.
However, there are two important exceptions.
\begin{itemize}
\item 
In structural inductions, if we consider a specific operator and decompose an expression $a$ by $a=\binop{a_1}{,\ldots,a_n}$ we usually introduce implicitly the new auxiliary identifier $a_1$.
\item
In inductions on the definition of reduction, if we consider a specific axiom or structural rule which requires a syntactic pattern we usually introduce implicitly the new auxiliary identifiers necessary for this pattern.
\end{itemize}
\end{remark}
\begin{remark}[Display of reduction relations]%
While we state all proofs involving reduction textually, we sometimes in addition illustrate reduction relations using diagrams:
\[\begin{tikzcd}
    a\ar{r}{}\ar{d}[swap]{} & b\ar{d}{}\\
    c\ar{r}{}           &{}d
\end{tikzcd}
\]
\end{remark}
\begin{remark}[Display of deductive trees]%
While we state all profs textually, we sometimes in addition illustrate the deductive trees in the following style. 
\[
\inferrule*[left=rule]
  {premise_1\quad
	\inferrule*[left=I.H.]{premise_2\quad premise_3}{conclusion_1}}
{conclusion_2}
\]
In inductive proofs we use $I.H.$ to denote the inductive hypothesis.
We sometimes stack multiple arguments (indicated by $ARGi$) vertically as follows.
\[
\inferrule*[left=rule1]
  {\inferrule*[left=rule2,right=ARG1]{premise_1\quad premise_2}{conclusion_2}\\ \\ \\
	\inferrule*[left=rule3,right=ARG2]{premise_3\quad premise_4}{conclusion_3}}
{conclusion_1}
\]
\end{remark}
\section{Basic properties of reduction}%
\label{rd.basic}
We begin with some basic properties of substitution and its relation to reduction.
\begin{law}[Basic properties of substitution]%
\label{basic.sub}
For all $a,b,c$ and $x,y$:
\begin{itemize}
\item[$i$:]
If $x\notin\free(a)$ then $a\gsub{x}{b}=a$.
\item[$ii$:]
If $x\neq y$ and $x\notin\free(c)$ then $a\gsub{x}{b}\gsub{y}{c}= a\gsub{y}{c}\gsub{x}{b\gsub{y}{c}}$.
\item[$iii$:]
If $x\neq y$, $x\notin\free(c), y\notin\free(b)$ then $a\gsub{x}{b}\gsub{y}{c}= a\gsub{y}{c}\gsub{x}{b}$.
\end{itemize}
\end{law}
\begin{proof}
$\;$
\begin{itemize}
\item[$i$:]
Property follows by structural induction on $a$.
\item[$ii$:]
Property follows by structural induction on $a$ using the definition of substitution:
For $a=\prim$ the property is obvious. If $a=z$ we have to distinguish three cases:
\begin{itemize}
\item $z=x$. Then obviously
\begin{eqnarray*}
&&z\gsub{z}{b}\gsub{y}{c}\\
&=&b\gsub{y}{c}\\
&=&z\gsub{z}{b\gsub{y}{c}}\\
&=&z\gsub{y}{c}\gsub{z}{b\gsub{y}{c}}
\end{eqnarray*}
\item $z=y$.
Then obviously
\begin{eqnarray*}
&&z\gsub{x}{b}\gsub{z}{c}\\
&=&z\gsub{z}{c}\\
&=&c\\
&=&c\gsub{x}{b\gsub{z}{c}}\qquad\qquad\text{(by part $i$ since $x\notin\free(c))$}\\
&=&z\gsub{z}{c}\gsub{x}{b\gsub{z}{c}}
\end{eqnarray*}
\item $z\neq x,y$.
Then obviously
\begin{eqnarray*}
&&z\gsub{x}{b}\gsub{y}{c}\\
&=&z\gsub{y}{c}\\
&=&z\\
&=&z\gsub{x}{b\gsub{z}{c}}\\
&=&z\gsub{y}{c}\gsub{x}{b\gsub{y}{c}}
\end{eqnarray*}
\end{itemize}
If $a=\binop{a_1}{\ldots a_n}$ then the property follows from the definition of substitution and the inductive hypothesis.
Finally if $a=\binbop{z}{a_1}{\ldots,a_n}$ then thanks to our assumption in Section~\ref{basic.definitions} about appropriate renaming in order to avoid name clashes during substitutions, we may assume that $z\neq x,y$ and therefore the property follows from the definition of substitution and the inductive hypothesis.
\item[$iii$:]
We can argue as follows:
\begin{eqnarray*}
&&a\gsub{x}{b}\gsub{y}{c}\\
&=&\quad\text{(by $ii$)}\\
&&a\gsub{y}{c}\gsub{x}{b\gsub{y}{c}}\\
&=&\quad\text{(since $y\notin\free(b)$)}\\
&&a\gsub{y}{c}\gsub{x}{b}
\end{eqnarray*}
\qedhere
\end{itemize}
\end{proof}
\begin{law}[Substitution and reduction]%
\label{rd.sub}
For all $a,b,c$ and $x$:
\begin{itemize}
\item[$i$:]$a\rd b$ implies $a\gsub{x}{c}\rd b\gsub{x}{c}$
\item[$ii$:]$a\rd b$ implies $c\gsub{x}{a}\rd c\gsub{x}{b}$
\item[$iii$:]$a\rd b$ implies $\free(b)\subseteq\free(a)$ 
\end{itemize}
\end{law}
\begin{proof}
$\;$
\begin{itemize}
\item[$i$:]
It is obviously sufficient to show the property for single-step reduction.
The proof is by induction on the definition of single-step reduction.
Both for the inductive base and the inductive steps it follows directly from the definition of substitution, for example in case of the axiom $\beta_1$ we have
\begin{eqnarray*}
&&([y:a_1]a_2\:a_3)\gsub{x}{c}\\
&=&([y:a_1\gsub{x}{c}]a_2\gsub{x}{c}\:a_3\gsub{x}{c})\\
&\srd&a_2\gsub{x}{c}\gsub{y}{a_3\gsub{x}{c}}\\
&=&a_2\gsub{y}{a_3}\gsub{x}{c}\qquad\text{(by Law~\ref{basic.sub}($ii$) since $y\notin\free(c)$)}
\end{eqnarray*}
Note that thanks to our assumption in Section~\ref{basic.definitions} about avoidance of name clashes due to renaming we can assume $y\neq x$.
A similar reasoning applies for the other axioms and the structural rules.
\item[$ii$:]
Property follows by structural induction on $c$.
If $c=\prim$ the property is obvious.
If $c=x$ then 
\[
c\gsub{x}{a}=a\rd b=c\gsub{x}{b},
\]
if $c=y\neq x$ then $c\gsub{x}{a}=c=c\gsub{x}{b}$. If $c=\binop{c_1}{\ldots c_n}$ then obviously
\begin{eqnarray*}
&&\binop{c_1}{\ldots c_n}\gsub{x}{a}\\
&=&\binop{c_1\gsub{x}{a}}{\ldots c_n\gsub{x}{a}}\\
&\rd&\binop{c_1\gsub{x}{b}}{\ldots c_n\gsub{x}{b}}\qquad\text{(inductive hyp., definition of $\rd$)}\\
&=&\binop{c_1}{\ldots c_n}\gsub{x}{b}
\end{eqnarray*}
A similar can be made for the case $c=\binbop{y}{c_1}{\ldots,c_n}$.
\item[$iii$:] 
The proof is by a straightforward induction on the definition of reduction.
\end{itemize}
\end{proof}
\noindent
Next we show basic decomposition properties of reduction. The first three properties are (strong) decomposition properties of the operators which are constructors, in the sense that they do not work on other operators. The remaining properties are (weak) decomposition properties of the other operators.
\begin{law}[Reduction decomposition]%
\label{rd.decomp}  
For all $a_1,\ldots,a_n,b,b_1,\ldots,b_n$ and $x$:
\begin{itemize}
\item[$i$:]
$\binop{a_1}{a_2}\rd b$, where $\binop{a_1}{a_2}$ is not an application implies $b=\binop{b_1}{b_2}$ where $a_1\rd b_1$ and $a_2\rd b_2$.
\item[$ii$:]
$\binbop{x}{a_1}{\ldots,a_n}\rd b$ implies $b=\binbop{x}{b_1}{\ldots,b_n}$, $a_i\rd b_i$, for $1\leq i\leq n$.
\item[$iii$:]
$a.i\rd\binbop{x}{b_1}{\ldots,b_n}$, $i=1,2$, implies $a\rd\prsumop{c_1}{c_2}$ or $a\rd\prdef{y}{c_1}{c_2}{c_3}$, for some $c_1, c_2,c_3$ with $c_i\rd\binbop{x}{b_1}{\ldots,b_n}$.
\item[$iv$:]
$a.i\rd\binop{a_1}{a_2}$, $i=1,2$, where $\binop{a_1}{a_2}$ is a sum, product or injection implies $a\rd\prsumopd{c_1}{c_2}$ or $a\rd\prdef{y}{c_1}{c_2}{c_3}$, for some $c_1, c_2,c_3$ with $c_i\rd\binop{a_1}{a_2}$.
\item[$v$:]
$\myneg a\rd[x:b_1]b_2$ implies $a\rd[x!c_1]c_2$ for some $c_1, c_2$ where $c_1\rd b_1$ and $\myneg c_2\rd b_2$.

$\myneg a\rd[b_1,b_2]$ implies $a\rd[c_1+c_2]$ for some $c_1, c_2$ where $\myneg c_1\rd b_1$ and $\myneg c_2\rd b_2$.

$\myneg a\rd[x!b_1]b_2$ implies $a\rd[x:c_1]c_2$ for some $c_1, c_2$ where $c_1\rd b_1$ and $\myneg c_2\rd b_2$.

$\myneg a\rd[b_1+b_2]$ implies $a\rd[c_1,c_2]$ for some $c_1, c_2$ where $\myneg c_1\rd b_1$ and $\myneg c_2\rd b_2$.

$\myneg a\rd\prdef{x}{b_1}{b_2}{b_3}$ implies $a\rd\prdef{x}{b_1}{b_2}{b_3}$.

$\myneg a\rd\case{b_1}{b_2}$ implies $a\rd\case{b_1}{b_2}$.

$\myneg a\rd\injl{b_1}{b_2}$ implies $a\rd\injl{b_1}{b_2}$.

$\myneg a\rd\injr{b_1}{b_2}$ implies $a\rd\injr{b_1}{b_2}$.
\item[$vi$:]
$(a_1\,a_2)\rd\binbop{x}{b_1}{\ldots,b_n}$ implies, for some $c_1, c_2, c_3, c_4$, one of the following cases
\begin{itemize}
\item $a_1\rd\binbopd{y}{c_1}{c_2}$, $a_2\rd c_3$ and $c_2\gsub{y}{c_3}\rd\binbop{x}{b_1}{\ldots, b_n}$, or
\item $a_1\rd\case{c_1}{c_2}$ and we have that either $a_2\rd\injl{c_3}{c_4}$ and $(c_1\,c_3)\rd\binbop{x}{b_1}{\ldots,b_n}$
 or $a_2\rd \injr{c_3}{c_4}$ and $(c_2\,c_4)\rd\binbop{x}{b_1}{\ldots,b_n}$.
\end{itemize}
\item[$vii$:]
$(a_1\,a_2)\rd\binop{b_1}{\ldots,b_n}$ where $\binop{b_1}{\ldots,b_n}$ is not an application implies, for some $c_1, c_2, c_3, c_4$, one of the following cases
\begin{itemize}
\item $a_1\rd\binbopd{y}{c_1}{c_2}$, $a_2\rd c_3$, and $c_2\gsub{x}{c_3}\rd\binop{b_1}{\ldots,b_n}$, or
\item $a_1\rd\case{c_1}{c_2}$ and either $a_2\rd\injl{c_3}{c_4}$ and $(c_1\,c_3)\rd\binop{b_1}{\ldots,b_n}$
 or $a_2\rd\injr{c_3}{c_4}$ and $(c_2\,c_4)\rd\binop{b_1}{\ldots,b_n}$.
\end{itemize}
\end{itemize}
\end{law}
\begin{proof}
$\;$
\begin{largeitemize}
\item[$i$,$ii$:]
Obvious, as none of the reduction axioms has any of these unary, binary, or ternary operators in outermost position.
\item[$iii$,$iv$:]
These properties are true since the only way to remove a projection is via a projection axiom.
\item[$v$:]
The only way to remove an outer negation is by one of the negation axioms $\nu_i$. We consider these cases in turn;
\begin{itemize}
\item If axiom $\nu_1$ is applied we have $a=\myneg a'$ and $\myneg\myneg a'\srd a'\rd\binbop{x}{b_1}{b_2}$ or $a'\rd\prsumop{b_1}{b_2}$.
In case of $a'\rd[x:b_1]b_2$ we then know that $a=\myneg a'\rd\myneg[x:b_1]b_2\srd[x!b_1]\myneg b_2$ which implies the proposition.
Similar for the other cases.
\item The property follows directly if one of the axioms $\nu_2$ to $\nu_5$ is applied.
\item Application of one of the axioms $\nu_6$ to $\nu_{10}$ yields the remaining cases.
\end{itemize}
\item[$vi$,$vii$:]
These properties are true since the only way to remove the outer application is via one of the axioms $\beta_i$, $1\leq i\leq 4$.
\qedhere
\end{largeitemize}
\end{proof}
\section{Confluence properties}%
\label{confl}
\subsection{Overview of the confluence proof}%
Well-known confluence proofs for untyped $\lambda$-calculus could be used, \eg~\cite{Bar:93} or~\cite{Takahashi95}(using parallel reduction) could be adapted to include the operators of \dcalc.
Due to the significant number of reduction axioms of~\dcalc, we use an alternative approach using explicit substitutions and an auxiliary relation of \emph{reduction with explicit substitution} which has detailed substitution steps on the basis of a definitional environment (this approach was basically already adopted in the Automath project~\cite{deBruijn80}) and which comprises sequences of negation-related reduction-steps into single steps.
The underlying idea is that reduction with explicit substitution can be shown to be directly confluent which implies its confluence.
We then show that this implies confluence of $\srd$.
There are several approaches to reduction with explicit substitutions, \eg~\cite{ACCL91,AK2010}.
Furthermore, there is a significant body of recent recent work in this context, however, as explicit substitution is not the main focus of this article we do not give an overview here.
However, we should note that, in general, confluence of calculi with explicit substitutions is proved by using confluence of the
underlying calculus without explicit substitutions. 
Here it is the other way around: confluence of reduction with explicit substitution is used to show confluence of $\srd$.
The approach introduced below introduces a definitional environment as part of the reduction relation 
to explicitly unfold single substitution instances and then discard substitution expressions when all instances are unfolded. 
This approach, as far as basic lambda calculus operators are concerned, is essentially equivalent to the system $\Lambda_{sub}$ 
which has been defined using substitution~\cite{Milner2007} or placeholders~\cite{KC2008} to indicate particular occurrences to be substituted.
Both approaches slightly differ from ours as as they duplicate the substituted expression on the right-hand side of the $\beta$-rule thus violating direct confluence.
\subsection{Explicit substitution, negation-reduction, reduction with explicit substitution}%
We first define explicit substitutions. 
\begin{definition}[Expressions with substitution]
\nomenclature[fBasic03]{$\dexps$}{set of expressions with substitution}
\nomenclature[bSets5]{$\mathbf{a}$, $\mathbf{b}$, $\mathbf{c}$, $\mathbf{d}$, $\ldots$}{expressions with substitution}
\nomenclature[eCalc10]{$[x\mydef\mathbf{a}]\mathbf{b}$}{internalized substitution}
\index{substitution!internalized}
The set $\dexps$ of \emph{expressions with substitution} is an extension of the set $\dexp$ of expressions adding
a substitution operator.
\begin{eqnarray*}
\dexps&\!::=\!&\underbrace{\{\prim\}\,\mid\,\cdots\,\mid\,\myneg\dexps}_{\text{(see Definition~\ref{expression})}}\,\mid\,[\dvar\mydef\dexps]\dexps
\end{eqnarray*}
Expressions with substitution will be denoted by $\mathbf{a},\mathbf{b},\mathbf{c},\mathbf{d},\ldots$. 
$[x\mydef\mathbf{a}]\mathbf{b}$ is an \emph{internalized substitution}.
As indicated by its name, the purpose of $[x\mydef\mathbf{a}]\mathbf{b}$ is to internalize the substitution function.
\end{definition}
\noindent\noindent
The function computing free variables (Definition~\ref{free}) is extended so as to treat internalized substitutions identical to abstractions.
\begin{eqnarray*}
\free([x\mydef\mathbf{a}]\mathbf{b})&=&\free(\mathbf{a})\union(\free(\mathbf{b})\!\setminus\!\{x\})
\end{eqnarray*}
Similar for $\alpha$-conversion (Definition~\ref{alpha}).
\[
\frac{y\notin\free(\mathbf{b})}{[x\mydef\mathbf{a}]\mathbf{b}\;=_{\alpha}\;[y\mydef\mathbf{a}]\mathbf{b}\gsub{x}{y}}
\]
As for expressions we will write variables as strings but always assume appropriate renaming of bound variables in order to avoid name clashes.

In order to define the $\mrd$-reduction relation we need an auxiliary reduction relation which comprises application sequences of axioms $\nu_1$, $\ldots$, $\nu_5$ in a restricted context.
\begin{definition}[Negation reduction]
\nomenclature[gRel06]{$\mathbf{a}\snur\mathbf{b}$}{single-step negation-reduction}
\nomenclature[gRel06]{$\mathbf{a}\nurn{n}\mathbf{b}$}{$n$-step negation-reduction}
\nomenclature[gRel06]{$\mathbf{a}\nur\mathbf{b}$}{negation-reduction}
\nomenclature[gRel06]{$\mathbf{a}\nurp\mathbf{b}$}{non-empty negation-reduction}
Single-step \emph{negation-reduction} $\mathbf{a}\snur\mathbf{b}$ is the smallest relation on expressions with explicit substitution satisfying the axiom and the inference rules of Table~\ref{nurd.rules}.
\begin{table}[!htb]
\fbox{
\begin{minipage}{0.96\textwidth}
\begin{center}
\begin{tabular}{@{$\;$}l@{}r@{$\;$}c@{$\;$}ll@{}r@{$\;$}c@{$\;$}l}
$\;$\\[-3mm]
$\mathit{(\nu_1)}$&$\;\myneg\myneg\mathbf{a}$&$\snur$&$\mathbf{a}$\\ 
$\mathit{(\nu_2)}$&$\;\myneg[\mathbf{a},\mathbf{b}]$&$\snur$&$[\myneg\mathbf{a}+\myneg\mathbf{b}]$&$\mathit{(\nu_3)}$&$\;\myneg[\mathbf{a}+\mathbf{b}]$&$\snur$&$[\myneg\mathbf{a},\myneg\mathbf{b}]$\\
$\mathit{(\nu_4)}$&$\;\myneg[x:\mathbf{a}]\mathbf{b}$&$\snur$&$[x!\mathbf{a}]\myneg\mathbf{b}$&$\mathit{(\nu_5)}$&$\;\myneg[x!\mathbf{a}]\mathbf{b}$&$\snur$&$[x:\mathbf{a}]\myneg\mathbf{b}$
\end{tabular}
\end{center}
\begin{align*}
\\[-6mm]
\qquad(\prsumop{\_}{\_}_1)\quad\frac{\mathbf{a}_1\snur\mathbf{a}_2}{\prsumop{\mathbf{a}_1}{\mathbf{b}}\snur\prsumop{\mathbf{a}_2}{\mathbf{b}}}
\qquad(\prsumop{\_}{\_}_2)\quad\frac{\mathbf{b}_1\snur\mathbf{b}_2}{\prsumop{\mathbf{a}}{\mathbf{b}_1}\snur\prsumop{\mathbf{a}}{\mathbf{b}_2}}
\end{align*}
\begin{align*}
\\[-8mm]
(\binbop{\_}{\_}{\_}_2)\quad\frac{\mathbf{b}_1\snur\mathbf{b}_2}{\binbop{x}{\mathbf{a}}{\mathbf{b}_1}\snur\binbop{x}{\mathbf{a}}{\mathbf{b}_2}}
\qquad
(\myneg{\_}_1)\quad\frac{\mathbf{a}_1\snur\mathbf{a}_2}{\myneg\mathbf{a}_1\snur\myneg\mathbf{a}_2}
\qquad
\end{align*}
\end{minipage}
}\caption{Axioms and rules for $\nur$-reduction.\label{nurd.rules}}
\end{table}
$n$-step negation-reduction $\mathbf{a}\nurn{n}\mathbf{b}$ ($n\geq 0$), negation-reduction $\mathbf{a}\nur\mathbf{b}$, and non-empty negation-reduction $\mathbf{a}\nurp\mathbf{b}$ are defined as follows:
\begin{eqnarray*}
\mathbf{a}\nurn{n}\mathbf{b}&:=&\exists\mathbf{b}_1,\cdots\mathbf{b}_{n-1}:\mathbf{a}\snur\mathbf{b}_1,\ldots,\mathbf{b}_{n-1}\snur\mathbf{b}.\\
\mathbf{a}\nur\mathbf{b}&:=&\exists k\geq 0:\mathbf{a}\nurn{k}\mathbf{b}\\
\mathbf{a}\nurp\mathbf{b}&:=&\exists k>0:\mathbf{a}\nurn{k}\mathbf{b} \\
\mathbf{a}\nuro\mathbf{b}&:=&\mathbf{a}=\mathbf{b}\;\text{or}\;\mathbf{a}\snur\mathbf{b}
\end{eqnarray*}
\end{definition}
\noindent
In order to define the $\mrd$-reduction relation on expression with substitution we need to introduce the notion of environments, which are used to record the definitions which are currently valid for a $\mrd$-reduction step.
\begin{definition}[Environment]
\nomenclature[bSets4]{$E, E_1, E_2, \cdots$}{environments}
\label{environment.ext}
\index{environment}
\emph{Environments}, denoted by $E$, $E_1$, $E_2$, etc.~are finite sequences of definitions $(x_1\mydef\mathbf{a}_1,\ldots ,x_n\mydef\mathbf{a}_n)$, where $x_i$ are variables, $x_i\neq x_j$.
The lookup of an variable in an environment is defined by $E(x)=\mathbf{a}_i$. 
$E,x\mydef\mathbf{a}$ denotes the extension of $E$ on the right by a definition $x\mydef\mathbf{a}$.
$E_1,E_2$ denotes the concatenation of two environments. 
The empty environment is written as $()$.
\end{definition}
\begin{definition}[Single-step reduction with explicit substitution]
\nomenclature[gRel06]{$E\sgv\mathbf{a}\smrd\mathbf{b}$}{single-step reduction with explicit substitution}
\emph{Single-step reduction reduction with explicit substitution} $E\sgv\mathbf{a}\smrd\mathbf{b}$ is the smallest relation on expressions with explicit substitution satisfying the axiom and the inference rules of Table~\ref{mred.rules}.
\begin{table}[!htb]
\fbox{
\begin{minipage}{0.96\textwidth}
\begin{center}
\begin{tabular}{@{$\;$}l@{}r@{$\;$}c@{$\;$}l@{$\;$}l@{}r@{$\;$}c@{$\;$}l}
$\;$\\[-3mm]
$\mathit{(\beta_1^{\mu})}$&$E\sgv([x:\mathbf{a}]\mathbf{b}\,\mathbf{c})$&$\smrd$&$[x\!\mydef\!\mathbf{c}]\mathbf{b}$&$\mathit{(\beta_2^{\mu})}$&$E\sgv([x!\mathbf{a}]\mathbf{b}\,\mathbf{c})$&$\smrd$&$[x\mydef\mathbf{c}]\mathbf{b}$\\
$\mathit{(\beta_3)}$&$E\sgv(\case{\mathbf{a}\!}{\!\mathbf{b}}\,\injl{\mathbf{c}}{\mathbf{d}})$&$\smrd$&$(\mathbf{a}\,\mathbf{c})$&$\mathit{(\beta_4)}$&$E\sgv(\case{\mathbf{a}\!}{\!\mathbf{b}}\,\injr{\mathbf{c}}{\mathbf{d}})$&$\smrd$&$(\mathbf{b}\,\mathbf{d})$\\[2mm]
$\it{(use)}$&$E\sgv x$&$\smrd$&$\mathbf{a}$&\multicolumn{4}{l}{if $E(x)=\mathbf{a}$}\\
$\it{(rem)}$&$E\sgv[x\mydef\mathbf{a}]\mathbf{b}$&$\smrd$&$\mathbf{b}$&\multicolumn{4}{l}{if $x\notin\free(\mathbf{b})$}\\[2mm]
$\mathit{(\pi_1)}$&$E\sgv\pleft{\prdef{x}{\mathbf{a}}{\mathbf{b}\!}{\!\mathbf{c}}}$&$\smrd$&$\mathbf{a}$&$\mathit{(\pi_2)}$&$E\sgv\pright{\prdef{x}{\mathbf{a}}{\mathbf{b}\!}{\!\mathbf{c}}}$&$\smrd$&$\mathbf{b}$\\
$\mathit{(\pi_3)}$&$E\sgv\pleft{[\mathbf{a},\mathbf{b}]}$&$\smrd$&$\mathbf{a}$&$\mathit{(\pi_4)}$&$E\sgv\pright{[\mathbf{a},\mathbf{b}]}$&$\smrd$&$\mathbf{b}$\\
$\mathit{(\pi_5)}$&$E\sgv\pleft{[\mathbf{a}+\mathbf{b}]}$&$\smrd$&$\mathbf{a}$&$\mathit{(\pi_6)}$&$E\sgv\pright{[\mathbf{a}+\mathbf{b}]}$&$\smrd$&$\mathbf{b}$\\[2mm]
$\mathit{(\nu_6)}$&$E\sgv\myneg\prim$&$\smrd$&$\prim$&$\mathit{(\nu_7)}$&$E\sgv\myneg\prdef{x}{\mathbf{a}}{\mathbf{b}\!}{\!\mathbf{c}}$&$\smrd$&$\prdef{x}{\mathbf{a}}{\mathbf{b}\!}{\!\mathbf{c}}$\\
$\mathit{(\nu_8)}$&$E\sgv\myneg\injl{\mathbf{a}}{\mathbf{b}}$&$\smrd$&$\injl{\mathbf{a}}{\mathbf{b}}$&$\mathit{(\nu_9)}$&$E\sgv\myneg\injr{\mathbf{a}}{\mathbf{b}}$&$\smrd$&$\injr{\mathbf{a}}{\mathbf{b}}$\\
$\mathit{(\nu_{10})}$&$E\sgv\myneg\case{\mathbf{a}}{\mathbf{b}}$&$\smrd$&$\case{\mathbf{a}}{\mathbf{b}}$
\end{tabular}
\end{center}
\begin{align*}
\mathit{(\nu)}\quad&\frac{\mathbf{a}\nurp\mathbf{b}}{E\gv\mathbf{a}\smrd\mathbf{b}}
\end{align*}
\begin{align*}
\\[-8mm]
\mathit{(\oplus{\overbrace{(\_,\ldots,\_)}^{n}}_i)}\quad&\frac{E\gv\mathbf{a}_i\smrd\mathbf{b}_i}{E\gv\binop{\mathbf{a}_1,\ldots,\mathbf{a}_i}{\ldots,\mathbf{a}_n}\smrd \binop{\mathbf{a}_1,\ldots,\mathbf{b}_i}{\ldots,\mathbf{a}_n}}\\
\mathit{(\oplus_x{\overbrace{(\_,\ldots,\_)}^{n}}_i)}\quad&\frac{E\gv\mathbf{a}_i\smrd\mathbf{b}_i}{E\gv\binbop{x}{\mathbf{a}_1,\ldots,\mathbf{a}_i}{\ldots,\mathbf{a}_n}\smrd \binbop{x}{\mathbf{a}_1,\ldots,\mathbf{b}_i}{\ldots,\mathbf{a}_n}}
\end{align*}
\begin{align*}
\\[-8mm]
\mathit{(L_{\smydef})}\quad&\!\frac{E\gv\mathbf{a}\smrd\mathbf{b}}{E\gv[x\mydef\mathbf{a}]\mathbf{c}\smrd[x\mydef\mathbf{b}]\mathbf{c}}&\quad
\mathit{(R_{\smydef})}\quad&\!\frac{E,x\mydef\mathbf{a}\gv\mathbf{b}\smrd\mathbf{c}}{E\gv[x\mydef\mathbf{a}]\mathbf{b}\smrd[x\mydef\mathbf{a}]\mathbf{c}}\\[-4mm]
\end{align*}
\end{minipage}
}\caption{Axioms and rules for single-step reduction with explicit substitution.\label{mred.rules}}
\end{table}
Compared to (conventional) reduction, the axiom $\beta$ has been decomposed into three axioms:
\begin{itemize}
\item $\beta^{\mu}_1$ and $\beta^{\mu}_2$ are reformulation of $\beta_1$ and $\beta_2$ using internalized substitution 
\item \emph{use} is unfolding single usages of definitions 
\item \emph{rem} is removing a definition without usage
\end{itemize} 
The axioms $\nu_1,\ldots,\nu_5$, which are not directly confluent \eg\ for $\myneg\myneg[a,b]$, have been removed and replaced by the rule $\nu$.
Furthermore there are two more structural rules $(L_{\smydef})$ and $(R_{\smydef})$ related to substitutions. 
Note that the rule $(R_{\smydef})$ is pushing a definition onto the environment $E$ when evaluating the body of a definition.
\end{definition}
\begin{definition}[Reduction with explicit substitution]
\nomenclature[gRel08]{$E\sgv\mathbf{a}\mrd\mathbf{b}$}{reduction with explicit substitution}
\nomenclature[gRel09]{$E\sgv\mathbf{a}\mrdr\mathbf{b}$}{reduction ($\smrd$)to common expression}
reduction  with explicit substitution $E\sgv\mathbf{a}\mrd\mathbf{b}$ of $\mathbf{a}$ to $\mathbf{b}$ is defined as the reflexive and transitive closure of $E\sgv\mathbf{a}\smrd\mathbf{b}$.
If two expressions $\mathbf{a}$ and $\mathbf{b}$ $\mrd$-reduce to a common expression we write it as $\mathbf{a}\mrdr\mathbf{b}$. 
\end{definition}
\begin{definition}[$\mrd$-Reduction notations]
\nomenclature[gRel10]{$E\sgv\mathbf{a}\mrdn{01}\mathbf{b}$}{zero-or-one-step reduction with explicit substitution}
\nomenclature[gRel11]{$E\sgv\mathbf{a}\mrdn{n}\mathbf{b}$}{$n$-step reduction with explicit substitution}
Zero-or-one-step reduction with explicit substitution and $n$-step reduction with explicit substitution are defined as follows
\begin{eqnarray*}
E\sgv\mathbf{a}\mrdn{01}\mathbf{b}&:=&E\sgv\mathbf{a}\smrd\mathbf{b}\;\vee\;\mathbf{a}=\mathbf{b} \\
E\sgv\mathbf{a}\mrdn{n}\mathbf{b}&:=&\exists\mathbf{b}_1,\cdots\mathbf{b}_{n-1}:E\sgv\mathbf{a}\smrd\mathbf{b}_1,\ldots,E\sgv\mathbf{b}_{n-1}\smrd\mathbf{b}.
\end{eqnarray*}
\end{definition}
\begin{remark}[Avoidance of name clashes through appropriate renaming]
Note that renaming is necessary to prepare use of the axiom \emph{use}: 
For example when reducing $y\mydef x\sgv [x:\prim][y,x]$ using $\mrd$, $[x:\prim][y,x]$ needs to be renamed to $[z:\prim][y,z]$ before substituting $y$ by $x$. 
\end{remark}
\noindent
We end this subsection with two straightforward properties of reduction with explicit substitution.
\begin{law}[Reduction implies reduction with explicit substitution]
\label{rd.mrd}
For all $a,b$: $a\rd b$ implies $()\sgv a\mrd b$.
\end{law}
\begin{proof}
Proof is by induction on the definition of $a\rd b$.
\begin{itemize}
\item 
Use of the axiom $\beta_1$ and $\beta_2$:
In case of $\beta_1$ we can show that 
\begin{eqnarray*}
()\gv&&([x:c]b\:a)\\
&\smrd&\quad\text{(by $\beta_1^{\mu}$)}\\
&&[x\mydef a]b\\
&\mrdn{n}&\quad\text{(by induction; with $n$ of free occurrences of $x$ in $b$)}\\
&&[x\mydef a](b\gsub{x}{a})\\
&\smrd&\quad\text{(by $\it{rem}$)}\\
&&b\gsub{x}{a}
\end{eqnarray*}
Similar for the axiom $\beta_2$. 
\item
The axioms $\beta_3$, $\beta_4$, $\pi_1$ to $\pi_6$, and $\nu_6$ to $\nu_{10}$ also exist for $\mrd$-reduction. 
\item
The other axioms of $\rd$ are implied by the rule $\nu$.
\item 
The structural rules of $\rd$ are included in the corresponding rules for $\mrd$ (always taking $E=()$).
\qedhere
\end{itemize}
\end{proof}
\subsection{Confluence of reduction with explicit substitution}%
\begin{remark}[Sketch of confluence proof]
First we show that negation-reduction is confluent and commutes with reduction with explicit substitution.
Based on these results, by induction on expressions with substitution one can establish direct confluence of $\smrd$, i.e. $E\sgv\mathbf{a}\smrd\mathbf{b}$ and $E\sgv\mathbf{a}\smrd\mathbf{c}$ imply $E\sgv\mathbf{b}\mrdn{01}\mathbf{d}$ and $E\sgv\mathbf{c}\mrdn{01}\mathbf{d}$ for some $\mathbf{d}$. Confluence follows by two subsequent inductions. 
\end{remark}
\noindent
As a first property we show confluence of $\snur$.
We begin with elementary properties of negation-reduction.
\begin{law}[Elementary properties of negation-reduction]
\label{nurd.basic}
For all $\mathbf{a},\mathbf{a}_1,\mathbf{a}_2,\mathbf{b},\mathbf{b}_1,\mathbf{b}_2$ and $x$, the following properties are satisfied:
\begin{itemize}
\item[$i$:]
$\prsumop{\mathbf{a}_1}{\mathbf{a}_2}\snur\mathbf{b}$ implies $b=\prsumop{\mathbf{c}_1}{\mathbf{c}_2}$, $\mathbf{a}_1 \nuro\mathbf{c}_1$, and $\mathbf{a}_2\nuro\mathbf{c}_2$, for some $\mathbf{c}_1$, $\mathbf{c}_2$.
\item[$ii$:]
$\binbop{x}{\mathbf{a}_1}{\mathbf{a}_2}\snur\mathbf{b}$ implies $\mathbf{b}=\binbop{x}{\mathbf{a}_1}{\mathbf{c}_2}$, $\mathbf{a}_2\snur\mathbf{c}_2$ for some $\mathbf{c}_2$.
\item[$iii$:] 
$\myneg\mathbf{a}\snur[\mathbf{b}_1,\mathbf{b}_2]$ implies $\mathbf{a}=[\mathbf{c}_1+\mathbf{c}_2]$, $\myneg\mathbf{c}_1\nuro\mathbf{b}_1$, $\myneg\mathbf{c}_2\nuro\mathbf{b}_2$ for some $\mathbf{c}_1$, $\mathbf{c}_2$.
\item[$iv$:] 
$\myneg\mathbf{a}\snur[\mathbf{b}_1+\mathbf{b}_2]$ implies $\mathbf{a}=[\mathbf{c}_1,\mathbf{c}_2]$, $\myneg\mathbf{c}_1\nuro\mathbf{b}_1$, $\myneg\mathbf{c}_2\nuro\mathbf{b}_2$ for some $\mathbf{c}_1$, $\mathbf{c}_2$.
\item[$v$:] 
$\myneg\mathbf{a}\snur[x:\mathbf{b}_1]\mathbf{b}_2$ implies $\mathbf{a}=[x!\mathbf{b}_1]\mathbf{c}_2$, $\myneg \mathbf{c}_2\snur\mathbf{b}_2$ for some $\mathbf{c}_2$.
\item[$vi$:] 
$\myneg\mathbf{a}\snur[x!\mathbf{b}_1]\mathbf{b}_2$ implies $\mathbf{a}=[x:\mathbf{b}_1]\mathbf{c}_2$, $\myneg\mathbf{c}_2\snur\mathbf{b}_2$ for some $\mathbf{c}_2$.
\item[$vii$:] 
$\myneg\mathbf{a}\snur\myneg\mathbf{b}$ implies $\mathbf{a}\snur\mathbf{b}$.
\end{itemize}
\end{law}
\begin{proof}
Parts $i$, $ii$, $iii$, $iv$, $v$, and $vi$ are direct consequences of the definition of negation-reduction.
They can be shown by induction on the definition of negation-reduction.
Part $vii$ is slightly more complex as there are two cases for
$\myneg\mathbf{a}\snur\myneg\mathbf{b}$:
\begin{itemize}
\item
Use of the axiom $\nu_1$, i.e.~$\mathbf{a}=\myneg\myneg\mathbf{b}$.
The proposition follows since $\mathbf{a}\snur\mathbf{b}$.
\item
Use of structural rule $\myneg{\_}_1$, i.e.~$\mathbf{a}\snur\mathbf{b}$. which also means we are done.
\qedhere
\end{itemize}
\end{proof}
\begin{law}[Strong normalisation of negation-reduction]
\index{strong normalisation!of negation-reduction}
\label{nurd.sn}
There is no infinite sequence $\mathbf{a}_1\snur\mathbf{a}_2\snur \ldots$.
\end{law}
\begin{proof} 
One way to see this is to define a weight $\negwt(\mathbf{a})$ such that
\[
\mathbf{a}\snur\mathbf{b}\quad\text{implies}\quad \negwt(\mathbf{b})<\negwt(\mathbf{a})
\]
A possible definition is given below: 
\begin{eqnarray*}
\negwt(\prim)&=&1\\
\negwt(x)&=&1\\
\negwt(\binop{\mathbf{a}_1,\ldots}{\mathbf{a}_n})&=&
\begin{cases}
(\negwt(\mathbf{a}_1)+1)^2\quad\text{if}\;n=1,\\
\qquad\qquad\qquad\qquad\qquad \binop{\mathbf{a}_1,\ldots}{\mathbf{a}_n}=\myneg\mathbf{a}_1\\
\negwt(\mathbf{a}_1)+\ldots+\negwt(\mathbf{a}_n)+1\\
\qquad\qquad\qquad\qquad\quad\qquad\text{otherwise}
\end{cases}\\
\negwt(\binbop{x}{\mathbf{a}_1,\ldots}{\mathbf{a}_n})&=&\negwt(\mathbf{a}_1)+\ldots+\negwt(\mathbf{a}_n)+1
\end{eqnarray*}
\qedhere
\end{proof}
\noindent
For the confluence proof of $\snur$ we will use induction on the size of expressions.
\begin{definition}[Size of expression]
\nomenclature[eCalc11]{$\sz(\mathbf{a})$}{Size of expression}
\index{size!of expression}
\label{ind.size}
The size of an expression is defined as follows: 
\begin{eqnarray*}
\sz(\prim)&=&1\\
\sz(x)&=&1\\
\sz(\binop{\mathbf{a}_1}{\ldots,\mathbf{a}_n})&=&\sz(\mathbf{a}_1)+\ldots+\sz(\mathbf{a}_n)+1\\
\sz(\binbop{x}{\mathbf{a}_1}{\ldots,\mathbf{a}_n})&=&\sz(\mathbf{a}_1)+\ldots+\sz(\mathbf{a}_n)+1
\end{eqnarray*}
Obviously $\sz(\mathbf{a})>0$ for all expressions $\mathbf{a}$.
\end{definition}
\begin{law}[Confluence of $\snur$]
\index{confluence!of single-step negation-reduction}
\label{nurd.confl}
For all $\mathbf{a},\mathbf{b},\mathbf{c}$: $\mathbf{a}\nur\mathbf{b}$ and $\mathbf{a}\nur\mathbf{c}$ imply $\mathbf{b}\nur\mathbf{d}$, and $\mathbf{c}\nur\mathbf{d}$ for some $\mathbf{d}$. 
\end{law}
\begin{proof}
The proof is by induction on the size of expressions and systematic investigation of critical pairs:
The case $\sz(\mathbf{a})=1$ is obviously true.
Consider an expression $\mathbf{a}$ with $\sz(\mathbf{a})=n>0$ and assume confluence for all expressions $\mathbf{b}$ with $\sz(\mathbf{b})<\sz(\mathbf{a})$.
 
First we establish local confluence from $a$, i.e.~$\mathbf{a}\snur\mathbf{b}$ and $\mathbf{a}\snur\mathbf{c}$ imply $\mathbf{b}\nur\mathbf{d}$, and $\mathbf{c}\nur\mathbf{d}$ for some $\mathbf{d}$. 
The following cases can be distinguished:
\begin{itemize}
\item
$\mathbf{a}=\prsumop{\mathbf{a}_1}{\mathbf{a}_2}$: Proposition follows from Law~\ref{nurd.basic}($i$) and the inductive hypothesis.
\item
$\mathbf{a}=\binbop{x}{\mathbf{a}_1}{\mathbf{a}_2}$: Proposition follows from Law~\ref{nurd.basic}($ii$) and the inductive hypothesis.
\item
$\mathbf{a}=\myneg\mathbf{a}_1$:
We have $E\sgv \myneg\mathbf{a}_1\snur\mathbf{b}$ and $E\sgv \myneg\mathbf{a}_1\snur\mathbf{c}$.
A somewhat clumsy but technically straightforward proof goes by systematic cases distinction on $\mathbf{b}$ and $\mathbf{c}$:
\begin{itemize}
\item
$\mathbf{b}=\myneg\mathbf{b}_1$, $\mathbf{c}=[\mathbf{c}_2,\mathbf{c}_3]$. 

By Law~\ref{nurd.basic}($vii$) we know that $\mathbf{a}_1\snur\mathbf{b}_1$.
By Law~\ref{nurd.basic}($iii$) we know that $\mathbf{a}_1=[\mathbf{a}_2+\mathbf{a}_3]$ for some $\mathbf{a}_2$, $\mathbf{a}_3$ where $\myneg\mathbf{a}_2\nurp\mathbf{c}_2$ and $\myneg\mathbf{a}_3\nurp\mathbf{c}_3$.
By inductive hypothesis applied to $\mathbf{a}_1$, there is an expression $\mathbf{e}$ such that $\mathbf{b}_1\nur\mathbf{e}$ and $[\mathbf{a}_2+\mathbf{a}_3]\nur\mathbf{e}$.
Hence $\mathbf{e}=[\mathbf{e}_2+\mathbf{e}_3]$ for some $\mathbf{e}_2$, $\mathbf{e}_3$ where $\mathbf{a}_2\nur\mathbf{e}_2$ and $\mathbf{a}_3\nur\mathbf{e}_3$. 
This can be summarized graphically as follows:
\[\begin{tikzcd}
    \mathbf{a}_1=[\mathbf{a}_2,\mathbf{a}_3]\arrow[r, "{\myneg}" description]\ar[rd, swap, "\stackrel{*}{\myneg}" description] &\mathbf{b}_1\ar{d}{\stackrel{*}{\myneg}}\\
                                      &\mathbf{e}=[\mathbf{e}_2,\mathbf{e}_3]
\end{tikzcd}
\]
Hence $\myneg\mathbf{a}_2\nur\myneg\mathbf{e}_2$ and $\myneg\mathbf{a}_3\nur\myneg\mathbf{e}_3$.
By inductive hypothesis applied to $\myneg\mathbf{a}_2$ and $\myneg\mathbf{a}_3$ (note that $\sz(\myneg\mathbf{a}_2),\sz(\myneg\mathbf{a}_3)<\sz(\mathbf{a})$) there are $\mathbf{d}_2$ and $\mathbf{d}_3$ where $\myneg\mathbf{e}_2\nur\mathbf{d}_2$ and $\mathbf{c}_2\nur\mathbf{d}_2$ as well as $\myneg\mathbf{e}_3\nur\mathbf{d}_3$ and $\mathbf{c}_3\nur\mathbf{d}_3$.
This can be summarized graphically as follows:
\[\begin{tikzcd}
   \myneg \mathbf{a}_2\ar[r, "\stackrel{*}{\myneg}" description]\ar{d}[swap]{\stackrel{*}{\myneg}} &\mathbf{c}_2\ar{d}{\stackrel{*}{\myneg}}\\
    \myneg \mathbf{e}_2\ar[r, "\stackrel{*}{\myneg}" description]           &{}\mathbf{d}_2
\end{tikzcd}
\qquad
\begin{tikzcd}
    \myneg \mathbf{a}_3\ar[r, "\stackrel{*}{\myneg}" description]\ar{d}[swap]{\stackrel{*}{\myneg}} &\mathbf{c}_3\ar{d}{\stackrel{*}{\myneg}}\\
    \myneg\mathbf{e}_3\ar[r, "\stackrel{*}{\myneg}" description]          &{}\mathbf{d}_3
\end{tikzcd}
\]
Hence we can define $\mathbf{d}=[\mathbf{d}_2,\mathbf{d}_3]$ where  $\mathbf{c}=[\mathbf{c}_2,\mathbf{c}_3]\nur\mathbf{d}$ and $\mathbf{b}=\myneg\mathbf{b}_1\nur\myneg \mathbf{e}=\myneg[\mathbf{e}_2+\mathbf{e}_3]\nur[\myneg\mathbf{e}_2,\myneg\mathbf{e}_3]\nur[\mathbf{d}_2,\mathbf{d}_3]=\mathbf{d}$. 
This can be summarized graphically as follows:
\[\begin{tikzcd}
    \mathbf{a}=\myneg \mathbf{a}_1=\myneg[\mathbf{a}_2+\mathbf{a}_3]\arrow[r, "{\myneg}" description]\arrow[d, "{\myneg}"]& \mathbf{b}=\myneg\mathbf{b}_1\ar{d}{\stackrel{*}{\myneg}}\\
    \mathbf{c}=[\mathbf{c}_2,\mathbf{c}_3]\ar[r, "\stackrel{*}{\myneg}" description]     &\mathbf{d}=[\mathbf{d}_2,\mathbf{d}_3]
\end{tikzcd}
\]
\item
$\mathbf{b}=\myneg\mathbf{b}_1$, $\mathbf{c}=[\mathbf{c}_2+\mathbf{c}_3]$. Symmetric to $\mathbf{c}=[\mathbf{c}_2,\mathbf{c}_3]$ using Law~\ref{nurd.basic}($iv$)
\item
$\mathbf{b}=\myneg\mathbf{b}_1$, $\mathbf{c}=[x:\mathbf{c}_2]\mathbf{c}_3$. By Law~\ref{nurd.basic}($vii$) we know that $\mathbf{a}_1\nur\mathbf{b}_1$.
By Law~\ref{nurd.basic}($v$) we know that $\mathbf{a}_1=[x!\mathbf{a}_2]\mathbf{a}_3$ for some $\mathbf{a}_2$, $\mathbf{a}_3$ where $\mathbf{a}_2=\mathbf{c}_2$ and $\myneg \mathbf{a}_3\nur\mathbf{c}_3$.
By inductive hypothesis applied to $\mathbf{a}_1$, there is an $\mathbf{e}$ such that $\mathbf{b}_1\nur\mathbf{e}$ and $[x!\mathbf{a}_2]\mathbf{a}_3\nur\mathbf{e}$.
Hence $\mathbf{e}=[x!\mathbf{a}_2]\mathbf{e}_3$ for some $\mathbf{e}_3$ where $\mathbf{a}_3\nur\mathbf{e}_3$.
This can be summarized graphically as follows:
\[\begin{tikzcd}
    \mathbf{a}_1=[x!\mathbf{a}_2]\mathbf{a}_3\arrow[r, "{\myneg}" description]\ar[rd, swap, "\stackrel{*}{\myneg}" description] &\mathbf{b}_1\ar{d}{\stackrel{*}{\myneg}}\\
                                      &\mathbf{e}=[x!\mathbf{a}_2]\mathbf{e}_3
\end{tikzcd}
\]
Hence also $\myneg\mathbf{a}_3\nur\myneg\mathbf{e}_3$.
By inductive hypothesis applied to $\myneg\mathbf{a}_3$ (note that $\sz(\myneg\mathbf{a}_3)<\sz(\mathbf{a})$) there is an expression $\mathbf{d}_3$ where $\myneg\mathbf{e}_3\nur\mathbf{d}_3$ and $\mathbf{c}_3\nur\mathbf{d}_3$.

Hence we can define $\mathbf{d}=[x:\mathbf{a}_2]\mathbf{d}_3$ where  $\mathbf{c}=[x:\mathbf{a}_2]\mathbf{c}_3\nur\mathbf{d}$ and
\[\mathbf{b}=\myneg\mathbf{b}_1\nur\myneg\mathbf{e}=\myneg[x!\mathbf{a}_2]\mathbf{e}_3\nur[x:\mathbf{a}_2]\myneg\mathbf{e}_3\nur\mathbf{d}
\]
This can be summarized graphically as follows:
\[\begin{tikzcd}
    \mathbf{a}=\myneg \mathbf{a}_1=\myneg[x!\mathbf{a}_2]\mathbf{a}_3\arrow[r,"{\myneg}" description]\arrow[d, "{\myneg}"]& \mathbf{b}=\myneg\mathbf{b}_1\ar{d}{\stackrel{*}{\myneg}}\\
    \mathbf{c}=[x:\mathbf{a}_2]\mathbf{c}_3\ar{r}{\stackrel{*}{\myneg}}           &\mathbf{d}=[x:\mathbf{a}_2]\mathbf{d}_3
\end{tikzcd}
\]
\item
$\mathbf{b}=\myneg\mathbf{b}_1$, $\mathbf{c}=[x!\mathbf{c}_2]\mathbf{c}_3$. Symmetric to case $\mathbf{c}=[x:\mathbf{c}_2]\mathbf{c}_3$ using Law\ref{nurd.basic}($vi$).
\item
$\mathbf{b}=\myneg\mathbf{b}_1$, $\mathbf{c}=\myneg\mathbf{c}_1$. By Law~\ref{nurd.basic}($vii$) we know that $\mathbf{a}_1\nur\mathbf{b}_1$ and $\mathbf{a}_1\nur\mathbf{c}_1$.
By inductive hypothesis applied to $\mathbf{a}_1$, there is an expression $\mathbf{d}$ such that $\mathbf{b}_1\nur\mathbf{d}$ and $\mathbf{c}_1\nur\mathbf{d}$.
Hence obviously $\mathbf{b}\nur\myneg\mathbf{d}$ and $\mathbf{c}\nur\myneg\mathbf{d}$.
\item
$\mathbf{c}=\myneg \mathbf{c}_1$, $\mathbf{b}$ is not a negation. Symmetric to previous cases where $\mathbf{b}$ was a negation and $\mathbf{c}$ was not a negation.
\item
$\mathbf{b}$ and $\mathbf{c}$ are both not negations. Obviously both $\mathbf{b}$ and $\mathbf{c}$ must be both either a product, a sum, or an abstraction.
\begin{itemize}
\item
In case $\mathbf{b}=[\mathbf{b}_2,\mathbf{b}_3]$ and $\mathbf{c}=[\mathbf{c}_2,\mathbf{c}_3]$, by Law~\ref{nurd.basic}($iii$) we know that $\mathbf{a}_1=[\mathbf{a}_2+\mathbf{a}_3]$ for some $\mathbf{a}_2$, $\mathbf{a}_3$ where $\myneg\mathbf{a}_2\nur\mathbf{b}_2$ and $\myneg\mathbf{a}_3\nur\mathbf{b}_3$ and also $\myneg\mathbf{a}_2\nur\mathbf{c}_2$ and $\myneg\mathbf{a}_3\nur\mathbf{c}_3$.
By inductive hypothesis applied to $\myneg\mathbf{a}_2$ and $\myneg\mathbf{a}_3$ (note that $\sz(\myneg\mathbf{a}_2)<\sz(\mathbf{a})$ and $\sz(\myneg\mathbf{a}_3)<\sz(\mathbf{a})$),
we know that  $\mathbf{b}_2\nur\mathbf{d}_2$ and $\mathbf{c}_2\nur\mathbf{d}_2$  as well as $\mathbf{b}_3\nur\mathbf{d}_3$ and $\mathbf{c}_3\nur\mathbf{d}_3$ for some $\mathbf{d}_2$ and $\mathbf{d}_3$.
Hence we can define $\mathbf{d}=[\mathbf{d}_2,\mathbf{d}_3]$ where  obviously $b\nur d$ and $\mathbf{c}\nur\mathbf{d}$.
This can be summarized graphically as follows:
\[\begin{tikzcd}
    \mathbf{a}=\myneg \mathbf{a}_1=\myneg[\mathbf{a}_2+\mathbf{a}_3]\arrow[r, "{\myneg}" description]\arrow[d, "{\myneg}"]&\mathbf{b}=[\mathbf{b}_2,\mathbf{b}_3]\ar{d}{\stackrel{*}{\myneg}}\\
    \mathbf{c}=[\mathbf{c}_2,\mathbf{c}_3]\ar[r, "\stackrel{*}{\myneg}" description]           &\mathbf{d}=[\mathbf{d}_2,\mathbf{d}_3]
\end{tikzcd}
\]
\item
The case $\mathbf{b}=[\mathbf{b}_2+\mathbf{b}_3]$, $\mathbf{c}=[\mathbf{c}_2+\mathbf{c}_3]$ is symmetric to the case $\mathbf{b}=[\mathbf{b}_2,\mathbf{b}_3]$, $\mathbf{c}=[\mathbf{c}_2,\mathbf{c}_3]$ using Law~\ref{nurd.basic}($iv$).
\item
In case $\mathbf{b}=[x:\mathbf{b}_2]\mathbf{b}_3$ and $\mathbf{c}=[x:\mathbf{c}_2]\mathbf{c}_3$, by Law~\ref{nurd.basic}($v$) we know that $\mathbf{a}_1=[x!\mathbf{a}_2]\mathbf{a}_3$  for some $\mathbf{a}_2$, $\mathbf{a}_3$ where $\mathbf{a}_2\nur\mathbf{b}_2$ and $\myneg\mathbf{a}_3\nur\mathbf{b}_3$ and $\mathbf{a}_2\nur\mathbf{c}_2$ and $\myneg \mathbf{a}_3\nur\mathbf{c}_3$.
By inductive hypothesis applied to $\mathbf{a}_2$ and $\myneg\mathbf{a}_3$ (note that $\sz(\myneg\mathbf{a}_3)<\sz(\mathbf{a})$),
we know $\mathbf{b}_2\nur\mathbf{d}_2$ and $\mathbf{c}_2\nur\mathbf{d}_2$  as well as $\mathbf{b}_3\nur\mathbf{d}_3$ and $\mathbf{c}_3\nur\mathbf{d}_3$ for some $\mathbf{d}_2$ and $\mathbf{d}_3$.
Hence we can define $\mathbf{d}=[x:\mathbf{d}_2]\mathbf{d}_3$ where $\mathbf{b}\nur\mathbf{d}$ and $\mathbf{c}\nur\mathbf{d}$.
This can be summarized graphically as follows:
\[\begin{tikzcd}
    \mathbf{a}=\myneg\mathbf{a}_1=\myneg[x!\mathbf{a}_2]\mathbf{a}_3\arrow[r, "{\myneg}" description]\arrow[d, "{\myneg}"]& \mathbf{b}=[x:\mathbf{b}_2]\mathbf{b}_3\ar{d}{\stackrel{*}{\myneg}}\\
    \mathbf{c}=[x:\mathbf{c}_2]\mathbf{c}_3\ar[r, "\stackrel{*}{\myneg}" description]   &\mathbf{d}=[x:\mathbf{d}_2]\mathbf{d}_3
\end{tikzcd}
\]
\item
The case $\mathbf{b}=[x!\mathbf{b}_2]\mathbf{b}_3$ and $\mathbf{c}=[x!\mathbf{c}_2]\mathbf{c}_3$ is symmetric to the case $\mathbf{b}=[x:\mathbf{b}_2]\mathbf{b}_3$ and $\mathbf{c}=[x:\mathbf{c}_2]\mathbf{c}_3$ using Law~\ref{nurd.basic}($vi$).
\end{itemize}
\end{itemize}
\item
In all other cases no axiom or structural rule of $\nur$ is matching.
\qedhere
\end{itemize}
\noindent
This completes the argument of local confluence.
Since $\nur$ is terminating, it is obviously also terminating for expressions $\mathbf{b}$ with $\sz(\mathbf{b})\leq n$.
Hence one can apply the diamond lemma~\cite{NEWMAN1942} to obtain confluence of $\snur$ for $\mathbf{a}$, which completes the inductive step.
\end{proof}
\begin{law}[Commutation of single-step reduction with explicit substitution and negation-reduction]
\label{nurd.comm}
For all $\mathbf{a}$, $\mathbf{b}$ and $\mathbf{c}$: $E\sgv\mathbf{a}\smrd\mathbf{b}$ and $\mathbf{a}\nur\mathbf{c}$ imply  $\mathbf{b}\nur\mathbf{d}$ and $E\sgv\mathbf{c}\mrdn{01}\mathbf{d}$ for some $\mathbf{d}$. 
\end{law}
\begin{proof}
First we prove by induction on $\mathbf{a}$ 
that for all $\mathbf{b}$ and $\mathbf{c}$, $E\sgv\mathbf{a}\smrd\mathbf{b}$ and $\mathbf{a}\snur\mathbf{c}$ imply  $\mathbf{b}\nur\mathbf{d}$ and $E\sgv\mathbf{c}\mrdn{01}\mathbf{d}$ for some $\mathbf{d}$.

Obviously we only have to consider those cases in which there exists a $\mathbf{c}$ such that $\mathbf{a}\snur\mathbf{c}$:
\begin{itemize}
\item
$\mathbf{a}=\prsumop{\mathbf{a}_1}{\mathbf{a}_2}$: $\mathbf{a}\snur\mathbf{c}$ obviously implies that $\mathbf{c}=\prsumop{\mathbf{c}_1}{\mathbf{c}_2}$ where either $\mathbf{a}_1\snur\mathbf{c}_1$ and $\mathbf{a}_2=\mathbf{c}_2$ or $\mathbf{a}_2\snur\mathbf{c}_2$ and $\mathbf{a}_1=\mathbf{c}_1$. 
Similarly, $E\sgv\mathbf{a}\smrd\mathbf{b}$ obviously implies that $\mathbf{b}=\prsumop{\mathbf{b}_1}{\mathbf{b}_2}$ where either $E\sgv\mathbf{a}_1\smrd \mathbf{b}_1$ and $\mathbf{a}_2=\mathbf{b}_2$ or $E\sgv\mathbf{a}_2\smrd\mathbf{b}_2$ and $\mathbf{a}_1=\mathbf{b}_1$.
Thus, there are four cases where in two of the cases the property follows directly and in the other two it follows from the inductive hypothesis. 
\item
$\mathbf{a}=\binbop{x}{\mathbf{a}_1}{\mathbf{a}_2}$: The proof is similar to the case $\mathbf{a}=\prsumop{\mathbf{a}_1}{\mathbf{a}_2}$.
\item
$\mathbf{a}=\myneg\mathbf{a}_1$: 
If $E\sgv \myneg\mathbf{a}_1\smrd\mathbf{b}$, due to the definition of $\mrd$-reduction, only the following two cases are possible:

The first case is the use of the rule $\nu$ on top-level, i.e.~$\myneg\mathbf{a}_1\nurp\mathbf{b}$.
By confluence of $\snur$ (Law~\ref{nurd.confl}) we know that there is an expression $\mathbf{d}$ such that $\mathbf{c}\nur\mathbf{d}$ and $\mathbf{b}\nur\mathbf{d}$.
We have $\mathbf{c}=\mathbf{d}$ or $\mathbf{c}\nurp\mathbf{d}$, and therefore obviously $E\sgv\mathbf{c}\mrdn{01}\mathbf{d}$.

The only other possible case is $\mathbf{b}=\myneg\mathbf{b}_1$ where $E\sgv\mathbf{a}_1\smrd\mathbf{b}_1$, for some $\mathbf{b}_1$.
From $\mathbf{a}=\myneg\mathbf{a}_1\snur\mathbf{c}$, by definition of $\nur$-reduction we know that at least one of the following cases must be true:
\begin{meditemize}
\item[$\nu_1$:]
$\mathbf{a}_1=\myneg\mathbf{a}_2$, $\mathbf{c}=\mathbf{a}_2$.
Hence $E\sgv\myneg\mathbf{a}_2\smrd\mathbf{b}_1$.
As argued above, there are two cases:
If $\myneg\mathbf{a}_2\nurp\mathbf{b}_1$ then obviously also $\myneg\mathbf{a}_1\nurp\mathbf{b}$.
This case has already been considered above.
Therefore $\mathbf{b}_1=\myneg\mathbf{d}$ where $E\sgv\mathbf{a}_2\smrd\mathbf{d}$, for some $\mathbf{d}$.
Obviously also $\mathbf{b}=\myneg\myneg\mathbf{d}\nur\mathbf{d}$.
\item[$\nu_2$:]
$\mathbf{a}_1=[\mathbf{a}_2,\mathbf{a}_3]$, $\mathbf{c}=[\myneg\mathbf{a}_2+\myneg\mathbf{a}_3]$ for some $\mathbf{a}_2$, $\mathbf{a}_3$.
Hence obviously $\mathbf{b}_1=[\mathbf{b}_2,\mathbf{b}_3]$ where $E\sgv\mathbf{a}_2\smrd\mathbf{b}_2$ and $\mathbf{a}_3=\mathbf{b}_3$  or vice versa.

In the first case we have $\mathbf{d}=[\myneg\mathbf{b}_2+\myneg\mathbf{a}_3]$ where $E\sgv\mathbf{c}=[\myneg\mathbf{a}_2+\myneg\mathbf{a}_3]\smrd\mathbf{d}$ and $\mathbf{b}=\myneg\mathbf{b}_1\snur\mathbf{d}$.
The second case runs analogously.
\item[$\nu_3$:]
$\mathbf{a}_1=[\mathbf{a}_2+\mathbf{a}_3]$, $c=[\myneg\mathbf{a}_2,\myneg\mathbf{a}_3]$ for some $\mathbf{a}_2$, $\mathbf{a}_3$.
This case is symmetric to the previous one.
\item[$\nu_4$:]
$\mathbf{a}_1=[x:\mathbf{a}_2]\mathbf{a}_3$, $c=[x!\mathbf{a}_2]\myneg\mathbf{a}_3$ for some $\mathbf{a}_2$, $\mathbf{a}_3$.
Hence obviously $\mathbf{b}_1=[x!\mathbf{b}_2]\mathbf{b}_3$ where $E\sgv\mathbf{a}_2\smrd\mathbf{b}_2$ and $\mathbf{a}_3=\mathbf{b}_3$ or $\mathbf{a}_2=\mathbf{b}_2$ and $E,x\mydef\mathbf{a}_2\sgv\mathbf{a}_3\smrd\mathbf{b}_3$.

In the first case we can infer that $\mathbf{d}=[x!\mathbf{b}_2]\myneg\mathbf{b}_3$ with $E\sgv \mathbf{c}=[x!\mathbf{a}_2]\myneg\mathbf{a}_3\smrd\mathbf{d}$ and $\mathbf{b}=\myneg\mathbf{b}_1\snur\mathbf{d}$.
The second case runs analogously.
\item[$\nu_5$:]
$\mathbf{a}_1=[x!\mathbf{a}_2]\mathbf{a}_3$, $\mathbf{c}=[x:\mathbf{a}_2]\myneg\mathbf{a}_3$ for some $\mathbf{a}_2$, $\mathbf{a}_3$.
This case is symmetric to the previous one.´
\item[$(\myneg\_)_1$:]
$\mathbf{c}=\myneg\mathbf{c}_1$ where $\mathbf{a}_1\snur\mathbf{c}_1$ for some $\mathbf{c}$.
By inductive hypothesis there is an expression $\mathbf{d}$ such that $\mathbf{b}_1\nur\mathbf{d}$ and $E\sgv\mathbf{c}_1\mrdn{01}\mathbf{d}$.
Hence obviously $\mathbf{b}\nur\myneg\mathbf{d}$ and $E\sgv\mathbf{c}=\myneg\mathbf{c}_1\mrdn{01}\myneg\mathbf{d}$.
\end{meditemize}
\end{itemize}
The next step is to prove the main property by induction on the length $n$ of $\nur$-reduction $\mathbf{a}\nurn{n}\mathbf{c}$.
\begin{itemize}
\item
In case of $n=0$ the property is trivial.
\item
Let  $E\sgv\mathbf{a}\smrd\mathbf{b}$ and $\mathbf{a}\nurn{n}\mathbf{c}'\snur\mathbf{c}$.
By inductive hypothesis we know there is an expression $\mathbf{d}'$ such that
$E\sgv\mathbf{c}'\mrdn{01}\mathbf{d}'$ and $\mathbf{b}\nur\mathbf{d}'$.
This situation can be graphically summarized as follows (leaving out the environment $E$):

\[\begin{tikzcd}
    \mathbf{a}\arrow[r,"\stackrel{n}{\myneg}" description]\arrow[d,"{:=}" description]&\mathbf{c}'\arrow[r,"{\myneg}" description]\ar{d}{\stackrel{01}{:=}}&\mathbf{c}\\
    \mathbf{b}\ar[r,"\stackrel{*}{\myneg}" description]                           &\mathbf{d}'
\end{tikzcd}
\]
If $\mathbf{c}'=\mathbf{d}'$, we know that $\mathbf{b}\nur\mathbf{c}'$ and hence $\mathbf{b}\nur\mathbf{c}$. 
Hence $\mathbf{d}=\mathbf{c}$ where $\mathbf{b}\nur\mathbf{d}$ and $E\sgv\mathbf{c}\mrdn{01}\mathbf{d}$.
This situation can be graphically summarized as follows (leaving out the environment $E$):
\[\begin{tikzcd}[column sep= huge]
    \mathbf{a}\arrow[r,"\stackrel{n}{\myneg}" description]\arrow[d,"{:=}"]&\mathbf{d}'=\mathbf{c}'\arrow[r,"{\myneg}" description]&\mathbf{d}=\mathbf{c}\\
    \mathbf{b}\ar[ru, "\stackrel{*}{\myneg}" description]                             
\end{tikzcd}
\]
Otherwise $E\sgv\mathbf{c}'\smrd\mathbf{d}'$ and $\mathbf{c}'\snur\mathbf{c}$.
By the argument above we know there is an expression $\mathbf{d}$ such that $\mathbf{d}'\nur\mathbf{d}$ and $E\sgv\mathbf{c}\mrdn{01}\mathbf{d}$.
Hence $\mathbf{b}\nur\mathbf{d}'\nur\mathbf{d}$ and $E\sgv\mathbf{c}\mrdn{01}\mathbf{d}$ which completes the proof.
This situation can be graphically summarized as follows (leaving out the environment $E$):
\[\begin{tikzcd}
    \mathbf{a}\arrow[r,"\stackrel{n}{\myneg}" description]\arrow[d,"{:=}"]&\mathbf{c}'\arrow[r,"\stackrel{1}{\myneg}" description]\ar{d}{:=}&\mathbf{c}\arrow[d,"\stackrel{01}{:=}"]\\
    \mathbf{b}\ar[r, "\stackrel{*}{\myneg}" description]                       &\mathbf{d}'\ar[r,"\stackrel{*}{\myneg}" description]                               &\mathbf{d} 
\end{tikzcd}
\]
\qedhere
\end{itemize}
\end{proof}
\begin{law}[Direct confluence of $\smrd$]
\index{confluence!of single-step reduction with expli-cit substitution, direct}
\label{mrd.confl.dir}
For all $E$, $\mathbf{a}$, $\mathbf{b}$, $\mathbf{c}$: $E\sgv\mathbf{a}\smrd\mathbf{b}$ and $E\sgv\mathbf{a}\smrd\mathbf{c}$ imply $E\sgv\mathbf{b}\mrdn{01}\mathbf{d}$, and $E\sgv\mathbf{c}\mrdn{01}\mathbf{d}$ for some $\mathbf{d}$.
\end{law}
\begin{proof}
Proof is by induction on $\mathbf{a}$ 
with a systematic investigation of critical pairs.

Due to the definition of $\smrd$,  critical pairs of $E\sgv\mathbf{a}\smrd\mathbf{b}$ and $E\sgv\mathbf{a}\smrd\mathbf{c}$ where where at least one of the steps is using axiom $\nu$, on top-level, i.e.~where $E\sgv\mathbf{a}\nurp\mathbf{b}$ or $E\sgv\mathbf{a}\nurp\mathbf{c}$, can be resolved thanks to Laws~\ref{nurd.confl} and~\ref{nurd.comm}, hence they are excluded in the following case distinctions.
\begin{itemize}
\item 
$\mathbf{a}=\prim$: Cannot be the case, since there is no matching left-hand side 
\item 
$\mathbf{a}=x$: The only matching axiom is \emph{use}, the property follows since $E(x)$ is a function, i.e. $\mathbf{c}=\mathbf{b}$.
\item 
$\mathbf{a}=[x:\mathbf{a}_1]\mathbf{a}_2$: Basically the property follows since there are no critical pairs. 
More formally, the only matching rules are $([x:\_]\_)_1$ and $([x:\_]\_)_2$. Therefore, the following four cases must be considered:
\begin{itemize}
\item 
$\mathbf{b}=[x:\mathbf{b}_1]\mathbf{a}_2$, $\mathbf{c}=[x:\mathbf{c}_1]\mathbf{a}_2$, where $E\sgv\mathbf{a}_1\smrd\mathbf{b}_1$ and $E\sgv\mathbf{a}_1\smrd\mathbf{c}_1$:
By inductive hypothesis there is an expression $\mathbf{d}$ such that $E\sgv\mathbf{b}_1\mrdn{01}\mathbf{d}$ and $E\sgv\mathbf{c}_1\mrdn{01}\mathbf{d}$, hence obviously $E\sgv\mathbf{b}\mrdn{01}[x:\mathbf{d}]\mathbf{a}_2$ and $E\sgv\mathbf{c}\mrdn{01}[x:\mathbf{d}]\mathbf{a}_2$.
\item 
$\mathbf{b}=[x:\mathbf{a}_1]\mathbf{b}_2$, $\mathbf{c}=[x:\mathbf{a}_1]\mathbf{c}_2$, where $E\sgv\mathbf{a}_2\smrd\mathbf{b}_2$ and $E\sgv\mathbf{a}_2\smrd\mathbf{c}_2$: By the inductive hypothesis there is an expression $d$ such that $E\sgv\mathbf{b}_2\mrdn{01}\mathbf{d}$ and $E\sgv\mathbf{c}_2\mrdn{01}\mathbf{d}$, hence obviously $E\sgv\mathbf{b}\mrdn{01}[x:\mathbf{a}_1]\mathbf{d}$ and $E\sgv\mathbf{c}\mrdn{01}[x:\mathbf{a}_1]\mathbf{d}$.
\item 
$\mathbf{b}=[x:\mathbf{b}_1]\mathbf{a}_2$, $\mathbf{c}=[x:\mathbf{a}_1]\mathbf{c}_2$, where $E\sgv\mathbf{a}_1\smrd\mathbf{b}_1$ and $E\sgv\mathbf{a}_2\smrd\mathbf{c}_2$. Obviously the common single-step reduct is $[x:\mathbf{b}_1]\mathbf{c}_2$.
\item 
$\mathbf{b}=[x:\mathbf{a}_1]\mathbf{c}_2$, $x=[x:\mathbf{b}_1]\mathbf{a}_2$, where $E\sgv\mathbf{a}_1\smrd\mathbf{b}_1$ and $E\sgv\mathbf{a}_2\smrd\mathbf{c}_2$. Obviously the common single-step reduct is $[x:\mathbf{b}_1]\mathbf{c}_2$.
\end{itemize}
\item 
$\mathbf{a}=[x!\mathbf{a}_1]\mathbf{a}_2$: Similar to $[x:\mathbf{a}_1]\mathbf{a}_2$ as there are no critical pairs.
\item 
$\mathbf{a}=\prdef{x}{\mathbf{a}_1}{\mathbf{a}_2}{\mathbf{a}_3}$: Similar to $[x:\mathbf{a}_1]\mathbf{a}_2$ as there are no critical pairs.
\item 
$\mathbf{a}=[x\mydef\mathbf{a}_1]\mathbf{a}_2$: The matching axiom and rules are \emph{rem}, \emph{L$_{\smydef}$}, and \emph{R$_{\smydef}$}. The use of \emph{L$_{\smydef}$} versus \emph{R$_{\smydef}$} can be argued as in the previous three cases. The interesting cases are the use of \emph{rem}, versus \emph{L$_{\smydef}$} or \emph{R$_{\smydef}$}: Hence we may assume that $x\notin \free(\mathbf{a}_2)$ and need to consider the following cases:
\begin{itemize}
\item 
$\mathbf{b}=\mathbf{a}_2$ and $\mathbf{c}=[x\mydef\mathbf{b}_1]\mathbf{a}_2$ where $E\sgv\mathbf{a}_1\smrd\mathbf{b}_1$.
We have $\mathbf{d}=\mathbf{a}_2$ since $x\notin \free(\mathbf{a}_2)$ and $\mathbf{c}$ reduces in one-step to $\mathbf{a}_2$.
This situation can be graphically summarized as follows (leaving out the environment $E$):
\[\begin{tikzcd}
    \mathbf{a}=[x\mydef\mathbf{a}_1]\mathbf{a}_2\arrow[r,"{:=}" description]\arrow[d,"{:=}"]&\mathbf{d}=\mathbf{b}=\mathbf{a}_2\\
    \mathbf{c}=[x\mydef\mathbf{b}_1]\mathbf{a}_2\arrow[ru,"{:=}" description]                           
\end{tikzcd}
\]
\item 
$\mathbf{b}=\mathbf{a}_2$ and $\mathbf{c}=[x\mydef\mathbf{a}_1]\mathbf{c}_2$ where $E,x\mydef\mathbf{a}_1\sgv\mathbf{a}_2\smrd\mathbf{c}_2$: 
Similar to Law~\ref{rd.sub}($iii$), by structural induction on expressions with internalized substitutions one can show that $x\notin\free(\mathbf{a}_2)$ implies that $x\notin\free(\mathbf{c}_2)$ and therefore $E \sgv\mathbf{a}_2\smrd\mathbf{c}_2$.
Therefore $\mathbf{d}=\mathbf{c}_2$ where $E\sgv\mathbf{c}\smrd\mathbf{c}_2$ and $E\sgv\mathbf{b}\smrd\mathbf{c}_2$.  
This situation can be graphically summarized as follows (leaving out the environment $E$):
\[\begin{tikzcd}
    \mathbf{a}=[x\mydef\mathbf{a}_1]\mathbf{a}_2\arrow[r,"{:=}" description]\arrow[d,"{:=}"]&\mathbf{b}=\mathbf{a}_2\arrow[d,"{:=}"]\\
    \mathbf{c}=[x\mydef\mathbf{a}_1]\mathbf{c}_2\arrow[r,"{:=}" description]                      & \mathbf{d}=\mathbf{c}_2     
\end{tikzcd}
\]
\item 
The other two cases are symmetric.
\end{itemize}
\item 
$\mathbf{a}=(\mathbf{a}_1\,\mathbf{a}_2)$: The four matching axiom and two matching rules are $\beta_1^{\mu}$, $\beta_2^{\mu}$, $\beta_3$, $\beta_4$, ${\_(\_)}_1$, and ${\_(\_)}_2$. Several cases have to be considered: The use of ${(\_\,\_)}_1$ versus ${(\_\,\_)}_2$ can be argued similar as in the case of (universal) abstraction.
The simultaneous application of two different axioms on top-level is obviously not possible.
The interesting remaining cases are the usage of one of the four axioms versus one of the rules.

The first case is the use of $\beta_1^{\mu}$, i.e. $\mathbf{a}_1=[x:\mathbf{a}_3]\mathbf{a}_4$ and $\mathbf{b}=[x\mydef\mathbf{a}_2]\mathbf{a}_4$, versus one of the rules ${(\_\,\_)}_1$, and ${(\_\,\_)}_2$. 
Two cases need to be considered:
\begin{itemize}
\item 
Use of rule ${(\_\,\_)}_1$, i.e.~$\mathbf{c}=(\mathbf{c}_1\,\mathbf{a}_2)$ where $E\sgv\mathbf{a}_1=[x:\mathbf{a}_3]\mathbf{a}_4\smrd\mathbf{c}_1$:
By definition of $\smrd$, there are two cases:
\begin{itemize}
\item 
$\mathbf{c}_1=[x:\mathbf{c}_3]\mathbf{a}_4$ where $E\sgv\mathbf{a}_3\smrd\mathbf{c}_3$: This means that $E\sgv\mathbf{c}\smrd[x\mydef\mathbf{a}_2]\mathbf{a}_4=\mathbf{b}$, i.e.~$\mathbf{d}=\mathbf{b}$ is a single-step reduct of $\mathbf{c}$.
This situation can be graphically summarized as follows (leaving out the environment $E$):
\[\begin{tikzcd}
    \mathbf{a}=(\mathbf{a}_1\,\mathbf{a}_2)=([x:\mathbf{a}_3]\mathbf{a}_4\:\mathbf{a}_2)\arrow[r,"{:=}" description]\arrow[d,"{:=}"]&\mathbf{d}=\mathbf{b}=[x\mydef\mathbf{a}_2]\mathbf{a}_4\\
    \mathbf{c}=(\mathbf{c}_1\,\mathbf{a}_2)=([x:\mathbf{c}_3]\mathbf{a}_4\:\mathbf{a}_2)\arrow[ru,"{:=}" description]                     
\end{tikzcd}
\]
\item 
$\mathbf{c}_1= [x:\mathbf{a}_3]\mathbf{c}_4$ where $E\sgv\mathbf{a}_4\smrd\mathbf{c}_4$: 
By definition of $\smrd$ we know that also $E\sgv[x\mydef\mathbf{a}_2]\mathbf{a}_4\smrd[x\mydef\mathbf{a}_2]\mathbf{c}_4$. 
Hence $\mathbf{d}=[x\mydef\mathbf{a}_2]\mathbf{c}_4$ with $E\sgv\mathbf{b}=[x\mydef\mathbf{a}_2]\mathbf{a}_4\smrd[x\mydef\mathbf{a}_2]\mathbf{c}_4=\mathbf{d}$ and $E\sgv \mathbf{c}=(\mathbf{c}_1\,\mathbf{a}_2)=([x:\mathbf{a}_3]\mathbf{c}_4\:\mathbf{a}_2)\smrd[x\mydef\mathbf{a}_2]\mathbf{c}_4=\mathbf{d}$. 
This situation can be graphically summarized as follows (leaving out the environment $E$):
\[\begin{tikzcd}
    \mathbf{a}=(\mathbf{a}_1\,\mathbf{a}_2)=([x:\mathbf{a}_3]\mathbf{a}_4\:\mathbf{a}_2)\arrow[r,"{:=}" description]\arrow[d,"{:=}"]&\mathbf{b}=[x\mydef\mathbf{a}_2]\mathbf{a}_4\arrow[d,"{:=}"]\\
    \mathbf{c}=(\mathbf{c}_1\,\mathbf{a}_2)=([x:\mathbf{a}_3]\mathbf{c}_4\:\mathbf{a}_2)\arrow[r,"{:=}" description] & \mathbf{d}=[x\mydef\mathbf{a}_2]\mathbf{c}_4 
\end{tikzcd}
\]
\end{itemize}
\item 
Use of rule ${(\_\,\_)}_2$, i.e.~$\mathbf{c}=(\mathbf{a}_1\,\mathbf{c}_2)$ where $E\sgv\mathbf{a}_2\smrd\mathbf{c}_2$:
It follows that $\mathbf{d}=[x\mydef\mathbf{c}_2]\mathbf{a}_4$ since $E\sgv (\mathbf{a}_1\,\mathbf{c}_2)\smrd\mathbf{d}$ and $\mathbf{d}$ is a single-step reduct of $\mathbf{b}$.
This situation can be graphically summarized as follows (leaving out the environment $E$):
\[\begin{tikzcd}
    \mathbf{a}=(\mathbf{a}_1\,\mathbf{a}_2)=([x:\mathbf{a}_3]\mathbf{a}_4\:\mathbf{a}_2)\arrow[r,"{:=}" description]\arrow[d,"{:=}"]&\mathbf{b}=[x\mydef\mathbf{a}_2]\mathbf{a}_4\arrow[d,"{:=}"]\\
    \mathbf{c}=(\mathbf{a}_1\,\mathbf{c}_2)=([x:\mathbf{a}_3]\mathbf{a}_4\:\mathbf{c}_2)\arrow[r,"{:=}" description] &\mathbf{d}=[x\mydef\mathbf{c}_2]\mathbf{a}_4 
\end{tikzcd}
\]
\end{itemize}
The second case is the use of $\beta_2^{\mu}$, i.e. $\mathbf{a}_1=[x!\mathbf{a}_3]\mathbf{a}_4$ and $\mathbf{b}=[x\mydef\mathbf{a}_2]\mathbf{a}_4$, versus one of the rules ${(\_\,\_)}_1$, and ${(\_\,\_)}_2$. It can be argued in the same way as the first case.

The third case is the use of $\beta_3$, i.e.~$\mathbf{a}_1=\case{\mathbf{a}_3}{\mathbf{a}_4}$, $\mathbf{a}_2=\injl{\mathbf{a}_5}{\mathbf{a}_6}$ and $\mathbf{b}=(\mathbf{a}_3\,\mathbf{a}_5)$, versus one of the rules ${(\_\,\_)}_1$, and ${(\_\,\_)}_2$.
The property follows by an obvious case distinction on whether $E\sgv\case{\mathbf{a}_3}{\mathbf{a}_4}\smrd\mathbf{c}$ is reducing in $\mathbf{a}_3$ or $\mathbf{a}_4$, and similarly for $E\sgv\injl{\mathbf{a}_5}{\mathbf{a}_6}\smrd\mathbf{c}$. 

The fourth case is symmetric to the third one.
\item 
$\mathbf{a}=\prsumop{\mathbf{a}_1}{\mathbf{a}_2}$: As there are no critical pairs, these cases are shown similar to the cases of universal abstraction.
\item 
$\mathbf{a}=\pleft{\mathbf{a}_1}$: The interesting cases are the use of one of the axioms $\pi_1$, $\pi_3$, or $\pi_5$ versus the rule $(\pleft{\_})_1$.

In case of $\pi_1$ we have $\mathbf{a}_1=[x!\mathbf{a}_2]\mathbf{a}_3$, $\mathbf{b}=\mathbf{a}_2$, and $\mathbf{c}=\pleft{\mathbf{c}_1}$ where $E\sgv[x!\mathbf{a}_2]\mathbf{a}_3\smrd\mathbf{c}_1$. 
Obviously $\mathbf{c}_1=[x!\mathbf{c}_2]\mathbf{c}_3$ where either $E\sgv\mathbf{a}_2\smrd\mathbf{c}_2$ and $\mathbf{c}_3=\mathbf{a}_3$ or $\mathbf{c}_2=\mathbf{a}_2$ and $E\sgv\mathbf{a}_3\smrd\mathbf{c}_3$.
In the first case $\mathbf{d}=\mathbf{c}_2$, in the second case $\mathbf{d}=\mathbf{a}_2$.

The cases $\pi_3$ and $\pi_5$ can be argued in a similar style.
\item 
$\mathbf{a}=\pright{\mathbf{a}_1}$: Similar to $\pleft{}$ considering the use of one of the axioms $\pi_2$, $\pi_4$, or $\pi_6$ versus the rule ${(\pleft{\_})}_2$.
\item 
$\mathbf{a}=\injl{\mathbf{a}_1}{\mathbf{a}_2}$, $\mathbf{a}=\injr{\mathbf{a}_1}{\mathbf{a}_2}$, $\mathbf{a}=\case{\mathbf{a}_1}{\mathbf{a}_2}$: 
As there are no critical pairs, these cases are shown similar to the cases of universal abstraction.
\item 
$\mathbf{a}=\myneg\mathbf{a}_1$: Matching are the axioms $\nu_i$ with $i\in\{6,\ldots,10\}$ and the rule $\mathit{{(\myneg\_)}_1}$.
There are obviously no critical pairs among the axioms.
Parallel use of $\mathit{{(\myneg\_)}_1}$ and any one of the axioms obviously implies direct confluence.
\end{itemize}
\end{proof}
\begin{law}[Confluence of $\smrd$]
\label{mrd.confl}
For all $E,\mathbf{a},\mathbf{b},\mathbf{c}$: $E\sgv\mathbf{a}\mrd\mathbf{b}$ and $E\sgv\mathbf{a}\mrd\mathbf{c}$ implies $E\sgv\mathbf{b}\mrdr\mathbf{c}$.
\end{law}
\begin{proof}
Based on Law~\ref{mrd.confl.dir}, by induction on the number $n$ of transition steps one can show that  $E\sgv\mathbf{a}=\mathbf{b}_0\mrdn{n}\mathbf{b}_n=\mathbf{b}$ and $E\sgv\mathbf{a}\smrd\mathbf{c}$ implies $E\sgv\mathbf{b}\mrdn{01}\mathbf{d}$, and $E\sgv\mathbf{c}\mrdn{*}\mathbf{d}$ for some $\mathbf{d}$.
This situation can be graphically summarized as follows (leaving out the environment $E$):
\[\begin{tikzcd}
\mathbf{a}=\mathbf{b}_0\arrow[r,":=" description]\arrow[d,"\stackrel{1}{:=}"]&\mathbf{b}_1\arrow[r,"\stackrel{1}{:=}" description]\arrow[d,"\stackrel{01}{:=}"]&\cdots\arrow[r,"\stackrel{1}{:=}" description]      &\mathbf{b}_n=\mathbf{b}\arrow[d,"\stackrel{01}{:=}"]\\
\mathbf{c}=\mathbf{d}_0\arrow[r,"\stackrel{01}{:=}" description]    &\mathbf{d}_1\arrow[r,"\stackrel{01}{:=}" description]  &\cdots\arrow[r,"\stackrel{01}{:=}" description]&\mathbf{d}_n=\mathbf{d}   
\end{tikzcd}
\]
Using this intermediate result, by induction on the number of transition steps $n$ one can show for any $m$ that $E\sgv\mathbf{a}\mrdn{n}\mathbf{b}$ and $E\sgv\mathbf{a}\mrdn{*}\mathbf{c}$ implies $E\sgv\mathbf{b}\mrdn{*}\mathbf{d}$ and $E\sgv\mathbf{c}\mrdn{*}\mathbf{d}$ for some $\mathbf{d}$.
This proves confluence of $\smrd$.
This situation can be graphically summarized as follows (leaving out the environment $E$):
\[\begin{tikzcd}
\mathbf{a}=\mathbf{b}_0\arrow[r,"{:=}" description]\arrow[d,"\stackrel{*}{:=}"]&\mathbf{b}_1\arrow[r,"{:=}" description]\arrow[d,"\stackrel{*}{:=}"]&\cdots\arrow[r,"{:=}" description]      &\mathbf{b}_n=\mathbf{b}\arrow[d,"\stackrel{*}{:=}"]\\
\mathbf{c}=\mathbf{d}_0\arrow[r,"\stackrel{01}{:=}" description]                    &\mathbf{d}_1\arrow[r,"\stackrel{01}{:=}" description]                  &\cdots\arrow[r,"\stackrel{01}{:=}" description]&\mathbf{d}_n=\mathbf{d}   
\end{tikzcd}
\]
\end{proof}
\subsection{Confluence of single-step reduction}%
By Law~\ref{rd.mrd} we know that reduction implies reduction with explicit substitution. The reverse direction is obviously not true. However we can show that $E\sgv\mathbf{a}\mrd b$ implies $a'\rd b'$ where 
$a'$ and $b'$ result from $\mathbf{a}$ and $\mathbf{b}$ by maximal evaluation of definitions, i.e.~by maximal application of the axioms \emph{use} and \emph{rem}.
First, we introduce the relation $E\sgv\mathbf{a}\dmrd\mathbf{b}$ of \emph{definition evaluation}
\begin{definition}[Single-step definition-evaluation]%
\nomenclature[gRel07]{$E\sgv\mathbf{a}\dmrd\mathbf{b}$}{single-step definition-evaluation}
\index{definition!evaluation, single-step}
\emph{Single-step definition- evaluation} $E\sgv\mathbf{a}\dmrd\mathbf{b}$ is defined just like single-step reduction with explicit substitution $E\sgv\mathbf{a}\smrd\mathbf{b}$ but without the rule $\nu$ and without any axioms except (\emph{rem}) and (\emph{use}). 
\end{definition}
\begin{law}[Confluence of $\dmrd$]%
\label{mrd.confl.def}
$\dmrd$ is confluent.
\end{law}
\begin{proof}
y removing all the axiom cases except \emph{rem} and \emph{use} and the rule $\nu$ in the proof of Lemma~\ref{mrd.confl.dir} (where these axioms only interacted with the structural rules for internalized substitution) it can be turned into a proof of direct confluence of $\dmrd$.
The confluence of $\dmrd$
then follows as in the proof of Lemma~\ref{mrd.confl}.
\end{proof}
\begin{law}[Strong normalization of definition-evaluation]%
\label{mrd.sn.def}
There are no infinite chains  $E\sgv\mathbf{a}_1\dmrd\mathbf{a}_2 \dmrd\mathbf{a}_3 \dmrd \cdots$
\end{law}
\begin{proof}
Since all definitions are non-recursive the property is straightforward. 
One way to see this is to define a weight $W(E,\mathbf{a})$ such that
\[
E\gv\mathbf{a}\dmrd\mathbf{b}\quad\text{implies}\quad W(E,\mathbf{b})<W(E,\mathbf{a})
\]
A possible definition is given below: 
\begin{eqnarray*}
W(E,\prim)&=&1\\
W(E,x)&=&\begin{cases}
W(E,E(x))+1&\text{if}\;x\in\dom(E)\\
1&\text{otherwise}
\end{cases}\\
W(E,\binop{\mathbf{a}_1,\ldots}{\mathbf{a}_n})&=&W(E,\mathbf{a}_1)+\ldots+W(E,\mathbf{a}_n)\\
W(E,\binbop{x}{\mathbf{a}_1,\ldots}{\mathbf{a}_n})&=&W(E,\mathbf{a}_1)+\ldots+W(E,\mathbf{a}_n)\\
W(E,[x\mydef\mathbf{a}]\mathbf{b})&==&W(E,\mathbf{a})+W((E,x:=\mathbf{a}),\mathbf{b})+1
\end{eqnarray*}
\end{proof}
\nomenclature[fBasic030]{$\dnf{E}(a)$}{definitional normal forms}
\index{normal form!definitional}
\begin{definition}[Definitional normal form]%
For any $\mathbf{a}$, let $\dnf{E}(\mathbf{a})\in\dexp$ denote the \emph{definitional normal form}, i.e.~ the expression resulting from $\mathbf{a}$ by maximal application of definition evaluation steps under environment $E$. This definition is sound since due to Law~\ref{mrd.sn.def} the evaluation of definitions always terminates and by Law~\ref{mrd.confl.def} the maximal evaluation of definitions delivers a unique result.
We abbreviate $\dnf{()}\mathbf{a}$ as $\dnfe(\mathbf{a})$.
\end{definition}
\begin{law}[Decomposition properties of $\dnf{E}(\mathbf{a})$]
\label{dnf.decomp}
For all $E,x,\mathbf{a}_1,\ldots \mathbf{a}_n$:
\begin{itemize}
\item[$i$:]
$\dnf{E}(\binop{\mathbf{a}_1}{\ldots,\mathbf{a}_n})=\binop{\dnf{E}(\mathbf{a}_1)}{\ldots,\dnf{E}(\mathbf{a}_n)}$
\item[$ii$:]
$\dnf{E}(\binbop{x}{\mathbf{a}_1}{\ldots,\mathbf{a}_n})=\binbop{x}{\dnf{E}(\mathbf{a}_1)}{\ldots,\dnf{E}(\mathbf{a}_n)}$
\end{itemize}
\end{law}
\begin{proof}
The proof of $i$ and $ii$ is by induction on the length of an arbitrary definition evaluation $E\sgv\binop{\mathbf{a}_1}{\ldots,\mathbf{a}_n}\dmrd\dnf{E}(\binop{\mathbf{a}_1}{\ldots,\mathbf{a}_n})$.
It is obvious since the outer operation is never affected by definition evaluation. 
\end{proof}
\begin{law}[Substitution properties of $\dnf{E}(\mathbf{a})$]%
\label{dnf.sub}
For all $E,x,\mathbf{a},\mathbf{b}$:
\begin{itemize}
\item[$i$:]
$\dnf{E}([x\mydef\mathbf{a}]\mathbf{b})=\dnf{E}(\mathbf{a}\gsub{x}{\mathbf{b}})$
\item[$ii$:]
$\dnf{E}(\mathbf{b}\gsub{x}{\mathbf{a}})=\dnf{E}(\mathbf{b})\gsub{x}{\dnf{E}(\mathbf{a})}$
\item[$iii$:]
$\dnf{E,x\smydef\mathbf{a}}(\mathbf{b})=\dnf{E}(\mathbf{b})\gsub{x}{\dnf{E}(\mathbf{a})}$
\end{itemize}
\end{law}
\begin{proof}
$\;$
\begin{itemize}
\item[$i$:]
As stated in the proof of Law~\ref{rd.mrd} we have $E\sgv[x\mydef\mathbf{a}]\mathbf{b}\mrd\mathbf{b}\gsub{x}{\mathbf{a}}$.
The proposition follows from confluence of $\dmrd$ (Law~\ref{mrd.confl.def}) since $E\sgv[x\mydef\mathbf{a}]\mathbf{b}\dmrd\cdots\dmrd\dnf{E}([x\mydef\mathbf{a}]\mathbf{b})$
and $E\sgv\mathbf{b}\gsub{x}{\mathbf{a}}\dmrd\cdots\dmrd\dnf{E}(\mathbf{b}\gsub{x}{\mathbf{a}})$.
\item[$ii$:]
Induction on the definition of substitution using the Law~\ref{dnf.decomp}.
\item[$iii$:]
Proof by induction on $\mathbf{b}$.
\begin{itemize}
\item
$\mathbf{b}=y$. 
\begin{itemize}
\item 
If $x=y$ then $\dnf{E}(\mathbf{b})=\dnf{E}(y)=\dnf{E}(x)=x$ and therefore
\begin{eqnarray*}
\dnf{E,x\smydef\mathbf{a}}(\mathbf{b})&=&\dnf{E,x\smydef\mathbf{a}}(x)\\
&=&\dnf{E,x\smydef\mathbf{a}}(\mathbf{a})\\
&=&\dnf{E}(\mathbf{a})\\
&=&x\gsub{x}{\dnf{E}(\mathbf{a})}\\
&=&\dnf{E}(\mathbf{b})\gsub{x}{\dnf{E}(\mathbf{a})}
\end{eqnarray*}
\item
If $x\neq y$ then 
\begin{eqnarray*}
\dnf{E,x\smydef\mathbf{a}}(\mathbf{b})&=&\dnf{E,x\smydef\mathbf{a}}(y)\\
&=&\dnf{E}(y)\\
&=&\dnf{E}(y)\gsub{x}{\dnf{E}(\mathbf{a})}\\
&=&\dnf{E}(\mathbf{b})\gsub{x}{\dnf{E}(\mathbf{a})}
\end{eqnarray*}
\end{itemize}
\item
$\mathbf{b}=\binop{\mathbf{a}_1}{\ldots,\mathbf{a}_n}$ or $\mathbf{b}=\binbop{x}{\mathbf{a}_1}{\ldots,\mathbf{a}_n}$.
In these cases the property follows from the inductive hypothesis and the Law~\ref{dnf.decomp}.
\item
$\mathbf{b}=[y\mydef\mathbf{c}]\mathbf{d}$. In this case, one can argue as follows:
\begin{eqnarray*}
&&\dnf{E,x\smydef\mathbf{a}}([y\smydef\mathbf{c}]\mathbf{d})\\
&=&\quad\text{(by $i$ and $ii$)}\\
&&\dnf{E,x\smydef\mathbf{a}}(\mathbf{d})\gsub{y}{\dnf{E,x\smydef\mathbf{a}}(\mathbf{c})}\\
&=&\quad\text{(inductive hypothesis)}\\
&&\dnf{E}(\mathbf{d})\gsub{x}{\dnf{E}(\mathbf{a})}\gsub{y}{\dnf{E,x\smydef\mathbf{a}}(\mathbf{c})}\\
&=&\quad\text{(inductive hypothesis)}\\
&&\dnf{E}(\mathbf{d})\gsub{x}{\dnf{E}(\mathbf{a})}\gsub{y}{\dnf{E}(\mathbf{c})\gsub{x}{\mathbf{a}}}\\
&=&\quad\text{(Law~\ref{basic.sub}($ii$) since $y\notin\free(\dnf{E}(\mathbf{a}))$)}\\
&&\dnf{E}(\mathbf{d})\gsub{y}{\dnf{E}(\mathbf{c})}\gsub{x}{\dnf{E}(\mathbf{a})}\\
&=&\quad\text{(by $i$ and $ii$)}\\
&&\dnf{E}([y\smydef\mathbf{c}]\mathbf{d})\gsub{x}{\dnf{E}(\mathbf{a})}
\end{eqnarray*}
\end{itemize}
\end{itemize}
\end{proof}
\begin{law}[Embedding law for negation reduction]%
\label{nurd.rd}
For all $\mathbf{a},\mathbf{b},E$: $\mathbf{a}\nurn{n}\mathbf{b}$ implies $\dnf{E}(\mathbf{a})\rd\dnf{E}(\mathbf{b})$.
\end{law}
\begin{proof}
Intuitively, due to Law~\ref{dnf.decomp}, the effect of $\dnf{E}$ is to replace some variables by expressions, which does not affect the reduction $\dnf{E}(\mathbf{a})\rd\dnf{E}(\mathbf{b})$.
Furthermore, no axiom of negation reduction has a free variable as its left-hand side.
Since the axioms and inference rules of $\nur$-reduction are a subset of those of reduction, the property
follows by a straightforward induction on the definition of $\mathbf{a}\nurn{n}\mathbf{b}$.
\end{proof}
\begin{law}[Embedding law for reduction with explicit substitution]%
\label{mrd.rd}
For all $E,\mathbf{a},\mathbf{b}$: $E\sgv\mathbf{a}\mrd\mathbf{b}$ implies $\dnf{E}(\mathbf{a})\rd\dnf{E}(\mathbf{b})$.
As a consequence, for all $a,b$, $()\sgv a\mrdr b$ implies $a\rdr b$.
\end{law}
\begin{proof}
First we show by induction on single-step reduction with explicit substitution that $E\sgv\mathbf{a}\smrd\mathbf{b}$ implies $\dnf{E}(\mathbf{a})\rd\dnf{E}(\mathbf{b})$.
The following cases have to be considered:
\begin{itemize}
\item 
The axioms \emph{use} or \emph{rem} were used: Obviously $\dnf{E}(\mathbf{a})=\dnf{E}(\mathbf{b})$.
\item 
The axiom $\beta_1^{\mu}$ was used, i.e. $\mathbf{a}=([x:\mathbf{a}_1]\mathbf{a}_2\:\mathbf{a}_3)$ and $\mathbf{b}=[x\mydef\mathbf{a}_3]\mathbf{a}_2$ 
We can argue as follows:
\begin{eqnarray*}
&&\dnf{E}(([x\!:\!\mathbf{a}_1]\mathbf{a}_2\:\mathbf{a}_3))\\
&=&\quad\text{(Law~\ref{dnf.decomp}($i$))}\\
&&([x\!:\!\dnf{E}(\mathbf{a}_1)]\dnf{E}(\mathbf{a}_2)\:\dnf{E}(\mathbf{a}_3))\\
&\srd&\quad\text{(axiom $\beta_1$)}\\
&&\dnf{E}(\mathbf{a}_2)\gsub{x}{\dnf{E}(\mathbf{a}_3)}\\
&=&\quad\text{(Law~\ref{dnf.sub}($ii$))}\\
&&\dnf{E}(\mathbf{a}_2\gsub{x}{\mathbf{a}_3})
\end{eqnarray*}
\item 
The axiom $\beta_2^{\mu}$ was used, i.e. $\mathbf{a}=([x!\mathbf{a}_1]\mathbf{a}_2\:\mathbf{a}_3)$ and $\mathbf{b}=[x\mydef \mathbf{a}_3]\mathbf{a}_2$ .
This case can be shown similar to case $\beta_1^{\mu}$.
\item 
Any other axiom was used to reduce $E\sgv\mathbf{a}\smrd\mathbf{b}$: In these cases obviously $\dnf{E}(\mathbf{a})\srd\dnf{E}(\mathbf{b})$.
\item
The rule $\nu$ was used to reduce $E\sgv\mathbf{a}\smrd\mathbf{b}$, i.e.~$\mathbf{a}\nurp\mathbf{b}$. 
By Law~\ref{nurd.rd}, $\dnf{E}(\mathbf{a})\rd\dnf{E}(\mathbf{b})$.
\item
A structural rule $\mathit{(\oplus{\overbrace{(\_,\ldots,\_)}^{n}}_i)}$ was used, i.e.
$\mathbf{a}=\binop{\mathbf{a}_1,\ldots,\mathbf{a}_i}{\ldots,\mathbf{a}_n}$, $\mathbf{b}=\binop{\mathbf{a}_1,\ldots,\mathbf{b}_i}{\ldots,\mathbf{a}_n}$ where $E\sgv\mathbf{a}_i\srd\mathbf{b}_i$.
By inductive hypothesis we know that $\dnf{E}(\mathbf{a}_i)\srd\dnf{E}(\mathbf{b}_i)$. Furthermore, by Law~\ref{dnf.decomp}($i$), we know that $\dnf{E}(\mathbf{a})=\binop{\dnf{E}(\mathbf{a}_1),\ldots,\dnf{E}(\mathbf{a}_i)}{\ldots,\dnf{E}(\mathbf{a}_n)}$ and $\dnf{E}(\mathbf{b})=\binop{\dnf{E}(\mathbf{a}_1),\ldots,\dnf{E}(\mathbf{b}_i)}{\ldots,\dnf{E}(\mathbf{a}_n)}$. Hence we know that $\dnf{E}(\mathbf{a})\rd\dnf{E}(\mathbf{b})$.
\item
A structural rule $\mathit{(\oplus_x{\overbrace{(\_,\ldots,\_)}^{n}}_i)}$ was used. This case can be shown like the previous one using Law~\ref{dnf.decomp}($ii$).
\item
The rule $\mathit{(L_{\smydef})}$ was used, i.e. $\mathbf{a}=[x\mydef\mathbf{a}_1]\mathbf{a}_2$ and $\mathbf{b}=[x\mydef\mathbf{b}_1]\mathbf{a}_2$, where $E\gv\mathbf{a}_1\smrd\mathbf{b}_1$.
By inductive hypothesis $\dnf{E}(\mathbf{a}_1)\rd\dnf{E}(\mathbf{b}_1)$. 
Hence, we can argue as follows:
\begin{eqnarray*}
&&\dnf{E}([x\mydef\mathbf{a}_1]\mathbf{a}_2)\\
&=&\quad\text{(Law~\ref{dnf.sub}($i,ii$))}\\
&&\dnf{E}(\mathbf{a}_2)\gsub{x}{\dnf{E}(\mathbf{a}_1)}\\
&\rd&\quad\text{(inductive hypothesis and Law~\ref{rd.sub}($ii$))}\\
&&\dnf{E}(\mathbf{a}_2)\gsub{x}{\dnf{E}(\mathbf{b}_1)}\\
&=&\quad\text{(Law~\ref{dnf.sub}($i,ii$))}\\
&&\dnf{E}([x\mydef\mathbf{b}_1]\mathbf{a}_2)
\end{eqnarray*}
\item
The rule $\mathit{(R_{\smydef})}$ was used, i.e. $\mathbf{a}=[x\mydef\mathbf{a}_1]\mathbf{a}_2$ and $\mathbf{b}=[x\mydef\mathbf{a}_1]\mathbf{b}_2$, where $E,x\mydef\mathbf{a}_1\gv\mathbf{a}_2\smrd\mathbf{b}_2$.
By inductive hypothesis $\dnf{E,x\mydef\mathbf{a}_1}(\mathbf{a}_2)\rd\dnf{E,x\mydef\mathbf{a}_1}(\mathbf{a}_2)$. 
Hence, we can argue as follows:
\begin{eqnarray*}
&&\dnf{E}([x\mydef\mathbf{a}_1]\mathbf{a}_2)\\
&=&\quad\text{(Law~\ref{dnf.sub}($iii$))}\\
&&\dnf{E,x\mydef\mathbf{a}_1}(\mathbf{a}_2)\\
&\rd&\quad\text{(inductive hypothesis)}\\
&&\dnf{E,x\mydef\mathbf{a}_1}(\mathbf{b}_2)\\
&=&\quad\text{(Law~\ref{dnf.sub}($iii$))}\\
&&\dnf{E,x\mydef\mathbf{a}_1}(\mathbf{b}_2)
\end{eqnarray*}
\end{itemize}

Now we turn to the main proposition.
Obviously, we have $E\sgv\mathbf{a}\mrdn{n}\mathbf{b}$ for some $n$. 
The proof is by induction on $n$. For $n=0$ the property is trivial. For $n>0$ assume  $E\sgv\mathbf{a}\smrd\mathbf{c}\mrdn{n-1}\mathbf{b}$. By inductive hypothesis $\dnf{E}(\mathbf{c})\rd\dnf{E}(\mathbf{b})$. By the previous property we know that $\dnf{E}(\mathbf{a})\rd\dnf{E}(\mathbf{c})$ which implies the proposition.

For the immediate consequence, for any $a,b$, assume there is an expression $\mathbf{c}$ such that $()\sgv a\mrd\mathbf{c}$ and
$()\sgv b\mrd\mathbf{c}$. We have just shown that also $\dnfe(a)\rd\dnfe(c)$ and $\dnfe(b)\rd\dnfe(c)$. The property follows since obviously $a=\dnfe(a)$ and $b=\dnfe(b)$. Hence $a\rdr b$.
\end{proof}
\begin{law}[Confluence of $\srd$]%
\index{confluence!of single-step reduction}
\label{rd.confl}
For all $a,b,c$:
$a\rd b$ and $a\rd c$ implies $b\rdr c$.
\end{law}
\begin{proof}
By Law~\ref{rd.mrd} we know that $\sgv a\mrd b$ and $\sgv a\mrd c$. Due to confluence of $\smrd$ (Law~\ref{mrd.confl}) we know that $\sgv b\mrdr c$.
By Law~\ref{mrd.rd} we obtain $b\rdr c$.
\end{proof}
\begin{law}[Common reduct preserved by reduction]%
\label{rdr.rd}
For all $a,b,c,d$: $a\rdr b$, $a\rd c$, and $b\rd d$ implies $c\rdr d$.
\end{law}
\begin{proof}
Direct consequence of Law~\ref{rd.confl}.
\end{proof}
\noindent
We can now show the well-known Church-Rosser property which allows us to consider $a\eqv b$ and $a\rdr b$ as equivalent.
\begin{law}[Church-Rosser property]%
\index{Church-Rosser property}
\label{cr}
For all $a,b$: $a\eqv b$ implies $a\rdr b$
\end{law}
\begin{proof}
Induction on the definition of congruence as the symmetric and transitive closure of reduction. 
In the base case and the inductive symmetry case the property is straightforward.
In the inductive transitive case we have transitivity of $a\rdr b$ due to Law~\ref{rd.confl}.
\end{proof}
\noindent
As an immediate consequence of confluence we show some basic properties of congruence.
\begin{law}[Substitution and congruence]%
\label{eqv.sub}
For all $x,a,b,c$: 
\begin{itemize}
\item[$i$:]$a\eqv b$ implies $a\gsub{x}{c}\eqv b\gsub{x}{c}$
\item[$ii$:]$a\eqv b$ implies $c\gsub{x}{a}\eqv c\gsub{x}{b}$
\end{itemize}
\end{law}
\begin{proof}
$\;$
\begin{itemize}
\item[$i$:]
By Law~\ref{cr} we know that $a\rd d$ and $b\rd d$ for some $d$.
By Law~\ref{rd.sub}($i$) we know that $a\gsub{x}{c}\rd d\gsub{x}{c}$ and $b\gsub{x}{c}\rd d\gsub{x}{c}$.
This implies $a\gsub{x}{c}\eqv b\gsub{x}{c}$.
\item[$ii$:]
Similar to part $i$.
\qedhere
\end{itemize}
\end{proof}
\begin{law}[Basic properties of congruence]%
\label{congr.basic}
For all $x,a,b,c,d$: 
\begin{itemize}
\item[$i$:]
$\binbop{x}{a_1}{\ldots,a_n}\eqv\binbop{y}{b_1}{\ldots,b_n}$ iff $x=y$ and $a_i\eqv b_i$ for $i=1,\ldots,n$.
\item[$ii$:]$\prsumop{a}{b}\eqv\prsumop{c}{d}$
iff $a\eqv c$ and $b\eqv d$.
\end{itemize}
\end{law}
\begin{proof}
$\;$
\begin{itemize}
\item[$i$:]Follows from Laws~\ref{cr} and~\ref{rd.decomp}($ii$).
\item[$ii$:]Follows from Laws~\ref{cr} and~\ref{rd.decomp}($i$).
\qedhere
\end{itemize}
\end{proof}
\section{Basic properties of typing}%
\label{typing.basics}
We frequently show properties about typing relations by induction on the definition of typing.
\begin{definition}[Induction on the definition of typing]
\index{induction!on definition of typing}
Properties which are proven for all $\Gamma$, $a$, and $b$ such that $\Gamma\sgv a:b$ (denoted here by $P(\Gamma,a,b)$) can be shown by shown by \emph{induction on the definition of typing} if the inductive base corresponds to the typing axiom and the inductive step corresponds to the inference rules of typing.
\end{definition} 
\noindent
We begin by a generalization of the weakening rule for typing.
\begin{law}[Context weakening]
\label{type.weak}
For all $\Gamma_1,\Gamma_2,x,a,b,c$: 
$(\Gamma_1,\Gamma_2)\sgv a:b$ and $\Gamma_1\sgv c$ imply $(\Gamma_1,x:c,\Gamma_2)\sgv a:b$.
\end{law}
\begin{proof}
Proof by induction on the definition of $(\Gamma_1,\Gamma_2)\sgv a:b$.
\begin{largeitemize}
\item[\ax:]
This case is not possible.
\item[\mystart:]
If $\Gamma_2=()$ then we have $\Gamma_1=(\Gamma_1',y:b)$ and $\Gamma_1',y:b\gv y:b$, for some $\Gamma_1'$, where $y=a$ and $\Gamma_1'\sgv b$.
Since $\Gamma_1\sgv c$, by rule \weak~we obtain $(\Gamma_1,x:c,\Gamma_2)\sgv a:b$.
This can be graphically illustrated as follows:
\[
\inferrule*[left=\normalfont{\weak}]
{{\Gamma_1',y:b\sgv y:b}\quad{\Gamma_1',y:b\sgv c}}
{\Gamma_1',y:b,x:c\gv y:b} 
\]
If $\Gamma_2=(\Gamma_2',y:d)$  for some $\Gamma_2'$, $y$, and $d$, then we have
$\Gamma_1,\Gamma_2',y:d\sgv y:d$ where $y=a$, $d=b$, and $(\Gamma_1,\Gamma_2')\sgv d$. 
Since $\Gamma_1\sgv c$, by inductive hypothesis $(\Gamma_1,x:c,\Gamma_2')\sgv d$.
Hence by rule \weak~$(\Gamma_1,x:c,\Gamma_2',y:d)\sgv y:d$ which is the same as $(\Gamma_1,x:c,\Gamma_2)\sgv a:b$. 
This can be graphically illustrated as follows:
\[
\inferrule*[left=\normalfont{\weak}]
  {\Gamma_1,\Gamma_2',y:d\sgv y:d\quad
	\inferrule*[left=I.H.]{\Gamma_1,\Gamma_2'\sgv d\quad \Gamma_1\sgv c}{\Gamma_1,x:c,\Gamma_2'\sgv d}}
{\Gamma_1,x:c,\Gamma_2',y:d\gv y:d}
\]
\item[\weak:]
If $\Gamma_2=()$ then we have $\Gamma_1\sgv a:b$. Since $\Gamma_1\sgv c$, by rule \weak~also $(\Gamma_1,x:c)\sgv a:b$ which is the same as $(\Gamma_1,x:c,\Gamma_2)\sgv a:b$.
This can be graphically illustrated as follows: 
\[
\inferrule*[left=\normalfont{\weak}]{\Gamma_1\sgv a:b\quad\Gamma_1\sgv c}{\Gamma_1,x:c\gv a:b}
\]
If $\Gamma_2=(\Gamma_2',y:d)$ for some $\Gamma_2'$, $y$, and $d$ then we have $(\Gamma_1,\Gamma_2')\sgv a:b$ and $(\Gamma_1,\Gamma_2')\sgv d$.
Since $\Gamma_1\sgv c$, by inductive hypothesis $(\Gamma_1,x:c,\Gamma_2')\sgv a:b$ as well as $(\Gamma_1,x:c,\Gamma_2')\sgv d$.
Hence by rule \weak~$(\Gamma_1,x:c,\Gamma_2',y:d)\sgv a:b$ which is the same as $(\Gamma_1,x:c,\Gamma_2)\sgv a:b$. 
This can be graphically illustrated as follows:
\[
\inferrule*[left=\normalfont{\weak}]{
  \inferrule*[left=I.H.]{\Gamma_1,\Gamma_2'\sgv a:b\quad\Gamma_1\sgv c}{\Gamma_1,x:c,\Gamma_2'\gv a:b}
	  \quad
	\inferrule*[left=I.H.]{\Gamma_1,\Gamma_2'\sgv d\quad\Gamma_1\sgv c}{\Gamma_1,x:c,\Gamma_2'\gv d}}
{\Gamma_1,x:c,\Gamma_2',y:d\gv a:b}
\]
\item[\conv:]
We have $(\Gamma_1,\Gamma_2)\sgv a:b$ where $(\Gamma_1,\Gamma_2)\sgv b$, $(\Gamma_1,\Gamma_2)\sgv a:b'$ for some $b'$ with $b'\eqv b$.
Since $\Gamma_1\sgv c$, by inductive hypothesis we know that $(\Gamma_1,x:c,\Gamma_2)\sgv b$ and $(\Gamma_1,x:c,\Gamma_2)\sgv a:b'$.
Hence by rule \conv~$(\Gamma_1,x:c,\Gamma_2)\sgv a:b$.
This can be graphically illustrated as follows:
\[
\inferrule*[left=\normalfont{\conv}]{
	\inferrule*[left=I.H.]{\Gamma_1,\Gamma_2\sgv a:b'\quad\Gamma_1\sgv c}{\Gamma_1,x:c,\Gamma_2\gv a:b'}
	\quad b'\eqv b\quad
  \inferrule*[left=I.H.]{\Gamma_1,\Gamma_2\sgv b\quad\Gamma_1\sgv c}{\Gamma_1,x:c,\Gamma_2\gv b}  
}
{\Gamma_1,x:c,\Gamma_2\gv a:b}
\]
\item[\absu:]
We have $a=[y:a_1]a_2$ and $b=[y:a_1]b_2$ for some $a_1$, $a_2$, and $b_2$, where $(\Gamma_1,\Gamma_2,y:a_1)\sgv a_2:b_2$.
Since $\Gamma_1\sgv c$, by inductive hypothesis we know that $(\Gamma_1,x:c,\Gamma_2,y:a_1)\sgv a_2:b_2$.
Hence by rule \absu~$(\Gamma_1,x:c,\Gamma_2)\sgv a:b$.
This can be graphically illustrated as follows:
\[
\inferrule*[left=\normalfont{\absu}]{
	\inferrule*[left=I.H.]{\Gamma_1,\Gamma_2,y:a_1\sgv a_2:b_2\quad\Gamma_1\sgv c}{\Gamma_1,x:c,\Gamma_2,y:a_1\gv a_2:b_2}}
{\Gamma_1,x:c,\Gamma_2\gv[y:a_1]a_2:[y:a_1]b_2}
\]
\item[\abse,\pdef:]
Analogous to case \absu.
\item[$\ldots$]
All remaining cases run analogous to case \conv.
\qedhere
\end{largeitemize}
\end{proof}
\noindent
In a similar style we show a generalization property about extraction of a typing from a context. 
\begin{law}[Context extraction]
\label{type.xtrct}
For all $\Gamma_1,\Gamma_2,a,b$: 
$(\Gamma_1,x:a,\Gamma_2)\sgv b$ implies $\Gamma_1\sgv a$
\end{law}
\begin{proof}
$(\Gamma_1,x:a,\Gamma_2)\sgv b$ means that there is an expression $c$ where $(\Gamma_1,x:a,\Gamma_2)\sgv b:c$.
The property that $(\Gamma_1,x:a,\Gamma_2)\sgv b:c$ implies $\Gamma_1\sgv a$ will be shown by induction on the definition of $(\Gamma_1,x:a,\Gamma_2)\sgv b:c$:
\begin{meditemize}
\item[\ax:]
This case is clearly not applicable. 
\item[\mystart:]
We either have $\Gamma_2=()$ in which case $b=x$ and $c=a$ and obviously $\Gamma_1\sgv a:c$ or we have $\Gamma_2=(\Gamma_2',y:c)$ for some $\Gamma_2'$ and $y$ where $y=b$ and $(\Gamma_1,x:a,\Gamma_2')\sgv c:d$ for some $d$. By inductive hypothesis $\Gamma_1\sgv a$.
\item[\weak:]
We either have $\Gamma_2=()$ in which case $\Gamma_1\sgv a$ or we have $\Gamma_2=(\Gamma_2',y:d)$ for some $\Gamma_2'$, $y$, and $d$ where $(\Gamma_1,x:a,\Gamma_2')\sgv d$. By inductive hypothesis $\Gamma_1\sgv a$.
\item[\absu:]
For the rule \absu~we have $b=[y:b_1]b_2$, $c=[y:b_1]c_2$ for some $b_1$, $b_2$, and $c_2$ where $(\Gamma_1,x:a,\Gamma_2,y:b_1)\sgv b_2:c_2$.
By inductive hypothesis $(\Gamma_1,x:a,\Gamma_2,y:b_1)\sgv c_2:d_2$ for some $d_2$.
Hence by rule \absu~we have $(\Gamma_1,x:a,\Gamma_2)\sgv c:[y:b_1]d_2$ which means $(\Gamma_1,x:a,\Gamma_2)\sgv c$.
\item[\abse:]
This case can be shown in a similar way.
\item[$\ldots$]
All other rules have at least on precondition not altering the context of the conclusion.
The property follows directly from the inductive hypothesis applied to this precondition as in the previous case.
\qedhere
\end{meditemize}
\end{proof}
\noindent
Next, we note a generalization of the \mystart~rule of typing.
\begin{law}[Start property]
\index{uniqueness!of types}
\label{start.confl}
For all $\Gamma,x,a,b$: 
$\Gamma,x:a\sgv x:b$ implies $a\eqv b$.
\end{law}
\begin{proof}
The proof is by induction on $\Gamma,x:a\sgv x:b$.
Obviously only the laws \mystart~and \conv~are applicable. 
In case of \mystart, we directly obtain $a=b$.
In case of \conv, we know that $\Gamma,x:a\sgv x:c$ for some $c$ where $c\eqv b$.
By inductive hypothesis $a\eqv c$ hence obviously $a\eqv b$.
\end{proof}
\noindent
Next we show several decomposition properties of typing.
\begin{law}[Typing decomposition]
\label{type.decomp}
For all $\Gamma,x,a_1,a_2,a_3,b$: 
\begin{itemize}
\item[$i$:]
$\Gamma\sgv\binbop{x}{a_1}{a_2}:b$ implies $b\eqv[x:a_1]c$ for some $c$ where $(\Gamma,x:a_1)\sgv a_2:c$.
\item[$ii$:]
$\Gamma\sgv\binbop{x}{a_1}{a_2}:[x:a_1]b$ implies $b\eqv c$ for some $c$ where $(\Gamma,x:a_1)\sgv a_2:c$ (easy consequence of $i$ and Law~\ref{cr}).
\item[$iii$:]
$\Gamma\sgv\injl{a_1}{a_2}:b$ implies $b\eqv [c+a_2]$ for some $c$ where $\Gamma\sgv a_1:c$ and $\Gamma\sgv a_2$.
$\Gamma\sgv\injr{a_1}{a_2}:b$ implies $b\eqv[a_1+c]$ for some $c$ where $\Gamma\sgv a_2:c$ and $\Gamma\sgv a_1$.
\item[$iv$:]
$\Gamma\sgv\prsumop{a_1}{a_2}:b$ implies $b\eqv[c_1,c_2]$ for some $c_1$, $c_2$ where $\Gamma\sgv a_1:c_1$ and $\Gamma\sgv a_2:c_2$.
\item[$v$:]
$\Gamma\sgv\myneg a_1:b$ implies $b\eqv c$ for some $c$ where $\Gamma\sgv a_1:c$. 
\item[$vi$:]
$\sgv\prim:b$ implies $b\eqv\prim$. 
\item[$vii$:]
$\Gamma\sgv (a_1\,a_2):b$ implies $b\eqv c_2\gsub{x}{a_2}$ where $\Gamma\sgv a_1:[x:c_1]c_2$ and $\Gamma\sgv a_2:c_1$ for some $c_1$, $c_2$. 
\item[$viii$:]
$\Gamma\sgv\pleft{a_1}:b$ implies $b\eqv c_1$ where $\Gamma\sgv a_1:[x!c_1]c_2$ or $\Gamma\sgv a_1:[c_1,c_2]$ for some $c_1$, $c_2$. 
\item[$ix$:]
$\Gamma\sgv\pright{a_1}:b$ implies, for some $c_1$, $c_2$, that either $b\eqv c_2$ where $\Gamma\sgv a_1:[c_1,c_2]$ or $b\eqv c_2\gsub{x}{\pleft{a_1}}$ where $\Gamma\sgv a_1:[x!c_1]c_2$. 
\item[$x$:]
$\Gamma\sgv\prdef{x}{a_1}{a_2}{a_3}:b$ implies $b\eqv[x!c]a_3$ for some $c$ where $\Gamma\sgv a_1:c$, $\Gamma\sgv a_2:a_3\gsub{x}{a_1}$, and $(\Gamma,x:c)\sgv a_3$. 
\item[$xi$:]
$\Gamma\sgv\case{a_1}{a_2}:b$ implies $b\eqv[x:[c_1+c_2]]c$ for some $c_1$, $c_2$, and $c$ where $\Gamma\sgv a_1:[x:c_1]c$, $\Gamma\sgv a_2:[x:c_2]c$ and $\Gamma\sgv c$. 
\end{itemize}
\end{law}
\begin{proof}
$\;$
\begin{itemize}
\item[$i$:]
Proof by induction on the definition of $\Gamma\sgv\binbop{x}{a_1}{a_2}:b$.
Only the rules \weak, \conv, \absu, and \abse~are relevant as all other rules in their conclusion type to an expression that cannot be equivalent to an abstraction. 
\begin{itemize}
\item
For the rule \weak~we have $\Gamma=(\Gamma',y:d)$, for some $\Gamma'$, $y$, and $d$ where $\Gamma'\sgv\binbop{x}{a_1}{a_2}:b$ as well as $\Gamma'\sgv d$.
By inductive hypothesis $(\Gamma',x:a_1)\sgv a_2:c$ for some $c$ where $b\eqv[x:a_1]c$.
By Law~\ref{type.weak} we can infer that $(\Gamma',y:d,x:a_1)\sgv a_2:c$ which can be rewritten as $(\Gamma,x:a_1)\sgv a_2:c$.
This can be graphically illustrated as follows:
\[
\inferrule*[left=\ref{type.weak}]
{	\inferrule*[left=I.H.]
  {\Gamma'\sgv\binbop{x}{a_1}{a_2}:b}
  {\Gamma',x:a_1\sgv a_2:c\;\text{where}\;b\eqv[x:a_1]c}
	\quad\Gamma'\sgv d
}
{\Gamma',y:d,x:a_1\gv a_2:c}
\]
\item
For the rule \conv~we have $\Gamma\sgv\binbop{x}{a_1}{a_2}:d$ for some $d$ where $d\eqv b$ and $\Gamma\sgv d$.
By inductive hypothesis $(\Gamma,x:a_1)\sgv a_2:c$ for some $c$ where $d\eqv[x:a_1]c$.
By symmetry and transitivity of $\eqv$ we obtain $c\eqv[x:a_1]c$. 
\item
The typing rules for abstractions \absu~and \abse~directly imply $b=[x:a_1]c$ for some $c$ where $(\Gamma,x:a_1)\sgv a_2:c$.
\end{itemize}
\item[$ii$:]
By part $i$ we obtain $[x:a_1]b\eqv[x:a_1]c$ for some $c$.
By Law~\ref{cr} this implies $[x:a_1]b\rdr[x:a_1]c$.
By Law~\ref{rd.decomp}($ii$) this implies $b\rdr c$ which implies the proposition.
\item[$iii$:]
Similar to part $i$, replacing the uses of \absu~or \abse~by \injll~or \injlr.
\item[$iv$:]
Similar to part $i$, replacing the uses of \absu~or \abse~by \bprod~or \bsum.
\item[$v$:]
Similar to part $i$, replacing the uses of \absu~or \abse~by \negate.
\item[$vi$:]
Proof by induction on the definition of typing.
Only the rule \conv~and the axiom \ax~are relevant as all other rules use a non-empty context in their conclusion or type to an expression that cannot be equivalent to the primitive constant. 
\begin{itemize}
\item
The case of the axiom \ax~is trivial.
\item
For the rule \conv~we have $\sgv\prim:a$ for some $a$ where $a\eqv b$ and $\sgv a$.
By inductive hypothesis $a\eqv\prim$ hence also $b\eqv\prim$.
\end{itemize}
\item[$vii$:]
Similar to part $i$, replacing the uses of \absu~or \abse~by \appl.
\item[$viii$:]
Similar to part $i$, replacing the uses of \absu~or \abse~by \prl~or \chin.
\item[$ix$:]
Similar to part $i$, replacing the uses of \absu~or \abse~by \prr~or \chba.
In case of \prr, one obtains $a_2\eqv c_2$ where $\Gamma\sgv a:[c_1,c_2]$ for some $c_1$, $c_2$. 
In case of \chba, one obtains $a_2\eqv c_2\gsub{x}{\pleft{a_1}}$ where $\Gamma\sgv a:[x!c_1]c_2$ for some $c_1$, $c_2$. 
\item[$x$:]
Similar to part $i$, replacing the uses of \absu~or \abse~by \pdef.
\item[$xi$:]
Similar to part $i$, replacing the uses of \absu~or \abse~by \cased.
\qedhere
\end{itemize}
\end{proof}
\noindent
We conclude this section with some basic decomposition properties of validity.
\begin{law}[Validity decomposition]
\label{val.decomp}
For all $\Gamma,x,a_1,\ldots,a_n$: 
\begin{itemize}
\item[$i$:]
$\Gamma\sgv\binbop{x}{a_1}{a_2}$ implies $\Gamma\sgv a_1$ and $(\Gamma,x:a_1)\sgv a_2$
\item[$ii$:]
$\Gamma\sgv\binop{a_1}{,\ldots a_n}$ implies $\Gamma\sgv a_1$, $\ldots$, $\Gamma\sgv a_n$. 
\item[$iii$:]
$\Gamma\sgv\prdef{x}{a_1}{a_2}{a_3}$ implies $\Gamma\sgv a_2$ and $\Gamma\sgv a_1:b$ for some $b$ where $(\Gamma,x:b)\sgv a_3$.
\end{itemize}
\end{law}
\begin{proof}
$\;$
\begin{itemize}
\item[$i$:]
Proof by induction on the definition of $\Gamma\sgv\binbop{x}{a_1}{a_2}:b$.
Only the rules \weak, \conv, and \absu~or \abse~are relevant as all other rules in their conclusion type to an expression that cannot be an abstraction. 
For the rules \weak~and \conv~the proposition follows from the inductive hypothesis.
The typing rules for abstractions \absu~or \abse~directly imply $(\Gamma,x:a)\sgv b$.
$\Gamma\sgv a$ then follows from Law~\ref{type.xtrct}.
\item[$ii$:]
For any operation $\binop{a_1}{,\ldots a_n}$ only the rules \emph{weak}, \emph{conv} and the corresponding unique rule introducing $\binop{a_1}{,\ldots a_n}$ are relevant.
For the rules \weak~and \conv~the proposition follows from the inductive hypothesis.
For the rule introducing $\binop{a_1}{,\ldots a_n}$ the property is directly provided by one of the antecedents.
\item[$iii$:]
Similar argument as for part $i$, using the rule \pdef~instead of \absu~or \abse.
\qedhere
\end{itemize}
\end{proof}
\section{Substitution properties of typing}%
\label{typing.sub}
A central prerequisite to the proof of closure of reduction and typing against validity is the following substitution property of typing.
In order to state the property, we need the following auxiliary definition.
\begin{definition}[Context substitution]%
\nomenclature[eCalc12]{$\Gamma\gsub{x}{b}$}{substitution in context}
\index{substitution!free variables in context}
The substitution function (Definition~\ref{substitute}) is extended to contexts $\Gamma\gsub{x}{a}$, where $a$ is an expression, as follows:
\begin{eqnarray*}
()\gsub{x}{a}&=&()\\
(y:b,\Gamma)\gsub{x}{a}&=&
\begin{cases}
(y:b\gsub{x}{a},\Gamma)&\text{if}\;x=y\\
(y:b\gsub{x}{a},\Gamma\gsub{x}{a})&\text{otherwise}
\end{cases}
\end{eqnarray*}
\end{definition}
\begin{law}[Substitution and typing]%
\label{type.sub}
Assume that $\Gamma_a=(\Gamma_1,x:a,\Gamma_2)$ and $\Gamma_b=(\Gamma_1,\Gamma_2\gsub{x}{b})$ for some $\Gamma_1,\Gamma_2,x,a,b$ where $\Gamma_1\sgv b:a$.
For all $c,d$: If $\Gamma_a\sgv c:d$ then $\Gamma_b\sgv c\gsub{x}{b}:d\gsub{x}{b}$. 
\end{law}
\begin{proof}
Let $\Gamma_a$ and $\Gamma_b$ as defined above and assume $\Gamma_1\sgv b:a$.
The proof that $\Gamma_a\sgv c:d$ implies $\Gamma_b\sgv c\gsub{x}{b}:d\gsub{x}{b}$ is by induction on the definition of $\Gamma_a\sgv c:d$.
We take a look at each typing rule:
\begin{meditemize}
\item[\ax:]
Trivial since $\Gamma_a=()$ is impossible.
\item[\mystart:] 
We have $c=y$ and $\Gamma_a=\Gamma_a',y:d$ for some $\Gamma_a'$, $y$, and $d$ where $(\Gamma_a',y:d)=(\Gamma_1,x:a,\Gamma_2)$.
There are two cases: 
\begin{itemize}
\item
$\Gamma_2=()$ and therefore $\Gamma_1=\Gamma_a'$, $y=x$ and $d=a$:
We can therefore argue as follows:
\begin{eqnarray*}
\Gamma_b=\Gamma_a'\gv
           &&  c\gsub{x}{b}\\
           &=& x\gsub{x}{b}\\
					  &=&b\\
						&:&a\qquad\qquad\text{($\Gamma_1\sgv b:a$ is assumed)}\\
						&=&a\gsub{x}{b}\\
						&=&d\gsub{x}{b}
\end{eqnarray*}
\item
$\Gamma_2=(\Gamma_2',y:d)$: Hence $x\neq y$ and $\Gamma_2\gsub{x}{b}(y)=d\gsub{x}{b}$. 
We can therefore argue as follows:
\begin{eqnarray*}
\Gamma_b=(\Gamma_1,\Gamma_2\gsub{x}{b})\gv
           &&  c\gsub{x}{b}\\
           &=& y\gsub{x}{b}\\
					  &=&y\\
						&:&\Gamma_2\gsub{x}{b}(y)\\
						&=&d\gsub{x}{b}
\end{eqnarray*}
\end{itemize}
\item[\weak:]
We have $\Gamma_a=(\Gamma_a',y:e)$ for some $\Gamma_a'$, $y$, and $e$ where $\Gamma_a'\sgv c:d$ and $\Gamma_a'\sgv e:f$ for some $f$.
There are two cases:
\begin{itemize}
\item
$\Gamma_2=()$ and therefore $x=y$: Hence $\Gamma_b=\Gamma_a'=\Gamma_1$ and $x\notin\free(c)\cup\free(d)$ which means that $c=c\gsub{x}{b}$ and $d=d\gsub{x}{b}$. 
Hence $\Gamma_b\sgv c\gsub{x}{b}:d\gsub{x}{b}$.
\item
$\Gamma_2=(\Gamma_2',y:e)$ for some $\Gamma_2'$.
Hence $x\neq y$, $\Gamma_a'=(\Gamma_1,\Gamma_2')$, and $\Gamma_b=(\Gamma_1,\Gamma_2'\gsub{x}{b},y:e\gsub{x}{b})$.
By inductive hypothesis we know that $(\Gamma_1,\Gamma_2'\gsub{x}{b})\sgv c\gsub{x}{b}:d\gsub{x}{b}$ and $(\Gamma_1,\Gamma_2'\gsub{x}{b})\sgv e\gsub{x}{b}:f\gsub{x}{b}$.
Hence by rule \weak~we know that $(\Gamma_1,\Gamma_2'\gsub{x}{b},y:e\gsub{x}{b})\sgv c\gsub{x}{b}:d\gsub{x}{b}$ which means that $\Gamma_b\sgv c\gsub{x}{b}:d\gsub{x}{b}$. 
This can be graphically illustrated as follows:
\[
\inferrule*[left=\normalfont{\weak}]{
  \inferrule*[left=I.H.,right=arg1]{\Gamma_1,\Gamma_2'\sgv c:d}{\Gamma_1,\Gamma_2'\gsub{x}{b}\gv c\gsub{x}{b}:d\gsub{x}{b}}
	  \\\\
	\inferrule*[left=I.H.,right=arg2]{\Gamma_1,\Gamma_2'\sgv e:f}{\Gamma_1,\Gamma_2'\gsub{x}{b}\gv e\gsub{x}{b}:f\gsub{x}{b}}}
{\Gamma_1,\Gamma_2'\gsub{x}{b},y:e\gsub{x}{b}\gv c\gsub{x}{b}:d\gsub{x}{b}}
\]
\end{itemize} 
\item[\conv:]
We have $\Gamma_a\sgv c:d'$ for some $d'$ where $d'\eqv d$ and $\Gamma_a\sgv d':e$ for some $e$.
By inductive hypothesis  $\Gamma_b\sgv c\gsub{x}{b}:d'\gsub{x}{b}$ and $\Gamma_b\sgv d'\gsub{x}{b}:e\gsub{x}{b}$.
By Law~\ref{eqv.sub}($i$) we can infer that $d'\gsub{x}{b}\eqv d\gsub{x}{b}$.
Therefore we can apply the rule \conv~to obtain $\Gamma_b\sgv c\gsub{x}{b}:d\gsub{x}{b}$.
This can be graphically illustrated as follows:
\[
\inferrule*[left=\normalfont{\conv}]{
		\inferrule*[left=\ref{eqv.sub}($i$)]{d'\eqv d}{d'\gsub{x}{b}\eqv d\gsub{x}{b}}\\\\
  \inferrule*[left=I.H.]{\Gamma_a\sgv c:d'}{\Gamma_b\sgv c\gsub{x}{b}:d'\gsub{x}{b}}\quad
	\inferrule*[left=I.H.]{\Gamma_a\sgv d':e}{\Gamma_b\sgv d'\gsub{x}{b}:e\gsub{x}{b}}}
{\Gamma_b\gv c\gsub{x}{b}:d\gsub{x}{b}}
\]
\item[\absu:]
We have $c=[y:c_1]c_2$ and $d=[y:c_1]d_2$ for some $y$, $c_1$, $c_2$, and $d_2$ where $(\Gamma_a,y:c_1)\sgv c_2:d_2$. 
By inductive hypothesis it follows that $(\Gamma_b,y:c_1\gsub{x}{b})\sgv c_2\gsub{x}{b}:d_2\gsub{x}{b}$.
Hence by \absu~we obtain $\Gamma_b\sgv [y:c_1\gsub{x}{b}]c_2\gsub{x}{b}:[y:c_1\gsub{x}{b}]d_2\gsub{x}{b}$.
By definition of substitution it follows that $\Gamma_b\sgv c\gsub{x}{b}:d\gsub{x}{b}$. 
\item[\abse:]
Similar to case \absu.
\item[\appl:] 
We have $c=(c_1\:c_2)$, $d=d_2\gsub{y}{c_2}$ for some $y$, $c_1$, $c_2$, and $d_2$ where $\Gamma_a\sgv (c_1\,c_2):d$, $\Gamma_a\sgv c_1:[y:d_1]d_2$, and $\Gamma_a\sgv c_2:d_1$. 
Obviously, we can assume that $x\neq y$.
We need to show that $\Gamma_b\sgv (c_1\,c_2)\gsub{x}{b}:d_2\gsub{y}{c_2}\gsub{x}{b}$.
We have:
\begin{eqnarray*}
\Gamma_b\gv &&c_1\gsub{x}{b}\\
&:&\quad\text{(inductive hypothesis)}\\
&&([y:d_1]d_2)\gsub{x}{b}\\
&=&\quad\text{(definition of substitution)}\\
&&[y:d_1\gsub{x}{b}](d_2\gsub{x}{b})
\end{eqnarray*}
Furthermore, since $\Gamma_a\sgv c_2:d_1$, by inductive hypothesis we know that $\Gamma_b\sgv c_2\gsub{x}{b}:d_1\gsub{x}{b}$.
Hence by typing rule \appl~we obtain
\[\Gamma_b\sgv c_1\gsub{x}{b}(c_2\gsub{x}{b}):(d_2\gsub{x}{b})\gsub{y}{c_2\gsub{x}{b}}\qquad (\ast)
\] 
We can now argue as follows:
\begin{eqnarray*}
\Gamma_b\gv &&(c_1\,c_2)\gsub{x}{b}\\
&=&\quad\text{(definition of substitution)}\\
&&(c_1\gsub{x}{b}\:c_2\gsub{x}{b})\\
&:&\quad\text{(using $\ast$)}\\
&&(d_2\gsub{x}{b})\gsub{y}{c_2\gsub{x}{b}}\\
&=&\quad\text{(Law~\ref{basic.sub}($ii$))}\\
&&d_2\gsub{y}{c_2}\gsub{x}{b}
\end{eqnarray*}
\item[\pdef:] 
We have $c=\prdef{y}{c_1}{c_2}{c_3}$ and  $d=[y!d_1]c_3$ for some $y$, $c_1$, $c_2$, $c_3$, and $d_1$ where $\Gamma_a\sgv c_1:d_1$, $\Gamma_a\sgv c_2:c_3\gsub{y}{c_1}$, and $(\Gamma_a,y:d_1)\sgv c_3:d_3$ for some $d_3$.
We need to show that $\Gamma_b\sgv c\gsub{x}{b}:d\gsub{x}{b}$. In order to apply rule \pdef, we need to show its premises:
\begin{itemize}
\item
By inductive hypothesis we have $\Gamma_b\sgv c_1\gsub{x}{b}:d_1\gsub{x}{b}$.
\item
We have
\begin{eqnarray*}
\Gamma_b\gv &&c_2\gsub{x}{b}\\
&:&\quad\text{(inductive hypothesis)}\\
&&c_3\gsub{y}{c_1}\gsub{x}{b}\\
&=&\quad\text{(by Law~\ref{basic.sub}($ii$), since $y\notin\free(b)$)}\\
&&c_3\gsub{x}{b}\gsub{y}{c_1\gsub{x}{b}}
\end{eqnarray*}
\item
By inductive hypothesis $(\Gamma_b,y:d_1\gsub{x}{b})\sgv c_3\gsub{x}{b}:d_3\gsub{x}{b}$.
\end{itemize}
Hence we may apply rule \pdef~to obtain 
\[\Gamma_b\sgv\prdef{y}{c_1\gsub{x}{b}}{c_2\gsub{x}{b}}{c_3\gsub{x}{b}}:[y!d_1\gsub{x}{b}](c_3\gsub{x}{b})\]
which by def.~of substitution is equivalent to $\Gamma_b\sgv c\gsub{x}{b}:d\gsub{x}{b}$.
\item[\chin:]
We have $c=\pleft{c_1}$ and $\Gamma_a\sgv c_1:[y!d]d_2$ for some $y$, $c_1$, and $d_2$.
By inductive hypothesis and definition of substitution we obtain  $\Gamma_b\sgv c_1\gsub{x}{b}:[y!d\gsub{x}{b}](d_2\gsub{x}{b})$.
By law \chin~we can infer that $\Gamma_b\sgv(\pleft{c_1})\gsub{x}{b}:d\gsub{x}{b}$.
\item[\chba:]
We have $c=\pright{c_1}$ and $d=d_2\gsub{y}{\pleft{c_1}}$ for some $y$, $c_1$, and $d_2$ where $\Gamma_a\sgv c_1:[y!d_1]d_2$ for some $d_1$.
Obviously we may assume $x\neq y$.
By inductive hypothesis and definition of substitution we obtain
\[\Gamma_b\sgv c_1\gsub{x}{b}:[y!d_1\gsub{x}{b}](d_2\gsub{x}{b})\qquad(\ast)
\]
We can argue as follows:
\begin{eqnarray*}
\Gamma_b\gv &&(\pright{c_1})\gsub{x}{b}\\
&=&\quad\text{(definition of substitution)}\\
&&\pright{(c_1\gsub{x}{b})}\\
&:&\quad\text{(typing rule \chba~using $\ast$)}\\
&&d_2\gsub{x}{b}\gsub{y}{\pleft{(c_1\gsub{x}{b})}}\\
&=&\quad\text{(definition of substitution)}\\
&&d_2\gsub{x}{b}\gsub{y}{\pleft{c_1}\gsub{x}{b})}\\
&=&\quad\text{(Law~\ref{basic.sub}($ii$))}\\
&&d_2\gsub{y}{\pleft{c_1}}\gsub{x}{b}
\end{eqnarray*}
\item[\bprod:]
We have $c=[c_1,c_2]$ and $d=[d_1,d_2]$ for some $c_1$, $c_2$, $d_1$, and $d_2$ where $\Gamma_a\sgv c_1:d_1$ and $\Gamma_a\sgv c_2:d_2$.
By inductive hypothesis $\Gamma_b\sgv c_1\gsub{x}{b}:d_1\gsub{x}{b}$ and $\Gamma_b\sgv c_2\gsub{x}{b}:d_2\gsub{x}{b}$. 
Hence by definition of substitution and by typing rule \bprod~we know that $\Gamma_b\sgv c\gsub{x}{b}:d\gsub{x}{b}$.
\item[\bsum:]
Similar to case \bprod.
\item[\prl:]
We have $c=\pleft{c_1}$ where $\Gamma_a\sgv c_1:[d,d_1]$ for some $c_1$ and $d_1$.
By inductive hypothesis and definition of substitution $\Gamma_b\sgv c_1\gsub{x}{b}:[d\gsub{x}{b},d_1\gsub{x}{b}]$. 
Hence by definition of substitution and by typing rule \prl~we know that $\Gamma_b\sgv(\pleft{c_1})\gsub{x}{b}:d\gsub{x}{b}$. 
\item[\prr:]
Similar to case \prl. 
\item[\injll:]
We have $c=\injl{c_1}{d_2}$ and $d=[d_1+d_2]$ for some $c_1$, $d_1$,  and $d_2$ where $\Gamma_a\sgv c_1:d_1$ and $\Gamma_a\sgv d_2:e_2$ for some $e_2$.
By inductive hypothesis $\Gamma_b\sgv c_1\gsub{x}{b}:d_1\gsub{x}{b}$ and $\Gamma_b\sgv d_2\gsub{x}{b}:e_2\gsub{x}{b}$. 
Hence by definition of substitution and by typing rule \injll~we know that $\Gamma_b\sgv c\gsub{x}{b}:d\gsub{x}{b}$.
\item[\injlr:]
Similar to case \injll.
\item[\cased:] 
We have $c=\case{c_1}{c_2}$ and $d=[y:[d_1+d_2]]e$ for some $c_1$, $c_2$, $y$, $d_1$, $d_2$, and $e$ where $\Gamma_a\sgv c_1:[y:d_1]e$, $\Gamma_a\sgv c_2:[y:d_2]e$, and $\Gamma_a\sgv e:f$ for some $f$.

By inductive hypothesis and definition of substitution we obtain $\Gamma_b\sgv c_1\gsub{x}{b}:[y:d_1\gsub{x}{b}]e\gsub{x}{b}$. 
Similarly, $\Gamma_b\sgv c_2\gsub{x}{b}:[y:d_2\gsub{x}{b}]e\gsub{x}{b}$, and $\Gamma_a\sgv d\gsub{x}{b}$.
This implies $\Gamma_b\sgv c\gsub{x}{b}:d\gsub{x}{b}$.
\item[\negate:]
We have $c=\myneg c_1$ where $\Gamma_a\sgv c_1:d$ for some $c_1$.
By inductive hypothesis we know that $\Gamma_b\sgv c_1\gsub{x}{b}:d\gsub{x}{b}$.
Hence by definition of typing and substitution $\Gamma_b\sgv c\gsub{x}{b}:d\gsub{x}{b}$.
\qedhere
\end{meditemize}
\end{proof}
\begin{law}[Substitution and validity]
\label{val.sub}
Assume that $\Gamma_a=(\Gamma_1,x:a,\Gamma_2)$ and $\Gamma_b=(\Gamma_1,(\Gamma_2\gsub{x}{b}))$ for some $\Gamma_1,\Gamma_2,x,a,b$ where $\Gamma_1\sgv b:a$.
For all $c$: $\Gamma_a\sgv c$ implies $\Gamma_b\sgv c\gsub{x}{b}$.
\end{law}
\begin{proof} 
Assume that $\Gamma_a\sgv c:d$ for some $d$. Hence by Law~\ref{type.sub} we know that $\Gamma_b\sgv c\gsub{x}{b}:e$ for some $e$.
Hence $\Gamma_b\sgv c\gsub{x}{b}$.
\end{proof}
\section{Closure properties of validity}%
\label{closure}
It is relatively straightforward to see that validity is closed against typing
\begin{law}[Valid expressions have valid types]%
\label{valid.type}
For all $\Gamma,a,b$:
If $\Gamma\sgv a:b$  then $\Gamma\sgv b$.
\end{law}
\begin{proof}
By induction on the definition of $\Gamma\sgv a:b$ we show that $\Gamma\sgv a:b$ implies $\Gamma\sgv b$.
We take a look at each typing rule:
\begin{meditemize}
\item[\ax:] 
Trivial.
\item[\mystart:] 
We have $x=a$ and $\Gamma=(\Gamma',a:b)$ for some $\Gamma'$ where $\Gamma'\sgv b:c$ for some $c$.
Hence we can apply rule \weak~to obtain $\Gamma\sgv b:c$ hence $\Gamma\sgv b$.
\item[\weak:] 
We have $\Gamma=(\Gamma',x:c)$ for some $\Gamma'$, $x$, and $c$ where $\Gamma'\sgv a:b$ and $\Gamma'\sgv c:d$ for some $d$.
By inductive hypothesis $\Gamma'\sgv b$.
Hence we can apply rule \weak~and the definition of validity to obtain $\Gamma=(\Gamma',x:c)\sgv b$.
\item[\conv:]
Obvious since $\Gamma\sgv b$ follows from the rules premise.
\item[\absu:] 
We have $a=[x:a_1]a_2$ and $b=[x:a_1]b_2$ for some $x$, $a_1$, $a_2$, and $b_2$ where $\Gamma,x:a_1\sgv a_2:b_2$. 
By inductive hypothesis we know that $(\Gamma,x:a_1)\sgv b_2$ which implies $\Gamma\sgv b$ by definition of validity and typing.
\item[\abse:]
Similar to case \absu.
\item[\appl:] 
We have $a=(a_1\,a_2)$ and $b=b_2\gsub{x}{a_2}$ for some $x$, $a_1$, $a_2$, and $b_2$ where $\Gamma\sgv a_1:[x:b_1]b_2$ for some $b_1$ and $\Gamma\sgv a_2:b_1$.
From the inductive hypothesis we know that $\Gamma\sgv[x:b_1]b_2$.
By Law~\ref{val.decomp}($i$) this implies $\Gamma,x:b_1\sgv b_2$. 
From $\Gamma,x:b_1\sgv b_2$, by validity substitution (Law~\ref{val.sub}, note that $\Gamma\sgv a_2:b_1$) we can infer that $\Gamma\sgv b_2\gsub{x}{a_2}=b$.
\item[\pdef:]
We have $a=\prdef{x}{a_1}{a_2}{a_3}$ and $b=[x!b_1]a_3$ for some $x$, $a_1$, $a_2$, $a_3$, and $b_1$ where $\Gamma\sgv a_1:b_1$ and $(\Gamma,x:b_1)\sgv a_3:a_4$ for some $a_4$.
By inductive hypothesis $\Gamma\sgv b_1$.
Hence by typing law \abse~and the definition of validity we obtain $\Gamma\sgv b=[x!b_1]a_3:[x:b_1]a_3$ which is equivalent to $\Gamma\sgv b$.
\item[\chin:]
We have $a=\pleft{a_1}$ where $\Gamma\sgv a_1:[x!b]b_1$ for some $x$, $a_1$, and $b_1$.
By inductive hypothesis we know that $\Gamma\sgv[x!b]b_1$.
By Law~\ref{val.decomp}($i$) this implies $\Gamma\sgv b$.
\item[\chba:]
We have $a=\pright{a_1}$ and $b=b_2\gsub{x}{\pleft{a_1}}$ for some $x$, $a_1$, and $b_2$ where $\Gamma\sgv a_1:[x!b_1]b_2$ for some $b_1$.
By inductive hypothesis we know that $\Gamma\sgv[x!b_1]b_2$.
By Law~\ref{val.decomp}($i$) this implies $\Gamma,x:b_1\sgv b_2$.
By definition of typing we have $\Gamma\sgv\pleft{a_1}:b_1$.
From $\Gamma,x:b_1\sgv b_2$, by validity substitution (Law~\ref{val.sub}, note that $\Gamma\sgv\pleft{a_1}:b_1$) we know that $\Gamma\sgv b=b_2\gsub{x}{\pleft{a_1}}$.
\item[\bprod:]
We have $a=[a_1,a_2]$ and $b=[b_1,b_2]$ for some $a_1$, $a_2$, $b_1$, and $b_2$ where $\Gamma\sgv a_1:b_1$ and $\Gamma\sgv a_2:b_2$.
By inductive hypothesis we know that $\Gamma\sgv b_1$ and $\Gamma\sgv b_2$. Hence $\Gamma\sgv b$.
\item[\bsum:]
Similar to case \bprod.
\item[\prl:]
We have $a=\pleft{a_1}$ where $\Gamma\sgv a_1:[b,b_2]$ for some $a_1$ and $b_2$.
By inductive hypothesis we know that $\Gamma\sgv[b,b_2]$.  
By Law~\ref{val.decomp}($ii$) this implies $\Gamma\sgv b$.
\item[\prr:]
Similar to case \prl. 
\item[\injll:]
Similar to case \bprod.
\item[\injlr:]
Similar to case \bprod.
\item[\cased:] 
We have $a=\case{a_1}{a_2}$ and $b=[x:[b_1+b_2]]b_3$  for some $a_1$, $a_2$, $b_1$, $b_2$ and $b_3$  where $\Gamma\sgv a_1:[x:b_1]b_3$, $\Gamma\sgv a_2:[x:b_1]b_3$, and $\Gamma\sgv b_3$.
By inductive hypothesis $\Gamma\sgv [x:b_1]b_3$ and $\Gamma\sgv [x:b_2]b_3$ hence by Law~\ref{val.decomp}($i$) we know that 
$\Gamma\sgv b_1]$ and $\Gamma\sgv b_2$.
By definition of typing and validity it follows that $\Gamma\sgv b=[x:[b_1+b_2]]b_3$
%
%
\item[\negate:]
This case follows directly from the inductive hypothesis and the definition of typing.
\qedhere
\end{meditemize}
\end{proof}
\noindent
Next we show an auxiliary result about the closure of typing against valid equivalent substitutions in the typing context.
\begin{law}[Context equivalence and typing]
\label{eqv.env}
Let $\Gamma_a=(\Gamma_1,x:a,\Gamma_2)$ and $\Gamma_b=(\Gamma_1,x:b,\Gamma_2)$ for some $\Gamma_1,\Gamma_2,x,a,b$ where $a\eqv b$ and $\Gamma_1\sgv b$:
For all $c,d$: If $\Gamma_a\sgv c:d$ then $\Gamma_b\sgv c:d$.
\end{law}
\begin{proof}
Let $\Gamma_a$ and $\Gamma_b$ where $a\eqv b$ and $\Gamma_1\sgv b$ as defined above :
The property is shown by induction on the definition of $(\Gamma_1,x:a,\Gamma_2)\sgv c:d$.
We take a look at each typing rule:
\begin{meditemize}
\item[\ax:] 
Trivial since $\Gamma_a=()$ is impossible.
\item[\mystart:] 
We have $\Gamma_a=(\Gamma_1,x:a,\Gamma_2)=(\Gamma_3,y:d)$ and $c=y$ for some $y$ and $\Gamma_3$ where $\Gamma_3\sgv d$.
There are two cases: 
\begin{itemize}
\item If $x=y$ then $\Gamma_a\sgv x:a$, i.e.~$d=a$, $\Gamma_2=()$, $\Gamma_1=\Gamma_3$, $\Gamma_a=(\Gamma_1,x:a)$, $\Gamma_b=(\Gamma_1,x:b)$,  and $\Gamma_1\sgv a$.
Since $\Gamma_1\sgv b$, by rule \weak~it follows that $\Gamma_b\sgv a$.
By rule \mystart~we obtain $\Gamma_b\sgv x:b$.
Hence by rule \conv~it follows that $\Gamma_b\sgv x:a$.
This can be graphically illustrated as follows:
\[
\inferrule*[left=\normalfont{\conv}]{
		\inferrule*[left=\normalfont{\mystart}]{\Gamma_1\sgv b}{\Gamma_1,x:b\gv x:b}
		\quad b\eqv a\quad
	\inferrule*[left=\normalfont{\weak}]{\Gamma_1\sgv a\quad\Gamma_1\sgv b}{\Gamma_1,x:b\sgv a}}
{\Gamma_b=(\Gamma_1,x:b)\gv x:a}
\]
\item
If $x\neq y$ then we have $\Gamma_2=(\Gamma_4,y:d)$ for some $\Gamma_4$ where $\Gamma_3=(\Gamma_1,x:a,\Gamma_4)$. 
Since $\Gamma_1,x:a,\Gamma_4\sgv d$, by inductive hypothesis we know that also $\Gamma_1,x:b,\Gamma_4\sgv d$ and therefore, by rule \mystart, we obtain $\Gamma_b=(\Gamma_1,x:b,\Gamma_2)\sgv y:d$.
\end{itemize}
\item[\weak:]
We have $\Gamma_a=(\Gamma_1,x:a,\Gamma_2)=(\Gamma_3,y:e)$ for some $y$ and $\Gamma_3$ where $\Gamma_3\sgv c:d$ and $\Gamma_3\sgv e$.
There are two cases:
\begin{itemize}
\item
If $x=y$ then $\Gamma_2=()$, $e=a$, and $\Gamma_1=\Gamma_3$.
From $\Gamma_1\sgv c:d$ and $\Gamma_1\sgv b$ by rule \weak~it follows that $\Gamma_b=(\Gamma_1,x:b)\sgv c:d$.
\item
If $x\neq y$ then $\Gamma_2=(\Gamma_4,y:e)$ for some $\Gamma_4$.
Hence $\Gamma_3=(\Gamma_1,x:a,\Gamma_4)$.
By inductive hypothesis $(\Gamma_1,x:b,\Gamma_4)\sgv c:d$.
Similarly by inductive hypothesis $(\Gamma_1,x:b,\Gamma_4)\sgv e$.
Hence, by rule \weak~it follows that $\Gamma_b=(\Gamma_1,x:b,\Gamma_2)=(\Gamma_1,x:b,\Gamma_d,y:e)\sgv c:d$.
\end{itemize} 
\item[\conv:] 
We have $\Gamma_a\sgv c:d_1$ for some $d_1$ where $d_1\eqv d$ and $\Gamma_a\sgv d$.
By inductive hypothesis $\Gamma_b\sgv c:d_1$ and $\Gamma_b\sgv d$.
Hence, since $d_1\eqv d$ by rule \conv~we obtain $\Gamma_b\sgv c:d$.
\item[\absu:] 
We have $c=[y:c_1]c_2$ and $d=[y:c_1]d_2$ for some $y$, $c_1$, $c_2$, and $d_2$ where $(\Gamma_a,y:c_1)\sgv c_2:d_2$. 
Obviously, we may assume $x\neq y$.
By inductive hypothesis we obtain $(\Gamma_b,y:c_1)\sgv c_2:d_2$.
By typing rule \absu~we obtain $\Gamma_b\sgv c=[y:c_1]c_2:[y:c_1]d_2=d$. 
\item[\abse:]
Similar to case \absu.
\item[\appl:]
We have $c=(c_1\,c_2)$ and $d=d_2\gsub{y}{c_2}$ for some $y$, $c_1$, $c_2$, and $d_2$ for some $x$ and $d_2$ where $\Gamma_a\sgv c_1:[y:d_1]d_2$ and $\Gamma_a\sgv c_2:d_1$ for some $d_1$. 
By inductive hypothesis $\Gamma_b\sgv c_1:[y:d_1]d_2$ as well as $\Gamma_b\sgv c_2:d_1$. 
By typing rule \appl~we obtain $\Gamma_b\sgv c:d$.
\item[\pdef:] 
We have $c=\prdef{y}{c_1}{c_2}{c_3}$ and $d=[y!d_1]c_3$ for some $y$, $c_1$, $c_2$, $c_3$, and $d_1$ where $\Gamma_a\sgv c_1:d_1$, $\Gamma_a\sgv c_2:c_3\gsub{y}{c_1}$, and $(\Gamma_a,y:d_1)\sgv c_3$.
Obviously, we may assume $x\neq y$.
By inductive hypothesis $\Gamma_b\sgv c_1:d_1$,  $\Gamma_b\sgv c_2:c_3\gsub{y}{c_1}$, and  $(\Gamma_b,y:d_1)\sgv c_3$. 
By typing rule \pdef~we obtain $\Gamma_b\sgv c:[y!d_1]c_3$.
\item[\chin:]
We have $c=\pleft{c_1}$ where $\Gamma_a\sgv c_1:[y!d]d_1$ for some $y$, $c_1$, $d_1$, and $d$.
By inductive hypothesis $\Gamma_b\sgv c_1:[y!d]d_1$.
By typing rule \chin~we obtain $\Gamma_b\sgv c=\pleft{c_1}:d$.
\item[\chba:]
We have $c=\pright{c_1}$ and $d=d_2\gsub{y}{\pleft{c_1}}$ for some $y$, $c_1$, and $d_2$ where  $\Gamma_a\sgv c_1:[y!d_1]d_2$ for some $d_1$.
Obviously, we may assume $x\neq y$.
By inductive hypothesis $\Gamma_b\sgv c_1:[y!d_1]d_2$. 
By typing rule \chin~we obtain $\Gamma_b\sgv c=\pright{c_1}:d_2\gsub{y}{\pleft{c_1}}=d$.
\item[\bprod:]
We have $c=[c_1,c_2]$ and $d=[d_1,d_2]$ for some $c_1$, $c_2$, $d_1$, and $d_2$ where $\Gamma_a\sgv c_1:d_1$ and $\Gamma_a\sgv c_2:d_2$. 
By inductive hypothesis $\Gamma_b\sgv c_1:d_1$ and $\Gamma_b\sgv c_2:d_2$ which by typing rule \bprod~implies that $\Gamma_b\sgv c=[c_1,c_2]:[d_1,d_2]=d$.
\item[\bsum:]
Similar to case \bprod.
\item[\prl:]
We have $c=\pleft{c_1}$ where  $\Gamma_a\sgv c_1:[d,d_2]$ for some $c_1$ and $d_2$. 
By inductive hypothesis $\Gamma_b\sgv c_1:[d,d_2]$.
By typing rule \prl~we obtain $\Gamma_b\sgv c=\pleft{c}:d$.
\item[\prr:]
Similar to case \prl.
\item[\injll:]
Similar to case \bprod.
\item[\injlr:]
Similar to case \bprod.
\item[\cased:] 
We have $c=\case{c_1}{c_2}$ and $d=[y:[d_1+d_2]]e$ for some $c_1$, $c_2$, $y$, $d_1$, $d_2$, and $e$ where $\Gamma_a\sgv c_1:[y:d_1]e$, $\Gamma_a\sgv c_2:[y:d_2]e$, and $\Gamma_a\sgv e$.
By inductive hypothesis $\Gamma_b\sgv c_1:[y:d_1]e$, $\Gamma_b\sgv c_2:[y:d_2]e$, and $\Gamma_b\sgv e$.
By typing rule \cased~we obtain $\Gamma_b\sgv c=\case{c_1}{c_2}:d$. 
\item[\negate:]
We have $c=\myneg c_1$ and $\Gamma_a\sgv c_1:d$ for some $c_1$.
By inductive hypothesis $\Gamma_b\sgv c_1:d$. 
By typing rule \negate~we obtain $\Gamma_b\sgv c=\myneg c_1:d$.
\qedhere
\end{meditemize}
\end{proof}
\noindent
We can now show the preservation of types under a reduction step.
\begin{law}[Preservation of types under reduction step]
\label{strd.type}
For all $\Gamma,a,b,c$:
$\Gamma\sgv a:c$ and $a\srd b$ imply $\Gamma\sgv b:c$.
\end{law}
\begin{proof}
Proof by induction on the definition of $a\srd b$.
We take a look at each axiom and structural rule of reduction.
Note that by Law~\ref{valid.type} we have $\Gamma\sgv c$.
This means that in order to show $\Gamma\sgv b:c$, it is sufficient to show $\Gamma\sgv b:b_1$ for some $b_1$ where $b_1\eqv c$.
$\Gamma\sgv b:c$ then follows by applying rule \conv. 
\begin{hugeitemize}
\item[$\beta_1$:] 
We have $a=([x:a_1]a_2)(a_3)$ and $b=a_2\gsub{x}{a_3}$ and $\Gamma\sgv([x:a_1]a_2)(a_3):c$ where $\Gamma\sgv a_3:a_1$.
The following typings can be derived:
\begin{eqnarray*}
&&\Gamma\sgv([x:a_1]a_2\:a_3):c\\
&\Rightarrow&\quad\text{(Law~\ref{type.decomp}($vii$))}\\
&&\Gamma\sgv[x:a_1]a_2:[x:d]e\quad\text{where $c\eqv e\gsub{x}{a_3}$, $\Gamma\sgv a_3:d$}\\
&\Rightarrow&\quad\text{(Law~\ref{type.decomp}($ii$))}\\
&&(\Gamma,x:a_1)\sgv a_2:b_2\quad\text{where $[x:d]e\eqv[x:a_1]b_2$}\\
&\Rightarrow&\quad\text{(Law~\ref{type.sub}, since $\Gamma\sgv a_3:a_1$)}\\
&&\Gamma\sgv a_2\gsub{x}{a_3}:b_2\gsub{x}{a_3}
\end{eqnarray*}
The following equivalences can be derived.
From $[x:d]e\eqv[x:a_1]b_2$, by Law~\ref{congr.basic}($i$) it follows that $e\eqv b_2$.
We can now argue as follows:
\begin{eqnarray*}
&&b_2\gsub{x}{a_3}\\
&\eqv&\quad\text{(by Law~\ref{eqv.sub}($i$) since $e\eqv b_2$)}\\
&&e\gsub{x}{a_3}\\
&\eqv&\quad\text{(by Law~\ref{type.decomp}($vii$) since $\Gamma\sgv([x:a_1]a_2\:a_3):c$)}\\
&&c
\end{eqnarray*}
Therefore we can apply rule \conv~to derive $\Gamma\sgv b=a_2\gsub{x}{a_3}:c$. 
This can be graphically illustrated as follows:
\[
\inferrule*[left=\normalfont{\conv}]{\Gamma\sgv a_2\gsub{x}{a_3}:b_2\gsub{x}{a_3}\qquad b_2\gsub{x}{a_3}\eqv c\qquad\Gamma\sgv c}{\Gamma\gv a_2\gsub{x}{a_3}:c}
\]
\item[$\beta_2$:]
The proof of this case is similar to the case $\beta_1$. 
\item[$\beta_3$:]
In this case we have $a=(\case{a_1}{a_2}\,\injl{a_3}{a_4})$, $b=(a_1\,a_3)$, and $\Gamma\sgv(\case{a_1}{a_2}\,\injl{a_3}{a_4}):c$.
From the latter, by Law~\ref{type.decomp}($xi$) we know that $c\eqv d$ for some $d$ where $\Gamma\sgv d$ and $\Gamma\sgv a_1:[x:c_1]d$, $\Gamma\sgv a_2:[x:c_2]d$, and $\Gamma\sgv\injl{a_3}{a_4}:[c_1+a_4]$ for some $c_1$ and $c_2$.

From $\Gamma\sgv\injl{a_3}{a_4}:[c_1+a_4]$, by Law~\ref{type.decomp}($iii$) we know that $\Gamma\sgv a_3:c_1$.
Hence from $\Gamma\sgv a_1:[x:c_1]d$, by rule \appl~we obtain $\Gamma\sgv (a_1\,a_3):d\gsub{x}{a_3}$.
Since $x\notin\free(d)$ this is equivalent to $\Gamma\sgv (a_1\,a_3):d$.
Since $d\eqv c$ and $\Gamma\sgv c$, by rule \conv~we can derive $\Gamma\sgv b=(a_1\,a_3):c$. 
\item[$\beta_4$:]
The proof of this case is similar to the case $\beta_3$. 
\item[$\pi_1$:]
We have $a=\pleft{\prdef{x}{a_1}{a_2}{a_3}}$, $b=a_1$, and $\Gamma\sgv\pleft{\prdef{x}{a_1}{a_2}{a_3}}:c$.
From the latter, by Law~\ref{type.decomp}($viii$) we know that $c\eqv b_1$ for some $b_1$ where $\Gamma\sgv\prdef{x}{a_1}{a_2}{a_3}:[x!b_1]b_2$ for some $b_2$ 
(the case $\Gamma\sgv\prdef{x}{a_1}{a_2}{a_3}:[b_1,b_2]$ is obviously impossible).
By Law~\ref{type.decomp}($x$) this implies that $[x!b_1]b_2\eqv[x!d]a_3$ for some $d$ where $\Gamma\sgv a_1:d$.
By basic properties of congruence (Law~\ref{congr.basic}($i$)) we know that $b_1\eqv d$.

Hence, since $\Gamma\sgv a_1:d$, $d\eqv b_1\eqv c$, and $\Gamma\sgv c$ we can apply rule \conv~to obtain $\Gamma\sgv b=a_1:c$ 
\item[$\pi_2$:]
We have $a=\pright{\prdef{x}{a_1}{a_2}{a_3}}$, $b=a_2$, and $\Gamma\sgv\pright{\prdef{x}{a_1}{a_2}{a_3}}:c$.
From the latter, by Law~\ref{type.decomp}($ix$) we know that $c\eqv b_2\gsub{x}{a_2}$ for some $b_2$ where $\Gamma\sgv\prdef{x}{a_1}{a_2}{a_3}:[x!b_1]b_2$
for some $b_1$
(the case $\Gamma\sgv\prdef{x}{a_1}{a_2}{a_3}:[b_1,b_2]$ is obviously impossible).
By Law~\ref{type.decomp}($x$) this implies that $[x!b_1]b_2\eqv[x!d]a_3$ for some $d$ where $\Gamma\sgv a_2:a_3\gsub{x}{a_1}$.

By basic properties of congruence (Law~\ref{congr.basic}($i$)) we have $b_2\eqv a_3$.
Hence by Law~\ref{eqv.sub}($i$) we know that $b_2\gsub{x}{a_2}\eqv a_3\gsub{x}{a_2}$.

Hence, since $\Gamma\sgv a_2:a_3\gsub{x}{a_1}$, $a_3\gsub{x}{a_2}\eqv b_2\gsub{x}{a_2}\eqv c$, and $\Gamma\sgv c$ we can apply rule \conv~to obtain $\Gamma\sgv b=a_2:c$. 
\item[$\pi_3$:]
We know that $a=\pleft{[a_1,a_2]}$ and $b=a_1$ where $\Gamma\sgv\pleft{[a_1,a_2]}:c$.
By Law~\ref{type.decomp}($viii$) we know that $c\eqv b_1$ for some $b_1$ where $\Gamma\sgv[a_1,a_2]:[b_1,b_2]$ for some $b_2$
(the case $\Gamma\sgv[a_1,a_2]:[x!b_1]b_2$ is obviously impossible).
By Law~\ref{type.decomp}($iv$) this implies that $[b_1,b_2]\eqv[d_1,d_2]$ for some $d_1$, $d_2$ where $\Gamma\sgv a_1:d_1$ and $\Gamma\sgv a_2:d_2$.
By basic properties of congruence (Law~\ref{congr.basic}($ii$)) we know that $b_1\eqv d_1$.

Hence, since $\Gamma\sgv a_1:d_1$, $d_1\eqv b_1\eqv c$, and $\Gamma\sgv c$ we can apply rule \conv~to obtain $\Gamma\sgv a_1:c$ 
\item[$\pi_4,\pi_5,\pi_6$:]
Similar to case $\pi_3$.
\item[$\nu_1$:] 
We have $a=\myneg\myneg a_1$ and $b=a_1$ and $\Gamma\sgv\myneg\myneg a_1:c$. 
By Law~\ref{type.decomp}($v$) (applied twice) we know that $c\eqv c_1$ for some $c_1$ where $\Gamma\sgv a_1:c_1$. 
Hence we can apply rule \conv~to infer that $\Gamma\sgv b=a_1:c$.
\item[$\nu_2$:] 
We have $a=\myneg[a_1,a_2]$, $b=[\myneg a_1+\myneg a_2]$, and $\Gamma\sgv\myneg[a_1,a_2]:c$.
By Law~\ref{type.decomp}($v$,$iv$) we know that $c\eqv[c_1,c_2]$ where  $\Gamma\sgv a_1:c_1$ and $\Gamma\sgv a_2:c_2$.
Hence $\Gamma\sgv\myneg a_1:c_1$ and $\Gamma\sgv\myneg a_2:c_2$ and therefore $\Gamma\sgv[\myneg a_1+\myneg a_2]:[c_1,c_2]$.
Hence we can apply rule \conv~to infer that $\Gamma\sgv b=[\myneg a_1+\myneg a_2]:c$.
\item[$\nu_3$:] 
Similar (symmetric) to case $\nu_3$.
\item[$\nu_4$:] 
We have $a=\myneg[x:a_1]a_2$, $b=[x!a_1]\myneg a_2$, and $\Gamma\sgv\myneg[x:a_1]a_2:c$.
By Law~\ref{type.decomp}($v$) we know that  $c\eqv d$ for some $d$ where $\Gamma\sgv[x:a_1]a_2:d$.
By rule \conv~we obtain $\Gamma\sgv[x:a_1]a_2:c$. 
From this, by Law~\ref{type.decomp}($i$) we know that $\Gamma\sgv[x:a_1]a_2:[x:a_1]c_2$ for some $c_2$ where $c\eqv[x:a_1]c_2$.
Hence by Law~\ref{type.decomp}($ii$) we know that $(\Gamma,x:a_1)\sgv a_2:c_3$ for some $c_3$ where $c_2\eqv c_3$.
Hence $c\eqv[x:a_1]c_2\eqv[x:a_1]c_3$.
By definition of typing $(\Gamma,x:a_1)\sgv \myneg a_2:c_3$ and therefore also $\Gamma\sgv[x!a_1]\myneg a_2:[x:a_1]c_3$.

Hence, since $[x:a_1]c_3\eqv[x:a_1]c_2\eqv c$ and $\Gamma\sgv c$ we can apply rule \conv~to infer that $\Gamma\sgv b=[x!a_1]\myneg a_2:c$.
\item[$\nu_5$:] 
Similar (symmetric) to case $\nu_4$.
\item[$\nu_6$:] 
We have $a=\myneg\prim$, $b=\prim$, and $\Gamma\sgv\myneg\prim:c$.
By Law~\ref{type.decomp}($v$) we know that $c\eqv d$ where $\Gamma\sgv \prim:d$.
Hence we can apply rule \conv~to infer that $\Gamma\sgv b=\prim:c$.
\item[$\nu_7,\ldots,\nu_{10}$:] 
Similar to case $\nu_6$.
\item[$\binbop{x}{\_}{\_}_1$:]
We have $a=\binbop{x}{a_1}{a_3}$, $b=\binbop{x}{a_2}{a_3}$ where $a_1\srd a_2$, and $\Gamma\sgv\binbop{x}{a_1}{a_3}:c$.
By Law~\ref{type.decomp}($i$), $c\eqv[x:a_1]c_3$ for some $c_3$ where $(\Gamma,x:a_1)\sgv a_3:c_3$. 
By Law~\ref{type.xtrct} this implies that $\Gamma\sgv a_1$ and therefore by inductive hypothesis we know that $\Gamma\sgv a_2$.
Therefore we can apply Law~\ref{eqv.env} which implies that $(\Gamma,x:a_2)\sgv a_3:c_3$. 
By definition of typing $\Gamma\sgv \binbop{x}{a_2}{a_3}:[x:a_2]c_3$.
Since $[x:a_2]c_3\eqv c\eqv[x:a_1]c_3$, we can apply typing rule \conv~to derive $\Gamma\sgv b=\binbop{x}{a_2}{a_3}:c$.
\item[$\binbop{x}{\_}{\_}_2$:]
We have $a=\binbop{x}{a_1}{a_2}$, $b=\binbop{x}{a_1}{a_3}$ where $a_2\srd a_3$ and $\Gamma\sgv\binbop{x}{a_1}{a_2}:c$. 
By Law~\ref{type.decomp}($i$), $c\eqv[x:a_1]c_2$ for some $c_2$ where $(\Gamma,x:a_1)\sgv a_2:c_2$. 
By inductive hypothesis, this implies  $(\Gamma,x:a_1)\sgv a_3:c_2$ and hence by definition of typing we obtain $\Gamma\sgv \binbop{x}{a_1}{a_3}:[x:a_1]c_2$.
Since $[x:a_1]c_2\eqv c$, by typing rule \conv~it follows that $\Gamma\sgv b=\binbop{x}{a_1}{a_3}:c$.
\item[${\prdef{x}{\_}{\_}{\_}}_1$:]
We have $a=\prdef{x}{a_1}{a_3}{a_4}$, $b=\prdef{x}{a_2}{a_3}{a_4}$ where $a_1\srd a_2$, and $\Gamma\sgv\prdef{x}{a_2}{a_3}{a_4}:c$.
By Law~\ref{type.decomp}($x$), $c\eqv[x!c_1]a_4$ for some $c_1$ where $\Gamma\sgv a_1:c_1$, $\Gamma\sgv a_3:a_4\gsub{x}{a_1}$, and $(\Gamma,x:c_1)\sgv a_4$.
By inductive hypothesis $\Gamma\sgv a_2:c_1$.
From $a_1\srd a_2$, by Law~\ref{rd.sub}($ii$) it follows that $a_4\gsub{x}{a_1}\eqv a_4\gsub{x}{a_2}$.
From $(\Gamma,x:c_1)\sgv a_4$ and $\Gamma\sgv a_2:c_1$, by Law~\ref{type.sub} it follows that $(\Gamma,x:c_1)\sgv a_4\gsub{x}{a_2}$.

Hence from $\Gamma\sgv a_3:a_4\gsub{x}{a_1}$, $a_4\gsub{x}{a_1}\eqv a_4\gsub{x}{a_2}$, and $(\Gamma,x:c_1)\sgv a_4\gsub{x}{a_2}$, by rule \conv~it follows that
$\Gamma\sgv a_3:a_4\gsub{x}{a_2}$.
Therefore, by definition of typing $\Gamma\sgv\prdef{x}{a_2}{a_3}{a_4}:[x!c_1]a_4$.
Since $[x!c_1]a_4\eqv c$, by typing rule \conv~it follows that $\Gamma\sgv b=\prdef{x}{a_2}{a_3}{a_4}:c$.
\item[${\prdef{x}{\_}{\_}{\_}}_2$:]
We have $a=\prdef{x}{a_1}{a_2}{a_4}$, $b=\prdef{x}{a_1}{a_3}{a_4}$ where $a_2\srd a_3$, and $\Gamma\sgv\prdef{x}{a_1}{a_3}{a_4}:c$.
By Law~\ref{type.decomp}($x$), $c\eqv[x!c_1]a_4$ for some $c_1$ where $\Gamma\sgv a_1:c_1$, $\Gamma\sgv a_2:a_4\gsub{x}{a_1}$, and $(\Gamma,x:c_1)\sgv a_4$.
By inductive hypothesis $\Gamma\sgv a_3:a_4\gsub{x}{a_1}$ and hence by definition of typing $\Gamma\sgv\prdef{x}{a_1}{a_3}{a_4}:[x!c_1]a_4$.
Since $[x!c_1]a_4\eqv c$, by typing rule \conv~it follows that $\Gamma\sgv b=\prdef{x}{a_1}{a_3}{a_4}:c$.
\item[${\prdef{x}{\_}{\_}{\_}}_3$:]
We have $a=\prdef{x}{a_1}{a_2}{a_3}$, $b=\prdef{x}{a_1}{a_2}{a_4}$ where $a_3\srd a_4$, and $\Gamma\sgv\prdef{x}{a_1}{a_2}{a_3}:c$.
By Law~\ref{type.decomp}($x$), $c\eqv[x!c_1]a_3$ for some $c_1$ where $\Gamma\sgv a_1:c_1$, $\Gamma\sgv a_2:a_3\gsub{x}{a_1}$, and $(\Gamma,x:c_1)\sgv a_3$.
By inductive hypothesis $(\Gamma,x:c_1)\sgv a_4$.  

Since $\Gamma\sgv a_1:c_1$, by Law~\ref{type.sub} we know that $\Gamma\sgv a_4\gsub{x}{a_1}$.  
By Law~\ref{rd.sub}($i$) we have $a_3\gsub{x}{a_1}\rd a_4\gsub{x}{a_1}$ hence, from $\Gamma\sgv a_2:a_3\gsub{x}{a_1}$ we can apply typing rule \conv~to infer that $\Gamma\sgv a_2:a_4\gsub{x}{a_1}$.
Therefore, by definition of typing $\Gamma\sgv \prdef{x}{a_1}{a_2}{a_4}:[x!c_1]a_4$. 
Since $[x!c_1]a_4\eqv[x!c_1]a_3\eqv c$, by typing rule \conv~it follows that $\Gamma\sgv b=\prdef{x}{a_1}{a_2}{a_4}:c$.
\item[${(\_\,\_)}_1$:]
We have $a=(a_1\,a_3)$, $b=(a_2\,a_3)$ where $a_1\srd a_2$. 
By~\ref{type.decomp}($vii$), $c\eqv c_2\gsub{x}{a_3}$ for some $c_2$ where $\Gamma\sgv a:c_2\gsub{x}{a_3}$, $\Gamma\sgv a_1:[x:c_1]c_2$ for some $c_1$, and $\Gamma\sgv a_3:c_1$.
By inductive hypothesis $\Gamma\sgv a_2:[x:c_1]c_2$.
Hence by definition of typing $\Gamma\sgv (a_2\,a_3):c_2\gsub{x}{a_3}$.
Since $c_2\gsub{x}{a_3}\eqv c$, by typing rule \conv~it follows that $\Gamma\sgv b=(a_2\,a_3):c$.
\item[${(\_\,\_)}_2$:]
$a=(a_1\,a_2)$ and $b=(a_1\,a_3)$ where $a_2\srd a_3$.
By Law~\ref{type.decomp}($vii$), $c\eqv c_2\gsub{x}{a_2}$ for some $c_2$ where $\Gamma\sgv a:c_2\gsub{x}{a_2}$, $\Gamma\sgv a_1:[x:c_1]c_2$ for some $c_1$, and $\Gamma\sgv a_2:c_1$.
By inductive hypothesis $\Gamma\sgv a_3:c_1$.
Hence by definition of typing $\Gamma\sgv (a_1\,a_3):c_2\gsub{x}{a_3}$.

By Law~\ref{rd.sub}($ii$) we know that $c_2\gsub{x}{a_2}\eqv c_2\gsub{x}{a_3}$.
Therefore $c_2\gsub{x}{a_3}\eqv c_2\gsub{x}{a_2}\eqv c$ and by typing rule \conv~it follows that $\Gamma\sgv b=(a_1,a_3):c$.
\item[$(\pleft{\_})_1$:]
We have $a=\pleft{a_1}$ and $b=\pleft{a_2}$ where $a_1\srd a_2$.
By Law~\ref{type.decomp}($viii$), $c\eqv c_1$ for some $c_1$ where either $\Gamma\sgv a_1:[x!c_1]c_2$ or $\Gamma\sgv a_1:[c_1,c_2]$, for some $c_2$.

In the first case, by inductive hypothesis we know that $\Gamma\sgv a_2:[x!c_1]c_2$.
In the second case, by inductive hypothesis we know that $\Gamma\sgv a_2:[c_1,c_2]$.
In both cases, by definition of typing $\Gamma\sgv\pleft{a_2}:c_1$.
Since $c_1\eqv c$, by typing rule \conv~it follows that $\Gamma\sgv b=\pleft{a_2}:c$.
\item[$(\pright{\_})_1$:]
We have $a=\pright{a_1}$ and $b=\pright{a_2}$ where $a_1\srd a_2$.

By Law~\ref{type.decomp}($ix$), for some $c_1$, $c_2$ either $c\eqv c_2\gsub{x}{\pleft{a_1}}$ and $\Gamma\sgv a_1:[x!c_1]c_2$
or $c\eqv c_2$ and $\Gamma\sgv a_1:[c_1,c_2]$.

In the first case, by inductive hypothesis we know that $\Gamma\sgv a_2:[x!c_1]c_2$.
Hence, by definition of typing $\Gamma\sgv\pright{a_2}:c_2\gsub{x}{\pleft{a_2}}$.
By Law~\ref{rd.sub}($ii$) we know that $c_2\gsub{x}{\pleft{a_1}}\eqv c_2\gsub{x}{\pleft{a_2}}$.
Since $c_2\gsub{x}{\pleft{a_2}}\eqv c$, by typing rule \conv~it follows that $\Gamma\sgv b=\pright{a_2}:c$.

In the second case, by inductive hypothesis we know that $\Gamma\sgv a_2:[c_1,c_2]$.
By definition of typing $\Gamma\sgv\pright{a_2}:c_2$.
Since $c_2\eqv c$, by typing rule \conv~it follows that $\Gamma\sgv b=\pright{a_2}:c$.
\item[${[\_,\_]}_1$:]
We have $a=[a_1,a_3]$ and $b=[a_2,a_3]$ where $a_1\srd a_2$.
By Law~\ref{type.decomp}($iv$) we know that $c\eqv[c_1,c_3]$ for some $c_1$, $c_3$ where $\Gamma\sgv a_1:c_1$ and $\Gamma\sgv a_3:c_3$.
By inductive hypothesis we know that $\Gamma\sgv a_2:c_1$ hence by definition of typing $\Gamma\sgv :[c_1,c_3]$.
Since $[c_1,c_3]\eqv c$ by typing rule \conv~it follows that $\Gamma\sgv b=[a_2,a_3]:c$. 
\item[${[\_,\_]}_2$:]
Similar to case ${[\_,\_]}_1$.
\item[${[\_+\_]}_i$:]
$i=1,2$, similar to case ${[\_,\_]}_1$.
\item[${\injl{\_}{\_}}_1$:]
Similar to case ${[\_,\_]}_1$ using Law~\ref{type.decomp}($iii$) instead.
\item[${\injl{\_}{\_}}_2$:]
We have $a=\injl{a_1}{a_2}$ and $b=\injl{a_1}{a_3}$ where $a_2\srd a_3$.
By Law~\ref{type.decomp}($iii$) we know that $c\eqv[c_1+a_2]$ for some $c_1$ where $\Gamma\sgv a_1:c_1$ and $\Gamma\sgv a_2$.

By inductive hypothesis we know that $\Gamma\sgv a_3$.
Hence by definition of typing $\Gamma\sgv \injl{a_1}{a_3}:[c_1+a_3]$.
Since obviously $[c_1+a_3]\eqv[c_1+a_2]$ we have $[c_1+a_3]\eqv c$ and by typing rule \conv~it follows that $\Gamma\sgv b=\injl{a_1}{a_3}:c$.
\item[${\injr{\_}{\_}}_i$:]
Similar to case ${\injl{\_}{\_}}_i$, $i=1,2$.
\item[${\case{\_}{\_}}_1$:]
We have $a=\case{a_1}{a_3}$, $b=\case{a_2}{a_3}$ where $a_1\srd a_2$.
By Law~\ref{type.decomp} ($xi$) we know that $c\eqv[x:[c_1+c_3]]c_4$ for some $c_1$, $c_3$, $c_4$ where $\Gamma\sgv a_1:[x:c_1]c_4$, $\Gamma\sgv a_3:[x:c_3]c_4$, and $\Gamma\sgv c_4$.
By inductive hypothesis it follows that $\Gamma\sgv a_2:[x:c_1]c_4$.
Hence by definition of typing $\Gamma\sgv \case{a_2}{a_3}:[x:[c_1+c_3]]c_4$.
Since $[x:[c_1+c_3]]c_4\eqv c$ by typing rule \conv~it follows that $\Gamma\sgv b=\case{a_2}{a_3}:c$.
\item[${\case{\_}{\_}}_2$:]
Similar to case $(\case{\_}{\_}\_)_1$.
\item[$(\myneg\_)_1$:]
We have $a=\myneg a_1$ and $b=\myneg a_2$ where $a_1\srd a_2$.
By Law~\ref{type.decomp}($v$), we know that $c\eqv d$ for some $d$ where $\Gamma\sgv a_1:d$.
By inductive hypothesis we know that $\Gamma\sgv a_2:d$ hence by definition of typing $\Gamma\sgv \myneg a_2:d$
Since $d\eqv c$ by typing rule \conv~it follows that $\Gamma\sgv b=\myneg a_2:c$.
%
%
\qedhere
\end{hugeitemize}
\end{proof}
\noindent
A simple inductive argument extends this property to general reduction.
This property is often referred to as \emph{subject reduction}.
\begin{law}[Subject reduction: Types are preserved under reduction]%
\index{reduction!subject}
\label{rd.type}
For all $\Gamma,a,b,c$:
$a\rd b$ and $\Gamma\sgv a:c$ imply that $\Gamma\sgv b:c$. 
\end{law}
\begin{proof}
Proof by induction on the number of reduction steps in $a\rdn{n} b$.
\begin{itemize}
\item $n=0$: We have $a=b$, and hence the proposition trivially holds.
\item $n>0$: 
We have $a\srd a'\rdn{n-1} b$ for some $a'$. 
By Law~\ref{strd.type} (preservation of type under single reduction step) we know that $\Gamma\sgv a':c$. 
By inductive hypothesis we know that $\Gamma\sgv b:c$.
\qedhere
\end{itemize}
\end{proof}
\noindent
Subject reduction can be reformulated using the validity notation without types.
\begin{law}[Valid expressions are closed against reduction]%
\label{valid.rd}
For all $\Gamma,a,b$:
$\Gamma\sgv a$ and $a\rd b$ implies $\Gamma\sgv b$ 
\end{law}
\begin{proof}
Direct consequence of Law~\ref{rd.type}.
\end{proof}
\noindent
Subject reduction can be easily extended to types.
\begin{law}[Elements are preserved under type reduction]%
\label{type.rd}
For all $\Gamma,a,b,c$:
$\Gamma\sgv a:b$ and $b\rd c$ implies $\Gamma\sgv a:c$.
\end{law}
\begin{proof}
By Law~\ref{valid.type} $\Gamma\sgv b$.
By Law~\ref{valid.rd} $\Gamma\sgv c$.
The proposition then follows from rule \conv.
\end{proof}
%
%
\noindent
We conclude this section with a conjecture about type hierarchies.
\begin{remark}[Conjecture about bounded type hierarchy]%
Repeated application of Law~\ref{valid.type} allows for constructing an infinite sequence of typings. 
However, we believe that this typing sequence eventually terminates, in the sense that at some point the type is congruent to the expression. 
We do not present a detailed proof of this property, the idea is to define a \emph{type-height} of a valid expression, \eg\ with $\prim$ having type-height $0$ and
$[x:\prim]x$ having type-height $1$ and so on, and to show several properties related to type-height most importantly that typing decreases the type-height and typing of an expression with type-height $0$ leads to a congruent expression.
\end{remark}
\section{Uniqueness of types}%
Before we show uniqueness of types, we need a property about the effect of removing unnecessary variables from a context.  
First we need the following lemma.
\begin{law}[Basic context contraction]
\label{contract.basic}
For all $\Gamma_1,\Gamma_2,x,a,b,c$:
$(\Gamma_1,x:c,\Gamma_2)\sgv a:b$ where $a=\prim$ or $a=y$, for some $y$, and $x\notin\free([\Gamma_2]a)$ implies 
 $(\Gamma_1,\Gamma_2)\sgv a:b'$  and $b\rd b'$ for some $b'$.
\end{law}
Note that in this lemma $x\notin\free([\Gamma_2]a)$ does not necessarily imply $x\notin\free([\Gamma_2]b)$, for example take  $x:\prim\sgv\prim:([y:\prim]\prim\,x)$, and hence the reduction of the type to some expression where $x$ is not free, in the example $([y:\prim]\prim\,x)\rd \prim$, is necessary.
\begin{proof}
Induction on the definition of typing.
Since $a=\prim$ or $a=y$ the only applicable rules are \weak\ and \conv. 
The rule \mystart\ is not applicable due to the condition about free occurrences of variables.

In case of rule \weak\ we have $(\Gamma_1,x:c,\Gamma_2)=(\Gamma,z:d)$ for some $\Gamma$, $z$, and $d$ where
$\Gamma\sgv a:b$ and $\Gamma\sgv d$. There are two cases:
\begin{itemize}
\item $z=x$: 
Hence $a=c$ and $\Gamma =\Gamma_1$.
Hence $b'=b$ and $\Gamma\sgv a:b$.
\item $z\neq x$ and hence $z\in\dom(\Gamma_2)$: 
Hence $\Gamma = (\Gamma_1,x:c,\Gamma_3)$ for some $\Gamma_3$.
By inductive hypothesis $\Gamma_1,\Gamma_3\sgv a:b'$ where $b\rd b'$ for some $b'$
and also $\Gamma_1,\Gamma_3\sgv d$. 
Hence by rule \weak\ we have $\Gamma_1,\Gamma_3,z:d\sgv a:b'$.
\end{itemize}

In case of rule \conv\ we have $(\Gamma_1,x:c,\Gamma_2)\sgv a:d$ where $b\eqv d$ and $(\Gamma_1,x:c,\Gamma_2)\sgv a:d$.
By inductive hypothesis we have $(\Gamma_1,\Gamma_2)\sgv a:d'$ where $d\rd d'$.
By Laws \ref{cr} and \ref{type.rd} we know that $(\Gamma_1,\Gamma_2)\sgv a:b\rdr d'$. 
\end{proof}

\begin{law}[Context contraction]
\label{context.reduction2}
For all $\Gamma_1,\Gamma_2,x,a,b,c$:
$(\Gamma_1,x:c,\Gamma_2)\sgv a:b$ and $x\notin\free([\Gamma_2]a)$ implies 
 $(\Gamma_1,\Gamma_2)\sgv a:b'$  and $b\rd b'$ for some $b'$.
\end{law}
Note that in this lemma $x\notin\free([\Gamma_2]a)$ does not necessarily imply $x\notin\free([\Gamma_2]b)$, for example take  $x:\prim\sgv\prim:([y:\prim]\prim\,x)$, and hence the reduction of the type to some expression where $x$ is not free, in the example $([y:\prim]\prim\,x)\rd \prim$, is necessary.
\begin{proof}
Induction on the definition of $(\Gamma_1,x:c,\Gamma_2)\sgv a:b$.
\begin{meditemize}
\item[\ax:] 
Trivial since the premise is false.
\item[\mystart:]
We have $a=y$ and $(\Gamma_1,x:c,\Gamma_2)=(\Gamma,y:b)$ for some $\Gamma$ where $\Gamma\sgv b$.
If $\Gamma_2=()$ then $y=x$ which would violate the free variable condition.
Otherwise $\Gamma_2=(\Gamma_3,y:b)$ for some $\Gamma_3$ and therefore $\Gamma=(\Gamma_1,x:c,\Gamma_3)$
By inductive hypothesis $\Gamma_1,\Gamma_3\sgv b$ and therefore by rule \mystart\ $\Gamma_1,\Gamma_2\sgv y:b$.
\item[\weak:]
In case of rule \weak\ we have $(\Gamma_1,x:c,\Gamma_2)=(\Gamma,y:d)$ for some $\Gamma$, $y$, and $d$ where
$\Gamma\sgv a:b$ and $\Gamma\sgv d$. There are two cases:
\begin{itemize}
\item $y=x$: 
Hence $a=c$ and $\Gamma =\Gamma_1$.
Hence $b'=b$ and $\Gamma\sgv a:b$.
\item $y\neq x$ and hence $y\in\dom(\Gamma_2)$: 
Hence $\Gamma = (\Gamma_1,x:c,\Gamma_3)$ for some $\Gamma_3$.
By inductive hypothesis $\Gamma_1,\Gamma_3\sgv a:b'$ where $b\rd b'$ for some $b'$
and also $\Gamma_1,\Gamma_3\sgv d$. 
Hence by rule \weak\ we have $\Gamma_1,\Gamma_3,y:d\sgv a:b'$.
\end{itemize}
\item[\conv:]
In case of rule \conv\ we have $(\Gamma_1,x:c,\Gamma_2)\sgv a:d$ for some $d$  where $b\eqv d$ and $(\Gamma_1,x:c,\Gamma_2)\sgv a:d$.
By inductive hypothesis we have $(\Gamma_1,\Gamma_2)\sgv a:d'$ for some $d'$ where $d\rd d'$.
By Laws \ref{cr} and \ref{type.rd} we know that $(\Gamma_1,\Gamma_2)\sgv a:b\rdr d'$. 
\item[\absu], \abse:
We have$a=\binbop{y}{a_1}{a_2}$, $b=[y:a_1]b_2$ for some $y$ where $y\neq x$ and $(\Gamma_1,x:c,\Gamma_2,y:a_1)\sgv a_2:b_2$ for some $b_2$.
Since $x\notin\free([\Gamma_2]\binbop{y}{a_1}{a_2})$ we know that $x\notin\free([\Gamma_2,y:a_1]a_2)$.
Therefore, by inductive hypothesis $(\Gamma_1,\Gamma_2,y:a_1)\sgv a_2:b_2'$ where $b_2\rd b_2'$ for some $b_2'$.
Hence $(\Gamma_1,\Gamma_2)\sgv\binbop{x}{a_1}{a_2}:[y:a_1]b_2'$. The property obviously follows with $b'=[y:a_1]b_2$.
\item[\appl:]
We have $a=(a_1\:a_2)$, $b= c_2\gsub{y}{a_2}$, $(\Gamma_1,x:c,\Gamma_2)\sgv a_1:[y:c_1]c_2$ for some $y$, $c_1$, $c_2$ where $y\neq x$, and $(\Gamma_1,x:c,\Gamma_2)\sgv a_2:c_1$.  
Since $x\notin\free([\Gamma_2]a_1)\cup\free([\Gamma_2]a_2)$ by inductive hypothesis 
$(\Gamma_1,\Gamma_2)\sgv a_1:d_1$ and  $(\Gamma_1,\Gamma_2)\sgv a_2:d_2$ for some $d_1$, $d_2$ where $[y:c_1]c_2\rd d_1$ and $c_1\rd d_2$.
By Law \ref{rd.decomp}($i$) we know that $d_1=[y:d_3]d_4$ where $c_1\rd d_3$ and $c_2\rd d_4$, for some $d_3$ and $d_4$.
Since $d_2\eqv d_3$, by Laws \ref{cr} and \ref{type.rd} we know that $(\Gamma_1,\Gamma_2)\sgv a_2:(d_2\rdr d_3)$ and $(\Gamma_1,\Gamma_2)\sgv a_1:[y:(d_2\rdr d_3)]d_4$.
Hence by rule \appl\ $(\Gamma_1,\Gamma_2)\sgv (a_1\:a_2):d_4\gsub{y}{a_2}$.
By Law \ref{rd.sub}($i$) we know that $c_2\gsub{y}{a_2}\rd d_4\gsub{y}{a_2}$ which implies the proposition with $b'=d_4\gsub{y}{a_2}$.
\item[\pdef:]
We have $a=\prdef{y}{a_1}{a_2}{a_3}$, $b\eqv[y!d]a_3$, for some $d$, $(\Gamma_1,x:c,\Gamma_2)\sgv a_1:d$, $(\Gamma_1,x:c,\Gamma_2)\sgv a_2:a_3\gsub{y}{a_1}$, and $(\Gamma_1,x:c,\Gamma_2,y:d)\sgv a_3$.
Since $x\notin\free([\Gamma_2]a_1)\cup\free([\Gamma_2]a_2\cup\free([\Gamma_2]a_2)$ by inductive hypothesis
$(\Gamma_1,\Gamma_2)\sgv a_1:d'$ for some $d'$ where $d\rd d'$, $(\Gamma_1,\Gamma_2)\sgv a_2:e$ for some $e$ where $a_3\gsub{y}{a_1}\rd e$, and 
$(\Gamma_1,\Gamma_2,y:d)\sgv a_3$.

By Law \ref{valid.type} we know that $(\Gamma_1,\Gamma_2)\sgv d'$.
Hence by Law \ref{eqv.env} from $(\Gamma_1,\Gamma_2,y:d)\sgv a_3$ we can conclude that $(\Gamma_1,\Gamma_2,y:d')\sgv a_3$. 
Since $(\Gamma_1,\Gamma_2)\sgv a_1:d'$ by Law~\ref{type.sub} this implies $(\Gamma_1,\Gamma_2)\sgv a_3\gsub{y}{a_1}$.
Since $(\Gamma_1,\Gamma_2)\sgv a_2:e$ and $e\eqv a_3\gsub{y}{a_1}$ by rule \conv\ we obtain $(\Gamma_1,\Gamma_2)\sgv a_2:a_3\gsub{y}{a_1}$.
Hence by definition of typing $\Gamma_1,\Gamma_2\sgv a:[y!d']a_3$ which implies $\Gamma_1,\Gamma_2\sgv a:b'$ where $b\rd b'=[y!d']a_3$.
\item[\chin:]
We have $a=\pleft{a_1}$ where $(\Gamma_1,x:c,\Gamma_2)\sgv a_1:[y!b]d_2$ for some $y\neq x$, $d_2$.
Since $x\notin\free(a_1)$, by inductive hypothesis $(\Gamma_1,\Gamma_2)\sgv a_1:e$ where $[y!b]d_2\rd e$ for some $e$.
By Law \ref{rd.decomp}($i$) we know that $e=[y!e_1]e_2$ where $b\rd e_1$ and $d_2\rd e_2$ for some $e_1$ and $e_2$.
By Law \ref{type.rd} and by definition of typing this implies $(\Gamma_1,\Gamma_2)\sgv \pleft{a_1}:e_1$.
Since $b\rd e_1$ by Law \ref{type.rd} we know that $(\Gamma_1,\Gamma_2)\sgv\pleft{a_1}:b'$ where $b'=e_1$.
\item[\chba:]
We have $a=\pright{a_1}$ and $b=d_2\gsub{y}{\pleft{a_1}}$  where $(\Gamma_1,x:c,\Gamma_2)\sgv a_1:[y!d_1]d_2$ for some $y\neq x$, $d_1$, $d_2$.
Since $x\notin\free(a_1)$, by inductive hypothesis $(\Gamma_1,\Gamma_2)\sgv a_1:e$ where $[y!d_1]d_2\rd e$ for some $e$.
By Law \ref{rd.decomp}($i$) we know that $e=[e_1,e_2]$ where $d_1\rd e_1$ and $d_2\rd e_2$ for some $e_1$ and $e_2$.
By Law \ref{type.rd} and by definition of typing this implies $(\Gamma_1,\Gamma_2)\sgv\pright{a_1}:e_2$.
Since $b=d_2\gsub{y}{\pleft{a_1}}$ by Laws \ref{sub.any}($i$) and \ref{type.rd} we know that $(\Gamma_1,\Gamma_2)\sgv\pright{a_1}:b'$ where $b'=e_2\gsub{y}{\pleft{a_1}}$.
\item[\bprod] and \bsum:
We have $a=\prsumop{a_1}{a_2}$, $b=[d_1,d_2]$ for some $d_1$, $d_2$ where $(\Gamma_1,x:c,\Gamma_2)\sgv a_1:d_1$ and $(\Gamma_1,x:c,\Gamma_2)\sgv a_2:d_2$.
Since $x\notin\free([\Gamma_2]\prsumop{a_1}{a_2})$ we know that $x\notin\free([\Gamma_2]a_1)\cup\free([\Gamma_2]a_2)$.
By inductive hypothesis $(\Gamma_1,\Gamma_2)\sgv a_1:e_1$ for some $e_1$ where $d_1\rd e_1$, and $(\Gamma_1,\Gamma_2)\sgv a_2:e_2$ for some $e_2$ where $d_2\rd e_2$.
By Law \ref{type.rd} and by definition of typing this implies $(\Gamma_1,\Gamma_2)\sgv\prsumop{a_1}{a_2}:[e_1,e_2]$.
Since $b=[e_1,e_2]$,  by Law \ref{type.rd} we know that $(\Gamma_1,\Gamma_2)\sgv\prsumop{a_1}{a_2}:b'$ where $b'=[e_1,e_2]$.  
\item[\prl:]
We have $a=\pleft{a_1}$  where $(\Gamma_1,x:c,\Gamma_2)\sgv a_1:[b,d_2]$ for some $d_2$.
Since $x\notin\free(a_1)$, by inductive hypothesis $(\Gamma_1,\Gamma_2)\sgv a_1:e$ where $[b,d_2]\rd e$ for some $e$.
By Law \ref{rd.decomp}($iv$) we know that $e=[e_1,e_2]$ where $b\rd e_1$ and $d_2\rd e_2$ for some $e_1$ and $e_2$.
By Law \ref{type.rd} and by definition of typing this implies $(\Gamma_1,\Gamma_2)\sgv \pleft{a_1}:e_1$.
Since $b\rd e_1$ by Law \ref{type.rd} we know that $(\Gamma_1,\Gamma_2)\sgv\pleft{a_1}:b'$ where $b'=e_1$.
\item[\prr:]
Similar to case \prl.
\item[\injll:]
We have $a=\injl{a_1}{a_2}$ and $b=[d,a_2]$ for some $d$ where $(\Gamma_1,x:c,\Gamma_2)\sgv a_1:d$ and $(\Gamma_1,x:c,\Gamma_2)\sgv a_2$.
Since $x\notin\free([\Gamma_2]\injl{a_1}{a_2})$ we know that $x\notin\free([\Gamma_2]a_1)\cup\free([\Gamma_2]a_2)$.
By inductive hypothesis $(\Gamma_1,\Gamma_2)\sgv a_1:e$ for some $e$ where $d\rd e$, and $(\Gamma_1,\Gamma_2)\sgv a_2$.
By Law \ref{type.rd} and by definition of typing this implies  $(\Gamma_1,\Gamma_2)\sgv\injl{a_1}{a_2}:[e+a_2]$.
Since $b=[e,a_2]$,  by Law \ref{type.rd} we know that $(\Gamma_1,\Gamma_2)\sgv\injl{a_1}{a_2}:b'$ where $b'=b\rdr[e,a_2]$. 
\item[\injlr:]
Similar to case \injll.
\item[\cased:]
We have $a=\case{a_1}{a_2}$, $b=[y:[d_1+d_2]]d$ for some $d_1$, $d_2$, and $d$ where $(\Gamma_1,x:c,\Gamma_2)\sgv a_1:[y:d_1]d$, $(\Gamma_1,x:c,\Gamma_2)\sgv a_2:[y:d_2]d$ and $(\Gamma_1,x:c,\Gamma_2)\sgv d$. Since $x\notin\free([\Gamma_2]\case{a_1}{a_2})$ we know that $x\notin\free([\Gamma_2]a_1)\cup\free([\Gamma_2]a_2)$.

By inductive hypothesis, $(\Gamma_1,\Gamma_2)\sgv a_1:e$ for some $e$ where $[y:d_1]d\rd e$, $(\Gamma_1,\Gamma_2)\sgv a_2:f$ for some $f$ where $[y:d_2]d\rd f$, 
and $(\Gamma_1,\Gamma_2)\sgv d$.
By Law \ref{rd.decomp}($iv$) we know that $e=[y!e_1]e_2$ where $d_1\rd e_1$ and $d\rd e_2$ for some $e_1$ and $e_2$.
By Law \ref{rd.decomp}($iv$) we also know that $f=[y!f_1]f_2$ where $d_2\rd f_1$ and $d\rd f_2$ for some $f_1$ and $f_2$.

Since $d\eqv e_2 \eqv f_2$ and by Laws \ref{cr} and \ref{type.rd} we know that $(\Gamma_1,\Gamma_2)\sgv a_1:[y!e_1]g$ 
and $(\Gamma_1,\Gamma_2)\sgv a_2:[y!f_1]g$ where $g=(d\rdr e_2\rdr f_2)$.
Since $(\Gamma_1,\Gamma_2)\sgv d$, by Law \ref{valid.rd} we obtain $(\Gamma_1,\Gamma_2)\sgv g$.

By Law \ref{type.rd} and by definition of typing this implies $(\Gamma_1,\Gamma_2)\sgv a:[y:[e_1+f_1]]g$.
Since $b=[y:[e_1+f_1]]g$,  by Law \ref{type.rd} we know that $(\Gamma_1,\Gamma_2)\sgv a:b'$ where $b'=[y:[e_1+f_1]]g$.  
\item[\negate:]
We have $a=\myneg a_1$ and $(\Gamma_1,x:c,\Gamma_2)\sgv a_1:b$. 
Since $x\notin\free([\Gamma_2]\myneg a_1)$ we know that $x\notin\free([\Gamma_2]a_1)$.
By inductive hypothesis, $(\Gamma_1,\Gamma_2)\sgv a_1:d$ for some $d$ with $b\rd d$.
By Law \ref{type.rd} and by definition of typing this implies $(\Gamma_1,\Gamma_2)\sgv\myneg a_1:d$.
Since $b\rd e$, by Laws \ref{cr} and \ref{type.rd} we know that $(\Gamma_1,\Gamma_2)\sgv a:b'$ where $b'=e$.
\end{meditemize}
\end{proof}
\begin{law}[Context contraction]
\label{context.reduction}
For all $\Gamma_1,\Gamma_2,x,a,b,c$:
$(\Gamma_1,x:c,\Gamma_2)\sgv a:b$ and $x\notin\free([\Gamma_2]a)$ implies 
 $(\Gamma_1,\Gamma_2)\sgv a:b'$ and $b\rd b'$ for some $b'$.
\end{law}
\begin{proof}
Proof is by induction on $a$.
\begin{itemize}
\item $a=\prim$, $a=y$: 
The proposition follows from Law~\ref{contract.basic}. 
\item $a=\binbop{x}{a_1}{a_2}$:
By Law \ref{type.decomp}($i$) $(\Gamma_1,x:c,\Gamma_2)\sgv\binbop{y}{a_1}{a_2}:b$ implies $b\eqv[y:a_1]d$ for some $d$ where $(\Gamma_1,x:c,\Gamma_2,y:a_1)\sgv a_2:d$.
Since $x\notin\free([\Gamma_2]\binbop{x}{a_1}{a_2})$ we know that $x\notin\free([\Gamma_2,y:a_1]a_2)$.
Therefore, by inductive hypothesis $(\Gamma_1,\Gamma_2,y:a_1)\sgv a_2:d'$ where $d\rd d'$ for some $d'$.
By definition of typing this implies $(\Gamma_1,\Gamma_2)\sgv\binbop{y}{a_1}{a_2}:[y:a_1]d'$.
Since $b\eqv[y:a_1]d'$ by Laws \ref{cr} and \ref{type.rd} we know that $(\Gamma_1,\Gamma_2)\sgv\binbop{y}{a_1}{a_2}:b'$ where $b'=b\rdr[y:a_1]d'$.
\item $a=(a_1\:a_2)$: 
By Law~\ref{type.decomp}($vii$) we have  $b\eqv c_2\gsub{y}{a_2}$ where $y\neq x$, $(\Gamma_1,x:c,\Gamma_2)\sgv a_1:[y:c_1]c_2$ and $(\Gamma_1,x:c,\Gamma_2)\sgv a_2:c_1$ for some $c_1$, $c_2$. 
Since $x\notin\free([\Gamma_2](a_1\,a_2))$ we know that $x\notin\free([\Gamma_2]a_1)\cup\free([\Gamma_2]a_2)$.
By inductive hypothesis $(\Gamma_1,\Gamma_2)\sgv a_1:d$ where $[y:c_1]c_2\rd d$, for some $d$, and $(\Gamma_1,\Gamma_2)\sgv a_2:e$ where $c_1\rd e$, for some $e$. 

By Law \ref{rd.decomp}($i$) we know that $d=[y:d_1]d_2$ where $c_1\rd d_1$ and $c_2\rd d_2$, for some $d_1$ and $d_2$.
Since $e\eqv d_1$, by Law \ref{cr} and \ref{type.rd} we know that $(\Gamma_1,\Gamma_2)\sgv a_2:(e\rdr d_1)$.
Since $[y:d_1]d_2\rd [y:e\rdr d_1]d_2$, by Law \ref{type.rd} we know that $(\Gamma_1,\Gamma_2)\sgv a_1:[y:(e\rdr d_1)]d_2$.

This implies $(\Gamma_1,\Gamma_2)\sgv a:d_2\gsub{y}{a_2}$.
Since $b\eqv c_2\gsub{y}{a_2}\eqv d_2\gsub{y}{a_2}$, by Law \ref{cr} and \ref{type.rd} we obtain  $(\Gamma_1,\Gamma_2)\sgv a:b'$ where $b'=b\rdr d_2\gsub{y}{a_2}$.
\item $a=\prdef{y}{a_1}{a_2}{a_3}$: 
By Law~\ref{type.decomp}($x$) we have $b\eqv[y!d]a_3$ for some $d$ where $(\Gamma_1,x:c,\Gamma_2)\sgv a_1:d$, $(\Gamma_1,x:c,\Gamma_2)\sgv a_2:a_3\gsub{y}{a_1}$, and $(\Gamma_1,x:c,\Gamma_2,y:d)\sgv a_3$.

Since $x\notin\free([\Gamma_2]\prdef{y}{a_1}{a_2}{a_3})$ we know that $x\notin\free([\Gamma_2]a_1)\cup\free([\Gamma_2]a_2)\cup\free([\Gamma_2]a_3)$.
By inductive hypothesis $(\Gamma_1,\Gamma_2)\sgv a_1:d'$ where $d\rd d'$ for some $d'$ and $(\Gamma_1,\Gamma_2)\sgv a_2:e$ where $a_3\gsub{y}{a_1}\rd e$ for some $e$.

By Law \ref{valid.type} we know that $(\Gamma_1,\Gamma_2)\sgv d'$. Since by Law \ref{type.xtrct} we have $\Gamma_1\sgv c$, by Law \ref{type.weak} we know that $(\Gamma_1,x:c,\Gamma_2)\sgv d'$.
Hence by Law \ref{eqv.env} from $(\Gamma_1,x:c,\Gamma_2,y:d)\sgv a_3$ we can conclude that $(\Gamma_1,x:c,\Gamma_2,y:d')\sgv a_3$. 
From that, since $x\notin\free([\Gamma_2,y:d']a_3)$, by inductive hypothesis  $(\Gamma_1,\Gamma_2,y:d')\sgv a_3$.
Since $(\Gamma_1,\Gamma_2)\sgv a_1:d'$ by Law~\ref{type.sub} this implies $(\Gamma_1,\Gamma_2)\sgv a_3\gsub{y}{a_1}$.

Since $(\Gamma_1,\Gamma_2)\sgv a_2:e$ and $e\eqv a_3\gsub{y}{a_1}$ by rule \conv\ we obtain $(\Gamma_1,\Gamma_2)\sgv a_2:a_3\gsub{y}{a_1}$.
Hence by definition of typing $\Gamma_1,\Gamma_2\sgv a:[y!d']a_3$ which implies $\Gamma_1,\Gamma_2\sgv a:b'$ where $b\rd b'=[y!d']a_3$.
\item $a=\pleft{a_1}$:
By Law~\ref{type.decomp}($viii$) we have $b\eqv d_1$ where either $(\Gamma_1,x:c,\Gamma_2)\sgv a_1:[y!d_1]d_2$ or $(\Gamma_1,x:c,\Gamma_2)\sgv a_1:[d_1,d_2]$ for some $d_1$, $d_2$.

In the first case, since $x\notin\free(a_1)$, by inductive hypothesis $(\Gamma_1,\Gamma_2)\sgv a_1:e$ where $[y!d_1]d_2\rd e$ for some $e$.
By Law \ref{rd.decomp}($i$) we know that $e=[y!e_1]e_2$ where $d_1\rd e_1$ and $d_2\rd e_2$ for some $e_1$ and $e_2$.
Therefore, by definition of typing,  $(\Gamma_1,\Gamma_2)\sgv \pleft{a_1}:e_1$.
Since $b\eqv d_1$ by Laws \ref{cr} and \ref{type.rd} we know that $(\Gamma_1,\Gamma_2)\sgv\pleft{a_1}:b'$ where $b'=b\rdr d_1$.

In the second case, since $x\notin\free(a_1)$, by inductive hypothesis $(\Gamma_1,\Gamma_2)\sgv a_1:e$ where $[d_1,d_2]\rd e$ for some $e$.
By Law \ref{rd.decomp}($iv$) we know that $e=[e_1,e_2]$ where $d_1\rd e_1$ and $d_2\rd e_2$ for some $e_1$ and $e_2$.
Therefore, by definition of typing,  $(\Gamma_1,\Gamma_2)\sgv \pleft{a_1}:e_1$.
Since $b\eqv e_1$ by Laws \ref{cr} and \ref{type.rd} we know that $(\Gamma_1,\Gamma_2)\sgv\pleft{a_1}:b'$ where $b'=b\rdr e_1$.
\item $a=\pright{a_1}$: 
By Law~\ref{type.decomp}($ix$) we have, for some $d_1$, $d_2$, that either $b\eqv d_2$ where $(\Gamma_1,x:c,\Gamma_2)\sgv a_1:[d_1,d_2]$ 
or $b\eqv d_2\gsub{y}{\pleft{a_1}}$ where $(\Gamma_1,x:c,\Gamma_2)\sgv a_1:[y!d_1]d_2$. 

In the first case, since $x\notin\free(a_1)$, by inductive hypothesis $(\Gamma_1,\Gamma_2)\sgv a_1:e$ where $[y!d_1]d_2\rd e$ for some $e$.
By Law \ref{rd.decomp}($iv$) we know that $e=[y!e_1]e_2$ where $d_1\rd e_1$ and $d_2\rd e_2$ for some $e_1$ and $e_2$.
Therefore, by definition of typing,  $(\Gamma_1,\Gamma_2)\sgv \pright{a_1}:e_2\gsub{y}{\pleft{a_1}}$.
Since $b\eqv e_2\gsub{y}{\pleft{a_1}}$ by Laws \ref{cr} and \ref{type.rd} we know that $(\Gamma_1,\Gamma_2)\sgv\pleft{a_1}:b'$ where $b'=b\rdr e_2\gsub{y}{\pleft{a_1}}$.

In the second case, since $x\notin\free(a_1)$, by inductive hypothesis $(\Gamma_1,\Gamma_2)\sgv a_1:e$ where $[y!d_1]d_2\rd e$ for some $e$.
By Law \ref{rd.decomp}($i$) we know that $e=[y!e_1]e_2$ where $d_1\rd e_1$ and $d_2\rd e_2$ for some $e_1$ and $e_2$.
Therefore, by definition of typing,  $(\Gamma_1,\Gamma_2)\sgv \pleft{a_1}:e_1$.
Since $b\eqv d_1$ by Laws \ref{cr} and \ref{type.rd} we know that $(\Gamma_1,\Gamma_2)\sgv\pleft{a_1}:b'$ where $b'=b\rdr d_1$.
\item $a=\prsumop{a_1}{a_2}$:
By Law \ref{type.decomp}($iv$) we know that $b\eqv[d_1,d_2]$ for some $d_1$, $d_2$ where $(\Gamma_1,x:c,\Gamma_2)\sgv a_1:d_1$ and $(\Gamma_1,x:c,\Gamma_2)\sgv a_2:d_2$.
Since $x\notin\free([\Gamma_2]\prsumop{a_1}{a_2})$ we know that $x\notin\free([\Gamma_2]a_1)\cup\free([\Gamma_2]a_2)$.
By inductive hypothesis $(\Gamma_1,\Gamma_2)\sgv a_1:e_1$ for some $e_1$ where $d_1\rd e_1$, and $(\Gamma_1,\Gamma_2)\sgv a_2:e_2$ for some $e_2$ where $d_2\rd e_2$.

By definition of typing $(\Gamma_1,\Gamma_2)\sgv\prsumop{a_1}{a_2}:[e_1,e_2]$.
Since $b\eqv[e_1,e_2]$,  by Laws \ref{cr} and \ref{type.rd} we know that $(\Gamma_1,\Gamma_2)\sgv\prsumop{a_1}{a_2}:b'$ where $b'=b\rdr[e_1,e_2]$.  
\item $a=\injl{a_1}{a_2}$:
By Law \ref{type.decomp}($iii$) we know that $b\eqv [d,a_2]$ for some $d$ where $(\Gamma_1,x:c,\Gamma_2)\sgv a_1:d$ and $(\Gamma_1,x:c,\Gamma_2)\sgv a_2$.
Since $x\notin\free([\Gamma_2]\injl{a_1}{a_2})$ we know that $x\notin\free([\Gamma_2]a_1)\cup\free([\Gamma_2]a_2)$.
By inductive hypothesis $(\Gamma_1,\Gamma_2)\sgv a_1:e$ for some $e$ where $d\rd e$, and $(\Gamma_1,\Gamma_2)\sgv a_2$.

By definition of typing $(\Gamma_1,\Gamma_2)\sgv\injl{a_1}{a_2}:[e+a_2]$.
Since $b\eqv[e,a_2]$,  by Laws \ref{cr} and \ref{type.rd} we know that $(\Gamma_1,\Gamma_2)\sgv\injl{a_1}{a_2}:b'$ where $b'=b\rdr[e,a_2]$. 
\item $a=\injr{a_1}{a_2}$:
By Law \ref{type.decomp}($iii$) we know that $b\eqv[a_1,d]$ for some $d$ where $(\Gamma_1,x:c,\Gamma_2)\sgv a_2:d$ and $(\Gamma_1,x:c,\Gamma_2)\sgv a_1$.
Since $x\notin\free([\Gamma_2]\injr{a_1}{a_2})$ we know that $x\notin\free([\Gamma_2]a_1)\cup\free([\Gamma_2]a_2)$.
By inductive hypothesis $(\Gamma_1,\Gamma_2)\sgv a_2:e$ for some $e$ where $d\rd e$, and $(\Gamma_1,\Gamma_2)\sgv a_1$.

By definition of typing $(\Gamma_1,\Gamma_2)\sgv\injr{a_1}{a_2}:[a_1+e]$.
Since $b\eqv[a_1,e]$,  by Laws \ref{cr} and \ref{type.rd} we know that $(\Gamma_1,\Gamma_2)\sgv\injr{a_1}{a_2}:b'$ where $b'=b\rdr[a_1,e]$. 
\item $a=\case{a_1}{a_2}$.
By Law \ref{type.decomp}($xi$) we know that $b\eqv[y:[d_1+d_2]]d$ for some $d_1$, $d_2$, and $d$ where $(\Gamma_1,x:c,\Gamma_2)\sgv a_1:[y:d_1]d$, $(\Gamma_1,x:c,\Gamma_2)\sgv a_2:[y:d_2]d$ and $(\Gamma_1,x:c,\Gamma_2)\sgv d$. 
Since $x\notin\free([\Gamma_2]\case{a_1}{a_2})$ we know that $x\notin\free([\Gamma_2]a_1)\cup\free([\Gamma_2]a_2)$.

By inductive hypothesis, $(\Gamma_1,\Gamma_2)\sgv a_1:e$ for some $e$ where $[y:d_1]d\rd e$, $(\Gamma_1,\Gamma_2)\sgv a_2:f$ for some $f$ where $[y:d_2]d\rd f$, 
and $(\Gamma_1,\Gamma_2)\sgv d$.
By Law \ref{rd.decomp}($iv$) we know that $e=[y!e_1]e_2$ where $d_1\rd e_1$ and $d\rd e_2$ for some $e_1$ and $e_2$.
By Law \ref{rd.decomp}($iv$) we also know that $f=[y!f_1]f_2$ where $d_2\rd f_1$ and $d\rd f_2$ for some $f_1$ and $f_2$.

Since $d\eqv e_2 \eqv f_2$ and by Laws \ref{cr} and \ref{type.rd} we know that $(\Gamma_1,\Gamma_2)\sgv a_1:[y!e_1]g$ 
and $(\Gamma_1,\Gamma_2)\sgv a_2:[y!f_1]g$ where $g=(d\rdr e_2\rdr f_2)$.
Since $(\Gamma_1,\Gamma_2)\sgv d$, by Law \ref{valid.rd} we obtain $(\Gamma_1,\Gamma_2)\sgv g$.

Therefore, by definition of typing  $(\Gamma_1,\Gamma_2)\sgv a:[y:[e_1+f_1]]g$.
Since $b\eqv[y:[e_1+f_1]]g$,  by Laws \ref{cr} and \ref{type.rd} we know that $(\Gamma_1,\Gamma_2)\sgv a:b'$ where $b'=b\rdr[y:[e_1+f_1]]g$.  
\item $a=\myneg a_1$.
By Law \ref{type.decomp}($v$) we know that $b\eqv d$ for some $d$ where $(\Gamma_1,x:c,\Gamma_2)\sgv a_1:d$. 
Since $x\notin\free([\Gamma_2]\myneg a_1)$ we know that $x\notin\free([\Gamma_2]a_1)$.
By inductive hypothesis, $(\Gamma_1,\Gamma_2)\sgv a_1:e$ for some $e$ with $d\rd e$.

By definition of typing $(\Gamma_1,\Gamma_2)\sgv \myneg a_1:e$.
Since $b\eqv e$, by Laws \ref{cr} and \ref{type.rd} we know that $(\Gamma_1,\Gamma_2)\sgv a:b'$ where $b'=b\rdr e$.
\qedhere
\end{itemize}
\end{proof}
\noindent
We can now show uniqueness of types. Most cases are straightforward, except for the weakening rule where we need the context contraction result above.
\begin{law}[Uniqueness of types]
\index{uniqueness!of types}
\label{type.confl}
For all $\Gamma,a,b,c$:
$\Gamma\sgv a:b$ and $\Gamma\sgv a:c$ implies $b\eqv c$.
\end{law}
\begin{remark}[Sketch of proof]
Using induction on the definition of $\Gamma\sgv a:b$, the property is relatively straightforward since, except for the rules \emph{weak} and \emph{conv}, there is exactly one typing rule for each construct. 
\end{remark}
\begin{proof}
Proof by induction on the definition of $\Gamma\sgv a:b$.
We look at each typing rule of $\Gamma\sgv a:b$ in turn. In each case we have to show that if also $\Gamma\sgv a:c$ then $b\eqv c$.
\begin{meditemize}
\item[\ax:]
This means $a=b=\prim$ and $\Gamma=()$. Let $\sgv\prim:c$.
By Law~\ref{type.decomp}($vi$) we know that $c\eqv\prim$ hence $b\eqv c$.
\item[\mystart:] 
We have $a=x$, $\Gamma=(\Gamma',x:b)$ for some $\Gamma'$ and $x$ where $(\Gamma',x:b)\sgv x:b$ and $\Gamma'\sgv b:d$ for some $d$:
Let $(\Gamma',x:b)\sgv x:c$.
By Law~\ref{start.confl} we know that $b\eqv c$. 
\item[\weak:]
We have $\Gamma=(\Gamma',x:d)$ for some $\Gamma'$ and $x$ where $(\Gamma',x:d)\sgv a:b$, $\Gamma'\sgv a:b$, and $\Gamma'\sgv d:e$ for some $e$.
Let $(\Gamma',x:d)\sgv a:c$. 
Since $x\notin\free(a)$, by Law~\ref{context.reduction} we know that $\Gamma'\sgv a:c'$ for some $c'$ with $c\rd c'$.
By inductive hypothesis $b\eqv c'$ which implies $b\eqv c$.
\item[\conv:] 
We have $\Gamma\sgv a:b$ and $\Gamma\sgv a:d$ for some $d$ where $d\eqv b$.
Let $\Gamma\sgv a:c$. 
By inductive hypothesis $b\eqv d\eqv c$.
\item[\absu:] 
We have $a=[x:a_1]a_2$ and $b=[x:a_1]b_2$ for some $x$, $a_1$, $a_2$, and $b_2$ where $\Gamma\sgv[x:a_1]a_2:[y:a_1]b_2$ and $(\Gamma,y:a_1)\sgv a_2:b_2$. 
Let $\Gamma\sgv[x:a_1]a_2:c$.
By Law~\ref{type.decomp}($i$) we know that $c\eqv[x:a_1]c_2$ for some $c_2$ where $\Gamma\sgv a_1$ and $(\Gamma,x:a_1)\sgv a_2:c_2$.
By inductive hypothesis applied on $(\Gamma,x:a_1)\sgv a_2:b_2$ we obtain $b_2\eqv c_2$.
Hence $b=[x:a_1]b_2\eqv[x:a_1]c_2=c$ .
\item[\abse:] 
Similar to case \absu.
\item[\appl:] 
We have $a=(a_1\,a_2)$ and $b=b_2\gsub{x}{a_2}$ for some $x$, $a_1$, $a_2$, and $b_2$ where $\Gamma\sgv (a_1\,a_2):b_2\gsub{x}{a_2}$,  $\Gamma\sgv a_1:[x:b_1]b_2$, and $\Gamma\sgv a_2:b_1$.
Let $\Gamma\sgv(a_1\,a_2):c$.
By Law~\ref{type.decomp}($vii$) we know that $c\eqv c_2\gsub{y}{a_2}$ for some $c_1$, $c_2$ where $\Gamma\sgv a_1:[y:c_1]c_2$ and $\Gamma\sgv a_2:c_1$.
By inductive hypothesis applied to $\Gamma\sgv a_1:[x:b_1]b_2$ it follows that $[x:b_1]b_2\eqv [y:c_1]c_2$.
Hence obviously $x=y$ and by basic properties of congruence (Law~\ref{congr.basic}($i$)) it follows that $b_2\eqv c_2$. 
Using Law~\ref{eqv.sub}($i$) we can argue $b\eqv b_2\gsub{x}{a_2}\eqv c_2\gsub{y}{a_2}\eqv c$.
\item[\pdef:]
We have $a=\prdef{x}{a_1}{a_2}{a_3}$ and $b=[x!b_1]a_3$ for some $x$, $a_1$, $a_2$, $a_3$, and $b_1$ where $\Gamma\sgv\prdef{x}{a_1}{a_2}{a_3}:[x!b_1]a_3$, $\Gamma\sgv a_1:b_1$, $\Gamma\sgv a_2:a_3\gsub{x}{a_1}$, and $(\Gamma,x:b_1)\sgv a_3$.
Let $\Gamma\sgv\prdef{x}{a_1}{a_2}{a_3}:c$.
By Law~\ref{type.decomp}($x$) we can infer that $c\eqv[x!c_1]a_3$ for some $c_1$ where $\Gamma\sgv a_1:c_1$ and $\Gamma\sgv a_2:a_3\gsub{x}{a_1}$.
By inductive hypothesis applied to $\Gamma\sgv a_1:b_1$ we know that $b_1\eqv c_1$.
Using Law~\ref{congr.basic}($i$) we can argue $b=[x!b_1]a_3\eqv[x!c_1]a_3\eqv c$. 
\item[\chin:]
We have $a=\pleft{a_1}$ for some $a_1$ where $\Gamma\sgv\pleft{a_1}:b$ and $\Gamma\sgv a_1:[x!b]b_2$ for some $x$ and $b_2$. 
Let $\Gamma\sgv\pleft{a_1}:c$.
By Law~\ref{type.decomp}($viii$) we know that $c\eqv c_1$ where either $\Gamma\sgv a_1:[x!c_1]c_2$ or $\Gamma\sgv a_1:[c_1,c_2]$ for some $c_1$ and $c_2$.

The case $\Gamma\sgv a_1:[c_1,c_2]$ is not possible since the inductive hypothesis applied to $\Gamma\sgv a_1:[x!b]b_2$ would imply $[x!b]b_2\eqv[c_1,c_2]$ which is obviously false.

Therefore we are left with the case $\Gamma\sgv a_1:[x!c_1]c_2$.
By inductive hypothesis applied to $\Gamma\sgv a_1:[x!b]b_2$ we know that $[x!b]b_2\eqv[x!c_1]c_2$.
Using elementary properties of congruence (Law~\ref{congr.basic}($i$)) we can argue $b\eqv c_1\eqv c$. 
\item[\chba:]
We have $a=\pright{a_1}$ and $b=b_2\gsub{x}{\pleft{a_1}}$ for some $a_1$, $x$, and $b_2$ where $\Gamma\sgv \pright{a_1}:b_2\gsub{x}{\pleft{a_1}}$ and $\Gamma\sgv a_1:[x!b_1]b_2$. 
Let $\Gamma\sgv \pright{a_1}:c$. 
By Law~\ref{type.decomp}($ix$) we know that either $c\eqv c_2\gsub{x}{\pleft{a_1}}$ for some $c_2$ where $\Gamma\sgv a_1:[x!c_1]c_2$ for some $c_1$
or $c\eqv[c_1,c_2]$ for some $c_1$, $c_2$ where $\Gamma\sgv a_1:[c_1,c_2]$.

The case $\Gamma\sgv a_1:[c_1,c_2]$ is not possible since the inductive hypothesis applied to $\Gamma\sgv a_1:[x!b_1]b_2$ would imply $[x!b_2]b_2\eqv[c_1,c_2]$ which is obviously false.

Therefore we are left with the case $\Gamma\sgv a_1:[x!c_1]c_2$.
By inductive hypothesis applied to $\Gamma\sgv a_1:[x!b_1]b_2$ we know that $[x!b_1]b_2\eqv[x!c_1]c_2$.
By elementary properties of congruence (Law~\ref{congr.basic}($i$))we obtain $b_2\eqv c_2$.
Using  elementary properties of congruence~(Law~\ref{eqv.sub}($i$)) we can argue $b=b_2\gsub{x}{\pleft{a_1}}\eqv c_2\gsub{x}{\pleft{a_1}}\eqv c$.
\item[\bprod:]
We have $a=[a_1,a_2]$ and $b=[b_1,b_2]$ for some $a_1$, $a_2$, $b_1$, and $b_2$ where $\Gamma\sgv[a_1,a_2]:[b_1,b_2]$, $\Gamma\sgv a_1:b_1$, and $\Gamma\sgv a_2:b_2$.
Let $\Gamma\sgv[a_1,a_2]:c$.
By Law~\ref{type.decomp}($iv$) we know that $c\eqv[c_1,c_2]$ for some $c_1$, $c_2$ where $\Gamma\sgv a_1:c_1$ and $\Gamma\sgv a_2:c_2$. 
By inductive hypothesis applied to $\Gamma\sgv a_1:b_1$ and $\Gamma\sgv a_2:b_2$ we obtain $b_1\eqv c_1$ and $b_2\eqv c_2$.
Using elementary properties of congruence (Law~\ref{congr.basic}($ii$)) we can argue $b=[b_1,b_2]\eqv[c_1,c_2]\eqv c$.
\item[\bsum:]
Similar to case \bprod.
\item[\prl:]
We have $a=\pleft{a_1}$ for some $a_1$ where $\Gamma\sgv\pleft{a_1}:b$ and $\Gamma\sgv a_1:[b,b_2]$ for some $x$ and $b_2$. 
Let $\Gamma\sgv\pleft{a_1}:c$.
By Law~\ref{type.decomp}($viii$) we know that $c\eqv c_1$ where $\Gamma\sgv a_1:[x!c_1]c_2$ or $\Gamma\sgv a_1:[c_1,c_2]$ for some $c_1$ and $c_2$.

The case $\Gamma\sgv a_1:[x!c_1]c_2$ is not possible since the inductive hypothesis applied to $\Gamma\sgv a_1:[b,b_2]$ would imply $[b,c_2]\eqv[x!c_1]c_2$ which is obviously false.

Therefore we are left with the case $\Gamma\sgv a_1:[c_1,c_2]$. 
By inductive hypothesis applied to $\Gamma\sgv a_1:[b,b_2]$ we know that $[b,b_2]\eqv[c_1,c_2]$.
Using elementary properties of congruence (Law~\ref{congr.basic}($ii$)) we can argue $b\eqv c_1\eqv c$. 
\item[\prr:]
Similar to case \prl.
\item[\injll:]
We have $a=\injl{a_1}{a_2}$ and $b=[b_1+a_2]$ for some $a_1$, $a_2$, and $b_1$ where $\Gamma\sgv\injl{a_1}{a_2}:[b_1+a_2]$, $\Gamma\sgv a_1:b_1$ and $\Gamma\sgv a_2$.
Let $\Gamma\sgv\injl{a_1}{a_2}:c$.
By Law~\ref{type.decomp}($iii$) we know that  $c\eqv[c_1+a_2]$ for some $c_1$ where $\Gamma\sgv a_1:c_1$.
By inductive hypothesis applied to $\Gamma\sgv a_1:b_1$ we know that $b_1\eqv c_1$.
Using elementary properties of congruence (Law~\ref{congr.basic}($ii$)) we obtain $b=[b_1+a_2]\eqv[c_1+a_2]\eqv c$.
\item[\injlr:]
Similar to case \injll.
\item[\cased:]
We have $a=\case{a_1}{a_2}$ and $b=[x:[b_1+b_2]]b_3$ for some $b_1$, $b_2$, and $b_3$ where $\Gamma\sgv\case{a_1}{a_2}:[x:[b_1+b_2]]b_3$, $\Gamma\sgv a_1:[x:b_1]b_3$, $\Gamma\sgv a_2:[x:b_2]b_3$, and $\Gamma\sgv b_3$. 
Let $\Gamma\sgv\case{a_1}{a_2}:c$.
By Law~\ref{type.decomp}($xi$) we know that $c\eqv[x:[c_1+c_2]]b_3'$ for some $c_1$, $c_2$, and $b_3'$ where $\Gamma\sgv a_1:[y:c_1]b_3'$ and $\Gamma\sgv a_2:[y:c_2]b_3'$.
By inductive hypothesis applied to $\Gamma\sgv a_1:[x:b_1]b_3$ and $\Gamma\sgv a_2:[x:b_2]b_3$ we know that $[x:b_1]b_3\eqv[x:c_1]b_3'$ and $[x:b_2]b_3\eqv[x:c_2]b_3'$.
By elementary properties of congruence (Law~\ref{congr.basic}($i,ii$)) we can obtain $b=[x:[b_1+b_2]]b_3\eqv[x:[c_1+c_2]]b_3'\eqv c$.
\item[\negate:]
We have $a=\myneg a_1$ for some $a_1$ where $\Gamma\sgv\myneg a_1:b$ and $\Gamma\sgv a_1:b$.
Let $\Gamma\sgv\myneg a_1:c$.
By Law~\ref{type.decomp}($v$) we know that $c\eqv c_1$ for some $c_1$ where $\Gamma\sgv a_1:c_1$. 
By inductive hypothesis $b\eqv c_1$ which implies $b\eqv c$.
\qedhere
\end{meditemize} 
\end{proof}
\noindent
A direct consequence of uniqueness of types is commutation of typing and congruence.
\begin{law}[Commutation of typing and congruence]
\label{congr.typ}
For all $\Gamma,a,b,c,d$: 
$\Gamma\sgv a:b$, $a\eqv c$, and $\Gamma\sgv c:d$ imply $b\eqv d$.
\end {law}
\begin{proof}
By Law~\ref{cr} we know that $a\rd c'$ and $c\rd c'$, for some $c'$. 
By Law~\ref{rd.type} we know that $\Gamma\sgv c':b$ and $\Gamma\sgv c':d$.
By Law~\ref{type.confl} we know that $b\eqv d$.
\end{proof}
\section{Strong normalization}%
\label{strong.normalization}
\subsection{Overview}%
Due to the Church-Rosser property (Law~\ref{cr}) we know that if a reduction terminates, the result will be unique.
In this section we will prove termination of all reduction sequences of valid expressions, this property is usually referred to as \emph{strong normalization}.

The idea for the proof of strong normalization of valid expressions in \dcalc\ is to classify expressions according to structural properties in order to make an inductive argument work. 
For this purpose we define structural skeletons called \emph{norms} as the subset of expressions built from the primitive $\prim$ and the product operation only and we define an partial function assigning norms to expressions. 
The domain of this norming function is the set of \emph{normable} expressions.
Norms are a reconstruction of a concept of \emph{simple types}, consisting of the atomic types and product types, within \dcalc\footnote{The construction of norms inside \dcalc~is due to convenient reuse of existing structures and definitions, norms could also be introduced as a separate mathematical structure}. 
Analogously to simple types, norms provide a handle to classify valid expressions into different degrees of structural complexity.
This can be used as a basis for making inductive arguments of strong normalization work.  
The idea of the strong normalisation argument is first to prove that all valid expressions are normable and then to prove that all normable expressions are strongly normalizable.

The good news is that in \dcalc\ we are not dealing with unconstrained parametric types as for example in System F (see \eg\ \cite{girard1989proofs}), 
and therefore we will be able to use more elementary methods to show strong normalisation as used for the simply typed lambda calculus. 
A common such method is to define a notion of \emph{reducible} expressions satisfying certain \emph{reducibility conditions} suitable for inductive arguments both on type structure and on reduction length, to prove that all reducible expressions satisfy certain \emph{reducibility properties} including strong normalisation, and then to prove that that all typable expressions are reducible (\cite{tait1967intensional}\cite{Girard72}).

We will basically adopt this idea, but unfortunately, common definitions of reducibility (\eg\ \cite{girard1989proofs}) cannot obviously be adapted to include the reduction of negations.
A more suitable basis for our purposes is to use the notion of \emph{computable expressions} as defined in language theoretical studies of Automath~\cite{vDaalen77}, see also~\cite{VANBENTHEMJUTTING1994}. 
This approach is basically extended here to cover additional operators, including negation.
 
\begin{figure}[!htb]
\setlength{\unitlength}{1mm}
\begin{picture}(100, 80)
\put(50,40){\circle{16}}
\put(45,40){$\Gamma\sgv a$}
\put(28,58){$\Gamma\sngv a$}
\put(28,53){$\text{or equivalently}\;a\in\ce_{\Gamma}$}
\put(50,40){\circle{64}}
\put(50,40){\circle{80}}
\put(45,75){$a\in\sn{}$}
\end{picture}
\caption{Relation between validity, normability, computability, and strong normalization\label{sn.proof}}
\end{figure}
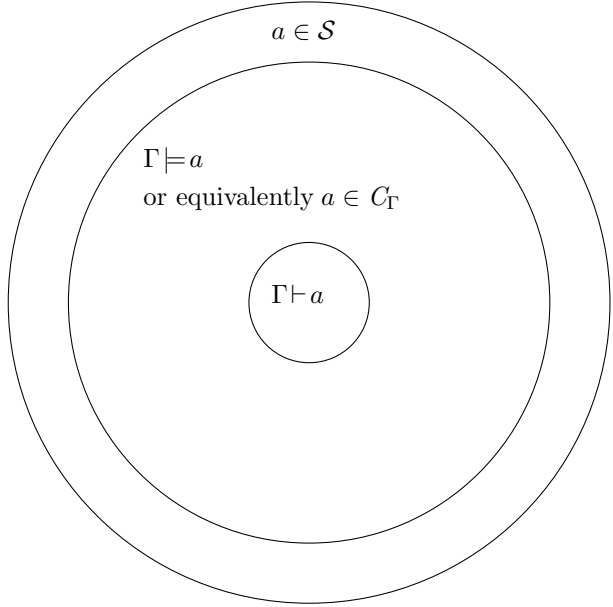

We motivate the basic idea of the proof (precise definitions can be found in the upcoming sections):
Consider the following condition necessary to establish strong normalization for an application $(a\,b)$ in the context of an inductive proof
(where $\sn{}$ denotes the set of strongly normalizable expressions):
\begin{itemize}
\item
If $a\rd[x:c_1]c_2\in\sn{}$ and $b\in\sn{}$ then $c_2\gsub{x}{b}\in\sn{}$.
\end{itemize}
Similarly, the following condition is necessary to establish strong normalization for a negation $\myneg a$ in the context of an inductive proof.
\begin{itemize}
\item
If $a\rd[x:c_1]c_2\in\sn{}$ then $\myneg c_2\in\sn{}$.
\end{itemize}
\noindent
These and other properties inspire the definition of the set of \emph{computable} expressions $C_{\Gamma}$ which 
are normable, strongly normalizable, and satisfy the property that $a,b\in C_{\Gamma}$ implies $\myneg a,(a\,b)\in C_{\Gamma}$. 
Unfortunately, the closure properties of computable expressions cannot be extended to abstractions.
Instead, we need to prove the stronger property
that normable expressions are computable under any substitution of their free variables to computable expressions.
This implies that all normable expressions are computable and therefore that normability and computability are equivalent notions. 
The logical relations between the various notions are graphically illustrated in Figure~\ref{sn.proof}.
\subsection{Basic properties of strong normalization}%
\label{sn.base}
We begin with some basic definitions and properties related to strong normalization. 
\begin{definition}[Strongly normalizable expressions]
\index{expression!strongly normalizable}
\nomenclature[fBasic05]{$\sn{}$}{strongly normalizable expressions}
\label{sn}
The set of \emph{strongly normalizable expressions} is denoted by $\sn{}$.
An expression $a$ is in $\sn{}$ iff there is no infinite sequence of one-step reductions $a\srd a_1\srd a_2\srd a_3\ldots$ 
\end{definition}
\begin{law}[Basic properties of strongly normalizable expressions]
\label{sn.basic}
For all $a,b,c,x$:
\begin{itemize}
\item[$i$:] 
$a\gsub{x}{b}\in\sn{}$ and $b\rd c$ imply $a\gsub{x}{c}\in\sn{}$
\item[$ii$:] 
$a,b\in\sn{}$ implies $\binbop{x}{a}{b}\in\sn{}$.
\item[$iii$:] 
$a\in\sn{}$ implies $\pleft{a}$, $\pright{a}\in\sn{}$.
\item[$iv$:]
$a,b,c\in\sn{}$ implies $\prdef{x}{a}{b}{c}\in\sn{}$. 
\item[$v$:] 
$a,b\in\sn{}$ implies $[a,b]$, $[a+b]$, $\injl{a}{b}$, $\injr{a}{b}$, $\case{a}{b}\in\sn{}$.
\end{itemize}
\end{law}
\begin{proof}
$\;$
\begin{itemize}
\item[$i$:]
Proof by contradiction:
Assume that $a\gsub{x}{b}\in\sn{}$, $b\rd c$, and $a\gsub{x}{c}\notin\sn{}$, i.e.~there is an infinite reduction sequence from $a\gsub{x}{c}$.
Since $b\rd c$ by Law~\ref{rd.sub}($ii$) we have $a\gsub{x}{b}\rd a\gsub{x}{c}$ hence $a\gsub{x}{b}\notin\sn{}$.
This is obviously a contradiction therefore $a\gsub{x}{c}\in\sn{}$.
\item[$ii$:]
Proof by contradiction:
Assume $a,b\in\sn{}$ and assume that there is an infinite reduction sequence from $\binbop{x}{a}{b}$. 
Using basic properties of reduction (Law~\ref{rd.decomp}($ii$)) one can show that each element of any finite prefix of this sequence has the form $\binbop{x}{a_i}{b_i}$ where $a\rd a_i$ and $b\rd b_i$.
The proof is by induction on the length $n$ of this prefix sequence, 
Hence obviously $a\notin\sn{}$ or $b\notin\sn{}$ which is a contradiction to $a,b\in\sn{}$, therefore $\binbop{x}{a}{b}\in\sn{}$.
\item[$iii$:]
We show the property for $\pleft{a}$ (the proof of $\pright{a}$ runs similar).

Proof by contradiction:
Assume $a\in\sn{}$ and assume there is an infinite reduction sequence $\pleft{a}\srd b_1\srd b_2\srd\ldots$. 
If $b_i=\pleft{b_i'}$, where $a\rdn{i} b_i'$ for all $i>0$, then we could construct an infinite reduction sequence from $a$ which would contradict our assumption. 
Hence there must be some index $j$ where $\pleft{a}\srd\pleft{a_1}\rdn{j-2}\pleft{a_{j-1}}\srd b_j$ and $b_j$ is not a left projection.
An analysis of the definition of $\rd$ shows that one of the axioms $\pi_1$, $\pi_3$, or $\pi_5$ must have been used do reduce $\pleft{a_{j-1}}\srd a'$.

In case of $\pi_1$ we know that $b_j=\pleft{\prdef{x}{b_{j+1}}{c}{d}}$ for some $c$ and $d$.
We can therefore write the sequence as $\pleft{a}\srd\pleft{a_1}\rdn{j-2}\pleft{a_{j-1}}\srd\pleft{\prdef{x}{b_{j+1}}{c}{d}}\srd b_{j+1}\srd\ldots$ where $a\srd a_1\rdn{j-2}a_{j-1}\srd\prdef{x}{b_{j+1}}{c}{d}$.
We can thus construct an infinite sequence from $a$ as $a\rdn{j}\prdef{x}{b_{j+1}}{c}{d}\srd\prdef{x}{b_{j+2}}{c}{d}\srd\ldots$ which would contradict our assumption.

In case of $\pi_3$ we know that $b_j=\pleft{[b_{j+1},c]}$ for some $c$.
We can therefore write the sequence as $\pleft{a}\srd\pleft{a_1}\rdn{j-2}\pleft{a_{j-1}}\srd\pleft{[b_{j+1},c]}\srd b_{j+1}\srd\ldots$ where $a\srd a_1\rdn{j-2}a_{j-1}\srd[b_{j+1},c]$.
We can thus construct an infinite sequence from $a$ as $a\rdn{j}[b_{j+1},c]\srd[b_{j+2},c]\srd\ldots$ which would contradict our assumption.

A similar argument can be made for the axiom $\pi_5$.
\item[$iv$:]
The proof is similar to case $ii$ using Law~\ref{rd.decomp}($ii$). 
\item[$v$:] 
The proof is similar to case $ii$ using Law~\ref{rd.decomp}($i$).
\qedhere
\end{itemize}
\end{proof}
\noindent
Next we show conditions under which application and negation are strongly normalizing. 
\begin{law}[Strong normalization conditions]
\label{sn.cond}
For all $a,b$:
\begin{itemize}
\item[$i$:]
For all $a,b\in\sn{}$: $(a\,b)\in\sn{}$ if for any $c_1$, $c_2$, $d_1$, $d_2$, and $x$, the following conditions are satisfied:
\begin{itemize}
\item[$C_1$:] $a\rd\binbop{x}{c_1}{c_2}$ implies that  $c_2\gsub{x}{b}\in\sn{}$
\item[$C_2$:] $a\rd\case{c_1}{c_2}$ and $b\rd\injl{d_1}{d_2}$ implies $(c_1\,d_1)\in\sn{}$
\item[$C_3$:] $a\rd\case{c_1}{c_2}$ and $b\rd\injr{d_1}{d_2}$ implies $(c_2\,d_2)\in\sn{}$
\end{itemize}
\item[$ii$:]
For all $a\in\sn{}$: $\myneg a\in\sn{}$ if for any $x$, $b$, and $c$, the following conditions are satisfied:
\begin{itemize}
\item[$C_1$:] $a\rd\prsumop{b}{c}$ implies $\myneg b$, $\myneg c\in\sn{}$ 
\item[$C_2$:] $a\rd\binbop{x}{b}{c}$ implies $\myneg c\in\sn{}$
\end{itemize} 
\end{itemize}
\end{law}
\begin{proof}
$\;$
\begin{itemize}
\item[$i$:]
Proof by contradiction:
Assume $a,b\in\sn{}$ as well as conditions $C_1$, $C_2$, and $C_3$ of $i$ and assume there is an infinite reduction $(a\,b)\srd e_1\srd e_2\srd\ldots$ for some $e_1$, $e_2$, $\ldots$.
Since $a,b\in\sn{}$, it is impossible that for all $i$ we have $e_i=(a_i\,b_i)$ where $a\rd a_i$ and $b\rd b_i$. 
An analysis of the definition of $\rd$ shows that therefore there must be a $j$ where one of the axioms $\beta_1$ to $\beta_4$ has been used (on-top-level) to reduce $e_j$ to $e_{j+1}$.

In case of $\beta_1$ and $\beta_2$ there must be a $j$ such that $e_j=(\binbop{x}{c_1}{c_2}\:b_j)$ for some $c_1$ and $c_2$, and $e_{j+1}=c_2\gsub{x}{b_j}$ where $a\rd\binbop{x}{c_1}{c_2}$ and $b\rd b_j$.
From $C_1$ we obtain $c_2\gsub{x}{b}\in\sn{}$.
Since $b\rd b'$ by Law~\ref{sn.basic}($i$) we know that $c_2\gsub{x}{b_j}\in\sn{}$ hence any reduction from $e_{j+1}=c_2\gsub{x}{b_j}$ must terminate.
This implies that the assumed infinite reduction from $(a\,b)$ also terminates which contradicts our assumption.

In case of the axiom $\beta_3$ there must be a $j$ such that  $e_j=(\case{c_1}{c_2}\;b_j)$ for some $c_1$ and $c_2$, $b_j=\injl{d_1}{d_2}$ for some $d_1$ and $d_2$, and $e_{j+1}=(c_1\,d_1)$ where
$a\rd\case{c_1}{c_2}$ and $b\rd\injl{d_1}{d_2}$.
From $C_2$ we obtain $c_1(d_1)\in\sn{}$ hence any reduction from $e_{j+1}=(c_1\,d_1)$ must terminate.
This implies that the assumed infinite reduction from $(a\,b)$ also terminates which contradicts our assumption.

A similar argument can be made for the axiom $\beta_4$.
\item[$ii$:] 
Assume $a,b\in\sn{}$ as well as conditions $C_1$ and $C_2$ of $ii$ and assume there is an infinite reduction $\myneg a\srd e_1\srd e_2\srd\ldots$ for some $e_1$, $e_2$, $\ldots$.
Since $a,b\in\sn{}$, it is impossible that for all $i$ we have $e_i=\myneg a_i$ where $a\rd a_i$.
An analysis of the definition of $\rd$ shows that therefore there must be a $j$ where one of the axioms $\nu_1$ to $\nu_{10}$ has been used (on-top-level) to reduce $e_j$ to $e_{j+1}$. We will go through the different axioms one-by-one.
\begin{hugeitemize}
\item[$\nu_1$:]
We have $e_j=\myneg\myneg b$ and $e_{j+1}=b$ for some $b$ where $a\rd b$.
Since $a\in\sn{}$ we obviously have $e_{j+1}=b\in\sn{}$ and therefore the reduction from $e_{j+1}$ must terminate.
\item[$\nu_2$:]
We have $e_j=\myneg[b,c]$ and $e_{j+1}=[\myneg b+\myneg c]$ for some $b$ and $c$ where $a\rd[b,c]$.
From $C_1$ we know that $\myneg b$, $\myneg c\in\sn{}$.
Hence by Law~\ref{sn.basic}($ii$) we know that $e_{j+1}=[\myneg b+\myneg c]\in\sn{}$ and therefore the reduction from $e_{j+1}$ must terminate.
\item[$\nu_3$:]
We have $e_j=\myneg[b+c]$ and $e_{j+1}=[\myneg b,\myneg c]$ for some $b$ and $c$ where $a\rd[b+c]$.
This argument for this case is similar to the one for $\nu_2$.
\item[$\nu_4$:]
We have $e_j=\myneg[x:b]c$ and $e_{j+1}=[x!b]\myneg c$ for some $b$ and $c$ where $a\rd[x:b]c$.
From $C_2$ we know that $\myneg c\in\sn{}$. 
Since $b\in\sn{}$, by Law~\ref{sn.basic}($ii$) we know that $e_{j+1}=[x!b]\myneg c\in\sn{}$ and therefore the reduction from $e_{j+1}$ must terminate.
\item[$\nu_5$:]
We have $e_j=\myneg[x!b]c$ and $e_{j+1}=[x:b]\myneg c$ for some $b$ and $c$ where $a\rd[x!b]c$.
This argument for this case is similar to the one for $\nu_4$.
\item[$\nu_i$,$6\leq i\leq 10$:]
In all these cases we have $e_j=\myneg b$ and $e_{j+1}=b$ for some $b$ where $a\rd b$
Since $a\in\sn{}$ we obviously have $e_{j+1}=b\in\sn{}$ and therefore the reduction from $e_{j+1}$ must terminate.
\end{hugeitemize}
In all cases we have derived that all reduction from $e_{j+1}$ must terminate.
This implies that the assumed infinite reduction from $\myneg a$ also terminates which contradicts our assumption.
\qedhere
\end{itemize}
\end{proof} 
\subsection{Norms and normable expressions}%
\label{norming}
Norms are a subset of expressions representing structural skeletons of expressions.
Norms play an important role to classify expressions in the course of the proof of strong normalization.
\begin{definition}[Norm]
\nomenclature[fBasic03]{$\dnrm$}{set of norms}
\nomenclature[bSets5]{$\bar{a}$, $\bar{b}$, $\bar{c}$, $\ldots$}{norms}
The set of norms $\dnrm$ is generated by the following rules
\begin{eqnarray*}
\dnrm&\!::=\!&\{\prim\}\,\mid\,[\dnrm,\dnrm]
\end{eqnarray*}
Obviously norms are a form of binary trees and $\dnrm\subset\dexp$.
We will use the notation $\bar{a}$, $\bar{b}$, $\bar{c}$, $\ldots$ to denote norms.
\end{definition}
\begin{definition}[Norming]
\index{norming}
\nomenclature[jNorm13]{$\nrm{\Gamma}{a}$}{norming}
The partial \emph{norming} function $\nrm{\Gamma}{a}$ defines for some expression $a$ the norm of $a$ under a context $\Gamma$.
It is defined by the equations in Table~\ref{norm.def}. 
\begin{table}[!htb]
\fbox{
\begin{minipage}{0.96\textwidth}
\begin{eqnarray*}
\\[-8mm]
\nrm{\Gamma}{\prim}&=&\prim\\
\nrm{\Gamma}{x}&=&\nrm{\Gamma}{\Gamma(x)}\quad\text{if}\;\Gamma(x)\;\text{is defined}\\
\nrm{\Gamma}{\binbop{x}{a}{b}}&=&[\nrm{\Gamma}{a},\nrm{\Gamma,x:a}{b}]\\
\nrm{\Gamma}{(a\,b)}&=&\bar{c}\quad\text{if}\;\nrm{\Gamma}{a}=[\nrm{\Gamma}{b},\bar{c}]\\
\nrm{\Gamma}{\prdef{x}{a}{b}{c}}&=&[\nrm{\Gamma}{a},\nrm{\Gamma}{b}]\quad\text{if}\;\nrm{\Gamma}{b}=\nrm{\Gamma,x:a}{c}\\
\nrm{\Gamma}{\prsumop{a}{b}}&=&[\nrm{\Gamma}{a},\nrm{\Gamma}{b}]\\
\nrm{\Gamma}{\pleft{a}}&=&\bar{a}\quad\text{if}\;\nrm{\Gamma}{a}=[\bar{a},\bar{b}]\\
\nrm{\Gamma}{\pright{a}}&=&\bar{b}\quad\text{if}\;\nrm{\Gamma}{a}=[\bar{a},\bar{b}]\\
\nrm{\Gamma}{\injl{a}{b}}&=&[\nrm{\Gamma}{a},\nrm{\Gamma}{b}]\\
\nrm{\Gamma}{\injr{a}{b}}&=&[\nrm{\Gamma}{a},\nrm{\Gamma}{b}]\\
\nrm{\Gamma}{\case{a}{b}}&=&[[\bar{a},\bar{b}],\bar{c}]\quad\text{if}\;\nrm{\Gamma}{a}=[\bar{a},\bar{c}],\;\nrm{\Gamma}{b}=[\bar{b},\bar{c}]\\
\nrm{\Gamma}{\myneg a}&=&\nrm{\Gamma}{a}
\end{eqnarray*}
\end{minipage}
}
\caption{Norming\label{norm.def}}
\end{table}
The partial norming function is well-defined, in the sense that one can show by structural induction on $a$ that, if defined, $\nrm{\Gamma}{a}$ is unique. 
\end{definition}
\begin{definition}[Normable expression]
\label{normable}
\index{expression!normable}
\nomenclature[jNorm13]{$\Gamma\sngv\, a$}{normable expression}
An expression $a$ is normable relative under context $\Gamma$ iff $\nrm{\Gamma}{a}$ is defined. 
This is written as $\Gamma\sngv a$.
Similarly to typing we use the notation $\Gamma\sngv a_1,\ldots,a_n$ as an abbreviation for $\Gamma\sngv a_1$, $\ldots$, $\Gamma\sngv a_n$.
Also we write $\sngv a$ if $\Gamma=()$.
\end{definition}
\begin{remark}[Examples]
There are valid and invalid normable expressions and there are strongly normalisable expression which are neither valid nor normable.
We present a few examples. 
We will show later that all valid expressions are normable (Law~\ref{val.nrm}) and that
all normable expressions are strongly normalizable (Laws~\ref{nrm.ce} and~\ref{sn.valid}).
Let $\Gamma =(p,q\!:\!\prim,\;z\!:\![x\!:\!p][y\!:\!q]\prim,\;w\!:\![x\!:\!\prim]x)$. 
Consider the expression $[x:p](z\,x)$:
\begin{itemize}
\item We have $\Gamma\sgv[x\!:\!p](z\,x)$.
\item We have $\Gamma\sngv[x\!:\!p](z\,x)$ since $\nrm{\Gamma}{[x\!:\!p](z\,x)}=[\prim,\nrm{\Gamma,x:p}{(z\,x)}]=[\prim,[\prim,\prim]]$. The latter equality is true since $\nrm{\Gamma,x:p}{z}=\nrm{\Gamma,x:p}{[x\!:\!p][y\!:\!q]\prim}=[\prim,[\prim,\prim]]$ and $\nrm{\Gamma,x:p}{x}=\nrm{\Gamma,x:p}{p}=\prim$.
\end{itemize}
Consider the expression $[x\!:\!p](z\,p)$:
\begin{itemize}
\item We do not have $\Gamma\sgv[x:p](z\,p)$ since we do not have $\Gamma\sgv p:p$.
\item We have $\Gamma\sngv[x:p](z\,p)$ since $\nrm{\Gamma}{[x:p](z\,p)}=[\prim,[\prim,\prim]]$ and $\nrm{\Gamma,x:p}{p}=\prim$ 
\end{itemize}
As a third example consider the expression $r=[x:[y:\prim]y](x\,x)$: 
\begin{itemize}
\item Obviously $r\in\sn{}$.
\item We do not have $()\sgv r$ since the application $(x\,x)$ cannot be typed.
\item We do not have $\sngv r$: The definition of norming yields  $\nrm{}{[x\!:\![y\!:\!\prim]y](x\,x)}=[[\prim,\prim],\nrm{x:[y:\prim]y}{(x\,x)}]$. 
However the expression $\nrm{x:[y:\prim]y}{(x\,x)}$ is not defined since $\nrm{x:[y:\prim]y}{x}=[\prim,\prim]$ which implies
$\nrm{x:[y:\prim]y}{x}\neq[\nrm{x:[y:\prim]y}{x},\bar{a}]$ for any $\bar{a}$.
Hence the norming condition for application is violated.
\end{itemize}
\end{remark}
\subsection{Properties of normable expressions}%
We show several basic properties of normable expressions culminating in the property that all valid expressions are normable. 
Some of these properties and proofs are structurally similar to the corresponding ones for valid expressions.
However, due to the simplicity of norms, the proofs are much shorter. 
%
%
\begin{law}[Norm equality in context]
\label{nrm.eq}
Let $\Gamma_a=(\Gamma_1,x:a,\Gamma_2)$ and $\Gamma_b=(\Gamma_1,x:b,\Gamma_2)$ for some $\Gamma_1,\Gamma_2,x,a,b$. For all $c$: If $\Gamma_1\sngv a,b$, $\nrm{\Gamma_1}{a}=\nrm{\Gamma_1}{b}$, and  $\Gamma_a\sngv c$ then  $\Gamma_b\sngv c$ and $\nrm{\Gamma_a}{c}=\nrm{\Gamma_b}{c}$.
\end{law}
\begin{proof}
The straightforward proof is by structural induction on $c$.
The most interesting case $c$ being an variable. There are two cases:
\begin{itemize}
\item
If $c=x$ then obviously $\nrm{\Gamma_a}{c}=\nrm{\Gamma_1}{a}=\nrm{\Gamma_1}{b}=\nrm{\Gamma_b}{c}$.
\item
If $c=y\neq x$ then $\Gamma_a(y)=\Gamma_b(y)$ and the property follows from the inductive hypothesis.
\qedhere
\end{itemize}
\end{proof}
\begin{law}[Substitution and norming]
\label{nrm.sub}
Let $\Gamma_a=(\Gamma_1,x:a,\Gamma_2)$ and $\Gamma_b=(\Gamma_1,\Gamma_2\gsub{x}{b})$ for some $\Gamma_1,\Gamma_2,x,a,b$. For all $c$: If $\Gamma_1\sngv a,b$, $\nrm{\Gamma_1}{a}=\nrm{\Gamma_1}{b}$, and  $\Gamma_a\sngv c$ then  $\Gamma_b\sngv c\gsub{x}{b}$ and $\nrm{\Gamma_a}{c}=\nrm{\Gamma_b}{c\gsub{x}{b}}$.
\end{law}
\begin{proof}
Proof is by structural induction on $c$.
\begin{giantitemize}
\item[$c=\prim$:] 
Obviously $c\gsub{x}{b}=c=\prim$, hence $\Gamma_b\sngv c\gsub{x}{b}$ and $\nrm{\Gamma_a}{c}=\prim=\nrm{\Gamma_b}{c}$.
\item[$c=y$:] 
If $y=x$ then obviously $c\gsub{x}{b}=b$ and therefore $\nrm{\Gamma_a}{x}=\nrm{\Gamma_1}{a}=\nrm{\Gamma_1}{b}=\nrm{\Gamma_b}{c\gsub{x}{b}}$ which also implies
 $\Gamma_b\sngv c\gsub{x}{b}$.

If $x\neq y$ then $c\gsub{x}{b}=y$. There are two subcases:
\begin{itemize}
\item
$\Gamma_1(y)$ is defined. $\Gamma_1(y)=d$. Obviously $\nrm{\Gamma_a}{y}=d=\nrm{\Gamma_b}{y}$.
\item
$\Gamma_1(y)$ is not defined and hence $\Gamma_2(y)=d$, for some $d$.
This obviously implies $(\Gamma_2\gsub{x}{b})(y)=d\gsub{x}{b}$.  
We can argue as follows:
\begin{eqnarray*}
\nrm{\Gamma_a}{y}
&=&\quad\text{(definition of norming)}\\
&&\nrm{\Gamma_a}{\Gamma_a(y)}\\
&=&\quad\text{(see above)}\\
&&\nrm{\Gamma_a}{d}\\
&=&\quad\text{(inductive hypothesis)}\\
&&\nrm{\Gamma_b}{d\gsub{x}{b}}\\
&=&\quad\text{(see above)}\\
&&\nrm{\Gamma_b}{(\Gamma_2\gsub{x}{b})(y)}\\
&=&\quad\text{(definition of lookup of variable)}\\
&&\nrm{\Gamma_b}{\Gamma_b(y)}\\
&=&\quad\text{(definition of norming)}\\
&&\nrm{\Gamma_b}{y}
\end{eqnarray*}
\end{itemize}
The arguments in both cases also imply $\Gamma_b\sngv c\gsub{x}{b}$.
\item[$c=\binbop{y}{c_1}{c_2}$:]
As usual we may assume $y\neq x$.
We can argue as follows:
\begin{eqnarray*}
&&\nrm{\Gamma_a}{\binbop{y}{c_1}{c_2}}\\
&=&\quad\text{(definition of norming)}\\
&&[\nrm{\Gamma_a}{c_1},\nrm{\Gamma_a,y:c_1}{c_2}]\\
&=&\quad\text{(inductive hypothesis)}\\
&&[\nrm{\Gamma_b}{c_1\gsub{x}{b}},\nrm{\Gamma_b,y:c_1\gsub{x}{b}}{c_2\gsub{x}{b}}]\\
&=&\quad\text{(definition of norming)}\\
&&\nrm{\Gamma_b}{\binbop{y}{c_1\gsub{x}{b}}{c_2\gsub{x}{b}}}\\
&=&\quad\text{(definition of substitution)}\\
&&\nrm{\Gamma_b}{\binbop{y}{c_1}{c_2}\gsub{x}{b}}
\end{eqnarray*}
This argument also implies $\Gamma_b\sngv c\gsub{x}{b}$.
\item[$c\!=\!\prdef{y}{c_1}{c_2\!}{\!c_3}$:]
As usual we may assume $y\neq x$.
As in the case of $c=\binbop{y}{c_1}{c_2}$ the proposition follows from the inductive hypothesis and the definition of substitution:
\begin{eqnarray*}
&&\nrm{\Gamma_a}{\prdef{y}{c_1}{c_2}{c_3}}\\
&=&\quad\text{(definition of norming)}\\
&&[\nrm{\Gamma_a}{c_1},\nrm{\Gamma_a}{c_2}]\\
&&\quad\text{(where $\nrm{\Gamma_a}{c_2}=\nrm{\Gamma_a,x:c_1}{c_3}$)}\\
&=&\quad\text{(inductive hypothesis)}\\
&&[\nrm{\Gamma_b}{c_1\gsub{x}{b}},\nrm{\Gamma_b}{c_2\gsub{x}{b}}]\\
&=&\quad\text{(definition of norming)}\\
&&\nrm{\Gamma_b}{\binbop{y}{c_1\gsub{x}{b}}{c_2\gsub{x}{b}}}\\
&=&\quad\text{(since the ind. hyp. implies}\\
&&\quad\text{$\nrm{\Gamma_b}{c_2\gsub{x}{b}}=\nrm{\Gamma_b,x:c_1}{c_3}\gsub{x}{b}$)}\\
&&\nrm{\Gamma_b}{\prdef{y}{c_1\gsub{x}{b}}{c_2\gsub{x}{b}}{c_3\gsub{x}{b}}}\\
&=&\quad\text{(definition of substitution)}\\
&&\nrm{\Gamma_b}{\prdef{y}{c_1}{c_2}{c_3}\gsub{x}{b}}
\end{eqnarray*}
This argument also implies $\Gamma_b\sngv c\gsub{x}{b}$.
\item[$c=(c_1\,c_2)$:]
We have $\nrm{\Gamma_a}{(c_1\,c_2)}=\bar{c}$ where $\nrm{\Gamma_a}{c_1}=[\nrm{\Gamma_a}{c_2},\bar{c}]$. 
By inductive hypothesis $\nrm{\Gamma_a}{c_1}=\nrm{\Gamma_b}{c_1\gsub{x}{b}}$ and $\nrm{\Gamma_a}{c_2}=\nrm{\Gamma_b}{c_2\gsub{x}{b}}$. 
Hence $\nrm{\Gamma_b}{c_1\gsub{x}{b}}=[\nrm{\Gamma_b}{c_2\gsub{x}{b}},\bar{c}]$.
This implies $\nrm{\Gamma_b}{(c_1\,c_2)\gsub{x}{b}}=\bar{c}$ and therefore also $\Gamma_b\sngv c\gsub{x}{b}$.
\item[$c=\pleft{c_1}$:]
We have $\nrm{\Gamma_a}{c}=\bar{c}_2$ where $\nrm{\Gamma_a}{c_1}=[\bar{c}_2,\bar{c}_3]$.
By inductive hypothesis $\nrm{\Gamma_a}{c_1}=\nrm{\Gamma_b}{c_1\gsub{x}{b}}$.
Hence $\nrm{\Gamma_b}{c\gsub{x}{b}}=\bar{c}_2$ and therefore also $\Gamma_b\sngv c\gsub{x}{b}$.
\item[$c=\pright{c_1}$:]
Similar (symmetric) to case $c=\pleft{c_1}$.
\item[$c=[c_1,c_2{]}$:]
We have $\nrm{\Gamma_a}{c}=[\nrm{\Gamma_a}{c_1},\nrm{\Gamma_a}{c_2}]$. The proposition follows from the inductive hypothesis and the definition of substitution.
\item[$c=[c_1+c_2{]}$:]
Similar (symmetric) to case $c=[c_1,c_2]$. 
\item[$c=\injl{c_1}{c_2}$:]
Similar (symmetric) to case $c=[c_1,c_2]$. 
\item[$c=\injr{c_1}{c_2}$:]
Similar (symmetric) to case $c=[c_1,c_2]$. 
\item[$c=\case{c_1}{c_2}$:] 
We have $\nrm{\Gamma_a}{c}=[[\bar{c}_1,\bar{c}_2],\bar{d}]$ where $\nrm{\Gamma_a}{c_1}=[\bar{c}_1,\bar{d}]$ as well as $\nrm{\Gamma_a}{c_2}=[\bar{c}_2,\bar{d}]$.
By inductive hypothesis $\nrm{\Gamma_b}{c_1\gsub{x}{b}}=[\bar{c}_1,\bar{d}]$ and $\nrm{\Gamma_b}{c_2\gsub{x}{b}}=[\bar{c}_2,\bar{d}]$.
Hence obviously $\nrm{\Gamma_b}{c\gsub{x}{b}}=[[\bar{c}_1,\bar{c}_2],\bar{d}]$ and therefore also $\Gamma_b\sngv c\gsub{x}{b}$.
\item[$c=\myneg c_1$:]
The proposition follows from the inductive hypothesis and the definition of substitution.
\qedhere
\end{giantitemize}
\end{proof}
\begin{law}[Reduction preserves norms]
\label{nrm.rd}
For all $\Gamma,a,b$:
$\Gamma\sngv a$ and $a\rd b$ implies $\nrm{\Gamma}{a}=\nrm{\Gamma}{b}$
\end{law}
\begin{proof}
It is obviously sufficient to show the property for single-step reduction which we do here by induction on the definition of single-step reduction.
We begin with the axioms:
\begin{meditemize}
\item[$\beta_1$:] 
We have $a=([x:a_1]a_2\:a_3)$, $b=a_2\gsub{x}{a_3}$, and $\nrm{\Gamma}{([x:a_1]a_2\:a_3)}=\nrm{\Gamma,x:a_1}{a_2}$ where $\nrm{\Gamma}{a_1}=\nrm{\Gamma}{a_3}$.
Therefore by Law~\ref{nrm.sub} we know that $\nrm{\Gamma,x:a_1}{a_2}=\nrm{\Gamma}{a_2\gsub{x}{a_3}}$ which implies the proposition.
\item[$\beta_2$:]
Similar to case $\beta_1$.
\item[$\beta_3$:]
We have $a=(\case{a_1}{a_2}\,\injl{a_3}{a_4})$ and $b=(a_1\,a_3)$.
By definition of norming $\nrm{\Gamma}{\injl{a_3}{a_4}}=[\nrm{\Gamma}{a_3},\nrm{\Gamma}{a_4}]$ and
$\nrm{\Gamma}{\case{a_1}{a_2}}=[[\nrm{\Gamma}{a_3},\nrm{\Gamma}{a_4}],\bar{c}]$ for some $\bar{c}$ where
$\nrm{\Gamma}{a_1}=[\nrm{\Gamma}{a_3},\bar{c}]$ as well as $\nrm{\Gamma}{a_2}=[\nrm{\Gamma}{a_4},\bar{c}]$.
Hence $\nrm{\Gamma}{a}=\bar{c}=\nrm{\Gamma}{(a_1\,a_3)}=\nrm{\Gamma}{b}$.
\item[$\beta_4$:]
Similar to case $\beta_3$.
\item[$\pi_1$:]
We have $a=\pleft{\prdef{x}{a_1}{a_2}{a_3}}$ and $b=a_1$.
By definition of norming we directly obtain $\nrm{\Gamma}{a}=\nrm{\Gamma}{a_1}=\nrm{\Gamma}{b}$.
\item[$\pi_2$:]
Similar to case $\pi_1$.
\item[$\pi_3$:]
We have $a=\pleft{[a_1,a_2]}$ and $b=a_1$.
By definition of norming we directly obtain $\nrm{\Gamma}{a}=\nrm{\Gamma}{a_1}=\nrm{\Gamma}{b}$.
\item[$\pi_4$:]
Similar to case $\pi_3$.
\item[$\pi_5$:]
We have $a=\pleft{[a_1+a_2]}$ and $b=a_1$.
By definition of norming we directly obtain $\nrm{\Gamma}{a}=\nrm{\Gamma}{a_1}=\nrm{\Gamma}{b}$.
\item[$\pi_6$:]
Similar to case $\pi_3$.
\item[$\nu_1$]
We have  $a=\myneg\myneg a_1$ and $b=a_1$.
By definition of norming we directly obtain $\nrm{\Gamma}{a}=\nrm{\Gamma}{a_1}=\nrm{\Gamma}{b}$.
\item[$\nu_2$:] 
We have $a=\myneg[x:a_1]a_2$ and $b=[x!a_1]\myneg a_2$.
By definition of norming we have $\nrm{\Gamma}{\myneg[x:a_1]a_2}=[\nrm{\Gamma}{a_1},\nrm{\Gamma,x:a_1}{a_2}]=\nrm{\Gamma}{[x!a_1]\myneg a_2}$.
\item[$\nu_3$:] 
Similar to case $\nu_2$.
\item[$\nu_4$:] 
We have $a=\myneg[a_1,a_2]$ and $b=[\myneg a_1,\myneg a_2]$.
By definition of norming we have $\nrm{\Gamma}{\myneg[a_1,a_2]}=[\nrm{\Gamma}{a_1},\nrm{\Gamma}{a_2}]=\nrm{\Gamma}{[\myneg a_1,\myneg a_2]}$
\item[$\nu_5$:] 
Similar to case $\nu_4$.
\item[$\nu_6$:] 
We have $a=\myneg\prim$ and $b=\prim$.
By definition of norming we directly obtain $\nrm{\Gamma}{a}=\prim=\nrm{\Gamma}{b}$.
\item[$\nu_i$:] 
$i>6$: Similar to case $\nu_6$.
%
\end{meditemize}
Next, we turn to the structural rules.
\begin{hugeitemize}
\item[$\binbop{x}{\_}{\_}_1$:]
We have $a=\binbop{x}{a_1}{a_3}$, $b=\binbop{x}{a_2}{a_3}$, and $a_1\srd a_2$.
By definition of norming $\nrm{\Gamma}{a}=[\nrm{\Gamma}{a_1},\nrm{\Gamma,x:a_1}{a_3}]$.
By inductive hypothesis $\nrm{\Gamma}{a_1}=\nrm{\Gamma}{a_2}$. 
Hence we can argue that $\nrm{\Gamma}{a}=[\nrm{\Gamma}{a_1},\nrm{\Gamma,x:a_1}{a_3}]$.
By Law~\ref{nrm.eq} we know that $\nrm{\Gamma,x:a_1}{a_3}\!=\nrm{\Gamma,x:a_2}{a_3}$ hence $\nrm{\Gamma}{a}=\nrm{\Gamma}{b}$.
\item[$\binbop{x}{\_}{\_}_2$:]
We have $a=\binbop{x}{a_1}{a_2}$, $b=\binbop{x}{a_1}{a_3}$, and $a_2\srd a_3$.
By definition of norming $\nrm{\Gamma}{a}=[\nrm{\Gamma}{a_1},\nrm{\Gamma,x:a_1}{a_2}]$.
By inductive hypothesis $\nrm{\Gamma,x:a_1}{a_2}=\nrm{\Gamma,x:a_1}{a_3}$. 
Hence we can argue that $\nrm{\Gamma}{a}=[\nrm{\Gamma}{a_1},\nrm{\Gamma,x:a_1}{a_2}]=[\nrm{\Gamma}{a_1},\nrm{\Gamma,x:a_1}{a_3}]=\nrm{\Gamma}{b}$.
\item[${\prdef{x}{\_}{\_}{\_}}_1$:]
We have $a=\prdef{x}{a_1}{a_3}{a_4}$, $b=\prdef{x}{a_2}{a_3}{a_4}$, and $a_1\srd a_2$.
By definition of norming $\nrm{\Gamma}{a}=[\nrm{\Gamma}{a_1},\nrm{\Gamma}{a_3}]$ where $\nrm{\Gamma}{a_4}=\nrm{\Gamma,x:a_1}{a_4}$.
By inductive hypothesis $\nrm{\Gamma}{a_1}=\nrm{\Gamma}{a_2}$.
By Law~\ref{nrm.eq} we know that $\nrm{\Gamma,x:a_1}{a_4}=\nrm{\Gamma,x:a_2}{a_4}$.
Hence we can argue that $\nrm{\Gamma}{a}=[\nrm{\Gamma}{a_1},\nrm{\Gamma}{a_3}]=[\nrm{\Gamma}{a_2},\nrm{\Gamma}{a_3}]=\nrm{\Gamma}{b}$.
\item[${\prdef{x}{\_}{\_}{\_}}_2$:]
We have $a=\prdef{x}{a_1}{a_2}{a_4}$, $b=\prdef{x}{a_1}{a_3}{a_4}$, and $a_2\srd a_3$.
By definition of norming $\nrm{\Gamma}{a}=[\nrm{\Gamma}{a_1},\nrm{\Gamma}{a_2}]$ where $\nrm{\Gamma}{a_2}=\nrm{\Gamma,x:a_1}{a_4}$.
By inductive hypothesis $\nrm{\Gamma}{a_2}=\nrm{\Gamma}{a_3}$.
Hence we can argue that $\nrm{\Gamma}{a}=[\nrm{\Gamma}{a_1},\nrm{\Gamma}{a_2}]=[\nrm{\Gamma}{a_1},\nrm{\Gamma}{a_3}]=\nrm{\Gamma}{b}$.
\item[${\prdef{x}{\_}{\_}{\_}}_3$:]
We have $a=\prdef{x}{a_1}{a_2}{a_3}$, $b=\prdef{x}{a_1}{a_2}{a_4}$, and $a_3\srd a_4$.
By definition of norming $\nrm{\Gamma}{a}=[\nrm{\Gamma}{a_1},\nrm{\Gamma}{a_2}]$ where $\nrm{\Gamma}{a_2}=\nrm{\Gamma,x:a_1}{a_3}$.
By inductive hypothesis $\nrm{\Gamma,x:a_1}{a_3}=\nrm{\Gamma,x:a_1}{a_4}$.
Hence we can argue that $\nrm{\Gamma}{a}=[\nrm{\Gamma}{a_1},\nrm{\Gamma}{a_2}]=\nrm{\Gamma}{b}$.
\item[$\mathit{(\oplus{\overbrace{(\_,\ldots,\_)}^{n}}_i)}$:]
We have $a=\oplus(a_1,\ldots,a_i,\ldots,a_n)$ and $b=\oplus(a_1,\ldots,b_i,\ldots,a_n)$ where $a_i\srd b_i$.
From $\Gamma\sngv a$ it obviously follows that $\Gamma\sngv a_1$.
By inductive hypothesis $\nrm{\Gamma}{a_1}=\nrm{\Gamma}{b_1}$.
By definition of norming this implies $\nrm{\Gamma}{a}=\nrm{\Gamma}{b}$.
\qedhere
\end{hugeitemize}
\end{proof}
\begin{law}[Context extension]
\label{norm.ext}
For all $\Gamma_1,\Gamma_2,x,a,b$ where $x\notin\dom(\Gamma_1,\Gamma_2)$:
$\Gamma_1,\Gamma_2\sngv a$ implies $(\Gamma_1,x:b,\Gamma_2)\sngv a$ and $\nrm{\Gamma_1,\Gamma_2}{a}=\nrm{\Gamma_1,x:b,\Gamma_2}{a}$.  
\end{law}
\begin{proof}
Obviosly $\Gamma_1,\Gamma_2\sngv a$ implies that $\free([\Gamma_1,\Gamma_2]a)=\emptyset$. 
Therefore the additional declaration $x:b$ will never be used when evaluating $\nrm{\Gamma_1,x:b,\Gamma_2}{a}$.
Hence the successful evaluation of $\nrm{\Gamma_1,\Gamma_2}{a}$ can be easily transformed into an evaluation of $\nrm{\Gamma_1,x:b,\Gamma_2}{a}$ with identical result.
\end{proof}
\begin{law}[Typing implies normability and preserves norm]
\label{nrm.dtyp}
For all $\Gamma,a,b$:
If $\Gamma\sgv a:b$ then $\sngv[\Gamma]a,[\Gamma]b$ and $\nrm{\Gamma}{a}=\nrm{\Gamma}{b}$.
\end{law}
\begin{proof}
Proof by induction on the definition of $\Gamma\sgv a:b$.
Note that $\sngv[\Gamma]a$ implies $\Gamma\sngv a$ but not vice-versa.
The stronger conclusion $\sngv[\Gamma]a$ is needed \eg\ in the rule \absu.
\begin{meditemize}
\item[\ax:] 
Obvious, since $a=b=\prim$ and $\Gamma=()$.
\item[\mystart:] 
We have $a=x$, $\Gamma=(\Gamma',x:b)$ for some $\Gamma'$ and $x$, and $\Gamma'\sgv b:c$ for some $c$.
By inductive hypothesis $\sngv[\Gamma']b$ and hence obviously $\Gamma'\sngv b$.
Since $x\notin\free(b)$, by Law~\ref{norm.ext} we know that  $\Gamma\sngv b$ and $\nrm{\Gamma}{b}=\nrm{\Gamma'}{b}$.
Hence, by definition of norming $\sngv[\Gamma]x$, where $\nrm{\Gamma}{x}=\nrm{\Gamma}{b}$.
\item[\weak:]
We have $\Gamma=(\Gamma',x:c)$ for some $\Gamma'$, $x$, and $c$ where $\Gamma'\sgv c$ and $\Gamma'\sgv a:b$. 
By inductive hypothesis $\sngv[\Gamma']a,[\Gamma']b,[\Gamma']c$ (obviously this implies $\Gamma'\sngv a,b,c$) and $\nrm{\Gamma'}{a}=\nrm{\Gamma'}{b}$.
Since $x\notin\free(a)\cup\free(b)$, by Law~\ref{norm.ext} we know that $\Gamma\sngv a$ and $\nrm{\Gamma}{a}=\nrm{\Gamma'}{a}$ as well as
$\Gamma\sngv b$ and $\nrm{\Gamma}{b}=\nrm{\Gamma'}{b}$.
Hence by definition of norming $\sngv[\Gamma]a$,$[\Gamma]b$ and $\nrm{\Gamma}{a}=\nrm{\Gamma}{b}$.
\item[\conv:]
We have $\Gamma\sgv a:b$ where $\Gamma\sgv a:c$ , $c\eqv b$, and $\Gamma\sgv b:d$ for some $c$ and $d$ (note that we are here using the rule with $b$ and $c$ exchanged).
By inductive hypothesis  $\sngv[\Gamma]a,[\Gamma]b,[\Gamma]c$ (obviously this implies $\Gamma\sngv a,b,c$) and $\nrm{\Gamma}{a}=\nrm{\Gamma}{c}$. 
Hence by Laws~\ref{rd.confl} and~\ref{nrm.rd} we know that $\nrm{\Gamma}{c}=\nrm{\Gamma}{b}$ which implies $\nrm{\Gamma}{a}=\nrm{\Gamma}{b}$.
\item[\absu:]
We have $a=[x:c]a_1$ and $b=[x:c]a_2$, for some $c$, $a_1$, and $a_2$ where $(\Gamma,x:c)\sgv a_1:a_2$. 
By inductive hypothesis we know that $\sngv[\Gamma,x:c]a_1,[\Gamma,x:c]a_2$ and $\nrm{\Gamma,x:c}{a_1}=\nrm{\Gamma,x:c}{a_2}$. 
Obviously this implies $\sngv[\Gamma][x:c]a_1$ and $[\Gamma][x:c]a_2$.  
Furthermore, \eg~$\sngv[\Gamma][x:c]a_1$ implies $\sngv[\Gamma]c$ and therefore
by definition of norming $\nrm{\Gamma}{a}=[\nrm{\Gamma}{c},\nrm{\Gamma,x:c}{a_1}]=[\nrm{\Gamma}{c},\nrm{\Gamma,x:c}{a_2}]=\nrm{\Gamma}{b}$.
\item[\abse:] 
Similar to case \absu.
\item[\appl:] 
We have $a=(a_1\,a_2)$ and $b=c_2\gsub{x}{a_2}$, for some $a_1$, $a_2$, $x$, and $c_2$ where $\Gamma\sgv a_1:[x:c_1]c_2$ and $\Gamma\sgv a_2:c_1$ for some $c_1$.
By inductive hypothesis we know that $\sngv[\Gamma]a_1,[\Gamma]a_2,[\Gamma][x:c_1]c_2,[\Gamma]c_1$ (obviously this implies $\Gamma\sngv a_1,a_2,[x:c_1]c_2,c_1$) and
$\nrm{\Gamma}{[x:c_1]c_2}=\nrm{\Gamma}{a_1}$ as well as $\nrm{\Gamma}{c_1}=\nrm{\Gamma}{a_2}$.
Hence $\nrm{\Gamma}{a_1}=[\nrm{\Gamma}{a_2},\nrm{\Gamma,x:c_1}{c_2}]$ and therefore by definition of norming $\nrm{\Gamma}{(a_1\,a_2)}=\nrm{\Gamma,x:c_1}{c_2}$ which implies $\Gamma\sngv a$ and (obviously) $\sngv[\Gamma]a$.
Since $\nrm{\Gamma}{a_2}=\nrm{\Gamma}{c_1}$ we can apply Law~\ref{nrm.sub} to obtain $\nrm{\Gamma,x:c_1}{c_2}=\nrm{\Gamma}{c_2\gsub{x}{a_2}}$.
Hence $\Gamma\sngv b$ and (obviously) $\sngv[\Gamma]b$ and $\nrm{\Gamma}{a}=\nrm{\Gamma}{(a_1\,a_2)}=\nrm{\Gamma}{c_2\gsub{x}{a_2}}=\nrm{\Gamma}{b}$.
\item[\pdef:] 
We have $a=\prdef{x}{a_1}{a_2}{a_3}$ and  $b=[x!b_1]a_3$, for some $x$, $a_1$, $a_2$, $a_3$, and $b_1$ where $\Gamma\sgv a_1:b_1$, $\Gamma\sgv a_2:a_3\gsub{x}{a_1}$, and $(\Gamma,x:b_1)\sgv a_3:a_4$.

By inductive hypothesis we know that $\sngv[\Gamma]a_1$, $[\Gamma]b_1$, $[\Gamma]a_2$, $[\Gamma]a_3\gsub{x}{a_1}$, $[\Gamma,x:b_1]a_3$, $[\Gamma,x:b_1]a_4$, and therefore $\Gamma\sngv a_1,b_1,a_2,a_3\gsub{x}{a_1}$ and $(\Gamma,x:b_1)\sngv a_3,a_4$, where $\nrm{\Gamma}{a_1}=\nrm{\Gamma}{b_1}$ and $\nrm{\Gamma}{a_2}=\nrm{\Gamma}{a_3\gsub{x}{a_1}}$.

Since $\nrm{\Gamma}{a_1}=\nrm{\Gamma}{b_1}$, by Law~\ref{nrm.sub} we know that $\nrm{\Gamma,x:b_1}{a_3}=\nrm{\Gamma}{a_3\gsub{x}{a_1}}$.
By Law~\ref{nrm.eq} we know that $\nrm{\Gamma,x:b_1}{a_3}=\nrm{\Gamma,x:a_1}{a_3}$ and therefore we obtain $\nrm{\Gamma}{a_2}=\nrm{\Gamma,x:a_1}{a_3}$.
Therefore, by definition of norming we know that $\Gamma\sngv a$ and $\nrm{\Gamma}{a}=[\nrm{\Gamma}{a_1},\nrm{\Gamma}{a_2}]$.

By definition of norming $\Gamma\sngv b$ and (obviously) $\sngv[\Gamma]b$ and $\nrm{\Gamma}{b}=[\nrm{\Gamma}{b_1},\nrm{\Gamma,x:b_1}{a_3}]=[\nrm{\Gamma}{a_1},\nrm{\Gamma,x:a_1}{a_3}]=[\nrm{\Gamma}{a_1},\nrm{\Gamma}{a_2}]=\nrm{\Gamma}{a}$.
\item[\chin:]
We have $a=\pleft{a_1}$ where $\Gamma\sgv a_1:[x!b]a_2$ for some $a_2$. 
By inductive hypothesis we know that $\sngv[\Gamma]a_1,[\Gamma][x!b]a_2$, and therefore$\Gamma\sngv a_1,[x!b]a_2$, and $\nrm{\Gamma}{a_1}=\nrm{\Gamma}{[x!b]a_2}$.
$[\Gamma][x!b]a_2$ obviously implies $\sngv[\Gamma]b$ and $\Gamma\sngv b$.
Therefore, by definition of norming $\nrm{\Gamma}{[x!b]a_2}=[\nrm{\Gamma}{b},\nrm{\Gamma,x:b}{a_2}]$ and $\nrm{\Gamma}{\pleft{a_1}}=\nrm{\Gamma}{b}$.
Therefore $\Gamma\sngv a$ and (obviously) $\sngv[\Gamma]a$ and $\nrm{\Gamma}{a}=\nrm{\Gamma}{b}$.
\item[\chba:]
We have $a=\pright{a_1}$ and $b=a_3\gsub{x}{\pleft{a_1}}$ for some $a_1$, $x$,  and$a_3$ where $\Gamma\sgv a_1:[x!a_2]a_3$ for some $a_2$.
By inductive hypothesis we know that $\sngv [\Gamma]a_1$, $[\Gamma][x!a_2]a_3$, and therefore $\Gamma\sngv a_1,[x!a_2]a_3$, and $\nrm{\Gamma}{a_1}=\nrm{\Gamma}{[x!a_2]a_3}$.
$[\Gamma][x!a_2]a_3$ obviously implies $\Gamma\sngv a_2$.
By definition of the norming we have $\nrm{\Gamma}{[x!a_2]a_3}=[\nrm{\Gamma}{a_2},\nrm{\Gamma,x:a_2}{a_3}]$, $\nrm{\Gamma}{\pleft{a_1}}=\nrm{\Gamma}{a_2}$, and $\nrm{\Gamma}{\pright{a_1}}=\nrm{\Gamma,x:a_2}{a_3}$.
Hence we can apply Law~\ref{nrm.sub} to obtain $\nrm{\Gamma,x:a_2}{a_3}=\nrm{\Gamma}{a_3\gsub{x}{\pleft{a_1}}}$.
Hence $\Gamma\sngv a,b$ and (obviously) $\sngv[\Gamma]a,[\Gamma]b$, where $\nrm{\Gamma}{a}=\nrm{\Gamma,x:a_2}{a_3}=\nrm{\Gamma}{a_3\gsub{x}{\pleft{a_1}}}=\nrm{\Gamma}{b}$.
\item[\bprod:]
We have $a=[a_1,a_2]$ and $b=[b_1,b_2]$ for some $a_1$, $a_2$, $b_1$, and $b_2$ where $\Gamma\sgv a_1:b_1$ and $\Gamma\sgv a_2:b_2$.
By inductive hypothesis we know that $\sngv[\Gamma]a_1,[\Gamma]b_1$, and therefore $\Gamma\sngv a_1,b_1$, and $\nrm{\Gamma}{a_1}=\nrm{\Gamma}{b_1}$, as well as $\sngv[\Gamma]a_2,[\Gamma]b_2$, and therefore $\Gamma\sngv a_2,b_2$, and $\nrm{\Gamma}{a_2}=\nrm{\Gamma}{b_2}$.
Hence $\Gamma\sngv a$ and (obviously) $\sngv[\Gamma]a$ and $\nrm{\Gamma}{a}=[\nrm{\Gamma}{a_1},\nrm{\Gamma}{a_2}]$.
Similarly $\Gamma\sngv b$ and (obviously) $\sngv[\Gamma]b$ and $\nrm{\Gamma}{b}=[\nrm{\Gamma}{b_1},\nrm{\Gamma}{b_2}]$.
Hence $\nrm{\Gamma}{a}=[\nrm{\Gamma}{a_1},\nrm{\Gamma}{a_2}]=[\nrm{\Gamma}{b_1},\nrm{\Gamma}{b_2}]=\nrm{\Gamma}{b}$.
\item[\bsum:]
Similar to case \bprod.
\item[\prl:]
We have $a=\pleft{a_1}$ where $\Gamma\sgv a_1:[b,b_2]$ for some $b_2$.
By inductive hypothesis we know that $\sngv[\Gamma]a_1$, $[\Gamma][b,b_2]$, and therefore $\Gamma\sngv a_1,[b,b_2]$, and $\nrm{\Gamma}{a_1}=\nrm{\Gamma}{[b,b_2]}=[\nrm{\Gamma}{b},\nrm{\Gamma}{b_2}]$.
Hence $\sngv[\Gamma]a$, $[\Gamma]b$ and $\nrm{\Gamma}{a}=\nrm{\Gamma}{b}$.
\item[\prr:]
Similar (symmetric) to case \prl. 
\item[\injll:]
We have $a=\injl{a_1}{a_2}$ and $b=[b_1,a_2]$ where $\Gamma\sgv a_1:b_1$ and $\Gamma\sgv a_2:c_2$ for some $c_2$.
By inductive hypothesis we know that $\sngv[\Gamma]a_1$, $[\Gamma]b_1$, and $[\Gamma]a_2$, and therefore $\Gamma\sngv a_1,b_1,a_2$, and $\nrm{\Gamma}{a_1}=\nrm{\Gamma}{b_1}$.
Hence $\Gamma\sngv a$ and (obviously) $\sngv[\Gamma]a$ and $\nrm{\Gamma}{a}=[\nrm{\Gamma}{a_1},\nrm{\Gamma}{a_2}]$.
Similarly $\Gamma\sngv b$ and (obviously) $\sngv[\Gamma]a$ and $\nrm{\Gamma}{b}=[\nrm{\Gamma}{b_1},\nrm{\Gamma}{a_2}]$.
Hence $\nrm{\Gamma}{a}=[\nrm{\Gamma}{a_1},\nrm{\Gamma}{a_2}]=[\nrm{\Gamma}{b_1},\nrm{\Gamma}{a_2}]=\nrm{\Gamma}{b}$.
\item[\injlr:]
Similar to case \injll.
\item[\cased:] 
We have $a=\case{a_1}{a_2}$ and $b=[z:[b_1+b_2]]b_3$ for some $a_1$, $a_2$, $b_1$, $b_2$, $b_3$, and $z$ where $\Gamma\sgv a_1:[x:b_1]b_3$, $\Gamma\sgv a_2:[y:b_2]b_3$, and $\Gamma\sgv b_3:c$ for some $x$, $y$, and $c$. 
By inductive hypothesis we know that $\sngv[\Gamma]a_1$, $[\Gamma]a_2$, $[\Gamma][x:b_1]b_3$, $[\Gamma]y:b_2]b_3$, $[\Gamma]b_3$, and therefore $\Gamma\sngv a_1,a_2,[x:b_1]b_3,[y:b_2]b_3,b_3$.
This obviously implies $\sngv[\Gamma]b_1$, $[\Gamma]b_2$ as well as $\Gamma\sngv b_1, b_2$.

$\Gamma\sgv b_3:c$ obviously implies $x,y,z\notin\free(b_3)$. Hence for any $d$ we have $\nrm{\Gamma,x:d}{b_3}=\nrm{\Gamma}{b_3}$
Therefore and by inductive hypothesis and the definition of norming we know that $\nrm{\Gamma}{a_1}=\nrm{\Gamma}{[x:b_1]b_3}=[\nrm{\Gamma}{b_1},\nrm{\Gamma}{b_3}]$ and $\nrm{\Gamma}{a_2}=\nrm{\Gamma}{[y:b_2]b_3}=[\nrm{\Gamma}{b_2},\nrm{\Gamma}{b_3}]$.

Hence  $\sngv[\Gamma]a$, $[\Gamma]b$ and $\nrm{\Gamma}{a}=[[\nrm{\Gamma}{b_1},\nrm{\Gamma}{b_2}],\nrm{\Gamma}{b_3}]=\nrm{\Gamma}{[z:[b_1+b_2]]b_3}=\nrm{\Gamma}{b}$.
\item[\negate:]
We have $a=\myneg a_1$ for some $a_1$ where $\Gamma\sgv a_1:b$. 
By inductive hypothesis we know that $\sngv[\Gamma]a_1$ ,$b$, and therefore $\Gamma\sngv a_1,b$, and $\nrm{\Gamma}{a_1}=\nrm{\Gamma}{b}$.
Obviously $\sngv[\Gamma]a$ and $\Gamma\sngv a$ and by definition of norming we know that $\nrm{\Gamma}{a}=\nrm{\Gamma}{a_1}=\nrm{\Gamma}{b}$.
\qedhere
\end{meditemize}
\end{proof}
\begin{law}[Valid expressions are normable]
\label{val.nrm}
For all $\Gamma,a$:
$\Gamma\sgv a$ implies $\Gamma\sngv a$.
\end{law}
\begin{proof}
$\Gamma\sgv a$ means that $\Gamma\sgv a:b$, for some $b$.
By Law~\ref{nrm.dtyp} this implies  $\Gamma\sngv a$.
\end{proof}
\subsection{Computable expressions}%
\label{compexp}
First we introduce a simple induction principle that we will use several times in this section.
\begin{definition}[Induction on the size of norms]
\label{induction.norm}
\index{induction!on size of norm}
The \emph{size} of a norm $\bar{a}$ is defined as the number of primitive constants $\prim$ it contains.
A property $P(\Gamma,\bar{a})$ is shown by \emph{norm-induction} iff for all $\bar{b}$ we know that:
If $P(\Gamma,\bar{c})$ for all $\bar{c}$ of size strictly smaller that $\bar{b}$ then $P(\Gamma,\bar{b})$.
\end{definition}
\begin{remark}[Use of norm-induction in proofs]
The clause defining norm-induction can be reformulated into a more convenient form for its use in proofs.
\begin{itemize}
\item {\bf Inductive base:}
$P(\Gamma,\prim)$.
\item {\bf Inductive step:}
For all $\bar{b},\bar{c}$:
If $P(\Gamma,\bar{a})$ for all $\bar{a}$ of size strictly smaller than the size of $[\bar{b},\bar{c}]$ then $P(\Gamma,[\bar{b},\bar{c}])$.
\end{itemize}
\end{remark}
\noindent 
Computable expressions are organized according to norm structure and satisfy strong normalization conditions (Law~\ref{sn.cond}).
\begin{definition}[Computable expressions]%
\label{ce}
\nomenclature[fBasic07]{$\ce_{\Gamma}(\bar{a})$}{computable expressions of norm $\bar{a}$}
\index{expression!computable of norm}
The set of \emph{computable expressions} of norm $\bar{a}$ under context $\Gamma$ is denoted by $\ce_{\Gamma}(\bar{a})$.
$a\in\ce_{\Gamma}(\bar{a})$ iff $a\in\sn{}$, $\Gamma\sngv a$ where $\nrm{\Gamma}{a}=\bar{a}$, and if $\bar{a}=[\bar{b},\bar{c}]$ for some $\bar{b}$ and $\bar{c}$ then
the following \emph{computability conditions} are satisfied:
\begin{itemize}[align=left]
\item[$(\alpha)$:]
For all $x$, $b$, $c$: If $a\rd\binbop{x}{b}{c}$ or $\myneg a\rd\binbop{x}{b}{c}$ then $c\in\ce_{\Gamma,x:b}(\bar{c})$ and $c\gsub{x}{d}\in\ce_{\Gamma}(\bar{c})$ for any $d\in\ce_{\Gamma}(\bar{b})$. 
\item[$(\beta)$:]
For all $b$, $c$: If $a\rd\prsumop{b}{c}$ or $\myneg a\rd\prsumop{b}{c}$ then $b\in\ce_{\Gamma}(\bar{b})$ and $c\in\ce_{\Gamma}(\bar{c})$.
\item[$(\gamma)$:]
For all $x$, $b$, $c$, $d$: $a\rd\prdef{x}{b}{c}{d}$ implies both $b\in\ce_{\Gamma}(\bar{b})$ and $c\in\ce_{\Gamma}(\bar{c})$, 
$a\rd\injl{b}{d}$ implies $b\in\ce_{\Gamma}(\bar{b})$, and $a\rd\injr{d}{c}$ implies $c\in\ce_{\Gamma}(\bar{c})$.
\item[$(\delta)$:]
If $\bar{b}=[\bar{b}_1,\bar{b}_2]$, for some $\bar{b}_1$ and $\bar{b}_2$, then, for all $b_1$, $b_2$:
$a\rd\case{b_1}{b_2}$ implies both $b_1\in\ce_{\Gamma}([\bar{b}_1,\bar{c}])$ and $b_2\in\ce_{\Gamma}([\bar{b}_2,\bar{c}])$.
\end{itemize}
\end{definition}
\begin{remark}[Motivation for the computability conditions]%
The conditions $\alpha$, $\beta$, $\gamma$, and $\delta$ are motivated by the strong normalization condition for applications (Law~\ref{sn.cond}($i$)) and negations (Law~\ref{sn.cond}($ii$)) and by the need for a monotonicity argument of computability with respect to injections and case distinctions.
\end{remark}
\noindent
Furthermore the recursion in the above definition eventually terminates which is a core argument of this proof of strong normalisation.
\begin{law}[The definition of computable expressions terminates]%
For all $\Gamma$ and $\bar{a}$: If $\:$ $\Gamma\sgv\bar{a}$ then the definition of $\ce_{\Gamma}(\bar{a})$ terminates.
\end{law}
\begin{proof}
Proof by norm-induction on $\bar{a}$.
$\Gamma\sngv a$ means that $\nrm{\Gamma}{a}$ is defined.

{\bf Inductive base:} In case of $\bar{a}=\prim$, the conditions $\alpha$, $\beta$, $\gamma$, and $\delta$ are trivially satisfied and the definition $\ce_{\Gamma}(\bar{a})$ obviously terminates.

{\bf Inductive step:} Let $\bar{a}=[\bar{b},\bar{c}]$ for some $\bar{b},\bar{c}$.
In case of conditions $\alpha$, $\beta$, and $\gamma$,
by inductive hypothesis the definitions of $\ce_{\Gamma}(\bar{b})$ and $\ce_{\Gamma}(\bar{c})$ terminate.
Therefore, by construction of  the conditions $\alpha$, $\beta$, and $\gamma$, obviously the definition of $\ce_{\Gamma}(\bar{a})$ terminates.
In case of conditions $\delta$, $\bar{b}=[\bar{b}_1,\bar{b}_2]$,
by inductive hypothesis, the definitions of $\ce_{\Gamma}([\bar{b}_1,\bar{c}])$ and $\ce_{\Gamma}([\bar{b}_1,\bar{c}])$ terminate
(the sizes of $[\bar{b}_1,\bar{c}]$ and $[\bar{b}_2,\bar{c}]$ are both strictly smaller than the size of $\bar{a}$).
Therefore, by construction of the condition $\gamma$, the definition of $\ce_{\Gamma}(\bar{a})$ terminates.
\end{proof}
\noindent 
We begin with some basic properties of computable expressions.
\begin{law}[Basic properties of computable expressions]%
\label{ce.basic}
For all $\Gamma,\Gamma_1,\Gamma_2$, $a$, $a_1$, $a_2$, $b$, $x$:
\begin{itemize}
\item[$i$:]
$a\in\ce_{\Gamma}(\nrm{\Gamma}{a})$ implies $a\in\sn{}$.
\item[$ii$:]
$\Gamma\sgv x$ implies $x\in\ce_{\Gamma}(\nrm{\Gamma}{x})$.
\item[$iii$:]
$\prim\in\ce_{\Gamma}(\prim)$.
\item[$iv$:]
$\Gamma_1\sngv a_1$, $a\in\ce_{\Gamma_1,x:a_1,\Gamma_2}(\nrm{\Gamma}{a})$, and $a_1\rd a_2$ imply $a\in\ce_{\Gamma_1,x:a_2,\Gamma_2}(\nrm{\Gamma}{a})$.
\item[$v$:]
$a\in\ce_{\Gamma}(\nrm{\Gamma}{a})$ and $a\rd b$ imply $b\in\ce_{\Gamma}(\nrm{\Gamma}{a})$.
\end{itemize}
\end{law}
\begin{proof}
$\;$
\begin{itemize}
\item[$i$:] 
By definition of computable expressions $a\in\ce_{\Gamma}(\nrm{\Gamma}{a})$ implies that $a\in\sn{}$.
\item[$ii$:] 
Obviously $x\in\sn{}$. From $\Gamma\sgv x$,  by Law~\ref{val.nrm} we obtain $\Gamma\sngv x$.
The computability conditions are trivially satisfied.
\item[$iii$:]
Obviously $\prim\in\sn{}$ and $\nrm{\Gamma}{\prim}=\prim$.
The computability conditions are trivially satisfied.
\item[$iv$:]
In general we can argue as follows:
Assume $\Gamma_1\sngv a_1$ and $a\in\ce_{\Gamma_1,x:a_1,\Gamma_2}(\bar{a})$ where $\bar{a}=\nrm{\Gamma_1,x:a_1,\Gamma_2}{a}$ and $a_1\rd a_2$. 
By Law~\ref{nrm.rd} we obtain $\Gamma_1\sngv a_2$ and $\nrm{\Gamma_1}{a_1}=\nrm{\Gamma_1}{a_2}$.
We need to show that $a\in\ce_{\Gamma_1,x:a_2,\Gamma_2}(\bar{a})$. 
From $a\in\ce_{\Gamma_1,x:a_1,\Gamma_2}(\bar{a})$ we know that $(\Gamma_1,x:a_1,\Gamma_2)\sngv a$.
Since $\nrm{\Gamma_1}{a_1}=\nrm{\Gamma_1}{a_2}$, by Law~\ref{nrm.eq} we obtain $(\Gamma_1,x:a_2,\Gamma_2)\sngv a$ and $\nrm{\Gamma_1,x:a_1,\Gamma_2}{a}=\nrm{\Gamma_1,x:a_2,\Gamma_2}{a}$.

We now show this part by norm-induction on $\bar{a}$.
More precisely, by norm-induction on $\bar{a}$ we show that for all $\Gamma_1$, $\Gamma_2$, $a$, $a_1$, $a_2$, $x$ and $\bar{a}$, that $\Gamma_1\sngv a_1$, $a\in\ce_{\Gamma_1,x:a_1,\Gamma_2}(\bar{a})$, and $a_1\rd a_2$ imply $a\in\ce_{\Gamma_1,x:a_2,\Gamma_2}(\bar{a})$.
The argument above takes care of all aspects of this proof except for the computability conditions.

{\bf Inductive base:}
In case of $\bar{a}=\prim$, the computability conditions become trivial and given the argument above we are finished.

{\bf Inductive step:}
Let $\bar{a}=[\bar{b},\bar{c}]$ for some $\bar{b}$ and $\bar{c}$.
We additionally have to show the computability conditions for $a$ under context $\Gamma_1,x:a_2,\Gamma_2$:
\begin{meditemize}
\item[($\alpha$):]
Let $a\rd\binbop{y}{b}{c}$ or $\myneg a\rd\binbop{y}{b}{c}$, for some $y$, $b$, and $c$.
From computability condition $\alpha$ for $a$ under context $\Gamma_1,x:a_1,\Gamma_2$ we know that $c\in\ce_{\Gamma_1,x:a_1,\Gamma_2,y:b}(\bar{c})$ and that $c\gsub{y}{d}\in\ce_{\Gamma_1,x:a_1,\Gamma_2}(\bar{c})$ for any $d\in\ce_{\Gamma_1,x:a_1,\Gamma_2}(\bar{b})$.
By inductive hypothesis this implies $c\in\ce_{\Gamma_1,x:a_2,\Gamma_2,y:b}(\bar{c})$ and $c\gsub{y}{d}\in\ce_{\Gamma_1,x:a_2,\Gamma_2}(\bar{c})$.
Hence the condition is satisfied. 
\item[($\beta$):] 
A similar argument as for condition $\alpha$ can be used.%
\item[($\gamma$):] 
A similar argument as for condition $\alpha$ can be used.%
\item[($\delta$):] 
A similar argument as for condition $\alpha$ can be used.%
\end{meditemize}
\item[$v$:]
We have that $a\in\ce_{\Gamma}(\bar{a})$ where $\nrm{\Gamma}{a}=\bar{a}$.
By $i$ we know that $a\in\sn{}$.
Obviously $b\in\sn{}$ and by Law~\ref{nrm.rd} we obtain $\nrm{\Gamma}{b}=\nrm{\Gamma}{a}$.  
We have to show that $b\in\ce_{\Gamma}(\bar{a})$: 
It is easy to prove the computability conditions, since from $b\rd c$ we can always infer $a\rd c$ and hence use the corresponding condition from the assumption $a\in\ce_{\Gamma}(\bar{a})$.
\qedhere
\end{itemize}
\end{proof}
\noindent
The closure of computable expressions against negation is shown first due to its frequent use in other monotonicity arguments.
\begin{law}[Computable expressions are closed against negation]%
\label{ce.mon.neg}
For all $\Gamma, a$:
$a\in\ce_{\Gamma}(\nrm{\Gamma}{a})$ implies $\myneg a\in\ce_{\Gamma}(\nrm{\Gamma}{a})$.
\end{law}
\begin{proof}
Assume that $a\in\ce_{\Gamma}(\nrm{\Gamma}{a})$.
Let $\bar{a}:=\nrm{\Gamma}{a}$.
By definition of norming obviously $\nrm{\Gamma}{\myneg a}=\bar{a}$.
We show that $a\in\ce_{\Gamma}(\bar{a})$ implies $\myneg a\in\ce_{\Gamma}(\bar{a})$ by norm-induction on $\bar{a}$.

{\bf Inductive base:} We have $\bar{a}=\prim$, therefore the computability conditions become trivial and it remains to show
that $\myneg a\in\sn{}$.
Since $a\in\sn{}$, according to Law~\ref{sn.cond}($ii$), we need to show that for any $x,b,c$:
\begin{itemize}
\item[$(C_1)$] $a\rd\prsumop{b}{c}$ implies $\myneg b$, $\myneg c\in\sn{}$ 
\item[$(C_2)$] $a\rd\binbop{x}{b}{c}$ implies $\myneg c\in\sn{}$
\end{itemize} 
Both conditions are satisfied since $a\rd\prsumop{b}{c}$ as well as $a\rd\binbop{x}{b}{c}$, by definition of norming and Law~\ref{nrm.rd},
imply that $\bar{a}\neq\prim$.

{\bf Inductive step:} Let $\bar{a}=[\bar{b},\bar{c}]$ for some $\bar{b}$ and $\bar{c}$.
First, we show that $\myneg a\in\sn{}$.
Since $a\in\sn{}$, according to Law~\ref{sn.cond}($ii$), for any $x,b,c$, we need to show conditions $(C_1)$ amd $(C_2)$:
\begin{itemize}
\item[$(C_1)$]
Let $a\rd\prsumop{b}{c}$.
Since $a\in\ce_{\Gamma}(\bar{a})$, by computability condition $\beta$ we know that $b\in\ce_{\Gamma}(\bar{b})$ and $c\in\ce_{\Gamma}(\bar{c})$. By inductive hypothesis $\myneg b\in\ce_{\Gamma}(\bar{b})$ and $\myneg c\in\ce_{\Gamma}(\bar{c})$ and therefore by Law~\ref{ce.basic}($i$),$\myneg b,\myneg c\in\sn{}$.
\item[$(C_2)$]
Let $a\rd\binbop{x}{b}{c}$. 
Since $a\in\ce_{\Gamma}(\bar{a})$, by computability condition $\alpha$  we know that $c\in\ce_{\Gamma,x:b}(\bar{c})$.
By inductive hypothesis $\myneg c\in\ce_{\Gamma,x:b}(\bar{c})$ and therefore by Law~\ref{ce.basic}($i$), $\myneg c\in\sn{}$.
\end{itemize} 
Hence $\myneg a\in\sn{}$.

\noindent
To finish the proof we have to show the computability conditions for $\myneg a$:
\begin{itemize}[align=left]
\item[($\alpha$):]
We have to consider four cases:

{\bf Cases 1 and 2:}
$\myneg a\rd[x:a_1]a_2$ or $\myneg a\rd[x!a_1]a_2$ for some $x$, $a_1$, and $a_2$.
By Law~\ref{rd.decomp}($v$) we know that $a\rd\binbop{x}{a_1'}{a_2'}$ for some $a_1'$ and $a_2'$ where $a_1'\rd a_1$ and $\myneg a_2'\rd a_2$ .

The first part of $\alpha$ can be argued as follows:
\begin{eqnarray*}
&&a\in\ce_{\Gamma}([\bar{b},\bar{c}])\\
&\Rightarrow&\quad\text{(by $\alpha$, since $a\rd\binbop{x}{a_1'}{a_2'}$ )}\\
&&a_2'\in\ce_{\Gamma,x:a_1'}(\bar{c})\\
&\Rightarrow&\quad\text{(inductive hypothesis)}\\
&&\myneg a_2'\in\ce_{\Gamma,x:a_1'}(\bar{c})\\
&\Rightarrow&\quad\text{(by Law~\ref{ce.basic}($v$), since $\myneg a_2'\rd a_2$)}\\
&&a_2\in\ce_{\Gamma,x:a_1'}(\bar{c})\\
&\Rightarrow&\quad\text{(by Law~\ref{ce.basic}($iv$), since $a_1'\rd a_1$)}\\
&&a_2\in\ce_{\Gamma,x:a_1}(\bar{c})
\end{eqnarray*}
For the second clause, for any $d\in\ce_{\Gamma}(\bar{c})$, we can argue as follows: 
\begin{eqnarray*}
&&a\in\ce_{\Gamma}([\bar{b},\bar{c}])\\
&\Rightarrow&\quad\text{(by $\alpha$)}\\
&&a_2'\gsub{x}{d}\in\ce_{\Gamma}(\bar{c})\\
&\Rightarrow&\quad\text{(inductive hypothesis)}\\
&&\myneg(a_2'\gsub{x}{d})\in\ce_{\Gamma}(\bar{c})\\
&\Rightarrow&\quad\text{(definition of substitution)}\\
&&(\myneg a_2')\gsub{x}{d}\in\ce_{\Gamma}(\bar{c})\\
&\Rightarrow&\quad\text{(by Laws~\ref{ce.basic}($v$) and~\ref{rd.sub}($i$), since $\myneg a_2'\rd a_2$)}\\
&&a_2\gsub{x}{d}\in\ce_{\Gamma}(\bar{c})
\end{eqnarray*}

{\bf Cases 3 and 4:}
$\myneg\myneg a\rd[x:a_1]a_2$ or $\myneg\myneg a\rd[x!a_1]a_2$ for some $x$, $a_1$, and $a_2$.
By Law~\ref{rd.decomp}($v$) we know that $\myneg a\rd\binbop{x}{a_1'}{a_2'}$ for some $a_1'$ and $a_2'$ where  $a_1'\rd a_1$ and $\myneg a_2'\rd a_2$. 
Applying Law~\ref{rd.decomp}($v$) again we know that $a\rd\binbopd{x}{a_1''}{a_2''}$ for some $a_1''$ and $a_2''$ where $a_1''\rd a_1'$ and $\myneg a_2''\rd a_2'$. This means that $a_1''\rd a_1$ and $\myneg\myneg a_2''\rd a_2$.

The first part of $\alpha$ can then be argued as follows:
\begin{eqnarray*}
&&a\in\ce_{\Gamma}([\bar{b},\bar{c}])\\
&\Rightarrow&\quad\text{(by $\alpha$, since $a\rd\binbopd{x}{a_1''}{a_2''}$ )}\\
&&a_2''\in\ce_{\Gamma,x:a_1''}(\bar{c})\\
&\Rightarrow&\quad\text{(inductive hypothesis, applied twice)}\\
&&\myneg\myneg a_2''\in\ce_{\Gamma,x:a_1''}(\bar{c})\\
&\Rightarrow&\quad\text{(by Law~\ref{ce.basic}($v$), since $\myneg\myneg a_2''\rd a_2$)}\\
&&a_2\in\ce_{\Gamma,x:a_1''}(\bar{c})\\
&\Rightarrow&\quad\text{(by Law~\ref{ce.basic}($vi$), since $a_1''\rd a_1$)}\\
&&a_2\in\ce_{\Gamma,x:a_1}(\bar{c})
\end{eqnarray*}
For the second clause, for any $d\in\ce_{\Gamma}(\bar{c})$, we can argue as follows: 
\begin{eqnarray*}
&&a\in\ce_{\Gamma}([\bar{b},\bar{c}])\\
&\Rightarrow&\quad\text{(by $\alpha$)}\\
&&a_2''\gsub{x}{d}\in\ce_{\Gamma}(\bar{c})\\
&\Rightarrow&\quad\text{(inductive hypothesis, applied twice)}\\
&&\myneg\myneg(a_2''\gsub{x}{d})\in\ce_{\Gamma}(\bar{c})\\
&\Rightarrow&\quad\text{(definition of substitution)}\\
&&(\myneg\myneg a_2')\gsub{x}{d}\in\ce_{\Gamma}(\bar{c})\\
&\Rightarrow&\quad\text{(by Laws~\ref{ce.basic}($v$) and~\ref{rd.sub}($i$), since $\myneg\myneg a_2'\rd a_2$)}\\
&&a_2\gsub{x}{d}\in\ce_{\Gamma}(\bar{c})
\end{eqnarray*}
\item[($\beta$):]
We have to consider four cases:

{\bf Cases 1,2:}
$\myneg a\rd[a_1,a_2]$ or $\myneg a\rd[a_1+a_2]$ for some $a_1$ and $a_2$.
By Law~\ref{rd.decomp}($v$) we know that $a\rd\prsumop{a_1'}{a_2'}$ for some $a_1'$ and $a_2'$ where $\myneg a_1'\rd a_1$ and $\myneg a_2'\rd a_2$.

Since $a\in\ce_{\Gamma}(\bar{a})$ and $a\rd\prsumop{a_1'}{a_2'}$ by computability condition $\alpha$ we obtain $a_1'\in\ce_{\Gamma}(\bar{b})$ anf $a_2'\in\ce_{\Gamma}(\bar{c})$.
By inductive hypotheses we know that also $\myneg a_1'\in\ce_{\Gamma}(\bar{b})$ and $\myneg a_2'\in\ce_{\Gamma}(\bar{c})$.
By Law~\ref{ce.basic}($v$) we get $a_1\in\ce_{\Gamma}(\bar{b})$ and $a_2\in\ce_{\Gamma}(\bar{c})$.

{\bf Cases 3,4:}
$\myneg\myneg a\rd[a_1,a_2]$ or $\myneg\myneg a\rd[a_1+a_2]$ for some $a_1$ and $a_2$.
By Law~\ref{rd.decomp}($v$) we know that $\myneg a\rd\prsumop{a_1'}{a_2'}$ for some $a_1'$ and $a_2'$ where $\myneg a_1'\rd a_1$ and $\myneg a_2'\rd a_2$ . 
Applying Law~\ref{rd.decomp}($v$) again we know that $a\rd\prsumopd{a_1''}{a_2''}$ for some $a_1''$ and $a_2''$ where $\myneg a_1''\rd a_1'$ and $\myneg a_2''\rd a_2'$. This means that $\myneg\myneg a_1''\rd a_1$ and $\myneg\myneg a_2''\rd a_2$.

Since $a\in\ce_{\Gamma}(\bar{a})$ and $a\rd\prsumopd{a_1''}{a_2'''}$ by computability condition $\alpha$ we obtain $a_1''\in\ce_{\Gamma}(\bar{b})$ anf $a_2''\in\ce_{\Gamma}(\bar{c})$.
By inductive hypotheses (applied twice)  we know that also $\myneg\myneg a_1'\in\ce_{\Gamma}(\bar{b})$ and $\myneg\myneg a_2'\in\ce_{\Gamma}(\bar{c})$.
By Law~\ref{ce.basic}($v$) we get $a_1\in\ce_{\Gamma}(\bar{b})$ and $a_2\in\ce_{\Gamma}(\bar{c})$.
\item[($\gamma$):]
We have to consider three cases:

{\bf Case 1:}
$\myneg a\rd\prdef{x}{a_1}{a_2}{a_3}$ for some $x$, $a_1$, $a_2$, $a_3$.
By Law~\ref{rd.decomp}($v$) we know that $a\rd\prdef{x}{a_1}{a_2}{a_3}$. 
Since $a\in\ce_{\Gamma}([\bar{b},\bar{c}])$, by computability condition $\beta$, we know that $a_1\in\ce_{\Gamma}(\bar{b})$, $a_2\in\ce_{\Gamma}(\bar{c})$.

{\bf Cases 2,3:}
$\myneg a\rd\injl{a_1}{a_2}$ or $\myneg a\rd\injr{a_1}{a_2}$ for some $a_1$ and $a_2$.
By Law~\ref{rd.decomp}($v$) we know that $a\rd\injl{a_1}{a_2}$ or $a\rd\injr{a_1}{a_2}$.
Since $a\in\ce_{\Gamma}([\bar{b},\bar{c}])$, by computability condition $\beta$, we know that $a_1\in\ce_{\Gamma}(\bar{b})$ or $a_2\in\ce_{\Gamma}(\bar{c})$, respectively. 
\item[($\delta$):]
Let $\bar{b}=[\bar{b}_1,\bar{b}_2]$ for some $\bar{b}_1$ and $\bar{b}_2$.
Let $\myneg a\rd\case{a_1}{a_2}$ for some $a_1$ and $a_2$.
By Law~\ref{rd.decomp}($v$) we know that $a\rd\case{a_1}{a_2}$.
By computability condition $\delta$ we know that $a_1\in\ce_{\Gamma}([\bar{b}_1,\bar{c}])$ and $a_2\in\ce_{\Gamma}([\bar{b}_2,\bar{c}])$.
\qedhere
\end{itemize}
\end{proof}
\noindent
Closure of computability against application has the most involved proof of monotonicity. 
\begin{law}[Closure of computable expressions against application]
\label{ce.mon.appl}
For all $\Gamma$, $a$, $b$:
$\Gamma\sngv (a\,b)$, $a\in\ce_{\Gamma}(\nrm{\Gamma}{a})$, and $b\in\ce_{\Gamma}(\nrm{\Gamma}{b})$ implies $\nrm{\Gamma}{a}=[\nrm{\Gamma}{b},\bar{c}]$ and $(a\,b)\in\ce_{\Gamma}(\bar{c})$ for some $\bar{c}$.
\end{law}
\noindent
The large size of the proof comes from many repetitions of similar arguments when showing the computability conditions of application.
\begin{proof}
Assume that $\Gamma\sngv(a\,b)$, $a\in\ce_{\Gamma}(\nrm{\Gamma}{a})$, and $b\in\ce_{\Gamma}(\nrm{\Gamma}{b})$.
Let $\bar{a}=\nrm{\Gamma}{a}$ and $\bar{b}=\nrm{\Gamma}{b}$.
$\Gamma\sngv(a\,b)$ implies that $\bar{a}=[\bar{b},\bar{c}]$ for some $\bar{c}$.
By induction on $\bar{a}$ we will show that $a\in\ce_{\Gamma}(\nrm{\Gamma}{a})$ and $b\in\ce_{\Gamma}(\nrm{\Gamma}{b})$ implies $(a\,b)\in\ce_{\Gamma}(\bar{c})$. 

{\bf Inductive base:} Trivial since $\bar{a}\neq\prim$. 

{\bf Inductive step:} 
We first need to show that $(a\,b)\in\sn{}$.
By Law~\ref{sn.cond}($i$) we have to show conditions $(C_1)$ and $(C_2)$. For any $x$, $b_1$, $c_1$, $c_2$, $d_1$, $d_2$:
\begin{itemize}
\item[$(C_1)$]
Let $a\rd\binbop{x}{b_1}{c_1}$. 
Since $a\in\ce_{\Gamma}([\bar{b},\bar{c}])$, by computability condition $\alpha$, for any $d\in\ce_{\Gamma}(\bar{b})$ we know that $c_1\gsub{x}{d}\in\ce_{\Gamma}(\bar{c})$.
Hence also $c_1\gsub{x}{b}\in\ce_{\Gamma}(\bar{c})$.
By Law~\ref{ce.basic}($i$) this implies $c_1\gsub{x}{b}\in\sn{}$ hence condition $(C_1)$ is satisfied.
\item[$(C_2)$]
Let $a\rd\case{c_1}{c_2}$ and $b\rd\injl{d_1}{d_2}$.
From $b\rd\injl{d_1}{d_2}$, by Law~\ref{nrm.rd} and by definition of norming we know that $\bar{b}=[\bar{d}_1,\bar{d}_2]$ and hence $\bar{a}=[[\bar{d}_1,\bar{d}_2],\bar{c}]$.
Since $b\in\ce_{\Gamma}([\bar{d}_1,\bar{d}_2])$ by computability condition $\beta$ we know that $d_1\in\ce_{\Gamma}(\bar{d}_1)$.
Since $a\in\ce_{\Gamma}([[\bar{d}_1,\bar{d}_2],\bar{c}])$ by computability condition $\delta$ we know that 
$c_1\in\ce_{\Gamma}([\bar{d}_1,\bar{c}])$.

By inductive hypothesis (the size of $[\bar{d}_1,\bar{c}]$ is strictly smaller than that of $\bar{a}$), we know that $(c_1\,d_1)\in\ce_{\Gamma}(\bar{c})$.
By definition of computability therefore $(c_1\,d_1)\in\sn{}$.
\item[$(C_3)$]: Proof is similar to $(C_2)$.
\end{itemize}
Therefore by Law~\ref{sn.cond}($i$) we know that $(a\,b)\in\sn{}$.
It remains to show the computability conditions for $(a\,b)$. 
Let $\bar{c}=[\bar{d},\bar{e}]$ for some $\bar{d}$ and $\bar{e}$:
\begin{itemize}[align=left]
\item[($\alpha$):]
If $(a\,b)\rd\binbop{x}{a_1}{a_2}$ or $\myneg(a\:b)\rd\binbop{x}{a_1}{a_2}$, for some $a_1$ and $a_2$ then, since $\Gamma\sngv(a\,b)$, by Law~\ref{nrm.rd} we have 
\[\nrm{\Gamma}{\binbop{x}{a_1}{a_2}}=[\nrm{\Gamma}{a_1},\nrm{\Gamma,x:a_1}{a_2}]=[\bar{d},\bar{e}]
\]
Therefore $\nrm{\Gamma}{a_1}=\bar{d}$ and $\nrm{\Gamma,x:a_1}{a_2}=\bar{e}$.
We have to show that $a_2\in\ce_{\Gamma,x:a_1}(\bar{e})$ and $a_2\gsub{x}{d}\in\ce_{\Gamma}(\bar{e})$ for any $d\in\ce_{\Gamma}(\bar{d})$.

We have to distinguish two cases:

{\bf Case $\alpha$1:} If $(a\,b)\rd\binbop{x}{a_1}{a_2}$ then by Law~\ref{rd.decomp}($vi$) we know that there are two cases: 
\begin{itemize}
\item
{\bf Case $\alpha$1.1:} $a\rd[y:a_3]a_4$, $b\rd b'$, and $a_4\gsub{y}{b'}\rd\binbop{x}{a_1}{a_2}$ for some $y$, $a_3$, $a_4$, and $b'$.
We can argue as follows:
\begin{eqnarray*}
&&a\in\ce_{\Gamma}([\bar{b},\bar{c}])\\
&\Rightarrow&\text{(by $\alpha$, since by Law~\ref{ce.basic}($v$) we know that $b'\in\ce_{\Gamma}(\bar{b})$)}\\
&&a_4\gsub{x}{b'}\in\ce_{\Gamma}(\bar{c})\\
&\Rightarrow&\text{(by Law~\ref{ce.basic}($v$), since $a_4\gsub{y}{b'}\rd\binbop{x}{a_1}{a_2}$)}\\
&&\binbop{x}{a_1}{a_2}\in\ce_{\Gamma}(\bar{c})
\end{eqnarray*}
\item 
{\bf Case $\alpha$1.2:} $a\rd\case{c_1}{c_2}$ and either $b\rd\injl{b_1}{b_2}$ and $c_1(b_1)\rd\binbop{x}{a_1}{a_2}$ or $b\rd\injr{b_1}{b_2}$ and $c_2(b_2)\rd\binbop{x}{a_1}{a_2}$ for some $c_1$, $c_2$, $b_1$, and $b_2$. 

This means that $\bar{b}=[\bar{b}_1,\bar{b}_2]$ for some $\bar{b}_1$ and $\bar{b}_2$ where $\nrm{\Gamma}{c_1}=[\bar{b}_1,\bar{c}]$ and $\nrm{\Gamma}{c_2}=[\bar{b}_2,\bar{c}]$.
Since $\Gamma\sngv(a\,b)$ we also know that $\nrm{\Gamma}{b_1}=\bar{b}_1$ and  $\nrm{\Gamma}{b_2}=\bar{b}_2$.
We will show the first case, the proof of the second case is similar.

From $b\rd\injl{b_1}{b_2}$, by computability condition $\gamma$ we know that $b_1\in\ce_{\Gamma}(\bar{b}_1)$.
From $a\rd\case{c_1}{c_2}$, by computability condition $\delta$ we know that $c_1\in\ce_{\Gamma}([\bar{b}_1,\bar{c}])$ and $c_2\in\ce_{\Gamma}([\bar{b}_2,\bar{c}])$.
Since the size of $[\bar{b}_1,\bar{c}]$ is strictly smaller than the size of $\bar{a}$ we can apply the inductive hypothesis to obtain
$(c_1\,c_3)\in\ce_{\Gamma}(\bar{c})$.
Hence, since $(c_1\,b_1)\rd\binbop{x}{a_1}{a_2}$, by Law~\ref{ce.basic}($v$) we have $\binbop{x}{a_1}{a_2}\in\ce_{\Gamma}(\bar{c})$.
\end{itemize}
Therefore in both cases we have shown $\binbop{x}{a_1}{a_2}\in\ce_{\Gamma}(\bar{c})$.
From computability condition $\alpha$ we obtain $a_2\in\ce_{\Gamma,x:a_1}(\bar{e})$ and $a_2\gsub{x}{d}\in\ce_{\Gamma}(\bar{e})$.

{\bf Case $\alpha$2:} If $\myneg(a\,b)\rd\binbop{x}{a_1}{a_2}$ then we need to consider two subcases:
\begin{itemize}
\item
{\bf Case $\alpha$2.1:} $\binbop{x}{a_1}{a_2}=[x:a_1]a_2$: 
By Law~\ref{rd.decomp}($v$) we know that $(a\,b)\rd[x!a_1']a_2'$ for some $a_1'$ and $a_2'$ where $a_1'\rd a_1$ and $\myneg a_2'\rd a_2$.

Therefore, by Law~\ref{rd.decomp}($vi$) we know that there are two cases:
\begin{itemize}
\item
{\bf Case $\alpha$2.1.1:}
$a\rd[y:a_3]a_4$ and $b\rd b'$ and $a_4\gsub{y}{b'}\rd[x!a_1']a_2'$ for some $y$, $a_3$, $a_4$, and $b'$.
As in case {\bf $\alpha$1.1} we can show that $[x!a_1']a_2'\in\ce_{\Gamma}(\bar{c})$ 
\item
{\bf Case $\alpha$2.1.2:}
$a\rd\case{c_1}{c_2}$ and either $b\rd\injl{b_1}{b_2}$ and $(c_1\,b_1)\rd[x!a_1']a_2'$ or $b\rd\injr{b_1}{b_2}$ and $c_2(b_2)\rd[x!a_1']a_2'$ for some $c_1$, $c_2$, $b_1$, and $b_2$.
As in case {\bf $\alpha$1.2} we can show that $[x!a_1']a_2'\in\ce_{\Gamma}(\bar{c})$.
\end{itemize}
Hence in all cases we have $[x!a_1']a_2'\in\ce_{\Gamma}(\bar{c})$.
We can now show the first part of $\alpha$ as follows:
\begin{eqnarray*}
&&[x!a_1']a_2'\in\ce_{\Gamma}(\bar{c})\\
&\Rightarrow&\text{(by $\alpha$)}\\
&&a_2'\in\ce_{\Gamma,x:a_1}(\bar{e})\\
&\Rightarrow&\text{(by Law~\ref{ce.mon.neg})}\\
&&\myneg a_2'\in\ce_{\Gamma,x:a_1}(\bar{e})\\
&\Rightarrow&\text{(by Law~\ref{ce.basic}($v$), since $\myneg a_2'\rd a_2$)}\\
&&a_2\in\ce_{\Gamma,x:a_1}(\bar{e})
\end{eqnarray*}
Let $d\in\ce_{\Gamma}(\bar{d})$. We can show the second part of $\alpha$ as follows:
\begin{eqnarray*}
&&[x!a_1']a_2'\in\ce_{\Gamma}(\bar{c})\\
&\Rightarrow&\text{(by $\alpha$, since $d\in\ce_{\Gamma}(\bar{d})$)}\\
&&a_2'\gsub{x}{d}\in\ce_{\Gamma}(\bar{e})\\
&\Rightarrow&\text{(by Law~\ref{ce.mon.neg})}\\
&&\myneg(a_2'\gsub{x}{d})\in\ce_{\Gamma,x:a_1}(\bar{e})\\
&\Rightarrow&\text{(definition of substitution )}\\
&&(\myneg a_2')\gsub{x}{d}\in\ce_{\Gamma,x:a_1}(\bar{e})\\
&\Rightarrow&\text{(by Law~\ref{ce.basic}($v$) and~\ref{rd.sub}($i$), since $\myneg a_2'\rd a_2$)}\\
&&a_2\gsub{x}{d}\in\ce_{\Gamma,x:a_1}(\bar{e})
\end{eqnarray*}
\item
{\bf Case $\alpha$2.2:} $\binbop{x}{a_1}{a_2}=[x!a_1]a_2$: This can be shown in a similar (symmetric) way as case {\bf $\alpha$2.1}.
\end{itemize}
\item[($\beta$):]
We have to distinguish two cases:

{\bf Case $\beta$1}: If $(a\,b)\rd\prsumop{a_1}{a_2}$ for some $a_1$ and $a_2$, then, since $\Gamma\sngv(a\,b)$, by Law~\ref{nrm.rd} we have 
\[\nrm{\Gamma}{(a\,b)}=[\nrm{\Gamma}{a_1},\nrm{\Gamma}{a_2}]=[\bar{d},\bar{e}]
\]
By Law~\ref{rd.decomp}($vii$) we know that there are two cases:
\begin{itemize}
\item
{\bf Case $\beta$1.1}:
$a\rd\binbop{x}{a_3}{a_4}$, $b\rd b'$, and $a_4\gsub{x}{b'}\rd\prsumop{a_1}{a_2}$ for some $x$, $a_3$, $a_4$, and $b'$.
By the same argument as in case {\bf $\alpha$1.1} we can show that $\prsumop{a_1}{a_2}\in\ce_{\Gamma}(\bar{c})$.
\item
{\bf Case $\beta$1.2}:
$a\rd\case{c_1}{c_2}$ and either $b\rd\injl{b_1}{b_2}$ and $(c_1\,b_1)\rd\prsumop{a_1}{a_2}$ or $b\rd\injr{b_1}{b_2}$ and $(c_2\,b_2)\rd\prsumop{a_1}{a_2}$ for some $c_1$, $c_2$, $b_1$, and $b_2$.
By the same argument as in case {\bf $\alpha$1.2} we can show that $\prsumop{a_1}{a_2}\in\ce_{\Gamma}(\bar{c})$.
\end{itemize}
Hence in both cases we have $\prsumop{a_1}{a_2}\in\ce_{\Gamma}(\bar{c})$.
By computability condition $\beta$ we obtain $a_1\in\ce_{\Gamma}(\bar{d})$ and $a_2\in\ce_{\Gamma}(\bar{e})$.

{\bf Case $\beta$2}: If $\myneg(a\,b)\rd\prsumop{a_1}{a_2}$ for some $a_1$ and $a_2$, then, since obviously $\Gamma\sngv\myneg(a\,b)$, by Law~\ref{nrm.rd} we have 
\[\nrm{\Gamma}{\myneg(a\,b)}=[\nrm{\Gamma}{a_1},\nrm{\Gamma}{a_2}]=[\bar{d},\bar{e}]
\]
We need to consider two subcases:
\begin{itemize}
\item
{\bf Case $\beta$2.1:} If $\myneg(a\,b)\rd[a_1,a_2]$ then by Law~\ref{rd.decomp}($v$) we know that $(a\,b)\rd[a_1'+a_2']$ where $\myneg a_1'\rd a_1$ and $\myneg a_2'\rd a_2$ for some $a_1'$ and $a_2'$.
Therefore by elementary properties of reduction (Law~\ref{rd.decomp}($vii$)) there are two cases
\begin{itemize}
\item
{\bf Case $\beta$2.1.1:}
$a\rd\binbop{x}{a_3}{a_4}$, $b\rd b'$, and $\myneg a_4\gsub{x}{b'}\rd[b_1+b_2]$ for some $x$, $a_3$, $a_4$, and $b'$.
By the same argument as in case {\bf $\alpha$1.1} we can show that $[a_1'+a_2']\in\ce_{\Gamma}(\bar{c})$.
\item
{\bf Case $\beta$2.1.2:}
$a\rd\case{c_1}{c_2}$ and either $b\rd\injl{b_1}{b_2}$ and $(c_1\,b_1)\rd[a_1'+a_2']$ or $b\rd\injr{b_1}{b_2}$ and $(c_2\,b_2)\rd[a_1'+a_2']$ for some $c_1$, $c_2$, $b_1$, and $b_2$.
By the same argument as in case {\bf $\alpha$1.2} we can show that $[a_1'+a_2']\in\ce_{\Gamma}(\bar{c})$. 
\end{itemize}
Hence in all cases we have $[a_1'+a_2']\in\ce_{\Gamma}(\bar{c})$.
By Law~\ref{ce.mon.neg} we have $\myneg[a_1'+a_2']\in\ce_{\Gamma}(\bar{c})$ and hence by Law~\ref{ce.basic}($v$) we have $[\myneg a_1',\myneg a_2']\in\ce_{\Gamma}(\bar{c})$ and therefore by definition of computable expressions (condition $\beta$) we obtain
 we have $\myneg a_1'\in\ce_{\Gamma}(\bar{d})$ and $\myneg a_2'\in\ce_{\Gamma}(\bar{e})$.
Since $\myneg a_1'\rd a_1$ and $\myneg a_2'\rd a_2$ by Law~\ref{ce.basic}($v$) we have $a_1\in\ce_{\Gamma}(\bar{d})$ and $a_2\in\ce_{\Gamma}(\bar{e})$.
\item
{\bf Case $\beta$2.2:}
$\myneg(a\,b)\rd[a_1+a_2]$: This can be shown in a similar (symmetric) way as case {\bf $\beta$2.1}.
\end{itemize}
\item[($\gamma$):]
We have to distinguish three cases:

{\bf Case $\gamma$1:}
If $(a\,b)\rd\prdef{x}{a_1}{a_2}{c}$ for some $a_1$ , $a_2$, and $c$ then, since $\Gamma\sngv(a\:b)$, by Law~\ref{nrm.rd} we have 
\[\nrm{\Gamma}{(a\,b)}=[\nrm{\Gamma}{a_1},\nrm{\Gamma}{a_2}]=[\bar{d},\bar{e}]
\]
By Law~\ref{rd.decomp}($vi$) we know that there are two cases
\begin{itemize}
\item
{\bf Case $\gamma$1.1:}
$a\rd\binbop{y}{a_3}{a_4}$, $b\rd b'$, and $\myneg a_4\gsub{y}{b'}\rd\prdef{x}{a_1}{a_2}{c}$ for some $y$, $a_3$, $a_4$, and $b'$.
By the same argument as in case {\bf $\alpha$1.1} we can show that $\prdef{x}{a_1}{a_2}{c}\in\ce_{\Gamma}(\bar{c})$. 
\item
{\bf Case $\gamma$1.2:}
$a\rd\case{c_1}{c_2}$ and either $b\rd\injl{b_1}{b_2}$ and $c_1(b_1)\rd\prdef{x}{a_1}{a_2}{c}$ or $b\rd\injr{b_1}{b_2}$ and $c_2(b_2)\rd\prdef{x}{a_1}{a_2}{c}$ for some $c_1$, $c_2$, $b_1$, and $b_2$.
By the same argument as in case {\bf $\alpha$1.2} we can show that $\prdef{x}{a_1}{a_2}{c}\in\ce_{\Gamma}(\bar{c})$. 
\end{itemize}
Hence in all cases $\prdef{x}{a_1}{a_2}{c}\in\ce_{\Gamma}(\bar{c})$.
By computability condition $\gamma$ we know that $a_1\in\ce_{\Gamma}(\bar{d})$ and $a_2\in\ce_{\Gamma}(\bar{e})$.

{\bf Case $\gamma$2:}
If $(a\,b)\rd\injl{a_1}{c}$ for some $a_1$ and $c$, then, since $\Gamma\sngv(a\:b)$, by Law~\ref{nrm.rd} we have 
\[\nrm{\Gamma}{(a\:b)}=[\nrm{\Gamma}{a_1},\nrm{\Gamma}{c}]=[\bar{d},\bar{e}]
\]
By Law~\ref{rd.decomp}($vii$) we know that there are two cases 
\begin{itemize}
\item
$a\rd\binbop{y}{a_3}{a_4}$, $b\rd b'$, and $a_4\gsub{y}{b'}\rd a_4\gsub{y}{b'}\rd\injl{a_1}{c}$.
By the same argument as in case {\bf $\alpha$1.1} we can show that  $\injl{a_1}{c}\in\ce_{\Gamma}(\bar{c})$.
\item
$a\rd\case{c_1}{c_2}$ and either $b\rd\injl{b_1}{b_2}$ and $c_1(b_1)\rd\injl{a_1}{c}$ or $b\rd\injr{b_1}{b_2}$ and $c_2(b_2)\rd\injr{a_1}{c}$ for some $c_1$, $c_2$, $b_1$, and $b_2$.
By the same argument as in case {\bf $\alpha$1.2} we can show that  $\injl{a_1}{c}\in\ce_{\Gamma}(\bar{c})$. 
\end{itemize}
Hence in all cases we have $\injl{a_1}{c}\in\ce_{\Gamma}(\bar{c})$.
By computability condition $\beta$ we obtain $a_1\in\ce_{\Gamma}(\bar{d})$.

{\bf Case $\gamma$3}: $\myneg(a\,b)\rd\injr{c}{a_2}$ for some $a_2$ and $c$:
This case is shown in a similar (symmetric) way as case {\bf $\gamma$2}.
\item[($\delta$):]
Let $\bar{d}=[\bar{d}_1,\bar{d}_2]$ for some $\bar{d}_1$ and $\bar{d}_2$:
$(a\,b)\rd\case{a_1}{a_2}$ for some $a_1$ and $a_2$ then, since $\Gamma\sngv(a\:b)$, by Law~\ref{nrm.rd} we have 
\[\nrm{\Gamma}{\case{a_1}{a_2}}=[\bar{d},\bar{e}]
\]
where $\nrm{\Gamma}{a_1}=[\bar{d}_1,\bar{e}]$ and $\nrm{\Gamma}{a_2}=[\bar{d}_2,\bar{e}]$.
By Law~\ref{rd.decomp}($vii$) we know that there are two cases 
\begin{itemize}
\item
{\bf Case $\delta$1:}
$a\rd[y:a_3]a_4$, $b\rd b'$, and $a_4\gsub{y}{b'}\rd a_4\gsub{y}{b'}\rd\case{a_1}{a_2}$.
By the same argument as in case {\bf $\alpha$1.1} we can show that $\case{a_1}{a_2}\in\ce_{\Gamma}(\bar{c})$.
\item
{\bf Case $\delta$2:}
$a\rd\case{c_1}{c_2}$ and either $b\rd\injl{b_1}{b_2}$ and $(c_1\,b_1)\rd\case{a_1}{a_2}$ or $b\rd\injr{b_1}{b_2}$, and $(c_2\,b_2)\rd\case{a_1}{a_2}$ for some $c_1$, $c_2$, $b_1$, and $b_2$.
By the same argument as in case {\bf $\alpha$1.2} we can show that $\case{a_1}{a_2}\in\ce_{\Gamma}(\bar{c})$. 
\end{itemize}
Hence in all cases we have $\case{a_1}{a_2}\in\ce_{\Gamma}(\bar{c})$.
By computability condition $\beta$ we obtain $a_1\in\ce_{\Gamma}([\bar{d}_1,\bar{e}])$ and $a_2\in\ce_{\Gamma}([\bar{d}_2,\bar{e}])$.
\qedhere
\end{itemize}
\end{proof}
\noindent
We now show all the other closure properties of computable expressions, some of which require normability of the operator.
\begin{law}[Closure properties of computable expressions]
\label{ce.mon}
For all $\Gamma$, $x$, $a$, $b$, $c$, $d$, $\bar{a}$, $\bar{b}$, $\bar{c}$:
\begin{itemize}
\item[$i$:]
$\Gamma\sngv\prdef{x}{a}{b}{c}$, $a\in\ce_{\Gamma}(\nrm{\gamma}{a})$, $b\in\ce_{\Gamma}(\nrm{\gamma}{b})$, and $c\in\ce_{\Gamma,x:a}(\nrm{\gamma}{b})$ implies $\prdef{x}{a}{b}{c}\in\ce_{\Gamma}([\nrm{\gamma}{a},\nrm{\gamma}{b}])$.
\item[$ii$:]
$a\in\ce_{\Gamma}(\nrm{\Gamma}{a})$ and $b\in\ce_{\Gamma}(\nrm{\Gamma}{b})$ implies $\prsumop{a}{b},\injl{a}{b},\injr{a}{b}\in\ce_{\Gamma}([\nrm{\Gamma}{a},\nrm{\Gamma}{b}])$.
\item[$iii$:]
$\Gamma\sngv\pleft{a}$ and $a\in\ce_{\Gamma}([\bar{b},\bar{c}])$ implies $\pleft{a}\in\ce_{\Gamma}(\bar{b})$.
\item[$iv$:]
$\Gamma\sngv\pright{a}$ and $a\in\ce_{\Gamma}([\bar{b},\bar{c}])$ implies $\pright{a}\in\ce_{\Gamma}(\bar{c})$.
\item[$v$:]
$a\in\ce_{\Gamma}([\bar{a},\bar{c}])$ and $b\in\ce_{\Gamma}([\bar{b},\bar{c}])$ implies $\case{a}{b}\in\ce_{\Gamma}([[\bar{a},\bar{b}],\bar{c}])$. 
\end{itemize}
\end{law}

\begin{proof}
For all $\Gamma$, $x$, $a$, $b$, $c$, $d$, $\bar{a}$, $\bar{b}$, $\bar{c}$:
\begin{itemize}
\item[$i$:] 
Since $\Gamma\sngv\prdef{x}{a}{b}{c}$ by definition of computable expressions we know that $\nrm{\Gamma}{\prdef{x}{a}{b}{c}}=[\nrm{\Gamma}{a},\nrm{\Gamma}{b}]$ where $\nrm{\Gamma,x:a}{c}=\nrm{\Gamma}{b}$.
Let $\bar{a}=\nrm{\Gamma}{a}$ and $\bar{b}=\nrm{\Gamma}{b}$.
Assume $a\in\ce_{\Gamma}(\bar{a})$, $b\in\ce_{\Gamma}(\bar{b})$, and $c\in\ce_{\Gamma,x:a}(\bar{b})$. 
We have to show $\prdef{x}{a}{b}{c}\in\ce_{\Gamma}([\bar{a},\bar{b}])$.

Since by definition of computable expressions $a,b,c\in\sn{}$, by Law~\ref{sn.basic}($iv$) we have $\prdef{x}{a}{b}{c}\in\sn{}$.
It remains to show the computability conditions:
By Law~\ref{rd.decomp}($ii$), $\prdef{x}{a}{b}{c}\rd d$ and $\myneg\prdef{x}{a}{b}{c}\rd d$ each imply $d=\prdef{x}{a'}{b'}{c'}$ for some $a'$, $b'$, and $c'$.
Therefore, the computability conditions $\alpha$, $\beta$, and $\delta$ are trivially satisfied.
The condition $\gamma$ is trivially satisfied except for the case $\prdef{x}{a}{b}{c}\rd\prdef{x}{a'}{b'}{c'}$ for some $a'$, $b'$, and $c'$.
By Law~\ref{rd.decomp}($ii$) we know that $a\rd a'$, $b\rd b'$, and $c\rd c'$.
By Law~\ref{ce.basic}($v$) we know that $a'\in\ce_{\Gamma}(\bar{a})$ and $b'\in\ce_{\Gamma}(\bar{b})$
\item[$ii$:]
Let $a\in\ce_{\Gamma}(\nrm{\Gamma}{a})$, and $b\in\ce_{\Gamma}(\nrm{\Gamma}{b})$. 
We show that $[a,b],[a+b],\injl{a}{b},\injr{a}{b}\in\ce_{\Gamma}([\nrm{\Gamma}{a},\nrm{\Gamma}{b}])$.
Obviously  $\Gamma\sngv[a,b],[a+b],\injl{a}{b},\injr{a}{b}$ and $[a,b],[a+b],\injl{a}{b},\injr{a}{b}\in\sn{}$. 

First we turn to the computability conditions for $[a,b]$:
Since products or negated products always reduce to product or sums, the conditions $\alpha$, $\gamma$, and $\delta$ are trivially satisfied. 
It remains to show the condition $\beta$:
Let $\Gamma\sgv[a,b]\rd\prsumop{a_1}{a_2}$  or $\Gamma\sgv\myneg[a,b]\rd\prsumop{a_1}{a_2}$ for some $a_1$ and $a_2$. 
By Law~\ref{rd.decomp}($i$,$v$) we know that $a\rd a_1$ or $\myneg a\rd a_1$ and $b\rd a_2$ or $\myneg b\rd a_2$.
By Law~\ref{ce.mon.neg} we know that $\myneg a\in\ce_{\Gamma}(\nrm{\Gamma}{a})$ and $\myneg b\in\ce_{\Gamma}(\nrm{\Gamma}{b})$.
Hence, by Law~\ref{ce.basic}($v$) in each case we know that $a_1\in\ce_{\Gamma}(\nrm{\Gamma}{a})$ and $a_2\in\ce_{\Gamma}(\nrm{\Gamma}{b})$.

The proof of the computability conditions for $[a+b]$ is similar.

Next we show the computability conditions for $\injl{a}{b}$ and $\injr{a}{b}$.
Since injections or negated injections always reduce to injections, the conditions $\alpha$, $\beta$, $\gamma$ (partially), and $\delta$ are trivially satisfied. 
It remains to show the non-trivial cases of the condition $\gamma$:
Let $\Gamma\sgv\injl{a}{b}\rd\injl{a_1}{a_2}$ for some $a_1$ and $a_2$. 
By Law~\ref{rd.decomp}($i$) we know that $a\rd a_1$ and $a\rd a_2$.
The required conclusions $a_1\in\ce_{\Gamma}(\nrm{\Gamma}{a})$ and $a_2\in\ce_{\Gamma}(\nrm{\Gamma}{b})$ follow by Law~\ref{ce.basic}($v$). 
\item[$iii$:]
Since $\Gamma\sngv\pleft{a}$ by definition of computable expressions we know that $\nrm{\Gamma}{\pleft{a}}=\bar{b}$ where
$\nrm{\Gamma}{a}=[\bar{b},\bar{c}]$.
Assume $a\in\ce_{\Gamma}([\bar{b},\bar{c}])$. We have to show $\pleft{a}\in\ce_{\Gamma}(\bar{b})$.

Since $a\in\sn{}$ by Law~\ref{sn.basic}($iii$) we know that $\pleft{a}\in\sn{}$.
It remains to show the computability conditions for $\pleft{a}$.
Let $\bar{b}=[\bar{d},\bar{e}]$ for some $\bar{d}$ and $\bar{e}$. 
\begin{itemize}[align=left]
\item[($\alpha$):]
Assume $\pleft{a}\rd\binbop{x}{a_1}{a_2}$ or $\myneg(\pleft{a})\rd\binbop{x}{a_1}{a_2}$ for some $a_1$ and $a_2$.
By Law~\ref{nrm.rd} we have 
\[\nrm{\Gamma}{\pleft{a}}=[\nrm{\Gamma}{a_1},\nrm{\Gamma}{a_2}]=[\bar{d},\bar{e}]
\]
If $\pleft{a}\rd\binbop{x}{a_1}{a_2}$, by Law~\ref{rd.decomp}($iii$) we have $a\rd\prdef{y}{a_3}{a_4}{a_5}$ or $a\rd\prsumop{a_3}{a_4}$, for some $y$, $a_3$, $a_4$, and $a_5$, where $a_3\rd\binbop{x}{a_1}{a_2}$.

If $\myneg(\pleft{a})\rd\binbop{x}{a_1}{a_2}$, Law~\ref{rd.decomp}($v$), we have $\pleft{a}\rd\binbopd{x}{a_1'}{a_2'}$ where $a_1'\rd a_1$ and $\myneg a_2'\rd a_2$ for some $a_1'$ and $a_2'$.
By Law~\ref{rd.decomp}($iii$) we know that $a\rd\prdef{y}{a_3}{a_4}{a_5}$ or $a\rd\prsumop{a_3}{a_4}$, for some $y$, $a_3$, $a_4$, and $a_5$, where $a_3\rd\binbopd{x}{a_1'}{a_2'}$.

In both cases, by Law~\ref{ce.basic}($v$) we can conclude that $\prdef{y}{a_3}{a_4}{a_5}\in\ce_{\Gamma}(\bar{a})$ or $\prsumop{a_3}{a_4}\in\ce_{\Gamma}(\bar{a})$.
By computability condition $\beta$ and $\gamma$ we have $a_3\in\ce_{\Gamma}(\bar{b})$.
Since $a_3\rd\binbop{x}{a_1}{a_2}$ or $a_3\rd\binbopd{x}{a_1'}{a_2'}$, by Law~\ref{ce.basic}($v$) we know that $\binbop{x}{a_1}{a_2}\in\ce_{\Gamma}(\bar{b})$ or $\binbopd{x}{a_1'}{a_2'}\in\ce_{\Gamma}(\bar{b})$.

If $\binbop{x}{a_1}{a_2}\in\ce_{\Gamma}(\bar{b})$, by computability condition $\alpha$ we obtain $a_2\in\ce_{\Gamma,x:a_1}(\bar{d})$ and $a_2\gsub{x}{d}\in\ce_{\Gamma}(\bar{e})$ for any $d\in\ce_{\Gamma}(\bar{d})$.

If $\binbopd{x}{a_1'}{a_2'}\in\ce_{\Gamma}(\bar{b})$, by computability condition $\alpha$ we obtain $a_2'\in\ce_{\Gamma,x:a_1}(\bar{d})$.

Hence, by Law~\ref{ce.mon.neg} we know that $\myneg a_2'\in\ce_{\Gamma,x:a_1}(\bar{d})$ and therefore by Law~\ref{ce.basic}($v$) we know that
$a_2\in\ce_{\Gamma,x:a_1}(\bar{d})$. 

Furthermore $a_2'\gsub{x}{d}\in\ce_{\Gamma}(\bar{e})$ for any $d\in\ce_{\Gamma}(\bar{d})$.
Hence by Law~\ref{ce.mon.neg} and definition of substitution we obtain $(\myneg a_2')\gsub{x}{d}\in\ce_{\Gamma}(\bar{e})$.c
Since  $\myneg a_2'\rd a_2$, by Laws~\ref{rd.sub}($i$) and~\ref{ce.basic}($v$) we know that $a_2\gsub{x}{d}\in\ce_{\Gamma}(\bar{e})$.
\item[($\beta$):]
Assume $\pleft{a}\rd\prsumop{a_1}{a_2}$ or $\myneg(\pleft{a})\rd\prsumop{a_1}{a_2}$.
By Law~\ref{nrm.rd} we have 
\[\nrm{\Gamma}{\pleft{a}}=[\nrm{\Gamma}{a_1},\nrm{\Gamma}{a_2}]=[\bar{d},\bar{e}]
\]
If $\pleft{a}\rd\prsumop{a_1}{a_2}$, by Law~\ref{rd.decomp}($iv$) we have $a\rd\prdef{y}{a_3}{a_4}{a_5}$ or $a\rd\prsumopd{a_3}{a_4}$, for some $y$, $a_3$, $a_4$, and $a_5$, where $a_3\rd\prsumop{a_1}{a_2}$.

If $\myneg(\pleft{a})\rd\prsumop{a_1}{a_2}$, Law~\ref{rd.decomp}($v$), we have $\pleft{a}\rd\prsumopd{a_1'}{a_2'}$ where $\myneg a_1'\rd a_1$ and $\myneg a_2'\rd a_2$ for some $a_1'$ and $a_2'$.
By Law~\ref{rd.decomp}($iii$) we know that $a\rd\prdef{y}{a_3}{a_4}{a_5}$ or $a\rd\prsumopdd{a_3}{a_4}$, for some $y$, $a_3$, $a_4$, and $a_5$, where $a_3\rd\prsumopd{a_1'}{a_2'}\rd\prsumopd{a_1}{a_2}$.

In both cases, by Law~\ref{ce.basic}($v$) we can conclude that $\prdef{y}{a_3}{a_4}{a_5}\in\ce_{\Gamma}(\bar{a})$ or $\prsumopd{a_3}{a_4}\in\ce_{\Gamma}(\bar{a})$.
By computability condition $\beta$ and $\gamma$ we have $a_3\in\ce_{\Gamma}(\bar{b})$.
Since $a_3\rd\prsumop{a_1}{a_2}$ or $a_3\rd\prsumopd{a_1}{a_2}$, by Law~\ref{ce.basic}($v$) we know that $\prsumop{a_1}{a_2}\in\ce_{\Gamma}(\bar{b})$ or $\prsumopd{a_1'}{a_2'}\in\ce_{\Gamma}(\bar{b})$.

Hence, by computability condition $\beta$ we obtain $a_2\in\ce_{\Gamma}(\bar{d})$ and $a_3\in\ce_{\Gamma}(\bar{e})$ or $a_2'\in\ce_{\Gamma}(\bar{d})$ and $a_3'\in\ce_{\Gamma}(\bar{e})$. 
In the second case by Law~\ref{ce.mon.neg} and definition of substitution we obtain $\myneg a_2'\in\ce_{\Gamma}(\bar{d})$ and $\myneg a_3'\in\ce_{\Gamma}(\bar{e})$.
Since  $\myneg a_2'\rd a_2$ and $\myneg a_3'\rd a_2$, by Law~\ref{ce.basic}($v$) we know that $a_2\in\ce_{\Gamma}(\bar{d})$ and $a_3\in\ce_{\Gamma}(\bar{e})$.
\item[($\gamma$):]
We have to consider three cases:

{\bf Case 1}:
Assume that $\pleft{a}\rd\prdef{x}{a_1}{a_2}{c}$ for some $c$.
By Law~\ref{nrm.rd} we have 
\[\nrm{\Gamma}{\pleft{a}}=[\nrm{\Gamma}{a_1},\nrm{\Gamma}{a_2}]=[\bar{d},\bar{e}]
\]
By Law~\ref{rd.decomp}($iii$) we have $a\rd\prdef{y}{a_3}{a_4}{a_5}$ or $a\rd\prsumop{a_3}{a_4}$, for some $y$, $a_3$, $a_4$, and $a_5$, where $a_3\rd\prdef{x}{a_1}{a_2}{c}$.
By Law~\ref{ce.basic}($v$) we have $\prdef{y}{a_3}{a_4}{a_5}\in\ce_{\Gamma}(\bar{a})$ or $\prsumop{a_3}{a_4}\in\ce_{\Gamma}(\bar{a})$.
By computability condition $\beta$ we have $a_3\in\ce_{\Gamma}(\bar{b})$.
Since $\pleft{a}\rd\prdef{x}{a_1}{a_2}{c}$, by Law~\ref{ce.basic}($v$) we have $\prdef{x}{a_1}{a_2}{c}\in\ce_{\Gamma}(\bar{b})$.
By computability condition $\gamma$ we obtain $a_1\in\ce_{\Gamma}(\bar{d})$ and $a_2\in\ce_{\Gamma}(\bar{e})$.

{\bf Case 2}:
Assume that $\pleft{a}\rd\injl{a_1}{c}$ for some $a_1$ and $c$.
By Law~\ref{nrm.rd} we have 
\[\nrm{\Gamma}{\pleft{a}}=[\nrm{\Gamma}{a_1},\nrm{\Gamma}{c}]=[\bar{d},\bar{e}]
\]
By Law~\ref{rd.decomp}($iv$) we have $a\rd\prdef{y}{a_3}{a_4}{a_5}$ or $a\rd\prsumop{a_3}{a_4}$, for some $y$, $a_3$, $a_4$, and $a_5$, where $a_3\rd\injl{a_1}{c}$.
By Law~\ref{ce.basic}($v$) we have $\prdef{y}{a_3}{a_4}{a_5}\in\ce_{\Gamma}(\bar{a})$ or $\prsumop{a_3}{a_4}\in\ce_{\Gamma}(\bar{a})$.
By computability condition $\beta$ we have $a_3\in\ce_{\Gamma}(\bar{b})$.
Since $\pleft{a}\rd\injl{a_1}{c}$, by Law~\ref{ce.basic}($v$) we have $\injl{a_1}{c}\in\ce_{\Gamma}(\bar{b})$.
By computability condition $\gamma$ we obtain $a_1\in\ce_{\Gamma}(\bar{d})$. 

{\bf Case 3}:
Assume that $\pleft{a}\rd\injl{a_1}{c}$ or $\pleft{a}\rd\injr{c}{a_2}$ for some $a_2$, and $c$.
The proof is similar to the one for the case 2.
\item[($\delta$):]
Let $\bar{d}=[\bar{d}_1,\bar{d}_2]$ for some $\bar{d}_1$ and $\bar{d}_2$.
Assume that $\pleft{a}\rd\case{a_1}{a_2}$.
By Law~\ref{nrm.rd} we have 
\[\nrm{\Gamma}{\pleft{a}}=[[\bar{d}_1,\bar{d}_2],\bar{e}]
\]
where $\nrm{\Gamma}{a_1}=[\bar{d}_1,\bar{e}]$ and $\nrm{\Gamma}{a_2}=[\bar{d}_2,\bar{e}]$.
By Law~\ref{rd.decomp}($iii$) we have $a\rd\prdef{y}{a_3}{a_4}{a_5}$ or $a\rd\prsumop{a_3}{a_4}$, for some $y$, $a_3$, $a_4$, and $a_5$, where $a_3\rd\case{a_1}{a_2}$.
By Law~\ref{ce.basic}($v$) we have $\prdef{y}{a_3}{a_4}{a_5}\in\ce_{\Gamma}(\bar{a})$ or $\prsumop{a_3}{a_4}\in\ce_{\Gamma}(\bar{a})$.
By computability condition $\beta$ we have $a_3\in\ce_{\Gamma}(\bar{b})$.
Since $a_3\rd\case{a_1}{a_2}$, by Law~\ref{ce.basic}($v$) we have $\case{a_1}{a_2}\in\ce_{\Gamma}(\bar{b})$.
By computability condition $\delta$ we obtain $a_1\in\ce_{\Gamma}([\bar{d}_1,\bar{e}])$ and $a_2\in\ce_{\Gamma}([\bar{d}_2,\bar{e}])$.
\end{itemize}
\item[$iv$:]
Since $\Gamma\sngv\pright{a}$ by definition of computable expressions we know that $\nrm{\Gamma}{\pright{a}}=\bar{c}$ where
$\nrm{\Gamma}{a}=[\bar{b},\bar{c}]$.
Assume $a\in\ce_{\Gamma}([\bar{b},\bar{c}])$.
We have to show $\pright{a}\in\ce_{\Gamma}(\bar{c})$.

Since $a\in\sn{}$ by Law~\ref{sn.basic}($iii$) we know that $\pright{a}\in\sn{}$.
It remains to show the computability conditions for $\pright{a}$.
The proof is very similar to the one for $\pleft{a}$: 
When having established, in each of the various sub-cases, that 
$\prdef{y}{a_3}{a_4}{a_5}\in\ce_{\Gamma}(\bar{a})$ or $\prsumop{a_3}{a_4}\in\ce_{\Gamma}(\bar{a})$, then
by computability condition $\beta$ we infer $a_4\in\ce_{\Gamma}(\bar{c})$ and proceed in each case with $a_4$ instead of $a_3$.
\item[$v$:]
Assume that $a\in\ce_{\Gamma}([\bar{a},\bar{c}])$ and $b\in\ce_{\Gamma}([\bar{b},\bar{c}])$.
We will show that $\case{a}{b}\in\ce_{\Gamma}([[\bar{a},\bar{b}],\bar{c}])$.
Obviously $\Gamma\sngv\case{a}{b}$ and $\case{a}{b}\in\sn{}$. 
Since case distinction or negated case distinctions always reduce to distinctions, the conditions $\alpha$, $\beta$, and $\gamma$ are trivially satisfied. 

It remains to show the condition $\delta$:
Assume that $\case{a}{b}\rd\case{a_1}{a_2}$.
By Law~\ref{rd.decomp}($i$) we know that $a\rd a_1$ and $a\rd a_2$.
The required conclusions $a_1\in\ce_{\Gamma}([\bar{a},\bar{c}])$ and $a_2\in\ce_{\Gamma}([\bar{b},\bar{c}])$ follow by Laws~\ref{ce.basic}($v$).  
\qedhere
\end{itemize}
\end{proof}
\noindent
Note that abstraction is missing from the properties of Law~\ref{ce.mon} since from $a\in\ce_{\Gamma,x:b}(\bar{a})$ we can not conclude that $a\gsub{x}{c}\in\ce_{\Gamma}$ for any $c\in\ce_{\Gamma}(\nrm{\Gamma}{b})$.
This is made precise by the following lemma.
\begin{law}[Abstraction closure for computable expressions]
\label{ce.abs}
For all $\Gamma$, $x$, $a$, and $b$ where $\Gamma\sngv\binbop{x}{a}{b}$:
If $a\in\ce_{\Gamma}(\nrm{\Gamma}{a})$, $b\in\ce_{\Gamma,x:a}(\nrm{\Gamma,x:a}{b})$, and for all $c$ with $c\in\ce_{\Gamma}(\nrm{\Gamma}{a})$ we have $b\gsub{x}{c}\in\ce_{\Gamma}(\nrm{\Gamma,x:a}{b})$ (this last assumption about substitution is crucial)
then $\binbop{x}{a}{b}\in\ce_{\Gamma}(\nrm{\Gamma}{\binbop{x}{a}{b}})$.
\end{law} 
\begin{proof}
Let $\bar{a}=\nrm{\Gamma}{a}$ and $\bar{b}=\nrm{\Gamma,x:a}{b}$.
From $a\in\ce_{\Gamma}(\bar{a})$ and $b\in\ce_{\Gamma,x:a}(\bar{b})$, 
by Law~\ref{ce.basic}($i$) we have $a,b\in\sn{}$.
By Law~\ref{sn.basic}($ii$) this implies $[x:a]b\in\sn{}$. 

To show that $\binbop{x}{a}{b}\in\ce_{\Gamma}([\bar{a},\bar{b}])$, 
it remains to show the computability conditions for $\binbop{x}{a}{b}$: 
\begin{meditemize}
\item[($\alpha$):]
By Law~\ref{rd.decomp}($ii$), only the cases $[x:a]b\rd[x:c]d$, $[x!a]b\rd[x!c]d$, $\myneg[x:a]b\rd[x!c]d$, and $\myneg[x!a]b\rd[x:c]d$ for some $c$ and $d$ are possible. 
By Law~\ref{rd.decomp}($ii$,$v$) we know that $a\rd c$ and either $b\rd d$ or $\myneg b\rd d$.  
We have to show the following properties:
\begin{itemize}
\item[$i$:]
We have to show $d\in\ce_{\Gamma,x:c}(\bar{b})$:
From the assumption $b\in\ce_{\Gamma,x:a}(\bar{b})$ either by directly using Law~\ref{ce.basic}($v$) or by first applying Law~\ref{ce.mon.neg} we obtain $d\in\ce_{\Gamma,x:a}(\bar{b})$. 
Since $a\rd c$, by Law~\ref{ce.basic}($iv$) we obtain that $d\in\ce_{\Gamma,x:c}(\bar{b})$.
\item[$ii$:]
Let $e\in\ce_{\Gamma}(\bar{a})$. 
We have to show that $d\gsub{x}{e}\in\ce_{\Gamma}(\bar{b})$ which follows from the assumption about substitution (by instantiating $c$ to $e$).
\end{itemize}
\item[($\beta$):]
The computability conditions $\beta$ is trivially satisfied since (negated) abstractions reduce to abstractions only. 
\item[($\gamma$):]
The computability conditions $\beta$ is trivially satisfied since (negated) abstractions reduce to abstractions only. 
\item[($\delta$):]
The computability conditions $\beta$ is trivially satisfied since (negated) abstractions reduce to abstractions only. 
\qedhere
\end{meditemize}
\end{proof}
\subsection{Computability law}%
To prove computability of all normable expressions, due to of Law~\ref{ce.abs}, we need to prove the stronger property
that normable expressions are computable under any substitution of their free variables to computable expressions.
First we need to extend the notion of substitution.
\begin{definition}[Extended substitution]%
The substitution operation $a\gsub{x}{b}$ to replace free occurrences of $x$ in $a$ by $b$ can be extended as follows:
Given sequences of pairwise disjoint variables $X=(x_1,\ldots,x_n)$ and expressions $B=(b_1,\ldots,b_n)$ where $n\geq 0$, a \emph{substitution function} $\sigma_{X,B}$ is defined on expressions and contexts as follows:
\begin{eqnarray*}
\sigma_{X,B}(a)&=&a\gsub{x_1}{b_1}\cdots\gsub{x_n}{b_n}\\[4mm]
\sigma_{X,B}(())&=&()\\
\sigma_{X,B}(x:a,\Gamma)&=&
\begin{cases}
(x:\sigma_{X,B}(a),\sigma_{X,B}(\Gamma))&\text{if}\;x\neq x_i\\
\sigma_{X,B}(\Gamma)&\text{otherwise}
\end{cases}
\end{eqnarray*}
If $x\neq x_i$ we write $\sigma_{X,B}\gsub{x}{b}$ for $\sigma_{(x_1,\ldots,x_n,x),(b_1,\ldots,b_n,b)}$.
\end{definition}
\begin{definition}[Norm-matching substitution]
\label{norm.match.sub}
A substitution $\sigma_{X,B}$  where $X=(x_1,\ldots,x_n)$ and $B=(b_1,\ldots,b_n)$ is called  \emph{norm matching \wrt\ }$\Gamma$ iff $\Gamma=(\Gamma_0,x_1:a_1,\Gamma_1\ldots x_n:a_n,\Gamma_n)$,
for some $\Gamma_0$ and $a_i$, $\Gamma_i$ where $1\leq i\leq n$ and furthermore
for all these $i$ we have, with $\sigma$ abbreviating $\sigma_{X,B}$:
\[
\sigma(\Gamma)\sngv \sigma(a_i),\quad\sigma(\Gamma)\sngv\sigma(b_i)\quad\text{and}\quad\nrm{\sigma(\Gamma)}{\sigma(a_i)}=\nrm{\sigma(\Gamma)}{\sigma(b_i)}
\] 
\end{definition}
\noindent
Norm-matching substitutions indeed preserve norms:
\begin{law}[Norm preservation of norm-matching substitutions]%
\label{nrm.sub.nrm}
Let $\sigma_{X,B}$ be norm-matching \wrt\ $\Gamma$, then for all $\Gamma$ and $a$ we have $\nrm{\sigma_{X,B}(\Gamma)}{\sigma_{X,B}(a)}=\nrm{\Gamma}{a}$.
\end{law}
\begin{proof}
Proof is by induction on the number $k$ of variables in $X$:
If $k=0$ the property is obviously true. 

If $k=n+1$ where $n\geq 0$ consider $\sigma=\sigma_{X,B}$ where $X=(x_1,\ldots,x_n,x)$, $B=(b_1,\ldots,b_n,b)$. 
Since $\sigma$ is norm-matching \wrt\ $\Gamma$, it can be written as $\Gamma =(\Gamma',x:a,\Gamma_{n+1})$ where $\Gamma'=(\Gamma_0,x_1:a_1,\Gamma_1\ldots x_n:a_n,\Gamma_n)$ for some $a',a_1,\ldots,a_n$.
Let $\sigma'=\sigma_{(x_1,\ldots,x_n),(b_1,\ldots,b_n)}$.
Obviously $\sigma'$ is norm-matching \wrt\ $\Gamma'$ and we can argue that: 
\begin{eqnarray*}
\nrm{\Gamma'}{a'}
&=&\quad\text{(inductive hypothesis)}\\
&&\nrm{\sigma'(\Gamma')}{\sigma'(a')}\\
&=&\quad\text{($\sigma'$ is norm-matching)}\\
&&\nrm{\sigma'(\Gamma')}{\sigma'(b)}\\
&=&\quad\text{(inductive hypothesis)}\\
&&\nrm{\Gamma'}{b}
\end{eqnarray*}
We can now show that $\sigma$ is norm-matching \wrt\ $\Gamma$: 
\begin{eqnarray*}
\nrm{\Gamma}{a}&=&\quad\text{(rewriting $\Gamma$)}\\
&&\nrm{\Gamma',x:a,\Gamma_{n+1}}{a}\\
&=&\quad\text{(by Law~\ref{nrm.sub} since $\nrm{\Gamma'}{a'}=\nrm{\Gamma'}{b}$)}\\
&&\nrm{\Gamma',(\Gamma_{n+1}\gsub{x}{b})}{a\gsub{x}{b}}\\
&=&\quad\text{(inductive hypothesis)}\\
&&\nrm{\sigma'(\Gamma',\Gamma_{n+1}\gsub{x}{b})}{\sigma'(a\gsub{x}{b})}\\
&=&\quad\text{(definition of $\sigma'$)}\\
&&\nrm{\sigma(\Gamma',x:a,\Gamma_{n+1})}{\sigma(a)}\\
&=&\quad\text{(rewriting $\Gamma$)}\\
&&\nrm{\sigma(\Gamma)}{\sigma(a)}
\end{eqnarray*}
\end{proof}
\noindent
As a consequence of Law~\ref{nrm.sub.nrm}, for any norm-matching substitution $\sigma$ we have $\Gamma\sngv a$ if and only if $\sigma(\Gamma)\sngv\sigma(a)$.
\begin{law}[Normability implies computability of all norm-matching substitutions to computable expressions]
\label{nrm.cesub}
For all $\Gamma$ and $a$: If $\Gamma\sngv a$ then for any norm-matching substitution $\sigma_{X,B}$ \wrt\ $\Gamma$ 
with $\sigma_{X,B}(x_i)\in\ce_{\sigma_{X,B}(\Gamma)}(\nrm{\Gamma}{b_i})$, for $1\leq i\leq n$, we have $\sigma_{X,B}(a)\in\ce_{\sigma_{X,B}(\Gamma)}(\nrm{\Gamma}{a})$.
\end{law} 
\begin{proof}
Let $\bar{a}=\nrm{\Gamma}{a}$ and $\bar{b}_i=\nrm{\Gamma}{b_i}$, for $1\leq i\leq n$.
Let $\sigma=\sigma_{X,B}$ be a norm matching substitution \wrt\ $\Gamma$ with $\sigma(x_i)\in\ce_{\sigma(\Gamma)}(\bar{b}_i)$.
The proof is by induction on $a$:
\begin{itemize}
\item
$a=\prim$: 
Since $\sigma(\prim)=\prim$ the property follows from Law~\ref{ce.basic}($iii$).
\item
$a=x:$ 
We have $\Gamma\sngv x$ and $\bar{a}=\nrm{\Gamma}{x}$.
There are two cases:
\begin{itemize}
\item
$x=x_i$, for some $i$.
Obviously $\sigma(x)=\sigma(b_i)$.
We know that $\sigma(b_i)\in\ce_{\sigma(\Gamma)}(\bar{b}_i)$.
\begin{eqnarray*}
\bar{a}&=&\nrm{\Gamma}{x}\\
&=&\quad\text{(Law~\ref{nrm.sub.nrm})}\\
&&\nrm{\sigma(\Gamma)}{\sigma(x)}\\
&=&\quad\text{(definition of substitution)}\\
&&\nrm{\sigma(\Gamma)}{\sigma(b_i)}\\
&=&\quad\text{(Law~\ref{nrm.sub.nrm})}\\
&&\nrm{\Gamma}{b_i}\\
&=&\quad\text{(rewriting)}\\
&&\bar{b}_i
\end{eqnarray*}
Hence $\sigma(x)=\sigma(b_i)\in\ce_{\sigma(\Gamma)}(\bar{a})$.
\item
If $x\neq x_i$ then $\sigma(x)=x$ and by Law~\ref{ce.basic}($ii$) we have $\sigma(x)\in\ce_{\sigma(\Gamma)}(\bar{c})$.
\end{itemize}
\item
$a=\binbop{x}{b}{c}$:
From $\Gamma\sngv\binbop{x}{b}{c}$ by definition of norming we obtain $\Gamma\sngv b$ and $\Gamma,x:b\sngv c$.
Let $\bar{b}=\nrm{\Gamma}{b}$ and $\bar{c}=\nrm{\Gamma,x:b}{c}$.
Applying the inductive hypothesis with the empty substitution $\sigma_{(),()}$, which is obviously norm-matching \wrt\ any context, we obtain
$b\in\ce_{\Gamma}(\bar{b})$ and $c\in\ce_{\Gamma,x:b}(\bar{c})$.

By Law~\ref{nrm.sub.nrm} we know that $\sigma(\nrm{\sigma(\Gamma)}{b})=\nrm{\Gamma}{b}=\bar{b}$.
By inductive hypothesis $\sigma(b)\in\ce_{\sigma(\Gamma)}(\bar{b})$.
and $c\in\ce_{\Gamma,x:b}(\bar{c})$.

Consider a $d$ where $d\in\ce_{\Gamma}(\bar{b})$.
In order to apply Law~\ref{ce.abs} we have to show that $\sigma(c)\gsub{x}{d}\in\ce_{\sigma(\Gamma)}(\bar{c})$.
If we define $X'=(x_1,\ldots,x_n,x)$, $B'=(b_1,\ldots,b_n,d)$, and $\sigma'=\sigma\gsub{x}{d}$
then obviously $\sigma'$ is norm-matching w.r.t~$(\Gamma,x:b)$ and substitutes to computable expressions.
Therefore by inductive hypothesis for $c$ we know that $\sigma'(c)=\sigma(c\gsub{x}{d})\in\ce_{\sigma(\Gamma)}(\bar{c})$.

Hence by Law~\ref{ce.abs} it follows that $\binbop{x}{\sigma(b)}{\sigma(c)}\in\ce_{\sigma(\Gamma)}([\bar{b},\bar{c}])$ which by definition of substitution is equivalent to $\sigma(\binbop{x}{b}{c})\in\ce_{\sigma(\Gamma)}([\bar{b},\bar{c}])$.
\item
$a=\prdef{x}{a_1}{a_2}{a_3}$:
From $\Gamma\sngv=\prdef{x}{a_1}{a_2}{a_3}$ by definition of norming we obtain $\Gamma\sngv a_1,a_2$, $\Gamma,x:a_1\sngv a_3$, and $\nrm{\Gamma}{a_2}=\nrm{\Gamma,x:a_1}{a_3}$.
Let $\bar{a}_1=\nrm{\Gamma}{a_1}$, $\bar{a}_2=\nrm{\Gamma}{a_2}$.

By inductive hypothesis with $\sigma$ we know that $\sigma_{X,B}(a_1)\in\ce_{\sigma(\Gamma)}(\bar{a}_1)$, $\sigma(a_2)\in\ce_{\sigma(\Gamma)}(\bar{a}_2)$, and
$\sigma_{X,B}(a_3)\in\ce_{\sigma(\Gamma,x:a_1)}(\bar{a}_2)=\ce_{(\sigma(\Gamma),x:\sigma(a_1))}(\bar{a}_2)$.

By Law~\ref{ce.mon}($i$) we then obtain  $\prdef{x}{\sigma(a_1)}{\sigma(a_2)}{\sigma(a_3)}\in\ce_{\sigma(\Gamma)}([\bar{a}_1,\bar{a}_2])$.
By definition of substitution this is equivalent to $\sigma(\prdef{x}{a_1}{a_2}{a_3})\in\ce_{\sigma(\Gamma)}([\bar{a}_1,\bar{a}_2])$.
\item
$a=\binop{a_1}{\ldots,a_n}$:
We have $\Gamma\sngv a$ and $\Gamma\sngv a_i$.
Let $\bar{a}=\nrm{\Gamma}{a}$, $\bar{a}_i=\nrm{\Gamma}{a_i}$.
By inductive hypothesis $\sigma(a_i)\in\ce_{\sigma(\Gamma)}(\bar{a}_i)$. 
By Laws~\ref{ce.mon.neg}, \ref{ce.mon.appl}, and the various cases of Law~\ref{ce.mon} we obtain $\binop{\sigma(a_1)}{\ldots,\sigma(a_n)}\in\ce_{\sigma(\Gamma)}(\bar{a})$.
By definition of substitution this is equivalent to $\sigma(\binop{a_1}{\ldots,a_n})\in\ce_{\sigma(\Gamma)}(\bar{a})$.
\qedhere
\end{itemize}
\end{proof}
\begin{law}[Computability law: Normability implies computability]
\label{nrm.ce}
For all $\Gamma$ and $a$: If $\Gamma\sngv a$ then $a\in\ce_{\Gamma}(\nrm{\Gamma}{a})$.
\end{law} 
\begin{proof}
Follows from Law~\ref{nrm.cesub} when taking the empty substitution $\sigma_{(),()}$ which obviously satisfies the required properties. 
\end{proof}
\subsection{Putting things together}%
\begin{law}[Strong normalization of valid expressions]%
\label{sn.valid}
For all $\Gamma$ and $a$: $\Gamma\sgv a$ implies $a\in\sn{}$.
\end{law} 
\begin{proof}
Assume $\Gamma\sgv a$.
By Law~\ref{val.nrm} this implies $\Gamma\sngv a$. 
By Law~\ref{nrm.ce} this implies $a\in\ce_{\Gamma}(\nrm{\Gamma}{a})$. 
By Law~\ref{ce.basic}($i$) this implies $a\in\sn{}$.
\end{proof}
\begin{definition}[Normal form]%
\index{expression!normal form of}
\nomenclature[fBasic06]{$\nf(a)$}{normal forms}
If $a\in\sn{}$ then $\nf(a)$ denotes the unique expression to which $a$ is maximally reducible.
Note that this definition is well-founded due to confluence of $\srd$ (Law~\ref{rd.confl}). 
Furthermore, due to strong normalisation (Law~\ref{sn.valid}), $\Gamma\sgv a$ implies that $\nf(a)$ exists.
\end{definition}
\begin{law}[Valid normal form and typing]%
\label{vnf.type}
For all $\Gamma$, $a$, and $b$: If $\Gamma\sgv a$ and $\Gamma\sgv a:b$ then $\Gamma\sgv\nf(a):b$
\end{law}
\begin{proof}
By Law~\ref{sn.valid} we know that $a\in\sn{}$ and hence that $\nf(a)$ exists. 
Obviously $a\rd\nf(a)$.
By Law~\ref{rd.type} this implies $\Gamma\sgv\nf(a):b$.
\end{proof}
\section{Decidability of the typing relation}%
\begin{law}[Decidability of the typing relation]
\label{decide.dtyp}
For any expression $a$ and context $\Gamma$ there is a terminating algorithm such that $\Gamma\sgv a$ iff the algorithm is not failing but computing an expression $b$ with $\Gamma\sgv a:b$. 
\end{law}
\begin{proof}
The algorithm to attempt to compute a type $b$ of $a$ is recursive on $a$ under the context $\Gamma$.
We outline the steps of this algorithm including its termination argument which is based on the weight function $W(E,a)$ from the proof of Law~\ref{mrd.sn.def} with $E=()$ and extended to contexts by the law
\[
W((),x_1:a_1,\ldots,x_n:a_n)=W((),a_1)+\ldots+W((),a_n)
\]
summing up the number of syntactic constructions in $\Gamma$.
\begin{giantitemize}
\item[$a=\prim$:] 
Obviously $b=\prim$.
\item[$a=x$:] 
If $\Gamma(x)$ is defined, i.e.~$\Gamma=(\Gamma_1,x:b,\Gamma_2)$ then try to compute a type of $b$ under $\Gamma_1$. 
The recursion will terminate since the weight of $\Gamma_1$ and $a$ is smaller than the weight of $\Gamma$ and $x$.
If the algorithm is successful then the result is $b$. In all other cases the algorithm fails.
\item[$a=\binbop{x}{a_1}{a_2}$:] 
First try to compute a type of $a_1$ under $\Gamma$. 
If the algorithm is successful then try to compute a type of $a_2$ under $(\Gamma,x:a_1)$. 
The recursion will terminate since the weight of $(\Gamma,x:a_1)$ and $a_2$ is smaller than the weight of $\Gamma$ and $\binbop{x}{a_1}{a_2}$. 
If the algorithm is successful and delivers a result $b_2$ then the result is $[x:a_1]b_2$. 
In all other cases the algorithm fails.
\item[$a=(a_1\,a_2)$:]
The algorithm is recursively applied to both $a_1$ and $a_2$ under $\Gamma$. 
If both applications are successful, we obtain expressions $b_1$ and $b_2$ where $\Gamma\sgv a_1:b_1$, and $\Gamma\sgv a_2:b_2$.
Due to strong normalization we can now check if $\nf(b_1)=[x:\nf(b_2)]d$ for some $d$.
If the check is positive, the result is $d\gsub{x}{a_2}$.  
In all other cases the algorithm fails.
\item[$a=\pleft{a_1}$:]
The algorithm is recursively applied to $a_1$ under $\Gamma$. 
If the applications is successful, we obtain an expression $b_1$ with $\Gamma\sgv a_1:b_1$.
Due to strong normalization we can now check if $\nf(b_1)=[c,d]$ or $\nf(b_1)=\prdef{x}{c}{d}{e}$ for some $c$, $d$, and $e$. 
If check is positive, the result is $c$.  
In all other cases the algorithm fails.
\item[$a=\pright{a_1}$:]
This case is similar to $\pleft{a_1}$.
\item[$\ldots$:]  All other cases are straightforward constructions along the lines of the previous cases.
\end{giantitemize}
This shows that the algorithm always delivers a correct result if it terminates. On the other hand, if $\Gamma\sgv a:b$ for some $b$, then one can show by a straightforward induction on the definition of $\Gamma\sgv a:b$ that our algorithm will compute a typing $\Gamma\sgv a:b'$ where $b\eqv b'$, hence the algorithm is also complete. We go through some crucial typing rules:
\begin{meditemize}
\item[\ax:] 
Obviously $b'=b=\prim$.
\item[\mystart:] 
We have $a=x$ and $\Gamma=(\Gamma',x:b)$ and $\Gamma'\sgv b:c$ for some $c$.
By inductive hypothesis our algorithm will compute a typing $\Gamma\sgv b:c'$ where $c\eqv c'$. 
Hence our algorithm will compute $b$ as type of $x$, i.e.~$b'=b$.
\item[\weak:]
We have $\Gamma=(\Gamma',x:c)$ where $\Gamma'\sgv c:d$ and $\Gamma'\sgv a:b$ for some $d$. 
By inductive hypothesis our algorithm will compute a typing $\Gamma'\sgv a:b'$ where $b\eqv b'$.
Since $x\notin\free(a)$, obviously our algorithm will also compute $\Gamma\sgv a:b'$ where $b\eqv b'$.
\item[\conv:]
We have $\Gamma\sgv a:b$ where $c\eqv b$ and $\Gamma\sgv c:d$ for some $c$ and $d$.
By inductive hypothesis our algorithm will compute typings $\Gamma\sgv a:b'$ where $b\eqv b'$.
Obviously $b'\eqv c$ which implies the proposition.
\item[\absu:] 
$a=[x:c]c_1$: We have $b=[x:c]c_2$ where $(\Gamma,x:c)\sgv c_1:c_2$ and  $\Gamma\sgv c:d$ for some $x$, $c$, $c_1$, $c_2$. 
By inductive hypothesis our algorithm will compute typings $(\Gamma,x:c)\sgv c_1:c_2'$ where $c_2\eqv c_2'$ and $\Gamma\sgv c:d'$ where $d\eqv d'$.
Hence it will compute $\Gamma\sgv a:[x:c]c_2'$ which implies the proposition since $[x:c]c_2'\eqv b$.
\item[$\ldots$] 
All other cases are straightforward constructions along the lines of the previous cases.
\qedhere
\end{meditemize}
\end{proof}
\section{Normal forms and consistency}%
\label{regular}
Due to confluence and the strong normalization result for valid expressions it is often sufficient to consider the normal form $\nf(a)$ instead of the expression $a$ itself when proving properties about expressions of \dcalc.
In Chapter~\ref{intro}, we have informally developed a stepwise characterization of the normal forms of valid expressions of \dcalc.
In this section, we study this more rigorously and use a characterization of valid normal forms to show consistency of \dcalc.
\begin{definition}[Valid normal forms]%
\index{normal forms!set of valid}
\index{dead ends!set of}
\nomenclature[fBasic12]{$\nf$}{valid normal forms}
\nomenclature[fBasic12]{$\de$}{dead ends}
\label{vnf}
The set of \emph{valid normal forms} is a subset of $\dexp$ and denoted by $\nf$.
The recursive characterization of valid normal forms in Table~\ref{vnf.set} also uses the auxiliary set of \emph{dead ends} denoted by $\de$.
\begin{table}[!htb]
\fbox{
\begin{minipage}{0.96\textwidth}
\begin{eqnarray*}
\\[-9mm]
\nf&=&\{\prim\}\;\union\;\{[x:a]b,[x!a]b,\prdef{x}{a}{b}{c}\sth a,b,c\in\nf\}\\
&&\;\union\;\{[a,b],[a+b],\injl{a}{b},\injr{a}{b},\case{a}{b}\sth a,b\in\nf\}\;\union\;\de\\
\de&=&\{x\sth x\in\dvar\}\;\union\;\{(a\,b),\pleft a, \pright a,\case{b}{c}(a)\sth a\in\de,b,c\in\nf\}\\
&&\;\union\;\{\myneg a\sth a\in\de,a\;\text{is not a negation}\}
\end{eqnarray*}
\end{minipage}
}
\caption{Valid normal forms\label{vnf.set}}
\end{table}
\end{definition}
\noindent
The following law shows that $\nf$ indeed characterizes the normal forms of valid expressions.
\begin{law}[Normal forms of valid expressions]%
\label{vnf.basic}
For all $a$ where $\sgv a$ we have that $a\in\nf$ iff $\nf(a)=a$.
\end{law}
\begin{proof}
We will prove the more general property that for all $a$ and $\Gamma$ with $\Gamma\sgv a$ we have that $a\in\nf$ iff $\nf(a)=a$.
Obviously, by construction, all elements of $\nf$ are irreducible.
For the reverse direction assume $\Gamma\sgv a$ and $\nf(a)=a$.
The proof of $a\in\nf$ is by induction on $a$. 
\begin{itemize}
\item 
If $a=x$ or $a=\prim$ then obviously $a\in\nf$.
\item 
If $a=[x:a_1]a_2$ then by Law~\ref{val.decomp}($i$) we know that $\Gamma\sgv a_1$ and $(\Gamma,x:a_1)\sgv a_2$. By inductive hypothesis $a_1, a_2\in\nf$. Hence by Law~\ref{sn.basic}($iii$) we know that $[x:a_1]a_2\in\nf$.
\item
The cases of $a$ being an existential abstraction, a product, a sum, a protected definition, a case distinction, or an injection, can be shown in a similar way.
\item 
If $a=(a_1\,a_2)$ then obviously $\Gamma\sgv a_1,a_2$ and therefore by inductive hypothesis $a_1,a_2\in\nf$.
We proceed by a case distinction on $a_1$:
\begin{itemize}
\item
Since $\Gamma\sgv (a_1\,a_2)$, $a_1$ cannot be the primitive constant, an injection, a protected definition, a sum, or a product.
\item
Since $\nf(a)=a$, $a_1$ cannot be a universal or existential abstraction.
\item
If $a_1=x$ then obviously $a_1\in\de$ and since $a_2\in\nf$ we have $a\in\de\subseteq\nf$.
\item
If $a_1=(a_3\,a_4)$, $a_1=\pleft{a_3}$, $a_1=\pright{a_3}$, or $a_1=\case{a_3}{a_4}$ then $a_1\in\nf$ (by definition of $\nf$) obviously implies $a_1\in\de$.
Hence $a\in\de\subseteq\nf$. 
\item
If $a_1=\myneg a_3$ then since $\nf(a_1)=a_1$ we know that $a_3$ can only be a variable.
Obviously $a_1\in\de$. Obviously this implies $a\in\de\subseteq\nf$. 
\end{itemize}
\item 
If $a=\pleft{a_1}$ then obviously $\Gamma\sgv a_1$ and therefore by inductive hypothesis $a_1\in\nf$.
We proceed by a case distinction on $a_1$:
\begin{itemize}
\item
Since $\Gamma\sgv\pleft{a_1}$, $a_1$ cannot be the primitive constant, a universal or existential abstraction, an injection, or a case distinction.
\item
Since $\nf(a)=a$, $a_1$ cannot be a protected definition, a product, or a sum.
\item
If $a_1=x$ then obviously $a_1\in\de$ and therefore $a\in\de\subseteq\nf$.
\item
If $a_1=(a_2\,a_3)$, $a_1=\pleft{a_2}$, $a_1=\pright{a_2}$, or $a_1=\case{a_2}{a_3}$ then $a_1\in\nf$ (by definition of $\nf$) obviously implies $a_1\in\de$.
Hence $a\in\de\subseteq\nf$. 
\item
If $a_1=\myneg a_2$ then since $\nf(a_1)=a_1$ we know that $a_2$ can only be an variable.
Obviously $a_1\in\de$. Obviously this implies $a\in\de\subseteq\nf$. 
\end{itemize}
\item 
The case $a=\pright{a_1}$ can be treated analogously to the previous case.
\item 
If $a=\myneg a_1$ then obviously $\Gamma\sgv a_1$ and therefore by inductive hypothesis $a_1\in\nf$.
Since $\nf(a)=a$, we know that $\nf(a_1)=a_1$ and that $a_1$ can only be an variable.
Obviously $a\in\de\subseteq\nf$. 
\qedhere
\end{itemize}
\end{proof}
\noindent
We need a couple of easy lemmas for the consistency proof.
First we note a property which motivated the construction of dead ends.
\begin{law}[Dead ends contain free variables]%
\label{nf.basic}
For all $a$: $\free(a)=\emptyset$ implies $a\notin\de$
\end{law}
\begin{proof}
The property is obvious by definition of $\de$. 
\end{proof}
\begin{law}[Valid normal forms of universal abstraction type]%
\label{nf.regular}
For all $x$, $a$, and $b$: If $a\in\nf$ and $\sgv a:[x:\prim]b$ then there is some $c\in\nf$ such that $a=\binbop{x}{\prim}{c}$ and $x:\prim\sgv c:b$.
\end{law}
\begin{proof}
Since $\free(a)=\emptyset$, by Law~\ref{nf.basic} we know that $a\notin\de$.
From $\sgv a:[x:\prim]b$, by Laws~\ref{valid.type} and~\ref{val.decomp}($i$) we know that $x:\prim\sgv b$.
Therefore we need to check the following remaining cases
\begin{itemize}
\item 
$a$ is the primitive constant, a protected definition, an injection, a sum, a case distinction, or a product, and $\sgv a:[x:\prim]b$: 
By definition of typing and properties of reduction this cannot be the case.
\item 
$a=\binbop{x}{a_1}{a_2}$. 
By Law~\ref{val.decomp}($i$) we know that $\sgv a_1$ and $x:a_1\sgv a_2$.
From  $\sgv\binbop{x}{a_1}{a_2}:[x:\prim]b$,  by Law~\ref{type.decomp}($ii$) we have $x:a_1\sgv a_2:c'$ for some $c'$ where $[x:a_1]c'\eqv[x:\prim]b$.
Hence $a_1\eqv \prim$ and $c'\eqv b$.

Therefore, since  $x:a_1\sgv a_2:c'$ and $a_1\eqv\prim$, by Law~\ref{eqv.env} we know that $x:\prim\sgv a_2:c'$.
Similarly since $c'\eqv b$, by typing rule \conv~we know that $x:\prim\sgv a_2:b$.
Hence we have shown the property with $c=a_2$.
\qedhere
\end{itemize}
\end{proof}
\noindent
We also need the following strengthening of Law~\ref{type.decomp}($ii$).
\begin{law}[Abstraction property]%
\label{abs.decomp}
For all $\Gamma$, $x$, $a$, $b$, and $c$: If $\Gamma\sgv\binbop{x}{a}{b}:[x:a]c$ then $\Gamma,x:a\sgv b:c$.
\end{law}
\begin{proof}
By Law~\ref{type.decomp}($ii$) we know that $\Gamma,x:a\sgv b:d$ for some $d$ with $c\eqv d$.
By Law~\ref{valid.type} we know that $\sgv [x:a]c$.
By Law~\ref{val.decomp}($i$) we know that $x:a\sgv c$.
Hence by rule \conv~we can infer that $\Gamma,x:a\sgv b:c$.
\end{proof}

\begin{law}[$\prim$-declaration property]%
\label{primdec}
For all $a$ and $b$: $x:\prim\sgv a:b$ implies that $b\neqv x$.
\end{law}
\begin{proof}
Du to Law~\ref{rd.type} (subject reduction) we may assume that $a\in\nf$.
The various cases of Law~\ref{type.decomp} imply that $b\neqv x$ in case $a$ is an abstraction, a sum, a product, a protected definition, a case distinction, or an injection. 
Hence by definition of $\nf$, it remains to look at the case $a\in\de$.
By definition of $\de$, since $x:\prim\sgv a$, $a$ can only be an variable $x$ or a negated variable $\myneg x$ and obviously $b\eqv\prim$.
The property follows since $\prim\neqv x$ and $\prim\neqv\myneg x$.
\end{proof}
\begin{law}[Consistency of \dcalc]%
\label{cons}
There is no expression $a$ such that $\sgv a:[x:\prim]x$. 
\end{law}
\begin{proof}
Assume that there is an expression $a$ with $\sgv a:[x:\prim]x$. 
By Theorem~\ref{sn.valid} (strong normalization) we know that there is a normal form $a':=\nf(a)$ with $a\rd a'$. 
By Lemma~\ref{valid.rd} we know that $\sgv a'$. 
By Theorem~\ref{rd.type} (subject reduction) we know that $\sgv a':[x:\prim]x$. 

By Lemma~\ref{vnf.basic} we know that $a'\in\nf$, hence, by Lemma~\ref{nf.regular} there is an expression $c$ where $\sgv\binbop{x}{\prim}{c}:[x:\prim]x$.
By Lemma~\ref{abs.decomp} this implies $x:\prim\sgv c:x$. 
By Lemma~\ref{primdec} this implies that $x\neqv x$.
Thus we have inferred a contradiction and therefore the proposition is true.
\end{proof}
\begin{remark}[Limitations of the consistency result]%
\label{cons.negation}
Law~\ref{cons} shows that there is no inherent flaw in the typing mechanism of \dcalc\ by which one could prove anything from nothing.
Note that the consistency result is limited to empty contexts, hence it does not cover the case of using negation or casting axioms (see Appendix~\ref{axioms}).
\end{remark}
\begin{remark}[Extension of the empty type]
It is an open issue if we can generalize the empty type $[x:\prim]x$ to $[x:a]x$ for any $a$ with $\sgv a$.
\end{remark}
\chapter{Comparison to other systems and possible extensions}
\label{related}
Due to the proposed systems use of $\lambda$-structured types, \dcalc\ falls outside the scope of PTS (see \eg\ \cite{Bar:93}).
In Section~\ref{concepts} we have indicated the differences between the core of~\dcalc\ and PTS\@.

Due to its origins from $\lambda^{\lambda}$ and its use of a reflexive typing axiom \dcalc\ does not use the concept of dependent product and it does not use a typing relation that can be interpreted as set membership.
Instead \dcalc\ introduces a number of operators which can be functionally interpreted by untyped $\lambda$-expressions by stripping of the type tags and negations, by interpreting sums, products, and protected definitions as binary pairs and both universal and existential abstraction as $\lambda$-abstraction (see Appendix~\ref{semantics}). The typing rules of \dcalc~then induce a relation between untyped $\lambda$-expressions.

Logically, the typing relation of \dcalc\ is restricted in the sense that additional axioms are required to obtain the complete set of negation properties.

In this chapter we discuss the use of \dcalc\ as a logic and then sketch several extensions of \dcalc, including paradoxical ones.
This will also illustrate the relation of \dcalc\ to other systems.
\section{Logical interpretation}%
\label{related.hol}
\dcalc\ is treating proofs and formulas uniformly as typed $\lambda$-expressions, and allows each of its operators to be used on both sides of the typing relation.
A subset of the operators of \dcalc, if used as types, can be associated with common logical predicates and connectors:
\begin{eqnarray*}
\prim&\simeq&\text{primitive constant}\\
(\ldots(x\,a_1)\,\ldots a_n)&\simeq&\text{atomic formula}\\{}
[x:a]b&\simeq&\text{universal quantification}\\{}
[x!a]b&\simeq&\text{existential quantification}\\{}
[a,b]&\simeq&\text{conjunction}\\{}
[a+b]&\simeq&\text{disjunction}\\{}
\myneg a&\simeq&\text{negation}
\end{eqnarray*}
In Section~\ref{ex.logic} we have shown that based on the type system of \dcalc\ many logical properties of these connectors can be derived without further assumptions. 
Furthermore, based on a strong normalization result we have shown (Law~\ref{cons}) that \dcalc\ is consistent in the sense that the type $[x:\prim]x$ is empty in \dcalc\ under the empty context.
In this sense, \dcalc\ can be seen as a (higher-order) logic where typing can be interpreted as a deduction typing to the proposition it has deduced~\cite{Howard69}. 

However, in order to have the complete properties of classical negation additional axioms have to be assumed and
we could not show consistency of the type system under these axioms by means of strong normalization.
Similarly, formalizations of mathematical structures in  \dcalc\ were introduced axiomatically.
In this sense the expressive power of \dcalc\ is limited and each axiomatization has to be checked carefully for consistency.

Furthermore, there are two important pragmatic issues which differ from common approaches: 
\begin{itemize}
\item
First, inference systems for higher-order logic based on typed-$\lambda$-calculus such as~\cite{church40}\cite{Paulson1989} typically make a distinction between the type of propositions 
and one or more types of \emph{individuals}. In \dcalc, one the one hand there is no such distinction, all such types must either be $\prim$ itself or declared using $\prim$.
On the other hand, due to the restricted formation rules which serve to ensure consistency as well as uniqueness of types, in \dcalc, $\prim$ does not allow to quantify over all propositions of \dcalc\ and additional axioms schemes must be used when reasoning with formulas of complex structure.
\item
Second, \dcalc\ has several operators which are not common logical connectors:
\begin{eqnarray*}
\prdef{x}{a}{b}{c}&\simeq&\text{protected definition}\\{}
\pleft{a},\pright{a}&\simeq&\text{projections}\\{}
\case{a}{b}&\simeq&\text{case distinction}\\{}
\injl{a}{b}, \injr{a}{b}&\simeq&\text{left and right injection}
\end{eqnarray*}
However, these operators have meaningful type-roles for defining functions over propositions. 
\end{itemize}
Note also that there is a strong relation between left projection $\pleft{a}$ on a deduction $a$ and Hilberts $\epsilon$-operator $\epsilon x.P$ on a formula $P$ as sometimes used in higher-order logic~\cite{church40,Paulson1989} with a law like:
\[
\forall x.(P \Rightarrow P\gsub{x}{\epsilon x.P})
\]
In a classical logical setting this is obviously implied by
\[
(\exists x.P) \Rightarrow P\gsub{x}{\epsilon x.P}
\]
The latter property can be approximated in \dcalc\ by
\[
[y:[x!\prim](P\,x)](P\,\pleft{y})
\]
and actually is a law since
\[
[y:[x!\prim](P\,x)]\pright{y}\;:\;[y:[x!\prim](P\,x)](P\,\pleft{y})
\]
This illustrates again how existential abstraction and the projection operators together embody a strong axiom of choice.

Finally there is no equality operator in \dcalc, a notion of equality is defined indirectly only through equality of expressions modulo reduction
\section{Negation}%
\label{related.negation}
In Section~\ref{negation} we have explained that the direct encoding of negation of $a$ as $[a\fun\ff]$, where $\ff$ abbreviates $[x:\prim]x$, is not possible due to the use of $\lambda$-structured types. 
This is a major obstacle to completely internalize negation properties into \dcalc, since encoding negation in the above style is very common in logic.

In \dcalc\, rather than defining negation by implication to falsehood, 
negation is incorporated by defining a subset of the equivalence laws of negation as equalities ($\eqv$).
The purpose is to have unique formal forms with respect to negation in order to simplify deductions.
A direct isomorphism between $\myneg\myneg a$ and $a$ has been advocated in~\cite{Munch2014}.
De-Morgan-style laws for propositional operators have been used to define an involutive negation in a type language~\cite{Barbanera:1996}.

As shown in Section~\ref{ex.logic}, additional axioms schemes must be assumed to have the full set of properties of logical negation.
While this is adequate when assuming computationally irrelevant proofs, it leaves the issues of consistency of the axiomatic extensions, i.e.~the question if typing with axioms $?\sgv a:b$\footnote{See Appendix~\ref{axioms}} is consistent.
We briefly discuss this drawback of \dcalc\ and why it seems inevitable.

Several approaches have been proposed to internalize classical reasoning into $\lambda$-calculus.
$\lambda\mu$-calculus is adding classical reasoning to $\lambda$-calculus by additional control operators~\cite{Parigot1992,Parigot2000,DBLP:conf/icfp/CurienH00,kesner_et_al:LIPIcs:2017:7713}.
Control operators give explicit control over the context in which an expression is evaluated. 
On the one hand, extending \dcalc\ by control operators for negation, obliterating the need for negation axioms, would extend its deductive means without clear necessity.
On the other hand, we are not aware of a lambda-typed system with control operators for negation in which one could translate \dcalc\ with negation axioms.

A major obstacle towards the use of control operators in \dcalc\ seems to be due to the use of a case distinction operator. 
To illustrate this issue in a more concrete way, suppose we would, along the lines of~\cite{barg97a}, define an negation introduction operator $[x:_{\mu}a,b]$ by means of the following typing rule:
\[
\frac{\Gamma,x:a\gv b:[c,\myneg c]}{\Gamma\gv[x:_{\mu}a,b]:\myneg a}
\]
For any $a$ with $\Gamma\gv a$ we could then prove:
\[
\Gamma\gv[x:_{\mu}[\myneg a,a],x]:[a+\myneg a]
\]
However, in order to normalize expressions so as to prove consistency of the system based on normal forms, we would have the need to define reduction rules for $[x:_{\mu}a,b]$. 
First of all, it not possible to reduce a negation introduction as an argument of a case distinction, i.e.~when normalizing a function with a sum domain.  
\[
(\case{c}{d}\,[x:_{\mu}a,b])\;\rd\;?
\]
Symmetrically, the same problem would arise when normalizing a function with an existential abstraction (intuitively an infinite sum) as domain.
\[
\pleft{[x:_{\mu}[y!c]d,b]}\;\rd\;? \qquad\pright{[x:_{\mu}[y!c]d,b]}\;\rd\;? 
\]
This phenomenon seems to be a consequence of Lafont's critical pairs (\cite{girard1989proofs}).
Note that \dcalc\ is using a simplified setting as we do not deal with general cut elimination, which would correspond to a function composition operator, but only a more specific elimination operator where a function is applied to an argument. 

Note also that there are fundamentally different approaches towards classical logic in type system, \eg\ in~\cite{Luo2007} a modular approach is proposed where various logical foundations can be combined with type systems.
\section{Paradoxical extension by subsumptive subtyping}%
\label{paradox.subtyping}
In Section~\ref{typecasting} we have mentioned that extending the core of \dcalc\ (Table~\ref{dcalculus.core}) by an subsumptive subtyping mechanism would lead to an inconsistent system.  
We present this argument here in more detail.

There is a well-known system with the property $type:type$ which is inconsistent~\cite{Girard72}. 
This system, referred to as $\lambda*$ in~\cite{Bar:93} (Section~5.5), can be presented as a pure type system $(S=*,A=\{*:*\};R=\{(*,*)\})$ as follows.
\begin{align*}
\it{(axioms)}\;\;&\frac{}{\sgv *:*}\\
\it{(start)}\;\;&\frac{\Gamma\sgv A:*}{\Gamma,x:A\sgv x:A}\\
\it{(weakening)}\;\;&\frac{\Gamma\sgv A:B\quad\Gamma\sgv C:*}{\Gamma,x:C\sgv A:B}\\
\it{(product)}\;\;&\frac{\Gamma\sgv A:*\quad\Gamma,x:A\sgv B:C}{\Gamma\sgv(\Pi x:A.B):*}\\
\it{(application)}\;\;&\frac{\Gamma\sgv F:(\Pi x:A.B)\quad\Gamma\sgv a:A}{\Gamma\sgv(Fa):B\gsub{x}{a}}\\
\it{(abstraction)}\;\;&\frac{\Gamma,x:A\sgv b:B\quad\Gamma\sgv(\Pi x:A.B):*}{\Gamma\sgv(\lambda x:A.b):(\Pi x:A.B)}\\
\it{(conversion)}\;\;&\frac{\Gamma\sgv A:B\quad\Gamma\sgv B':*\quad B\eqv B'}{\Gamma\sgv a:B'}
\end{align*}
Here $A$, $B$, $C$ range over types and $a$, $b$, $c$ range over $\lambda$-expressions. $x$, $y$, $z$ range over variables of both.

\noindent
When replacing $(\Pi x:A.B)$ and $\lambda x:A.b$ by (universal) abstraction $[x:a]b$ and writing $\prim$ for $*$ the above rules rewrite as
\begin{align*}
\it{(axioms)}\;\;&\frac{}{\sgv \prim:\prim}\\
\it{(start)}\;\;&\frac{\Gamma\sgv a:\prim}{\Gamma,x:a\sgv x:a}\\
\it{(weakening)}\;\;&\frac{\Gamma\sgv a:b\quad\Gamma\sgv c:\prim}{\Gamma,x:c\sgv a:b}\\
\it{(product)}\;\;&\frac{\Gamma\sgv a:\prim\quad\Gamma,x:a\sgv b:c}{\Gamma\sgv[x:a]b:\prim}\\
\it{(application)}\;\;&\frac{\Gamma\sgv c:[x:a]b\quad\Gamma\sgv d:a}{\Gamma\sgv (c\,d):b\gsub{x}{d}}\\
\it{(abstraction)}\;\;&\frac{\Gamma,x:a\sgv b:c\quad\Gamma\sgv [x:a]c:\prim}{\Gamma\sgv [x:a]b:[x:a]c}\\
\it{(conversion)}\;\;&\frac{\Gamma\sgv a:b\quad\Gamma\sgv b':\prim\quad b\eqv b'}{\Gamma\sgv a:b'}
\end{align*}
These rules would all be true in an extension of core of \dcalc\ with a subsumptive subtyping rule and axiom as defined in Section~\ref{typecasting}: 
\[
\frac{\Gamma\gv a:b\qquad b\leqslant c\qquad \Gamma\gv c:d}{\Gamma\gv a:c}
\qquad\frac{}{a\leqslant\prim}
\]
For example consider the rule \emph{product}: Assume that $\Gamma\sgv a:\prim$ and $\Gamma,x:a\sgv b:c$. By typing rule \emph{abs$_U$} we can infer $\Gamma\sgv[x:a]b:[x:a]c$ and then, since $[x:a]c\leqslant\prim$ by the type inclusion rule we obtain $\Gamma\sgv[x:a]b:\prim$.

Hence such a system (and thus obviously~\dcalc) would be inconsistent in the sense that there would be an expression $a$ such that $\sgv a:[x:\prim]x$.
\section{Paradoxical extension with casting operators}%
\label{paradox.casting}
In Section~\ref{typecasting} we have mentioned that extending the core of \dcalc\ with $\prim$-casting and cast introduction and elimination operators, $\cast$, $\castin{a}b$, and  $\castout{a}b$ where  
\[
\frac{\Gamma\sgv a:b}{\Gamma\sgv\cast a:\prim}
\qquad
\frac{\Gamma\sgv a:b}{\Gamma\sgv\castin{b}a:\cast b}
\qquad
\frac{\Gamma\sgv a:\cast b}{\Gamma\sgv\castout{b}a:b}
\]
as well as a reduction axiom for canceling out pairs of cast introduction and elimination, i.e.
\[
\castout{c}\castin{b}a\rd a
\]
would lead to an inconsistent system.
To motivate this consider the following reconstruction of $\lambda*$ (see Section~\ref{paradox.subtyping}) using $\prim$-casting operations:
\begin{align*}
\it{(axioms)}\;\;&\frac{}{\sgv \prim:\prim}\\
\it{(start)}\;\;&\frac{\Gamma\sgv a:\prim}{\Gamma,x:a\sgv x:a}\\
\it{(weakening)}\;\;&\frac{\Gamma\sgv a:b\quad\Gamma\sgv c:\prim}{\Gamma,x:c\sgv a:b}\\
\it{(product)}\;\;&\frac{\Gamma\sgv a:\prim\quad\Gamma,x:a\sgv b:c}{\Gamma\sgv\cast[x:a]b:\prim}\\
\it{(application)}\;\;&\frac{\Gamma\sgv c:\cast[x:a]b\quad\Gamma\sgv d:a}{\Gamma\sgv(\castout{[x:a]b}{c}\;d):b\gsub{x}{d}}\\
\it{(abstraction)}\;\;&\frac{\Gamma,x:a\sgv b:c\quad\Gamma\sgv\cast[x:a]c:\prim}{\Gamma\sgv\castin{[x:a]c}[x:a]b:\cast[x:a]c}\\
\it{(conversion)}\;\;&\frac{\Gamma\sgv a:b\quad\Gamma\sgv b':\prim\quad b\eqv b'}{\Gamma\sgv a:b'}
\end{align*}
This motivates the following mapping $\PTStoD_{\Gamma}(A)$ from a PTS expression $A$ and a PTS-context $\Gamma$ where $\Gamma\sgv A:B$, for some $B$, to an expression in \dcalc\ with extended $\prim$-casting:
\begin{eqnarray*}
\PTStoD_{\Gamma}(*)&=&\prim\\
\PTStoD_{\Gamma}(x)&=&x\\
\PTStoD_{\Gamma}(\Pi x:A.B)&=&\cast [x:\PTStoD_{\Gamma}(A)]\PTStoD_{\Gamma,x:A}(B)\\
\PTStoD_{\Gamma}(\lambda x:A.B)&=&\castin{c}[x:\PTStoD_{\Gamma}(A)]\PTStoD_{\Gamma,x:A}(B)\;\;\text{where}\;\;\PTStoD(\Gamma)\sgv \PTStoD_{\Gamma}(\lambda x:A.B):c\\
\PTStoD_{\Gamma}((A\,B))&=&(\castout{c}\PTStoD_{\Gamma}(A)\;\PTStoD_{\Gamma}(B))\;\;\text{where}\;\;\PTStoD(\Gamma)\sgv \PTStoD_{\Gamma}(A):\cast c
\end{eqnarray*}
where the mapping $\PTStoD(\Gamma)$ from PTS-contexts to contexts in \dcalc\ is recursively defined as follows:
\begin{eqnarray*}
\PTStoD(\Gamma)&=&\PTStoD_{()}(\Gamma)\\
\PTStoD_{\Gamma}(())&=&()\\
\PTStoD_{\Gamma}(x:A,\Gamma')&=&(x:\PTStoD_{\Gamma}(A),\PTStoD_{\Gamma,x:A}(\Gamma'))
\end{eqnarray*}
It is straightforward to show that this translation preserves reduction in $\lambda*$ and therefore typing.
\[
\frac{\Gamma\sgv A:B\quad A\brd{n}C}{\PTStoD_{\Gamma}(A)\rdn{m}\PTStoD_{\Gamma}(C)\;\text{where}\; m\geq n}\qquad\frac{\Gamma\gv A:B}{\PTStoD(\Gamma)\gv\PTStoD_{\Gamma}(A):\PTStoD_{\Gamma}(B)}
\]
Here $A\brd{n}C$ denotes a $n$-step beta reduction in a PTS.
The crucial step of the $\beta$-reduction preservation proof is using extended $\prim$-casting:
\begin{eqnarray*}
\PTStoD_{\Gamma}((\lambda x:A.B\;C))&=&(\castout{d}\PTStoD_{\Gamma}(\lambda x:A.B)\;\PTStoD_{\Gamma}(C))\\
&=&(\castout{d}\castin{d}[x:\PTStoD_{\Gamma}(A)]\PTStoD_{\Gamma,x:A}(B)\;\PTStoD_{\Gamma}(C))\\
&\srd&([x:\PTStoD_{\Gamma}(A)]\PTStoD_{\Gamma,x:A}(B)\;\PTStoD_{\Gamma}(C))\\
&\srd&\PTStoD_{\Gamma,x:A}(B)\gsub{x}{\PTStoD_{\Gamma}(C)}\\
&=&\PTStoD_{\Gamma}(B\gsub{x}{C})
\end{eqnarray*}
where $\PTStoD(\Gamma)\sgv\PTStoD_{\Gamma}(\lambda x:A.B):\cast d$ for some $d$.

Hence a (nonterminating) proof $A$ of $\Pi x:*.x$ in $\lambda *$ could be transformed into a (nonterminating) proof of $\cast[x:\prim]x$, and therefore obviously also of $[x:\prim]x$, in \dcalc\ with extended casting rules.
\section{Relaxing uniqueness of types}%
\label{related.unique}
Note that uniqueness of types (\ref{type.confl}) was not needed in the strong normalisation proof but only the weaker property~\ref{val.nrm}.

Protected definitions $\prdef{x}{a}{c}{d}$ carry a type tag $d$ allowed to use $x$ in order to ensure uniqueness of types.
Law~\ref{val.nrm} would be retained if we remove the type tag and the binding of $x$ from protected definitions: 
\[
\frac{\Gamma\sgv a: b\quad\Gamma\sgv c:d\gsub{x}{a}\quad\Gamma,x:b\sgv d:e}{\Gamma\sgv[\_\!\doteq\!a,c]:[x!b]d}
\]
However, this may exponentially increase the type variants of a protected definition, \eg:
\[
\frac{\Gamma\sgv a:b\quad\Gamma\sgv c:[d\fun d]\quad\Gamma,x:b\sgv d:e}{\Gamma\sgv [\_\!\doteq\!a,c]:[x!b][d\fun d]}
\quad
\frac{\Gamma\sgv a:b\quad\Gamma\sgv c:[d\fun d]\quad\Gamma,x:b\sgv d:e}{\Gamma\sgv [\_\!\doteq\!a,c]:[x!b][x\fun d]}
\]
\[
\frac{\Gamma\sgv a:b\quad\Gamma\sgv c:[d\fun d]\quad\Gamma,x:b\sgv d:e}{\Gamma\sgv [\_\!\doteq\!a,c]:[x!b][x\fun d]}
\quad
\frac{\Gamma\sgv a:b\quad\Gamma\sgv c:[d\fun d]\quad\Gamma,x:b\sgv d:e}{\Gamma\sgv [\_\!\doteq\!a,c]:[x!b][x\fun x]}
\]
See also the discussion about the typing rule for existential abstraction in Section~\ref{existential2}.
In case of universal abstractions the situation is different:
Adding the following type rule for universal abstractions
\[
\frac{\Gamma,x:a\sgv b:\prim}{\Gamma\sgv [x:a]b:\prim}
\]
would violate both~\ref{type.confl} and~\ref{val.nrm} and together with $\sgv \prim:\prim$ result in a paradoxical system. 
\section{Lack of extensionality}%
As indicated in Section~\ref{overview}, \dcalc\ introduces universal and existential abstractions (and similarly products and sums) as functions that are semantically distinct but have equivalent behavior on arguments.
This naturally precludes adding axioms of extensionality as their sole purpose is to deny such distinctions.

While this restriction can be criticized on a theoretical level, from a pragmatic point of view it seems less relevant as in formalizations one can always replace a function reference $x$ where \eg\ $x:[y:\prim]\prim$ or $x:[y!\prim]\prim$ by $[y:\prim](x\,y)$.
\section{Abbreviation systems}%
\label{related.definition}
Complex systems of abbreviations spanning over several conceptual or abstraction levels play a major conceptual role in mathematical work.
A multitude of proposals for incorporating definitions into typed $\lambda$-calculi have been made, \eg\ \cite{deBruijn94}~\cite{Guidi09}~\cite{KAMAREDDINE1999}. 
Support for definitional extensions for systems closely related to Automath's $\Lambda$ have been investigated in~\cite{PdG91}. 
Note that definitional extensions would allow for a relaxation in the type rule for existential abstractions:
\[
\pdefm\qquad\frac{\Gamma\sgv a: b\quad\Gamma\sgv c:d\gsub{x}{a}\quad\Gamma,x:b\sgv d:e}{\Gamma\sgv\prdef{x}{a}{c}{d}:[x!b]d}
\]
Since $c$ may use $a$ in its type, it could also use an abbreviation $x$ of $a$. Hence it seems highly plausible that also $c$ itself may use an abbreviation $x$ for $a$. 
In a calculus that includes definitions this dependency could be modelled. In \dcalc, this is not possible and maximally unfolded expressions must be used. 

While support for definitional extensions is undoubtedly important, in our setting, they have not been necessary to formulate \dcalc.
Note that in other settings this might be different and abbreviation systems become indispensable, \eg\ \cite{Luo2003}.
In our case, these concepts must of course play a major role in any practically useful approach for formal deductions based on \dcalc. 
\section{Inductively defined datatypes and functions}%
\dcalc\ has computationally-irrelevant proofs, i.e.~it is not possible to extract for example primitive recursive functions from valid expressions.
This constitutes a major difference to well-known typing systems such as Coq~\cite{Bertot2010}.
Therefore in \dcalc, it is not possible to extract programs from proofs or to support verification of programs using (recursively-defined) custom datatypes.
However, one could argue to include generic mechanisms for axiomatizing well-grounded inductive definitions as this may relieve the notational burden of specifying constructors and induction principles, \eg\ instead of the declarations of $\nats$, $0$, $s$, $S_1$, $S_2$, and $ind$ in Section~\ref{examples.nats}, one could introduce a shorthand notation such as
\[
{\bf Inductive}\;\nats : \it{Sort} := 0 : \nats\;|\;S : [\nats\fun\nats]
\] 
However, note that the equality relation used to state injectivity of constructors on natural numbers was not predefined but axiomatized. 
Therefore one would need an extended shorthand notation to also define injectivity with respect to a equality relation $e:Equality$ 
\[
{\bf Inductive}(e:Equality)\;\nats : \it{Sort} := 0 : \nats\;|\;S : [\nats\fun\nats]
\] 
where all laws involving equality would be stated using the following definition: 
\[()\!\!=\!\!():=\pright e
\]
Again, we see such generic notational extensions and specific support for inductive proofs as part of practically useful languages for formal deductions based on \dcalc. 
\appendix
\chapter{Typing and validity with axioms for negation and casting}
\label{axioms}
We can define a weakening of the typing relation, which assumes a finite number of declarations instantiating axiom schemes for negation and casting, as follows:
\begin{definition}[Typing and validity with axioms]
\nomenclature[gRel04]{$?,\Gamma\sgv a:b$}{typing with axioms}%
\index{typing!with axioms}
\nomenclature[gRel05]{$?,\Gamma\sgv a$}{validity with axioms}
\index{validity!with axioms}
In Chapter~\ref{examples}, we have been using the following axiom schemes:
\begin{eqnarray*}
\negax^+_{a,b}&:&[[a+b]\fun[\myneg a\fun b]]\\
\negax^-_{a,b}&:&[[\myneg a\fun b]\fun[a+b]]\\
\cast_a&:&[a\fun\prim]\\
\castin{a}&:&[x:a][x\fun \cast_a x]\\
\castout{a}&:&[x:a][\cast_a x\fun x]\\
\dcastin{a,b}&:&[x:a;y:[x\fun b];z:x][y(z)\fun y(\castout{a}(x,\castin{a}(x,z)))]\\
\dcastout{a,b}&:&[x:a;y:[x\fun b];z:x][y(\castout{a}(x,\castin{a}(x,z)))\fun y(z)]
\end{eqnarray*}
where we assumed $\negax^+_{a,b},\negax^-_{a,b},\cast_a,...\in\dvar$ form an infinite subset of variables  $\mathit{I_{Ax}}$ indexed over expressions.

Formally, typing with axioms $?,\Gamma\gv a:b$ under the above axiom scheme could be defined so as to require that $\free([\Gamma]b)=\emptyset$ and that there is a context $\Gamma'$ consisting of (a finite set of) declarations of variables from $\mathit{I_{Ax}}$ with type-mappings as defined above, such that $\Gamma',\Gamma\sgv a:b$.

Similar to typing $?\sgv a:b$ abbreviates $?,()\sgv a:b$.
We also use the notion of validity with axioms, defined by
\begin{eqnarray*}
?,\Gamma\gv a&:=&\exists b:\;\;?,\Gamma\sgv a:b
\end{eqnarray*}
If $\Gamma=()$ we just write $?\sgv a:b$.
\end{definition}
\chapter{Mapping to untyped lambda-calculus}
\label{semantics}
This chapter defines tentative steps towards a set-theoretical semantics of \dcalc.
As we do not make a distinction between $\lambda$ and $\Pi$ we do not take the aproach to interpret the typing relation as set inclusion in additive domains as done in the semantic of classical Automath~\cite{BARENDREGT1983127}.
Rather, we take a two-level approach: the typing relation will be interpreted as a ternary relation involving untyped $\lambda$-calculus expressions which are then themselves interpreted \eg\ in reflexive domains as usual.
Obviously the semantic interpretations should reflect intended properties of~\dcalc. 
To this end we define a semantic space and some fundamental properties our semantic interpretations are to satisfy. 
We define two straightforward mappings from \dcalc\ expressions into the semantic space and discuss their properties. 
Both interpretations are non-trivial but both abstract from some of the properties of \dcalc. 
We briefly discuss how more granular semantics spaces and corresponding mappings could recover more detailed properties of~\dcalc.  
\section{The semantic space}
\subsection{Lambda-calculus}
\index{lambda-expressions!untyped}
\nomenclature[bSets7]{$t,t_1,t_2,\cdots$}{untyped lambda-expressions}
\nomenclature[kCalc01]{$D$}{lambda-expressions}
\index{reduction!of lambda-expressions}
\nomenclature[kCalc02]{$t_1\brd{}t_2$}{reduction on lambda-expressions}
\index{equivalence!of lambda-expressions}
\nomenclature[kCalc020]{$t_1\eq t_2$}{equivalence on lambda-expressions}
We use $\lambda$-expressions with the usual notations $\lambda x.t$ for $\lambda$-abstraction and $(t_1\;t_2)$ for application. 
We use the notation $\brd{}$ for $\beta$-contraction and $\eq$ for $\beta$-contraction-induced congruence,
Furthermore, to avoid notational clutter we often use the following notations:
\begin{eqnarray*}
\lambda x_1\cdots x_n.t&\equiv&\lambda x_1\cdots.\lambda x_n.t\\
(t\;t_1\;t_2 \cdots t_n)&\equiv&(\cdots((t\;t_1)\;t_2) \cdots t_n)
\end{eqnarray*}
We also often omit the brackets around the outermost applications when expression are sufficiently disambiguated.
We assume a special constant $\primfun$.
\subsection{A rudimentary semantic structure}
\begin{definition}[$\lambda$-Context]
\index{lambda-context}
\nomenclature[kCalc03]{$C$}{lambda context}
A $\lambda$-context $C$ is a sequence $(x_1:t_1,\cdots,x_n:t_n)$ where $x_i\neq x_j$ and $t_i\in D$.
The lookup $C(x)$ is defined in an obvious way.
\end{definition}
\noindent
We now characterize a subset of triples $(C,t_1,t_2)$.  
\begin{definition}[d-Structure]\index{lambda-context}
\nomenclature[kCalc04]{$D_s$}{d-structure}
\index{d-structure}
A d-structure $D_s$ is a set of triples $(C,t_1,t_2)$ where $C$ is a $\lambda$-context, $t_1,t_2\in D$ and which respects the following rules
\begin{itemize}
\item[$i$:]$((),\primfun,\primfun)\in D_s$.
\item[$ii$:]$(C,t_1,t_2)\in D_s$ implies $((C,x:t_1),x,t_1)\in D_s$.
\item[$iii$:]$(C,t_1,t_2),(C,t_3,t_4)\in D_s$ implies $((C,x:t_3),t_1,t_2)\in D_s$.
\item[$iv$:]$(C,t_1,t_2)\in D_s$, $t_2\eq t_3$, and $(C,t_3,t_4)\in D_s$ imply $(C,t_1,t_3)\in D_s$.
\item[$v$:]$(C,t_1,t_2)\in D_s$, $t_1\eq t_3$, and $(C,t_3,t_4)\in D_s$ imply $(C,t_3,t_2)\in D_s$.
\item[$vi$:]$(C,t_1,t_2),(C,t_3,t_4)\in D_s$ imply $(C,\lambda x.(x\;t_1\;t_2),\lambda x.(x\;t_3\;t_4))\in D_s$.
\item[$vii$:]$(C,t_1,t_2)\in D_s$ imply $(C,t_1\;\lambda xy.x,t_2\;\lambda xy.x),(C,t_1\;\lambda xy.y,t_2\;\lambda xy.y)\in D_s$.
\end{itemize}
\end{definition}
\noindent
Note that due to the first condition $\emptyset$ is not a d-structure.
\section{A type-stripping interpretation}
\subsection{Mapping to functional content}
\begin{definition}[Mapping to $D$]
\nomenclature[kCalc05]{$\ssem$}{type-stripping semantic interpretation}
\index{semantic interpretation!type-stripping}
The function $\ssem$ translates a \dcalc\ expression into $D$ (Table~\ref{sem1.def}).
\begin{table}[!htb]
\fbox{
\begin{minipage}{0.96\textwidth}
\begin{eqnarray*}
\\[-8mm]
\ssem(\prim)&=&\primfun\\
\ssem(x)&=&x\\
\ssem([x:a]b)=\ssem([x!a]b)&=&\lambda x.\ssem(b)\\
\ssem((a\,b))&=&\ssem(a)\;\ssem(b)\\
\ssem(\prdef{x}{a}{b}{c})=\ssem([a,b])=\ssem([a+b])&=&\lambda x.(x\;\ssem(a)\;\ssem(b))\\ 
\ssem(\pleft{a})&=&\ssem(a)\;\lambda xy.x\\ 
\ssem(\pright{a})&=&\ssem(a)\;\lambda xy.y\\
\ssem(\case{a}{b})&=&\lambda x.(x\;\ssem(a)\;\ssem(b))\\ 
\ssem(\injl{a}{b})&=&\lambda xy.(x\;\ssem(a))\\
\ssem(\injr{a}{b})&=&\lambda xy.(y\;\ssem(b))\\
\ssem(\myneg a)&=&\ssem(a)
\end{eqnarray*} 
\end{minipage}
}
\caption{Type-stripping mapping to $D$\label{sem1.def}}
\end{table}
For a context $\Gamma$, 
$\ssem(\Gamma)$ is defined by applying $\ssem$ to all expressions in $\Gamma$ to yield a $\lambda$-context, i.e.~$\ssem()=()$ and $\ssem(x:a, \Gamma)=(x:\ssem(a),\ssem(\Gamma))$. 
\end{definition}
\begin{remark}[Examples]
Consider the expression $[x:\prim][y:x]y$ which is of type $[x:\prim][y:x]x$. It is interpreted as $\lambda\,xy.y$ of type $\lambda\,xy.x$.
As a slightly more involved example consider the derivation of the modus-ponens rule 
\[[p:\prim][q:\prim][x:p][y:[z:p]q](y\,x)
\]
which is of type
\[[p,q:\prim][x:p][y:[z:p]q]q
\]
Is interpreted as $\lambda\,pqxy.(y\,x)$ of type $\lambda\,pqxy.q$. 
\end{remark}
\begin{law}[Reduction law]
\label{sem1.rd}
$a\rd b$ implies $\ssem(a)\brd{}\ssem(b)$. 
\end{law}
\begin{proof}
The proposition is shown by induction on the definition of reduction.
\begin{itemize}
\item{$\beta_1$}:
\begin{eqnarray*} 
\ssem(([x:a]b\;c))
&=&(\ssem([x:a]b)\,\ssem(c))\\
&=&((\lambda x.\ssem(b))\,\ssem(c))\\
&\brd{}&\ssem(b)\gsub{x}{\ssem(c)}\\
&=&\ssem(b\gsub{x}{c})
\end{eqnarray*}
\item{$\beta_2$}: Similar
\item{$\beta_3$}:
\begin{eqnarray*} 
\ssem((\case{a}{b}\,\injl{c}{d}))
&=&(\lambda x.(x\;\ssem(a)\;\ssem(b))\;\lambda xy.(x\;\ssem(c)))\\
&\brd{}&\lambda xy.(x\;\ssem(c))\;\ssem(a)\;\ssem(b)\\
&\brd{}&\lambda y.(\ssem(a)\;\ssem(c))\;\ssem(b)\\
&\brd{}&\ssem(a)\;\ssem(c)\\
&=&\ssem((a\,c))
\end{eqnarray*}
\item{$\beta_4$}: Similar
\item{$\pi_1$}:
\begin{eqnarray*} 
\ssem(\pleft{\prdef{x}{a}{b}{c}})
&=&\ssem(\prdef{x}{a}{b}{c})\;\lambda xy.x\\
&=&\lambda x.(x\;\ssem(a)\;\ssem(b))\;\lambda xy.x\\
&\brd{}&\lambda xy.x\;\ssem(a)\;\ssem(b)\\
&\brd{}&\lambda y.\ssem(a)\;\ssem(b)\\
&\brd{}&\ssem(a)
\end{eqnarray*}
\item{$\pi_2$, $\ldots$, $\pi_6$}: Similar
\item{$\nu_1$}:
\begin{eqnarray*} 
\ssem(\myneg\myneg a)
&=&\ssem(\myneg a)\\
&=&\ssem(a)
\end{eqnarray*}
\item{$\nu_2$}:
\begin{eqnarray*} 
\ssem(\myneg[x:a]b)
&=&\ssem(\myneg[x:a]b)\\
&=&\lambda x.\ssem(b)\\
&=&\lambda x.\ssem(\myneg b)\\
&=&\ssem([x!a]\myneg b)
\end{eqnarray*}
\item{$\nu_3,\ldots,\nu_{10},\kappa$}: Similar
\item{$\mathit{(\oplus{\overbrace{(\_,\ldots,\_)}^{n}}_i)}$}:
If $a_i\srd b_i$ then by inductive hypothesis $\ssem(a_i)\brd{}\ssem(b_i)$ and therefore obviously
\[
\ssem(\binop{a_1,\ldots,a_i}{\ldots,a_n}\brd{}\ssem(\binop{a_1,\ldots,b_i}{\ldots,a_n})
\]
\item{$\mathit{(\oplus_x{\overbrace{(\_,\ldots,\_)}^{n}}_i)}$}: Similar
\qedhere
\end{itemize}
\end{proof}
\begin{law}[Irreducibility of semantic range]
\label{sem1.irred}
$\Gamma\sgv a$ then if $a$ is irreducible in \dcalc\ then $\ssem(a)$ is irreducible in $D$.
\end{law}
\begin{proof}
Let $\snf$ be the set of semantic normal forms, i.e.\ the set of all $\ssem(a)$ where $a\in\nf$.
By induction on the structure of \dcalc\ expressions one can calculate that $\snf$ can be characterized as follows where $\sde$ denotes the set of semantic dead ends.
\begin{eqnarray*}
\snf&=&\{\primfun\}\\
&&\union\;\{\lambda x.t\sth t\in\snf\}\\ 
&&\union\;\{\lambda x.(x\;t_1\;t_2)\sth t_1,t_2\in\snf\}\\ 
&&\union\;\{\lambda xy.(x\;t),\lambda xy.(y\;t)\sth t\in\snf\}\\ 
&&\union\;\sde\\
\sde&=&\{x\sth x\;\text{variable}\}\\  
&&\union\;\{t_1\;t_2\sth t_1\in\sde,t_2\in\snf\}\\
&&\union\;\{(t\;\lambda xy.x),(t\;\lambda xy.y)\sth t\in\sde\}
\end{eqnarray*}
\noindent
By induction on the structure of $\lambda$-expressions one can calculate that $\snf$ are normal forms of $D$. This implies the proposition.
\end{proof}
\begin{law}[Congruence law]
\label{sem1.eqv}
$\Gamma\sgv a,b$ and $a\eqv b$ imply $\ssem(a)\eq\ssem(b)$. 
\end{law}
\begin{proof}
Due to strong normalization, $a$ and $b$ have unique normal forms $a'$ and $b'$. If $a\eqv b$ then due to Law~\ref{cr} we know that $a'=b'$ and by Law~\ref{sem1.rd} that
$\ssem(a)\eq\ssem(b)$.  Note that $\ssem(a)=\ssem(b)$ does not imply $a=b$ since the latter may contain inequivalent type assignments.
\end{proof}
\begin{law}[Validity law]
\label{sem1.valid}
$\Gamma\sgv a$ implies $\ssem(a)$ is normalizing and has a unique normal form.
\end{law}
\begin{proof}
By Law~\ref{sn.valid} we have $a\in\sn{}$. Let $a'=\nf(a)$. Due to Law~\ref{sem1.rd} we have
$\ssem(a)\eq\ssem(a')$. By Law~\ref{sem1.irred} $\ssem(a')$ is irreducible.
Due to confluence of extended $\lambda$-calculus, this normal form is unique.
\end{proof}
\subsection{Typing relation}
\begin{definition}[Induced semantic typing relation]
\nomenclature[kCalc06]{$C\sdgv t_1:t_2$}{semantic typing induced by type-stripping interpretation}
\index{semantic typing!type-stripping interpretation}
\label{sem1.semtype}
The \emph{semantic typing relation} $C\sdgv t_1:t_2$ induced by $\ssem$ is defined as follows:
\begin{eqnarray*}
C\sdgv t_1:t_2&\equiv&\exists\;\Gamma,a,b:\Gamma\sgv a:b\;\wedge\;\ssem(\Gamma)=C\;\wedge\;\ssem(a)=t_1\;\wedge\;\ssem(b)=t_2
\end{eqnarray*}
\end{definition}
\begin{law}[Semantic typing relation does not depend on reduction]
\label{sem1.semtypered}
If $C\sdgv t_1:t_2$, $t_1\eq t_3$, and $t_2\eq t_4$ then $C\sdgv t_3:t_4$.
\end{law}
\begin{proof}
Follows from the definition of semantic typing
\end{proof}
It is interesting to investigate the properties of the semantic typing relation.
Since the $\lambda$-expressions have been stripped of type information we cannot expect to get a property such as 
\[\frac{C\sdgv t_1:\lambda x.t_2}{C\sdgv (t_1\,t):t_2\gsub{x}{t}}
\]
A counterexample would be $C=()$, $t_1=\lambda x.x$, $t_2=\primfun$, and $t=(\primfun\,\primfun)$.
Since $\sgv[x:\prim]x:[x:\prim]\prim$ we know that $\sdgv\lambda x.x:\lambda x.\primfun$ which means $C\sdgv t_1:\lambda x.t_2$.
However $(t_1\,t)\eq(\primfun\,\primfun)$ and it is easy to see there is no valid expression $a$ such that $\ssem(a)=(\primfun\,\primfun)$.
Hence it is not possible that $C\sdgv (t_1\,t):\primfun$.
 
Nevertheless, we can show the semantic typing relation is a d-structure.
\begin{law}[Properties of the semantic typing relation]
\label{sem1.semtypeprop}
The relation $C\sdgv t_1:t_2$ is a d-structure. 
\end{law}
\begin{proof}
We have to show the properties of d-structures:
\begin{itemize}
\item[$i$:]
$()\sdgv \primfun:\primfun$ follows from axiom \emph{ax} and the definition of $\ssem$.
\item[$ii$:]
Assume that $C\sdgv t_1:t_2$. Hence $\Gamma\sgv a:b$ where $\ssem(\Gamma)=C$, $\ssem(a)=t_1$, and $\ssem(b)=t_2$.
Hence by rule \emph{start} we know that $\Gamma,x:a\sgv x:a$. The proposition follows by definition of $\ssem$.
\item[$iii$:]
Follows from rule \emph{weak} and the definition of $\ssem$.
\item[$iv$:]
Follows from rule \emph{conv} and the definition of $\ssem$.
\item[$v$:]
Follows from Laws~\ref{cr} and~\ref{rd.type} and the definition of $\ssem$.
\item[$vi$:]
Follows from rule \emph{prod} and the definition of $\ssem$.
\item[$vii$:]
Follows from rule \emph{neg} and the definition of $\ssem$.
\qedhere
\end{itemize}
\end{proof}
\begin{remark}[Limitation of type-stripping interpretation]
Law~\ref{cons} states that there is no valid expression $a$ with $\sgv a:[x:\prim]x$.
Obviously $\ssem([x:\prim]x)=\lambda x.x$.
One would like to extend the consistency result to semantic typing, i.e.~there is no $t$ such that $\sdgv t:\lambda x.x$.
However, due to type stripping, this does not hold under some reasonable additional assumptions: 
In remark~\ref{cons.negation}, assuming some additional axiom schemes for cast introduction and elimination, we have constructed an expression $a$ with $\sgv a:[x:[y:\prim]y]x$.
Since $\ssem(x:[y:\prim]y]x)=\lambda x.x$ we have $\sgv \ssem(a):\lambda x.x$.
In this sense the semantic typing relation is logically limited.
\end{remark}
\section{A type-encoding interpretation}
\subsection{Mapping to annotated functional content}
The basic idea is to encode an abstraction $[x:a]b$ as a pair $\langle a,\lambda x.b\rangle$ and to encode application $(a\:b)$ as $(a.2\;b)$.
We do not need to assume new operators, because pairs and projections are already available.
\begin{definition}[Mapping to $D$]
\nomenclature[kCalc07]{$\sem$}{type-encoding semantic interpretation}
\index{semantic interpretation!type-encoding}
The function $\sem$ translates a \dcalc\ expression into $D$ (Table~\ref{sem2.def}).
\begin{table}[!htb]
\fbox{
\begin{minipage}{0.96\textwidth}
\begin{eqnarray*}
\\[-8mm]
\sem(\prim)&=&\primfun\\
\sem(x)&=&x\\
\sem([x:a]b)=\sem([x!a]b)&=&\lambda y.(y\;\sem(a)\;\lambda x.\sem(b))\\
\sem((a\,b))&=&\sem(a)\;\lambda xy.y\;\sem(b)\\
\sem(\prdef{x}{a}{b}{c})=\ssem([a,b])=\sem([a+b])&=&\lambda x.(x\;\sem(a)\;\sem(b))\\ 
\sem(\pleft{a})&=&\sem(a)\;\lambda xy.x\\ 
\sem(\pright{a})&=&\sem(a)\;\lambda xy.y\\
\sem(\case{a}{b})&=&\lambda x.(x\;\sem(a)\;\sem(b))\\ 
\sem(\injl{a}{b})&=&\lambda xy.(x\;\lambda xy.y\:\sem(a))\\
\sem(\injr{a}{b})&=&\lambda xy.(y\;\lambda xy.y\;\sem(b))\\
\sem(\myneg a)&=&\sem(a)
\end{eqnarray*} 
\end{minipage}
}
\caption{Type-encoding mapping to $D$\label{sem2.def}}
\end{table}
For a context $\Gamma$, 
$\sem(\Gamma)$ is defined by applying $\sem$ to all expressions in $\Gamma$ to yield a $\lambda$-context, i.e.~$\sem()=()$ and $\sem(x:a, \Gamma)=(x:\sem(a),\sem(\Gamma))$. 
\end{definition}
\begin{remark}[Examples]
Consider the expression $[x:\prim][y:x]y$ which is of type $[x:\prim][y:x]x$. These two expressions are interpreted as follows:
\begin{eqnarray*}
\sem([x:\prim][y:x]y)&=&\lambda z.(z\;\primfun\;\lambda x.(\lambda z.(z\;x\;\lambda y.y)))\\
\sem([x:\prim][y:x]x)&=&\lambda z.(z\;\primfun\;\lambda x.(\lambda z.(z\;x\;\lambda y.y)))
\end{eqnarray*}
\noindent
As another example consider the derivation of the modus ponens rule $[p:\prim][q:\prim][x:p][y:[z:p]q](y\,x)$ which is of type $[p:\prim][q:\prim][x:p][y:[z:p]q]q$.
The expression $\sem([p\!:\!\prim][q\!:\!\prim][x\!:\!p][y\!:\![z\!:\!p]q](y\,x))$ is interpreted as follows:
\begin{eqnarray*}
&&\lambda w.(w\;\primfun\;\\
&&\;\;\lambda p.(\\
&&\;\;\;\;\lambda w.(w\;\primfun\;\\
&&\;\;\;\;\;\;\lambda q.(\\
&&\;\;\;\;\;\;\;\;\lambda w.(w\;\primfun\;\\
&&\;\;\;\;\;\;\;\;\;\;\lambda x.(\\
&&\;\;\;\;\;\;\;\;\;\;\;\;\lambda w.(w\;p\;\\
&&\;\;\;\;\;\;\;\;\;\;\;\;\;\;\lambda y.(\\
&&\;\;\;\;\;\;\;\;\;\;\;\;\;\;\;\;\lambda w.(w\;\lambda w.(w\;p\;\lambda z.q)\;(y\;\lambda xy.y\;x))))))))))
\end{eqnarray*}
The expression $\sem([p:\prim][q:\prim][x:p][y:[z:p]q]q)$ is interpreted as follows:
\begin{eqnarray*}
&&\lambda w.(w\;\primfun\;\\
&&\;\;\lambda p.(\\
&&\;\;\;\;\lambda w.(w\;\primfun\;\\
&&\;\;\;\;\;\;\lambda q.(\\
&&\;\;\;\;\;\;\;\;\lambda w.(w\;\primfun\;\\
&&\;\;\;\;\;\;\;\;\;\;\lambda x.(\\
&&\;\;\;\;\;\;\;\;\;\;\;\;\lambda w.(w\;p\;\\
&&\;\;\;\;\;\;\;\;\;\;\;\;\;\;\lambda y.(\\
&&\;\;\;\;\;\;\;\;\;\;\;\;\;\;\;\;\lambda w.(w\;\lambda w.(w\;p\;\lambda z.q)\;q)))))))))
\end{eqnarray*}
\end{remark}
\noindent
Several laws of the type stripping interpretation are also satisfied by the type-encoding variant.
\begin{law}[Reduction law]
\label{sem2.rd}
$a\rd b$ implies $\sem(a)\brd{}\sem(b)$. 
\end{law}
\begin{proof}
The proof is similar as for Law~\ref{sem1.rd}. The only deviation is the handling of the $\beta$-reduction axioms, \eg:
\begin{itemize}
\item$\beta_1$:
\begin{eqnarray*} 
\sem(([x:a]b\;c))&=&\sem([x:a]b)\;\lambda xy.y\;\sem(c)\\
&=&\lambda y.(y\;\sem(a)\;\lambda x.\sem(b))\;\lambda xy.y\;\sem(c)\\
&\brd{}&\lambda xy.y\;\sem(a)\;\lambda x.\sem(b)\;\sem(c)\\
&\brd{}&\lambda y.y\;\lambda x.\sem(b)\;\sem(c)\\
&\brd{}&\lambda x.\sem(b)\;\sem(c)\\
&\brd{}&\sem(b)\gsub{x}{\sem(c)}\\
&=&\sem(b\gsub{x}{c})
\end{eqnarray*}
\end{itemize}
\end{proof}
\begin{law}[Irreducibility of semantic range]
\label{sem2.irred}
If $\Gamma\sgv a$ and $a$ is irreducible in \dcalc\ then $\sem(a)$ is irreducible in $D$.
\end{law}
\begin{proof}
The proof is similar to the one for Law~\ref{sem1.irred}.
The sets $\snf'$ and $\sde'$ are slightly different due to the different $\lambda$-encoding.

\begin{eqnarray*}
\snf'&=&\{\primfun\}\\
&&\union\;\lambda y.(y\;t_1\;\lambda x.t_2)\sth t_1,t_2\in\snf'\}\\
&&\union\;\{\lambda x.(x\;t_1\;t_2)\sth t_1,t_2\in\snf'\}\\ 
&&\union\;\{\lambda xy.(x\;\lambda xy.y\:t),\lambda xy.(y\;\lambda xy.y\;t)\sth t\in\snf'\}\\ 
&&\union\;\sde'\\
\sde'&=&\{x\sth x\;\text{variable}\}\\  
&&\union\;\{t_1\;\lambda xy.y\;t_2\sth t_1\in\sde',t_2\in\snf'\}\\
&&\union\;\{(t\;\lambda xy.x),(t\;\lambda xy.y)\sth t\in\sde'\}
\end{eqnarray*}

\end{proof}
\begin{law}[Congruence law]
\label{sem2.eqv}
$\Gamma\sgv a,b$ implies that $a\eqv b$ implies $\sem(a)\eq\sem(b)$. 
\end{law}
\begin{proof}
The proof is similar to the one for Law~\ref{sem1.eqv}.
\end{proof}
\begin{law}[Validity law]
\label{sem2.valid}
$\Gamma\sgv a$ implies $\sem(a)$ is normalizing and has a unique normal form.
\end{law}
\begin{proof}
The proof is similar to the one for Law~\ref{sem1.valid}.
\end{proof}
\noindent
Note that the following law does not hold for the type-stripping interpretation.
\begin{law}[Uniqueness law]
\label{sem2.unique}
If $\sem(a)=\sem(b)$ then $\Gamma\sgv a:c$ iff $\Gamma\sgv b:c$. Similarly if $\sem(\Gamma)=\sem(\Gamma')=C$ then $\Gamma\sgv a:b$ iff $\Gamma'\sgv a:b$
\end{law}
\begin{proof}
Follows from the definition of $\sem$ and uniqueness of types (Law~\ref{type.confl}).
\qedhere
\end{proof}
\subsection{Typing relation}
\begin{definition}[Induced semantic typing relation]
\nomenclature[kCalc08]{$C\sgv t_1:t_2$}{semantic typing induced by type-encoding interpretation}
\index{semantic typing!type-encoding interpretation}
\label{sem2.semtype}
The \emph{semantic typing relation} $C\sgv t_1:t_2$ is defined as follows:
\begin{eqnarray*}
C\sgv t_1:t_2&\equiv&\exists\;\Gamma,a,b:\Gamma\sgv a:b\;\wedge\;\sem(\Gamma)=C\;\wedge\;\sem(a)=t_1\;\wedge\;\sem(b)=t_1
\end{eqnarray*}
\end{definition}
%
%
\begin{law}[Semantic typing relation does not depend on reduction]
\label{sem2.semtypered}
If $C\sgv t_1:t_2$, $t_1\eq t_3$, and $t_2\eq t_4$ then $C\sgv t_3:t_4$.
\end{law}
\begin{proof}
Follows from the definition of semantic typing
\end{proof}
\noindent
The following structural properties can be shown:
\begin{law}[Properties of the semantic typing relation]
\label{sem2.semtypeprop}
$\;$
\begin{itemize}
\item[$i:$] 
The relation $C\sdgv t_1:t_2$ is a d-structure
\item[$ii$:]
If $C,x:t\sgv t_1:t_2$ then $C\sgv\lambda y.(y\,t\,\lambda x.t_1):\lambda y.(y\,t\,\lambda x.t_1)$
\item[$iii$:] 
If $C\sgv t_1:\lambda y.(y\,t_3\,\lambda x.t_2)$ and $C\sgv t_4:t_3$ then $C\sgv(t_1\,\lambda xy.y\,t_4):t_2\gsub{x}{t_4}$
\end{itemize}
\end{law}
\begin{proof}
$\;$
\begin{itemize}
\item[$i$:]
The proof of the properties of d-structures is as for Law~\ref{sem1.semtypeprop}.
\item[$ii:$]
$C,x:t\sgv t_1:t_2$ implies that $\Gamma,x:a\sgv b_1:b_2$ where $\sem(\Gamma)=C$, $\sem(a)=t$, $\sem(b_1)=t_1$, and $\sem(b_2)=t_2$. 
Hence $\Gamma\sgv[x:a]b_1:[x:a]b_2$ which implies $C\sgv\lambda y.(y\,t\,\lambda x.t_1):\lambda y.(y\,t\,\lambda x.t_1)$.
\item[$iii$:]
$C\sgv t_1:\lambda y.(y\,t_3\,\lambda x.t_2)$ implies that $\Gamma\sgv b_1:b_2$ where $\sem(\Gamma)=C$, $\sem(b_1)=t_1$ and $\sem(b_2)=\lambda y.(y\,t_3\,\lambda x.t_2)$. 
Due to basic properties of typing we may assume that $b_2\in\nf$. 
By definition of $\sem$ this is only possible if $b_2=[x:c_1]c_2$ where $\sem(c_1)=t_3$ and $\sem(c_2)=t_2$. 
Since $C\sgv t_4:t_3$ we know that there is some $b_3$ such that $\sem(b_3)=t_4$ and $\Gamma'\sgv b_3:c_1$ where $\sem(\Gamma')=C$.
By Law~\ref{sem2.unique} and by definition of typing we know that $\Gamma\sgv b_1(b_3):c_2\gsub{x}{b_3}$. 
This implies $C\sgv(t_1\,\lambda xy.y\,t_4):t_2\gsub{x}{t_4}$. 
\qedhere
\end{itemize}
\end{proof}
\begin{remark}[Limitations of type-encoding interpretation]
The type-encoding interpretation is already capturing some but not quite all aspects of \dcalc.
On the positive side it allows for richer properties of the semantic structure as shown in Law~\ref{sem2.semtypeprop}.

On the negative side negation and casting are still interpreted as neutral operations which is clearly undesirable.
Similarly, the type encoding interpretation is identifying existential and universal abstraction, hence it cannot mirror the consistency result of \dcalc.
By adding more variables to $D$-spaces and adaptation of the semantic mapping one could increase the details of the semantics to a point where it distinctively models all operations and properties of \dcalc\ in detail.
\end{remark}
\chapter{\dcalct\ as a logical framework}
\label{formalization}
In this chapter we sketch how \dcalc\ can be used to present logic in such a way that provability of a formula in the original logic reduces to a type inhabitation problem in \dcalc. Note that \dcalc\ as a logical framework plays the role of a meta-logic, therefore while \dcalc\ itself is a classical logic, this does not prevent its use for modeling non-classical ones.  
\section{Adequate formalization}
Assume that some logical system $T$ is formalized as a pair $(\wff,\vwff)$ where $\wff$ is the set of well-formed formulas of that theory and $\vwff$ is the subset of $\wff$ consisting of those formulas which are considered as valid within the system. The goal is to adequately formalize that system in \dcalc. This can be achieved through the declarations and definitions of a context $\Gamma_T$ such that the following proposition becomes true.
\[
\text{For all }f\in\wff:f\in\vwff\text{ iff there is some expression $a$ such that }\Gamma_T\sgv a:\alpha(f)
\]
where $\alpha$ is an injective function from $\wff$ to expressions in \dcalc.

The direction from left to right in the above proposition establishes the completeness of the formalization, i.e. $\Gamma_T$ allows deriving everything which is valid in \emph{T}. This direction is usually not difficult to prove, especially if $\Gamma_T$ contains axioms and inference rules which are direct translations of those used in \emph{T}.

The direction from right-to-left establishes the correctness of the formalization, i.e. every formula $\alpha(f)$ derived under $\Gamma_T$ is valid in \emph{T}. The problem with the proof of this direction is that the derivation of $\alpha(f)$ may be an arbitrary expression. A proof by structural induction on valid expressions is difficult because many such expressions are equivalent modulo reduction. However due to Law~\ref{vnf.type} of \dcalc\ the above proposition can be simplified as follows:
\[
\text{For all }f\in\wff:f\in\vwff\text{ iff there is some $a\in\nf$ such that }\Gamma_T\sgv a:\alpha(f)
\]
This allows for proving correctness by induction on the recursive structure of normal forms.
\section{Example: Minimal logic}
Formulas $\mathbb{F}$ of minimal logic are build from the constants $T,F$ and the binary function $\Rightarrow$.
The set of valid formulas can be defined \eg\ by a sequent logic ($\Gamma$ denotes a list of formulas). 
\[
\frac{}{\Gamma, A\gv A}
\qquad
\frac{\Gamma, A\gv B}{\Gamma \gv A\Rightarrow B}
\qquad
\frac{\Gamma\gv A\quad \Gamma \gv A\Rightarrow B}{\Gamma \gv B}
\]
A formula $A$ is valid iff $\gv A$. 
In \dcalc, minimal logic can be axiomatized by the context $\minimal$ defined in Section~\ref{minimallogic}.
Next, we define the following injective mapping from minimal logic formulas to \dcalc\ expressions
\begin{eqnarray*}
\alpha(T)&=&t\\
\alpha(F)&=&f\\
\alpha(A \Rightarrow B)&=&((I\,\alpha(A))\,\alpha(B))
\end{eqnarray*}
Obviously $a\in ran(\alpha)$ implies that $\minimal\sgv a:F$. 
To prove adequate formalization we have to show that for any $A\in\mathbb{F}$:
\[
\sgv A\;\text{iff there is some $a\in\nf$ such that}\;\minimal\sgv a:\alpha(A)
\]
It is straightforward to see the direction from left-to-right as any derivation of valid formulas in minimal logic can be directly formalized in $\minimal$.  
In the following sections, we consider the direction from right-to-left, i.e.~no invalid formulas of minimal logic can be derived from $\minimal$. 

\noindent
Consider the set of expressions $a$ with $\minimal\sgv a:\alpha(A)$ for some $A$.
\begin{eqnarray*}
\mathcal{P}&:=&\{a\in\nf\mid \minimal\sgv a:b, b\in ran(\alpha)\} 
\end{eqnarray*}
Our goal is to find a recursive characterization of $\mathcal{P}$ that is, based on the particular context $\minimal$, simpler than that of $\nf$.
To this end, let $a\in\mathcal{P}$. Since $a\in\nf$ we have several case to consider. The following ones can be excluded:
\begin{itemize}
\item $a=[x_1:a_1]a_2$: We would have $\minimal\sgv a:[x:a_1]b_2$, where $\minimal,x:a_1\sgv a_2:b_2$ and $[x:a_1]b_2\in ran(\alpha)$.
This implies $\minimal\sgv [x:a_1]b_2:F$ which is impossible.
\item $a=[x_1!a_1]a_2$, $a=\prdef{x_1}{a_1}{a_2}{a_3}$: 
For similar reasons as in the previous case, this case is impossible.
\item $a=[a_1,a_2]$:
We would have $\minimal\sgv a:[b_1,b_2]$ where $[b_1,b_2]\in ran(\alpha)$.
This implies $\minimal\sgv[b_1,b_2]:F$ which is impossible.
\item $a=[a_1+a_2]$, $a=\injl{a_1}{a_2}$, $a=\injr{a_1}{a_2}$
For similar reasons as in the previous case, this case is impossible.
\end{itemize}
By definition of $\nf$ this implies
\begin{eqnarray*}
\mathcal{P}&=&\{a\in\de\mid \minimal\sgv a:b, b\in ran(\alpha)\}
\end{eqnarray*} 
This completes the first step of the characterization of formula derivations.

Given the context $\minimal$ we can specialize the result $a\in\de$ as follows:  We know that for any $a$ with $\minimal\sgv a:b$ and $b\in ran(\alpha)$ we have $a\in\mathcal{M}_0$ where the set $\mathcal{M}_0$ is characterized as follows:
\begin{eqnarray*}
\mathcal{M}_0&=&\{x\sth x\in\{F,t,f,I,i,o\}\}\\  
&&\union\;\{(a_1\,a_2)\sth a_1\in\mathcal{M}_0,a_2\in\nf\}\\
&&\union\;\{\pleft{a},\pright{a}\sth a\in\mathcal{M}_0\}\\
&&\union\;\{\case{a_1}{a_2}\sth a_1,a_2\in\mathcal{M}_0\}\\
&&\union\;\{\myneg a_1\sth a_1\in\mathcal{M}_0,a_1\;\text{is not a negation}\}
\end{eqnarray*}
Obviously we have $\mathcal{M}_0\subseteq\de$. Let us now take a closer look at the various operations characterizing the set $\mathcal{M}_0$:

In the characterization of $\mathcal{M}_0$, the case $x=F$ is not possible since $a=F$ would imply $\prim\in ran(\alpha)$ and all the other combinations would not be well-typed except for $a=\myneg F$. 
The case $a=\myneg F$ would imply $\prim\in ran(\alpha)$ for some $b$ which is impossible.
Similarly one can show that $x=t$, $x=f$, $x=I$ are not possible either since they would always lead to expressions in $\de$ whose types are not equivalent to any expression in $ran(\alpha)$. This leaves us with the cases $x\in\{i,o\}$ which completes the second step of the characterization of formula derivations.

We now take a closer look at the types of elements of $a\in\mathcal{M}_0$.
By structural induction one can show that for any $a\in\mathcal{M}_0$ exactly one of the following cases must be true ($n\geq 0$).
\begin{itemize}
\item $\minimal\sgv a:[x:F][y:F][[x\fun y]\fun ((I\,x)\,y)]$,
\item $\minimal\sgv a:[y:F][[b\fun y]\fun ((I\,b)\,y)]$ where $\minimal\sgv b:F$
\item $\minimal\sgv a:[[b\fun c]\fun ((I\,b)\,c)]$ where $\minimal\sgv b:F$ and $\minimal\sgv c:F$
\item $\minimal\sgv a:[x:F][y:F][((I\,x)\,y)\fun[x\fun y]]$
\item $\minimal\sgv a:[y:F][((I\,b)\,y)\fun[b\fun y]]$ where $\minimal\sgv b:F$
\item $\minimal\sgv a:[((I\,b)\,c)\fun[b\fun c]]$ where $\minimal\sgv b:F$ and $\minimal\sgv c:F$
\item $\minimal\sgv a:[b\fun c]$ where $\minimal\sgv b:F$ and $\minimal\sgv c:F$
\item $\minimal\sgv a:b$ where $\minimal\sgv b:F$
\end{itemize}
However this implies that $\minimal\sgv a:[b,c]$ is not possible and hence the cases $\pleft{a}$ and $\pright{a}$ are not possible.
Similar arguments can be made for  for $\minimal\sgv a:[b+c]$, $\minimal\sgv a:\injl{b}{c}$, $\minimal\sgv a:\injr{b}{c}$, $\minimal\sgv a:[x!b]c$, and $\minimal\sgv a:\prdef{x}{b}{c}{d}$, are not possible implying also that $\case{a_1}{a_2}$ is not possible.

This completes the third step of the characterization of formula derivations. We can summarize our results up to this point as follows:
\begin{eqnarray*}
\mathcal{P}&=&\{a\in\mathcal{M}_1\mid \minimal\sgv a:b, b\in ran(\alpha)\}
\end{eqnarray*}
where
\begin{eqnarray*}
\mathcal{M}_1&=&\{x\sth x\in\{i,o\}\}\;\union\;\{(a\,b)\sth b\in\nf,a\in\mathcal{M}_1\}\;\union\;\{\myneg a \sth a\in\mathcal{M}_1, a\neq\myneg b\}
\end{eqnarray*}
All together we can summarize that for a given formula $A$:
\begin{eqnarray*}
&&\{a\in\nf\sth\minimal\sgv a:\alpha(A)\}\\
\text{(see above)}&=&\{a\in\mathcal{M}_1\sth\minimal\sgv a:\alpha(A)\}
\end{eqnarray*}
This means that $\minimal\sgv a:\alpha(A)$ implies that $a\in\mathcal{M}_1$. 

It remains to show that only valid formulas are produced by $\mathcal{M}_1$. To this end we define the partial mapping $\beta$ from \dcalc\ to expressions in $\mathbb{F}$ extended by variables:
\begin{eqnarray*}
\beta(x)&=&
\begin{cases}
T&\text{if}\;x=t\\
F&\text{if}\;x=f\\
x&\text{othwerwise}
\end{cases}\\
\beta((a\,b))&=&\beta(c)\Rightarrow\beta(b)\quad\text{if}\;a=I(c)\\
\beta([x:a]b)&=&\begin{cases}\beta(b)&\text{if}\;a=F\\
\beta(a)\Rightarrow \beta(b)&\text{otherwise}\end{cases}
\end{eqnarray*}
Obviously we have $\beta(\alpha(A))=A$.
By structural induction on $a$ one can show that $\minimal\sgv a:b$ and $a\in\mathcal{M}_1$ implies that $\beta(b)$ is defined and $\sgv\beta(b)$.
Together it follows that $\minimal\sgv a:\alpha(A)$ implies that $\sgv A$.
\chapter{Bounded polymorphism: The system \texorpdfstring{\dcalcb}{\dcalcbt}}
\label{sub}
\section{Overview}
\label{sub.overview}
The limited range of elements of type $\prim$, resulting in the need for explicit casting axioms (see Section~\ref{ex.casting}) is a significant restriction of \dcalc.
In this chapter, the type concept of  \dcalc\ is extended by defining a partial ordering on functions based on an infinite set of primitive constants.
This ordering will then be used to introduce bounded polymorphism. The resulting system is referred to as \dcalcb.

To recall the issue, consider the following example where $\Gamma:=(N:\prim,0:N)$ and $x:=[p:[N\fun\prim]](p\,0)$. Here, under $\Gamma$, we can
apply $x$ to $[n:N]n$ but not \eg\ to $[n:N][m:N]m$, $[n:N][m!N]m$, $[n:N][m,m]$, or $[n:N][m+m]$ since $[m:N]m$, $[m!N]m$, $[m,m]$, and $[m+m]$ cannot be typed to $\prim$. 

Since in \dcalc\ a type is not a first class concept but the role of an element in the type relation, we define use a partial ordering of functions to introduce bounded polymorphism.
Unfortunately, to use $\prim$ as single top element in such an ordering leads to paradoxes, \eg\ in Section~\ref{paradox.subtyping} we have shown how a naive subsumptive subtyping law leads to a paradoxical system.

Here we will instead introduce constants $\any{n}$, where $n\geq 0$, with $\any{n}:\any{n}$ and a strict linear order $\any{0}<\any{1}<\ldots$. 
More precisely, the idea is to extend the type relation by first introducing the ordering of $\prim$-\emph{inclusion}, $\Gamma\sgv a<\any{n}$, where an element is strictly smaller than some $\any{n}$ if the element itself or a member of its type hierarchy is based on constants $\any{k}$ where $k<n$ only (see Table~\ref{sub.sinc.rules}). 
\begin{table}[!htb]
\fbox{
\begin{minipage}{0.96\textwidth}
\begin{align*}
\sstartm\;\;&\frac{\Gamma\sgv\any{m}\quad\Gamma\sgv\any{n}\quad m<n}{\Gamma\sgv\any{m}<\any{n}}&
\sstypm\;\;&\frac{\Gamma\sgv a:b\quad \Gamma\sgv b<\any{n}}{\Gamma\sgv a<\any{n}}\\
\ssabsum\;\;&\frac{\Gamma\sgv a<\any{n}\quad\Gamma,x:a\sgv b<\any{n}}{\Gamma\sgv\binbop{x}{a}{b}<\any{n}}&
\ssbprodm\;\;&\frac{\Gamma\sgv a<\any{n}\quad\Gamma\sgv b<\any{n}}{\Gamma\sgv\prsumop{a}{b}<\any{n}}\\
\ssinjlm\;\;&\frac{\Gamma\sgv a<\any{n}\quad\Gamma\sgv b<\any{n}}{\Gamma\sgv\injl{a}{b}<\any{n}}&
\ssinjrm\;\;&\frac{\Gamma\sgv a<\any{n}\quad\Gamma\sgv b<\any{n}}{\Gamma\sgv\injr{a}{b}<\any{n}}\\[-11mm]
\end{align*}
\begin{align*}
\sspdefm\;\;&\frac{\Gamma\sgv a<\any{n}\quad\Gamma\sgv b<\any{n}\quad\Gamma\sgv c<\any{n}\quad\Gamma,x:b\sgv d<\any{n}}{\Gamma\sgv\prdef{x}{a^b}{c}{d}<\any{n}}
\end{align*}
\end{minipage}
}
\caption{Rules for $\prim$-inclusion.\label{sub.sinc.rules}}
\end{table}

The relation of (general) \emph{inclusion} is then introduced as the reflexive embedding of $\prim$-inclusion on valid expressions that is monotonic with respect to products/sums and abstractions (see Table~\ref{sub.inc.rules}). 
We will show that $\prim$-inclusion is a strict (modulo congruence) partial ordering (Laws~\ref{sub.sinc.trans} and~\ref{sub.sinc.irrefl}) and that inclusion is a partial ordering (Law~\ref{sub.leq.por}).
\begin{table}[!htb]
\fbox{
\begin{minipage}{0.96\textwidth}
\begin{align*}
\sreflm\;\;&\frac{\Gamma\sgv a}{\Gamma\sgv a\leq a}&
\sembedm\;\;&\frac{\Gamma\sgv a<\any{n}}{\Gamma\sgv a\leq\any{n}}\\
\sabsm\;\;&\frac{\Gamma,x:a\sgv b\leq c}{\Gamma\sgv\binbop{x}{a}{b}\leq\binbop{x}{a}{c}}&
\sbprodm\;\;&\frac{\Gamma\sgv a\leq b\quad c\leq d}{\Gamma\sgv\prsumop{a}{c}\leq\prsumop{b}{d}}
\end{align*}
\end{minipage}
}
\caption{Rules for inclusion.\label{sub.inc.rules}}
\end{table}

Finally, as an extension of the existing definition of typing (see Table~\ref{typ.rules}) the axiom is \ax\ is extended to primitive constants $\any{n}$ and a new inclusion rule \sinc\ is added
which, when interpreting expressions as types, effectively introduces bounded polymorphism (see Table \ref{sub.typ.rules}).
\begin{table}[!htb]
\fbox{
\begin{minipage}{0.96\textwidth}
\begin{align*}
\axm\;\;&\frac{n\geq 0}{\sgv\any{n}:\any{n}}&
\sincm\;\;&\frac{\Gamma\sgv a:b\quad \Gamma\sgv b\leq c}{\Gamma\sgv a:c}
\end{align*}
\end{minipage}
}
\caption{Rules for primitive constants and bounded polymorphism.\label{sub.typ.rules}}
\end{table}

Note that the rules of the type relation, of inclusion, and of $\any{}$-inclusion are interdependent as the relations reference each other.
Hence properties have to be established by arguments simultaneously considering all three relations.

Furthermore this approach obviously precludes a notion of uniqueness of types \wrt\ $\beta$-conversion, \eg\ \cite{lungu_et_al:LIPIcs:2018:9849}. 
However, in this approach one can extend the propositions-as-types analogy by viewing the partial order of inclusion logically as implication .
We will show that there is a (unique) minimal - and hence logically strongest - type among all the types of an element thus recovering an alternative notion of uniqueness of types.

When looking at the rules for $\any{}$-inclusion (Table~\ref{sub.sinc.rules}), it is obvious there is no $a$ such that $\Gamma\sgv a<\any{0}$, and therefore 
from the rules of inclusion (Table~\ref{sub.inc.rules}) it obviously follows that $\Gamma\sgv b\leq\any{0}$ implies $b=\any{0}$. 
This similarity of $\any{0}$ with the constant $\any{}$ of \dcalc\ justifies the convention to write $\prim$ for $\any{0}$.

We also introduce the notion of $\lambda$-\emph{inclusion} $\Gamma\sgv a\incl b$ which is
an extension of inclusion by congruence
\nomenclature[lgaRel05]{$\Gamma\sgv a\incl b$}{$\lambda$-inclusion}%
\index{inclusion!lambda}
\begin{eqnarray*}
\Gamma\sgv a\incl b&\text{iff}&\Gamma\sgv b,\;\; 
                     a\eqv a',\;\Gamma\sgv a'\leq b',\;\text{and}\;b'\eqv b\;\text{for some $a'$, $b'$}
\end{eqnarray*}
Note the validity condition on $b$ which ensures that \sinc\ generalizes to $\lambda$-inclusion (Law \ref{sub.incl.prop}($vii$)).
Informally speaking, for $n=0$, $1$, $\ldots$ the inclusion $\sgv a\incl\any{n}$ characterizes an increasing set of expressions $a$:
\begin{itemize}
\item[] $\;\sgv a\incl\any{0}$: $a\eqv\any{0}$. 
\item[] $\;\sgv a\incl\any{1}$: $a\eqv\any{1}$ or a member of its type hierarchy is based on type $\any{0}$ only.   
\item[] $\;\sgv a\incl\any{2}$: $a\eqv\any{2}$ or a member of its type hierarchy is based on types $\any{0}$, $\any{1}$ only.
\item[] $\;\ldots$
\end{itemize}
\noindent
By the \emph{type hierarchy} of a given expression $a$ we refer to sequences $a$, $b$, $c$, $\ldots$, where $\sgv a:b$, $\sgv b:c$, $\ldots$.
The role of primitive constants as upper bounds for types is subject of Law \ref{sub.inc.upper}.

Using constant $\any{1}$, we can rewrite the above example as $\Gamma:=(N:\prim,0:N)$ and $f:=[x:N_1](x\,0)$ where $N_1:=[N\fun\any{1}]$.
Obviously $\Gamma\sgv f:[N_1\fun\any{1}]$, $\Gamma\sgv[n:N]n:[N\fun N]$, and $\Gamma\sgv[N\fun N]\leq N_2$ as well as $\Gamma\sgv[n:N][m:N]n:[N\fun[N\fun N]]$ and $\Gamma\sgv[N\fun[N\fun N]]\leq N_1$.
Hence, by rule \sinc~we know that both $\Gamma\sgv[n:N]n:N_1$ and $\Gamma\sgv[n:N][m:N]n:N_1$.
Under $\Gamma$, $f$ can be applied both to $[n:N]n$ and $[n:N][m:N]n$ where $\Gamma\sgv (f\:[n:N]n):N$ as well as $\Gamma\sgv (f\:[n:N][m:N]n):[N\fun N]$ and
\[
(f\:[n:N]n)\,\eqv\,([n:N]n\;0)\,\eqv\,0
\]  
\[
(f\:[n:N][m:N]n)\,\eqv\,([n:N][m:N]n\;0)\,\eqv\,[m:N]0
\]
Note that the situation is similar when using $f$ and its arguments in a type-role, i.e.~with $\Gamma:=(N:\prim,0:N,r:f,x:[n:N]n,y:[n:N][m:N]n)$ we have $\Gamma\sgv (r\,x):(f\,x)$ and 
 $\Gamma\sgv (r\,y):(f\,y)$.

An example of a function over $\any{2}$ can be illustrated as follows: First, recall that $\Gamma\sgv f:[N_1\fun\any{1}]$. 
Let $u_1:=[p:N_1][n:N](p\,n)$ and $u_2:=[p:N_1][n:N][m:N](p\,n)$.
We have $\Gamma\sgv u_1:[N_1\fun[N\fun\any{1}]]$ and $\Gamma\sgv u_2:[N_1\fun[N\fun[N\fun\any{1}]]]$.  
A function accepting both $u_1$ and $u_2$ as arguments would  therefore need $N_2:=[N_1\fun[N\fun\any{2}]]$ as type of its domain.
For example, we can define a parametric function $F:=[P:N_2][q:N_1]((P\;q)\;0)$.
We have $\Gamma\sgv F:[N_2\fun [N_1\fun N]]$ and
\[
(F\,u_1)\,\eqv\,[q\!:\!N_1](([p\!:\!N_1][n\!:\!N](p\,n)\,q)\,0)\,\eqv\,[q\!:\!N_1](q\,0)
\] 
\[
(F\,u_2)\,\eqv\,[q\!:\!N_1](([p\!:\!N_1][n:\!N][m:\!N](p\,n)\,q)\,0)\,\eqv\,[q\!:\!N_1][n\!:\!N](q\,0)
\] 
Again the situation is similar when using $F$ in a type role.
Further examples are described and discussed in Section~\ref{sub.ex.logic}.

Note that the type rule \sinc\ for inclusions violates uniqueness of types since
$\sgv[x:\any{}]\any{}:[x:\any{}]\any{}$ as well as $\sgv[x:\any{}]\any{}:\any{1}$.
However we will show (Law \ref{sub.min.min}) that each valid expression has a unique type that is minimal \wrt\ inclusion.

In the following section we describe the (significant) additions or changes to the definitions, examples, and proofs of \dcalc\ necessary for the additional primitive constants and the extended type relation.
We also show some new properties related to the ordering on functions introduced by \dcalcb (Section \ref{sub.order}).
\section{Comparison to pure type systems}
To compare \dcalcb\ with common type systems we will use an ad-hoc variation of pure type systems called PFTS ({\em pure function-based type system})
which is generalizing the definition of a subset of \dcalcb\ covering primitive constants, variables, universal abstractions, and application only. 
A PFTS specification is a triple ($C$, $A$, $O$) where $C=\{k_1, k_2,\ldots\}\subseteq\dexp$ is the set of {\em constants},  $A\subseteq C\times C$ is the set of {\em type axioms}, and $O\subseteq C\times C$ is the set of {\em ordering axioms}. PFTS has the typing rules shown in Table~\ref{sub.pts}.
\begin{table}[!htb]
\fbox{
\begin{minipage}{0.96\textwidth}
\begin{align*}
\\[-7mm]
\it{(axiom)}\;\;&\frac{}{()\gv k_1:k_2}\quad\text{where}\;(k_1,k_2)\in A\\
\it{(start)}\;\;&\frac{\Gamma\gv a:b}{\Gamma,x:a\gv x:a}\\
\it{(weakening)}\;\;&\frac{\Gamma\gv a:b\quad\Gamma\gv c:d}{\Gamma,x:c\gv a: d}\\
\it{(abstraction)}\;\;&\frac{\Gamma,x:a\gv b:c\quad\Gamma\gv[x:a]c: d}{\Gamma\gv [x:a]b:[x:a]c}\\
\it{(application)}\;\;&\frac{\Gamma\gv a:[x:b]c\quad\Gamma\gv d:b}{\Gamma\gv(a\,d):c\gsub{x}{d}}\\
\it{(conversion)}\;\;&\frac{\Gamma\gv a:b\quad b\eqv b'\quad\Gamma\gv b':c}{\Gamma\gv a:b'}\\
\it{(inclusion)}\;\;&\frac{\Gamma\gv a:b\quad\Gamma\gv b\leq c}{\Gamma\gv a:c}\\[3mm]
\it{(reflexivity)}\;\;&\frac{\Gamma\gv a:b}{\Gamma\gv a\leq a}\\
\it{(embedding)}\;\;&\frac{\Gamma\gv a<k}{\Gamma\gv a\leq k}\quad\text{where}\;k\in C\\
\it{(abstraction_{\leq})}\;\;&\frac{\Gamma,x:a\gv b\leq c}{\Gamma\gv[x:a]b\leq[x:a]c}\\[3mm]
\it{(ordering)}\;\;&\frac{\Gamma\gv k_1:a\quad\Gamma\gv k_2:b}{()\gv\Gamma\gv k_1<k_2}\quad\text{where}\;(k_1,k_2)\in O\\
\it{(abstraction_<)}\;\;&\frac{\Gamma\gv a<k\quad\Gamma,x:a\gv b< k}{\Gamma\gv[x:a]b< k}\quad\text{where}\;k\in C\\
\it{(type)}\;\;&\frac{\Gamma\gv a:b\quad\Gamma\gv b<k}{\Gamma\gv a<k}\quad\text{where}\;k\in C
\end{align*}
\end{minipage}
}
\caption{Pure function-based type system\label{sub.pts}}
\end{table}
Obviously, when restricted to expressions of PFTS, \dcalcb\ is a pure function-based type system with $C=\{\any{i}\mid 0\leq i\}$, $A=\{(\any{i},\any{i})\mid 0\leq i\}$, and
$O=\{(\any{i},\any{j})\mid 0\leq i<j\}$. 
One can also consider bounded variants of \dcalcb: Given a fixed index $n$, let \dcalc$^{(n)}_{\leq}$ denote the subsystem of \dcalcb\ which is built from constants $\any{0},\ldots,\any{n}$ only.
\dcalc$^{(n)}_{\leq}$ corresponds to the pure function-based type system with $C_n=\{\any{i}\mid 0\leq i\leq n\}$, $A_n=\{(\any{i},\any{i})\mid 0\leq i\leq n\}$ and $O_n=\{(\any{i},\any{j})\mid 0\leq i<j\leq n \}$.
For illustration, we enumerate some values:
\[
\begin{array}{llll}
\text{\dcalc$^{(0)}_{\leq}$}:&C_0=\{\any{0}\},&A_1=\{(\any{0},\any{0})\}
                  &O_0=\emptyset\\[1mm]
\text{\dcalc$^{(1)}_{\leq}$}:&C_1=C_0\cup\{\any{1}\},&A_1=A_0\cup\{(\any{1},\any{1})\}
                  &O_1=\{(\any{0},\any{1})\}\\[1mm]
\text{\dcalc$^{(2)}_{\leq}$}:&C_2=C_1\cup\{\any{2}\},&A_2=A_1\cup\{(\any{2},\any{2})\}
                  &O_2=O_1\cup\{(\any{0},\any{2}),(\any{1},\any{2})\}\\
\ldots
\end{array}
\]
\noindent
Obviously \dcalc$^{(0)}_{\leq}$ corresponds to \dcalc\ restricted to expressions of PFTS.
As another example, consider the PFTS
\[
C=\{\prim,\kappa\},\quad A=\{(\prim,\kappa)\},\quad O=\emptyset
\]
which corresponds to the system $\lambda^{\lambda}$ \cite{PdG93} which is very similar to \dcalc$^{(0)}_{\leq}$ as it has no ordering axioms.
One may extend $\lambda^{\lambda}$ with polymorphism as follows:
\[
C=\{\prim,\kappa\},\quad A=\{(\prim,\kappa)\},\quad O=\{(\prim,\kappa)\}
\]
The resulting system is similar to \dcalc$^{(1)}_{\leq}$ allowing for parametrization of functions build from $\prim$.
As an example of paradoxical system consider:
\[
C=\{\prim\},\quad A=\{(\prim,\prim)\},\quad O=\{(\prim,\prim)\}
\]
In this system with $x:=[y:\prim]y$ we have $\sgv x:[\prim\fun\prim]$ and $\sgv[\prim\fun\prim]<\prim$ and therefore by rule \sstyp\ we obtain $\sgv x<\prim$. 
Hence from $z:x\sgv z:x$, by rule \sinc\ we obtain $z:x\sgv z:\prim$.
Since also $z:x\sgv z:[y:\prim]y$ we can type the self application $z:x\sgv(z\,z):z$.
Therefore with  $u:=[z:x](z\,z)$ we obtain $\sgv u:[z:x]z$ and with a similar argument we can type the diverging element $\sgv(u\,u):x$. 

\begin{remark}
Note that a similar inconsistency would come up in variations pf PFTS removing $\any{}$-inclusion alltogether and replacing its use in the rules (\emph{type}) and (\emph{abstraction}$_<$) by inclusion. One would have $\,\sgv x\leq\prim$ and $\,\sgv \prim\fun\prim]\leq\prim$ and hence could type $\sgv(u\,u):I$. Note this typing is independent on the rule (\emph{axiom}) in the sense that it only needs the property $\,\sgv\prim$. 
\end{remark}

As for the relationship between PFTS and PTS, one should note that
PFTS is viewing terms and types as function-roles and therefore collapsing many kinds of systems which are differentiated with PTS.
For example, when comparing the systems of the $\lambda$-cube and PFTS, the most simple meaningful PFTS, the system \dcalc$^{(0)}_{\leq}$, is a system which already has parametrized terms and types both as roles of functions. This means that, in case of PFTS, the three axis of the $\lambda$ cube collapse into one.

When looking at the systems \dcalc$^{(0)}_{\leq}$, the lack of ordering axioms and hence bounded polymorphism puts it as the most distant one to the calculus of constructions~\cite{COQUAND198895} and most close one to the simply typed $\lambda$-calculus.
Similarly one may understand the systems \dcalc$^{(1)}_{\leq}$, \dcalc$^{(2)}_{\leq}$, $\ldots$ as successive parallel approximations of the functions of the calculus of constructions and its type universes.
\section{Definition of \texorpdfstring{\dcalcb}{\dcalcbt}}
\label{sub.definition}
\subsection{Basic definitions}	%
\label{sub.basic.definitions}
\nomenclature[lbCalc03]{$\any{n}$}{primitive constant of level $n$}
\index{primitive constant!of level $n$}
\nomenclature[lcBasic03]{$\dsuper$}{set of primitive constants}
\nomenclature[ldCalc07]{$\prdef{x}{a^b}{c}{d}$}{tagged protected definition}
\index{protected definition!tagged}
The set $\dexp$ of expressions is extended by adding a infinite set of primitive constants.
\begin{eqnarray*}
\dexp&\!::=\!&\underbrace{\{\prim\}\,\mid\,\cdots\,\mid\,\myneg\dexp}_{\text{(as in Definition~\ref{expression} except for protected definitions)}}\,\mid\,\dsuper\,\mid\,
\prdef{\dvar}{\dexp^{\dexp}}{\dexp}{\dexp}
\end{eqnarray*}
where $\any{0},\any{1},\cdots\in\dsuper$ are {\em primitive constants of level} $0$, $1$, $\cdots$.
While for succinctness, in the following definitions we will write $\any{0}$ for $\prim$,
for readability, in the examples and illustration we will use the notation $\prim$. 
Note the modification of protected definitions by a type tag $b$ of the substituted expression $a$.
As will be seen, the tag is needed in order to ensure minimal types.
We also use this extended notation for the proof of strong normalisation.
We also use the \emph{binary operator} $\prsuminjop{a}{b}$ to stand for $\prsumop{a}{b}$, $\injl{a}{b}$, or $\injr{a}{b}$.
\nomenclature[ldCalc07]{$\prsuminjop{a}{b}$}{binary operator}
\index{operator!binary}

\noindent
The definition of free variables and substitution are extended in an obvious way.
\begin{eqnarray*}
\free(\any{n})&=&\{\}\\
\any{n}\gsub{x}{b}&=&\any{n}\\
\prdef{x}{a^b}{c}{d}&=&\free(a)\cup\free(b)\cup\free(c)\cup(\free(d)\setminus\{x\})
\end{eqnarray*}
\subsection{Reduction and congruence}
\label{sub.reduction}
Single-step reduction (see Table~\ref{red.rules}) is extended by extending axiom $\nu_6$ to primitive constants $\any{n}$.
\begin{eqnarray*}
\mathit{(\nu_6)}\qquad \myneg\any{n}&\srd&\any{n}
\end{eqnarray*}
The structural rules for single-step reduction of protected definitions (Definition \ref{sred}) are adapted in an obvious way to type tags. 
The definitions of reduction and congruence remain unchanged.
\subsection{Typing with bounded polymorphism}
\label{sub.typing}
\begin{definition}[Typing with bounded polymorphism]%
\label{sub.inclusion}
\nomenclature[leRel05]{$\Gamma\sgv a\leq b$}{inclusion}%
\index{inclusion}
\nomenclature[lfRel05]{$\Gamma\sgv a<\any{n}$}{$\prim$-inclusion}%
\index{inclusion!$\any{}$}
\nomenclature[lgbRel05]{$\Gamma\sgv a\gincl b$}{$\lambda^*$-inclusion}%
\index{inclusion!lambda$^*$}
In the definition of typing (see Table~\ref{typ.rules}) the \ax\ is extended to primitive constants $\any{n}$ and a rule for inclusion is added (see Table~\ref{sub.typ.rules}). 
The rule \pdef\ is modified as follows. 
\[
\pdefm\quad\frac{\Gamma\sgv a:b\quad\Gamma\sgv c:d\gsub{x}{a}\quad\Gamma,x:b\sgv d:e}{\Gamma\sgv\prdef{x}{a^b}{c}{d}:[x!b]d}
\]
Furthermore the type relation $\Gamma\sgv a:b$ is now defined simultaneously with two new inclusion relations:
\begin{itemize}
\item
\emph{Inclusion} $\Gamma\sgv a\leq b$ is defined as the smallest relation generated by the inclusion rules in Table~\ref{sub.inc.rules}.
\item
\emph{$\prim$-inclusion} $\Gamma\sgv a<\any{n}$ is defined as the smallest relation generated by the inclusion rules in Table~\ref{sub.sinc.rules}.
\end{itemize}
\noindent 
We also introduce here the notion of $\lambda^*$-inclusion $\Gamma\sgv a\gincl b$ which is defined as the transitive closure of 
$\lambda$-\emph{inclusion}. We will later show (Law \ref{sub.incl.por}($iv$)) that it is equivalent to $\lambda$-inclusion.
\end{definition}
\begin{remark}[Examples of inclusions]
Some examples illustrate the two notions of inclusion. 
\begin{itemize}
\item $\sgv[y:\prim]y<\any{1}$ since after applying the \ssabsu\ and \sstyp\ the conditions of \sstart\ are obviously satisfied.
\item $z:\prim\sgv([x:\prim][x,\prim]\;z)\incl([x:\prim][x,\any{1}]\;z)$ since the condition on validity is obviously satisfied and 
\[
([x:\prim][x,\prim]\;z)\eqv[z,\prim],\;\;
z:\prim\sgv[z,\prim]\leq[z,\any{1}],\;\;
[z,\any{1}]\eqv([x:\prim][x,\any{1}]\;z)
\]
\end{itemize}
\end{remark}
\begin{remark}[Examples of typing with bounded polymorphism]
Some examples illustrate typing with bounded polymorphism: Let $z:=[\prim\fun\prim]$. We have $\sgv z:z$ and $\sgv z\leq\any{1}$ and hence by rule \sinc\ we have $\sgv z:\any{1}$.  
\begin{itemize}
\item  Let $\Gamma:=(y:[x:\any{1}][x\fun x])$. Since $\sgv z:\any{1}$ we have $\Gamma\sgv (y\;z):[z\fun z]$ 
\item  Let $t:=([x:\any{1}][x\fun\prim]\;z)$: Since $\sgv z:\any{1}$ we have $\sgv t:[z\fun\prim]$ and since $\sgv z:z$ we have $\sgv(t\;z):\prim$.
\end{itemize}
\end{remark}
\noindent
%
\section{Examples}
We describe changes that use the new primitive constants and bounded polymorphism. Note that casting axioms (see~\ref{ex.casting}, \ref{ex.casting2}) are not needed anymore.  
In general, the choice of primitive constants $\any{n}$ depends on the degree of functional complexity needed.
Note that in this section we are using the notational conventions from Section \ref{examples}.
\subsection{Basic logical properties}%
\label{sub.ex.logic}
We have presented some logical properties in Section~\ref{ex.logic} schematically, i.e.~ranging over meta-variables $a$, $b$, etc.
Using the primitive constants we can now present them completely formalized, i.e. as expressions in \dcalcb.

In this simple context, we are using the primitive constants $\any{1}$ and $\any{2}$ as follows:
\begin{itemize}
\item
$\any{2}$ is used as type of propositional identifiers used in logical axioms, \eg\ the infinite set of axioms $\negax^+_{a,b},\negax^-_{a,b}\in\dvar$
can be approximated by a single axiom.
\[
\it{neg}:[x,y:\any{2}][[[x+y]\fun[\myneg x\fun y]],[[x\fun y]\fun[\myneg x+y]]]
\]
\item
$\any{1}$ is used as
type of propositional identifiers used in derived logical laws, \eg\ one can now derive the principle of the excluded middle as follows
\begin{eqnarray*}
[x:\any{1}](\pright{\it{neg}(\myneg x,\myneg x)}([y:\myneg x]y))&:&[x:\any{1}][x+\myneg x]\qquad\quad(\it{tnd})\\{}
[x,y:\any{1}](\pleft{\it{neg}([\myneg x\!+\!x],y)}(\injl{\it{tnd}(x)}{y}))&:&[x,y:\any{1}][[x,\myneg x]\fun y]
\end{eqnarray*}
\noindent
Note that in the second example the argument $[\myneg x+x]$ of \emph{neg} is of type $[\any{1}+\any{1}]$ which is not included in $\any{1}$, necessitating the quantification of $x:\any{2}$ in \emph{neg}.
\end{itemize}
\noindent
In this setting, trivial laws can be approximated by the following formalization:
\begin{eqnarray*}
[x:\any{1}][y:x]y&:&[x:\any{1}][x\fun x]
\end{eqnarray*}
\noindent
Logical falsehood (see Section~\ref{ex.logic}) can be defined as the statement that that any proposition is true:
\begin{eqnarray*}
\ff&:=&[x:\any{1}]x
\end{eqnarray*}
\begin{remark}
Note that the quantification over $\any{2}$ for axiom \emph{neg} is not needed from a theoretical point of view since any use of a derived logical law can, by unfolding of its proof, ultimately be replaced by an expression using axioms on a concrete proposition $p:\any{1}$. Here we choose to have a separation for the purely pragmatic reasons to handle an independent intermediate layer of definition, which in a logical interpretation amounts to a layer of auxiliary laws. 
Other formalizations may use axioms ranging over constants $\any{3}$ or higher, allowing for additional levels of intermediate results.
The use of levels of intermediate lemmas is illustrated schematically in Table~\ref{sub.any.levels}.
\begin{table}[!htb]
\fbox{
\begin{minipage}{0.96\textwidth}
\begin{eqnarray*}
A&:& [x_1:\any{n}]\ldots[x_{k_n}:\any{n}]P_A(x_1\,\ldots x_{k_n})\\
L_{n-1}
&:& [x_1:\any{n-1}]\ldots[x_{k_{n-1}}:\any{n-1}]P_{n-1}(x_1\,\ldots x_{k_{n-1}})\\
&:=& d_{n-1}\;(\it{using}\;A)\\
&\ldots&\\
L_1
&:& [x_1:\any{1}]\ldots[x_{k_1}:\any{1}]P_1(x_1\,\ldots x_{k_1})\\
&:=& d_1\;(\it{using}\;A\;\it{and}\;L_2\;\it{to}\;L_{n-1})\\
\it{law}
&:& [x_1:\any{0}]\ldots[x_{k_0}:\any{0}]P(x_1\,\ldots x_{k_0})\\
&:=& d_0\;(\it{using}\;A\;\it{and}\;L_1\;\it{to}\;L_{n-1}) 
\end{eqnarray*} 
\end{minipage}
}
\caption{Intermediate logical propositions.\label{sub.any.levels}}
\end{table}
Here $P$, $P_1$, $\ldots$, $P_{n-1}$, and $P_A$, stand for propositions parametrized over elements of type $\any{i}$, for some $i$.
In a practical setting,  in an intermediate state of the proof of \emph{law}, some of the proofs $d_i$ may not yet be available, hence 
for the time being some of the elements \emph{law} and $L_i$ are declared rather than defined.
However this may lead to the necessity for raising the index $i$ of $\any{i}$ at each level so as to satisfy the type constraint of applications.
An example for the need to use at least $\any{2}$ has been given in the overview section (\ref{sub.overview}).
\end{remark}
\subsection{Type casting}%
\label{sub.ex.casting}
The example from Section~\ref{ex.casting} can now be rewritten for $\ff$ using $\any{1}$.
\[
x:\any{1}\gv[y:\ff]y(x):[\ff\fun x]
\]
\noindent
Therefore, the casting axiom are not needed anymore.
\subsection{Natural Numbers}
\label{sub.examples.nats}
The type of $\nats$ (see Section~\ref{examples.nats}) is extended to $\any{1}$ .
\begin{eqnarray*}
\it{Naturals}:=\quad(\quad \nats&:&\any{1}\\
0&:&\nats\\
s&:&[\nats\fun\nats]\\
\ldots&&)
\end{eqnarray*}
As a consequence,  as illustrated in the Section~\ref{sub.ex.sets}, instantiations to $\nats$ do not need casting anymore. 
Note that $\emph{even}$ is not included in $\any{1}$, however
\[\it{even}\;:=\;[n:\nats;m!\nats]n=2*m
\]
\[
\it{Naturals}\gv\it{even}\;:\;[\nats\fun[\nats\fun\prim]]\;\leq\;\any{2}
\]
\subsection{Equality}
\label{sub.equality}
Since natural numbers are of type $\any{1}$, the type of the type parameter of the equality function in Section~\ref{equality} is also extended to $\any{1}$.
Basic axioms about an equality congruence relation on expressions of equal type can be formalized as context \emph{Equality}:
\begin{eqnarray*}
\it{Equality}&:=&(\\
\noarg=_{\noarg}\noarg&:&[S:\any{1}][S;S\fun\prim],\\
E_1&:&[S:\any{1};x:S]x=_{S}x,\\
\ldots&&)
\end{eqnarray*}
\subsection{Sets}%
\label{sub.ex.sets}
Basic set operators can now be declared using $\any{1}$, and $\any{2}$.
Note that $\any{2}$ is used in the type of $\{\noarg\}_{\noarg}$ as the set building operator ranges over predicates such as \emph{even}, 
similar for the set membership axioms $I$ and $O$.
\begin{eqnarray*}
Sets&:=&(\\
\mathbb{P}&:&[\any{1}\fun\prim],\\
\noarg\in_{\noarg}\noarg&:&[S:\any{1}][S;\mathbb{P}(S)\fun\prim],\\
\{\noarg\}_{\noarg}&:&[S:\any{1}][[S\fun\any{2}]\fun\mathbb{P}(S)],\\
I&:&[S:\any{1};x:S;P:[S\fun\any{2}]][P(x)\fun x\in_{S}\{[y:S]P(y)\}_S],\\
O&:&[S:\any{1};x:S;P:[S\fun\any{2}]][x\in_{S}\{[y:S]P(y)\}_S\fun P(x)]\\
) 
\end{eqnarray*}
The formalization of sets (see Section~\ref{ex.sets}) can be simplified by using set comprehension without casting.
\begin{eqnarray*}
\emptyset&:=& [S:\any{1}]\{[S\fun\ff]\}_S\\
&&:\quad[S:\any{1}]\mathbb{P}(S)\\
\noarg\!\cup_{\noarg}\!\noarg&:=& [S:\any{1};A,B:\mathbb{P}(S)]\{[x:S][x\in_{S}A+x\in_{S}B]\}_S\\
&&:\quad[S:\any{1}][\mathbb{P}(S);\mathbb{P}(S)\fun \mathbb{P}(S)]\\
\it{Even}&:=&\{\it{even}\}_{\nats}\\
&&:\quad\mathbb{P}(\nats)
\end{eqnarray*}
Properties of individual elements can be deduced on the basis of the axiom $O$, for example
we can extract the property $P:=[n!\nats](x=_{\nats}2*n)$ of a member $x$ of {\it Even} without type casting as follows:
\begin{eqnarray*}
(x:\nats,asm:x\in_{\nats}\it{Even})&\sgv&O(\nats,x,\it{even},\it{asm})\;:\; P
\end{eqnarray*}
\noindent
With $\it{Int}:\prim$ (see Section~\ref{ex.sets}) we have $[n:\it{Int}](n\geq 0_I):[\it{Int}\fun\prim]\leq\any{1}$. Therefore, an alternative definition of naturals from integers can be given without casting:
\begin{eqnarray*}
(\nats\;:=\;\{[n:\it{Int}](n\geq 0_I)\}_{\it Int},\;\;0\;:=\ldots)
\end{eqnarray*}  
\section{Properties of \texorpdfstring{\dcalcb}{\dcalcbt}}
We go through the existing properties and discuss the necessity of changes or additions for the additional primitive constants and the inclusion relation.
For simplicity we often do not introduce a new numbering for these adapted properties but instead keep using the old numbering.
In other words, if we extend an existing property in a minor way and in the upcoming sections reference an such an existing property, we will use the old reference but always
mean to refer to its (adapted) version covering the extensions of \dcalcb. 
\subsection{Basic properties of reduction}
\label{sub.rd.basic}
The properties in Section~\ref{rd.basic} are obviously all valid also for primitive constants and tagged protected definition.
The proofs need to be adapted in very obvious ways due to adapted induction principles.
\subsection{Confluence properties}
\label{sub.confl}
The properties in Section~\ref{confl} are obviously all valid also for primitive constants and tagged protected definitions.
The extended reduction axiom $\nu_6$ does not create any new critical pairs and therefore 
proofs need to be adapted in very obvious ways only.
\subsection{Basic properties of typing and inclusion}
\label{sub.typing.basics}
The properties in Sections \ref{typing.basics}, \ref{typing.sub} and \ref{closure} have to be modified for the rules of \dcalcb.
Moreover some additional properties are needed and there are new dependencies in the proofs which require to reorganize the sequence of presentation.
\begin{law}[Free variables in inclusions and $\any{}$-inclusions.]
\label{sub.leq.free}
For all $\Gamma_1$, $\Gamma_2$, $x$, $a$, $b$, $c$:
If $\Gamma_1,x:c,\Gamma_2\sgv a\leq b$ and $x\notin{\free}([\Gamma_2]a)$ then $x\notin{\free}([\Gamma_2]b)$ and $\Gamma_1,\Gamma_2\sgv a\leq b$.
Similar for $\any{}$-inclusions.
\end{law}
\begin{proof}
A straightforward inspection of the inference rules of $\Gamma_1,x:c,\Gamma_2\sgv a\leq b$ shows that they cannot not introduce free occurrences of $x$ in $b$.
Similar for $\any{}$-inclusions.
\end{proof}
\noindent
The Law~\ref{type.sub} (substitution and typing) has to be generalized as follows: 
\begin{law}[Substitution and typing - extended by inclusion]%
\label{sub.type.sub}

Assume that $\Gamma_a=(\Gamma_1,x:a,\Gamma_2)$ and $\Gamma_b=(\Gamma_1,\Gamma_2\gsub{x}{b})$ for some $\Gamma_1,\Gamma_2,x,a,b$ where $\Gamma_1\sgv b:a$.
For all $c,d,n$: 
\begin{itemize}
\item[$i$:] 
If $\Gamma_a\sgv c<\any{n}$ then $\Gamma_b\sgv c\gsub{x}{b}<\any{n}$. 
\item[$ii$:]  
If $\Gamma_a\sgv c\leq d$ then $\Gamma_b\sgv c\gsub{x}{b}\leq d\gsub{x}{b}$. 
\item[$iii$:] 
If $\Gamma_a\sgv c:d$ then $\Gamma_b\sgv c\gsub{x}{b}: d\gsub{x}{b}$. 
\end{itemize}
\end{law}
\begin{proof}
The proof is by simultaneous induction on all three parts.
\begin{itemize}
\item[$i$:]
\begin{meditemize}
\item[\sstart:]  
We have $\Gamma_a\sgv\any{m}$, $\Gamma_a\sgv\any{n}$, and $m<n$.
Using the inductive hypothesis we obtain $\Gamma_b\sgv\any{m}$, $\Gamma_b\sgv\any{n}$ and therefore the proposition.
\item[\sstyp:]
We have $\Gamma_a\sgv c:e$ and  $\Gamma_a\sgv e<\any{n}$ for some $e$. 
By inductive hypothesis $\Gamma_b\sgv c\gsub{x}{b}:e\gsub{x}{b}$ and $\Gamma_b\sgv e\gsub{x}{b}<\any{n}$ hence $\Gamma_b\sgv c\gsub{x}{b}<\any{n}$. 
\item[\ssabsu:]
We have $c=\binbop{y}{c_1}{c_2}$ where $\Gamma_a\sgv c_1:\any{n}$ and $\Gamma_a,y:c_1\sgv c_2:\any{n}$ for some $y$, $c_1$, $c_2$ where we may assume $y\neq x$.
By inductive hypothesis $\Gamma_b\sgv c_1\gsub{x}{b}:\any{n}$ and $\Gamma_b,y:c_1\gsub{x}{b}\sgv c_2\gsub{x}{b}:\any{n}$ hence 
 $\Gamma_b\sgv c\gsub{x}{b}<\any{n}$.
\end{meditemize}
\noindent
Similar arguments can be made for rules \ssbprod, \ssinjl, \ssinjr, and \sspdef.
\item[$ii$:]
\begin{meditemize}
\item[\srefl:]  
We have $c=d$. If $\Gamma_a\sgv c$ then by inductive hypothesis we know that $\Gamma_b\sgv c\gsub{x}{b}$ which implies the proposition.
\item[\sembed:]
Follows directly from the inductive hypothesis.
\item[\sabs:]
We have $c=\binbop{y}{c_1}{c_2}$ and $d=\binbop{y}{c_1}{d_2}$ where $\Gamma_a\sgv c_1\leq d_1$ and $\Gamma_a,y:c_1\sgv c_2\leq d_2$ for some $y$, $c_1$, $c_2$, $d_2$ where we may assume $y\neq x$.
By inductive hypothesis $\Gamma_b,y:c_1\gsub{x}{b}\sgv c_2\gsub{x}{b}\leq d_2\gsub{x}{b}$ hence $\Gamma_b\sgv c\gsub{x}{b}\leq d\gsub{x}{b}$.
\item[\sbprod:]
Similar to previous case.
\end{meditemize}
\item[$iii$:]
The proof is by induction on the definition of  $\Gamma_a\sgv c:d$. All arguments in the proof of Law~\ref{type.sub} remain unchanged. The new typing rule can be shown as follows:
\begin{meditemize}
\item[\sinc:]
We have $\Gamma_a\sgv c:e$, $\Gamma_a\sgv e\leq d$.  
By inductive hypothesis  $\Gamma_b\sgv c\gsub{x}{b}:e\gsub{x}{b}$ and $\Gamma_b\sgv e\gsub{x}{b}\leq d\gsub{x}{b}$. 
Hence by rule \sinc~ we can infer that $\Gamma_b\sgv c\gsub{x}{b}: d\gsub{x}{b}$.  \qedhere
\end{meditemize}
\end{itemize}
\end{proof}
\noindent
The proof of the Law~\ref{val.sub} (substitution and validity) will use Law \ref{sub.type.sub} instead of Law~\ref{type.sub}.

\noindent
As a direct consequence of the rules of inclusion and $\any{}$-inclusion we note that both are relations on valid expressions. 
\begin{law}[Inclusion and validity]
\label{sub.incl.valid}
For all $\Gamma,a,b,n$: 
\begin{itemize}
\item[$i$:] 
$\Gamma\sgv a<\any{n}$ implies $\Gamma\sgv a$.
\item[$ii$:]  
$\Gamma\sgv a\leq b$ implies $\Gamma\sgv a,b$.
\end{itemize}
\end{law}
\begin{proof}
Obvious inductive arguments on the definition of $\any{}$-inclusion and inclusion.
\end{proof}
In the proof of Law~\ref{valid.type} (valid expressions have valid types) the case \sinc\ has to be added, all other cases can remain unchanged.
In case of \sinc, the proposition follows from Law~\ref{sub.incl.valid}.
\noindent
The properties of Section~\ref{typing.basics} must be generalized to the notion of typing with inclusion.
First, Law~\ref{type.weak}(context weakening) must be adapted as follows: 
\begin{law}[Context weakening - extended by inclusions]
\label{sub.type.weak}
For all $\Gamma_1,\Gamma_2,x,n,a,b,c$ where $\Gamma_1\sgv c$: 
\begin{itemize}
\item[$i$:] 
$(\Gamma_1,\Gamma_2)\sgv a<\any{n}$ implies $(\Gamma_1,x:c,\Gamma_2)\sgv a<\any{n}$.
\item[$ii$:]  
$(\Gamma_1,\Gamma_2)\sgv a\leq b$ implies $(\Gamma_1,x:c,\Gamma_2)\sgv a\leq b$.
\item[$iii$:] 
$(\Gamma_1,\Gamma_2)\sgv a:b$ implies $(\Gamma_1,x:c,\Gamma_2)\sgv a:b$.
\item[$iv$:]  
$(\Gamma_1,\Gamma_2)\sgv a\incl b$ implies $(\Gamma_1,x:c,\Gamma_2)\sgv a\incl b$.
\item[$v$:]  
$(\Gamma_1,\Gamma_2)\sgv a\gincl b$ implies $(\Gamma_1,x:c,\Gamma_2)\sgv a\gincl b$.
\end{itemize}
\end{law}
\begin{proof}
For Parts $i$, $ii$, $iii$ the proof is by simultaneous induction on all three parts.
For Part $iv$ the proof follows directly from the definition of $\lambda$-inclusion and Part $ii$.
For Part $v$ the proof is by induction on the definition of $\lambda^*$-inclusion using Part $iv$.
\end{proof}
\noindent 
The Law~\ref{type.xtrct} (context extraction) is not affected by the new inclusion relations and its proof does not need to be changed except for an obvious extension for rule \sinc.
The Law~\ref{val.decomp} (validity decomposition) whose proof uses Law~\ref{type.xtrct} only is not affected by the new inclusion relations.
It proof has to be extended by including the rule \sinc\ in an obvious way.

\noindent 
The Law~\ref{eqv.env} (context equivalence and typing) has to be extended by the inclusion relations.
Note that the proof of Law~\ref{eqv.env} did not depend on any auxiliary properties. 
The same will be case for the adapted version.
\begin{law}[Context equivalence and typing - extended by inclusions]
\label{sub.eqv.env}
Let $\Gamma_a=(\Gamma_1,x:a,\Gamma_2)$ and $\Gamma_b=(\Gamma_1,x:b,\Gamma_2)$ for some $\Gamma_1,\Gamma_2,x,a,b$ where $a\eqv b$ and $\Gamma_1\sgv b$:
For all $c,d,n$: 
\begin{itemize}
\item[$i$:]
If $\Gamma_a\sgv c<\any{n}$ then $\Gamma_b\sgv c<\any{n}$.
\item[$ii$:]
If $\Gamma_a\sgv c\leq d$ then $\Gamma_b\sgv c\leq d$.
\item[$iii$:]
If $\Gamma_a\sgv c:d$ then $\Gamma_b\sgv c:d$.
\end{itemize}
As a consequence $\Gamma_a\sgv c\incl d$ implies  $\Gamma_b\sgv c\incl d$ and similar for $\lambda^*$-inclusion.
\end{law}
\begin{proof}
The proof is by simultaneous induction on all three parts.
\begin{itemize}
\item[$i$:]
\begin{meditemize}
\item[\sstyp:]
We have $\Gamma_a\sgv a:c$ and  $\Gamma_a\sgv c\leq\any{n}$. 
By inductive hypothesis $\Gamma_b\sgv b:c$ and $\Gamma_b\sgv c\leq\any{n}$ and therefore $\Gamma_b\sgv b\leq\any{n}$. 
\end{meditemize}
Similar arguments can be made for rules \sstart, \ssabsu, \ssbprod, \ssinjl, \ssinjr, and \sspdef.
\item[$ii$:]
\begin{meditemize}
\item[\srefl:]  
Obvious.
\item[\sembed:]
Follows directly from the inductive hypothesis.
\item[\sabs:]
We have $c=\binbop{y}{c_1}{c_2}$ where $\Gamma_a\sgv c_1<\any{n}$ and $\Gamma_a,y:c_1\sgv c_2<\any{n}$.
By inductive hypothesis $\Gamma_b\sgv c_1\gsub{x}{b}<\any{n}$ and $\Gamma_b,y:c_1\gsub{x}{b}\sgv c_2\gsub{x}{b}<\any{n}$ hence  $\Gamma_b\sgv c\gsub{x}{b}<\any{n}$.
\item[\sbprod:]
Similar to previous case.
\end{meditemize}
\item[$iii$:]
In the proof of Law~\ref{eqv.env} the case \sinc\ has to be added. It follows from the inductive hypothesis.
\end{itemize}
The consequence is obvious since congruence does not depend on a context.\qedhere
\end{proof}
\noindent
As a prerequisite for other properties we need to show some basic properties for inclusion, $\lambda$-inclusion, and $\lambda^*$-inclusion. 
\begin{law}[Basic properties of inclusion]
\label{sub.leq.prop}  
For all $\Gamma$, $x$, $n$, $a$, $b$, $b_1$, $b_2$, $c$ where $x\notin\dom(\Gamma)$: 
\begin{itemize}
\item[$i$:]
$\Gamma\sgv\any{n}\leq a$ implies $a=\any{m}$ for some $m$ where either $m=n$ or $\Gamma\sgv\any{n}<\any{m}$. 
(In Law \ref{sub.any}($ii$)) we will show that also $m\geq n$).
\item[$ii$:]
If $\Gamma\sgv a\leq\binbop{x}{b}{c}$ then $a=\binbop{x}{b}{d}$ where $\Gamma,x:b\sgv d\leq c$ for some $d$.
\item[$iii$:]
If $\Gamma\sgv\binbop{x}{b}{c}\leq a$ then either $a=\binbop{x}{b}{d}$ and $\Gamma,x:b\sgv c\leq d$ for some $d$,
or $a=\any{m}$ for some $m$.
\item[$iv$:]
If $\Gamma\sgv a\leq\prsumop{b_1}{b_2}$ then $a=\prsumop{a_1}{a_2}$ and $\Gamma\sgv a_i\leq b_i$ for some $a_i$  where $i=1,2$.
\item[$v$:]
If $\Gamma\sgv\prsumop{b_1}{b_2}\leq a$ then either $a=\prsumop{a_1}{a_2}$ and $\Gamma\sgv a_i\leq b_i$ for some $a_i$ where $i=1,2$,
or $a=\any{m}$ for some $m$.
\end{itemize}
\end{law}
\begin{proof}
Simple consequences of the definition of inclusion. \qedhere
\end{proof}
\begin{law}[Basic properties of $\lambda$-inclusion]
\label{sub.incl.prop}
For all $\Gamma$, $a$, $b$, $c$: 
\begin{itemize}
\item[$i$:]
If $\Gamma\sgv a\incl b$ and $a\eqv a'$, $b\eqv b'$ some $a'$, $b'$ and $\Gamma\sgv b'$ then $\Gamma\sgv a'\incl b'$. 
\item[$ii$:]
If $\Gamma\sgv a\incl b$ where $a\eqv\any{n}$ for some $n$ then $b\eqv\any{m}$ for some $m$ where $\Gamma\sgv\any{n}\incl\any{m}$.
\item[$iii$:]
If $\Gamma\sgv a\incl\binbop{x}{b_1}{b_2}$ for some $x$, $b_1$, $b_2$
then $a\eqv\binbop{x}{b_1}{a_2}$ for some $a_2$ where $\Gamma\sgv\binbop{x}{b_1}{a_2}$ and $\Gamma,x:b_1\sgv a_2\incl b_2$.
\item[$iv$:]
If $\Gamma\sgv\binbop{x}{a_1}{a_2}$ and $\Gamma\sgv\binbop{x}{a_1}{a_2}\incl b$ for some $x$, $a_1$, $a_2$ then
either $b\eqv\binbop{x}{a_1}{b_2}$ for some $b_2$ where $\Gamma,x:a_1\sgv a_2\incl b_2$,
or $b\eqv\any{m}$ for some $m$ where $\Gamma\sgv\binbop{x}{a_1}{a_2}\incl\any{m}$.
\item[$v$:]
If $\Gamma\sgv a\incl\prsumop{b_1}{b_2}$ for some $b_1$, $b_2$ then $a\eqv\prsumop{a_1}{a_2}$ where $\Gamma\sgv\prsumop{a_1}{a_2}$ and $\Gamma\sgv a_i\incl b_i$ for some $a_i$ and $i=1,2$.
\item[$vi$:]
If $\Gamma\sgv\prsumop{a_1}{a_2}$ and $\Gamma\sgv\prsumop{a_1}{a_2}\incl b$ for some $a_1$, $a_2$ then either $a\eqv\prsumop{a_1}{a_2}$ and $\Gamma\sgv a_i\incl b_i$ for some $b_i$  where $i=1,2$,
or $a\eqv\any{m}$ for some $m$ where $\Gamma\sgv\prsumop{a_1}{a_2}\incl\any{m}$.
\item[$vii$:]
If $\Gamma\sgv a:b$ and $\Gamma\sgv b\incl c$ then  $\Gamma\sgv a:c$.
\end{itemize}
\end{law}
\begin{proof}
For all $\Gamma$, $a$, $b$, $c$: 

Part $i$ follows from the definition of $\lambda$-inclusion and Law \ref{cr}.

Part $ii$ can be seen as follows:
If $\Gamma\sgv a\incl b$ where $a\eqv\any{n}$ for some $n$ then $a\eqv a'$, $\Gamma\sgv a'\leq b'$, and $b'\eqv b$ for some $a'$, $b'$. 
By definition of inclusion one of the following five cases must be true: $b'=a'$, $b'=\any{m}$ for some $m$ where $\Gamma\sgv a'<\any{m}$,
$b'$ is an abstraction, $b'$ is a product, or $b'$ is a sum. 
In the first case we obtain $\any{n}\eqv a\eqv a'=b'\eqv\any{m}$ which by Law \ref{cr} implies $n=m$ which, since $\Gamma\sgv\any{m}$, implies the proposition.
In the second case we obtain $\Gamma\sgv a'\incl\any{m}$ and therefore by Part $i$ $\Gamma\sgv\any{n}\incl\any{m}$. 
If $b'$ is an abstraction then by Law \ref{sub.leq.prop}($ii$) we know that $a'$ is an abstraction which is impossible since $\any{n}\eqv a'$.
Similar for the other two cases.

Part $iii$ can be seen as follows: 
If $\Gamma\sgv a\incl\binbop{x}{b_1}{b_2}$ for some $b_1$, $b_2$ 
then by definition of $\lambda$-inclusion $\Gamma\sgv\binbop{x}{b_1}{b_2}$, $a\eqv a'$, $\Gamma\sgv a'\leq b'$ and $b'\eqv\binbop{x}{b_1}{b_2}$ for some $a'$, $b'$. 
By definition of inclusion one of the following four cases must be true: $b'=a'$ ($1$), $b'=\any{m}$ for some $m$ ($2$),
$a'=\binbop{x}{b_1'}{a_2}$ and $b'=\binbop{x}{b_1'}{b_2'}$ for some $b_1'$, $b_2'$, and $a_2$ where $\Gamma,x:b_1'\sgv a_2\leq b_2'$ ($3$), or 
$a'=\prsumop{a_1}{a_2}$ and $b'=\prsumop{b_1'}{b_2'}$ for some $a_1$, $a_2$, $b_1'$, and $b_2'$, where $\Gamma\sgv a_1\leq b_1'$, $\Gamma\sgv a_2\leq b_2'$ ($4$).

In case $1$ $a\eqv a'=b'\eqv\binbop{x}{b_1}{b_2}$ and since $\Gamma\sgv\binbop{x}{b_1}{b_2}$ 
by Law \ref{type.xtrct} $\Gamma,x:b_1\sgv b_2$ and hence obviously $\Gamma,x:b_1\sgv b_2\incl b_2$ which by Law \ref{cr} implies the proposition with $b_2=a_2$.
Case $2$ is impossible since by basic properties of reduction $b'=\any{m}\neqv\binbop{x}{b_1}{b_2}$.
In case $3$ $\binbop{x}{b_1}{b_2}\eqv b'=\binbop{x}{b_1'}{b_2'}$ hence by Law \ref{cr} and basic properties of reduction $b_1\eqv b_1'$ as well as $b_2\eqv b_2'$.
Hence $a\eqv\binbop{x}{b_1}{a_2}$.
Since by Law \ref{type.xtrct} $\Gamma\sgv b_1$ by Law \ref{sub.eqv.env} $\Gamma,x:b_1\sgv a_2\leq b_2'$ 
and hence since by Law \ref{val.decomp} $\Gamma,x:b_1\sgv b_2$, by Part $i$ we obtain $\Gamma,x:b_1\sgv a_2\incl b_2$. 
Case $4$ is impossible since by basic properties of reduction $b'\neqv\binbop{x}{b_1}{b_2}$.

Part $iv$ can be seen as follows: 
If $\Gamma\sgv\binbop{x}{a_1}{a_2}$ and $\Gamma\sgv\binbop{x}{a_1}{a_2}\incl b$ for some $a_1$, $a_2$ then
then by definition of $\lambda$-inclusion $\Gamma\sgv b$, $\binbop{x}{a_1}{a_2}\eqv a'$, $\Gamma\sgv a'\leq b'$ and $b'\eqv b$ for some $a'$, $b'$. 
By definition of inclusion one of the following four cases must be true: $b'=a'$ ($1$), $b'=\any{m}$ for some $m$ ($2$),
$a'=\binbop{x}{a_1'}{a_2'}$ and $b'=\binbop{x}{a_1'}{b_2}$ for some $a_1'$, $a_2'$, and $b_2$, where $\Gamma,x:a_1'\sgv a_2'\leq b_2$ ($3$), or 
$a'=\prsumop{a_1'}{a_2'}$ and $b'=\prsumop{b_1}{b_2}$ for some $a_1'$, $a_2'$, $b_1$, and $b_2$, where $\Gamma\sgv a_1'\leq b_1$, $\Gamma\sgv a_2'\leq b_2$ ($4$).

In case $1$ $\binbop{x}{a_1}{a_2}\eqv a'=b'\eqv b$  which by Law \ref{cr} implies $\binbop{x}{a_1}{a_2}\eqv b$.
Since $\Gamma\sgv\binbop{x}{a_1}{a_2}$ by Law \ref{val.decomp} $\Gamma,x:a_1\sgv a_2$ and hence $\Gamma,x:a_1\sgv a_2\incl a_2$. 
The property follows with $b_2=a_2$.
In case $2$ the property follows immediately.
In case $3$ $\binbop{x}{a_1}{a_2}\eqv\binbop{x}{a_1'}{a_2'}$ hence by Law \ref{cr} and basic properties of reduction $a_1\eqv a_1'$ as well as $a_2\eqv a_2'$.
Hence $b\eqv\binbop{x}{a_1}{b_2}$.
Since $\Gamma,x:a_1'\sgv a_2'\leq b_2$ and by Law \ref{val.decomp} $\Gamma\sgv a_1$, Law \ref{sub.eqv.env} implies $\Gamma,x:a_1\sgv a_2'\leq b_2$.
By Part $i$ $\Gamma,x:a_1\sgv a_2\incl b_2$.
Case $4$ is impossible since by basic properties of reduction $a'\neqv\prsumop{a_1}{a_2}$.

Part $v$ can be argued similar to Part $iii$.
Part $vi$ can be argued similar to Part $iv$.
In Part $vii$ we have $\Gamma\sgv a:b$ and $b\eqv b'$, $\Gamma\sgv b'\leq c'$, $c'\eqv c$, and $\Gamma\sgv c$ for some $b'$ and $c'$.  
By Law \ref{sub.incl.valid}($ii$) we know that $\Gamma\sgv b'$ and $\Gamma\sgv c'$ hence by rule \conv\ we know that $\Gamma\sgv a:b'$, hence by rule \sinc\ $\Gamma\sgv a:c'$ and by rule \conv\ $\Gamma\sgv a:c$.
\qedhere
\end{proof}
\begin{law}[Basic properties of $\lambda^*$-inclusion]
\label{sub.gincl.prop}
For all $\Gamma$, $a$, $b$, $c$: 
\begin{itemize}
\item[$i$:]
If $\Gamma\sgv a\gincl b$, $a\eqv a'$, $b\eqv b'$, and $\Gamma\sgv b'$ then $\Gamma\sgv a'\gincl b'$ for all $a'$, $b'$. 
\item[$ii$:]
If $\Gamma\sgv a\gincl b$ where $a\eqv\any{n}$ for some $n$ then $b\eqv\any{m}$ for some $m$ where $\Gamma\sgv\any{n}\gincl\any{m}$. 
\item[$iii$:]
If $\Gamma\sgv a\gincl\binbop{x}{b_1}{b_2}$ for some $x$, $b_1$, $b_2$
then $a\eqv\binbop{x}{b_1}{a_2}$ where $\Gamma\sgv\binbop{x}{b_1}{a_2}$ and $\Gamma,x:b_1\sgv a_2\gincl b_2$ for some $a_2$.
\item[$iv$:]
If $\Gamma\sgv\binbop{x}{a_1}{a_2}$ and $\Gamma\sgv\binbop{x}{a_1}{a_2}\gincl b$ for some $x$, $a_1$, $a_2$ then either
 $\Gamma,x:b_1\sgv a_2\gincl b_2$ for some $b_2$ where $\Gamma\sgv\binbop{x}{a_1}{b_2}\gincl b$.
or $b\eqv\any{m}$ for some $m$ where $\Gamma\sgv\binbop{x}{a_1}{a_2}\gincl\any{m}$.
\item[$v$:]
$\Gamma\sgv a\gincl\prsumop{b_1}{b_2}$ for some $b_1$, $b_2$ implies $a\eqv\prsumop{a_1}{a_2}$ where $\Gamma\sgv\prsumop{a_1}{a_2}$  and $\Gamma\sgv a_i\gincl b_i$ for some $a_i$ and $i=1,2$.
\item[$vi$:]
If $\Gamma\sgv\prsumop{a_1}{a_2}$ and $\Gamma\sgv\prsumop{a_1}{a_2}\gincl b$ for some $a_1$, $a_2$ 
then either $\Gamma\sgv a_i\gincl b_i$ for some $b_i$ and $i=1,2$,
or $b\eqv\any{m}$ for some $m$ where $\Gamma\sgv\prsumop{a_1}{a_2}\gincl\any{m}$.
\item[$vii$:]
If $\Gamma\sgv a:b$ and $\Gamma\sgv b\gincl c$ then  $\Gamma\sgv a:c$.
\end{itemize}
\end{law}
\begin{proof}
For all $\Gamma$, $a$, $b$, $c$: 

Part $i$ follows by an inductive argument on $\Gamma\sgv a\gincl b$.
For the base $\Gamma\sgv a\incl b$ the property corresponds to \ref{sub.incl.prop}($i$).
Otherwise assume $\Gamma\sgv a\gincl d$ and $\Gamma\sgv d\gincl b$ for some $d$ and $a\eqv a'$ and $b\eqv b'$ where $\Gamma\sgv b'$.
By inductive assumption on $\Gamma\sgv a\gincl d$ we obtain $\Gamma\sgv a'\gincl d$ and by inductive assumption on $\Gamma\sgv d\gincl b$ we obtain $\Gamma\sgv d\gincl b'$.
By definition of $\lambda^*$-inclusion this implies $\Gamma\sgv a'\gincl b'$.

Part $ii$ can be shown by induction on the definition of $\Gamma\sgv a\gincl b$.
If $\Gamma\sgv a\incl b$ and $a\eqv\any{n}$ for some $n$ then the proposition follows from \ref{sub.incl.prop}($ii$).
Otherwise assume $\Gamma\sgv a\gincl d$ and $\Gamma\sgv d\gincl b$ for some $d$ and $a\eqv\any{n}$ for some $n$.
By inductive hypothesis of $\Gamma\sgv a\gincl d$: $d\eqv\any{k}$ for some $k$ where $\Gamma\sgv\any{n}\gincl\any{k}$.
Hence by inductive hypothesis of $\Gamma\sgv d\gincl b$: $b\eqv\any{m}$ for some $m$ where $\Gamma\sgv\any{k}\gincl\any{m}$.
By definition of $\lambda^*$-inclusion $\Gamma\sgv\any{n}\gincl\any{m}$ which implies the proposition.

For Part $iii$ first note that the proposition can be reformulated equivalently as follows: If $\Gamma\sgv a\gincl b$ where $b\eqv\binbop{x}{b_1}{b_2}$ and $\Gamma\sgv\binbop{x}{b_1}{b_2}$ for some $b$, $b_1$, $b_2$ then $\ldots$ (as before).
This is because $\Gamma\sgv a\gincl\binbop{x}{b_1}{b_2}$ where $\Gamma\sgv\binbop{x}{b_1}{b_2}$ if and only if $\Gamma\sgv a\gincl b$ where $b\eqv\binbop{x}{b_1}{b_2}$ and $\Gamma\sgv\binbop{x}{b_1}{b_2}$. The direction from left to right is obvious the reverse direction follows from Part $i$.
 
The reformulated proposition can then be shown by induction on the definition of $\Gamma\sgv a\gincl b$: 
If $\Gamma\sgv a\incl b$, $b\eqv\binbop{x}{b_1}{b_2}$, and $\Gamma\sgv\binbop{x}{b_1}{b_2}$
then by Law \ref{cr} $\Gamma\sgv a\incl\binbop{x}{b_1}{b_2}$ and the proposition follows \ref{sub.incl.prop}($iii$).
Otherwise assume $\Gamma\sgv a\gincl d$, $\Gamma\sgv d\gincl b$ for some $d$, and $b\eqv\binbop{x}{b_1}{b_2}$ and $\Gamma\sgv\binbop{x}{b_1}{b_2}$.
By inductive hypothesis of $\Gamma\sgv d\gincl b$: $d\eqv\binbop{x}{b_1}{d_2}$ where $\Gamma\sgv\binbop{x}{b_1}{d_2}$ and $\Gamma,x:b_1\sgv d_2\gincl b_2$ for some $d_2$.
Hence by inductive hypothesis of $\Gamma\sgv a\gincl d$: $a\eqv\binbop{x}{b_1}{a_2}$ where $\Gamma\sgv\binbop{x}{b_1}{a_2}$ and $\Gamma,x:b_1\sgv a_2\gincl d_2$ for some $a_2$.
By definition of $\lambda^*$-inclusion $\Gamma,x:b_1\sgv a_2\gincl b_2$ which implies the proposition.

For Part $iv$ note that analogously to Part $iii$  the proposition can be reformulated equivalently as follows: If $\Gamma\sgv a\gincl b$ where $\Gamma\sgv a$, $a\eqv\binbop{x}{a_1}{a_2}$, and $\Gamma\sgv\binbop{x}{a_1}{a_2}$  for some $a$, $a_1$, $a_2$ then $\ldots$ (as before).
The argument runs analogously to Part $iii$.
The reformulated proposition can then be shown by induction on the definition of $\Gamma\sgv a\gincl b$:
If $\Gamma\sgv a\incl b$ and $a\eqv\binbop{x}{a_1}{a_2}$ for some $a_1$, $a_2$ then by Law \ref{sub.incl.prop}($i$) $\Gamma\sgv\binbop{x}{a_1}{a_2}\incl b$ and, since $\Gamma\sgv\binbop{x}{a_1}{a_2}$, the proposition follows from \ref{sub.incl.prop}($iv$).
Otherwise assume $\Gamma\sgv a\gincl d$ and $\Gamma\sgv d\gincl b$ for some $d$ where $\Gamma\sgv a$, $a\eqv\binbop{x}{a_1}{a_2}$ and $\Gamma\sgv\binbop{x}{a_1}{a_2}$.
By inductive hypothesis of $\Gamma\sgv a\gincl d$ there are two cases:
\begin{itemize}
\item
$d\eqv\binbop{x}{a_1}{d_2}$ and $\Gamma,x:a_1\sgv a_2\gincl d_2$ for some $d_2$.
Since $\Gamma\sgv d$ we can apply the inductive hypothesis of $\Gamma\sgv d\gincl b$ and obtain
either $b\eqv\binbop{x}{a_1}{b_2}$ and $\Gamma,x:a_1\sgv d_2\gincl b_2$ for some $d_2$,
or we obtain $b\eqv\any{m}$ for some $m$  where $\Gamma\sgv\binbop{x}{a_1}{d_2}\gincl\any{m}$. 
In the first case by transitivity of $\lambda^*$-inclusion we have $\Gamma,x:a_1\sgv a_2\gincl b_2$ which implies the proposition.
In the second case we have $b\eqv\any{m}$ for some $m$  where $\Gamma\sgv\binbop{x}{a_1}{d_2}\gincl\any{m}$ and 
since an obvious argument on $\lambda^*$-inclusion implies $\Gamma\sgv\binbop{x}{a_1}{a_2}\gincl\binbop{x}{a_1}{d_2}$ 
by transitivity of $\lambda^*$-inclusion $\Gamma\sgv\binbop{x}{a_1}{a_2}\gincl\any{m}$ which implies the proposition.
\item
$d\eqv\any{k}$ for some $k$ where $\Gamma\sgv\binbop{x}{a_1}{a_2}\gincl\any{k}$.
Since $\Gamma\sgv d\gincl b$ by Part $ii$ we know that $b\eqv\any{m}$ for some $m$ where $\Gamma\sgv\any{k}\gincl\any{m}$.
By transitivity of $\lambda^*$-inclusion $\Gamma\sgv\binbop{x}{b}{c}\gincl\any{m}$ which implies the proposition.
\end{itemize}

Part $v$ can be argued similar to Part $iii$ using \ref{sub.incl.prop}($v$).
Part $vi$ can be argued similar to Part $iv$ using \ref{sub.incl.prop}($vi$).
Part $vii$ can be shown by induction on the definition of $\lambda^*$-inclusion.
If $\Gamma\sgv b\incl c$ the proposition follows from \ref{sub.incl.prop}($vii$).
Otherwise assume $\Gamma\sgv b\gincl d$ and $\Gamma\sgv d\gincl c$ for some $d$.
By inductive assumption we obtain $\Gamma\sgv a:d$ and then $\Gamma\sgv a:c$. \qedhere
\end{proof}
\noindent
Due to the new rule \sinc, Law~\ref{type.decomp} (typing decomposition) needs adaptations in its statements and corresponding modifications of its proofs. 
\begin{law}[Typing decomposition - extended by inclusion]
\label{sub.type.decomp}
For all $\Gamma$, $x$, $n$, $a_1$, $\ldots$, $a_4$ $b$, $b_1$, $b_2$ the following statements are true: 
\begin{itemize}
\item[$i$:]
$\Gamma\sgv\any{n}:b$ implies $b\eqv\any{m}$ for some $m$.
\item[$ii$:]
$\Gamma\sgv x:b$ implies $x\in\dom(\Gamma)$ and $\Gamma\gv\Gamma(x)\gincl b$. 
\item[$iii$:]
$\Gamma\sgv\binbop{x}{a_1}{a_2}:b$ implies 
$\Gamma,x:a_1\sgv a_2:c$ and $\Gamma\sgv[x:a_1]c\gincl b$ for some $c$. 
\item[$iv$:]
$\Gamma\sgv\binbop{x}{a_1}{a_2}:[x:b_1]b_2$ implies $a_1\eqv b_1$ and $\Gamma,x:a_1\sgv a_2:b_2$.
\item[$v$:]
$\Gamma\sgv (a_1\,a_2):b$ implies $\Gamma\sgv a_1:[y:c_1]c_2$, $\Gamma\sgv a_2:c_1$, and $\Gamma\sgv c_2\gsub{y}{a_2}\gincl b$ for some $y$, $c_1$, $c_2$.  
\item[$vi$:]
$\Gamma\sgv\prsumop{a_1}{a_2}:b$ implies $\Gamma\sgv a_1:c_1$ and $\Gamma\sgv a_2:c_2$ for some $c_1$, $c_2$ and $\Gamma\sgv[c_1,c_2]\gincl b$. 
$\Gamma\sgv\prsumop{a_1}{a_2}:[b_1,b_2]$ implies $\Gamma\sgv a_1:b_1$ and $\Gamma\sgv a_2:b_2$.
\item[$vii$:]
$\Gamma\sgv\prdef{x}{a_1^{a_2}}{a_3}{a_4}:b$ implies $\Gamma\sgv a_1:a_2$, $\Gamma\sgv a_3:a_4\gsub{x}{a_1}$, and $(\Gamma,x:a_2)\sgv a_4$  where $\Gamma\sgv [x!a_2]a_3\gincl b$. 
$\Gamma\sgv\prdef{x}{a_1^{a_2}}{a_3}{a_4}:[x!b_1]b_2$ implies $a_2\eqv b_1$, $a_4\eqv b_2$, $\Gamma\sgv a_1:a_2$, $\Gamma\sgv a_3:a_4\gsub{x}{a_1}$, and $(\Gamma,x:a_2)\sgv a_4$.  
\item[$viii$:]
$\Gamma\sgv\pleft{a_1}:b$ implies either $\Gamma\sgv a_1:[y!c_1]c_2$ and $\Gamma\sgv c_1\gincl b$ or $\Gamma\sgv a_1:[c_1,c_2]$ and $\Gamma\sgv c_1\gincl b$  for some $y$, $c_1$, $c_2$. 
\item[$ix$:]
$\Gamma\sgv\pright{a_1}:b$ implies either $\Gamma\sgv a_1:[c_1,c_2]$ and $\Gamma\sgv c_2\gincl b$ 
or $\Gamma\sgv a_1:[y!c_1]c_2$ and $\Gamma\sgv c_2\gsub{y}{\pleft{a_1}}\gincl b$ for some $y$, $c_1$, $c_2$.
\item[$x$:]
$\Gamma\sgv\injl{a_1}{a_2}:b$ implies $\Gamma\sgv a_2$ and $\Gamma\sgv a_1:c$ for some $c$ and $\Gamma\sgv[c+a_2]\gincl b$. 
$\Gamma\sgv\injl{a_1}{a_2}:[b_1+b_2]$ implies $\Gamma\sgv a_1:b_1$.
Analogous properties are valid for $\Gamma\sgv\injr{a_1}{a_2}:b$.
\item[$xi$:]
$\Gamma\sgv\case{a_1}{a_2}:b$ implies $\Gamma\sgv a_1:[y:c_1]c$, $\Gamma\sgv a_2:[y:c_2]c$, and $\Gamma\sgv c$ for some $y$, $c_1$, $c_2$, $c$ where $\Gamma\sgv[y:[c_1+c_2]]c\gincl b$.
\item[$xii$:]
$\Gamma\sgv\myneg a_1:b$ implies $\Gamma\sgv a_1:b$. 
\end{itemize}
\end{law}
\noindent
In comparison to Law \ref{type.decomp} we have dropped Part $vi$ which will be shown at a later point (Law \ref{sub.any}($iii$)).
Furthermore we have added a new statement (Part $ii$) about the type of variables.
\begin{proof}
In these proofs we repeatedly make use of the fact that most typing rules of \dcalcb\ type a unique operator, \eg\ a variable, an abstraction, an application, $\ldots$.
The only exceptions are the rules \weak, \conv, and \sinc, which modify the type of an given operator.
Hence in an induction argument about a statement $\Gamma\sgv a:b$, where $a$ is an concrete operator of \dcalcb, it is sufficient to consider the unique introduction rule of the operator followed by use of the rules  \weak, \conv, and \sinc.
\begin{itemize}
\item[$i$:]
$\Gamma\sgv\any{n}:b$  can only be established by the introduction rules \ax, followed by a usage of the rules \weak, \conv, and \sinc.
The rule \ax\ establishes $\Gamma\sgv b\eqv\any{m}$ for some $m$, the rules \weak\ and \conv\ obviously preserve it.
For the rule \sinc\ the property is preserved thanks to Law \ref{sub.leq.prop}($i$).
\item[$ii$:]
$\Gamma\sgv x:b$ can only be established by the use of the introduction rule \mystart, followed by a usage of the rules \weak, \conv, and \sinc.
Rule \mystart\ obviously implies $x\in\dom(\Gamma)$ and $\Gamma\sgv\Gamma(x)\leq b$. 
The other rules obviously preserve these properties.
\item[$iii$:]
Proof by induction on definition of typing.
$\Gamma\sgv\binbop{x}{a_1}{a_2}:b$ can only be established by the use of either one of the introduction rules \absu~and \abse, followed by a usage of the rules \weak, \conv, and \sinc.
Use of \absu\ or \abse\ implies that $\Gamma,x:a_1\sgv a_2:c$ for some $c$. The property follows with $b=[x:a_1]c$ since
by Law \ref{valid.type} we know that $\Gamma\sgv b$ which implies $\Gamma\sgv b\gincl b$. 

Three rules are remaining:
\begin{meditemize}
\item[\weak:]
We have $\Gamma=(\Gamma',y:d)$, for some $\Gamma'$, $y$, and $d$ where $\Gamma'\sgv\binbop{x}{a_1}{a_2}:b$ as well as $\Gamma'\sgv d$.
By inductive hypothesis $\Gamma',x:a_1\sgv a_2:c$ where $\Gamma'\sgv[x:a_1]c\gincl b$ for some $c$.
By Law \ref{sub.type.weak} $\Gamma\sgv[x:a_1]c\gincl b$.
\item[\conv:]
We have $\Gamma\sgv\binbop{x}{a_1}{a_2}:d$ for some $d$ where $d\eqv b$ and $\Gamma\sgv b$.
By inductive hypothesis $\Gamma,x:a_1\sgv a_2:c$ for some $c$ where $\Gamma\sgv[x:a_1]c\gincl d$.  
Since obviously $\Gamma\sgv d\gincl b$, by definition of $\lambda^*$-inclusion we obtain $\Gamma\sgv[x:a_1]c\gincl b$.
\item[\sinc:]
We have $\Gamma\sgv\binbop{x}{a_1}{a_2}:d$ for some $d$ where $\Gamma\sgv d\leq b$.
By inductive hypothesis $(\Gamma,x:a_1)\sgv a_2:c$ for some $c$ where $\Gamma\sgv[x:a_1]c\gincl d$.
Since obviously $\Gamma\sgv d\gincl b$, by definition of $\lambda^*$-inclusion we obtain $\Gamma\sgv[x:a_1]c\gincl b$.
\end{meditemize}
\item[$iv$:]
Assume $\Gamma\sgv\binbop{x}{a_1}{a_2}:[x:b_1]b_2$.
By Part $iii$ we obtain $\Gamma,x:a_1\sgv a_2:d_2$ and $\Gamma\sgv[x:a_1]d_2\gincl[x:b_1]b_2$ for some $d_2$.
By Law \ref{sub.gincl.prop}($iii$) $[x:a_1]d_2\eqv[x:b_1]c_2$ where $\Gamma,x:b_1\sgv c_2\gincl b_2$ for some $c_2$.
Since $a_1\eqv b_1$ and by Law \ref{val.decomp} $\Gamma\sgv a_1$ by Law \ref{sub.eqv.env} $\Gamma,x:a_1\sgv c_2\gincl b_2$.
ince obviously $d_2\eqv c_2$ by Law \ref{sub.gincl.prop}($i$) $\Gamma,x:a_1\sgv d_2\gincl b_2$.
By Law \ref{sub.gincl.prop}($vii$) we have $\Gamma,x:a_1\sgv a_2:b_2$.
\item[$v$:]
Proof by induction on the definition of typing.
According to the remark at the beginning, only the rules \weak, \conv, \sinc, and the rule \appl~are relevant.
For the rule \appl\ we have $b=c_2\gsub{y}{a_2}$ where $\Gamma\sgv a_1:[y:c_1]c_2$ and $\Gamma\sgv a_2:c_1$ for some $y$, $c_1$, $c_2$. 
By Law \ref{valid.type} we know that $\Gamma\sgv[y:c_1]c_2$ and by Part $iii$ we know that
 $\Gamma,y:c_1\sgv c_2$. By Law \ref{sub.type.sub} $\Gamma\sgv b$ and therefore $\Gamma\sgv b\gincl b$.

Three other rules may follow.
\begin{meditemize}
\item[\weak:]
We have $\Gamma,z:d\sgv(a_1\,a_2):b$ where $\Gamma\sgv(a_1\,a_2):b$ and $\Gamma\sgv d$ for some $z$, $d$.
By inductive hypothesis $\Gamma\sgv a_1:[y:c_1]c_2$ and $\Gamma\sgv a_2:c_1$ for some $c_1$, $c_2$ where $\Gamma\sgv c_2\gsub{y}{a_2}\gincl b$.
By Law \ref{sub.type.weak} we obtain $\Gamma,z:d\sgv c_2\gsub{y}{a_2}\gincl b$.
\item[\conv:]
We have $\Gamma\sgv(a_1\,a_2):d$ where $d\eqv b$ and $\Gamma\sgv b$.
By inductive hypothesis $\Gamma\sgv a_1:[y:c_1]c_2$ and $\Gamma\sgv a_2:c_1$ for some $c_1$, $c_2$  where $\Gamma\sgv c_2\gsub{y}{a_2}\gincl d$.
Since obviously $\Gamma\sgv d\gincl b$, by transitivity of $\lambda^*$-inclusion we obtain $\Gamma\sgv c_2\gsub{y}{a_2}\gincl b$.
\item[\sinc:]
We have $\Gamma\sgv(a_1\,a_2):d$ where $\Gamma\sgv d\leq b$.
By inductive hypothesis $\Gamma\sgv a_1:[y:c_1]c_2$ and $\Gamma\sgv a_2:c_1$ for some $c_1$, $c_2$ where $\Gamma\sgv c_2\gsub{y}{a_2}\gincl d$.
Since obviously $\Gamma\sgv d\gincl b$, by transitivity of $\lambda^*$-inclusion we obtain $\Gamma\sgv c_2\gsub{y}{a_2}\gincl b$.
\end{meditemize}
\item[$vi$:]
The first property is shown similar to Part $iii$, replacing the uses of \absu~or \abse~by \bprod~or \bsum.
The second property is shown similar to Part $iv$, replacing the use of Law \ref{sub.gincl.prop}($iii$) by Law \ref{sub.gincl.prop}($v$).
\item[$vii$:]
The first property is shown similar to Part $iii$, replacing the uses of \absu~or \abse~by \pdef.
The second property is shown similar to Part $iv$.
\item[$viii$:]
Similar to Part $v$, using \prl\ or \chin\ instead of \appl.
\item[$ix$:]
Similar to Part $v$, using \prr\ or \chba\ instead of \appl.
\item[$x$:]
The first property is shown similar to Part $iii$, replacing the uses of \absu~or \abse~by \injll~or \injlr.
The second property is shown similar to Part $iv$, replacing the use of Law \ref{sub.gincl.prop}($iii$) by Law \ref{sub.gincl.prop}($v$).
\item[$xi$:]
Similar to Part $iii$ and $iv$, replacing the uses of \absu~or \abse~by \cased.
\item[$xii$:]
$\Gamma\sgv\myneg a_1:b$  can only be established by the introduction rules \negate, followed by a usage of the rules \weak, \conv, and \sinc.
The rule \negate\ establishes $\Gamma\sgv a_1:b$, the other rules obviously preserve it.  
\qedhere
\end{itemize}
\end{proof}
\noindent
Due rule \sinc, Law~\ref{start.confl} (start property) is not valid anymore
\subsection{Closure properties of validity}%
\label{sub.closure}
The property~\ref{strd.type} (preservation of types under reduction step) has to be extended by the inclusion relations as follows:
\begin{law}[Preservation of types under reduction step - extended by inclusion]
\label{sub.strd.type}
For all $\Gamma,n,a,b,c$:
\begin{itemize}
\item[$i$:]
$\Gamma\sgv a:c$ and $a\srd b$ imply $\Gamma\sgv b:c$.
\item[$ii$:]
$\Gamma\sgv a<\any{n}$ and $a\srd b$ imply $\Gamma\sgv b<\any{n}$.
\item[$iii$:]
$\Gamma\sgv a\leq c$ and $a\srd b$ imply $\Gamma\sgv b\leq c$.
\item[$iv$:]
$\Gamma\sgv a\leq b$ and $b\srd c$ imply $\Gamma\sgv a\leq c$.
\end{itemize}
\end{law}
\begin{proof} $\;$

\noindent {\bf Part} $i$:
In the proof of this Part several cases in the proof of \ref{strd.type} have to adapted.
\begin{hugeitemize}
\item[$\beta_1$:] 
We have $a=([x:a_1]a_2\:a_3)$, $b=a_2\gsub{x}{a_3}$, and $\Gamma\sgv([x:a_1]a_2\:a_3):c$.
By Law~\ref{sub.type.decomp}($v$) we know that $\Gamma\sgv[x:a_1]a_2:[x:c_1]c_2$ and $\Gamma\sgv a_3:c_1$ for some $c_1$ and $c_2$.
By Law~\ref{sub.type.decomp}($iv$) this implies $a_1\eqv c_1$ and $(\Gamma,x:a_1)\sgv a_2:c_2$.
By Law~\ref{type.xtrct} we know that $\Gamma\sgv a_1$ and therefore by rule \conv~it follows that $\Gamma\sgv a_3:a_1$ and 
further by Law~\ref{sub.type.sub} we know that $\Gamma\sgv a_2\gsub{x}{a_3}:c_2\gsub{x}{a_3}$. 
By Law~\ref{sub.type.decomp}($v$) $\Gamma\sgv c_2\gsub{x}{a_3}\gincl c$.
By Law \ref{sub.gincl.prop}($vii$) $\Gamma\sgv a_2\gsub{x}{a_3}:c$
\item[$\beta_2$:] Similar to $\beta_1$. 
\item[$\beta_3$:] 
We have $a=(\case{a_1}{a_2}\,\injl{a_3}{a_4})$, $b=(a_1\,a_3)$, and $\Gamma\sgv a:c$. 
By Law~\ref{sub.type.decomp}($v$) we know that 
$\Gamma\sgv\case{a_1}{a_2}:[x:c_1]c_2$, $\Gamma\sgv\injl{a_3}{a_4}:c_1$, and $\Gamma\sgv d\gsub{x}{\injl{a_3}{a_4}}\gincl c$ for some $c_1$, $c_2$.
By Law~\ref{sub.type.decomp}($xi$) we know that $c_1\eqv[b_1+b_2]$ for some $b_1$, $b_2$ where $\Gamma\sgv a_1:[x:b_1]c_2$ and $\Gamma\sgv c_2$. 
Obviously $c_2\gsub{x}{\injl{a_3}{a_4}}=c_2$.
Since obviously $\Gamma\sgv[b_1+b_2]$ by rule \conv\ we obtain $\Gamma\sgv\injl{a_3}{a_4}:[b_1+b_2]$ hence by Law \ref{sub.type.decomp}($x$), we know that $\Gamma\sgv a_3:b_1$.
Hence by rule \appl\ we obtain $\Gamma\sgv(a_1\,a_3):c_2\gsub{x}{a_3}=c_2$.
By Law \ref{sub.gincl.prop}($vii$) $\Gamma\sgv(a_1\,a_3):c$.
\item[$\beta_4$:]
Similar to $\beta_3$. 
\item[$\pi_1$:] 
We have $a=\pleft{\prdef{x}{a_1^{a_2}}{a_3}{a_4}}$, $b=a_1$, and $\Gamma\sgv a:c$.
By Law~\ref{sub.type.decomp}($viii$), $\Gamma\sgv a:c$ 
implies either $\Gamma\sgv\prdef{x}{a_1^{a_2}}{a_3}{a_4}:[x!c_1]c_2$ and $\Gamma\sgv c_1\gincl c$
or $\Gamma\sgv\prdef{x}{a_1^{a_2}}{a_3}{a_4}:[c_1,c_2]$ and $\Gamma\sgv c_1\gincl c$ for some $c_1$, $c_2$.
If $\Gamma\sgv\prdef{x}{a_1^{a_2}}{a_3}{a_4}:[c_1,c_2]$ then by Law~\ref{sub.type.decomp}($vii$)
$\Gamma\sgv a_1:d$, $\Gamma\sgv a_3:a_4\gsub{x}{a_1}$, and $(\Gamma,x:d)\sgv a_4$ for some $d$ where $\Gamma\sgv[x!d]a_4\gincl[c_1,c_2]$.
However, since $[x!d]a_4\neqv[c_1,c_2]$ due to Law~\ref{sub.gincl.prop}($v$) this cannot be the case.
Hence we must have $\Gamma\sgv\prdef{x}{a_1^{a_2}}{a_3}{a_4}:[x!c_1]c_2$ and then by Law~\ref{sub.type.decomp}($vii$)
$\Gamma\sgv a_1:b_1$ where $b_1\eqv c_1$. 
Using Laws~\ref{valid.type} and \ref{type.xtrct} we can infer that $\Gamma\sgv c_1, c$ hence by rule \conv~$\Gamma\sgv a_1:c_1$.
By Law \ref{sub.gincl.prop}($vii$) this implies $\Gamma\sgv a_1:c$.
\item[$\pi_2$:] 
Similar to $\pi_1$ using Law \ref{sub.type.decomp}($ix$, second case) instead of \ref{sub.type.decomp}($viii$). 
\item[$\pi_3$:]
Similar to $\pi_1$ using Law \ref{sub.type.decomp}($ix$, first case) instead of \ref{sub.type.decomp}($viii$). 
\item[$\pi_4$,]$\pi_5$, $\pi_6$: Similar to $\pi_3$. 
\item[$\nu_1$:] 
$a=\myneg\myneg a_1$, $b=a_1$, and $\Gamma\sgv a:c$ for some $c$.
Laws~\ref{valid.type} we know that $\Gamma\sgv c$.
By Law~\ref{sub.type.decomp}($xii$) applied twice $\Gamma\sgv b:c$.
\item[$\nu_2$:]
$a=\myneg[a_1,a_2]$, $b=[\myneg a_1+\myneg a_2]$, and $\Gamma\sgv a:c$ for some $c$.
Law~\ref{sub.type.decomp}($xii$) $\Gamma\sgv[a_1,a_2]:c$, hence by Law \ref{sub.type.decomp}($vi$)
$\Gamma\sgv a_i:c_i$ for some $c_1$, $c_2$ where $\Gamma\sgv[c_1,c_2]\gincl c$.
By rules of typing we know that $\Gamma\sgv[\myneg a_1+\myneg a_2]:[c_1,c_2]$ which implies the proposition.
\item[$\nu_3$:]
Similar to $\nu_2$. 
\item[$\nu_4$:]
Similar to $\nu_1$ using Law~\ref{sub.type.decomp}($xii$) and \ref{sub.type.decomp}($iii$).
\item[$\nu_5$:] Similar to $\nu_4$. 
\item[$\nu_6$]
$\nu_7$, $\nu_8$, $\nu_9$, $\nu_{10}$: Similar to $\nu_1$. 
\item[$\binbop{x}{\_}{\_}_1$:]
We have $a=\binbop{x}{a_1}{a_3}$, $b=\binbop{x}{a_2}{a_3}$ where $a_1\srd a_2$, and $\Gamma\sgv\binbop{x}{a_1}{a_3}:c$.
By Law~\ref{sub.type.decomp}($iii$) we know that $(\Gamma,x:a_1)\sgv a_3:c_2$ and $\Gamma\sgv[x:a_1]c_2\gincl c$ for some $c_2$.
By Law~\ref{type.xtrct} this implies that $\Gamma\sgv a_1$ and therefore by inductive hypothesis we know that $\Gamma\sgv a_2$. 
Hence we can apply Law~\ref{sub.eqv.env} to obtain $(\Gamma,x:a_2)\sgv a_3:c_2$.
By definition of typing $\Gamma\sgv\binbop{x}{a_2}{a_3}:[x:a_2]c_2$. 
By Law~\ref{valid.type} we know that $\Gamma\sgv c$, $\Gamma\sgv[x:a_2]c_2$, and $\Gamma\sgv[x:a_1]c_2$.
Since $[x:a_2]c_2\eqv[x:a_1]c_2$, by definition of $\lambda^*$-inclusion we know that $\Gamma\sgv[x:a_2]c_2\gincl[x:a_1]c_2$ and therefore we obtain $\Gamma\sgv[x:a_2]c_2\gincl c$.
Hence by Law \ref{sub.gincl.prop}($vii$) we know that $\Gamma\sgv b:c$.
\item[$\binbop{x}{\_}{\_}_2$:]
We have $a=\binbop{x}{a_1}{a_2}$, $b=\binbop{x}{a_1}{a_3}$ where $a_2\srd a_3$ and $\Gamma\sgv\binbop{x}{a_1}{a_2}:c$. 
By Law~\ref{valid.type} we know that $\Gamma\sgv c$.
By Law~\ref{sub.type.decomp}($iii$) we know that $(\Gamma,x:a_1)\sgv a_2:c_2$ and $\Gamma\sgv[x:a_1]c_2\gincl c$ for some $c_2$.
This implies that $(\Gamma,x:a_1)\sgv a_2$ and therefore by inductive hypothesis we know that $(\Gamma,x:a_1)\sgv a_3:c_2$. 
By definition of typing $\Gamma\sgv \binbop{x}{a_1}{a_3}:[x:a_1]c_2$. 
Hence by Law \ref{sub.gincl.prop}($vii$) we know that $\Gamma\sgv b:c$.
\item[${\prdef{x}{\_^{\_}}{\_}{\_}}_i$:] where $i=1,2,3,4$. 
Similar to the cases $\binbop{x}{\_}{\_}_1$ and $\binbop{x}{\_}{\_}_2$ using Law~\ref{sub.type.decomp}($vii$).
\item[${(\_\,\_)}_1$:]
We have $a=(a_1\,a_3)$, $b=(a_2\,a_3)$ where $a_1\srd a_2$ and $\Gamma\sgv(a_2\,a_3):c$.
By Law~\ref{valid.type} we know that $\Gamma\sgv c$.
By Law~\ref{sub.type.decomp}($v$) we know that $\Gamma\sgv a_1:[x:b_1]b_2$ $\Gamma\sgv a_3:b_1$ for some $b_1$, $b_2$.
By inductive hypothesis we know that $\Gamma\sgv a_2:[x:b_1]b_2$. 
Hence $\Gamma\sgv(a_2\:a_3):b_2\gsub{x}{a_3}$.
Further by Law~\ref{sub.type.decomp}($v$) we know that $\Gamma\sgv b_2\gsub{x}{a_3}\gincl c$.
Hence by Law \ref{sub.gincl.prop}($vii$) we know that $\Gamma\sgv b:c$.
\item[${(\_\,\_)}_2$:]
We have $a=(a_1\,a_2)$, $b=(a_1\,a_3)$ where $a_2\srd a_3$ and $\Gamma\sgv(a_2\,a_3):c$.
By Law~\ref{valid.type} we know that $\Gamma\sgv c$.
By Law~\ref{sub.type.decomp}($v$) we know that $\Gamma\sgv a_1:[x:b_1]b_2$ $\Gamma\sgv a_3:b_1$ for some $b_1$, $b_2$.
By inductive hypothesis we know that $\Gamma\sgv a_3:b_1$. 
Hence $\Gamma\sgv(a_1\:a_3):b_2\gsub{x}{a_3}$.
Further by Law~\ref{sub.type.decomp}($v$) we know that $\Gamma\sgv b_2\gsub{x}{a_3}\gincl c$.
Hence by Law \ref{sub.gincl.prop}($vii$) we know that $\Gamma\sgv b:c$.
\item[${(\_\,\_)}_2$:]
Similar to case ${(\_\,\_)}_1$.
\item[$(\pleft{\_})_1$:]
Similar to case ${(\_\,\_)}_1$ using Law~\ref{sub.type.decomp}($viii$).
\item[$(\pright{\_})_1$:]
Similar to case ${(\_\,\_)}_1$ using Law~\ref{sub.type.decomp}($ix$).
\item[${[\_,\_]}_1$:]
Similar to case $\binbop{x}{\_}{\_}_2$ using Law~\ref{sub.type.decomp}($vi$).
\item[${[\_,\_]}_2$:]
Similar to case ${[\_,\_]}_1$.
\item[${[\_+\_]}_i$:]
$i=1,2$, similar to case ${[\_,\_]}_1$.
\item[${\injl{\_}{\_}}_1$:]
Similar to case $\binbop{x}{\_}{\_}_2$ using Law~\ref{sub.type.decomp}($x$).
\item[${\injl{\_}{\_}}_2$:]
$i=1,2$, similar to case ${\injl{\_}{\_}}_1$:.
\item[${\injr{\_}{\_}}_i$:]
Similar to case ${\injl{\_}{\_}}_i$, $i=1,2$.
\item[${\case{\_}{\_}}_1$:]
Similar to case $\binbop{x}{\_}{\_}_2$ using Law~\ref{sub.type.decomp}($xi$).
\item[${\case{\_}{\_}}_2$:]
Similar to case $(\case{\_}{\_}\_)_1$.
\item[$(\myneg\_)_1$:]
Similar to case $\binbop{x}{\_}{\_}_2$ using Law~\ref{sub.type.decomp}($vii$).
\qedhere
\end{hugeitemize}
\noindent {\bf Part} $ii$:  The proof is by induction on the definition of $\any{}$-inclusion
\begin{meditemize}
\item[\sstart:] 
Obvious since $a=b$.
\item[\sstyp:]
We have $\Gamma\sgv a:c$ and  $\Gamma\sgv c<\any{n}$. 
By Part $i$ $\Gamma\sgv b:c$ and therefore $\Gamma\sgv b<\any{n}$. 
\item[\ssabsu:]
We have $a=\binbop{y}{a_1}{a_2}$ where $\Gamma\sgv a_1<\any{n}$ and $\Gamma,y:a_1\sgv a_2:\any{n}$ and $\binbop{y}{a_1}{a_2}\srd b$.
By Law~\ref{rd.decomp}($ii$) we know that $b=\binbop{x}{b_1}{b_2}$ where $a_1\rd b_1$ and $a_2\rd b_2$. 
By inductive hypothesis $\Gamma\sgv b_1<\any{n}$ as well as $\Gamma,x:a_1\sgv b_2<\any{n}$ and by Law~\ref{sub.eqv.env}($i$) we know that 
$\Gamma,x:b_1\sgv b_2<\any{n}$ hence  $\Gamma\sgv b<\any{n}$.  
\item[\ssbprod,]\ssinjl, \ssinjr, and \sspdef: similar to \ssabsu.
\end{meditemize}
\noindent {\bf Part} $iii$:  The proof is by induction on the definition of inclusion
\begin{meditemize}
\item[\srefl:]  
Follows directly from Part $i$.
\item[\sembed:]
Follows directly from Part $ii$.
\item[\sabs:]
We have $a=\binbop{y}{a_1}{a_2}$ and $c=\binbop{y}{c_1}{c_2}$ where $\Gamma\sgv a_1\leq c_1$ and $\Gamma,y:a_1\sgv a_2\leq c_2$ and $\binbop{y}{a_1}{a_2}\srd b$.
By Law~\ref{rd.decomp}($ii$) we know that $b=\binbop{x}{b_1}{b_2}$ where $a_1\srd b_1$ and $a_2\srd b_2$.
By inductive hypothesis $\Gamma\sgv b_1\leq c_1$ as well as $\Gamma,x:a_1\sgv b_2\leq c_2$ and, by Law~\ref{sub.eqv.env}($ii$), we know that 
$\Gamma,x:b_1\sgv b_2\leq c_2$ therefore in all of the above cases we obtain $\Gamma\sgv b\leq c$.  
\item[\sbprod:]
Similar to previous case.
\end{meditemize}

\noindent {\bf Part} $iv$: :  The proof is by induction on the definition of inclusion
\begin{meditemize}
\item[\srefl:]  
Follows directly from Part $i$.
\item[\sembed:]
Follows directly from Part $ii$.
\item[\sabs:]
We have $a=\binbop{y}{a_1}{a_2}$ and $b=\binbop{y}{b_1}{b_2}$ where $\Gamma\sgv a_1\leq b_1$ and $\Gamma,y:a_1\sgv a_2\leq b_2$ and $\binbop{y}{b_1}{b_2}\srd c$.
By Law~\ref{rd.decomp}($ii$) we know that $c=\binbop{x}{c_1}{c_2}$ where $b_1\srd c_1$ and $b_2\srd c_2$.
By inductive hypothesis $\Gamma\sgv a_1\leq c_1$ as well as $\Gamma,x:a_1\sgv a_2\leq c_2$ and therefore in all of the above cases we obtain $\Gamma\sgv a\leq c$.  
\item[\sbprod:]
Similar to previous case.\qedhere
\end{meditemize}
\end{proof}
\noindent
Law~\ref{rd.type} (subject reduction) has to be generalized analogously.
\begin{law}[Subject reduction: Types are preserved under reduction - extended by inclusion]%
\index{reduction!subject - with inclusion}
\label{sub.rd.type}
For all $\Gamma,n,a,b,c,d$:
\begin{itemize}
\item[$i$:]
$a\rd b$ and $\Gamma\sgv a:c$ imply that $\Gamma\sgv b:c$. 
\item[$ii$:]
$a\rd b$ and $\Gamma\sgv a<\any{n}$ imply that $\Gamma\sgv b<\any{n}$.  
\item[$iii$:]
$a\rd b$, $c\rd d$  and $\Gamma\sgv a\leq c$ imply that $\Gamma\sgv b\leq d$.
\end{itemize}
\end{law}
\begin{proof}
Obvious extension of Law \ref{sub.strd.type} to the new cases.
\end{proof}
\noindent
Law~\ref{type.rd}(Elements are preserved under type reduction) can remain unchanged.
As a first consequence of Law \ref{sub.rd.type} we can generalize and strengthen Law \ref{sub.type.decomp}($i$). 
\begin{law}[Typing and inclusion of $\any{}$]
\label{sub.any}
For all $\Gamma$, $b$, $n$:
\begin{itemize}
\item[$i:$]
$\Gamma\sgv\any{n}<\any{m}$ for some $m$ implies $n<m$. 
\item[$ii:$]
$\Gamma\sgv\any{n}\leq b$ implies $b\eqv\any{m}$ for some $m$ where $n\leq m$. 
\item[$iii:$]
$\Gamma\sgv\any{n}:b$ implies $b\eqv\any{m}$ for some $m$ where $n\leq m$. 
\end{itemize}
\end{law}
\begin{proof}
The proof is by simultaneous induction on all three relations.
For Part $i$, we only need to consider the rules \sstart\ and \sstyp:
\begin{meditemize}
\item[\sstart:] 
Obvious.
\item[\sstyp:] 
We have $\Gamma\sgv\any{n}:b$ and $\Gamma\sgv b<\any{m}$ for some $b$.
Note that the inductive hypothesis can be applied to the two relations for any such $b$.
By inductive hypothesis (for $b$) applied to $\Gamma\sgv\any{n}:b$ we know that $b\eqv\any{k}$ for some $k$ with $n\leq k$.
Due to Law \ref{cr} we know that $b\rd\any{k}$ hence by Law~\ref{sub.rd.type} we know that $\Gamma\sgv\any{n}:\any{k}$ and $\Gamma\sgv\any{k}<\any{m}$.
Hence by inductive hypothesis (for $\any{k}$) applied to $\Gamma\sgv\any{k}<\any{m}$ we obtain $k<m$.
Hence $n<m$.
\end{meditemize}
For Part $ii$, we only need to consider the rules \srefl\ and \sembed.
\begin{meditemize}
\item[\srefl:] 
Obvious.
\item[\sembed:] 
We have $b=\any{m}$ and  $\Gamma\sgv\any{n}<\any{m}$. 
By inductive hypothesis $n<m$.
\end{meditemize}
For Part $iii$, $\Gamma\sgv \any{n}:b$ can only be established by the use of \ax\ (where $b=\any{n}$), followed by use of the rules \weak, \conv, and \sinc.
\begin{meditemize}
\item[\weak:] 
Follows from the inductive hypothesis and Law \ref{type.weak}.
\item[\conv:] 
$\Gamma\sgv \any{n}:c$, $c\eqv b$, and $\Gamma\sgv b$ for some $c$.
By inductive hypothesis $c\eqv\any{m}$ for some $m$ where $n\leq m$. 
Due to Law \ref{cr} we know that $b\eqv\any{m}$.
\item[\sinc:]
We have $\Gamma\sgv\any{n}:c$ and $\Gamma\sgv c\leq b$ for some $c$.
By inductive hypothesis (for $c$ applied to $c\eqv\any{k}$ for some $k$ where $n\leq k$. 
Due to Law \ref{cr} we know that $c\rd\any{k}$ hence by Law~\ref{sub.rd.type} we know that $\Gamma\sgv\any{n}:\any{k}$ and $\Gamma\sgv\any{k}\leq b$.
Hence by inductive hypothesis (for $\any{k}$) $b\eqv\any{m}$ for some $m$ where $k\leq m$. Hence $n\leq m$. \qedhere
\end{meditemize}
\end{proof}
\subsection{Ordering properties}
\label{sub.order}
\noindent
First we show that $\any{}$-inclusion is transitive
\begin{law}[Transitivity of $\prim$-inclusion]
\label{sub.sinc.trans}
For all $\Gamma$, $a$, $n,m$: $\Gamma\sgv a<\any{n}$ and $\Gamma\sgv\any{n}<\any{m}$ imply $\Gamma\sgv a<\any{m}$.   
\end{law}
\begin{proof}
Proof by induction on the definition of $\Gamma\sgv a<\any{n}$.
\begin{meditemize}
\item[\sstart:] 
Using laws \ref{sub.any} and \ref{sub.incl.valid} this case is obvious.
\item[\sstyp:] 
We have $\Gamma\sgv a:b$ and $\Gamma\sgv b<\any{n}$. By inductive hypothesis $\Gamma\sgv b<\any{m}$. Hence by rule \sstyp\ we know that $\Gamma\sgv a<\any{m}$.
\end{meditemize}
Analogous arguments can be made for rules \ssabsu, \ssbprod, \ssinjl, \ssinjr, and \sspdef.
\qedhere
\end{proof}
\noindent
In order to show transitivity of inclusion, we first need to show a restricted version. 
\begin{law}[Restricted transitivity of inclusion]
\label{sub.leq.top}
For all $\Gamma,x,a,b,n$ where $x\notin\dom(\Gamma)$:
If $\Gamma\sgv a\leq b$ and $\Gamma\sgv b<\any{n}$ then $\Gamma\sgv a<\any{n}$.  
\end{law}
\begin{proof}
By induction on the definition of $\Gamma\sgv a\leq b$ we show that $\Gamma\sgv a\leq b$ and $\Gamma\sgv b<\any{n}$ implies $\Gamma\sgv a<\any{n}$.
\begin{meditemize}
\item[\srefl:] 
Obvious 
\item[\sembed:] 
We have $b=\any{m}$, for some $m$, where $\Gamma\sgv a\leq\any{m}$ and $\Gamma\sgv\any{m}<\any{n}$.
If $a=\any{m}$ we are obviously done, otherwise $\Gamma\sgv a<\any{m}$ and the property follows from Law \ref{sub.sinc.trans}.
\item[\sabs:]
We have $a=\binbop{x}{a_1}{a_2}$ and $b=\binbop{x}{a_1}{b_2}$ where $\Gamma,x:a_1\sgv a_2\leq b_2$.  
From $\Gamma\sgv\binbop{x}{a_1}{b_2}<\any{n}$ according to the definition of $\prim$-inclusion we either have 
$\Gamma,x:a_1\sgv b_2<\any{n}$ or we have $\Gamma\sgv\binbop{x}{a_1}{b_2}:c$ and $\Gamma\sgv c<\any{n}$ for some $c$.

In the first case, by inductive hypothesis $\Gamma,x:a_1\sgv a_2<\any{n}$.
Hence by rule \ssabsu\ we obtain $\Gamma\sgv a<\any{n}$.

In the second case by basic properties of typing (Law~\ref{sub.type.decomp}($iii$)) we know that $\Gamma,x:a_1\sgv b_2:c_2$ and $\Gamma\sgv[x:a_1]c_2\gincl c$ for some $c_2$.
Since by Law \ref{valid.type} $\Gamma\sgv[x:a_1]c_2$ we can apply Law \ref{sub.gincl.prop}($iv$) and obtain either $c\eqv[x:a_1]d$ and $\Gamma,x:a_1\sgv c_2\gincl d$ for some $d$,
or $c\eqv\any{m}$ for some $m$  where $\Gamma\sgv[x:a_1]c_2\gincl\any{m}$.
\begin{itemize}
\item
In the first case by Law \ref{sub.gincl.prop}($vii$) $\Gamma,x:a_1\sgv b_2:d$ and therefore $\Gamma\sgv\binbop{x}{a_1}{b_2}:[x:a_1]d$.
By rule \conv\ we have $\Gamma\sgv\binbop{x}{a_1}{b_2}:c$ hence by rule \sstyp\ $\Gamma\sgv\binbop{x}{a_1}{b_2}<\any{n}$.
\item
In the second case since $\Gamma\sgv\binbop{x}{a_1}{b_2}:[x:a_1]c_2$ by Law \ref{sub.gincl.prop}($vii$) we know that $\Gamma\sgv\binbop{x}{a_1}{b_2}:\any{m}$.
Obviously $c\rd\any{m}$ hence since $\Gamma\sgv c<\any{n}$ by Law \ref{sub.rd.type}($ii$) we know that $\Gamma\sgv \any{m}<\any{n}$.
Hence by rule \sstyp\ $\Gamma\sgv\binbop{x}{a_1}{b_2}<\any{n}$.
\end{itemize}
\item[\sbprod:]
The proof is similar to case \sabs\ using Law \ref{sub.gincl.prop}($v$,$vi$). \qedhere
\end{meditemize}
\end{proof}
\begin{law}[Inclusion is a partial ordering]
\label{sub.leq.por}
For all $\Gamma$, $a$, $b$: $\Gamma\sgv a\leq b$ is reflexive on valid expressions.
Furthermore it is transitive and antisymmetric.
\end{law}
\begin{proof}
Reflexivity is trivial. 
For transitivity,
by induction on the definition of $\Gamma\sgv a\leq b$ we show that $\Gamma\sgv a\leq b$ and $\Gamma\sgv b\leq c$ implies $\Gamma\sgv a\leq c$.
\begin{meditemize}
\item[\srefl:] 
Obvious 
\item[\sembed:] 
We have $b=\any{n}$, for some $n$, and $\Gamma\sgv a<\any{n}$ as well as $\Gamma\sgv\any{n}\leq c$. There are two cases:
If $c=\any{n}$ we are done.
Otherwise $c=\any{m}$ for some $m$ where $\Gamma\sgv\any{n}<\any{m}$. The proposition follows from Law \ref{sub.leq.top}.
\item[\sabs:]
We have $a=\binbop{x}{a_1}{a_2}$ and $b=\binbop{x}{a_1}{b_2}$ where $\Gamma,x:a_1\sgv a_2\leq b_2$.  
From $\Gamma\sgv\binbop{x}{a_1}{b_2}\leq c$, by Law~\ref{sub.leq.prop}($iii$) there are two cases:
Either $c=\any{n}$, $\Gamma\sgv\binbop{x}{a_1}{b_2}\leq\any{n}$, and thus obviously $\Gamma\sgv\binbop{x}{a_1}{b_2}<\any{n}$, and then by Law~\ref{sub.leq.top} we know that $\Gamma\sgv a=\binbop{x}{a_1}{a_2}<\any{n}=c$, 
or $c=\binbop{x}{a_1}{c_2}$ and $\Gamma,x:a_1\sgv b_2\leq c_2$ for some $c_2$, and then 
by inductive hypothesis $\Gamma,x:a_1\sgv a_2\leq c_2$ and therefore $\Gamma\sgv a\leq c$.
\item[\sbprod:]
The proof is similar to case \sabs\ using Law \ref{sub.leq.prop}($v$).
\end{meditemize}
For antisymmetry,
by induction on the definition of $\Gamma\sgv a\leq b$ we show that $\Gamma\sgv a\leq b$ and $\Gamma\sgv b\leq a$ implies $a=b$.
\begin{meditemize}
\item[\srefl:] 
Obvious 
\item[\sembed:] 
We have $b=\any{n}$, for some $n$, and $\Gamma\sgv a<\any{n}$ as well as $\Gamma\sgv\any{n}\leq a$.
From $\Gamma\sgv\any{n}\leq a$ we know that either $a=\any{n}$ or $a=\any{m}$ for some $m$ where $\Gamma\sgv\any{n}<\any{m}$.  
In the latter case, by Law \ref{sub.leq.top} $\Gamma\sgv\any{m}<\any{m}$ which due to Law \ref{sub.any} is not possible.
\item[\sabs:]
We have $a=\binbop{x}{a_1}{a_2}$ and $b=\binbop{x}{a_1}{b_2}$ where $\Gamma,x:a_1\sgv a_2\leq b_2$.  
From $\Gamma\sgv\binbop{x}{a_1}{b_2}\leq a$, by Law~\ref{sub.leq.prop}($iii$) we know that $\Gamma,x:a_1\sgv b_2\leq a_2$, and then 
by inductive hypothesis $a_2=b_2$ which implies $a=b$.
\item[\sbprod:]
The proof is similar to case \sabs\ using Law \ref{sub.leq.prop}($v$). \qedhere
\end{meditemize}
\end{proof}
\begin{law}[$\lambda$-Inclusion is a partial ordering]
\label{sub.incl.por}
For all $\Gamma$, $a$, $b$:
\begin{itemize}
\item[$i:$]
If $\Gamma\sgv b$ and $a\eqv b$ then $\Gamma\sgv a\incl b$.
\item[$ii:$]
The relation $\Gamma\sgv a\incl b$ is transitive.
\item[$iii$:] 
The relation $\Gamma\sgv a\incl b$ is antisymmetric modulo congruence.
\item[$iv$:] 
$\Gamma\sgv a\incl b$ if and only if $\Gamma\sgv a\gincl b$.
\end{itemize}
\end{law}
\begin{proof}
For all $\Gamma$, $a$, $b$:
\begin{itemize}
\item[$i:$]
By Law \ref{cr} there is a $c$ where $a\rd c$ and $b\rd c$.
By Law \ref{sub.rd.type}($i$) we know that $\Gamma\sgv c$ and therefore $\Gamma\sgv c\leq c$. 
Hence $\Gamma\sgv a\incl b$.
\item[$ii$:]
For transitivity assume $\Gamma\sgv a\incl b$ and $\Gamma\sgv b\incl c$ for some $c$. 
Hence by definition of $\lambda$-inclusion
$a\eqv a'$,  $\Gamma\sgv a'\leq b'$, $b'\eqv b$, $b\eqv b''$,  $\Gamma\sgv b''\leq c'$, $c'\eqv c$, and $\Gamma\sgv c$ for some $b'$, $b''$, $c'$.
By Law \ref{cr} we know that $b'\rd \hat{b}$ and $b''\rd \hat{b}$ for some $\hat{b}$.
Hence by Law \ref{sub.rd.type}($iii$) we know that $\Gamma\sgv a'\leq\hat{b}$ and $\Gamma\sgv\hat{b}\leq c'$.
By Law \ref{sub.leq.por} we know that $\Gamma\sgv a'\leq c'$ and therefore $\Gamma\sgv a\incl c$. 
\item[$iii$:]
For antisymmetry assume $\Gamma\sgv a\incl b$ and $\Gamma\sgv b\incl a$.
Hence by definition of $\lambda$-inclusion
$a\eqv a'$,  $\Gamma\sgv a'\leq b'$, $b'\eqv b$, $b\eqv b''$,  $\Gamma\sgv b''\leq a''$, $a''\eqv a$, $\Gamma\sgv a$ for some $b'$, $b''$, $a'$, $a''$.
By Law \ref{cr} we know that $b'\rd \hat{b}$ and $b''\rd \hat{b}$ for some $\hat{b}$ and that  $a'\rd \hat{a}$ and $a''\rd \hat{a}$ for some $\hat{a}$.  
Hence by Law \ref{sub.rd.type}($iii$) we know that $\Gamma\sgv \hat{a}\leq\hat{b}$ and $\Gamma\sgv\hat{b}\leq\hat{a}$.
By Law \ref{sub.leq.por} we know that $\hat{b}=\hat{a}$ and therefore $a\eqv b$. 
\item[$iv$:]
Follows from $ii$ since $\Gamma\sgv a\gincl b$ is defined as the transitive closure of $\Gamma\sgv a\incl b$. \qedhere
\end{itemize}
\end{proof}
\noindent
In order to show a property about upper bounds we need the following substitution lemma.
\begin{law}[Substitution property for $\any{}$-inclusion]
\label{sub.inc.sub}
For any $\Gamma_1$, $\Gamma_2$, $x$, $a_1$, and $a_2$ where $\Gamma_1\sgv a_2:a_1$, let $\Gamma=(\Gamma_1,x:a_1,\Gamma_2)$ and $\Gamma'=(\Gamma_1,\Gamma_2\gsub{x}{a_2})$:
For all $b$: If $\Gamma_1\sgv a_2<\any{n}$, $\ran(\Gamma)<\any{n}$, and $\Gamma\sgv b<\any{n}$ then $\ran(\Gamma')<\any{n}$ and $\Gamma'\sgv b\gsub{x}{a_2}<\any{n}$.
\end{law}
\begin{proof}
Structural induction on the definition of $\Gamma\sgv b<\any{n}$.
\begin{meditemize}
\item[\sstart:]
We have $b=\any{m}$ for some $m$ where $\Gamma\sgv\any{m}$ and $\Gamma\sgv\any{n}$.
By Law \ref{sub.type.sub} $\Gamma'\sgv\any{m}$ and $\Gamma'\sgv\any{n}$ which implies the proposition.
\item[\sstyp:]
We have $\Gamma\sgv b:c$ for some $c$ where $\Gamma\sgv c<\any{n}$.
By Law \ref{sub.type.sub} $\Gamma'\sgv b\gsub{x}{a_2}:c\gsub{x}{a_2}$ by inductive hypothesis $\ran(\Gamma')<\any{n}$  and $\Gamma'\sgv c\gsub{x}{a_2}<\any{n}$
hence $\Gamma'\sgv b\gsub{x}{a_2}<\any{n}$. 
\item[\ssabsu:]
We have $b=\binbop{y}{b_1}{b_2}$ where $\Gamma\sgv b_1<\any{n}$ and $\Gamma,y:b_1\sgv b_2<\any{n}$ and $\ran(\Gamma)<\any{n}$.
Hence $\ran(\Gamma,y:b_1)<\any{n}$.
By inductive hypothesis $\ran(\Gamma')<\any{n}$ and $\Gamma'\sgv b_1\gsub{x}{a_2}<\any{n}$. 
Similarly $\ran(\Gamma',y:b_1\gsub{x}{a_2})<\any{n}$ and $\Gamma',y:b_1\gsub{x}{a_2}\sgv b_2\gsub{x}{a_2}<\any{n}$.
Hence $\Gamma'\sgv b\gsub{x}{a_2}<\any{n}$.
\item[\ssbprod,]\ssinjl, \ssinjr, \sspdef:
Similar to \ssabsu. \qedhere
\end{meditemize}
\end{proof}
\noindent
The following property illustrates the power of the primitive constants $\any{n}$. 
Note that in this context the property is a final result in the sense that is not needed for the arguments of other subsequent properties. 
\begin{law}[Existence of upper bound]
\label{sub.inc.upper}
For all $\Gamma$, $a$, $b$: If  $\Gamma\sgv a:b$ then there is an $n$ such that $\Gamma\sgv b<\any{n}$ and $\ran(\Gamma)<\any{n}$.
Here $\ran(\Gamma)<\any{n}$ is an abbreviation for the condition that $\Gamma\sgv c<\any{n}$ for all $c$ in $\ran(\Gamma)$.
As a consequence every valid expression has an upper bound $\any{n}$ with respect to inclusion.
\end{law}
\begin{proof}
The straightforward proof is by induction on $\Gamma\sgv a:b$. We show a few cases below.
We write $\ran(\Gamma)<a$ as abbreviation for the condition that $\Gamma\sgv b<a$ for all $b$ in $\ran(\Gamma)$.
\begin{meditemize}%
\item[\ax:] 
Obvious
\item[\mystart:]
We have $a=x$ and $(\Gamma,x:b)\sgv x:b$ where $\Gamma\sgv b:c$ for some $x$ and $c$. 
By inductive hypothesis $\Gamma\sgv c<\any{n}$ and $\ran(\Gamma)<\any{n}$  for some $n$.
Hence by rule~\sstyp~we obtain $\Gamma\sgv b<\any{n}$. This implies $\ran(\Gamma,x:b)<\any{n}$. 
\item[\weak:]
We have $\Gamma,x:c\sgv a:b$ where $\Gamma\sgv a:b$ and $\Gamma\sgv c:d$ for some $x$, $c$ and $d$.
By inductive hypothesis $\Gamma\sgv b<\any{n}$ and $\ran(\Gamma)<\any{n}$ for some $n$.
Hence by rule \sstyp\ $\Gamma\sgv a<\any{n}$ and by Law \ref{sub.type.weak} $\Gamma,x:c\sgv a<\any{n}$.
Similarly by inductive hypothesis $\Gamma\sgv d<\any{m}$ and $\ran(\Gamma)<\any{m}$ for some $m$. 
The property follows for the maximum of $n$ and $m$ using Law \ref{sub.leq.top}. 
\item[\conv:]
We have $\Gamma\sgv a:b$ where $\Gamma\sgv a:c$, $c\eqv b$ and $\Gamma\sgv b:d$ for some $c$ and $d$.
By inductive hypothesis $\Gamma\sgv d<\any{n}$ and $\ran(\Gamma)<\any{n}$ for some $n$.
Hence by rule \sstyp~$\Gamma\sgv b<\any{n}$. 
\item[\absu:]
We have $a=[x:a_1]a_2$, $b=[x:a_1]b_2$ and $\Gamma\sgv[x:a_1]a_1:[x:a_1]b_2$ where $\Gamma,x:a_1\sgv a_2:b_2$ for some $x$, $a_1$, $a_2$, $b_1$, $b_2$.
By inductive hypothesis $\Gamma,x:a_1\sgv b_2<\any{n}$ and $\ran(\Gamma,x:a_1)<\any{n}$ for some $n$.
The property follows from the definition of $\any{}$-inclusion.
\item[\abse:]
Similar to \absu.
\item[\appl:]
We have  $a=(a_1\,a_2)$ and $b=b_2\gsub{x}{a_2}$ and $\Gamma\sgv(a_1\,a_2):b_2\gsub{x}{a_2}$ where $\Gamma\sgv a_1:[x:b_1]b_2$ and $\Gamma\sgv a_2:b_1$ for some $a_1$, $a_2$, $x$, $b_1$, and $b_2$.
By inductive hypothesis $\Gamma\sgv[x:b_1]b_2<\any{n}$ and $\ran(\Gamma)<\any{n}$ for some $n$.
Hence obviously $\Gamma\sgv b_1<\any{n}$ and $(\Gamma,x:b_1)\sgv b_2<\any{n}$. 
Hence by rule \sstyp~$\Gamma\sgv a_2<\any{n}$. 
Furthermore we obviously have $\ran(\Gamma,x:b_1)<\any{n}$. 
By Law~\ref{sub.inc.sub} we obtain $\Gamma\sgv b_2\gsub{x}{a_2}<\any{n}$.
\item[\pdef:]
Similar to \absu.
\item[\chin:]
We have $a=\pleft{a_1}$, $\Gamma\sgv a_1:[x!b]b_2$ for some $x$, $b_2$, and $\Gamma\sgv a:c$.
By inductive hypothesis $\Gamma\sgv[x!b]b_2<\any{n}$ and $\ran(\Gamma)<\any{n}$ for some $n$.
Hence obviously $\Gamma\sgv b<\any{n}$.
\item[\chba:]
We have $a=\pright{a_1}$, $\Gamma\sgv a_1:[x!b_1]b_2$,  $b=b_2\gsub{x}{\pleft{a_1}}$ for some $x$, $b_1$, $b_2$, and $\Gamma\sgv a:c$.
By inductive hypothesis $\Gamma\sgv[x!b_1]b_2<\any{n}$ and $\ran(\Gamma)<\any{n}$ for some $n$.
Obviously $\Gamma,x:b_1\sgv b_2<\any{n}$ and $\Gamma\sgv b_1<\any{n}$. 
Furthermore $\Gamma\sgv\pleft{a_1}:b_1$ hence by rule \sstyp\ $\Gamma\sgv \pleft{a_1}<\any{n}$ and therefore by Law~\ref{sub.inc.sub} we obtain 
$\Gamma\sgv b<\any{n}$.
\item[\bprod:]
Similar to \absu.
\item[\bsum:]
Similar to \absu.
\item[\prl:]
Similar to \chin.
\item[\prr:]
Similar to \chin.
\item[\injll:]
Similar to \absu.
\item[\injlr:]
Similar to \absu.
\item[\cased:]
We have $a=\case{a_1}{a_2}$, $b=[z:[b_1+b_2]]b_3$ where $\Gamma\sgv a:[x:b_1]b_3$, $\Gamma\sgv a:[x:b_2]b_3$, $\Gamma\sgv b_3$ for some $x$, $a_1$, $a_2$, $b_1$, $b_2$, $b_3$, and $\Gamma\sgv a:c$.  
By inductive hypothesis $\Gamma\sgv[x:b_1]b_3<\any{n}$, $\Gamma\sgv[x:b_2]b_3<\any{n}$, $\Gamma\sgv b_3<\any{n}$ and $\ran(\Gamma)<\any{n}$ for some $n$.
Hence obviously $\Gamma\sgv b<\any{n}$.
\item[\negate:]
Similar to \absu.
\end{meditemize}

\noindent
As a consequence using rules~\sstyp\ and~\sembed\ we can infer that $\Gamma\sgv a\leq\any{n}$. \qedhere
\end{proof}
\noindent We show the ``inverse'' to Law \ref{sub.any} which is  irreflexivity of $\prim$-inclusion modulo congruence.
\begin{law}[Irreflexivity of $\prim$-inclusion modulo congruence]
\label{sub.sinc.irrefl}
For all $\Gamma$, $a$, $n$, $m$: $\Gamma\sgv a<\any{n}$ implies $a\neqv\any{m}$ for $n\leq m$. 
\end{law}
\begin{proof}
By induction on the definition of $\Gamma\sgv a<\any{n}$.
\begin{meditemize}
\item[\sstart:] 
Obvious.
\item[\sstyp:] 
We have $\Gamma\sgv a:b$ and $\Gamma\sgv b<\any{n}$.  
Assume that $a\eqv\any{k}$ for some $k$ with $n\leq k$. 
By Law~\ref{cr} we know that $a\rd\any{k}$ and by Law~\ref{sub.rd.type}(subject reduction) we know that $\Gamma\sgv\any{k}:b$.
Hence by Law~\ref{sub.any}($iii$) we know that $b\eqv\any{l}$ for some $l$ with $k\leq l$.
However by inductive hypothesis $b\neqv\any{m}$ for any $m$ with $n\leq m$ which is a contradiction since $n\leq l$.
\end{meditemize}
Analogous arguments can be made for rules \ssabsu, \ssbprod, \ssinjl, \ssinjr, and \sspdef.\qedhere
\end{proof}
\subsection{Properties of typing}%
While there is no corresponding result for Law \ref{sub.inc.upper} on lower bounds we can show that the types of an expression have a minimal element.
\begin{definition}[Minimal type]
\label{sub.min.type}
\nomenclature[lzeNorm14]{$\mint{\Gamma}{a}$}{minimal type}%
\index{type!minimal}
For all $\Gamma$, $a$ with $\Gamma\sgv a$ we define the \emph{minimal type} $\mint{\Gamma}{a}$ recursively as follows:
$\mint{\Gamma}{\any{n}}=\any{n}$,
$\mint{\Gamma}{x}=\Gamma(x)$, 
$\mint{\Gamma}{\binbop{x}{a}{b}}=[x:a]\mint{\Gamma,x:a}{b}$, 
$\mint{\Gamma}{(a\:b)}=d\gsub{x}{b}$ where $\mint{\Gamma}{a}=[x:c]d$,
$\mint{\Gamma}{\prsumop{a}{b}}=\prsumop{\mint{\Gamma}{a}}{\mint{\Gamma}{b}}$, 
$\mint{\Gamma}{\prdef{y}{a^{b}}{c}{d}}=[x!b]d$,
$\mint{\Gamma}{\pleft{a}}=b$ where $\mint{\Gamma}{a}=[b,c]$,
$\mint{\Gamma}{\pright{a}}=c$ where $\mint{\Gamma}{a}=[b,c]$,
$\mint{\Gamma}{\injl{a}{b}}=[\mint{\Gamma}{a}+b]$,
$\mint{\Gamma}{\injr{a}{b}}=[a+\mint{\Gamma}{b}]$,
$\mint{\Gamma}{\case{a}{b}}=[[c+d]\fun e]$ where $\mint{\Gamma}{a}=[c\fun e]$, $\mint{\Gamma}{b}=[d\fun e]$, and
$\mint{\Gamma}{\myneg a}=\mint{\Gamma}{a}$.
\end{definition}
\begin{remark}
The function is completely defined on its domain since $\Gamma\sgv(a\:b)$ implies (Table \ref{sub.typ.rules}) $\Gamma\sgv a:[x:a_1]a_2$ for some $a_1$, $a_2$ and hence $\mint{\Gamma}{a}=[x:a_1]a_3$ for some $a_3$.
\end{remark}
\noindent
We will show that minimal types are indeed minimal. Before that we need to generalize \ref{sub.type.sub} to $\lambda$-inclusions.
\begin{law}[$\lambda$-Inclusion and substitution]
\label{sub.incl.sub}
For all $\Gamma$, $x$, $a$, $b$, $c$, $d$: $\Gamma,x:a\sgv b\incl c$ and $\Gamma\sgv d:a$ imply $\Gamma\sgv b\gsub{x}{d}\incl c\gsub{x}{d}$.
\end{law}
\begin{proof}
$\Gamma,x:a\sgv b\incl c$ implies $b\eqv b'$, $\Gamma,x:a\sgv b'\leq c'$, $c'\eqv c$, and $\Gamma\sgv c$ for some $b'$, $c'$. 
By \ref{sub.type.sub} $\Gamma\sgv b'\gsub{x}{d}\leq c'\gsub{x}{d}$ and $\Gamma\sgv c\gsub{x}{d}$.
Since $b\gsub{x}{d}\eqv b'\gsub{x}{d}$ and $c\gsub{x}{d}\eqv c'\gsub{x}{d}$ we have $\Gamma\sgv b\gsub{x}{d}\incl c\gsub{x}{d}$.
\end{proof}
\noindent
\begin{law}[Minimal types are minimal]
\label{sub.min.min}
For all $\Gamma$, $a$, $b$: $\Gamma\sgv a:b$ implies that $\Gamma\sgv a:\mint{\Gamma}{a}$ and $\Gamma\sgv\mint{\Gamma}{a}\incl b$.
\end{law}
\begin{proof}
Structural induction on $a$.
\begin{itemize}
\item$a=\any{n}$ for some $n$: By Law \ref{sub.any} $b\eqv\any{m}$ for some $m$ where $n\leq m$. 
If $n=m$ then by rule \srefl\ $\Gamma\sgv\any{n}\leq\any{n}$. If $n<m$ then
by rule \sstart, which is applicable since $b\rd\any{m}$ and by Law \ref{valid.type} $\Gamma\sgv b$ and therefore by Law \ref{sub.rd.type} $\Gamma\sgv\any{m}$, we obtain $\Gamma\sgv\any{n}<\any{m}$.
In both cases this implies $\Gamma\sgv\mint{\Gamma}{a}=\any{n}\incl b$.
\item$a=x$ for some $x$: By \ref{sub.type.decomp}($ii$) $\Gamma\sgv\Gamma(x)\incl b$. Hence $\Gamma\sgv\mint{\Gamma}{a}\incl b$.
\item$a=\binbop{x}{a_1}{a_2}$ for some $x$, $a_1$, $a_2$: By Law \ref{sub.type.decomp}($iii$) 
$\Gamma,x:a_1\sgv a_2:c$ and $\Gamma\sgv[x:a_1]c\incl b$ for some $c$. 
By inductive hypothesis $\Gamma,x:a_1\sgv\mint{\Gamma,x:a_1}{a_2}\incl c$ hence obviously $\Gamma\sgv\mint{\Gamma}{a}\incl[x:a_1]c$
and therefore by Law \ref{sub.incl.por} $\Gamma\sgv\mint{\Gamma}{a}\incl b$.
\item$a=(a_1\:a_2)$ for some $a_1$, $a_2$: By Law \ref{sub.type.decomp}($v$)
$\Gamma\sgv a_1:[x:c_1]c_2$ for some $x$, $c_1$, $c_2$ where $\Gamma\sgv a_2:c_1$ and $\Gamma\sgv c_2\gsub{x}{a_2}\incl b$.
By inductive hypothesis $\Gamma\sgv\mint{\Gamma}{a_1}\incl[x:c_1]c_2$. 
By Law \ref{sub.gincl.prop}($iii$) $\mint{\Gamma}{a_1}\eqv[x:c_1]c_3$ for some $c_3$ where $\Gamma,x:c_1\sgv c_3\incl c_2$.
By Law \ref{sub.incl.sub} $\Gamma\sgv c_3\gsub{x}{a_2}\incl c_2\gsub{x}{a_2}$ and
therefore by definition of minimal type and Law \ref{sub.incl.por} $\Gamma\sgv\mint{\Gamma}{a}\incl b$.
\item$a=\prsumop{a_1}{a_2}$ for some $a_1$, $a_2$: 
Similar to the case of abstraction using Law \ref{sub.type.decomp}($vi$).
\item$a=\prdef{y}{a_1^{a_2}}{a_3}{a_4}$ for some $a_1$, $a_2$, $a_3$, $a_4$:
Similar to the case of abstraction using Law \ref{sub.type.decomp}($vii$).
\item$a=\pleft{a_1}$ for some $a_1$: 
By Law \ref{sub.type.decomp}($viii$) there are two cases. In the first case the property follows
from Law \ref{sub.gincl.prop}($iii$) in the second case it follows from Laws 
\ref{sub.gincl.prop}($v$). 
\item$a=\pright{a_1}$ for some $a_1$: 
By Law \ref{sub.type.decomp}($ix$) there are two cases. In the first case the property follows
from Law \ref{sub.gincl.prop}($iii$) in the second case, similar to applications it follows from Laws \ref{sub.gincl.prop}($v$), \ref{sub.incl.sub}, and \ref{sub.incl.por}. 
\item$a=\injl{a_1}{a_2}$ for some $a_1$, $a_2$: 
Similar to the case of abstraction using Law \ref{sub.type.decomp}($x$).
\item$a=\injr{a_1}{a_2}$ for some $a_1$, $a_2$: 
Similar to the case of abstraction using Law \ref{sub.type.decomp}($x$).
\item$a=\case{a_1}{a_2}$ for some $a_1$, $a_2$: 
Similar to the case of abstraction using Law \ref{sub.type.decomp}($xi$) and
Law \ref{sub.incl.por}.
\item$a=\myneg a_1$ for some $a_1$: 
Follows from \ref{sub.type.decomp}($xi$) and the inductive hypothesis.
\qedhere
\end{itemize}
\end{proof}
\subsection{Norms and norming}
\label{sub.skeletons}
In the proof of strong normalization (see~\ref{strong.normalization})
we have introduced the {\em norm} $\nrm{\Gamma}{a}$ of an expression $a$ (\ref{norming}) as a structural characterization.  
While the notion of norms as defined in \ref{norming} amounted to a reconstruction of simple types based on the primitive constant and products only, the inclusion of the new primitive constants $\any{0}$, $\any{1}$, $\ldots$, and of the adapted type rule for applications will require several extensions to norms. First of all, the primitive constants itself will obviously themselves be norms.
Second we have to add a parametric mechanism to norms. As a motivation consider the expression $g:=[x:\any{1}][x\fun\any{0}]$ which is intended to allow for application with arguments of different structure, \eg\ $(g\,\any{0}\,\any{0})$ and $(g\,[\any{0}\fun\any{0}]\,[\any{0}\fun\any{0}])$. 
Finally, we can no longer norm abstractions to products norms (\ref{norming}), as these operators have different properties \wrt\ parametrization.
Since these extensions and their properties are complex they will be presented here in a dedicated new Section before
being used in an adapted proof of strong normalisation (see~\ref{sub.strong.normalization}).

In general, norms are expressions without elimination operators.
The set of norms introduced in the old strong normalisation proof (see~\ref{norming}) is extended to include the use of primitive constants, variables, abstractions, protected definitions with tags, sums, and injections. 
\nomenclature[lhSets5]{$\bar{\Gamma}$, $\bar{\Gamma}_1$, $\bar{\Gamma}_2$, $\ldots$}{norm contexts}
\begin{eqnarray*}
\dnrm&\!::=\!&\{\any{0}\}\,\mid\,\dsuper\,\mid\,\dvar\,\mid\,\binbop{\dvar}{\dnrm}{\dnrm}\,\mid\,\prdef{\dvar}{\dnrm^{\dnrm}}{\dnrm}{\dnrm}\,\mid\,\prsumop{\dnrm}{\dnrm}\,\mid\,\injl{\dnrm}{\dnrm}\,\mid\,\injr{\dnrm}{\dnrm}
\end{eqnarray*}
We recall the notation $\bar{a}$, $\bar{b}$, $\bar{c}$, $\ldots$ to denote norms.
Furthermore, contexts declaring norms only are denoted by $\bar{\Gamma}$, $\bar{\Gamma}_1$, $\bar{\Gamma}_2$, $\ldots$.
\begin{law}[Fundamental properties of norms]
\label{sub.nrm.fund}
Obviously norms are irreducible. This simplifies many laws, for example on norms congruence $\bar{a}\eqv\bar{b}$ is the same as equality $\bar{a}=\bar{b}$ and $\lambda$-inclusion $\Gamma\sgv\bar{a}\incl\bar{b}$ is the same as inclusion $\Gamma\sgv\bar{a}\leq\bar{b}$.
Furthermore all norms without free identifiers are valid and all types of norms are norms as well.  
In order not to clutter the following presentations we assume implicit use of this law.
\end{law}
\noindent
The goal is to assign a unique norm to all expressions satisfying certain structural properties.
Furthermore this assignment should have two key properties: Reduction of expression should not affect norm assignment (Law~\ref{nrm.rd}) and it should be possible to assign a norm to all valid expressions (Law~\ref{val.nrm}).

\subsubsection{Definition of norming}
\label{sub.norm.def}
The definition of norm assignment is technically complex since, as it turns out, when trying to achieve the above two properties, norming of applications cannot be defined in a compositional style. More precisely, when norming an application of a function to an argument the former cannot be normed in isolation but its norm also depends on its argument. 
To see this consider the expression $t:=[x:[\any{0}\fun\any{1}]](x\,\any{0})$. If we norm it to $[[\any{0}\fun\any{1}]\fun\any{1}]$ independently of it use, then the application $(t\,[\any{0}\fun\any{0}])$ 
would norm to $\any{1}$. However $(t\,[\any{0}\fun\any{0}])\rd([\any{0}\fun\any{0}]\,\any{0})\rd\any{0}$ which would violate the intended Law~\ref{nrm.rd}.
As a consequence we have to include the argument context of a function when computing its norm. A similar argument can be made for the projection context of a product or an existential abstraction. 
The machinery needed is provided by the following technical but straightforward definitions.
\begin{definition}[Elimination list]%
\label{sub.norm.elimination}
\index{norming!elimination list}
\index{norming!eliminator}
\nomenclature[liSets5]{$\Theta=(\theta_1,\ldots,\theta_n)$}{elimination list}
\nomenclature[ljSets5]{$\theta_1$, $\theta_1$, $\theta_3$, $\ldots$}{eliminators}
An {\em elimination list} $\Theta=(\theta_1,\ldots,\theta_n)$ where, for each $i$, the \emph{eliminator} $\theta_i\in\dexp\cup\{1,2\}\cup\{2_a\mid a\in\dexp\}$ is used to model pending eliminations related to applications and projections.
Similar to contexts, we write $(\theta,\Theta)$ or $(\Theta,\theta)$ for adding $\theta$ to the left or right of $\Theta$. 
Substitution $\Theta\gsub{x}{a}$ on elimination lists is defined as $(\theta_1\gsub{x}{a},\ldots,\theta_n\gsub{x}{a})$ where we extend substitution by $(2_b)\gsub{x}{a}=2_{b\gsub{x}{a}}$, $1\gsub{x}{a}=1$, and $2\gsub{x}{a}=2$.
\end{definition}
\noindent
In order to use pending elimination elements when defining norms we introduce an extension to contexts containing both definitions and declarations. 
\begin{definition}[Extended context]
\nomenclature[lkCalc001]{$x\odot b$}{declaration or definition}
\index{operator!declaration or definition}
\nomenclature[llSets5]{$\Lambda, \Lambda_1, \Lambda_2, \cdots$}{extended contexts}
\label{sub.environment}
\index{extended context}
\nomenclature[lmCalc04]{$\domdec(\Lambda)$}{declarations of extended context}
\index{extended context!declarations}
\nomenclature[lnCalc04]{$\domdef(\Lambda)$}{definitions of extended context}
\index{extended context!definitions}
\emph{Extended contexts}, denoted by $\Lambda$, $\Lambda_1$, $\Lambda_2$, etc.~are finite sequences $(x_1\odot_1 a_1$, $\ldots$, $x_n\odot_n a_n$ $)$, where $x_i$ are variables, $x_i\neq x_j$,
where each $x_i\odot_i a_i$ stands for a declaration $x_i:a_i$ or definition $x_i\mydef a_i$.
For an extended context $\Lambda$, $\domdec(\Lambda)$ denotes the set of all variables which are declared, $\domdef(\Lambda)$ denotes the set of all variables which are defined.

The lookup of an variable in an extended context is defined by $\Lambda(x)=a_i$. 
$\Lambda,x\mydef a$ denotes the extension of $\Lambda$ on the right by a definition $x\mydef a$.
$\Lambda_1,\Lambda_2$ denotes the concatenation of two extended contexts. 
The empty extended context is written as $()$.
Obviously, every context $\Gamma$ is an extended context.
\end{definition}
\nomenclature[lo1Rel05]{$\Gamma\sgv a\ginc b$}{generalized inclusion}%
\index{inclusion!generalized}
\noindent
Finally, we introduce the notion of {\em generalized inclusion}:
\[
\Gamma\sgv a\ginc b\quad\equiv\quad\Gamma\sgv a\leq b\quad\text{or}\quad\Gamma\sgv a:b
\]
The motivation for this notation is provided further below in the definition of norming (see example E.5).
We can now adapt the definition of norming (Section~\ref{normable}).
\begin{definition}[Norming]
\label{sub.genNormable}
\index{norming}
\index{norming!extended context}
\nomenclature[lpNorm14]{$\Lambda,\Theta\sngv a$}{normable expression}%
\nomenclature[lqNorm14]{$\nrm{\Lambda,\Theta}{a},\nrm{\Lambda}{a}$}{norming}%
\nomenclature[lrNorm14]{$\nrm{\Lambda_1}{\Lambda_2},\nrm{}{\Lambda}$}{context norming}%
To foster understanding we break up the definition into three parts.
Norming $\nrm{\Lambda,\Theta}{a}$ of an expression $a$ under extended context $\Lambda$, and elimination list $\Theta$ is a partial function defined in Tables~\ref{sub.schema.def},\ref{sub.schema.def2}, and~\ref{sub.schema.def3}.
\begin{table}[!htb]
\fbox{
\begin{minipage}{0.96\textwidth}
\begin{eqnarray*}
\\[-8mm]
\nrm{\Lambda,\Theta}{\any{n}}&=&\any{n}\\
\nrm{\Lambda,\Theta}{x}&=&
\begin{cases}
x&\text{if}\;x\!\in\!\domdec(\Lambda),\Theta=()\\
\!\nrm{\Lambda,\Theta}{\Lambda(x)}&\text{if}\;x\!\in\!\domdec(\Lambda),\Theta\neq()\;\text{or if}\;x\!\in\!\domdef(\Lambda)
\end{cases}\\
\nrm{\Lambda,\Theta}{\binbop{x}{a}{b}}&=&
\begin{cases}
\binbop{x}{\nrm{\Lambda,()}{a}}{\nrm{(\Lambda,x:a),()}{b}}&\text{if}\;\Theta=()\\
\binbop{x}{\nrm{\Lambda,()}{a}}{\nrm{(\Lambda,x\smydef c),\Theta'}{b}}&\text{if}\;\Theta=(c,\Theta')
\end{cases}\\
\nrm{\Lambda,\Theta}{(a\,b)}&=&\bar{c}\quad
\text{if}\;\nrm{\Lambda,(b,\Theta)}{a}=\binbop{x}{\bar{d}}{\bar{c}}\;\;\text{and}\;\nrm{}{\Lambda}\;\sgv\;\nrm{\Lambda}{b}\!\ginc\bar{d}
\end{eqnarray*}
\end{minipage}
}
\caption{Norming under extended context and elimination list (part one)\label{sub.schema.def}}
\end{table}
Tables~\ref{sub.schema.def} defines norming for the core subset of abstractions and applications only.
It illustrates the basic mechanism of how application arguments are pushed onto elimination lists when norming the function of an application and removed from elimination lists when norming the body of an abstraction.

Norming is defined in parallel with {\em context norming} $\nrm{\Lambda_1}{\Lambda_2}$ which is a partial function yielding a context declaring norms. It is defined in Table~\ref{sub.norming.context}.
\begin{table}[!htb]
\fbox{
\begin{minipage}{0.96\textwidth}
\begin{eqnarray*}
\\[-8mm]
\nrm{\Lambda_1}{()}&=&()\\
\nrm{\Lambda_1}{(x:a,\Lambda_2)}&=&(x:\nrm{\Lambda_1,()}{a},\nrm{\Lambda_1,x\odot a}{\Lambda_2})\\
\nrm{\Lambda_1}{(x\mydef a,\Lambda_2)}&=&\nrm{\Lambda_1,x\smydef a}{\Lambda_2}
\end{eqnarray*}
\end{minipage}
}
\caption{Context norming~\label{sub.norming.context}}
\end{table}
\noindent
An expression $a$ is {\em normable} under $\Lambda$ and $\Theta$ if $\nrm{\Lambda,\Theta}{a}$ is defined. 
This is written as $\Lambda,\Theta\sngv\,a$.
As for empty contexts $()$ we omit writing empty elimination list and empty extended contexts. For example, we write $\nrm{\Lambda}{a}$ as abbreviation for $\nrm{\Lambda,()}{a}$
and $\nrm{}{a}$ as abbreviation for $\nrm{(),()}{a}$.

Table~\ref{sub.schema.def2} is adding protected definitions and their associated projections also relevant existential abstractions, .
These rules are considerably more complex as existential abstractions are normed to existential abstractions and therefore projection mechanisms must be 
extended to existential abstractions as well. Similar to application, the law of projection is pushing left and right projections onto elimination lists.
Protected definitions are normed to protected definitions so as to be able to access its elements using projections.
Not that in contrast for to existential abstractions for protected definitions the expression $e$ in $2_e$ is not used as there is no need for instantiation of the extended environment.
\begin{table}[!htb]
\fbox{
\begin{minipage}{0.96\textwidth}
\begin{eqnarray*}
\\[-8mm]
\nrm{\Lambda,\Theta}{[x\sdef a]b}&=&
\begin{cases}
[x\sdef\nrm{\Lambda,\Theta'}{a}]\nrm{(\Lambda,x:a),()}{b}&\text{if}\;\Theta=(1,\Theta')\\
[x\sdef\nrm{\Lambda,()}{a}]\nrm{(\Lambda,x:=c),\Theta'}{b}&\text{if}\;\Theta\!=\!(2_c,\Theta')  
\end{cases}\\
\nrm{\Lambda,\Theta}{\prdef{x}{a^b\!}{c}{d}}&=&
\begin{cases}
\prdef{x}{\nrm{\Lambda,()}{a}^{\snrm{\Lambda,()}{b}}}{\!\nrm{\Lambda,()}{c}}{\nrm{(\Lambda,x:b),()}{d}}&\!\!\!\!\text{if}\;\Theta\!=\!()\\
\prdef{x}{\nrm{\Lambda,\Theta'}{a}^{\snrm{\Lambda,\Theta'}{b}}}{\!\!\nrm{\Lambda,()}{c}}{\nrm{\!(\Lambda,x:b),()}{d}}&\!\!\!\!\text{if}\;\Theta\!=\!(1,\!\Theta')\\
\prdef{x}{\nrm{\Lambda,()}{a}^{\snrm{\Lambda,()}{b}}}{\nrm{\Lambda,\Theta'}{c}}{\nrm{(\Lambda,x:b),\Theta'}{d}}&\!\!\!\!\text{if}\;\Theta\!=\!(2_e,\!\Theta')
\end{cases}\\
&&\;\text{if}\;\nrm{}{\Lambda}\;\sgv\;\nrm{\Lambda,()}{a}\!\ginc\nrm{\Lambda,()}{b}\\
&&\;\text{and}\;\nrm{}{\Lambda}\;\sgv\;\nrm{\Lambda,()}{c}\!\ginc\nrm{(\Lambda,x\smydef a),()}{d} \\[2mm]
\nrm{\Lambda,\Theta}{a.i}&=&
\begin{cases}
\bar{b}_1&\text{if}\;i=1,\;\nrm{\Lambda,(1,\Theta)}{a}=[x\sdef\bar{b}_1]\bar{b}_2\\
\bar{b}_2&\text{if}\;i=2,\;\nrm{\Lambda,(2_{\pleft{a}},\Theta)}{a}=[x\sdef\bar{b}_1]\bar{b}_2\\
\bar{b}_1&\text{if}\;i=1,\;\nrm{\Lambda,(1,\Theta)}{a}=\prdef{x}{\bar{b}_1^{\bar{b_2}}}{\bar{b}_3}{\bar{b}_4}\\
\bar{b}_3&\text{if}\;i=2,\;\nrm{\Lambda,(2_{\pleft{a}},\Theta)}{a}=\prdef{x}{\bar{b}_1^{\bar{b_2}}}{\bar{b}_3}{\bar{b}_4}
\end{cases}
\end{eqnarray*}
\end{minipage}
}
\caption{Norming under extended context and elimination list (part two)\label{sub.schema.def2}}
\end{table}
Note that projection is well defined assuming projection of its arguments is well defined and therefore either $\nrm{\Lambda,(2_{\pleft{a}},\Theta)}{a}=[x\sdef\bar{b}_1]\bar{b}_2$ 
or $\nrm{\Lambda,(2_{\pleft{a}},\Theta)}{a}=\prdef{x}{\bar{b}_1^{\bar{b_2}}}{\bar{b}_3}{\bar{b}_4}$.

Table~\ref{sub.schema.def3} is adding the finite counterparts of (existential) abstractions, case distinction, and negation.
The rules reflect the choices made for the infinite versions of product and sum.
\begin{table}[!htb]
\fbox{
\begin{minipage}{0.96\textwidth}
\begin{eqnarray*}
\\[-8mm]
\nrm{\Lambda,\Theta}{\prsuminjop{a}{b}}&=&
\begin{cases}
\prsuminjop{\nrm{\Lambda,()}{a}}{\nrm{\Lambda,()}{b}}&\text{if}\;\Theta=()\\
\prsuminjop{\nrm{\Lambda,\Theta'}{a}}{\nrm{\Lambda}{b}}&\text{if}\;\Theta=(1,\Theta')\\
\prsuminjop{\nrm{\Lambda}{a}}{\nrm{\Lambda,\Theta'}{b}}&\text{if}\;\Theta=(2,\Theta')
\end{cases}\\
\nrm{\Lambda,\Theta}{a.i}&=&\bar{b}_i\quad\text{if}\nrm{\Lambda,(i,\Theta)}{a}=\prsuminjop{\bar{b}_1}{\bar{b}_2}\\
\nrm{\Lambda,()}{\case{a}{b}}&=&[[\bar{c}+\bar{d}]\fun\bar{e}]\;\text{if}\;\nrm{\Lambda}{a}=[\bar{c}\fun\bar{e}],\nrm{\Lambda}{b}=[\bar{d}\fun\bar{e}]\\
\nrm{\Lambda,(c,\Theta)}{\case{a}{b}}&=&[[\bar{c}+\bar{d}]\fun\bar{e}]\;\text{if}\;\nrm{\Lambda,(\pleft{c},\Theta)}{a}=[\bar{c}\fun\bar{e}],\nrm{\Lambda,(\pright{c},\Theta)}{b}=[\bar{d}\fun\bar{e}]\\
\nrm{\Lambda,\Theta}{\myneg a}&=&\nrm{\Lambda,\Theta}{a} 
\end{eqnarray*}
\end{minipage}
}
\caption{Norming under extended context and elimination list (part three)\label{sub.schema.def3}}
\end{table}
Regarding well-definedness on projections a similar remark as for Table \ref{sub.schema.def2} applies. 
\end{definition}
\noindent
\begin{remark}[Examples of norms]
Some properties of norming are illustrated by basic examples: 
\begin{itemize}
\item[E.1] 
Let $\Gamma=(y:[x:\prim][x\fun x])$: The expression $\nrm{\Gamma}{(y\;[\prim\fun\prim])}$ is not defined 
since $\nrm{\Gamma,[\prim\fun\prim]}{y}=\nrm{\Gamma,[\prim\fun\prim]}{[x:\prim][x\fun x]}=[x:\prim]\nrm{\Gamma,x\smydef[\prim\fun\prim]}{[x\fun x]}$, $\nrm{\Gamma}{[\prim\fun\prim]}=[\prim\fun\prim]$,
and neither $\nrm{}{\Gamma}\gv[\prim\fun\prim]:\prim$ nor $\nrm{}{\Gamma}\gv[\prim\fun\prim]\leq\prim$.
\item[E.2] 
Let $\Gamma=(y:[x:\any{1}][x\fun x])$: $\nrm{\Gamma}{(y\;[\prim\fun\prim])}=[[\prim\fun\prim]\fun[\prim\fun\prim]]$
since $\nrm{\Gamma,[\prim\fun\prim]}{y}=\nrm{\Gamma,[\prim\fun\prim]}{[x:\any{1}][x\fun x]}=[x:\any{1}]\nrm{\Gamma,x\smydef[\prim\fun\prim]}{[x\fun x]}$
and $\nrm{}{\Gamma}\gv\nrm{\Gamma}{[\prim\fun\prim]}=[\prim\fun\prim]\leq\any{1}$.
\item[E.3]  
Let $u=[x:[\any{1}\fun\prim]](x\,x)$: The expression $\nrm{x:[\any{1}\sfun\prim]}{(x\,x)}$ and therefore also $\nrm{}{u}$ are undefined. 
When considering $(x\,x)$, first we obtain $\nrm{x:[\any{1}\sfun\prim],x}{x}=\nrm{x:[\any{1}\sfun\prim],x}{[\any{1}\fun\prim]}=[\any{1}\fun\prim]$ as the norm of the function $x$ in $(x\,x)$.
However, since neither $x:[\any{1}\fun\prim]\sgv x:\any{1}$ nor $x:[\any{1}\fun\prim]\sgv x\leq\any{1}$ the argument $x$ in $(x\,x)$ does not satisfy the required generalized inclusion constraint. 
Furthermore note that $\nrm{}{(u\,u)}$ is undefined too since $\nrm{u}{u}=[x:[\any{1}\fun\prim]]\nrm{x\smydef u}{(x\,x)}$ leading to an infinite recursion.
\item[E.4]
The non-compositional nature of norming can be illustrated as follows: 
Let $\Gamma=(x:[\prim\fun\any{1}],y:[\prim\fun\prim],)$, $t=[z:[\prim\fun\any{1}]](z\,\prim)$. We have $\nrm{\Gamma}{t}=[[\prim\fun\any{1}]\fun\any{1}]$.
Now consider the expressions $(t\;x)$ as well as $(t\;y)$.

We have $\nrm{\Gamma,x}{t}=[z:[\prim\fun\any{1}]]\nrm{\Gamma,z\smydef x}{(z\,\prim)}=[[\prim\fun\any{1}]\fun\any{1}]$
since $\nrm{\Gamma,z\smydef x,\prim}{z}=\nrm{\Gamma,z\smydef x,\prim}{x}=[\prim\fun\any{1}]$.
Hence $\nrm{\Gamma,}{(t\,x)}=\any{1}$.

We have $\nrm{\Gamma,y}{t}=[z:[\prim\fun\any{1}]]\nrm{\Gamma,z\smydef y}{(z\,\prim)}=[[\prim\fun\any{1}]\fun\prim]$
since $\nrm{\Gamma,z\smydef y,\prim}{z}=\nrm{\Gamma,z\smydef y,\prim}{x}=[\prim\fun\prim]$.
Hence $\nrm{\Gamma}{(t\,y)}=\prim$.

This illustrates that one cannot compute the norm of $(t\,y)$ compositionally, i.e.~by instantiating the norm of $t$ by the norm of $y$. 
The above situation can be presented in another way: If $\Gamma'=(\Gamma,z:[\prim\fun\any{1}])$ then 
$\nrm{\Gamma'}{(z\,\prim)\gsub{z}{[\prim\fun\prim]}}=\nrm{\Gamma'}{([\prim\fun\prim]\,\prim)}=\prim$
and  $\nrm{\Gamma'}{(z\,\prim)}\gsub{z}{[\prim\fun\prim]}=\nrm{\Gamma'}{([\prim\fun\any{1}]\,\prim)}\gsub{z}{[\prim\fun\prim]}=\any{1}$.
This example shows that the norm $\nrm{\Gamma}{a\gsub{x}{b}}$ of a substitution is not always equivalent to the substitution $\nrm{\Gamma}{a}\gsub{x}{\nrm{\Gamma}{b}}$ of a norm. 
\item[E.5]
So far we have always used the type condition when checking applications. The generalized inclusion condition is needed in the following example:
Let $\Gamma=(x:\prim,y:[z:\prim]z)$.
On the one hand, we have $\Gamma\sgv (y\,x):x$.
On the other hand, since $\nrm{\Gamma}{x}=x$, $\nrm{\Gamma,x}{y}=\nrm{\Gamma,x}{[z:\prim]z}=[z:\prim]\nrm{\Gamma,z\smydef x}{z}=[z:\prim]x$ and  $\nrm{}{\Gamma}=\Gamma\sgv x\ginc\prim$, we have
$\nrm{\Gamma}{(y\,x)}=x$.
As a consequence when norming \eg\ $([z:x]\prim\;(y\,x))]$ we cannot use typing for the application condition but rather have to use inclusion.
\item[E.6]
Let $\Gamma=(x:\prim,y:[z!x][[z\fun z],z])$.
We have $\nrm{\Gamma}{y}=[z!x][[z\fun z],z]$ and 
$\nrm{\Gamma}{\pleft{y}}=x$.
Furthermore since $\nrm{\Gamma,2_{\pleft{y}}}{y}
=\nrm{\Gamma,2_{\pleft{y}}}{[z!x][[z\fun z],z]}
=[z!x]\nrm{\Gamma,z:=\pleft{y}}{[[z\fun z],z]}
=[z!x][[x\fun x],x]$ we have $\nrm{\Gamma}{\pright{y}}=[[x\fun x],x]$
and therefore $\nrm{\Gamma}{\pleft{\pright{y}}}=[x\fun x]$ and $\nrm{\Gamma}{\pright{\pright{y}}}=x$. 
%
\item[E.7]
Let $\Gamma=(x:\prim)$ and $t=\prdef{y}{x^{\prim}}{[x\fun x]}{[y\fun\prim]}$.
We have  
$\nrm{}{\Gamma}\gv\nrm{\Gamma}{x}=x:\prim$ and
$\nrm{}{\Gamma}\gv\nrm{\Gamma}{[x\fun x]}=[x\fun x]:[x\fun\prim]$ and therefore
$\nrm{\Gamma}{t}=\prdef{y}{x^{\prim}}{[x\fun x]}{\nrm{\Gamma,y:\prim}{[y\fun\prim]}}=t$.
As a consequence we have
$\nrm{\Gamma}{\pleft{t}}=x$ and 
$\nrm{\Gamma}{\pright{t}}=[x\fun x]$.
\end{itemize}
\end{remark}
\begin{remark}[Norming of case distinctions]
\label{norm.cased}
While the above definition for norming case distinctions is satisfying the desired property that norms are preserved by reduction (Law \ref{sub.nrm.rd}). 
It does not satisfy the desired property that all valid expression are normable (Law \ref{sub.val.nrm}).
A very simple example is $a=[x:[\any{}\fun\any{}]][x?x]$. Obviously $\sgv a:[[\any{}+\any{}]\fun\any{}]$ but not $\sngv a$ since $\nrm{x:[\any{}\fun\any{}]}{x}=x$ is not an abstraction.
While this might be handled by a notion of norm-type, there is a more basic underlying issue.
To see this let
\[ a=([x:\any{1}][\any{}\fun x]\,\any{}), \;b=[\any{}\fun\any{1}] 
\]
Obviously $\sgv[a?b]$ since $a\rd[\any{}\fun\any{}]$ which implies $\sgv a:[\any{}\fun\any{1}]$, and $\sgv b:[\any{}\fun\any{1}]$, and therefore  $\sgv[a?b]:[[\any{}+\any{}]\fun\any{1}]$.
However $[a?b]$ is not normable since
\[
\nrm{}{a}=[\any{}\fun\any{}], \; \nrm{}{b}=[\any{}\fun\any{1}],
\]
Modifying the definition of norming by some notion of upper or lower bound of two norms would be in violation of axioms $\beta_3$, $\beta_4$.

\end{remark}
\subsubsection{Basic properties of norming}
\noindent
The following property ensures that norming of application does not lead to free identifiers.
\begin{law}[Norming and free variables]
\label{sub.nrm.basic}
For all $\Lambda$, $\Lambda_1$, $\Lambda_2$, $\Theta$, $x$, $a$, and $b$:
\begin{itemize}
\item[$i$:]
If $\Lambda=(\Lambda_1,x\mydef a,\Lambda_2)$ and $\Lambda\;\sngv\;\nrm{\Lambda,\Theta}{b}$ then $x\notin\free(\nrm{\Lambda,\Theta}{b})$.
\item[$ii$:]
If $\nrm{\Lambda,(b,\Theta)}{a}=\binbop{x}{\bar{c}}{\bar{d}}$ or $\nrm{\Lambda,(2_b,\Theta)}{a}=[x\sdef\bar{c}]\bar{d}$ for some  $\bar{c}$, $\bar{d}$
then $x\notin\free(\bar{d})$.
\end{itemize}
\end{law}
\begin{proof}
Consider arbitrary all $\Lambda$, $\Lambda_1$, $\Lambda_2$, $\Theta$, $x$, $a$, and $b$:

Part $i$ can be shown by induction on the definition of $\Lambda\sngv\nrm{\Lambda,\Theta}{b}$.
The proof is straightforward since definitions are always unfolded during norming.

For Part $ii$ the proof is a bit technical but not really difficult.
We need to define an auxiliary binary relation $E_{\Theta}(\bar{a},\bar{b})$ on norms: For all $\bar{a}$, $\bar{a}_1$, $\bar{a}_2$, $\bar{b}$, $\bar{c}$, $\bar{d}$, e, $x$: 
$E_{()}(\bar{a},\bar{a})$;
$E_{(e,\Theta)}(\binbop{x}{\bar{c}}{\bar{a}},\bar{d})$  if $E_{\Theta}(\bar{a},\bar{d})$;
$E_{(i,\Theta)}(\prsumop{\bar{a}_1}{\bar{a}_2},\bar{d})$ for $i=1,2$, if $E_{\Theta}(\bar{a}_i,\bar{d})$;
$E_{(1,\Theta)}([x\sdef\bar{a}_1]\bar{a}_2,\bar{d})$ and $E_{(1,\Theta)}(\prdef{x}{\bar{a}_1^{\bar{c}}}{\bar{a}_2}{\bar{d}})$ if $E_{\Theta}(\bar{a}_1,\bar{d})$;
$E_{(2_e,\Theta)}([x\sdef\bar{a}_1]\bar{a}_2,\bar{d})$ and $E_{(2_e,\Theta)}(\prdef{x}{\bar{a}_1^{\bar{c}}}{\bar{a}_2}{\bar{d}})$  if $E_{\Theta}(\bar{a}_2,\bar{d})$; 
$E_{\Theta}(\injl{\bar{a}_1}{\bar{a}_2},\bar{d})$ and $E_{\Theta}(\injr{\bar{a}_1}{\bar{a}_2},\bar{d})$ if $E_{\Theta}(\bar{a}_i,\bar{d})$ for $i=1,2$;

We then prove for all $\Lambda$, $\Theta_1$, $\Theta_2$, $a$, $b$ with $\Theta=(\Theta_1,b,\Theta_2)$ by induction on the definition of $\nrm{\Lambda,\Theta}{a}$ that $E_{\Theta_1}(\nrm{\Lambda,\Theta}{a},\binbop{x}{\bar{c}}{\bar{d}})$ for some $x$, $\bar{c}$, $\bar{d}$ implies $x\notin\free(\bar{d})$. 
\begin{itemize}
\item $a=\any{n}$: Trivial, since $\Theta\neq()$.
\item $a=y$: the case $\Theta=()$ is not possible, otherwise the proposition follows from the inductive hypothesis.
\item $a=\binbop{y}{a_1}{a_2}$ for some $y$, $a_1$, and $a_2$: There are several subcases for $\Theta$ in the definition of norming of abstractions:
\begin{itemize}
\item
$\Theta=()$: Not possible,
\item 
$\Theta=(b,\Theta_2)$, i.e.~$\Theta_1=()$ and $x=y$:  
Since $\nrm{\Lambda,\Theta}{\binbop{x}{a_1}{a_2}}$ = $\binbop{x}{\nrm{\Lambda}{a_1}}{\nrm{(\Lambda,x\smydef b),\Theta_2}{a_2}}$, 
by Part $i$ we know that $x\notin\free(\nrm{(\Lambda,y\smydef b),\Theta_2}{a_2})$.
\item
$\Theta=(c,\Theta_3,b,\Theta_2)$ for some $c$ and $\Theta_3$:
From $E_{c,\Theta_3}(\nrm{\Lambda,\Theta)}{a},\binbop{x}{\bar{c}}{\bar{d}})$ we can infer $E_{\Theta_3}(\nrm{(\Lambda,y\smydef c),(\Theta_3,b,\Theta_2)}{a_2},\binbop{x}{\bar{c}}{\bar{d}})$.
By inductive hypothesis $x\notin\free(\bar{d})$.
\item
$\Theta=(1,\Theta_3,b,\Theta_2)$ for some $\Theta_3$: Hence $a=[y\sdef a_1]a_2$.
From $E_{1,\Theta_3}(\nrm{\Lambda,\Theta}{a},\binbop{x}{\bar{c}}{\bar{d}})$ we can infer $E_{\Theta_3}(\nrm{(\Lambda,()}{a_1},\binbop{x}{\bar{c}}{\bar{d}})$.
By inductive hypothesis $x\notin\free(\bar{d})$.
\item
$\Theta_1=(2_c,\Theta_3,b,\Theta_2)$ for some $c$ and $\Theta_3$:
Hence $a=[y\sdef a_1]a_2$.
$E_{2_c,\Theta_3}(\nrm{\Lambda,\Theta}{a},\binbop{x}{\bar{c}}{\bar{d}})$ implies $E_{\Theta_3}(\nrm{(\Lambda,y\smydef c),(\Theta_3,b,\Theta_2)}{a_2},\binbop{x}{\bar{c}}{\bar{d}})$.
By inductive hypothesis $x\notin\free(\bar{d})$.
\end{itemize}
\item $a=(a_1\,a_2)$ for some $a_1$ and $a_2$: Let $\nrm{\Lambda,(\Theta_1,b,\Theta_2)}{(a_1\,a_2)}=\binbop{x}{\bar{c}}{\bar{d}}$. 
Obviously $\nrm{\Lambda,(a_2,\Theta_1,b,\Theta_2))}{a_1}=\binbop{y}{\bar{f}}{\binbop{x}{\bar{c}}{\bar{d}}}$ for some $\bar{f}$.
Hence from $E_{\Theta_1}(\nrm{\Lambda,(\Theta_1,b,\Theta_2)}{a},\binbop{x}{\bar{c}}{\bar{d}})$ we can infer $E_{(a_2,\Theta_1)}(\nrm{\Lambda,(a_2,\Theta_1,b,\Theta_2)}{a_1},\binbop{x}{\bar{c}}{\bar{d}})$. 
By inductive hypothesis $x\notin\free(\bar{d})$.
\item $a=\pleft{a_1}$, for some $a_1$. 
We know that $\nrm{\Lambda,(1,\Theta_1,b,\Theta_2)}{a_1}=\prsumop{\nrm{\Lambda,(\Theta_1,b,\Theta_2)}{a}}{\bar{b}_2}$ or $\nrm{\Lambda,(1,\Theta_1,b,\Theta_2)}{a_1}=[y\sdef\nrm{\Lambda,(\Theta_1,b,\Theta_2)}{a}]\bar{b}_2$.
From $E_{\Theta_1}(\nrm{\Lambda,(\Theta_1,b,\Theta_2)}{a},\binbop{x}{\bar{c}}{\bar{d}})$ we can infer $E_{(1,\Theta_1)}(\nrm{\Lambda,(1,\Theta_1,b,\Theta_2)}{a_1},\binbop{x}{\bar{c}}{\bar{d}})$.
By inductive hypothesis $x\notin\free(\bar{d})$.
\item 
The remaining cases of right-projection, protected definitions,  products, sums, injections, case distinctions, and negation can be shown in a similar style.
\end{itemize}
\noindent
The proposition follows with $\Theta_1=()$ since $E_{()}(\bar{a},\bar{a})$ for all $\bar{a}$. 

\noindent
We can prove in a similar way by induction on the definition of $\nrm{\Lambda,(\Theta_1,2_b,\Theta_2)}{a}$ that $E_{\Theta_1}(\nrm{\Lambda,(\Theta_1,2_b,\Theta_2)}{a},[x!\bar{c}]\bar{d})$ for some $\bar{c}$, $\bar{d}$ implies $x\notin\free(\bar{d})$. 
\qedhere
\end{proof}
%
\subsubsection{Properties of inclusion on norms}
\noindent
First, we note some decomposition properties of generalized inclusion on norms.
\begin{law}[Generalized inclusions on norms]
\label{sub.ginc.prop}
For all $\bar{\Gamma}$, $x$, $\bar{a}$,  $\bar{a}_1$,  $\bar{a}_2$, $\bar{b}$, $\bar{b}_1$, $\bar{b}_2$ where $\bar{a}\notin\dvar$: 
\begin{itemize}
\item[$i$:]
$\Gamma\sgv\any{n}\ginc\bar{b}$ for some $n$ implies $\bar{b}\eqv\any{m}$ for some $m$. 
\item[$ii$:]
$\bar{\Gamma}\sgv x\ginc\bar{b}$ implies $\bar{b}=\any{m}$ for some $m$, or $\bar{b}=x$, or $\bar{\Gamma}\sgv x:\bar{b}$. 
\item[$iii$:]
If $\bar{\Gamma}\sgv\bar{a}\ginc[x:\bar{b}_1]\bar{b}_2$ and $\bar{a}\notin\dvar$ then $\bar{a}=\binbop{x}{\bar{b}_1}{\bar{c}}$ for some $\bar{c}$ where $\bar{\Gamma},x:\bar{b}_1\sgv\bar{c}\ginc\bar{b}_2$.
\item[$iv$:]
If $\bar{\Gamma}\sgv[x:\bar{a}_1]\bar{a}_2\ginc\bar{b}$ then either $\bar{b}=\any{m}$ for some $m$ or $\bar{b}=[x:\bar{a}_1]\bar{c}$ for some $\bar{c}$ where $\bar{\Gamma},x:\bar{a}_1\sgv\bar{a}_2\ginc\bar{c}$.
\item[$v$:]
If $\bar{\Gamma}\sgv\bar{a}\ginc[\bar{b}_1,\bar{b}_2]$ and $\bar{a}\notin\dvar$ then $\bar{a}=\prsumop{\bar{c}_1}{\bar{c}_2}$ for some $\bar{c}_1$, $\bar{c}_2$
where $\bar{\Gamma}\sgv\bar{c}_i\ginc\bar{b}_i$. for $i=1,2$.
\item[$vi$:]
If $\bar{\Gamma}\sgv[\bar{a}_1,\bar{a}_2]\ginc\bar{b}$ then either $\bar{b}=\any{m}$ for some $m$ or $\bar{b}=[\bar{c}_1,\bar{c}_2]$ for some $\bar{c}_1$, $\bar{c}_2$ where $\bar{\Gamma}\gv\bar{a}_i\ginc\bar{c}_i$ for $i=1,2$.
\item[$vii$:]
If $\bar{\Gamma}\sgv\bar{a}\ginc[x\sdef\bar{b}_1]\bar{b}_2$ and $\bar{a}\notin\dvar$ then either $\bar{a}=[x\sdef\bar{b}_1]\bar{c}_2$ where $\bar{\Gamma},x:\bar{b}_1\sgv\bar{c}_2\leq\bar{b}_2$ 
or $\bar{\Gamma}\sgv\bar{a}:[x\sdef\bar{b}_1]\bar{b}_2$ and $\bar{a}=\prdef{x}{\bar{c}_1^{\bar{b}_1}}{\bar{c}_2}{\bar{b}_2}$ 
for some $\bar{c}_1$, $\bar{c}_2$. 
\item[$viii$:]
If $\bar{\Gamma}\sgv[x\sdef\bar{a}_1]\bar{a}_2\ginc\bar{b}$ then either $\bar{b}=\any{m}$ for some $m$ or $\bar{b}=\binbop{x}{\bar{a}_1}{\bar{c}}$ for some $\bar{c}$ where $\bar{\Gamma},x:\bar{a}_1\gv \bar{a}_2\ginc\bar{c}$.
\end{itemize}
\end{law}
\begin{proof}
For all $\bar{\Gamma}$, $x$,  $\bar{a}$,  $\bar{a}_1$,  $\bar{a}_2$,  $\bar{b}$,  $\bar{b}_1$, $\bar{b}_2$ where $\bar{a}\notin\dvar$:
\begin{itemize}
\item[$i:$]
If $\bar{\Gamma}\sgv\any{n}\leq\bar{b}$ for some $n$ the property follows Laws \ref{sub.leq.prop}($i$) and \ref{sub.any}$(i$). 
If $\bar{\Gamma}\sgv\any{n}:\bar{b}$ for some $n$ the property follows from Law \ref{sub.any}$(iii$). 
\item[$ii:$]
By definition of inclusion $\bar{\Gamma}\sgv x\leq\bar{b}$ can only be introduced by rules \srefl\ and \sembed. 
\item[$iii$:]
If $\bar{\Gamma}\sgv\bar{a}\leq[x:\bar{b}_1]\bar{b}_2$ the property follows from Law~\ref{sub.leq.prop}($ii$). 
If $\bar{\Gamma}\sgv\bar{a}:[x:\bar{b}_1]\bar{b}_2$ then since $\bar{a}\notin\dvar$ and Part $i$ it can only be an abstraction
and therefore by Law \ref{sub.type.decomp}($iv$) we know that $\bar{a}=[x:\bar{b}_1]\bar{c}$, for some $\bar{c}$ where $\bar{\Gamma},x:\bar{b}_1\sgv\bar{c}:\bar{b}_2$ which implies the proposition.
\item[$iv$:]
If $\bar{\Gamma}\sgv[x:\bar{a}_1]\bar{a}_2\leq\bar{b}$ the property follows from Law \ref{sub.leq.prop}($iii$).
Otherwise, if $\bar{\Gamma}\sgv[x:\bar{a}_1]\bar{a}_2:\bar{b}$ then by Laws \ref{sub.type.decomp}($iii$) and \ref{sub.nrm.fund} we know that
$\bar{\Gamma},x:\bar{a}_1\gv \bar{a}_2:\bar{d}$ for some $\bar{d}$ where $\bar{\Gamma}\sgv[x:\bar{a}_1]\bar{d}\leq\bar{b}$.
By Law \ref{sub.leq.prop}($iii$) we have $\bar{b}=\any{m}$ for some $m$ or $\bar{b}=[x:\bar{a}_1]\bar{c}$ for some $\bar{c}$ where $\bar{\Gamma},x:\bar{a}_1\gv\bar{d}\leq\bar{c}$.
The first case is obvious, in the second case by rule \sinc\ $\bar{\Gamma},x:\bar{a}_1\gv \bar{a}_2:\bar{c}$ which implies the proposition.
\item[$v$:]
$\bar{\Gamma}\sgv\bar{a}\ginc[\bar{b}_1,\bar{b}_2]$ means that either $\bar{\Gamma}\sgv\bar{a}\leq[\bar{b}_1,\bar{b}_2]$ or $\bar{\Gamma}\sgv\bar{a}:[\bar{b}_1,\bar{b}_2]$.
In the first case the property follows from Law~\ref{sub.leq.prop}($iv$). 
In the second case, since $\bar{a}\notin\dvar$, it can only be the primitive constant, an abstraction, a product, a sum, an injection, or a protected definition.
Using the various cases of Law~\ref{sub.type.decomp} we can exclude all of these cases except for products and sums (Law~\ref{sub.type.decomp}($vi$)) 
and therefore know that $\bar{a}=\prsumop{\bar{c}_1}{\bar{c}_2}$, for some $\bar{c}_i$ where $\bar{\Gamma}\sgv\bar{c}_i:\bar{b}_i$
which implies the proposition
\item[$vi$:]
Similar to Part $iv$ using Law \ref{sub.type.decomp}($vi$) and \ref{sub.leq.prop}($iii$)
\item[$vii$:]
If $\bar{\Gamma}\sgv\bar{a}\leq[x\sdef\bar{b}_1]\bar{b}_2$ the argument is similar to Part $iii$.
If $\bar{\Gamma}\sgv\bar{a}:[x\sdef\bar{b}_1]\bar{b}_2$ the argument is similar to $v$ excluding all cases but Law~\ref{sub.type.decomp}($vii$).
\item[$viii$:]
Similar to Part $iv$.
\qedhere
\end{itemize}
\end{proof}
\noindent
We introduce the notion of \emph{flexible inclusion} which essentially augments generalized inclusion on norms over the norm constructors.
This notion will be needed 
in Section \ref{sub.norm.typ} for understanding the relation between norming and typing.
\begin{definition}[Flexible inclusion]
\label{inclusion.finc}
\nomenclature[loRel05]{$\Gamma\sgv a\finc b$}{flexible inclusion}%
\index{inclusion!flexible}
\label{sub.inst.flex}
The definition of flexible inclusion is shown in Table~\ref{sub.finc.rules}. 
\begin{table}[!htb]
\fbox{
\begin{minipage}{0.96\textwidth}
\begin{align*}
\sfembedm\;\;&\frac{\bar{\Gamma}\sgv\bar{a}\ginc\bar{b}}{\bar{\Gamma}\sgv\bar{a}\finc\bar{b}}\\
\sfabsm\;\;&\frac{\bar{\Gamma},x:\bar{a}\sgv\bar{b}\finc\bar{c}}{\bar{\Gamma}\sgv\binbop{x}{\bar{a}}{\bar{b}}\finc[x:\bar{a}]\bar{c}}&
\sfabsem\;\;&\frac{\bar{\Gamma},x:\bar{a}\sgv\bar{b}\leq\bar{c}}{\bar{\Gamma}\sgv[x!\bar{a}]\bar{b}\finc[x!\bar{a}]\bar{c}}\\
\sfbprodm\;\;&\frac{\bar{\Gamma}\sgv\bar{a}\finc\bar{b}\quad\bar{c}\finc\bar{d}}{\bar{\Gamma}\sgv\prsumop{\bar{a}}{\bar{c}}\finc[\bar{b},\bar{d}]}&
\sfpdefm\;\;&\frac{}{\bar{\Gamma}\sgv\prdef{x}{\bar{a}^{\bar{b}}}{\bar{c}}{\bar{d}}\finc[x!\bar{b}]\bar{d}}\\
\finjllm\;\;&\frac{\bar{\Gamma}\sgv\bar{a}\finc\bar{b}}{\bar{\Gamma}\sgv\injl{\bar{a}}{\bar{c}}\finc[\bar{b}+\bar{c}]}&
\finjlrm\;\;&\frac{\bar{\Gamma}\sgv\bar{b}\finc\bar{c}}{\bar{\Gamma}\sgv\injr{\bar{a}}{\bar{b}}\finc[\bar{a}+\bar{c}]}
\end{align*}
\end{minipage}
}
\caption{Rules for flexible inclusion.\label{sub.finc.rules}}
\end{table}
\end{definition}
\noindent
Flexible inclusions can be extended by (generalized) inclusion. 
\begin{law}[Flexible inclusion and inclusion]
\label{sub.flex.incl}
For all $\bar{\Gamma}$, $\bar{a}$, $\bar{b}$, $\bar{c}$, $n$:
$\bar{\Gamma}\sgv\bar{a}\finc\bar{b}$ and $\bar{\Gamma}\sgv\bar{b}<\any{n}$ imply $\bar{\Gamma}\sgv\bar{a}<\any{n}$. 
$\bar{\Gamma}\sgv\bar{a}\finc\bar{b}$ and $\bar{\Gamma}\sgv\bar{b}\leq\bar{c}$ imply $\bar{\Gamma}\sgv\bar{a}\finc\bar{c}$. 
\end{law}
\begin{proof}
Both properties are shown by induction on the definition of $\bar{\Gamma}\sgv\bar{a}\finc\bar{b}$.

Consider the the first property: 
In case of rule \sfembed\ the property follows either from typing rule \sinc\ or, since norms are irreducible, from Law \ref{sub.leq.top}.
In case of rule \sfabs\ we have $\bar{a}=[x:\bar{a}_1]\bar{a}_2$ and $\bar{b}=[x:\bar{a}_1]\bar{b}_2$ where  $\bar{\Gamma},x:\bar{a}_1\sgv\bar{b}_1\finc\bar{b}_2$.
Since $\bar{\Gamma}\sgv[x:\bar{a}_1]\bar{b}_2<\any{n}$, obviously  $\bar{\Gamma},x:\bar{a}_1\sgv\bar{b}_2<\any{n}$.
By inductive hypothesis we know that $\bar{\Gamma},x:\bar{a}_1\sgv\bar{a}_2<\any{n}$ hence $\bar{\Gamma}\sgv\bar{a}<\any{n}$.
The other structural rules are shown in a similar style. 

Consider the second one: In case of rule \sfembed\ the property follows either from typing rule \sinc\ or from Law \ref{sub.leq.por}.
In case of rule \sfabs\ we have $\bar{a}=[x:\bar{a}_1]\bar{a}_2$ and $\bar{b}=[x:\bar{a}_1]\bar{b}_2$ where  $\bar{\Gamma},x:\bar{a}_1\sgv\bar{b}_1\finc\bar{b}_2$.
Since $\bar{\Gamma}\sgv[x:\bar{a}_1]\bar{b}_2\leq\bar{c}$, by Law \ref{sub.leq.prop}($iii$) there are two cases:
If $\bar{c}=[x:\bar{a}_1]\bar{c}_2$ where $\bar{\Gamma},x:\bar{a}_1\sgv\bar{b}_2\leq\bar{c}_2$ then by inductive hypothesis we know that  $\bar{\Gamma},x:\bar{a}_1\sgv\bar{a}_2\finc\bar{c}_2$ and hence
$\bar{\Gamma}\sgv\bar{a}\finc\bar{c}$.
If $\bar{c}=\any{n}$ for some $n$ then by the first property we know that $\bar{\Gamma}\sgv\bar{a}\finc\bar{c}$.
The other structural rules are shown in a similar style. 
\end{proof}
\noindent
For later use we show some decomposition properties of flexible inclusion.
\begin{law}[Flexible inclusion on norms]
\label{sub.flex.prop}
For all $\bar{\Gamma}$, $x$, $\bar{a}$,  $\bar{a}_1$,  $\bar{a}_2$,  $\bar{b}$, $\bar{b}_1$, $\bar{b}_2$ where $\bar{a}\notin\dvar$: 
\begin{itemize}
\item[$i$:]
If $\bar{\Gamma}\sgv\any{n}\finc\bar{b}$ for soem $n$ then $\bar{b}=\any{m}$ for some $m$ where $m\geq n$.
\item[$ii$:]
$\bar{\Gamma}\sgv x\finc\bar{b}$ implies $\bar{b}=\any{m}$ for some $m$, $\bar{b}=x$, or $\bar{\Gamma}\sgv x:\bar{b}$. 
\item[$iii$:]
If $\bar{\Gamma}\sgv\bar{a}\finc[x:\bar{b}_1]\bar{b}_2$ and $\bar{a}\notin\dvar$ then $\bar{a}=\binbop{x}{\bar{b}_1}{\bar{c}}$ for some $\bar{c}$ where $\bar{\Gamma},x:\bar{b}_1\sgv\bar{c}\finc\bar{b}_2$.
\item[$iv$:]
If $\bar{\Gamma}\sgv[x:\bar{a}_1]\bar{a}_2\finc\bar{b}$ then either $\bar{b}=\any{m}$ for some $m$ or $\bar{b}=[x:\bar{a}_1]\bar{c}$  for some $\bar{c}$ where $\bar{\Gamma},x:\bar{a}_1\gv \bar{a}_2\finc\bar{c}$.
\item[$v$:]
If $\bar{\Gamma}\sgv\bar{a}\finc[\bar{b}_1,\bar{b}_2]$ and $\bar{a}\notin\dvar$  then $\bar{a}=\prsumop{\bar{c}_1}{\bar{c}_2}$  for some $\bar{c}_1$, $\bar{c}_2$
where $\bar{\Gamma}\sgv\bar{c}_i\finc\bar{b}_i$. 
\item[$vi$:]
If $\bar{\Gamma}\sgv[\bar{a}_1,\bar{a}_2]\finc\bar{b}$ then either $\bar{b}=\any{m}$ for some $m$ or $\bar{b}=[\bar{c}_1,\bar{c}_2]$  for some $\bar{c}_1$, $\bar{c}_2$ where $\bar{\Gamma}\gv\bar{a}_i\finc\bar{c}_i$ for $i=1,2$.
\item[$vii$:]
If $\bar{\Gamma}\sgv\bar{a}\finc[x\sdef\bar{b}_1]\bar{b}_2$ and $\bar{a}\notin\dvar$  then either $\bar{a}=\prdef{x}{\bar{c}_1^{\bar{b}_1}}{\bar{c}_2}{\bar{b}_2}$ 
or $\bar{a}=[x\sdef\bar{b}_1]\bar{c}_2$ where $\bar{\Gamma},x:\bar{b}_1\sgv\bar{c}_2\leq\bar{b}_2$  for some $\bar{c}_1$, $\bar{c}_2$. 
\item[$viii$:]
If $\bar{\Gamma}\sgv[x\sdef\bar{a}_1]\bar{a}_2\finc\bar{b}$ then either $\bar{b}=\any{m}$ for some $m$ or $\bar{b}=\binbop{x}{\bar{a}_1}{\bar{c}}$ for some $\bar{c}$ where $\bar{\Gamma},x:\bar{a}_1\gv \bar{a}_2\finc\bar{c}$.
\end{itemize}
\end{law}
\begin{proof}
For all $\bar{\Gamma}$, $x$, $\bar{a}$,  $\bar{a}_1$,  $\bar{a}_2$,  $\bar{b}$, $\bar{b}_1$, $\bar{b}_2$ where $\bar{a}\notin\dvar$:
\begin{itemize}
\item[$i:$]
Proof by induction on the definition of $\bar{\Gamma}\sgv\any{n}\finc\bar{b}$.
For rule \sfembed\ the property follows from Law \ref{sub.ginc.prop}($i$).
For the other rules rules the property is trivial.
\item[$ii$:]
Follows from Law \ref{sub.ginc.prop}($ii$) and the structural rules of flexible inclusion.
\item[$iii$:]
For rule \sfembed\ the property follows from Law \ref{sub.ginc.prop}($iii$).
In case of the structural rule \sfabs\ the property follows immediately since $\bar{b}=\any{m}$ is not possible.
For the other rules the property is trivial.

\item[$iv$:]
For rule \sfembed\ the property follows from Law \ref{sub.ginc.prop}($iv$).
The other rules are analogous to Part $iii$.
\item[$v$:]
For rule \sfembed\ the property follows from Law \ref{sub.ginc.prop}($v$).
In case of the structural rule \sfbprod\ the property follows immediately since $\bar{b}=\any{m}$ is not possible.
For the other rules the property is trivial.
\item[$vi$:]
For rule \sfembed\ the property follows from Law \ref{sub.ginc.prop}($vi$).
The other rules are analogous to Part $v$.
\item[$vii$:]
For rule \sfembed\ the property follows from Law \ref{sub.ginc.prop}($vii$).
In case of the rules \sfpdef\ and \sfabse\ the property follows immediately.
For the other rules the property is trivial.
\item[$viii$:]
For rule \sfembed\ the property follows from Law \ref{sub.ginc.prop}($viii$).
The other rules are analogous to Part $iv$.
\qedhere
\end{itemize}
\end{proof}
\noindent
\begin{law}[Flexible inclusion is antisymmetric]%
\label{sub.flex.anti}
For all $\bar{\Gamma}$, $\bar{a}$, $\bar{b}$: $\bar{\Gamma}\sgv\bar{a}\finc\bar{b}$ and $\bar{\Gamma}\sgv\bar{b}\finc\bar{a}$ implies $\bar{a}=\bar{b}$.
\end{law}
\begin{proof}
We first show that $\bar{\Gamma}\sgv\bar{a}:\bar{b}$ and $\bar{\Gamma}\sgv\bar{b}\finc\bar{a}$ imply $\bar{a}=\bar{b}$
and that $\bar{\Gamma}\sgv\bar{a}:\bar{b}$ and  $\bar{\Gamma}\sgv\bar{b}\finc\bar{a}$ imply $\bar{a}=\bar{b}$.
The argument is by simultaneous induction on $\bar{\Gamma}\sgv\bar{a}:\bar{b}$ and $\bar{\Gamma}\sgv\bar{a}\leq\bar{b}$. 
We begin with $\bar{\Gamma}\sgv\bar{a}:\bar{b}$:
\begin{meditemize}
\item[\ax:]
Obviously $\bar{a}=\bar{b}=\any{m}$ for some $m$.
\item[\mystart:]
$\bar{\Gamma}=(\bar{\Gamma}',x:\bar{b})$, $\bar{a}=x$ for some $\bar{\Gamma}'$ and $x$.
On the other hand $\bar{\Gamma}\sgv\bar{b}\finc x$ which is obviously impossible.
\item[\weak:]
Follows from obvious weakening properties for typing and the inductive hypothesis. 
\item[\sinc:]
$\bar{\Gamma}\sgv\bar{a}:\bar{c}$ where $\bar{\Gamma}\sgv\bar{c}\leq\bar{b}$ for some $\bar{c}$ and $\bar{\Gamma}\sgv\bar{b}\finc\bar{a}$.
By an obvious argument $\bar{\Gamma}\sgv\bar{c}\finc\bar{a}$ hence by inductive hypothesis for $\bar{\Gamma}\sgv\bar{a}:\bar{c}$ we have $\bar{c}=\bar{a}$
and therefore by inductive hypothesis for $\bar{\Gamma}\sgv\bar{c}\leq\bar{b}$ we have $\bar{b}=\bar{a}$.
\item[\absu:]
Follows from the definition of typing and the inductive hypothesis.
\item[$\ldots$:]
The other structural cases follow from the definition of typing and the inductive hypothesis.
\end{meditemize}
\noindent
Next come $\bar{\Gamma}\sgv\bar{a}\leq\bar{b}$:
\begin{meditemize}
\item[\srefl:]
Obvious.
\item[\sembed:]
We have $\bar{b}=\any{n}$ for some $n$ where $\bar{\Gamma}\sgv\bar{a}<\any{n}$ and $\bar{\Gamma}\sgv\any{n}\finc\bar{a}$. By an obvious argument the latter implies that $\bar{a}=\any{m}$ for some $m\geq n$.
However $\bar{\Gamma}\sgv\any{m}<\any{n}$ obviously implies $m<n$. Hence this case is trivial.  
\item[\sabs:]
Follows from the definition of inclusion and the inductive hypothesis.
\item[$\ldots$:]
The other structural cases follow from the definition of inclusion and the inductive hypothesis.
\end{meditemize}
\noindent
The main property is shown by induction on $\bar{\Gamma}\sgv\bar{a}\finc\bar{b}$.
The above arguments have shown the base case \sfembed. The structural cases are straightforward.
\end{proof}
\noindent
\subsubsection{Norming and reduction}
The Law~\ref{nrm.eq}(norm equality in context) is also valid for the modified version of norming.
\begin{law}[Norm equality in context]
\label{sub.nrm.eq}
Let $\Lambda_a=(\Lambda_1,x\odot a,\Lambda_2)$ and $\Lambda_b=(\Lambda_1,x\odot b,\Lambda_2)$ for some $\Lambda_1,\Lambda_2,x,a,b$. For all $c$: If $\Lambda_1\sngv a$, $\Lambda_1\sngv b$, $\nrm{\Lambda_1}{a}=\nrm{\Lambda_1}{b}$, and $\Lambda_a\sngv c$ then $\Lambda_b\sngv c$ and $\nrm{\Lambda_a}{c}=\nrm{\Lambda_b}{c}$.
\end{law}
\begin{proof}
Straightforward induction on the definition of $\Lambda_a\sngv c$.
\end{proof}
\begin{law}[Norm equality in elimination list]
\label{sub.nrm.eq.elim}
For all $\Lambda$, $\Theta$, $a$, $b$, $c$: If $\Lambda\sngv a$, $\Lambda\sngv b$, $\nrm{\Lambda}{a}=\nrm{\Lambda}{b}$, and $\Lambda,(\Theta,a)\sngv c$ then $\Lambda,(\Theta,b)\sngv c$ 
and $\nrm{\Lambda,(\Theta,a)}{c}=\nrm{\Lambda,(\Theta,b)}{c}$.
\end{law}
\begin{proof}
Straightforward induction on the definition of $\Lambda,(a,\Theta)\sngv c$ using Law \ref{sub.nrm.eq} in case of abstractions 
\end{proof}

As discussed above in the example E.4 of norming, the Law~\ref{nrm.sub}(substitution and norming) is not valid anymore for norming.
However a restricted version can still be shown.
\begin{law}[Norm inner substitution property]
\label{sub.nrm.sub}
For all $x$, $a$, $b$, $\Lambda_1$,$\Lambda_2$, $\Theta$ let $\Lambda=(\Lambda_1,x\mydef a,\Lambda_2)$, $\Lambda'=(\Lambda_1,\Lambda_2\gsub{x}{a})$, $\Theta'=\Theta\gsub{x}{a}$.
If $\Lambda_1,x\mydef a\sngv \Lambda_2$, $\Lambda_1\sngv\Lambda_2\gsub{x}{a}$, and $\nrm{\Lambda_1,x\smydef a}{\Lambda_2}=\nrm{\Lambda_1}{\Lambda_2\gsub{x}{a}}$ 
then $\Lambda,\Theta\sngv b$ iff $\Lambda',\Theta'\sngv b\gsub{x}{a}$ and $\nrm{\Lambda,\Theta}{b}=\nrm{\Lambda',\Theta'}{b\gsub{x}{a}}$. 
\end{law}
\begin{proof}
Proof by induction on the definition of $\nrm{\Lambda,\Theta}{b}$.
Before that, note that the assumption $\nrm{\Lambda_1,x\smydef a}{\Lambda_2}=\nrm{\Lambda_1}{\Lambda_2\gsub{x}{a}}$ is equivalent to $\nrm{}{\Lambda}=\nrm{}{\Lambda}'$ since
$\nrm{}{\Lambda}=(\nrm{}{\Lambda_1},\nrm{\Lambda_1,x\smydef a}{\Lambda_2})$ and $(\nrm{}{\Lambda_1},\nrm{\Lambda_1}{\Lambda_2\gsub{x}{a}})=\nrm{}{\Lambda'}$. 
\begin{itemize}
\item
$b=\any{n}$: 
Obviously $\nrm{\Lambda,\Theta}{\any{n}}=\any{n}=\nrm{\Theta',\Lambda'}{\any{n}}=\nrm{\Theta',\Lambda'}{\any{n}\gsub{x}{a}}$.
\item
$b=x$: We have $x\in\domdef(\Lambda)$ and $\nrm{\Lambda,\Theta}{x}=\nrm{\Lambda,\Theta}{a}$.
Since obviously $\free(a)\cap{\dom}(\Lambda_2)=\emptyset$ we have
$\nrm{\Lambda,\Theta}{a}=\nrm{\Lambda',\Theta'}{a}=\nrm{\Lambda',\Theta'}{x\gsub{x}{a}}$.
\item
$b=y$ for some $y$ where $y\neq x$ and $y\gsub{x}{a}=y$. If $\Lambda,\Theta\sngv y$ we need to show that $\Lambda',\Theta'\sngv y$ and $\nrm{\Lambda,\Theta}{y}=\nrm{\Lambda',\Theta'}{y}$.
There are three subcases:
\begin{itemize}
\item
$y\in\domdec(\Lambda)$, $\Theta=()$: Obviously $\nrm{\Lambda}{y}=y=\nrm{\Lambda'}{y}$
\item
$y\in\domdec(\Lambda)$, $\Theta\neq()$: 
We have $\nrm{\Lambda,\Theta}{y}=\nrm{\Lambda,\Theta}{\Lambda(y)}$.
If $y\in{\domdec}(\Lambda_1)$ then $\Lambda'(y)=\Lambda_1(y)=\Lambda_1(y)\gsub{x}{a}=\Lambda(y)\gsub{x}{a}$.
Otherwise if $y\in{\domdec}(\Lambda_2)$ then $\Lambda'(y)=(\Lambda_2\gsub{x}{a})(y)=\Lambda_2(y)\gsub{x}{a}=\Lambda(y)\gsub{x}{a}$.
In both cases, by inductive hypothesis: $\nrm{\Lambda,\Theta}{\Lambda(y)}=\nrm{\Lambda',\Theta'}{\Lambda(y)\gsub{x}{a}}=\nrm{\Lambda',\Theta'}{\Lambda'(y)}=\nrm{\Lambda',\Theta'}{y}$.
\item
$y\in\domdef(\Lambda)$:
We have $\nrm{\Lambda,\Theta}{y}=\nrm{\Lambda,\Theta}{\Lambda(y)}$.
The cases $y\in{\dom}(\Lambda_1)$ and $y\in{\dom}(\Lambda_2)$ run similar to the previous case. 
In both cases, by inductive hypothesis: $\nrm{\Lambda,\Theta}{\Lambda(y)}=\nrm{\Lambda',\Theta'}{\Lambda(y)\gsub{x}{a}}=\nrm{\Lambda',\Theta'}{\Lambda'(y)}=\nrm{\Lambda',\Theta'}{y}$
\end{itemize}
Conversely if  $\Lambda',\Theta'\sngv y\gsub{x}{a}$ we need to show that $\Lambda,\Theta\sngv y$ and $\nrm{\Lambda,\Theta}{y}=\nrm{\Lambda',\Theta'}{y}$.
First note that $y\in\domdec(\Lambda)$ iff $y\in\domdec(\Lambda')$, $y\in\domdef(\Lambda)$ iff $y\in\domdef(\Lambda')$, and $\Theta'=()$ iff $\Theta=()$.
Hence we can consider the same three subcases: 
\begin{itemize}
\item
$y\in\domdec(\Lambda')$, $\Theta'=()$: Obviously $\nrm{\Lambda}{y}=y=\nrm{\Lambda'}{y}$
\item
$y\in\domdec(\Lambda')$, $\Theta'\neq()$: 
We have $\nrm{\Lambda',\Theta'}{y}=\nrm{\Lambda',\Theta'}{\Lambda'(y)}$.
If $y\in{\domdec}(\Lambda_1)$ then $\Lambda'(y)=\Lambda_1(y)=\Lambda_1(y)\gsub{x}{a}=\Lambda(y)\gsub{x}{a}$.
Otherwise if $y\in{\domdec}(\Lambda_2\gsub{x}{a})$ then $\Lambda'(y)=(\Lambda_2\gsub{x}{a})(y)=\Lambda_2(y)\gsub{x}{a}=\Lambda(y)\gsub{x}{a}$.
In both cases, by inductive hypothesis: $\nrm{\Lambda,\Theta}{\Lambda(y)}=\nrm{\Lambda',\Theta'}{\Lambda(y)\gsub{x}{a}}=\nrm{\Lambda',\Theta'}{\Lambda'(y)}=\nrm{\Lambda',\Theta'}{y}$.
\item
$y\in\domdef(\Lambda')$:
We have $\nrm{\Lambda',\Theta'}{y}=\nrm{\Lambda',\Theta'}{\Lambda'(y)}$.
The cases $y\in{\dom}(\Lambda_1)$ and $y\in{\dom}(\Lambda_2\gsub{x}{a})$ run similar to the previous case. 
In both cases, by inductive hypothesis: $\nrm{\Lambda,\Theta}{\Lambda(y)}=\nrm{\Lambda',\Theta'}{\Lambda(y)\gsub{x}{a}}=\nrm{\Lambda',\Theta'}{\Lambda'(y)}=\nrm{\Lambda',\Theta'}{y}$
\end{itemize}
\item
$b=\binbop{y}{b_1}{b_2}$ for some $y$, $b_1$, $b_2$ where we may assume $x\neq y$:
If $\Lambda,\Theta\sngv b$ there are three cases:
\begin{itemize}
\item $\Theta=()$:
We can calculate as follows:
\begin{eqnarray*}
&&\nrm{\Lambda}{\binbop{y}{b_1}{b_2}}\\
&=&\binbop{y}{\nrm{\Lambda}{b_1}}{\nrm{\Lambda,y:b_1}{b_2}}\\
&=_{(I.H.)}&\binbop{y}{\nrm{\Lambda'}{b_1\gsub{x}{a}}}{\nrm{\Lambda',y:b_1\gsub{x}{a}}{b_2\gsub{x}{a}}}\\
&=&\nrm{\Lambda'}{\binbop{y}{b_1\gsub{x}{a}}{b_2\gsub{x}{a}}}\\
&=&\nrm{\Lambda'}{(\binbop{y}{b_1}{b_2})\gsub{x}{a}}\\
\end{eqnarray*}
\noindent 
Note that the use of the inductive hypothesis on $b_2$ is justified since based on the inductive hypothesis on $b_1$ we know that
$\nrm{\Lambda_1,x\smydef a}{\Lambda_2,y:b_1}=(\nrm{\Lambda_1,x\smydef a}{\Lambda_2},y:\nrm{\Lambda}{b_1})
=(\nrm{\Lambda_1}{\Lambda_2\gsub{x}{a}},y:\nrm{\Lambda'}{b_1\gsub{x}{a}})
=\nrm{\Lambda_1}{(\Lambda_2,x:b_1)\gsub{x}{a}}$. 
This argument is also used in case $\Theta=(1,\Theta_1)$ below:
\item
$\Theta = (c,\Theta_1)$ for some $c$ and $\Theta_1$: Hence
$\Theta' = (c\gsub{x}{a},\Theta_1')$ where $\Theta_1'=\Theta_1\gsub{x}{a}$.
We can calculate as follows:
\begin{eqnarray*}
&&\nrm{\Lambda,\Theta}{\binbop{y}{b_1}{b_2}}\\
&=&\binbop{y}{\nrm{\Lambda}{b_1}}{\nrm{(\Lambda,y\smydef c),\Theta_1}{b_2}}\\
&=_{I.H.}&\binbop{y}{\nrm{\Lambda'}{b_1\gsub{x}{a}}}{\nrm{(\Lambda',y\smydef c\gsub{x}{a}),\Theta_1'}{b_2\gsub{x}{a}}}\\
&=&\nrm{\Lambda',(c\gsub{x}{a},\Theta_1')}{\binbop{y}{b_1\gsub{x}{a}}{b_2\gsub{x}{a}}}\\
&=&\nrm{\Lambda',(c\gsub{x}{a},\Theta_1')}{\binbop{y}{b_1}{b_2}\gsub{x}{a}}\\
&=&\nrm{\Lambda',\Theta'}{\binbop{y}{b_1}{b_2}\gsub{x}{a}}
\end{eqnarray*}
Note that the use of the inductive hypothesis on $b_2$ is justified since the definition $y\mydef c$ is removed by context norming.
\item
$\Theta = (1,\Theta_1)$ for some $\Theta_1$:  This means $\binbop{y}{b_1}{b_2}=[y!b_1]b_2$ and $\Theta' = (1,\Theta_1')$ where $\Theta_1'=\Theta_1\gsub{x}{a}$. We can calculate as follows:
\begin{eqnarray*}
&&\nrm{\Lambda,\Theta}{[y!b_1]b_2}\\
&=&[y!\nrm{\Lambda,\Theta_1}{b_1}]\nrm{\Lambda,y:b_1}{b_2}\\
&=_{I.H.}&[y!\nrm{\Lambda',\Theta_1'}{b_1\gsub{x}{a}}]\nrm{\Lambda',y:b_1\gsub{x}{a}}{b_2\gsub{x}{a}}\\
&=&\nrm{\Lambda',(1,\Theta_1')}{[y!b_1\gsub{x}{a}]b_2\gsub{x}{a}}\\
&=&\nrm{\Lambda',\Theta'}{([y!b_1]b_2)\gsub{x}{a}}
\end{eqnarray*}
\item
$\Theta = (2_c,\Theta_1)$ for some $c$ and $\Theta_1$: Hence $\binbop{y}{b_1}{b_2}=[y\sdef b_1]b_2$
$\Theta' = (2_{c\gsub{x}{a}},\Theta_1')$ where $\Theta_1'=\Theta_1\gsub{x}{a}$.
We can calculate as follows:
\begin{eqnarray*}
&&\nrm{\Lambda,\Theta}{[y\sdef b_1]b_2}\\
&=&[y\sdef\nrm{\Lambda}{b_1}]\nrm{(\Lambda,y\smydef c),\Theta_1}{b_2}\\
&=_{I.H.}&[y\sdef\nrm{\Lambda'}{b_1\gsub{x}{a}}]\nrm{(\Lambda',y\smydef c\gsub{x}{a}),\Theta_1'}{b_2\gsub{x}{a}}\\
&=&\nrm{\Lambda',(2_{c\gsub{x}{a}},\Theta_1')}{[y\sdef b_1\gsub{x}{a}](b_2\gsub{x}{a})}\\
&=&\nrm{\Lambda',(2_{c\gsub{x}{a}},\Theta_1')}{([y\sdef b_1]b_2)\gsub{x}{a}}\\
&=&\nrm{\Lambda',\Theta'}{([y\sdef b_1]b_2)\gsub{x}{a}}
\end{eqnarray*}
Note that the use of the inductive hypothesis on $b_2$ is justified since the definition $y\mydef c$ is removed by context norming.
\end{itemize}
Otherwise if $\Lambda',\Theta'\sngv b\gsub{x}{a}$ an analogous case distinction applies with $\Theta'$.
\item
$b=(b_1\,b_2)$ for some $b_1$, $b_2$:  If $\Lambda,\Theta\sngv b$ we have
$\nrm{\Lambda,\Theta}{(b_1\,b_2)}=\bar{c}_2$ for some $\bar{c}_2$
where $\nrm{\Lambda,(b_2,\Theta)}{b_1}=\binbop{y}{\bar{c}_1}{\bar{c}_2}$ for some $\bar{c}_1$ and $y$ where we may assume $y\neq x$. 
and where $\nrm{}{\Lambda}\gv\nrm{\Lambda}{b_2}\ginc\bar{c}_1$. 

From the inductive hypothesis we know that
$\Lambda',(b_2\gsub{x}{a},\Theta')\sngv b_1\gsub{x}{a}$, $\nrm{\Lambda,(b_2,\Theta)}{b_1}=\nrm{\Lambda',(b\gsub{x}{a},\Theta')}{b_1\gsub{x}{a}}$,
$\Lambda'\sngv b_2\gsub{x}{a}$, and $\nrm{\Lambda}{b_2}=\nrm{\Lambda'}{b_2\gsub{x}{a}}$.
We can calculate $\nrm{\Lambda',(b\gsub{x}{a},\Theta')}{b_1\gsub{x}{a}}=\nrm{\Lambda,(b_2,\Theta)}{b_1}=\binbop{y}{\bar{c}_1}{\bar{c}_2}$.
Note that this implies that $x\notin{\free}(\binbop{y}{\bar{c}_1}{\bar{c}_2})$.
We can calculate as follows:
\begin{eqnarray*}
&&\nrm{}{\Lambda}\gv\nrm{\Lambda}{b_2}\ginc\bar{c}_1\\
&\Rightarrow&\quad\text{(I.H.)}\\
 &&\nrm{}{\Lambda}\gv\nrm{\Lambda'}{b_2\gsub{x}{a}}\ginc\bar{c}_1\\
&\Rightarrow&\quad\text{($\nrm{}{\Lambda}=\nrm{}{\Lambda}'$)}\\
&&\nrm{}{\Lambda'}\gv\nrm{\Lambda'}{b_2\gsub{x}{a}}\ginc\bar{c}_1
\end{eqnarray*}
Therefore by definition of norming
$\nrm{\Lambda,\Theta}{(b_1\,b_2)}=\bar{c}_2=\nrm{\Lambda',\Theta'}{(b_1\,b_2)\gsub{x}{a}}$.

Otherwise if $\Lambda',\Theta'\sngv b\gsub{x}{a}$ we have
$\nrm{\Lambda',\Theta'}{(b_1\gsub{x}{a}\,b_2\gsub{x}{a})}=\bar{c}_2$ for some $\bar{c}_2$
where $\nrm{\Lambda',(b_2\gsub{x}{a},\Theta')}{b_1\gsub{x}{a}}=\binbop{y}{\bar{c}_1}{\bar{c}_2}$ for some $\bar{c}_1$ and $y$ where we may assume $y\neq x$. 
and where $\nrm{}{\Lambda'}\gv\nrm{\Lambda'}{b_2\gsub{x}{a}}\ginc\bar{c}_1$. 

From the inductive hypothesis we know that
$\Lambda,(b_2,\Theta)\sngv b_1$, $\nrm{\Lambda,(b_2,\Theta)}{b_1}=\nrm{\Lambda',(b\gsub{x}{a},\Theta')}{b_1\gsub{x}{a}}$,
$\Lambda\sngv b_2$, and $\nrm{\Lambda}{b_2}=\nrm{\Lambda'}{b_2\gsub{x}{a}}$.
We can calculate $\nrm{\Lambda',(b\gsub{x}{a},\Theta')}{b_1\gsub{x}{a}}=\nrm{\Lambda,(b_2,\Theta)}{b_1}=\binbop{y}{\bar{c}_1}{\bar{c}_2}$.
Note that this implies that $x\notin{\free}(\binbop{y}{\bar{c}_1}{\bar{c}_2})$.
We can calculate as follows:
\begin{eqnarray*}
&&\nrm{}{\Lambda'}\gv\nrm{\Lambda'}{b_2\gsub{x}{a}}\ginc\bar{c}_1\\
&\Rightarrow&\quad\text{(I.H.)}\\
 &&\nrm{}{\Lambda'}\gv\nrm{\Lambda}{b_2}\ginc\bar{c}_1\\
&\Rightarrow&\quad\text{($\nrm{}{\Lambda}=\nrm{}{\Lambda}'$)}\\
&&\nrm{}{\Lambda}\gv\nrm{\Lambda}{b_2}\ginc\bar{c}_1
\end{eqnarray*}
Therefore by definition of norming
$\nrm{\Lambda,\Theta}{(b_1\,b_2)}=\bar{c}_2=\nrm{\Lambda',\Theta'}{(b_1\,b_2)\gsub{x}{a}}$.
\item
$b=\prdef{y}{b_1^{b_2}}{b_3}{b_4}$ for some $y$, $b_1$, $b_2$, $b_3$, $b_4$ where we may assume $y\neq x$: 

Assume that $\Lambda,\Theta\sngv b$:
First we take a look at the two conditions.
For the first condition we can argue as follows:
\begin{eqnarray*}
&&\quad\text{(definition of norming)}\\
&&\nrm{}{\Lambda}\gv\nrm{\Lambda}{b_1}\ginc\nrm{\Lambda}{b_2}\\
&\Rightarrow&\quad\text{($\nrm{}{\Lambda}=\nrm{}{\Lambda}'$)}\\
&&\nrm{}{\Lambda'}\gv\nrm{\Lambda}{b_1}\ginc\nrm{\Lambda}{b_2}\\
&\Rightarrow&\quad\text{(I.H.)}\\
&&\nrm{}{\Lambda'}\gv\nrm{\Lambda'}{b_1\gsub{x}{a}}\ginc\nrm{\Lambda'}{b_2\gsub{x}{a}}
\end{eqnarray*}
For the second condition we can argue as follows:
\begin{eqnarray*}
&&\quad\text{(definition of norming)}\\
&&\nrm{}{\Lambda}\gv\nrm{\Lambda}{b_3}\ginc\nrm{\Lambda,y\smydef b_1}{b_4}\\
&\Rightarrow&\quad\text{($\nrm{}{\Lambda}=\nrm{}{\Lambda}'$)}\\
&&\nrm{}{\Lambda'}\gv\nrm{\Lambda}{b_3}\ginc\nrm{\Lambda,y\smydef b_1}{b_4}\\
&\Rightarrow&\quad\text{(I.H.)}\\
&&\nrm{}{\Lambda'}\gv\nrm{\Lambda'}{b_3\gsub{x}{a}}\ginc\nrm{\Lambda',y\smydef b_1\gsub{x}{a}}{b_4\gsub{x}{a}}
\end{eqnarray*}
Note that the use of the inductive hypothesis on $b_4$ is justified since the definition $y\mydef b_1$ is removed by context norming.
This argument can be repeated several times below.

\noindent
We have just shown that the conditions are also valid for the desired substitution of the protected definition $b$.
Next, we take a look at each of the three cases in the norming definition
\begin{itemize}
\item $\Theta=()$: 
We can calculate as follows:
\begin{eqnarray*}
&&\nrm{\Lambda}{\prdef{y}{b_1^{b_2}}{b_3}{b_4}}\\
&=&\prdef{y}{\nrm{\Lambda}{b_1}^{\snrm{\Lambda}{b_2}}}{\nrm{\Lambda}{b_3}}{\nrm{\Lambda,y\smydef b_1}{b_4}}\\
&=_{I.H.}&\prdef{y}{\nrm{\Lambda'}{b_1\gsub{x}{a}}^{\snrm{\Lambda'}{b_2\gsub{x}{a}}}}{\nrm{\Lambda'}{b_3\gsub{x}{a}}}{\nrm{\Lambda',y\smydef b_1\gsub{x}{a}}{b_4\gsub{x}{a}}}\\
&=&\nrm{\Lambda'}{\prdef{y}{b_1\gsub{x}{a}^{b_2\gsub{x}{a}}}{b_3\gsub{x}{a}}{b_4\gsub{x}{a}}}\\
&=&\nrm{\Lambda'}{\prdef{y}{b_1^{b_2}}{b_3}{b_4}\gsub{x}{a}}
\end{eqnarray*}
\item $\Theta=(1,\Theta_1)$ for some $\Theta_1$: 
We can calculate as follows:
\begin{eqnarray*}
\!\!\!&\!\!\!&\nrm{\Lambda,(1,\Theta_1)}{\prdef{y}{b_1^{b_2}}{b_3}{b_4}}\\
\!\!\!&=\!\!\!&\prdef{y}{\nrm{\Lambda,\Theta_1}{b_1}^{\snrm{\Lambda,\Theta_1}{b_2}}}{\nrm{\Lambda}{b_3}}{\nrm{\Lambda,y\smydef b_1}{b_4}}\\
\!\!\!&=_{I.H.}\!\!\!& \prdef{y}{\nrm{\Lambda',\Theta_1'}{b_1\gsub{x}{a}}^{\snrm{\Lambda',\Theta_1'}{b_2\gsub{x}{a}}}}{\nrm{\Lambda'}{b_3\gsub{x}{a}}}{\nrm{\Lambda',y\smydef b_1\gsub{x}{a}}{b_4\gsub{x}{a}}}\\
\!\!\!&=\!\!\!&\nrm{\Lambda',(1,\Theta_1')}{\prdef{y}{b_1\gsub{x}{a}^{b_2\gsub{x}{a}}}{b_3\gsub{x}{a}}{b_4\gsub{x}{a}}}\\
\!\!\!&=\!\!\!&\nrm{\Lambda',(1,\Theta_1')}{\prdef{y}{b_1^{b_2}}{b_3}{b_4}\gsub{x}{a}}
\end{eqnarray*}
\item $\Theta=(2_c,\Theta_1)$ for some $c$ and $\Theta_1$:
We can calculate as follows:
\begin{eqnarray*}
&&\nrm{\Lambda,(2_c,\Theta_1)}{\prdef{y}{b_1^{b_2}}{b_3}{b_4}}\\
&=&\prdef{y}{\nrm{\Lambda}{b_1}^{\snrm{\Lambda}{b_2}}}{\nrm{\Lambda,\Theta_1}{b_3}}{\nrm{\Lambda,y\smydef b_1,\Theta_1}{b_4}}\\
&=_{I.H.}&[y\mydef\nrm{\Lambda'}{b_1\gsub{x}{a}}^{\snrm{\Lambda'}{b_2\gsub{x}{a}}},\\
&&\nrm{\Lambda',\Theta_1'}{b_3\gsub{x}{a}}:\nrm{\Lambda',y\smydef b_1\gsub{x}{a},\Theta_1'}{b_4\gsub{x}{a}}]\\
&=&\nrm{\Lambda',(2_{c\gsub{x}{a}},\Theta_1')}{\prdef{y}{b_1\gsub{x}{a}^{b_2\gsub{x}{a}}}{b_3\gsub{x}{a}}{b_4\gsub{x}{a}}}\\
&=&\nrm{\Lambda',(2_{c\gsub{x}{a}},\Theta_1')}{\prdef{y}{b_1^{b_2}}{b_3}{b_4}\gsub{x}{a}}
\end{eqnarray*}
\end{itemize}
Otherwise if $\Lambda',\Theta'\sngv b\gsub{x}{a}$ an analogous argument applies. 
\item
$b=\pleft{b_1}$ for some $b_1$:

If $\Lambda,\Theta\sngv b$ there are several cases:
\begin{itemize}
\item
$\nrm{\Lambda,\Theta}{b_1}=\bar{c}_1$ where $\nrm{\Lambda,(1,\Theta)}{b_1}=[y!\bar{c}_1]\bar{c}_2$ for some $\bar{c}_1$ and $\bar{c}_2$.
By I.H. $\nrm{\Lambda,(1,\Theta)}{b_1}=\nrm{\Lambda',(1,\Theta')}{b_1\gsub{x}{a}}$ and therefore
$\nrm{\Lambda',\Theta'}{b\gsub{x}{a}}=\nrm{\Lambda',\Theta'}{\pleft{(b_1\gsub{x}{a})}}=\bar{c}_1$.
\item
$\nrm{\Lambda,\Theta}{b}=\bar{c}_1$ where $\nrm{\Lambda,(1,\Theta)}{b_1}=\prdef{y}{\bar{c}_1^{\bar{c}_2}}{\bar{c}_3}{\bar{c}_4}$ for some $y$, $\bar{c}_1$, $\bar{c}_2$, $\bar{c}_3$, $\bar{c}_4$ where we may assume $y\neq x$.
By I.H. we know that $\nrm{\Lambda,(1,\Theta)}{b_1}=\nrm{\Lambda',(1,\Theta')}{b_1\gsub{x}{a}}$ and therefore we know that
$\nrm{\Lambda',\Theta'}{b\gsub{x}{a}}=\nrm{\Lambda',\Theta'}{\pleft{(b_1\gsub{x}{a})}}=\bar{c}_1$.
\item
$\nrm{\Lambda,\Theta}{b}=\bar{c}_1$ where $\nrm{\Lambda,(1,\Theta)}{b_1}=\prsumop{\bar{c}_1}{\bar{c}_2}$ for some $\bar{c}_1$ and $\bar{c}_2$.
By I.H. $\nrm{\Lambda,(1,\Theta)}{b_1}=\nrm{\Lambda',(1,\Theta')}{b_1\gsub{x}{a}}$ and therefore
$\nrm{\Lambda',\Theta'}{b\gsub{x}{a}}=\nrm{\Lambda',\Theta'}{\pleft{(b_1\gsub{x}{a})}}=\bar{c}_1$.
\end{itemize}
Otherwise if $\Lambda',\Theta'\sngv b\gsub{x}{a}$ an analogous case distinction applies. 
\item
$b=\pright{b_1}$ for some $b_1$:
If $\Lambda,\Theta\sngv b$ there are several cases:
\begin{itemize}
\item
$\nrm{\Lambda,\Theta}{b}=\bar{c}_2$ where $\nrm{\Lambda,(2_{\pleft{b}},\Theta)}{b_1}=[y!\bar{c}_1]\bar{c}_2$ for some $\bar{c}_1$ and $\bar{c}_2$.
By I.H. $\nrm{\Lambda,(2_{\pleft{b}},\Theta)}{b_1}=\nrm{\Lambda',(2_{\pleft{b}\gsub{x}{a}},\Theta')}{b_1\gsub{x}{a}}
=\nrm{\Lambda',(2_{\pleft{b\gsub{x}{a}}},\Theta')}{b_1\gsub{x}{a}}$ 
and therefore we obtain
$\nrm{\Lambda',\Theta'}{\pright{b_1}\gsub{x}{a}}$ = $\nrm{\Lambda',\Theta'}{\pright{b_1\gsub{x}{a}}}$ = $\bar{c}_2$.
\item 
$\nrm{\Lambda,\Theta}{b}=\bar{c}_2$ where $\nrm{\Lambda,(2_c,\Theta)}{b_1}=\prdef{y}{\bar{c}_1^{\bar{c}_2}}{\bar{c}_3}{\bar{c}_4}$ for some $y$, $\bar{c}_1$, $\bar{c}_2$, $\bar{c}_3$, $\bar{c}_4$ where we may assume $y\neq x$.
By I.H. $\nrm{\Lambda,(2_{\pleft{b}},\Theta)}{b_1}=\nrm{\Lambda',(2_{\pleft{b}\gsub{x}{a}},\Theta')}{b_1\gsub{x}{a}}
=\nrm{\Lambda',(2_{\pleft{b\gsub{x}{a}}},\Theta')}{b_1\gsub{x}{a}}$ 
and therefore we obtain
$\nrm{\Lambda',\Theta'}{\pright{b_1}\gsub{x}{a}}$ = $\nrm{\Lambda',\Theta'}{\pright{b_1\gsub{x}{a}}}$ = $\bar{c}_2$.
\item 
$\nrm{\Lambda,\Theta}{b}=\bar{c}_2$ where $\nrm{\Lambda,(2,\Theta)}{b_1}=[\bar{c}_1,\bar{c}_2]$ for some $\bar{c}_1$ and $\bar{c}_2$.
By I.H. $\nrm{\Lambda,(2,\Theta)}{b_1}=\nrm{\Lambda',(2,\Theta')}{b_1\gsub{x}{a}}$ and therefore
$\nrm{\Lambda',\Theta'}{b\gsub{x}{a}}=\nrm{\Lambda',\Theta'}{\pright{(b_1\gsub{x}{a})}}=\bar{c}_1$.
\end{itemize}
Otherwise if $\Lambda',\Theta'\sngv b\gsub{x}{a}$ an analogous case distinction applies. 
\item
$b=\prsuminjop{b_1}{b_2}$, $\case{b_1}{b_2}$, or $\myneg{b_1}$ for some $b_1$, $b_2$: Follows from the definition of norming, substitution, and the inductive hypothesis. 
\qedhere
\end{itemize}
\end{proof}
\noindent
Law~\ref{nrm.rd}(reduction preserves norms) is also valid for the adapted notion of norms.
\begin{law}[Reduction preserves norms] 
\label{sub.nrm.rd}
For all $\Gamma,a,b$:
$\Gamma\sngv a$ and $a\rd b$ implies $\Gamma\sngv b$ and $\nrm{\Gamma}{a}=\nrm{\Gamma}{b}$
\end{law}
\begin{proof}
It is obviously sufficient to show the property for single-step reduction which we do here by induction on the definition of single-step reduction.
We begin with the axioms:
\begin{meditemize}
\item[$\beta_1$:] 
We have $a=([x:a_1]a_2\:a_3)$, $b=a_2\gsub{x}{a_3}$ for some $x$, $a_1$, $a_2$, $a_3$. By definition of norming 
$\nrm{\Gamma}{([x:a_1]a_2\:a_3)}=\nrm{\Gamma,x\smydef a_3}{a_2}$ 
since $\nrm{\Gamma,a_3}{[x:a_1]a_2}=[x:\nrm{\Gamma}{a_1}]\nrm{\Gamma,x\smydef a_3}{a_2}$.
By Law~\ref{sub.nrm.sub} (with $\Lambda_1=\Gamma$, $\Lambda_2=()$, $\Theta=()$) we know that $\nrm{\Gamma,x\smydef a_3}{a_2}=\nrm{\Gamma}{a_2\gsub{x}{a_3}}$ which implies the proposition.
\item[$\beta_2$:]
Similar to case $\beta_1$.
\item[$\beta_3$:]
We have $a=(\case{a_1}{a_2}\,\injl{a_3}{a_4})$ and $b=(a_1\,a_3)$ for some $a_1$, $a_2$, $a_3$, $a_4$.
By definition of norming $\nrm{\Gamma}{a}=\bar{c}$ for some $\bar{c}$ where
$\nrm{\Gamma,\injl{a_3}{\,a_4}}{\case{a_1}{a_2}}=[[\nrm{\Gamma}{a_3}+\nrm{\Gamma}{a_4}]\fun\bar{c}]$,
$\nrm{\Gamma,\pleft{\injl{a_3}{\,a_4}}}{a_1}=[\nrm{\Gamma}{a_3}\,\fun\bar{c}]$, and $\nrm{\Gamma,\pright{\injl{a_3}{\,a_4}}}{a_2}=[\nrm{\Gamma}{a_4}\,\fun\bar{c}]$.
By Law \ref{sub.nrm.eq.elim} $\nrm{\Gamma,\pleft{\injl{a_3}{\,a_4}}}{a_1}=\nrm{\Gamma,a_3}{a_1}$ which implies the proposition.
\item[$\beta_4$:]
Similar to case $\beta_3$.
\item[$\pi_i$:] $i=1,\ldots,6$.
Follows from the definition of norming.
\item[$\nu_i$:] $i=1,\ldots,10$.
Follows from the definition of norming.
\end{meditemize}
Next we turn to the structural rules
\begin{itemize}
\item $a=\binop{a_1,\ldots,b_i}{\ldots,a_n}$ for some $a_1$, $\ldots$, $a_n$ and some $b_i$.
Follows from the definition of norming.
\item $a=\binbop{x}{a_1}{a_2}$, $b=\binbop{x}{a_1'}{a_2}$ where $a_1\srd a_1'$ for some $a_1$, $a_1'$, $a_2$: 
By inductive hypothesis $\nrm{\Gamma}{a_1}=\nrm{\Gamma}{a_1'}$. We can calculate as follows 
\begin{eqnarray*}
&&\nrm{\Gamma}{\binbop{x}{a_1}{a_2}}\\
&=&\quad\text{(definition of norming)}\\
&&\binbop{x}{\nrm{\Gamma}{a_1}}{\nrm{\Gamma,x:a_1}{a_2}}\\
&=&\quad\text{(I.H.)}\\
&&\binbop{x}{\nrm{\Gamma}{a_1'}}{\nrm{\Gamma,x:a_1}{a_2}}\\
&=&\quad\text{(Law~\ref{sub.nrm.eq})}\\
&&\binbop{x}{\nrm{\Gamma}{a_1'}}{\nrm{\Gamma,x:a_1'}{a_2}}\\
&=&\quad\text{(definition of norming)}\\
&&\nrm{\Gamma}{\binbop{x}{a_1'}{a_2}}
\end{eqnarray*}
\item $a=\binbop{x}{a_1}{a_2}$, $b=\binbop{x}{a_1}{a_2'}$ where $a_2\srd a_2'$ for some $a_1$, $a_2$, $a_2'$: 
Follows from the inductive hypothesis and the definition of norming.
\item 
$a=\prdef{x}{a_1^{a_2}}{a_3}{a_4}$, $b=\prdef{x}{b_1^{b_2}}{b_3}{b_4}$ for some $a_1$, $\ldots$, $a_4$, $b_1$, $\ldots$, $b_4$ where $a_i\srd b_i$ for $i=1,\ldots,4$ and $a_i=b_i$ otherwise.
Similar to second case of $a=\binbop{x}{a_1}{a_2}$.
\qedhere
\end{itemize}
\end{proof}
\noindent
The Law~\ref{norm.ext}(context extension) remains valid for norms.
\subsubsection{Norming and typing}
\label{sub.norm.typ}
\noindent
Unfortunately, Law~\ref{nrm.dtyp}(Typing implies normability and preserves norms) is not valid anymore for norming and typing in \dcalcb. This can be illustrated by a simple example.
Let $\Gamma=(x:[\any{0}\fun\any{0}])$ and $t=[z:[\any{0}\fun\any{1}](z\,\any{0})$. Consider $\Gamma\sgv(t\,x)$, the typing relation yields $\Gamma\sgv(t\,x):\any{1}$.
Norming however which is defined using the argument context as described above yields $\nrm{\Gamma}{(t\,x)}=\any{0}$. Hence Law~\ref{nrm.dtyp} cannot be true for norming. 
Fortunately, an adapted property is satisfied: typability implies normability and the norm of an element is included (w.r.t flexible inclusion \ref{inclusion.finc}) in the norm of its type, \eg\ $\any{0}\finc\any{1}$ in the above example. 
\begin{remark}[Restricted Validity]
\nomenclature[ls1Rel04]{$\Gamma\sgv a\rtyp b$}{restricted typing}%
\nomenclature[ls2Rel04]{$\Gamma\sgv a\rsinc\any{n}$}{restricted $\any{}$-inclusion}%
\nomenclature[ls3Rel04]{$\Gamma\sgv a\rleq b$}{restricted inclusion}%
\nomenclature[ls3Rel04]{$\Gamma\sgv a\rincl b$}{restricted $\lambda$-inclusion}%
\nomenclature[ls3Rel04]{$\Gamma\sgv a\rginc b$}{restricted general inclusion}%
\nomenclature[ls3Rel04]{$\Gamma\rsgv a$}{restricted validity}%
\index{typing!restricted}
\index{inclusion!restricted}
\index{inclusion!restricted general}
\index{$\lambda$-inclusion!restricted}
\index{$\any{}$-inclusion!restricted}
\index{validity!restricted}
Due to the fact that not all typable case distinction are normable (See remark in \ref{norm.cased}) we consider a restricted version of typing $\Gamma\sgv a\rtyp b$ (and corresponding restricted versions $\Gamma\sgv a\rleq b$ of inclusion,  $\Gamma\sgv a\rincl b$ of $\lambda$-inclusion, $\Gamma\sgv a\rleq \any{n}$ of $\any{}$-inclusion, and of restricted validity $\Gamma\rsgv a$. All rules are defined as for the unrestricted version but for the 
rule \cased\ which is modified as follows:
\[
\casedrm\quad
\frac{\Gamma\sgv a\rtyp[x:c_1]d\quad\Gamma\sgv b\rtyp[y:c_2]d\quad\Gamma\sgv d\rtyp e\quad C_{\casedrm}		
}{\Gamma\sgv\case{a}{b}\rtyp[z:[c_1+c_2]]d
}
\]
where $C_{\casedrm}$ is defined by two conditions: 
\[
 \frac{\Gamma\ngv a\quad\Gamma\ngv b}{\Gamma\ngv[a?b]}
\]
and for all $\Theta$, $c$;
\[ \frac{\Gamma,(\pleft{c},\Theta)\ngv a\quad\Gamma,(\pright{c},\Theta)\ngv b}{\Gamma,\Theta\ngv[a?b]}
\]
All laws about typing, inclusion, and $\any{}$-inclusion remain valid for the restricted version.
Furthermore we introduce restricted general inclusion as follows:
\[
\Gamma\sgv a\rginc b\quad\equiv\quad\Gamma\sgv a\rleq b\;\text{or}\;\Gamma\sgv a\rtyp b
\]
All the properties for general inclusions remain true analogously for restricted geberal inclusions.
Furthermore, on norms obviously all the restricted variants of concepts are identical to the original version.
\end{remark}

\noindent
In order to show an adapted version of Law~\ref{nrm.dtyp} that is valid for norming in the context of bounded polymorphism we need to bridge the different definitional structures of typing and norming.
This is achieved by defining an {\em instantiation set} for each type judgement $\Gamma\sgv a\rtyp b$ and then prove some properties about this set which imply that $\nrm{}{\Gamma}\gv\nrm{\Gamma}{a}\finc\nrm{\Gamma}{b}$. In order to define instantiation sets we need a notion of typing for extended contexts.
\begin{definition}[Context typing]
\label{sub.inst}
\nomenclature[lsRel04]{$\Lambda:\Gamma$}{context typing}%
\index{context typing}
The notion of \emph{context typing} between an extended context $\Lambda$ and a context $\Gamma$, written as $\Lambda:\Gamma$, is defined as follows:
$\Lambda:\Gamma$ if and only if
\begin{itemize}
\item[$i$:]
$\dom(\Lambda)=\dom(\Gamma)$,
\item[$ii$:]
for all $x\in\domdec(\Lambda)$ we have $\Lambda(x)=\Gamma(x)$, and
\item[$iii$:]	 
for all $x\in\domdef(\Lambda)$ we have $\Gamma\gv\Lambda(x)\rtyp\Gamma(x)$.
\end{itemize}
\end{definition}
\begin{definition}[Instantiation sets]
\label{sub.typ.inst}
\nomenclature[ltRel04]{$\inst{\Gamma\sgv a:b}$}{instantiation set of a type judgement}%
\nomenclature[luRel04]{$\inst{\Gamma\sgv a\rleq b}$}{instantiation set of an inclusion}%
\nomenclature[lvRel04]{$\inst{\Gamma\sgv a\rsinc \any{n}}$}{instantiation set of an $\any{}$-inclusion}%
\index{instantiation set!of type judgement}
The function $\inst{\Gamma\gv a\rtyp b}$ assigns a set of pairs $(\Lambda,\Theta)$ to a typing $\Gamma\gv a\rtyp b$.
The function $\inst{\Gamma\gv a\rleq b}$ assigns a set of pairs $(\Lambda,\Theta)$ to an inclusion $\Gamma\gv a\rleq b$.
The function $\inst{\Gamma\gv a\rsinc \any{n}}$ assigns a set of pairs $(\Lambda,\Theta)$ to an $\any{}$-inclusion $\Gamma\gv a\rsinc \any{n}$.
The recursive definition of these functions is structured according to the inference rules of typing, inclusion, and $\any{}$-inclusion and refers to the identifiers used in these rules (see Tables~\ref{typ.rules}, \ref{sub.typ.rules}, \ref{sub.sinc.rules}, \ref{sub.inc.rules}):

\noindent $\inst{\Gamma\,\gv a\rtyp b}$:
\begin{meditemize}
\item[\ax:]$\inst{\,\gv\any{n}\rtyp \any{n}}$ consists of all pairs $((),\Theta)$.
\item[\mystart:]
$\inst{\Gamma,x:a\,\gv x\rtyp a}$ consists of all pairs $((\Lambda,x:a),\Theta)$ where $(\Lambda,\Theta)\in\inst{\Gamma\,\gv a\rtyp b}$
and all pairs $((\Lambda,x\mydef c),\Theta)$ for some $c$ with $\Gamma\gv c\rtyp a$ and $(\Lambda,\Theta)\in\inst{\Gamma\,\gv c\rtyp a}$.
\item[\weak:]
$\inst{\Gamma,x:c\,\gv a\rtyp b}$ consists of all pairs $((\Lambda,x:c),\Theta)$ and $((\Lambda,x\mydef d),\Theta)$ 
where $(\Lambda,\Theta)\in\inst{\Gamma\,\gv a\rtyp b}$ and $\Gamma\gv d\rtyp c$ for some $d$ with $(\Lambda,())\in\inst{\Gamma\,\gv d\rtyp c}$.
\item[\conv:]
$\inst{\Gamma\,\gv a\rtyp c}$ consists of all pairs $(\Lambda,\Theta)$ where $(\Lambda,\Theta)\in\inst{\Gamma\,\gv a\rtyp b}$ and $b\eqv c$.
\item[\sinc:]
$\inst{\Gamma\,\gv a\rtyp c}$ consists of all pairs $(\Lambda,\Theta)$ where $(\Lambda,\Theta)\in\inst{\Gamma\,\gv a\rtyp b}\cap\inst{\Gamma\,\gv b\rleq c}$. 
\item[\absu]$\!$and~\abse:
$\inst{\Gamma\,\gv\binbop{x}{a}{b}\rtyp [x:a]c}$ consists of all pairs $(\Lambda,())$ where $((\Lambda,x:a),())\in\inst{\Gamma,x:a\,\gv b\rtyp c}$
and of all pairs $(\Lambda,(d,\Theta))$  for some $d$  where $((\Lambda,x\mydef d),\Theta)\in\inst{\Gamma,x:a\,\gv b\rtyp c}$.
\item[\appl:] 
$\inst{\Gamma\,\gv(a\,b)\rtyp c\gsub{x}{b}}$ consists of all pairs $(\Lambda,\Theta)$ where $(\Lambda,(b,\Theta))\in\inst{\Gamma\,\gv a\rtyp [x:d]c}$ and 
$(\Lambda,())\in\inst{\Gamma\,\gv b\rtyp d}$. 
\item[\pdef:]
$\inst{\Gamma\,\gv\prdef{x}{a^b}{c}{d}\rtyp [x\sdef b]d}$ consists of all pairs $(\Lambda,())$ where $(\Lambda,())\in\inst{\Gamma\,\gv a\rtyp b}\cap\inst{\Gamma\,\gv c\rtyp d\gsub{x}{a}}$,
of all pairs $(\Lambda,(1,\Theta))$ where $(\Lambda,\Theta)\in\inst{\Gamma\,\gv a\rtyp b}$, 
and of all pairs $(\Lambda,(2_e,\Theta))$ where  $(\Lambda,\Theta)$ $\in$ $\inst{\Gamma\,\gv c\rtyp d\gsub{x}{a}}$. 
\item[\chin:]
$\inst{\Gamma\gv\pleft{a}\rtyp b}$ contains all pairs $(\Lambda,\Theta)$ where $(\Lambda,(1,\Theta))\in\inst{\Gamma\gv a\rtyp [x\sdef b]c}$.
\item[\prl:]
$\inst{\Gamma\gv\pleft{a}\rtyp b}$ contains all pairs $(\Lambda,\Theta)$ where $(\Lambda,(1,\Theta))\in\inst{\Gamma\gv a\rtyp [b,c]}$.
\item[\chin]$\!$and~\prl: 
$\inst{\Gamma\gv\pleft{a}\rtyp b}$ contains no other pairs than those defined by rule \chin\ and \prl. 
\item[\chba:]
$\inst{\Gamma\,\gv\pright{a}\rtyp c}$ consists of all pairs $(\Lambda,\Theta)$
where $c=d\gsub{x}{\pleft{a}}$, for some $d$ and $x$, and
$(\Lambda,(2_{\pleft{a}},\Theta))\in\inst{\Gamma\,\gv a\rtyp [x\sdef b]d}$.
\item[\prr:]
$\inst{\Gamma\,\gv\pright{a}\rtyp c}$ consists of all pairs $(\Lambda,\Theta)$ where $(\Lambda,(2,\Theta))\in\inst{\Gamma\,\gv a\rtyp [b,c]}$.
\item[\chba]$\!$and~\prr: 
$\inst{\Gamma\gv\pright{a}\rtyp b}$ contains no other pairs than those defined by rule \chba\ and \prr. 
\item[\bprod]$\!$and~\bsum:
$\inst{\Gamma\,\gv\prsumop{a}{b}\rtyp [c,d]}$ consists of all pairs $(\Lambda,())$ where $(\Lambda,())\in\inst{\Gamma\,\gv a\rtyp c}$ $\cap$ $\inst{\Gamma\,\gv b\rtyp d}$,
of all pairs  $(\Lambda,(1,\Theta))$ where  $(\Lambda,\Theta)\in\inst{\Gamma\,\gv a\rtyp c}$,
and of all pairs  $(\Lambda,(2,\Theta))$ where  $(\Lambda,\Theta)\in\inst{\Gamma\,\gv b\rtyp d}$,
\item[\injll:]
$\inst{\Gamma\,\gv\injl{a}{\,b}\rtyp [c+b]}$ consists of all pairs $(\Lambda,())$ where $(\Lambda,())\in\inst{\Gamma\,\gv a\rtyp c}$.
\item[\injlr:]
$\inst{\Gamma\,\gv\injr{{\;\;}a}{b}\rtyp [a+c]}$ consists of all pairs $(\Lambda,())$ where $(\Lambda,())\in\inst{\Gamma\,\gv b\rtyp c}$.
\item[\cased:]
$\inst{\Gamma\,\gv\case{a}{b}\rtyp [x:[c_1+c_2]]d}$ consists of
\begin{itemize}
\item all pairs $(\Lambda,())$ where $(\Lambda,())$ $\in$ $\inst{\Gamma\,\gv a\rtyp [x:c_1]d}$ $\cap\inst{\Gamma\,\gv b\rtyp [x:c_2]d}$,
\item and all pairs $(\Lambda,(e,\Theta))$ for some $e$  where $(\Lambda,(\pleft{e},\Theta))$ $\in$ $\inst{\Gamma\,\gv a\rtyp [x:c_1]d}$ and $(\Lambda,(\pright{e},\Theta))$ $\in$ $\inst{\Gamma\,\gv b\rtyp [x:c_2]d}$.
\end{itemize}
\item[\negate:]
$\inst{\Gamma\,\gv\myneg a\rtyp  b}$ consists of all pairs $(\Lambda,())$ where $(\Lambda,())\in\inst{\Gamma\,\gv a\rtyp b}$.
\end{meditemize}
\noindent $\inst{\Gamma\,\gv a\rleq b}$:
\begin{meditemize}
\item[\srefl:]
$\inst{\Gamma\,\gv a\rleq a}$ consists of all pairs $(\Lambda,\Theta)\in\inst{\Gamma\,\gv a\rtyp b}$ for some $b$ with $\Gamma\,\sgv a\rtyp b$.
\item[\sembed:]
$\inst{\Gamma\,\gv a\rleq\any{n}}$ consists of all pairs $(\Lambda,\Theta)\in\inst{\Gamma\,\gv a\rsinc \any{n}}$.
\item[\sabs:]
$\inst{\Gamma\,\gv\binbop{x}{a}{b}\rleq\binbop{x}{a}{c}}$ consists of all pairs $(\Lambda,())\in\inst{\Gamma,x:a\,\gv b\rleq c}$
and of all pairs $(\Lambda,(d,\Theta))$  for some $d$  where $((\Lambda,x\mydef d),\Theta)\in\inst{\Gamma,x:a\,\gv b\rleq c}$.
\item[\sbprod:]
$\inst{\Gamma\,\gv\prsumop{a}{c}\rleq\prsumop{b}{d}}$ 
consists of all pairs $(\Lambda,())$ where $(\Lambda,())\in\inst{\Gamma\gv a\rleq b}\cap\inst{\Gamma\,\gv c\rleq d}$,
and of all pairs  $(\Lambda,(1,\Theta))$ where  $(\Lambda,\Theta)\in\inst{\Gamma\,\gv a\rleq b}$,
and of all pairs  $(\Lambda,(2,\Theta))$ where  $(\Lambda,\Theta)\in\inst{\Gamma\,\gv c\rleq d}$.
\end{meditemize}
\noindent $\inst{\Gamma\,\gv a\rsinc \any{n}}$:
\begin{meditemize}
\item[\sstart:]
$\inst{\Gamma\,\gv\any{m}\rsinc \any{n}}$ consists of all pairs $(\Lambda,\Theta)\in\inst{\Gamma\gv\any{m}\rtyp a}\cap\inst{\Gamma\gv\any{n}\rtyp b}$.
\item[\sstyp:]
$\inst{\Gamma\,\gv a\rsinc \any{n}}$ consists of all pairs $(\Lambda,\Theta)$ with $(\Lambda,\Theta)\in\inst{\Gamma\,\gv a\rtyp b}\cap\inst{\Gamma\,\gv b\rsinc \any{n}}$.
\item[\ssbprod:]
$\inst{\Gamma\,\gv\prsumop{a}{b}\rsinc \any{n}}$ 
consists of all pairs $(\Lambda,())$ where we have $(\Lambda,())$ $\in$ $\inst{\Gamma\,\gv a\rsinc \any{n}}$ $\cap$ 
$\inst{\Gamma\,\gv b\rsinc \any{n}}$ ,
of all pairs  $(\Lambda,(1,\Theta))$ where  $(\Lambda,\Theta)\in\inst{\Gamma\,\gv a\rsinc \any{n}}$,
and of all pairs  $(\Lambda,(2,\Theta))$ where  $(\Lambda,\Theta)\in\inst{\Gamma\,\gv b\rsinc \any{n}}$.
\item[\ssabsu:]
$\inst{\Gamma\,\gv\binbop{x}{a}{b}\rsinc \any{n}}$ consists of all pairs $(\Lambda,())\in\inst{\Gamma,x\rtyp a\,\gv b\rsinc \any{n}}$
and of all pairs $(\Lambda,(d,\Theta))$  for some $d$  where $((\Lambda,x\mydef d),\Theta)\in\inst{\Gamma,x:a\,\gv b\rsinc \any{n}}$.
\item[\ssinjl:]
$\inst{\Gamma\,\gv\injl{a}{\,b}\rsinc \any{n}}$ are all pairs $(\Lambda,\!())$ where $(\Lambda,\!())\in\inst{\Gamma\,\gv a\rsinc \any{n}}\cap\inst{\Gamma\,\gv b\rsinc \any{n}}$.
\item[\ssinjr:]
$\inst{\Gamma\,\gv\injr{a}{\,b}\rsinc \any{n}}$ are all pairs $(\Lambda,\!())$ where $(\Lambda,\!())\in\inst{\Gamma\,\gv a\rsinc \any{n}}\cap\inst{\Gamma\,\gv b\rsinc \any{n}}$.
\item[\sspdef:]
$\inst{\Gamma\,\gv\prdef{x}{a^b}{c}{d}\rsinc \any{n}}$ 
consists of all pairs $(\Lambda,())$ where we have $(\Lambda,())\in\inst{\Gamma\,\gv a\rsinc \any{n}}\cap\inst{\Gamma\,\gv b\rsinc \any{n}}\cap\inst{\Gamma\,\gv c\rsinc \any{n}}$
and $((\Lambda,x:b),())\in\inst{\Gamma,x:b\,\gv d\rsinc \any{n}}$,
of all pairs $(\Lambda,(1,\Theta))$ where $(\Lambda,\Theta)$ $\in$ $\inst{\Gamma\,\gv a\rsinc \any{n}}$, 
and of all pairs $(\Lambda,(2_e,\Theta))$ where  $(\Lambda,\Theta)$ $\in$ $\inst{\Gamma\,\gv c\rsinc \any{n}}$. 
\end{meditemize}
\end{definition}
\begin{law}[Instantiation sets are well defined]
\label{sub.inst.def}
For all $\Gamma$, $a$, $b$, $n$: If $\Gamma\sgv a\rtyp b$ then $\inst{\Gamma\sgv a\rtyp b}$ is defined. Similarly for $\Gamma\gv a\rleq b$  and $\Gamma\gv a\rsinc\any{n}$. 
\end{law}
\begin{proof}
Note that the rules \mystart\ and \weak\ which extend typing contexts introduce recursive references to instantiations sets with expressions that do not appear in the premises of the corresponding rule.
Note that the additional references are always using the reduced context hence these references do not lead to an infinite recursion.
\end{proof}
\begin{remark}[Examples of instantiation sets]
Consider the type statement $\sgv([x:\prim]x\,\prim)\rtyp \prim$, the derivation steps of which can be graphically presented as follows.
\[
\inferrule*[left=\normalfont{\appl}]{
	\inferrule*[left=\normalfont{\absu}]{
	   \inferrule*[left=\normalfont{\mystart}]{
				\inferrule*[left=\normalfont{\ax}]{
				}
				{\sgv\prim\rtyp \prim}
			}
			{x:\prim\sgv x\rtyp \prim}
	}
	{\sgv[x:\prim]x\rtyp [x:\prim]\prim}
	\quad
	\inferrule*[left=\normalfont{\ax}]{
	}
	{\sgv\prim\rtyp \prim}
	}
{\sgv([x:\prim]x\;\prim)\rtyp \prim\gsub{x}{\prim}}
\]
This leads to the following successively calculated instantiation sets:
\begin{eqnarray*}
\inst{\,\gv\prim\rtyp \prim}&=&\{((),())\}\\
\inst{x:\prim\,\gv x\rtyp \prim}&=&\{((x:\prim),())\}\cup\{((x\mydef a),())\mid((),())\in\inst{\,\gv a\rtyp \prim}%
                                                                                                      \}\\
\inst{\,\gv[x:\prim]x\rtyp [x:\prim]\prim}&=&\{((),())\}\cup\{((),a)\mid((x\mydef a),())\in\inst{x:\prim\,\gv x\rtyp \prim}\}\\
&=&\{((),())\}\cup\{((),a)\mid((),())\in\inst{\,\gv a\rtyp \prim}\}\\
\inst{\,\gv([x:\prim]x\;\prim)\rtyp \prim\gsub{x}{\prim}}&=&\{((),())\mid((),(\prim,()))\in\inst{\Gamma\gv[x:\prim]x\rtyp [x:\prim]\prim}\}\\
&=&\{((),())\}
\end{eqnarray*} 
\end{remark}
\noindent
Next we need to show that instantiation sets correct and complete in the sense that they contain only pairs related by inclusions and that they contain all intended pairs. 
These arguments require an auxiliary operator.
\begin{definition}[Elimination function]
\label{sub.elim.op}
\nomenclature[lwNorm14]{$\bar{a}[\Theta]$}{Elimination function}%
The partial function $\bar{a}[\Theta]$ applies an elimination list $\Theta$ to a norm $\bar{a}$, it is recursively defined as follows: 
$\any{n}[\Theta]=\any{n}$, $\binbop{x}{\bar{a}}{\bar{b}}[c,\Theta]=\bar{b}[\Theta]$, 
$[x\sdef\bar{a}]\bar{b}[1,\Theta]=\bar{a}[\Theta]$, $[x\sdef\bar{a}]\bar{b}[2_c,\Theta]=\bar{b}[\Theta]$, 
$\prdef{x}{\bar{a}_1^{\bar{b}_1}}{\bar{a}_2}{\bar{b}_2}[1,\Theta]=\bar{a}_1[\Theta]$, 
$\prdef{x}{\bar{a}_1^{\bar{b}_1}}{\bar{a}_2}{\bar{b}_2}[2_c,\Theta]=\bar{a}_2[\Theta]$, and $\prsumop{\bar{a}_1}{\bar{a}_2}[i,\Theta]=\bar{a}_i[\Theta]$.
\end{definition}
\noindent
True prefixes of eliminations do not eliminate to variables.
\begin{law}[Basic elimination property]
\label{sub.elim.basic}
For all $\Lambda$ and $a$ let $\Theta=(\Theta_1,\Theta_2)$ for some $\Theta_1$ and $\Theta_2$ where $\Theta_2\neq()$:
If $\Lambda,\Theta\sngv a$ then $\nrm{\Lambda,\Theta}{a}[\Theta_1]$ is defined and $\nrm{\Lambda,\Theta}{a}[\Theta_1]\notin\dvar$.
\end{law}
\begin{proof}
Inductive argument on the definition of $\nrm{\Lambda,\Theta}{a}$.
\begin{itemize}
\item $a=\any{n}$ for some $n$: Obvious.
\item $a=x$ for some $x$: $\nrm{\Lambda,\Theta}{x}=\nrm{\Lambda,\Theta}{\Lambda(x)}$ reducing the property to the inductive hypothesis. Similar if $a$ is a negation.
\item $a=\binbop{x}{a_1}{a_2}$ for some $a_1$, $a_2$:  
Since $\Theta=()$ is not possible there are two cases: 
If $\Theta_1=()$ then obviously $\nrm{\Lambda,\Theta_2}{a}\notin\dvar$.
Otherwise $\Theta_1=(b,\Theta_1')$ for some $b,\Theta_1'$ and hence $\nrm{\Lambda,\Theta}{a}=\binbop{x}{\nrm{\Lambda}{a_1}}{\nrm{(\Lambda,x\smydef b),(\Theta_1',\Theta_2)}{a_2}}$.
By inductive hypothesis $\nrm{\Lambda,\Theta}{a}[\Theta_1]=\nrm{(\Lambda,x\smydef b),(\Theta_1',\Theta_2)}{a_2}[\Theta_1']\notin\dvar$.
\item $a=(a_1\;a_2)$ for some $a_1$, $a_2$: $\nrm{\Lambda,\Theta}{a}=\bar{d}$ where $\nrm{\Lambda,(a_2,\Theta)}{a_1}=[x:\bar{c}]\bar{d}$. 
By inductive hypothesis $\nrm{\Lambda,(a_2,\Theta)}{a_1}[a_2,\Theta_1]=\bar{d}[\Theta_1]\notin\dvar$. 
\item $a$ is a protected definition, a product, a sum, an injection, or a case distinction: Similar to the case of abstractions above.
\item $a=\pleft{a_1}$ for some $a_1$: $\nrm{\Lambda,\Theta}{a}=\bar{c}$ where $\nrm{\Lambda,(1,\Theta)}{a_1}=[\bar{c},\bar{d}]$ or $\nrm{\Lambda,(1,\Theta)}{a_1}=[x!\bar{c}]\bar{d}$ for some $\bar{c}$, $\bar{d}$. By inductive hypothesis $\nrm{\Lambda,(1,\Theta)}{a_1}[1,\Theta_1]=\bar{c}[\Theta_1]\notin\dvar$. 
\item $a=\pright{a_1}$ for some $a_1$: $\nrm{\Lambda,\Theta}{a}=\bar{d}$ where either $\nrm{\Lambda,(2,\Theta)}{a_1}=[\bar{c},\bar{d}]$ or 
$\nrm{\Lambda,\Theta}{a}=\bar{d}\gsub{x}{\pleft{a_1}}$ where $\nrm{\Lambda,(2_{\pleft{a_1}},\Theta)}{a_1}=[x!\bar{c}]\bar{d}$ for some $\bar{c}$, $\bar{d}$. 
By inductive hypothesis either $\nrm{\Lambda,(2,\Theta)}{a_1}[2,\Theta_1]=\bar{d}[\Theta_1]\notin\dvar$
or  $\nrm{\Lambda,(2_{\pleft{a_1}},\Theta)}{a_1}[2_{\pleft{a_1}},\Theta_1]=\bar{d}\gsub{x}{\pleft{a_1}}[\Theta_1]\notin\dvar$.
\qedhere
\end{itemize}
\end{proof}
\noindent
We can now show that that all elements of instantiations sets imply flexible inclusions.
\begin{law}[Instantiation sets and flexible inclusion]
\label{sub.flex.inst}
For all $\Lambda$, $\Theta$, $\Gamma$, $a$, $b$, $n$:
\begin{itemize}
\item[$i:$]
If $\Gamma\gv a\rtyp b$ and $(\Lambda,\Theta)\in\inst{\Gamma\gv a\rtyp b}$ then $\nrm{}{\Lambda}\gv\nrm{\Lambda,\Theta}{a}\finc\nrm{\Lambda,\Theta}{b}$.
\item[$ii:$]
If $\Gamma\gv a\rleq b$ and $(\Lambda,\Theta)\in\inst{\Gamma\gv a\rleq b}$ then $\nrm{}{\Lambda}\gv\nrm{\Lambda,\Theta}{a}\leq\nrm{\Lambda,\Theta}{b}$.
\item[$iii:$]
If $\Gamma\gv a\rsinc \any{n}$ and $(\Lambda,\Theta)\in\inst{\Gamma\gv a\rsinc \any{n}}$ then $\nrm{}{\Lambda}\gv\nrm{\Lambda,\Theta}{a}<\any{n}$.
\end{itemize}
Furthermore in all cases the respective normings are defined.
\end{law}
\begin{proof}
\noindent
All parts are shown simultaneously by induction on the definition of instantiation sets.

\noindent {\bf Part} $i$:
\begin{meditemize}
\item[\ax:] Obviously for any $n$ we have $\gv\nrm{\Theta}{\any{n}}\finc\nrm{\Theta}{\any{n}}$.
\item[\mystart:] 
According to the definition of $\inst{\Gamma,x:a\,\gv x\rtyp a}$ there are two cases:
In the first case we know that $\Lambda=(\Lambda',x:a)$ where $(\Lambda',\Theta)\in\inst{\Gamma\gv a\rtyp b}$ for some $b$.
By inductive hypothesis $\sngv\Lambda'$ and $(\Lambda',\Theta)\sngv a$ and therefore obviously $\sngv(\Lambda',x:a)$. 
If $\Theta=()$ then $\nrm{(\Lambda',x:a),\Theta}{x}=x$. 
Obviously $\nrm{}{(\Lambda',x:a)}=(\nrm{}{\Lambda'},x:\nrm{\Lambda'}{a})\gv x:\nrm{\Lambda'}{a}=\nrm{\Lambda',x:a}{a}$.
If $\Theta\neq()$ then $\nrm{(\Lambda',x:a),\Theta}{x}=\nrm{(\Lambda',x:a),\Theta}{a}$.
Either way, by definition of generalized inclusion $\nrm{}{\Lambda',x:a}\gv\nrm{(\Lambda',x:a),\Theta}{x}\ginc\nrm{(\Lambda',x:a),\Theta}{a}$.
By rule \sfembed~this implies $\nrm{}{\Lambda',x:a}\gv\nrm{(\Lambda',x:a),\Theta}{x}\finc\nrm{(\Lambda',x:a),\Theta}{a}$. 

In the second case we know that 
$\Lambda=(\Lambda',x\mydef c)$ where $(\Lambda',\Theta)\in\inst{\Gamma\gv c\rtyp a}$ for some $\Lambda'$, $\Theta$, $c$ with $\Gamma\sgv c\rtyp a$.
By inductive hypothesis $\sngv\Lambda'$, $(\Lambda',\Theta)\sngv c$, $(\Lambda',\Theta)\sngv a$, and $\nrm{}{\Lambda'}\gv\nrm{\Lambda,\Theta}{c}\finc\nrm{\Lambda,\Theta}{a}$.
Hence obviously $\sngv(\Lambda',x\mydef c)$. 
Since $\nrm{}{\Lambda',x\mydef c}=\nrm{}{\Lambda'}$, $\nrm{(\Lambda',x\smydef c),\Theta}{x}=\nrm{\Lambda',\Theta}{c}$, and $\nrm{(\Lambda',x\smydef c),\Theta}{a}=\nrm{\Lambda',\Theta}{a}$
we obtain $\nrm{}{\Lambda',x\mydef c}\gv\nrm{(\Lambda',x\smydef c),\Theta}{x}\finc\nrm{(\Lambda',x\smydef c),\Theta}{a}$.
\item[\weak:]
$\inst{\Gamma,x:c\,\gv a\rtyp b}$ consists of all pairs $((\Lambda,x:c),\Theta)$ and $((\Lambda,x\mydef d),\Theta)$ 
where $(\Lambda,\Theta)\in\inst{\Gamma\,\gv a\rtyp b}$ and $\Gamma\gv d\rtyp c$ for some $d$ with $(\Lambda,())\in\inst{\Gamma\,\gv d\rtyp c}$.
By inductive assumption we know that $\sngv\Lambda$, $\Lambda,\Theta\sngv a$,  $\Lambda,\Theta\sngv a$, and $\nrm{}{\Lambda}\gv\nrm{\Lambda,\Theta}{a}\finc\nrm{\Lambda,\Theta}{b}$. 
In the first case we have to show that $\nrm{}{\Lambda,x:c}\gv\nrm{(\Lambda,x:c),\Theta}{a}\finc\nrm{(\Lambda,x:a),\Theta}{b}$
and in the second case we have to show that $\nrm{}{\Lambda,x\mydef d}\gv\nrm{(\Lambda,x\smydef d),\Theta}{a}\finc\nrm{\Lambda,x\smydef d),\Theta}{b}$.
both of which obviously follows since $x\notin\free(a)\cup\free(b)$.
\item[\conv:]
Members of $\inst{\Gamma\gv a\rtyp c}$ are $(\Lambda,\Theta)$ where $(\Lambda,\Theta)\in\inst{\Gamma\gv a\rtyp b}$ with $b\eqv c$.
By inductive hypothesis $\nrm{}{\Lambda}\gv\nrm{\Lambda,\Theta}{a}\finc\nrm{\Lambda,\Theta}{b}$.
By Laws \ref{cr} and \ref{sub.nrm.rd} it follows that $\nrm{}{\Lambda}\gv\nrm{\Lambda,\Theta}{a}\finc\nrm{\Lambda,\Theta}{c}$.
\item[\sinc:]
Members of $\inst{\Gamma\gv a\rtyp c}$ are all pairs $(\Lambda,\Theta)$ where $(\Lambda,\Theta)\in\inst{\Gamma\gv a\rtyp b}\cap\inst{\Gamma\gv b\rleq  c}$.
By inductive hypothesis $\nrm{}{\Lambda}\gv\nrm{\Lambda,\Theta}{a}\finc\nrm{\Lambda,\Theta}{b}$ and $\nrm{}{\Lambda}\gv\nrm{\Lambda,\Theta}{b}\leq\nrm{\Lambda,\Theta}{c}$.
By Law \ref{sub.flex.incl}~it follows that $\nrm{}{\Lambda}\gv\nrm{\Lambda,\Theta}{a}\finc\nrm{\Lambda,\Theta}{c}$.
\item[\absu]and~\abse:
Members of $\inst{\Gamma\gv\binbop{x}{a}{b}\rtyp [x:a]c}$ consist of pairs $(\Lambda,())$ where $((\Lambda,x:a),())\in\inst{\Gamma,x:a\gv b\rtyp c}$
and pairs $(\Lambda,(d,\Theta))$, where $((\Lambda,x\mydef d),\Theta)\in\inst{\Gamma,x:a\gv b\rtyp c}$,$\Lambda\sngv a$, $\Lambda\sngv d$, and $\nrm{}{\Lambda}\gv\nrm{\Lambda}{d}\ginc\nrm{\Lambda}{a}$.

According to the definition of $\inst{\Gamma\gv[x:a]b\rtyp [x:a]c}$ there are two cases:
In the first case we have to show that $\nrm{}{\Lambda}\gv\nrm{\Lambda}{[x:a]b}\ginc\nrm{\Lambda}{[x:a]c}$.
By definition of generalized inclusion, norming, and typing this follows from $\nrm{}{\Lambda,x:a}\gv\nrm{\Lambda,x:a}{b}\ginc\nrm{\Lambda,x:a}{c}$ which corresponds to the inductive hypothesis.
In the second case we have to show that $\nrm{}{\Lambda,x\mydef d}=\nrm{}{\Lambda}\gv\nrm{\Lambda,(d,\Theta)}{[x:a]b}\ginc\nrm{\Lambda,(d,\Theta)}{[x:a]c}$
where $((\Lambda,x\mydef d),\Theta)\in\inst{\Gamma,x:a\,\gv b\rtyp c}$. The argument is similar to the first case.
In both cases the inductive assumption ensure that all normings are defined.
\item[\appl:] 
Members of $\inst{\Gamma\gv(a\,b)\rtyp c_2\gsub{x}{b}}$ are $(\Lambda,\Theta)$ where $\Lambda,(b,\Theta)\in\inst{\Gamma\gv a\rtyp [x:c_1]c_2}$ 
 and $(\Lambda,())\in\inst{\Gamma\,\gv b:c_1}$.  
We have to show that $\nrm{}{\Lambda}\gv\nrm{\Lambda,\Theta}{(a\,b)}\ginc\nrm{\Lambda,\Theta}{c_2\gsub{x}{b}}$.

By inductive hypothesis $\nrm{}{\Lambda}\gv\nrm{\Lambda,(b,\Theta)}{a}\ginc\nrm{\Lambda,(b,\Theta)}{[x:c_1]c_2}=[x:\nrm{\Lambda}{c_1}]\nrm{(\Lambda,x\smydef b),\Theta}{c_2}
=_\text{(Law~\ref{sub.nrm.sub})}[x:\nrm{\Lambda}{c_1}]\nrm{\Lambda,\Theta}{c_2\gsub{x}{b}}$. 
Using Law \ref{sub.elim.basic} with $\Theta_1=()$ and $\Theta_2=(b,\Theta)$ we can apply Law \ref{sub.ginc.prop}($iii$) to obtain
$\nrm{\Lambda,(b,\Theta)}{a}=[x:\nrm{\Lambda}{c_1}]\bar{d}$ for some $\bar{d}$ where $\nrm{}{\Lambda}\gv\bar{d}\ginc\nrm{\Lambda,\Theta}{c_2\gsub{x}{b}}$.
Since by inductive hypothesis $\nrm{}{\Lambda}\gv\nrm{\Lambda}{b}\ginc\nrm{\Lambda}{c_1}$, by definition of norming 
$\nrm{\Lambda,\Theta}{(a\,b)}=\bar{d}$. This implies 
$\nrm{}{\Lambda}\gv\nrm{\Lambda,\Theta}{(a\,b)}\ginc\nrm{\Lambda,\Theta}{c_2\gsub{x}{b}}$.
As in case \absu\ the inductive assumption ensure that all normings are defined.
\item[\pdef:]
Members of $\inst{\Gamma\,\gv\prdef{x}{a^b}{c}{d}\rtyp [x\sdef b]d}$ are pairs $(\Lambda,())$ where $(\Lambda,())\in\inst{\Gamma\,\gv a\rtyp b}\cap\inst{\Gamma\,\gv c\rtyp d\gsub{x}{a}}$,
pairs $(\Lambda,(1,\Theta))$ where $(\Lambda,\Theta)\in\inst{\Gamma\,\gv a\rtyp b}$, 
and all pairs $(\Lambda,$ $(2_e,$ $\Theta))$ where  $(\Lambda,\Theta)$ $\in$ $\inst{\Gamma\,\gv c\rtyp d\gsub{x}{a}}$. 

In the first case we have to show that $\nrm{}{\Lambda}\gv\nrm{\Lambda}{\prdef{x}{a^b}{c}{d}}\finc\nrm{\Lambda}{[x\sdef b]d}$.
We know that
$\nrm{\Lambda}{\prdef{x}{a^b}{c}{d}}
=\prdef{x}{\nrm{\Lambda}{a}^{\snrm{\Lambda}{b}}}{\nrm{\Lambda}{c}}{\nrm{\Lambda,x:b}{d}}$ and $\nrm{\Lambda}{[x\sdef b]d}=[x\sdef\snrm{\Lambda}{b}]\nrm{\Lambda,x:b}{d}$
Hence by rule \sfpdef~$\nrm{}{\Lambda}\gv\nrm{\Lambda}{\prdef{x}{a^b}{c}{d}}\finc\nrm{\Lambda}{[x\sdef b]d}$.

In the second case we have to show
$\nrm{}{\Lambda}\gv\nrm{\Lambda,(1,\Theta)}{\prdef{x}{a^b}{c}{d}}\finc\nrm{\Lambda,(1,\Theta)}{[x\sdef b]d}$.  
We know that $\nrm{\Lambda,(1,\Theta)}{\prdef{x}{a^b}{c}{d}}
=\prdef{x}{\nrm{\Lambda,\Theta}{a}^{\snrm{\Lambda,\Theta}{b}}}{\nrm{\Lambda}{c}}{\nrm{\Lambda,x:b}{d}}$ 
and $\nrm{\Lambda,(1,\Theta)}{[x\sdef b]d}=[x\sdef\snrm{\Lambda,\Theta}{b}]\nrm{\Lambda,x:b}{d}$.
Hence by rule \sfpdef~$\nrm{}{\Lambda}\gv\nrm{\Lambda,(1,\Theta)}{\prdef{x}{a^b}{c}{d}}\finc\nrm{\Lambda,(1,\Theta)}{[x\sdef b]d}$.  

In the third case we have to show
$\nrm{}{\Lambda}\gv\nrm{\Lambda,(2_e,\Theta)}{\prdef{x}{a^b}{c}{d}}\finc\nrm{\Lambda,(2_e,\Theta)}{[x\sdef b]d}$.  
We know that $\nrm{\Lambda,(2_e,\Theta)}{\prdef{x}{a^b}{c}{d}}
=\prdef{x}{\nrm{\Lambda}{a}^{\snrm{\Lambda}{b}}}{\nrm{\Lambda}{c}}{\nrm{(\Lambda,x:b),\Theta}{d}}$
and $\nrm{\Lambda,(2_e,\Theta)}{[x\sdef b]d}=[x\sdef\snrm{\Lambda}{b}]\nrm{(\Lambda,x:b),\Theta}{d}$.
Hence by rule \sfpdef~$\nrm{}{\Lambda}\gv\nrm{\Lambda,(2_e,\Theta)}{\prdef{x}{a^b}{c}{d}}\finc\nrm{\Lambda,(2_e,\Theta)}{[x\sdef b]d}$.  

As in case \absu\ the inductive assumption ensure that all normings are defined.
\item[\chin:]
Members of $\inst{\Gamma\gv\pleft{a}\rtyp b}$ considered here are pairs $(\Lambda,\Theta)$ where $(\Lambda,(1,\Theta))\in\inst{\Gamma\gv a\rtyp [x\sdef b]c}$.
By inductive hypothesis we know that $\nrm{}{\Lambda}\gv\nrm{\Lambda,(1,\Theta)}{a}\finc\nrm{\Lambda,(1,\Theta)}{[x\sdef b]c}=[x\sdef\nrm{\Lambda,\Theta}{b}]\nrm{\Lambda,x:b}{c}$. 
Due to Law \ref{sub.elim.basic} with $\Theta_1=()$ and $\Theta_2=(1,\Theta)$ we can apply Law~\ref{sub.flex.prop}($vii$) which implies two cases:

If $\nrm{\Lambda,(1,\Theta)}{a}=\prdef{x}{\bar{a}_1^{\snrm{\Lambda,\Theta}{b}}}{\bar{a}_3}{\nrm{\Lambda,x:b}{c}}$ 
for some $\bar{a}_1$, $\bar{a}_3$ then $\nrm{}{\Lambda}\gv\nrm{\Lambda,\Theta}{\pleft{a}}=\pleft{\nrm{\Lambda,(1,\Theta)}{a}}=\nrm{\Lambda,\Theta}{b}$. 
Similarly if $\nrm{\Lambda,(1,\Theta)}{a}=[x\sdef\nrm{\Lambda,\Theta}{b}]\bar{a}_2$ for some $\bar{a}_2$.
then $\nrm{}{\Lambda}\gv\nrm{\Lambda,\Theta}{\pleft{a}}=\pleft{\nrm{\Lambda,(1,\Theta)}{a}}=\nrm{\Lambda,\Theta}{b}$.

As in case \absu\ the inductive assumption ensure that all normings are defined.
\item[\chba:]
Members of $\inst{\Gamma\gv\pright{a}\rtyp d\gsub{x}{\pleft{a}}}$ considered are pairs $(\Lambda,\Theta)$ with $(\Lambda,(2_{\pleft{a}},\Theta))$ $\in$ $\inst{\Gamma\gv a\rtyp [x\sdef b]d}$.
By inductive hypothesis
$\nrm{}{\Lambda}\gv\nrm{\Lambda,(2_{\pleft{a}},\Theta)}{a}\finc\nrm{\Lambda,(2_{\pleft{a}},\Theta)}{[x\sdef b]d}$ = $[x\sdef\nrm{\Lambda}{b}]\nrm{(\Lambda,x\smydef\pleft{a}),\Theta}{d}$.
Due to Law \ref{sub.elim.basic} with $\Theta_1=()$ and $\Theta_2=(2_{\pleft{a}},\Theta)$ we can apply Law~\ref{sub.flex.prop}($vii$) which implies two cases: 

If $\nrm{\Lambda,(2_{\pleft{a}},\Theta)}{a}=\prdef{x}{\bar{a}_1^{\snrm{\Lambda,\Theta}{b}}}{\bar{a}_3}{\nrm{(\Lambda,x\smydef\pleft{a}),\Theta}{d}}$
for some $\bar{a}_1$, $\bar{a}_3$ 
then $\nrm{}{\Lambda}\gv\nrm{\Lambda,\Theta}{\pright{a}}=
\pright{\nrm{\Lambda,(2_{\pleft{a}},\Theta)}{a}}=\nrm{(\Lambda,x\smydef\pleft{a}),\Theta}{d}=_{\ref{sub.nrm.sub}}\nrm{\Lambda,\Theta}{d\gsub{x}{\pleft{a}}}$.
Otherwise if $\nrm{\Lambda,(2_{\pleft{a}},\Theta)}{a}=[x\sdef\nrm{\Lambda,\Theta}{b}]\bar{a}_2$ for some
$\bar{a}_2$ where $\nrm{}{\Lambda},x:\nrm{\Lambda}{b}\gv\bar{a}_2\leq\nrm{(\Lambda,x\smydef\pleft{a}),\Theta}{d}$.
Since $x\notin{\free}(\nrm{(\Lambda,x\smydef\pleft{a}),\Theta}{d})$ by Law \ref{sub.leq.free} $x\notin{\free}(\bar{a}_2)$
and therefore $\nrm{}{\Lambda}\gv\nrm{\Lambda,\Theta}{\pright{a}}=
\pright{\nrm{\Lambda,(2_{\pleft{a}},\Theta)}{a}}=\bar{a}_2\finc\nrm{(\Lambda,x\smydef\pleft{a}),\Theta}{d}=_{\ref{sub.nrm.sub}}\nrm{\Lambda,\Theta}{d\gsub{x}{\pleft{a}}}$.

As in case \absu\ the inductive assumption ensure that all normings are defined.
\item[\bprod]and \bsum:
Member of $\inst{\Gamma\gv\prsumop{a}{b}\rtyp [c,d]}$ are pairs $(\Lambda,())$ where $(\Lambda,())\in\inst{\Gamma\gv a\rtyp c}\cap\inst{\Gamma\gv b\rtyp c}$,
pairs  $(\Lambda,(1,\Theta))$ where  $(\Lambda,\Theta)\in\inst{\Gamma\gv a\rtyp c}$,
and also pairs $(\Lambda,(2,\Theta))$ where  $(\Lambda,\Theta)\in\inst{\Gamma\gv b\rtyp d}$.
In the first case, by inductive hypothesis $\nrm{}{\Lambda}\gv\nrm{\Lambda}{a}\finc\nrm{\Lambda}{c}$ and
$\nrm{}{\Lambda}\gv\nrm{\Lambda}{b}\finc\nrm{\Lambda}{d}$.
Hence obviously  $\nrm{}{\Lambda}\gv\nrm{\Lambda}{\prsumop{a}{b}}\finc\nrm{\Lambda}{[c,d]}$.
In the second case, by inductive hypothesis $\nrm{}{\Lambda,\Theta}\gv\nrm{\Lambda}{a}\finc\nrm{\Lambda,\Theta}{c}$. 
Hence obviously  $\nrm{}{\Lambda}\gv\nrm{\Lambda,(1,\Theta)}{\prsumop{a}{b}}=\prsumop{\nrm{\Lambda,\Theta}{a}}{\nrm{\Lambda}{b}}\finc\nrm{\Lambda,(1,\Theta)}{[a,b]}$.
The third case is analogous.
As in case \absu\ the inductive assumption ensure that all normings are defined.
\item[\prl:]
Members of $\inst{\Gamma\gv\pleft{a}\rtyp b}$ considered here are pairs $(\Lambda,\Theta)$ where $(\Lambda,(1,\Theta))\in\inst{\Gamma\gv a\rtyp [b,c]}$.
By inductive hypothesis we know that $\nrm{}{\Lambda}\gv\nrm{\Lambda,(1,\Theta)}{a}\finc\nrm{\Lambda,(1,\Theta)}{[b,c]}=[\nrm{\Lambda,\Theta}{b},\nrm{\Lambda}{c}]$. 
Due to Law \ref{sub.elim.basic} with $\Theta_1=()$ and $\Theta_2=(1,\Theta)$ by Law~\ref{sub.flex.prop}($v$) $\nrm{\Lambda,(1,\Theta)}{a}=[\bar{a}_1,\bar{a}_2]$ for some $\bar{a}_1$, $\bar{a}_2$ where $\nrm{}{\Lambda}\gv\bar{a}_1\finc\nrm{\Lambda,\Theta}{b}$.
Hence $\nrm{}{\Lambda}\gv\nrm{\Lambda,\Theta}{\pleft{a}}=\bar{a}_1\finc\nrm{\Lambda,\Theta}{b}$.
As in case \absu\ the inductive assumption ensure that all normings are defined.
\item[\prr:]
Analogous to \prl.
\item[\injll:]
Members of $\inst{\Gamma\gv\injl{a}{\,b}\rtyp [c+b]}$ are pairs $(\Lambda,())$ where $(\Lambda,())\in\inst{\Gamma\gv a\rtyp c}$.
By inductive hypothesis we know that $\nrm{}{\Lambda}\gv\nrm{\Lambda}{a}\finc\nrm{\Lambda}{c}$. 
By rule~\finjll~we have $\nrm{}{\Lambda}\gv\nrm{\Lambda}{\injl{a}{b}}=\injl{\nrm{\Lambda}{a}}{\nrm{\Lambda}{b}}\finc[\nrm{\Lambda}{c}+\nrm{\Lambda}{b}]=\nrm{\Lambda}{[c+b]}$.
As in case \absu\ the inductive assumption ensure that all normings are defined.
\item[\injlr:] 
Analogous to \injll.
\item[\cased:]
For pairs $(\Lambda,())\in\inst{\Gamma\,\gv\case{a}{b}\rtyp [x:[c_1+c_2]]d}$ obviously 
$(\Lambda,())\in\inst{\Gamma\,\gv a\rtyp [x:c_1]d}$ and $(\Lambda,())\in\inst{\Gamma\,\gv b\rtyp [x:c_2]d}$.
By inductive hypothesis $\nrm{}{\Lambda}\gv\nrm{\Lambda}{a}\finc[x:\nrm{\Lambda}{c_1}]\nrm{\Lambda}{d}$
and $\nrm{}{\Lambda}\gv\nrm{\Lambda}{b}\finc[x:\nrm{\Lambda}{c_2}]\nrm{\Lambda}{d}$.
The inductive assumption ensures that all normings are defined.
By rule \casedr\ $\Gamma\sngv[a?b]$ exists, hence the norms of $a$ and $b$ have equal range.
Therefore $\nrm{}{\Lambda}\gv\nrm{\Lambda}{\case{a}{b}}\finc\nrm{\Lambda}{[x:[c_1+c_2]]d}$.

For pairs $(\Lambda,(e,\Theta))\in\inst{\Gamma\,\gv\case{a}{b}\rtyp [x:[c_1+c_2]]d}$ we know that 
$(\Lambda,(\pleft{e},\Theta))$ $\in$ $\inst{\Gamma\,\gv a\rtyp [x:c_1]d}$ and $(\Lambda,(\pright{e},\Theta))\in\inst{\Gamma\,\gv b\rtyp [x:c_2]d}$.
By inductive hypothesis $\nrm{}{\Lambda}\gv\nrm{\Lambda,(\pleft{e},\Theta)}{a}\finc[x:\nrm{\Lambda}{c_1}]\nrm{\Lambda,\Theta}{d}$
and $\nrm{}{\Lambda}\gv\nrm{\Lambda,(\pright{e},\Theta)}{b}\finc[x:\nrm{\Lambda}{c_2}]\nrm{\Lambda,\Theta}{d}$.
The inductive assumption ensures that all normings are defined.
By rule \casedr\ $\Gamma,\Theta\sngv[a?b]$ exists, hence the norms of $a$ and $b$ have equal range.
Similar to the first case we can argue that $\nrm{}{\Lambda}\gv\nrm{\Lambda,(e,\Theta)}{\case{a}{b}}\finc\nrm{\Lambda,(e,\Theta)}{[x:[c_1+c_2]]d}$.
\item[\negate:]
Members of $\inst{\Gamma\gv\myneg a\rtyp  b}$ are pairs $(\Lambda,())$ where $(\Lambda,())\in\inst{\Gamma\gv a\rtyp b}$.
The property follows from the inductive hypothesis since $\nrm{\Lambda}{\myneg a}=\nrm{\Lambda}{a}$. 
As in case \absu\ the inductive assumption ensure that all normings are defined.
\end{meditemize}
\noindent {\bf Part} $ii$:
\begin{meditemize}
\item[\srefl:]
Members of $\inst{\Gamma\,\gv a\rleq  a}$ are pairs $(\Lambda,\Theta)\in\inst{\Gamma\,\gv a\rtyp b}$ for some $b$ with $\Gamma\,\sgv a\rtyp b$.
By inductive hypothesis $\Lambda,\Theta\sngv a$. Hence $\nrm{}{\Lambda}\gv\nrm{\Lambda,\Theta}{a}\leq\nrm{\Lambda,\Theta}{a}$.
\item[\sembed:]
Members of $\inst{\Gamma\,\gv a\rleq \any{n}}$ are pairs $(\Lambda,\Theta)\in\inst{\Gamma\,\gv a\rsinc \any{n}}$.
By inductive hypothesis $\nrm{}{\Lambda}\gv\nrm{\Lambda,\Theta}{a}\rsinc \any{n}$ hence $\nrm{}{\Lambda}\gv\nrm{\Lambda,\Theta}{a}\leq\any{n}$.
\item[\sabs:]
Members of $\inst{\Gamma\,\gv\binbop{x}{a}{b}\rleq \binbop{x}{a}{c}}$ are pairs $(\Lambda,()\in\inst{\Gamma,x:a\,\gv b\rleq  c}$
and pairs $(\Lambda,(d,\Theta))$  for some $d$  where $((\Lambda,x\mydef d),\Theta)\in\inst{\Gamma,x:a\,\gv b\rleq  c}$. 
The argument is similar to cases \absu\ and \abse.
\item[\sbprod:]
Members of  $\inst{\Gamma\,\gv\prsumop{a}{c}\rleq \prsumop{b}{d}}$ are pairs $(\Lambda,())$ where $(\Lambda,())\in\inst{\Gamma\gv a\rleq  b}\cap\inst{\Gamma\,\gv c\rleq  d}$,
 pairs  $(\Lambda,(1,\Theta))$ where  $(\Lambda,\Theta)\in\inst{\Gamma\,\gv a\rleq  b}$,
and pairs  $(\Lambda,(2,\Theta))$ where  $(\Lambda,\Theta)\in\inst{\Gamma\,\gv c\rleq  d}$.
The argument is similar to cases \bprod\ and \bsum.
\end{meditemize}
\noindent {\bf Part} $iii$:
\begin{meditemize}
\item[\sstart:]
Members of $\inst{\Gamma\,\gv\any{m}\rsinc \any{n}}$ are pairs $(\Lambda,\Theta)\in\inst{\Gamma\gv\any{m}\rtyp a}\cap\inst{\Gamma\gv\any{n}\rtyp b}$.
Obviously $\nrm{}{\Lambda}\gv\nrm{\Lambda,\Theta}{\any{m}}=\any{m}<\any{n}= \nrm{}{\Lambda}\gv\nrm{\Lambda,\Theta}{\any{n}}$.
\item[\sstyp:]
Members of $\inst{\Gamma\,\gv a\rsinc \any{n}}$ are pairs $(\Lambda,\Theta)$ with $(\Lambda,\Theta)\in\inst{\Gamma\,\gv a\rtyp b}\cap\inst{\Gamma\,\gv b\rsinc \any{n}}$.
By inductive hypothesis $\nrm{}{\Lambda}\gv\nrm{\Lambda}{a}\finc\nrm{\Lambda}{b}$ and $\nrm{}{\Lambda}\gv\nrm{\Lambda}{b}<\any{n}$.
By Law \ref{sub.flex.incl} we know that $\nrm{}{\Lambda}\gv\nrm{\Lambda}{a}<\any{n}$.
\item[\ssbprod:]
Members of  $\inst{\Gamma\,\gv\prsumop{a}{b}\rsinc \any{n}}$ 
are all pairs $(\Lambda,())$ where $(\Lambda,())\in\inst{\Gamma\,\gv a\rsinc \any{n}}$ $\cap$ $\inst{\Gamma\,\gv b\rsinc \any{n}}$,
pairs  $(\Lambda,(1,\Theta))$ where  $(\Lambda,\Theta)\in\inst{\Gamma\,\gv a\rsinc \any{n}}$,
and all pairs  $(\Lambda,(2,\Theta))$ where  $(\Lambda,\Theta)\in\inst{\Gamma\,\gv b\rsinc \any{n}}$.
The argument is similar to cases \bprod\ and \bsum.
\item[\ssabsu:]
Members of $\inst{\Gamma\,\gv\binbop{x}{a}{b}\rsinc \any{n}}$ are pairs $(\Lambda,()\in\inst{\Gamma,x:a\,\gv b\rsinc \any{n}}$
and $(\Lambda,(d,\Theta))$  for some $d$  where $((\Lambda,x\mydef d),\Theta)\in\inst{\Gamma,x:a\,\gv b\rsinc \any{n}}$.
The argument is similar to cases \absu\ and \abse.
\item[\ssinjl:]
Members of $\inst{\Gamma\,\gv\injl{a}{\,b}\rsinc \any{n}}$ are pairs $(\Lambda,())$ where $(\Lambda,())\in\inst{\Gamma\,\gv a\rsinc \any{n}}\cap\inst{\Gamma\,\gv b\rsinc \any{n}}$.
By inductive hypothesis $\nrm{}{\Lambda}\gv\nrm{\Lambda}{a}<\any{n}$ and $\nrm{}{\Lambda}\gv\nrm{\Lambda}{b}<\any{n}$.
Hence $\nrm{}{\Lambda}\gv\nrm{\Lambda}{\injl{a}{b}}<\any{n}$.
\item[\ssinjr:]
Similar to \ssinjl.
\item[\sspdef:]
Members of $\inst{\Gamma\,\gv\prdef{x}{a^b}{c}{d}\rsinc \any{n}}$ 
are pairs $(\Lambda,())$ where $(\Lambda,())\in\inst{\Gamma\,\gv a\rsinc \any{n}}$ $\cap$ $\inst{\Gamma\,\gv b\rsinc \any{n}}$ $\cap$ $\inst{\Gamma\,\gv c\rsinc \any{n}}$
and $((\Lambda,x:b),())\in\inst{\Gamma,x:b\,\gv d\rsinc \any{n}}$, 
pairs $(\Lambda,(1,\Theta))$ where $(\Lambda,\Theta)$ $\in$ $\inst{\Gamma\,\gv a\rsinc \any{n}}$, 
and $(\Lambda,(2_e,\Theta))$ where  $(\Lambda,\Theta)$ $\in$ $\inst{\Gamma\,\gv c\rsinc \any{n}}$. 

In the first case, by inductive hypothesis $\nrm{}{\Lambda}\gv\nrm{\Lambda}{a}<\any{n}$, $\nrm{}{\Lambda}\gv\nrm{\Lambda}{b}<\any{n}$, $\nrm{}{\Lambda}\gv\nrm{\Lambda}{c}<\any{n}$,
and $\nrm{}{\Lambda},x:\nrm{\Lambda}{b}\gv\nrm{\Lambda}{d}<\any{n}$ hence $\nrm{}{\Lambda}\gv\nrm{\Lambda}{\prdef{x}{a^b}{c}{d}}<\any{n}$.
The other two cases are similar to the second and third case of \pdef.
\qedhere
\end{meditemize}
\end{proof}
\noindent
In order to show that instantiation sets are complete (Law \ref{sub.inst.prop}) we need some properties involving elimination and generalized inclusion
\begin{law}[Elimination and flexible inclusion]
\label{sub.elim.prop}
For all $\Lambda$, $\Theta$, $a$, $b$, $n$:
\begin{itemize}
\item[$i$:]
If $\Lambda,\Theta\sngv (a\,b)$
then $\nrm{\Lambda,\Theta}{(a\,b)}\![\Theta]=\nrm{\Lambda,(b,\Theta)}{a}\![b,\Theta]$,
if $\Lambda,\Theta\sngv\pleft{a}$ 
then $\nrm{\Lambda,\Theta}{\pleft{a}}\![\Theta]= \nrm{\Lambda,(1,\Theta)}{a}\![1,\Theta]$, 
if $\Lambda,\Theta\sngv\pright{a}$ and $\nrm{\Lambda,(2,\Theta)}{a}=\prsumop{\bar{b}_1}{\bar{b}_2}$ 
then $\nrm{\Lambda,\Theta}{\pright{a}}\![\Theta]=\nrm{\Lambda,(2,\Theta)}{a}\![2,\Theta]$,
and if $\Lambda,\Theta\sngv\pright{a}$ and $\nrm{\Lambda,(2_{\pleft{a}},\Theta)}{a}=[x\sdef\bar{b}_1]\bar{b}_2$ 
then $\nrm{\Lambda,\Theta}{\pright{a}}\![\Theta]=\nrm{\Lambda,(2_{\pleft{a}},\Theta)}{a}\![2_{\pleft{a}},\Theta]$.
\item[$ii$:]
If $\bar{\Gamma}\gv\bar{a}<\any{n}$ and $\bar{a}[\Theta]$ is defined $\bar{\Gamma}\gv\bar{a}[\Theta]<\any{n}$. 
\item[$iii$:]
If $\bar{\Gamma}\gv\bar{a}\leq\bar{b}$ and $\bar{a}[\Theta]$ is defined then $\bar{b}[\Theta]$ is defined and $\bar{\Gamma}\gv\bar{a}[\Theta]\leq\bar{b}[\Theta]$. 
\item[$iv$:]
If $\bar{\Gamma}\gv\bar{a}:\bar{b}$ and $\bar{a}[\Theta]$ is defined then $\bar{b}[\Theta]$ is defined and $\bar{\Gamma}\gv\bar{a}[\Theta]:\bar{b}[\Theta]$. 
\item[$v$:]
If $\bar{\Gamma}\gv\bar{a}\finc\bar{b}$ and $\bar{a}[\Theta]$ is defined then $\bar{b}[\Theta]$ is defined and $\bar{\Gamma}\gv\bar{a}[\Theta]\finc\bar{b}[\Theta]$. 
\end{itemize}
\end{law}
\begin{proof}
For all $\Lambda$, $\Theta$, $a$, and $b$:
\begin{itemize}
\item[$i$:]
Consider arbitrary $\Lambda$, $\Theta$:
Assume $\Lambda,\Theta\sngv (a\,b)$. Hence $\Lambda,(b,\Theta)\sngv a$ and $\nrm{\Lambda,(b,\Theta)}{a}=\binbop{x}{\bar{c}}{\nrm{\Lambda,\Theta}{(a\,b)}}$ for some $\bar{c}$. 
Hence $\nrm{\Lambda,(b,\Theta)}{a}\![b,\Theta]=\nrm{\Lambda,\Theta}{(a\,b)}\![\Theta]$. The other statements are shown in a similar style.  
\item[$ii$:]
Interconnected inductive argument for Parts $ii$, $iii$, and $iv$. 
Consider the rules of $\bar{\Gamma}\gv\bar{a}<\any{n}$. 
For rule \sstart\ the proposition follows since $\any{n}[\Theta]=\any{n}$.
For rule \sstyp\ we have $\bar{\Gamma}\gv\bar{a}:\bar{b}$ and $\bar{\Gamma}\gv\bar{b}<\any{n}$.
The proposition follows from the inductive hypothesis for Parts $ii$ and $iv$.
For rule \ssabsu\ we have  $\bar{a}=\binbop{x}{\bar{a}_1}{\bar{a}_2}$ where 
$\bar{\Gamma}\gv\bar{a}_1<\any{n}$ and $\bar{\Gamma},x:\bar{a}_1\gv\bar{a}_2<\any{n}$. 
The proposition follows from the inductive hypothesis for Part $ii$.
Similar for the other structural rules.
\item[$iii$:]
Interconnected inductive argument for Parts $ii$, $iii$, and $iv$. 
Consider the rules of $\bar{\Gamma}\gv\bar{a}\leq\bar{b}$:
For rule \srefl\ the property is trivial, for rule \sembed\ it follows from the inductive assumption of Part $ii$.
For rule \sabs\ we have $\bar{a}=\binbop{x}{\bar{a}_1}{\bar{a}_2}$ and $\bar{b}=\binbop{x}{\bar{a}_1}{\bar{b}_2}$ where $\bar{\Gamma},x:\bar{a}_1\gv\bar{a}_2\leq\bar{b}_2$.
Assume $\bar{a}[\Theta]$ is defined, hence $\Theta=(c,\Theta')$ and $\bar{a}_2[\Theta']$ is defined.
By inductive hypothesis $\bar{b}_2[\Theta']$ is defined and $\bar{\Gamma},x:\bar{a}_1\gv\bar{a}_2[\Theta']\leq\bar{b}_2[\Theta']$.
Hence $\bar{\Gamma}\gv\bar{a}[\Theta]\leq\bar{b}[\Theta]$.
Similar for the rule \sbprod.
\item[$iv$:]
Interconnected inductive argument for Parts $ii$, $iii$, and $iv$. 
Consider the rules of $\bar{\Gamma}\gv\bar{a}:\bar{b}$.
We obviously only need to consider the rules for constructors.
For rule \ax\ the proposition follows since $\any{n}[\Theta]=\any{n}$.
Rule \mystart\ is trivial since $x[\Theta]$ is not defined.
Rule \weak\ follows directly from the inductive hypothesis.
Rule \conv\ is trivial since norms are irreducible.
The structural rule \absu\ is argued similar to rule \sabs\ in Part $iii$.
Similar for the other structural rules.
\item[$v$:]
Proof by induction on $\bar{\Gamma}\gv\bar{a}\finc\bar{b}$.
For the rule \sfembed\ the property follows form Parts $ii$ and $iii$.
The structural \sfabs, \sfbprod, $\ldots$ are shown similar to Part $iii$.
%
\qedhere
\end{itemize}
\end{proof}
\noindent
We can now show how elimination lists in instantiation sets can be extended.  
First we need a simple lemma about eliminators with equal norms.
\begin{law}[Norm equality in instantiations sets]
\label{sub.inst.equal}
For all $\Gamma$, $\Lambda$, $\Theta_1$, $\Theta_2$, $a$, $b$, $c$, $d$, $n$, and $S$ where $S=\inst{\Gamma\gv a\rtyp b}$,
$S=\inst{\Gamma\gv a\rleq  b}$, or $S=\inst{\Gamma\gv a\rsinc \any{n}}$: 
If $\Lambda\sngv c$, $\Lambda\sngv d$, and $\nrm{\Lambda}{c}=\nrm{\Lambda}{d}$ then $(\Lambda,(\Theta_1,c,\Theta_2))\in S$ if and only if $(\Lambda,(\Theta_1,d,\Theta_2))\in S$.
\end{law}
\begin{proof}
This is a direct consequence of the way instantiation sets are defined
The proof is by simultaneous induction on the three sets.
\end{proof}
\noindent
\begin{law}[Extension of instantiation sets]
\label{sub.inst.law}
For all $\Lambda$, $\Theta$, $\Gamma$, $a$, $b$, $n$: Assume $(\Lambda,\Theta)\in S$ where $S=\inst{\Gamma\gv a\rtyp b}$, $S=\inst{\Gamma\gv a\rleq  b}$, or $S=\inst{\Gamma\gv a\rsinc \any{n}}$. 
The following statements are true
\begin{itemize}
\item[$i$:]
For all $e$:
If $\nrm{}{\Lambda}\gv\nrm{\Lambda,\Theta}{a}\![\Theta]\finc[x:\bar{c}]\bar{d}$ and $\nrm{}{\Lambda}\gv\nrm{\Lambda}{e}\ginc\bar{c}$ for some $x$, $\bar{c}$, $\bar{d}$ then
$(\Lambda,(\Theta,e))\in S$ where, depending on the assumption, $S=\inst{\Gamma\gv a\rtyp b}$, $S=\inst{\Gamma\gv a\rleq  b}$, or $S=\inst{\Gamma\gv a\rsinc \any{n}}$. 
\item[$ii$:]
If
$\nrm{}{\Lambda}\gv\nrm{\Lambda,\Theta}{a}\![\Theta]\finc[\bar{c},\bar{d}]$ for some $\bar{c}$, $\bar{d}$
then $(\Lambda,(\Theta,i))\in S$ for $i=1,2$ where, depending on the assumption, $S=\inst{\Gamma\gv a\rtyp b}$, $S=\inst{\Gamma\gv a\rleq  b}$, or $S=\inst{\Gamma\gv a\rsinc \any{n}}$.  
\item[$iii$:]
If 
$\nrm{}{\Lambda}\gv\nrm{\Lambda,\Theta}{a}\![\Theta]\finc[x\sdef\bar{c}]\bar{d}$ for some $x$, $\bar{c}$, $\bar{d}$
then $(\Lambda,(\Theta,1))$, $(\Lambda,(\Theta,2_{\pleft{e}}))$ $\in S$ where, depending on the assumption, $S=\inst{\Gamma\gv a\rtyp b}$, $S=\inst{\Gamma\gv a\rleq  b}$, or $S=\inst{\Gamma\gv a\rsinc \any{n}}$.
\item[$iv$:]
For all $e$, $i=1,2$:
If $\nrm{}{\Lambda}\gv\nrm{\Lambda,\Theta}{a}\![\Theta]=\any{m}$ for some $m$ then
$(\Lambda,(\Theta,e))$, $(\Lambda,(\Theta,i))$, $(\Lambda,(\Theta,2_{\pleft{e}}))$ $\in S$ where, depending on the assumption, $S=\inst{\Gamma\gv a\rtyp b}$, $S=\inst{\Gamma\gv a\rleq  b}$, or $S=\inst{\Gamma\gv a\rsinc \any{n}}$.
\end{itemize}
\end{law}
\begin{proof}
All properties are shown in parallel by induction on the definition of instantiation sets.

\noindent $\inst{\Gamma\gv a\rtyp b}$:
\begin{meditemize}
\item[\ax:] 
For Parts $i$, $ii$, and $iii$ the proposition is trivial since all of the premises are false.
For Part $iv$, $\Lambda=()$ and $a=b=\any{m}$ for some $m$. 
Obviously $\nrm{\Theta}{\any{m}}\![\Theta]=\any{m}\![\Theta]=\any{m}$
and by definition of $\inst{\Gamma\gv\any{m}\rtyp \any{m}}$ $((),(\Theta,e))\in S$.
\item[\mystart:]
$\inst{\Gamma,x:a\,\gv x\rtyp a}$ consists of all pairs $((\Lambda,x:a),\Theta)$ where $(\Lambda,\Theta)\in\inst{\Gamma\,\gv a\rtyp b}$.
and all pairs  $((\Lambda,x\mydef c),\Theta)$ for some $c$ with $\Gamma\gv c\rtyp a$ and $(\Lambda,\Theta)\in\inst{\Gamma\,\gv c\rtyp a}$.

Assume that $(\Lambda,\Theta)\in\inst{\Gamma,x:a\gv x\rtyp a}$ and, for Part $i$, that $\nrm{}{\Lambda}\gv\nrm{\Lambda,\Theta}{x}\![\Theta]\finc[y:\bar{c}]\bar{d}$ for some $y$, $\bar{c}$, $\bar{d}$ 
where $\nrm{}{\Lambda}\gv\nrm{\Lambda}{e}\ginc\bar{c}$. Analogous assumptions can be made for Parts $ii$, $iii$, and $iv$. 
For Part $i$ we will show that $(\Lambda,(\Theta,e))\in\inst{\Gamma,x:a\gv x\rtyp a}$.
For Parts $ii$, $iii$, and $iv$ we show corresponding statements.
By definition of $\inst{\Gamma,x:a\gv x\rtyp a}$ there are two cases:

In the first case $\Lambda=(\Lambda',x:a)$ where $(\Lambda',\Theta)\in\inst{\Gamma\gv a\rtyp b}$. 
We consider Part $i$ first. If $\Theta=()$ then by definition of norming and the elimination function $\nrm{}{\Lambda}\gv\nrm{\Lambda}{x}\![()]=\nrm{\Lambda}{x}=x:\nrm{\Lambda'}{a}=\nrm{\Lambda'}{a}\![()]$.
$\nrm{}{\Lambda}\gv x\finc[y:\bar{c}]\bar{d}$ obviously implies $\nrm{}{\Lambda}\gv x\ginc[y:\bar{c}]\bar{d}$.
Since $\nrm{}{\Lambda}\gv x\leq[y:\bar{c}]\bar{d}$ would contradict Law \ref{sub.leq.prop}($ii$) we have $\nrm{}{\Lambda}\gv x:[y:\bar{c}]\bar{d}$.
By Law \ref{sub.type.decomp}($ii$) and since $x\notin{\free(a)}\cup{\free}([x:\bar{c}]\bar{d})$ $\nrm{}{\Lambda'}\gv\nrm{\Lambda'}{a}\leq[y:\bar{c}]\bar{d}$.
Hence by inductive hypothesis we obtain $(\Lambda',e)\in\inst{\Gamma\gv a\rtyp b}$ and therefore $(\Lambda,e)\in\inst{\Gamma,x:a\gv x\rtyp a}$.
If $\Theta\neq()$ then by definition of norming $\nrm{}{\Lambda'}\gv\nrm{\Lambda',\Theta}{x}\![\Theta]=\nrm{\Lambda',\Theta}{a}\![\Theta]$.
Hence we can apply the inductive hypothesis to obtain $(\Lambda',(\Theta,e))\in\inst{\Gamma\gv a\rtyp b}$ and therefore $(\Lambda,(\Theta,e))\in\inst{\Gamma,x:a\gv x\rtyp a}$.
This case be shown analogously for Parts $ii$ (using Law \ref{sub.leq.prop}($iv$)) and $iii$ (using Law \ref{sub.leq.prop}($ii$)).
For Part $iv$ assume that $\nrm{}{\Lambda}\gv\nrm{\Lambda,\Theta}{x}\![\Theta]=\any{m}$. 
If $\Theta=()$ then by definition of norming $\nrm{}{\Lambda}\gv\nrm{\Lambda}{x}\![()]=\nrm{\Lambda}{x}=x$ which cannot be the case.
If $\Theta\neq()$ then by definition of norming $\nrm{\Lambda',\Theta}{a}\![\Theta]=\any{m}$.
Hence by inductive hypothesis $(\Lambda,(\Theta,e))\in\inst{\Gamma,x:a\gv x\rtyp a}$.

In the second case $\Lambda=(\Lambda',x\mydef c)$ where $(\Lambda',\Theta)\in\inst{\Gamma\gv c\rtyp a}$.
We consider Part $i$ first.
By Law \ref{sub.flex.inst} we know that $\nrm{}{\Lambda'}\gv\nrm{\Lambda',\Theta}{c}\finc\nrm{\Lambda',\Theta}{a}$.
By definition of norming $\nrm{}{\Lambda}\gv\nrm{\Lambda',\Theta}{x}=\nrm{\Lambda',\Theta}{c}$.
Hence by inductive hypothesis $(\Lambda',(\Theta,e))\in\inst{\Gamma\gv c\rtyp a}$ and therefore by definition of instantiation sets
$(\Lambda,(\Theta,e))\in\inst{\Gamma,x:a\gv x\rtyp a}$. 
Similar arguments can be made for Parts $ii$, $iii$, and $iv$.
\item[\weak:]
$\inst{\Gamma,x:c\,\gv a\rtyp b}$ consists of all pairs $((\Lambda,x:c),\Theta)$ and $((\Lambda,x\mydef d),\Theta)$ 
where $(\Lambda,\Theta)\in\inst{\Gamma\,\gv a\rtyp b}$ and $\Gamma\gv d\rtyp c$ or some $d$ with $(\Lambda,())\in\inst{\Gamma\,\gv d\rtyp c}$.

Assume that $(\Lambda,\Theta)\in\inst{\Gamma,x:c\gv a\rtyp b}$ and, for Part $i$,  
that $\nrm{}{\Lambda}\gv\nrm{\Lambda,\Theta}{a}\![\Theta]\finc[y:\bar{c}]\bar{d}$ for some $y$, $\bar{c}$, and $\bar{d}$,
and where $\nrm{}{\Lambda}\gv\nrm{\Lambda}{e}\ginc\bar{c}$. Analogous assumptions can be made for Parts $ii$, $iii$, and $iv$. 
For Part $i$ we will show that $(\Gamma,(\Theta,e))\in\inst{\Gamma,x:c\gv a\rtyp b}$.
For Parts $ii$, $iii$, and $iv$ we show corresponding statements.
By definition of $\inst{\Gamma,x:c\gv a\rtyp b}$ there are two cases:

In the first case $\Lambda=(\Lambda',x:c)$ where $(\Lambda',\Theta)\in\inst{\Gamma\gv a\rtyp b}$
Hence by inductive hypothesis we know that $(\Lambda',(\Theta,e))\in\inst{\Gamma\gv a\rtyp b}$.
By definition of instantiation sets we know that $(\Lambda,(\Theta,e))\in\inst{\Gamma,x:c\gv a\rtyp b}$.
Similar arguments can be made for Parts $ii$, $iii$, and $iv$.

In the second case $\Lambda=(\Lambda',x\mydef d)$ where $(\Lambda,\Theta)\in\inst{\Gamma\gv a\rtyp b}$ and $(\Lambda,())\in\inst{\Gamma\,\gv d\rtyp c}$.
By inductive hypothesis we know that $(\Lambda',(\Theta,e))\in\inst{\Gamma\gv a\rtyp b}$.
By definition of instantiation sets it follows that $(\Lambda,(\Theta,e))\in\inst{\Gamma,x:c\gv a\rtyp b}$.
Similar arguments can be made for Parts $ii$, $iii$, and $iv$.
\item[\conv:]
$\inst{\Gamma\gv a\rtyp c}$ consists of all pairs $(\Lambda,\Theta)$ where $(\Lambda,\Theta)\in\inst{\Gamma\gv a\rtyp b}$ and $b\eqv c$.
For Part $i$ assume $\nrm{}{\Lambda}\gv\nrm{\Lambda,\Theta}{a}\![\Theta]\finc[x:\bar{c}]\bar{d}$ and $\nrm{}{\Lambda}\gv\nrm{\Lambda}{e}\ginc\bar{c}$ for some $x$, $\bar{c}$, $\bar{d}$.
By inductive hypothesis $(\Lambda,(\Theta,e))\in\inst{\Gamma\gv a\rtyp b}$. Since $b\eqv c$ this implies $(\Lambda,(\Theta,e))\in\inst{\Gamma\gv a\rtyp c}$.
Similar arguments can be made for Parts $ii$, $iii$, and $iv$.
\item[\sinc:]
$\inst{\Gamma\gv a\rtyp c}$ consists of all pairs $(\Lambda,\Theta)$ where $(\Lambda,\Theta)\in\inst{\Gamma\gv a\rtyp b}\cap\inst{\Gamma\gv b\rleq  c}$.

Consider Part $i$: Assume $\nrm{}{\Lambda}\gv\nrm{\Lambda,\Theta}{a}\![\Theta]\finc[x:\bar{c}]\bar{d}$ and $\nrm{}{\Lambda}\gv\nrm{\Lambda}{e}\ginc\bar{c}$ for some $e$, $x$, $\bar{c}$, $\bar{d}$. 
By inductive hypothesis we know that $(\Lambda,(\Theta,e))\in\inst{\Gamma\gv a\rtyp b}$ which by Law~\ref{sub.flex.inst} implies $\Lambda,\Theta\sngv b$ and
$\nrm{}{\Lambda}\gv\nrm{\Lambda,\Theta}{a}\finc\nrm{\Lambda,\Theta}{b}$.
By Law \ref{sub.elim.prop}($v$) $\nrm{\Lambda,\Theta}{b}\![\Theta]$ is defined and $\nrm{}{\Lambda}\gv\nrm{\Lambda,\Theta}{a}\![\Theta]\ginc\nrm{\Lambda,\Theta}{b}\![\Theta]$.
In order to show that the conditions for applying the inductive hypothesis for $\inst{\Gamma\gv b\rleq  c}$ are satisfied we make a case distinction:
\begin{itemize}
\item 
If $\nrm{\Lambda,\Theta}{a}\![\Theta]\in\dvar$ then by Law \ref{sub.flex.prop}($ii$) $\nrm{}{\Lambda}\gv \nrm{\Lambda,\Theta}{a}\![\Theta]:[y:\bar{c}]\bar{d}$.
Furthermore from 
$\nrm{}{\Lambda}\gv\nrm{\Lambda,\Theta}{a}\ginc\nrm{\Lambda,\Theta}{b}$ by law \ref{sub.flex.prop}($iv$) we know that
$\nrm{\Lambda,\Theta}{b}\![\Theta]=\any{k}$ for some $k$, or
$\nrm{}{\Lambda}\gv\nrm{\Lambda,\Theta}{a}\![\Theta]:\nrm{\Lambda,\Theta}{b}\![\Theta]$.
In the first case the conditions for applying the inductive hypothesis for $\inst{\Gamma\gv b\rleq  c}$ are directly satisfied.
In the second case by Law \ref{sub.min.min} $\nrm{}{\Lambda}\gv \nrm{\Lambda,\Theta}{a}\![\Theta]:\bar{f}$ for some $\bar{f}$
where $\nrm{}{\Lambda}\gv\bar{f}\leq[y:\bar{c}]\bar{d}$ and $\nrm{}{\Lambda}\gv\bar{f}\leq\nrm{\Lambda,\Theta}{b}\![\Theta]$.
Using Law \ref{sub.leq.prop}($ii$, $iii$) we can argue that $\nrm{\Lambda,\Theta}{b}\![\Theta]=[y:\bar{c}]\bar{d}'$ for some $\bar{d}'$ or $\nrm{\Lambda,\Theta}{b}\![\Theta]=\any{k}$ for some $k$.
\item
If $\nrm{\Lambda,\Theta}{a}\![\Theta]\notin\dvar$ by Law \ref{sub.flex.prop}($iii$) $\nrm{}{\Lambda}\gv\!\nrm{\Lambda,\Theta}{a}\![\Theta]=[x:\bar{c}]\bar{d}'$ for some $\bar{d}'$.
By Law \ref{sub.flex.prop}($iv$) either $\nrm{}{\Lambda}\gv\!\nrm{\Lambda,\Theta}{b}\![\Theta]=[x:\bar{c}]\bar{d}''$ for some $\bar{d}''$ or 
$\nrm{}{\Lambda}\gv\!\nrm{\Lambda,\Theta}{b}\![\Theta]=\any{k}$ for some $k$.
\end{itemize}
Hence in both cases the conditions for applying the inductive hypothesis for $\inst{\Gamma\gv b\rleq  c}$ are satisfied which implies $(\Lambda,(\Theta,e))\in\inst{\Gamma\gv b\rleq  c}$
and therefore $(\Lambda,(\Theta,e))\in\inst{\Gamma\gv a\rtyp c}$.

Similar arguments can be made for Parts $ii$ (using Laws \ref{sub.flex.prop}($v$,$vi$) and \ref{sub.leq.prop}($iv$,$v$)), and $iii$ (using Law \ref{sub.flex.prop}($vii$, $viii$)).

For Part $iv$ by Law \ref{sub.flex.prop}($i$) $\nrm{\Lambda,\Theta}{b}=\any{k}$ for some $k$ and hence $\nrm{\Lambda,\Theta}{b}\![\Theta]=\any{k}$ which means that the inductive hypothesis for  
$\inst{\Gamma\gv b\rleq  c}$ can be applied.
\item[\absu]and~\abse:
Assume that $(\Lambda,\Theta)\in\inst{\Gamma\gv\binbop{x}{a}{b}\rtyp [x:a]c}$ 
and, for Part $i$, that $\nrm{}{\Lambda}\gv\nrm{\Lambda,\Theta}{\binbop{x}{a}{b}}\![\Theta]\finc[y:\bar{c}]\bar{d}$ for some $y$, $\bar{c}$, and $\bar{d}$.
and some $e$ with $\nrm{}{\Lambda}\gv\nrm{\Lambda}{e}\ginc\bar{c}$.
We have to show that $(\Gamma,(\Theta,e))\in\inst{\Gamma\gv\binbop{x}{a}{b}\rtyp [x:a]c}$.
By definition of $\inst{\Gamma\gv\binbop{x}{a}{b}\rtyp [x:a]c}$ there are two cases:

In the first case $\Theta=()$ and $((\Lambda,x:a),())\in\inst{\Gamma,x:a\gv b\rtyp c}$. 
Obviously $\nrm{\Lambda}{\binbop{x}{a}{b}}[()]=\nrm{\Lambda}{\binbop{x}{a}{b}}=\binbop{x}{\nrm{\Lambda}{a}}{\nrm{\Lambda,x:a}{b}}$.
(Note that Parts $ii$, $iii$, $iv$ will be trivially valid at this point as their assumptions are false.)
From $\nrm{}{\Lambda}\gv\nrm{\Lambda}{\binbop{x}{a}{b}}\![()]\finc[y:\bar{c}]\bar{d}$ by Law \ref{sub.flex.prop}($iii$, $iv$) we know that $x=y$ and $\nrm{\Lambda}{a}=\bar{c}$. 
Furthermore $\Lambda\sngv a$, $\Lambda\sngv e$, and $\nrm{}{\Lambda}\gv\nrm{\Lambda}{e}\ginc\nrm{\Lambda}{a}$. 
Hence, by definition of instantiation sets we know that $((\Lambda,x\mydef e),())\in\inst{\Gamma,x:a\gv b\rtyp c}$ and therefore
$(\Lambda,e)\in\inst{\Gamma\gv\binbop{x}{a}{b}\rtyp [x:a]c}$.

In the second case $\Theta=(d,\Theta')$, for some $d$, and $((\Lambda,x\mydef d),\Theta')\in\inst{\Gamma,x:a\gv b\rtyp c}$ and $\nrm{}{\Lambda}\gv\nrm{\Lambda}{d}\ginc\nrm{\Lambda}{a}$. 
Clearly $\nrm{}{\Lambda}\gv\nrm{\Lambda,\Theta}{\binbop{x}{a}{b}}\![d,\Theta']$ 
=$\binbop{x}{\nrm{\Lambda}{a}}{\nrm{(\Lambda,x\smydef d),\Theta}{b})}\![d,\Theta']$ 
=$\nrm{\Lambda,x\mydef d}{b}\![\Theta']$.
By assumption $\nrm{\Lambda,x\mydef d}{b}\![\Theta']\finc[y:\bar{c}]\bar{d}$.
By inductive hypothesis for $\inst{\Gamma,x:a\gv b\rtyp c}$ $(\Lambda,x\mydef d,(\Theta',e))\in\inst{\Gamma,x:a\gv b\rtyp c}$.
Hence by definition of instantiation sets we have $(\Lambda,(d,\Theta',e))$ $\in$ $\inst{\Gamma\gv\binbop{x}{a}{b}\rtyp [x:a]c}$.
Note that the condition $\nrm{}{\Lambda}\gv\nrm{\Lambda}{d}\ginc\nrm{\Lambda}{a}$ was not needed.

Similar arguments can be made for Parts $ii$, $iii$, and $iv$.
\item[\appl:] 
$\inst{\Gamma\gv(a\,b)\rtyp c\gsub{x}{b}}$ consists of all pairs $(\Lambda,\Theta)$ where $(\Lambda,(b,\Theta))\in\inst{\Gamma\gv a\rtyp [x:d]c}$ and $\nrm{}{\Lambda}\gv\nrm{\Lambda}{b}\ginc\nrm{\Lambda}{d}$. 
Assume that $(\Lambda,\Theta)\in\inst{\Gamma\gv(a\,b)\rtyp c\gsub{x}{b}}$. 

For Part $i$ assume that $\nrm{}{\Lambda}\gv\nrm{\Lambda,\Theta}{(a\,b)}\![\Theta]\finc[y:\bar{c}]\bar{d}$ for some $y$, $\bar{c}$, and $\bar{d}$ and assume
some $e$ with $\nrm{}{\Lambda}\gv\nrm{\Lambda}{e}\ginc\bar{c}$.
We have to show that $(\Gamma,(\Theta,e))\in\inst{\Gamma\gv(a\,b)\rtyp c\gsub{x}{b}}$.
Since by Law \ref{sub.elim.prop}($i$)~$\nrm{\Lambda}{(a\,b)}\![\Theta]=\nrm{\Lambda}{a}\![b,\Theta]$,
by inductive hypothesis for $\inst{\Gamma\gv a\rtyp [x:c]d}$ we obtain
$(\Lambda,(b,\Theta,e))\in\inst{\Gamma\gv a\rtyp [x:d]c}$.
By definition of instantiation sets we obtaib $(\Lambda,(\Theta,e))$ $\in$ $\inst{\Gamma\gv(a\,b):c\gsub{x}{b}}$.

\noindent
For Part $ii$ assume that
$\nrm{}{\Lambda}\gv\nrm{\Lambda}{(a\,b)}\![\Theta]\finc[\bar{c},\bar{d}]$ for some $\bar{c}$ and $\bar{d}$. With a similar argument we can show that
$(\Lambda,(\Theta,1)),(\Lambda,(\Theta,2))\in\inst{\Gamma\gv(a\,b)\rtyp c\gsub{x}{b}}$. 

\noindent
Parts $iii$ and $iv$ can be shown in a similar way.
\item[\pdef:] 
Assume that $(\Lambda,\Theta)\in\inst{\Gamma\gv\prdef{x}{a^b}{c}{d}\rtyp [x\sdef b]d}$.
Members of $\inst{\Gamma\gv\prdef{x}{a^b}{c}{d}\rtyp [x\sdef b]d}$ are pairs $(\Lambda,())$ where $(\Lambda,())\in\inst{\Gamma\gv a\rtyp b}\cap\inst{\Gamma\gv c\rtyp d\gsub{x}{a}}$, 
or $(\Lambda,(1,\Theta))$ where $(\Lambda,\Theta)\in\inst{\Gamma,\Theta\gv a\rtyp b}$, 
or $(\Lambda,(2_e,\Theta))$ where $(\Lambda,\Theta)\in\inst{\Gamma,\Theta\gv c\rtyp d\gsub{x}{a}}$. 

If $\Theta=()$ then $\nrm{\Lambda}{\prdef{x}{a^b}{c}{d}}\![()]=\prdef{x}{\nrm{\Lambda}{a}^{\snrm{\Lambda}{b}}}{\nrm{\Lambda}{c}}{\nrm{\Lambda}{d}}$.
Hence Parts $i$, $ii$, and $iv$ become trivial.
For Part $iii$, if $(\Lambda,())\in\inst{\Gamma\gv a\rtyp b}$ and $(\Lambda,())\in\inst{\Gamma\gv c\rtyp d\gsub{x}{a}}$ then
by definition of $\inst{\Gamma\gv\prdef{x}{a^b}{c}{d}\rtyp [x\sdef b]d}$
we know for any $e$ that $(\Lambda,1),(\Lambda,e)\in\inst{\Gamma\gv\prdef{x}{a^b}{c}{d}\rtyp [x\sdef b]d}$.

If $\Theta=(1,\Theta')$ then $\nrm{\Lambda,\Theta}{\prdef{x}{a^b}{c}{d}}\![1,\Theta']=\nrm{\Lambda,\Theta}{a}\![\Theta']$.
For Part $i$ assume that $\nrm{}{\Lambda}\gv\nrm{\Lambda,\Theta}{a}\![\Theta']\finc[y:\bar{c}]\bar{d}$ for some $y$, $\bar{c}$, and $\bar{d}$ and assume $\nrm{}{\Lambda}\gv\nrm{\Lambda}{e}\ginc\bar{c}$, for some $e$.
By inductive hypothesis for $\inst{\Gamma\gv a\rtyp b}$ $(\Lambda,(\Theta',e)\in\inst{\Gamma\gv a\rtyp b}$, hence $(\Lambda,(1,\Theta',e))=(\Lambda,(\Theta,e))\in\inst{\Gamma\gv\prdef{x}{a^b}{c}{d}\rtyp [x\sdef b]d}$. 
The Parts $ii$, $iii$, $iv$ are shown analogously.

The case $\Theta=(2_e,\Theta')$ is shown analogously.
\item[\bprod]and \bsum:
Assume that $(\Lambda,\Theta)\in\inst{\Gamma\gv\prsumop{a}{b}\rtyp [c,d]}$.
Member of $\inst{\Gamma\gv\prsumop{a}{b}\rtyp [c,d]}$ are pairs $(\Lambda,())$ where $(\Lambda,())\in\inst{\Gamma\gv a\rtyp c}\cap\inst{\Gamma\gv b\rtyp c}$,
and pairs  $(\Lambda,(1,\Theta))$ where  $(\Lambda,\Theta)\in\inst{\Gamma\gv a\rtyp c}$,
and pairs $(\Lambda,(2,\Theta))$ where  $(\Lambda,\Theta)\in\inst{\Gamma\gv b\rtyp d}$.

If $\Theta=()$ then $\nrm{\Lambda}{\prsumop{a}{b}}\![()]=\prsumop{\nrm{\Lambda}{a}}{\nrm{\Lambda}{b}}$.
Hence Parts $i$, $iii$, and $iv$ become trivial.
For Part $ii$, since $(\Lambda,())\in\inst{\Gamma\gv a\rtyp c}\cap\inst{\Gamma\gv b\rtyp c}$, by definition of $\inst{\Gamma\gv\prsumop{a}{b}\rtyp [c,d]}$ 
we know that $(\Lambda,1)\in\inst{\Gamma\gv a\rtyp c}$, $(\Lambda,2)\in\inst{\Gamma\gv b\rtyp d}$.

If $\Theta=(1,\Theta')$ then $\nrm{\Lambda,\Theta}{\prsumop{a}{b}}\![\Theta]=\nrm{\Lambda,\Theta}{a}\![\Theta']$.
For Part $i$ assume that $\nrm{}{\Lambda}\gv\nrm{\Lambda,\Theta}{a}\![\Theta']\finc[x:\bar{c}]\bar{d}$ and $\nrm{}{\Lambda}\gv\nrm{\Lambda}{e}\ginc\bar{c}$, for some $e$
By inductive hypothesis $(\Lambda,(\Theta',e)\in\inst{\Gamma\gv a\rtyp c}$, hence $(\Lambda,(1,\Theta',e)=(\Lambda,(\Theta',e))\in\inst{\Gamma\gv\prsumop{a}{b}\rtyp [c,d]}$.
The Parts $ii$, $iii$, $iv$ are shown analogously.

The case $\Theta=(2,\Theta')$ is show analogously.
\item[\prl:]
Assume that $(\Lambda,\Theta)\in\inst{\Gamma\gv\pleft{a}\rtyp b}$ where $\Gamma\gv a\rtyp [b,c]$.
Members of $\inst{\Gamma\gv\pleft{a}\rtyp b}$ are pairs $(\Lambda,\Theta)$ where $(\Lambda,(1,\Theta))\in\inst{\Gamma\gv a\rtyp [x\sdef b]c}$ $\cup\inst{\Gamma\gv a\rtyp [b,c]}$.

For Part $i$ assume that
$\nrm{}{\Lambda}\gv\nrm{\Lambda,\Theta}{\pleft{a}}\![\Theta]\finc[x:\bar{c}]\bar{d}$ for some $x$, $\bar{c}$, and $\bar{d}$ and assume
some $e$ with $\nrm{}{\Lambda}\gv\nrm{\Lambda}{e}\ginc\bar{c}$.
We have to show that $(\Gamma,(\Theta,e))\in\inst{\Gamma\gv\pleft{a}\rtyp b}$.
Since by Law \ref{sub.elim.prop}($i$) $\nrm{\Lambda}{\pleft{a}}\![\Theta]=\nrm{\Lambda}{a}\![1,\Theta]$,
by inductive hypothesis for $\inst{\Gamma\gv a\rtyp [b,c]}$ we obtain
$(\Lambda,(1,\Theta,e))\in\inst{\Gamma\gv a\rtyp [b,c]}$.
By definition of instantiation sets $(\Lambda,(\Theta,e))\in\inst{\Gamma\gv\pleft{a}\rtyp b}$.

The Parts $ii$, $iii$, $iv$ are shown analogously.
\item[\prr:]
Similar to \prl.
\item[\chin:]
Proof is analogously to \prl. 
\item[\chba:]
Assume that $(\Lambda,\Theta)\in\inst{\Gamma\gv\pright{a}\rtyp c\gsub{x}{\pleft{a}}}$ where $\Gamma\gv a\rtyp [x\sdef b]c$.
Members of $\inst{\Gamma\gv\pright{a}\rtyp c\gsub{x}{\pleft{a}}}$ are pairs $(\Lambda,\Theta)$ where $(\Lambda,(2_{\pleft{a}},\Theta))\in\inst{\Gamma\gv a\rtyp [x\sdef b]c}$.

For Part $i$ assume that
$\nrm{}{\Lambda}\gv\nrm{\Lambda}{\pright{a}}\![\Theta]\finc[y:\bar{c}]\bar{d}$ for some $y$, $\bar{c}$, and $\bar{d}$ and assume
some $e$ with $\nrm{}{\Lambda}\gv\nrm{\Lambda}{e}\ginc\bar{c}$.
We have to show that $(\Gamma,(\Theta,e))\in\inst{\Gamma\gv\pright{a}\rtyp c\gsub{x}{\pleft{a}}}$.
Since by Law \ref{sub.elim.prop}($i$) $\nrm{\Lambda}{\pright{a}}\![\Theta]=\nrm{\Lambda}{a}\![(2_{\pleft{a}},\Theta)]$,
by inductive hypothesis for $\inst{\Gamma\gv a\rtyp [x\sdef b]c}$ we obtain
$(\Lambda,(2_{\pleft{a}},\Theta,e))\in\inst{\Gamma\gv a\rtyp [x\sdef b]c}$.
By definition of instantiation sets $(\Lambda,(\Theta,e))\in\inst{\Gamma\gv\pright{a}\rtyp c\gsub{x}{\pleft{a}}}$.

Parts $ii$, $iii$, $iv$ are shown analogously.
\item[\injll:]
Assume that $(\Lambda,\Theta)\in\inst{\Gamma\gv\injl{a}{\,b}\rtyp [c+b]}$.
Members of $\inst{\Gamma\gv\injl{a}{\,b}\rtyp [c+b]}$ are pairs $(\Lambda,())$ where $(\Lambda,())\in\inst{\Gamma\gv a\rtyp c}$, therefore $\Theta=()$.
By Law~\ref{sub.flex.inst} we know that $\nrm{}{\Lambda}\gv\nrm{\Lambda}{\injl{a}{\,b}}\finc\nrm{\Lambda}{[c+b]}=[\nrm{\Gamma}{c}+\nrm{\Gamma}{b}]$.
Hence as a consequence of Law \ref{sub.flex.prop} Parts $i$ to $iv$ are trivially true. 
\item[\injlr:]
Similar to \injll.
\item[\cased:]
Assume that $(\Lambda,\Theta)\in\inst{\Gamma\,\gv\case{a}{b}\rtyp [x:[c_1+c_2]]d}$ 
and, for Part $i$, that $\nrm{}{\Lambda}\gv\nrm{\Lambda,\Theta}{\case{a}{b}}\![\Theta]\finc[y:\bar{c}']\bar{d}'$ for some $y$, $\bar{c}'$, and $\bar{d}'$.
and some $e$ with $\nrm{}{\Lambda}\gv\nrm{\Lambda}{e}\ginc\bar{c}'$.
We have to show that $(\Gamma,(\Theta,e))\in\inst{\Gamma\,\gv\case{a}{b}\rtyp [x:[c_1+c_2]]d}$.
By definition of $\inst{\case{a}{b}\rtyp [x:[c_1+c_2]]d}$ there are two cases:

In the first case $\Theta=()$ and $(\Lambda,())\in\inst{\Gamma\,\gv\case{a}{b}\rtyp [x:[c_1+c_2]]d}$. 
This means $(\Lambda,())\in\inst{\Gamma\,\gv a\rtyp [x:c_1]d}$ and $(\Lambda,())$ $\in$ $\inst{\Gamma\,\gv b\rtyp [x:c_2]d}$.
Furthermore  $\nrm{\Lambda}{a}=[x:\nrm{\Lambda}{c_1}]\nrm{\Lambda}{d}$, $\nrm{\Lambda}{b}=[x:\nrm{\Lambda}{c_2}]\nrm{\Lambda}{d}$ and therefore $\nrm{\Lambda}{[a?b]}=[x:[\nrm{\Lambda}{c_1}+\nrm{\Lambda}{c_2}]\nrm{\Lambda}{d}$. Hence for this case Parts $ii$, $ii$ $iv$ will become trivial.
By inductive hypothesis $(\Lambda,e_1)\in\inst{\Gamma\,\gv a\rtyp [x:c_1]d}$ and $(\Lambda,e_2)\in\inst{\Gamma\,\gv b\rtyp [x:c_2]d}$ for some $e_1$, $e_2$ where $\Gamma\sgv  e_1\rtyp c_1$ and $\Gamma\sgv e_2\rtyp c_2$.
Let $e=[e_1+e_2]$. By Law \ref{sub.inst.equal} $(\Lambda,\pleft{e})\in\inst{\Gamma\,\gv a\rtyp [x:c_1]d}$ and $(\Lambda,\pright{e})\in\inst{\Gamma\,\gv b\rtyp [x:c_2]d}$
Hence $(\Lambda,e)\in\inst{\Gamma\,\gv\case{a}{b}\rtyp [x:[c_1+c_2]]d}$.

In the second case we have $\Theta=(f,\Theta')$ for some $f$ and $(\Lambda,(f,\Theta'))$ $\in$ $\inst{\Gamma\,\gv\case{a}{b}\rtyp [x:[c_1+c_2]]d}$.
Thus $(\Lambda,(\pleft{f},\Theta'))$ $\in$ $\inst{\Gamma\gv a\rtyp [x:c_1]d}$ and $(\Lambda,(\pright{f},\Theta'))$ $\in$ $\inst{\Gamma\gv b\rtyp [x:c_2]d}$. 
$\nrm{}{\Lambda}\gv\nrm{\Lambda}{f}\ginc\nrm{\Lambda}{[c_1+c_2]}$,
$\nrm{\Lambda,(f,\Theta')}{a}=[x:\nrm{\Lambda}{c_1}]\nrm{\Lambda,\Theta'}{d}$, and $\nrm{\Lambda,(f,\Theta')}{b}=[x:\nrm{\Lambda}{c_2}]\nrm{\Lambda,\Theta'}{d}$.
Since $\nrm{\Lambda,\Theta}{\case{a}{b}}\![\Theta]=\nrm{\Lambda,\Theta'}{d}[\Theta']$ we can apply the inductive hypothesis twice to obtain
$(\Lambda,(\pleft{f},\Theta',e))\in\inst{\Gamma\gv a\rtyp [x:c_1]d}$ and $(\Lambda,(\pright{f},\Theta',e))\in\inst{\Gamma\gv b\rtyp [x:c_2]d}$ and therefore
$(\Lambda,[\Theta,e))\in\inst{\Gamma\,\gv\case{a}{b}\rtyp [x:[c_1+c_2]]d}$.
Furthermore we know that Parts $ii$, $ii$ $iv$ will become trivial.
\item[\negate:]
Follows from the definition of $\inst{\Gamma\gv\myneg{a}\rtyp b}$ and the inductive assumptions of the four parts.
\end{meditemize}
\noindent $\inst{\Gamma\gv a\rleq  b}$:
\begin{meditemize}
\item[\srefl:]
Members of $\inst{\Gamma\,\gv a\rleq  a}$ are pairs $(\Lambda,\Theta)\in\inst{\Gamma\,\gv a\rtyp b}$ for some $b$ with $\Gamma\,\sgv a\rtyp b$.
Given the assumptions of Part $i$, by inductive hypothesis we know that $(\Lambda,(e,\Theta))\in\inst{\Gamma\,\gv a\rtyp b}$ hence
$(\Lambda,(e,\Theta))\in\inst{\Gamma\,\gv a\rleq  a}$.
Similar for Parts $ii$, $iii$, $iv$.
\item[\sembed:]
The argument is analogous to the above case \srefl.
\item[\sabs:]
Members of $\inst{\Gamma\,\gv\binbop{x}{a}{b}\rleq \binbop{x}{a}{c}}$ are all pairs $(\Lambda,()\in\inst{\Gamma,x:a\,\gv b\rleq  c}$
and all pairs $(\Lambda,(d,\Theta))$  for some $d$  where $((\Lambda,x\mydef d),\Theta)\in\inst{\Gamma,x:a\,\gv b\rleq  c}$, $\Lambda\sngv a$, $\Lambda\sngv d$, and
$\nrm{}{\Lambda}\gv\nrm{\Lambda}{d}\ginc\nrm{\Lambda}{a}$.
The argument is similar to cases \absu\ and \abse.
\item[\sbprod:]
Members of  $\inst{\Gamma\,\gv\prsumop{a}{c}\rleq \prsumop{b}{d}}$ are all pairs $(\Lambda,())$ where $(\Lambda,())\in\inst{\Gamma\gv a\rleq  b}\cap\inst{\Gamma\,\gv c\rleq  d}$,
and all pairs  $(\Lambda,(1,\Theta))$ where  $(\Lambda,\Theta)\in\inst{\Gamma\,\gv a\rleq  b}$,
and all pairs  $(\Lambda,(2,\Theta))$ where  $(\Lambda,\Theta)\in\inst{\Gamma\,\gv c\rleq  d}$.
The argument is similar to cases \bprod\ and \bsum.
\end{meditemize}
\noindent $\inst{\Gamma\gv a\rsinc \any{n}}$:
\begin{meditemize}
\item[\sstart:]
Members of $\inst{\Gamma\,\gv\any{m}\rsinc \any{n}}$ are all pairs $(\Lambda,\Theta)\in\inst{\Gamma\gv\any{m}\rtyp a}\cap\inst{\Gamma\gv\any{n}\rtyp b}$.
The premises of Parts $i$, $ii$, and $iii$ are obviously false.
For Part $iv$ since $\nrm{\Lambda,\Theta}{\any{m}}\![\Theta]=\any{m}$ by inductive hypothesis we know that $(\Lambda,(\Theta,d))\in\inst{\Gamma\gv\any{m}\rtyp a}$.
Similar for $\any{n}$ which implies the proposition.
\item[\sstyp:]
Members of $\inst{\Gamma\,\gv a\rsinc \any{n}}$ are pairs $(\Lambda,\Theta)$ with $(\Lambda,\Theta)\in\inst{\Gamma\,\gv a\rtyp b}\cap\inst{\Gamma\,\gv b\rsinc \any{n}}$.
By an argument analogous to the one in case \sinc\ we can show that the inductive hypothesis for $\inst{\Gamma\,\gv b\rsinc \any{n}}$ is applicable which implies the proposition.
\item[\ssabsu:]
$\inst{\Gamma\,\gv\binbop{x}{a}{b}\rsinc \any{n}}$ consists of all pairs $(\Lambda,())\in\inst{\Gamma,x:a\,\gv b\rsinc \any{n}}$
and of all pairs $(\Lambda,(d,\Theta))$  for some $d$  where $((\Lambda,x\mydef d),\Theta)\in\inst{\Gamma,x:a\,\gv b\rsinc \any{n}}$, $\Lambda\sngv a$, $\Lambda\sngv d$, and
$\nrm{}{\Lambda}\gv\nrm{\Lambda}{d}\ginc\nrm{\Lambda}{a}$.

Assume that $(\Lambda,\Theta)\in\inst{\Gamma\gv\binbop{x}{a}{b}\rsinc \any{n}}$ 
and, for Part $i$, that $\nrm{}{\Lambda}\gv\nrm{\Lambda,\Theta}{\binbop{x}{a}{b}}\![\Theta]\finc[y:\bar{c}]\bar{d}$ for some $y$, $\bar{c}$, and $\bar{d}$.
and some $e$ with $\nrm{}{\Lambda}\gv\nrm{\Lambda}{e}\ginc\bar{c}$.
We have to show that $(\Gamma,(\Theta,e))\in\inst{\Gamma\gv[x:a]b\rsinc \any{n}}$.
By definition of $\inst{\Gamma\gv\binbop{x}{a}{b}\rsinc \any{n}}$ there are two cases:

In the first case $\Theta=()$ and $((\Lambda,x:a),())\in\inst{\Gamma,x:a\gv b\rsinc \any{n}}$. 
Obviously $\nrm{\Lambda}{\binbop{x}{a}{b}}[()]=\nrm{\Lambda}{\binbop{x}{a}{b}}=\binbop{x}{\nrm{\Lambda}{a}}{\nrm{\Lambda,x:a}{b}}$.
(Note that Parts $ii$, $iii$, $iv$ will be trivially valid at this point as their assumptions are false.)
From $\nrm{}{\Lambda}\gv\nrm{\Lambda}{\binbop{x}{a}{b}}\![()]\finc[y:\bar{c}]\bar{d}$ by Law \ref{sub.flex.prop}($iii$,$iv$) we know that $x=y$ and $\nrm{\Lambda}{a}=\bar{c}$. 
Furthermore $\nrm{}{\Lambda}\gv\nrm{\Lambda}{e}\ginc\nrm{\Lambda}{a}$. 
Hence, by definition of instantiation sets we know that $((\Lambda,x\mydef e),())\in\inst{\Gamma,x:a\gv b\rsinc \any{n}}$ and therefore
$(\Lambda,e)\in\inst{\Gamma\gv\binbop{x}{a}{b}\rsinc \any{n}}$.

In the second case $\Theta=(d,\Theta')$, for some $d$, and $((\Lambda,x\mydef d),\Theta')\in\inst{\Gamma,x:a\gv b\rsinc \any{n}}$ and $\nrm{}{\Lambda}\gv\nrm{\Lambda}{d}\ginc\nrm{\Lambda}{a}$. 
Clearly $\nrm{}{\Lambda}\gv\nrm{\Lambda,\Theta}{\binbop{x}{a}{b}}\![d,\Theta']$ 
=$\binbop{x}{\nrm{\Lambda}{a}}{\nrm{(\Lambda,x\smydef d),\Theta}{b})}\![d,\Theta']$ 
=$\nrm{\Lambda,x\mydef d}{b}\![\Theta']$.
By assumption $\nrm{\Lambda,x\mydef d}{b}\![\Theta']\finc[y:\bar{c}]\bar{d}$.
By inductive hypothesis for $\inst{\Gamma,x:a\gv b\rtyp c}$ $(\Lambda,x\mydef d,(\Theta',e))\in\inst{\Gamma,x:a\gv b\rsinc \any{n}}$.
Hence by definition of instantiation sets we have $(\Lambda,(d,\Theta',e))$ $\in$ $\inst{\Gamma\gv\binbop{x}{a}{b}\rsinc \any{n}}$.
Note that the condition $\nrm{}{\Lambda}\gv\nrm{\Lambda}{d}\ginc\nrm{\Lambda}{a}$ was not needed.

The Parts $ii$, $iii$, $iv$ are shown analogously.
\item[\ssbprod:]
The argument is similar to cases \bprod\ and \bsum.

\item[\ssinjl:]
The argument is similar to cases \injll.
\item[\ssinjr:]
Similar to \ssinjl.
\item[\sspdef:]
Members of $\inst{\Gamma\,\gv\prdef{x}{a^b}{c}{d}\rsinc \any{n}}$ 
are pairs $(\Lambda,())$ where $(\Lambda,())\in\inst{\Gamma\,\gv a\rsinc \any{n}}$ $\cap$ $\inst{\Gamma\,\gv b\rsinc \any{n}}$ $\cap$ $\inst{\Gamma\,\gv c\rsinc \any{n}}$
and $((\Lambda,x:b),())\in\inst{\Gamma,x:b\,\gv d\rsinc \any{n}}$, 
pairs $(\Lambda,(1,\Theta))$ where $(\Lambda,\Theta)$ $\in$ $\inst{\Gamma\,\gv a\rsinc \any{n}}$, 
and $(\Lambda,(2_e,\Theta))$ where  $(\Lambda,\Theta)$ $\in$ $\inst{\Gamma\,\gv c\rsinc \any{n}}$. 
The argument is similar to case \ssabsu.
\qedhere
\end{meditemize}
\end{proof}
\begin{law}[Instantiation sets contain empty elimination]
\label{sub.inst.prop}
For all $\Gamma$, $\Lambda$, $n$, $a$, $b$ with $\Lambda:\Gamma$: 
\begin{itemize}
\item[$i$:]
If $\Gamma\gv a\rtyp b$ then $(\Lambda,())\in\inst{\Gamma\gv a\rtyp b}$.
\item[$ii$:]
If $\Gamma\gv a\rleq  b$ then $(\Lambda,())\in\inst{\Gamma\gv a\rleq  b}$.
\item[$iii$:]
If $\Gamma\gv a\rsinc \any{n}$ then $(\Lambda,())\in\inst{\Gamma\gv a\rsinc \any{n}}$.
\end{itemize}
\end{law}
\begin{proof}
All parts are shown simultaneously by induction on the definition of instantiation sets.

\noindent {\bf Part} $i$:
\begin{meditemize}
\item[\ax:] Obvious.
\item[\mystart:]
$\inst{\Gamma,x:a\,\gv x\rtyp a}$ consists of all pairs $((\Lambda,x:a),\Theta)$ where $(\Lambda,\Theta)\in\inst{\Gamma\,\gv a\rtyp b}$.
and all pairs  $((\Lambda,x\mydef c),\Theta)$ for some $c$ with $\Gamma\gv c\rtyp a$ and $(\Lambda,\Theta)\in\inst{\Gamma\,\gv c\rtyp a}$.

Consider some $\Lambda$ where $\Lambda:(\Gamma, x:a)$: Obviously $\Lambda=(\Lambda',x:a)$ or $\Lambda=(\Lambda',x\mydef c)$ where $\Lambda':\Gamma$ and 
$\Gamma\sgv c\rtyp a$.
In the first case we have to show that $((\Lambda',x:a),())\in\inst{\Gamma,x:a\,\gv x\rtyp a}$. 
By definition of instantiation sets and the inductive hypothesis we know that $(\Lambda',())\in\inst{\Gamma\gv a\rtyp b}$ where $\Lambda':\Gamma$.
By definition of instantiation sets $((\Lambda',x:a),())\in\inst{\Gamma,x:a\,\gv x\rtyp a}$.
In the second case we have to show that $((\Lambda',x\mydef b),())\in\inst{\Gamma,x:a\,\gv x\rtyp a}$. 
By definition of instantiation sets and the inductive hypothesis we know that $(\Lambda',())\in\inst{\Gamma\gv c:a}$ where $\Lambda':\Gamma$.
Hence by definition of instantiation sets $((\Lambda',x\mydef b),())\in\inst{\Gamma,x:a\,\gv x\rtyp a}$.
\item[\weak:]
$\inst{\Gamma,x:c\,\gv a\rtyp b}$ consists of all pairs $((\Lambda,x:c),\Theta)$ and $((\Lambda,x\mydef d),\Theta)$ 
where $(\Lambda,\Theta)\in\inst{\Gamma\,\gv a\rtyp b}$ and $\Gamma\gv d\rtyp c$ for some $d$ with $(\Lambda,())\in\inst{\Gamma\,\gv d\rtyp c}$.

Consider some $\Lambda$ where $\Lambda:(\Gamma, x:c)$: Obviously $\Lambda=(\Lambda',x:c)$ or $\Lambda=(\Lambda',x\mydef d)$ where $\Lambda':\Gamma$ and 
$\Gamma\gv d\rtyp c$.
The argument then goes as in case~\mystart.
\item[\conv:]
$\inst{\Gamma\gv a\rtyp c}$ consists of all pairs $(\Lambda,\Theta)$ where $(\Lambda,\Theta)\in\inst{\Gamma\gv a\rtyp b}$ and $b\eqv c$.
The proposition follows directly from the inductive hypothesis.
\item[\sinc:]
$\inst{\Gamma\gv a\rtyp c}$ consists of all  pairs $(\Lambda,\Theta)$ where $(\Lambda,\Theta)\in\inst{\Gamma\gv a\rtyp b}\cap\inst{\Gamma\gv b\rleq  c}$.
Consider some $\Lambda$ where $\Lambda:\Gamma$:
By inductive hypothesis $(\Lambda,())\in\inst{\Gamma\gv a\rtyp b}\cap\inst{\Gamma\gv b\rleq  c}$.
Therefore by definition of $\inst{\Gamma\gv a\rtyp c}$ we know that $(\Lambda,())\in\inst{\Gamma\gv a\rtyp c}$.
\item[\absu]and \abse:
$\inst{\Gamma\gv\binbop{x}{a}{b}\rtyp [x:a]c}$ includes all pairs $(\Lambda,())$ with $((\Lambda,x:a),())\in\inst{\Gamma,x:a\gv b\rtyp c}$.
Consider some $\Lambda$ where $\Lambda:\Gamma$:  By inductive hypothesis $(\Lambda',())\in\inst{\Gamma,x:a\gv b\rtyp c}$ for all $\Lambda':(\Gamma,x:a)$.
Obviously $(\Lambda,x:a):(\Gamma,x:a)$, hence $((\Lambda,x:a),())\in\inst{\Gamma,x:a\gv b\rtyp c}$. By definition of instantiation sets $(\Lambda,())\in\inst{\Gamma\gv\binbop{x}{a}{b}\rtyp [x:a]c}$.
\item[\appl:] 
Members of $\inst{\Gamma\gv(a\,b)\rtyp c\gsub{x}{b}}$ are pairs $(\Lambda,\Theta)$ where $(\Lambda,(b,\Theta))\in\inst{\Gamma\gv a\rtyp [x:d]c}$ and $\nrm{}{\Lambda}\gv\nrm{\Lambda}{b}\ginc\nrm{\Lambda}{d}$. 
We show that $(\Lambda,())\in\inst{\Gamma\gv(a\,b)\rtyp c\gsub{x}{b}}$ which follows
from $(\Lambda,b)\in\inst{\Gamma\gv a\rtyp [x:d]c}$.
By inductive hypothesis $(\Lambda,())\in\inst{\Gamma\gv a\rtyp [x:d]c}$.
By Law~\ref{sub.flex.inst} we know that $\nrm{}{\Lambda}\gv\nrm{\Lambda}{a}\finc\nrm{\Lambda}{[x:d]c}=[x:\nrm{\Lambda}{d}]\bar{e}$ for some $\bar{e}$. 
Since $\nrm{}{\Lambda}\gv\nrm{\Lambda}{b}\ginc\nrm{\Lambda}{d}$, by Law~\ref{sub.inst.law}($i$) with $\Theta=()$ we know that $(\Lambda,b)\in\inst{\Gamma\gv a\rtyp [x:d]c}$.
\item[\pdef:] 
Members of $\inst{\Gamma\gv\prdef{x}{a^b}{c}{d}\rtyp [x\sdef b]d}$ include all pairs $(\Lambda,())$ where $(\Lambda,())\in\inst{\Gamma\gv a\rtyp b}\cap\inst{\Gamma\gv c\rtyp d\gsub{x}{a}}$.
For any $\Lambda$ where $\Lambda:\Gamma$ by inductive hypothesis $(\Lambda',())\in\inst{\Gamma\gv a\rtyp b}$ for all $\Lambda':\Gamma$.
Hence $(\Lambda,())\in\inst{\Gamma\gv\prdef{x}{a^b}{c}{d}\rtyp [x\sdef b]d}$.
\item[\chin:] 
$\inst{\Gamma\gv\pleft{a}\rtyp b}$ contains $(\Lambda,\Theta)$ where $(\Lambda,(1,\Theta))\in\inst{\Gamma\gv a\rtyp [x\sdef b]c}$ $\cup\inst{\Gamma\gv a\rtyp [b,c]}$
By Law~\ref{sub.flex.inst} we know that $\nrm{}{\Lambda}\gv\nrm{\Lambda}{a}\finc\nrm{\Lambda}{[x\sdef b]c}=[x\sdef\nrm{\Lambda}{b}]\nrm{\Lambda,x:b}{c}$.
By Law~\ref{sub.inst.law}($iii$) with $\Theta=()$ we know that $(\Lambda,1)\in\inst{\Gamma\gv a\rtyp [x\sdef b]c}$.
By definition of instantiation sets this implies $(\Lambda,())\in\inst{\Gamma\gv\pleft{a}\rtyp b}$. 
\item[\chba:] 
Similar to case \chin.
\item[\bprod]and \bsum:
Member of $\inst{\Gamma\gv\prsumop{a}{b}\rtyp [c,d]}$ include all pairs $(\Lambda,())$ where $(\Lambda,())$ $\in$ $\inst{\Gamma\gv a\rtyp c}\cap\inst{\Gamma\gv b\rtyp c}$,
Consider some $\Lambda$ where $\Lambda:\Gamma$:  By inductive hypothesis $(\Lambda',())\in\inst{\Gamma\gv a\rtyp c}\cap\inst{\Gamma\gv b\rtyp c}$ for all $\Lambda':\Gamma$.
Hence $(\Lambda,())\in\inst{\Gamma\gv\prsumop{a}{b}\rtyp [c,d]}$.
\item[\prl] and \prr:  
Similar to case \chin using~\ref{sub.inst.law}($ii$).
\item[\cased:]
$\inst{\Gamma\,\gv\case{a}{b}\rtyp [x:[c_1+c_2]]d}$ includes all pairs $(\Lambda,())$ where $(\Lambda,())$ $\in$ $\inst{\Gamma\,\gv a\rtyp [x:c_1]d}$ $\cap\inst{\Gamma\,\gv b\rtyp [x:c_2]d}$.
Consider some $\Lambda'$ where $\Lambda':\Gamma$:  By inductive hypothesis $(\Lambda',())\in\inst{\Gamma\,\gv a\rtyp [x:c_1]d}$ and  $(\Lambda',())\in\inst{\Gamma\,\gv b\rtyp [x:c_2]d}$ 
for all $\Lambda':\Gamma$.
Hence $(\Lambda',())\in\inst{\Gamma\,\gv\case{a}{b}\rtyp [x:[c_1+c_2]]d}$.
\item[\negate:]
Follows from the inductive hypothesis.
\end{meditemize}
\noindent
{\bf Part} $ii$, $iii$: In all cases obvious consequence from the inductive hypothesis.
\end{proof}
\begin{law}[Expressions with restricted validity are normable]
\label{sub.val.nrm}
For all $\Gamma,a$:
$\Gamma\rsgv a$ implies $\Gamma\sngv a$.
\end{law}
\begin{proof}
$\Gamma\rsgv a$ means that $\Gamma\sgv a\rtyp b$, for some $b$.
Since $\Gamma:\Gamma$, by Law~\ref{sub.inst.prop}, this implies  $(\Gamma,())\in\inst{\Gamma\gv a\rtyp b}$.
By Law~\ref{sub.flex.inst}, this implies  $\nrm{}{\Gamma}\gv\nrm{\Gamma}{a}\ginc\nrm{\Gamma}{b}$.
Hence $\Gamma\sngv a$
\end{proof}
\subsection{Strong normalization}
\label{sub.strong.normalization}
Due to the addition of type tags, Part $iv$ of Law~\ref{sn.basic}(Basic properties of strongly normalizable expressions), has to be adapted as follows
\[
a,b,c\in\sn{}\quad\text{implies}\quad\prdef{x}{a^d}{b}{c}\in\sn{}. 
\]
The proof can be adapted in a straightforward manner. 

First, in order to use inductive arguments based on norming we need to define a well-founded order on norms. 
\begin{definition}[Norm order, Weak norm order]
The $\any{}$-{\emph{size}} $\asize{\bar{\Gamma}}{\bar{a}}{n}$ of a norm $\bar{a}$ under norm context $\bar{\Gamma}$ is recursively defined by 
$\asize{\bar{\Gamma}}{\any{m}}{n}=1$ if $n=m$ and $0$ otherwise, $\asize{\bar{\Gamma}}{x}{n}=\asize{\bar{\Gamma}}{\bar{\Gamma}(x)}{n}$ if $x\in{\dom}(\bar{\Gamma})$, 
$\asize{\bar{\Gamma}}{\prsuminjop{\bar{a}}{\bar{b}}}{n}=\asize{\bar{\Gamma}}{\bar{a}}{n}+\asize{\bar{\Gamma},x:\bar{a}}{\bar{b}}{n}$, and
$\asize{\bar{\Gamma}}{\binbop{x}{\bar{a}_1}{\bar{a}_2,\ldots,\bar{a}_m}}{n}=\asize{\bar{\Gamma}}{\bar{a}_1}{n}+\asize{\bar{\Gamma},x:\bar{a}_1}{\bar{a}_2}{n}$ + $\ldots$ 
+ $\asize{\bar{\Gamma},x:\bar{a}_1}{\bar{a}_m}{n}$.
The {\emph{size}} $S_{\bar{\Gamma}}(\bar{a})$ of a norm $\bar{a}$ under norm context $\bar{\Gamma}$ is defined as as the maximal number $n$ with $\asize{\bar{\Gamma}}{\bar{a}}{n}>0$.
For norm context $\bar{\Gamma}$ and norms $\bar{a}$, $\bar{b}$, {\em norm order} $\bar{\Gamma}\sgv\bar{a}\nord\bar{b}$  
is defined by the rules shown in Table \ref{sub.nord.rules}. 
\begin{table}[!htb]
\fbox{
\begin{minipage}{0.96\textwidth}
\emph{Norm order:}
\begin{align*}
\\[-7mm]
\ostartm\;\frac{S_{\bar{\Gamma}}(\bar{a})\rsinc n}{\bar{\Gamma}\sgv\bar{a}\nord\any{n}}
\quad\ovarm\;\frac{x\in\dom{(\Gamma)}\;\;\bar{\Gamma}\sgv\bar{\Gamma}(x)\nord\bar{b}}{\bar{\Gamma}\sgv x\nord\bar{b}}\\[-11mm]
\end{align*}
\begin{align*}
\mathit{(\otimes_{\nord,1})}\;\frac{\bar{\Gamma}\sgv\bar{a}_1\nord\bar{b}_1\quad\bar{\Gamma}\sgv\bar{a}_2\wnord\bar{b}_2}{\prsuminjop{\bar{a}_1}{\bar{a}_2}\nord\prsuminjop{\bar{b}_1}{\bar{b}_2}}
\quad
\mathit{(\otimes_{\nord,2})}\;\frac{\bar{\Gamma}\sgv\bar{a}_1\wnord\bar{b}_1\quad\bar{\Gamma}\sgv\bar{a}_2\nord\bar{b}_2}{\prsuminjop{\bar{a}_1}{\bar{a}_2}\nord\prsuminjop{\bar{b}_1}{\bar{b}_2}}\\[-11mm]
\end{align*}
\begin{align*}
\mathit{(\oplus_x{\overbrace{(\_,\ldots,\_)}^{n}}_{\nord,i})}\;\frac{\bar{\Gamma}\sgv\bar{a}\wnord\bar{b}_i\quad\bar{\Gamma}\sgv\bar{b}_j, j\neq i}{\bar{a}\nord \binbop{x}{\bar{b}_1,\ldots,\bar{b}_i}{\ldots,\bar{b}_n}}\\[-11mm]
\end{align*}
\begin{align*}
\mathit{(\oplus_x{\overbrace{(\_,\ldots,\_)}^{n}}_{\nord,n+1})}\;
\frac{\bar{\Gamma}\sgv\bar{a}_1\nord\bar{b}_1
\quad\bar{\Gamma},x:\bar{a}_1\sgv\bar{a}_i\wnord\bar{b}_i}{\binbop{x}{\bar{a}_1,\ldots,\bar{a}_i}{\ldots,\bar{a}_n}\nord \binbop{x}{\bar{b}_1,\ldots,\bar{b}_i}{\ldots,\bar{b}_n}}\\[-11mm]
\end{align*}
\begin{align*}
\mathit{(\oplus_x{\overbrace{(\_,\ldots,\_)}^{n}}_{\nord,n+i})}\;
\frac{\bar{\Gamma}\sgv\bar{a}_1\!\wnord\bar{b}_1
\;\;\bar{\Gamma}\!,x:\bar{a}_1\sgv\bar{a}_i\!\nord\bar{b}_i
\;\;\bar{\Gamma}\!,x:\bar{a}_1\sgv\bar{a}_j\!\wnord\bar{b}_j, j\neq i>1}{\binbop{x}{\bar{a}_1,\ldots,\bar{a}_i}{\ldots,\bar{a}_n}\nord \binbop{x}{\bar{b}_1,\ldots,\bar{b}_i}{\ldots,\bar{b}_n}}
\end{align*}
\end{minipage}
}
\caption{Norm order.\label{sub.nord.rules}}
\end{table}
The rules also introduce {\em weak norm order} $\bar{\Gamma}\sgv\bar{a}\wnord\bar{b}$ defined as follows:
\[
\bar{\Gamma}\sgv\bar{a}\wnord\bar{b}\;\text{if and only if}\;\bar{\Gamma}\sgv\bar{a}\nord\bar{b}\;\text{or}\;\bar{\Gamma}\sgv\bar{a},\bar{a}=\bar{b},
\] 
\noindent
Note that \eg\ $\otimes_{\nord,1}$ and $\otimes_{\nord,2}$ overlap, however they are consistent and in this formulation only two rules are needed. 
Furthermore the two rules obviously imply 
\begin{equation}
\label{sub.nord.bin}
\frac{\bar{\Gamma}\sgv\bar{a}\wnord\bar{c}\quad\bar{\Gamma},x:\bar{a}\sgv\bar{b}\wnord\bar{d}}{\bar{\Gamma}\sgv\binop{\bar{a}}{\bar{b}}\wnord\binop{\bar{c}}{\bar{d}}}
\end{equation}
Similarly, \eg\ $\oplus_x{(\_,\_)}_{\nord,3}$ and $\oplus_x{(\_,\_)}_{\nord,4}$ overlap, however they are consistent and in this formulation only two rules are needed. 
Furthermore the two rules obviously imply 
\begin{equation}
\label{sub.nord.abs}
\frac{\bar{\Gamma}\sgv\bar{a}\wnord\bar{c}\quad\bar{\Gamma},x:\bar{a}\sgv\bar{b}\wnord\bar{d}}{\bar{\Gamma}\sgv\binbop{x}{\bar{a}}{\bar{b}}\wnord\binbop{x}{\bar{c}}{\bar{d}}}
\end{equation}
For readability we sometimes write $\bar{\Gamma}\sgv\bar{a}\nord\bar{b}$ as $\bar{\Gamma}\sgv\bar{b}\rnord\bar{a}$.
\end{definition}
\noindent
We note some basic properties involving (weak) norm order.
\begin{law}[Norm order and $\any{}$-size]
\label{sub.norm.order.size}
For any $\bar{\Gamma}$, $\bar{a}$, $\bar{b}$:
If $\bar{\Gamma}\sgv\bar{a}\nord\bar{b}$ then 
there is a $k$ such that $\asize{\bar{\Gamma}}{\bar{a}}{k}<\asize{\bar{\Gamma}}{\bar{b}}{k}$ and
$\asize{\bar{\Gamma}}{\bar{a}}{i}=\asize{\bar{\Gamma}}{\bar{b}}{i}$ for all $i\geq k$. 
As a consequence $S_{\bar{\Gamma}}(\bar{a})\leq S_{\bar{\Gamma}}(\bar{b})$.
\end{law}
\begin{proof}
Induction on $\bar{\Gamma}\sgv\bar{a}\nord\bar{b}$.
The index $k$ which is obviously unique is called the {\em discriminator}.
In case of \ostart\ obviously $k=n$. 
For rule \ovar\ $k$ is the discriminator obtained from the inductive hypothesis.
For rule $\binbop{x}{\_}{\_}_{\nord,1}$ if $\bar{a}=\bar{b}_1$ $k$ is  $S_{\bar{\Gamma}}(\bar{b}_2)$ 
otherwise $k$ is the maximum of the discriminator of $\bar{\Gamma}\sgv\bar{a}\nord\bar{b}_1$ and $S_{\bar{\Gamma}}(\bar{b}_2)$.
Similar for $\binbop{x}{\_}{\_}_{\nord,2}$.
For rule $\binbop{x}{\_}{\_}_{\nord,3}$ by inductive hypothesis we know that 
there is a discriminator $k$ of $\bar{\Gamma}\sgv\bar{a}_1\nord\bar{b}_1$.
If $\bar{a}_2=\bar{b}_2$ then $k$ is obviously also the discriminator for the conclusion of the rule.
Otherwise, by inductive hypothesis we know that 
there is a discriminator $j$ for $\bar{\Gamma},x:\bar{a}_1\sgv\bar{a}_2\nord\bar{b}_2$ and
the desired discriminator is obviously the maximum of $k$ and $j$.  
Similar for $\binbop{x}{\_}{\_}_{\nord,4}$ and the other rules.

\noindent
For the consequence obviously $S_{\bar{\Gamma}}(\bar{a})<S_{\bar{\Gamma}}(\bar{b})$ if the discriminator equals the $\any{}$-size of $\bar{b}$ and equality otherwise. 
\end{proof}
\noindent
\begin{law}[Norm order size law]
\label{sub.norm.order.intro}
For any $\bar{\Gamma}$, $\bar{a}$, $\bar{b}$:
If $S_{\bar{\Gamma}}(\bar{a})<S_{\bar{\Gamma}}(\bar{b})$ then $\bar{\Gamma}\sgv\bar{a}\nord\bar{b}$.
\end{law}
\begin{proof}
Induction on the definition of $S_{\bar{\Gamma}}(\bar{b})$. We write $n=S_{\bar{\Gamma}}(\bar{b})$.
The case $\bar{b}=\any{m}$ for some $m$ is obvious by rule \ostart.
The case $\bar{b}=x$ for some $x$ follows from the inductive hypothesis and rule \ovar.
If $\bar{b}=[x:\bar{b}_1]\bar{b}_1$ for some $x$, $\bar{b}_1$, $\bar{b}_2$ then $n$ is the maximum of $m=S_{\bar{\Gamma}}(\bar{b}_1)$ and $k=S_{\bar{\Gamma},x:\bar{b}_1}(\bar{b}_2)$.
If $n=m$ then $S_{\bar{\Gamma}}(\bar{a})<m$ and hence $\bar{\Gamma}\sgv\bar{a}\nord\bar{b}_1$. 
By $\binbop{x}{\_}{\_}_{\nord,1}$ $\bar{\Gamma}\sgv\bar{a}\nord\bar{b}$. Similar for $n=k$.
This argument can be extended to the other constructors in an obvious way. 
\end{proof}
\noindent
We now come to the most important property of norm order.
\begin{law}[Norm order is well founded]
\label{sub.norm.order.term}
There is no infinite sequence of successively smaller norms w.r.t~norm order 
\end{law}
\begin{proof}
The basic idea is to consider infinite norm order sequences starting from $\bar{a}$ which satisfy a parametrized $\any{}$-size constraint $C^{(n)}_{\bar{\Gamma}}(\bar{a})$ inspired by law \ref{sub.norm.order.size}. 
Given an infinite chain $\bar{\Gamma}\sgv\bar{a}=\bar{a}_1\rnord\bar{a}_2\rnord\ldots$ we define 
\[
C^{(n)}_{\bar{\Gamma}}(\bar{a})\;\equiv\;\asize{\bar{\Gamma}}{\bar{a}}{j}=\asize{\bar{\Gamma}}{\bar{a}_i}{j}\;\text{for all}\;i\geq 1\;\text{and all}\;j\;\text{where}\;n<j\leq S_{\bar{\Gamma}}(\bar{a})
\]
We show by induction on $n$ that for all $\bar{\Gamma}$, $\bar{a}$ with $\bar{\Gamma}\sgv\bar{a}$ there is no infinite chain $\bar{\Gamma}\sgv\bar{a}=\bar{a}_1\rnord\bar{a}_2\rnord\ldots$ for which $C^{(n)}_{\bar{\Gamma}}(\bar{a})$ is valid. 
As a consequence this property holds for $n=S_{\bar{\Gamma}}(\bar{a})$, and since $C^{(S_{\bar{\Gamma}}(\bar{a}))}_{\bar{\Gamma}}(\bar{a})$ is trivially true, there are no infinite chains at all.

\noindent
The case $n=0$ is obvious since otherwise $\asize{\bar{\Gamma}}{\bar{a}}{j}=\asize{\bar{\Gamma}}{\bar{a}_2}{j}$ for all $j\leq S_{\bar{\Gamma}}(\bar{a})$ which violates law \ref{sub.norm.order.size}.
For the case $n>0$ assume an infinite chain $\bar{\Gamma}\sgv\bar{a}=\bar{a}_1\rnord\bar{a}_2\rnord\ldots$ 
with $C^{(n)}_{\bar{\Gamma}}(\bar{a})$, i.e.~$\asize{\bar{\Gamma}}{\bar{a}}{j}=\asize{\bar{\Gamma}}{\bar{a}_i}{j}$ for all $i\geq 1$ and all $j$ with $n<j\leq S(\bar{a})$.
Obviously $\asize{\bar{\Gamma}}{\bar{a}_i}{n}=\asize{\bar{\Gamma}}{\bar{a}}{n}$, for all $i\geq 1$ is not possible since otherwise
we would have an infinite sequence with $\asize{\bar{\Gamma}}{\bar{a}}{j}=\asize{\bar{\Gamma}}{\bar{a}_i}{j}$ for all $i\geq 1$ and all $j$ with $n-1<j\leq S(a)$, 
or in other words satisfying $C^{(n-1)}_{\bar{\Gamma}}(\bar{a})$, and therefore violating the inductive hypothesis.
Hence there is an index $i_0$ where $\asize{\bar{\Gamma}}{\bar{a}}{n}>\asize{\bar{\Gamma}}{\bar{a}_{i_0}}{n}$.
By repeating this argument on the infinite sequence ${\bar{\Gamma}}\sgv\bar{a}_{i_0}\rnord\bar{a}_{{i_0}+1}\rnord\ldots$ we eventually reach an index $i_h$ and infinite sequence $\bar{\Gamma}\sgv\bar{a}_{i_h}\rnord\bar{a}_{i_h+1}\rnord\ldots$ where $\asize{\bar{\Gamma}}{\bar{a}_{i_h}}{n}=0$.
However this means that $\asize{\bar{\Gamma}}{\bar{a}_i}{n}=0$, for all $i\geq i_h$,
or in other words satisfying $C^{(n-1)}_{\bar{\Gamma}}(\bar{a}_{i_h})$ which violates the inductive hypothesis. \qedhere
\end{proof}
\noindent
As a consequence of Law \ref{sub.norm.order.term}, the following \emph{norm-induction} principle is valid: 
\begin{law}[Norm induction]
\label{sub.norm.induction}
For all $\Gamma_1$ and $\bar{a}_1$ with ${\free}(\bar{a}_1)\subseteq{\dom}(\Gamma_1)$, a property $P(\Gamma_1,\bar{a}_1)$ is true 
if  $\nrm{}{\Gamma_1,\Gamma_2}\gv\bar{a}_1\rnord\bar{a}_2$ and $P((\Gamma_1,\Gamma_2),\bar{a}_2)$ for all $\bar{a}_2$ and $\Gamma_2$ implies $P(\Gamma_1,\bar{a}_1)$.
\end{law}
\begin{proof}
Note that the repeated use of the inductive assumption implies norm orderings $\nrm{}{\Gamma_1,\ldots,\Gamma_n}\gv\bar{a}_1\rnord\ldots\rnord\bar{a}_n$, for $n\geq 1$.
By Law \ref{sub.norm.order.term} the induction is well-founded.
\end{proof}
\noindent
Next, we note some basic properties involving (weak) norm order.
These properties will be used for the single purpose to show that, under certain conditions, computable expressions (Definition \ref{sub.compexp}) always exist (Law \ref{sub.comp.well}). 
They can be split into two parts: The first part is about norm order itself ending with Law \ref{sub.norm.order.basic}, the second part is about
norming and its relation to norm order, ending with Law \ref{sub.norm.order.sub}.
\begin{law}[Weakening/Strengthening of (weak) norm order]
\label{sub.norm.order.weak}
For any $\bar{\Gamma}$, $x$, $\bar{a}$, $\bar{b}$, $\bar{c}$: 
If $\bar{\Gamma}\sgv\bar{b}\nord\bar{c}$ then $\bar{\Gamma},x:\bar{a}\sgv\bar{b}\nord\bar{c}$.
If $\bar{\Gamma},x:\bar{a}\sgv\bar{b}\nord\bar{c}$ and $x\notin\free(\bar{b})$ then $x\notin\free(\bar{c})$  and $\bar{\Gamma}\sgv\bar{b}\nord\bar{c}$.
Similar for weak norm order.
\end{law}
\begin{proof}
Obvious inductive argument.
\end{proof}
\noindent
\begin{law}[Left decomposition of weak norm order]
\label{sub.norm.order.body}
For any $\bar{\Gamma}$, $x$, $\bar{a}$, $\bar{b}$, $\bar{c}$: 
If $\bar{\Gamma}\sgv\binbop{x}{\bar{a}}{\bar{b}}\wnord\bar{c}$ then $\bar{\Gamma}\sgv\bar{a}\nord\bar{c}$ and $\bar{\Gamma},x:\bar{a}\sgv\bar{b}\nord\bar{c}$.
Similar for the other constructors.
\end{law}
\begin{proof}
We show the case of universal abstractions $\bar{\Gamma}\sgv[x:\bar{a}]\bar{b}$, the other case can be argued analogously.
If $[x:\bar{a}]\bar{b}=\bar{c}$, $\bar{\Gamma}\sgv\bar{c}$ the property follows directly form rules $\binbop{x}{\_}{\_}_{\nord,1}$, $\binbop{x}{\_}{\_}_{\nord,2}$, and law \ref{sub.norm.order.weak}. 
It is therefore sufficient to show that if
$\bar{\Gamma}\sgv[x:\bar{a}]\bar{b}\nord\bar{c}$ then $\bar{\Gamma}\sgv\bar{a}\nord\bar{c}$ and $\bar{\Gamma},x:\bar{a}\sgv\bar{b}\nord\bar{c}$.
The proof is by induction on $\bar{\Gamma}\sgv[x:\bar{a}]\bar{b}\nord\bar{c}$:
For rule \ostart\ it follows from an obvious property of $\any{}$-size.
Rule \ovar\ is trivial.
For rule $\binbop{x}{\_}{\_}_{\nord,1}$ we have $\bar{c}=[x:\bar{c}_1]\bar{c}_2$ and $\bar{\Gamma}\sgv[x:\bar{a}]\bar{b}\wnord\bar{c}_1$ for some $\bar{c}_1$, $\bar{c}_2$. 
If $\bar{\Gamma}\sgv[x:\bar{a}]\bar{b}\nord\bar{c}_1$ then by inductive hypothesis $\bar{\Gamma}\sgv\bar{a}\nord\bar{c}_1$ and $\bar{\Gamma},x:\bar{a}\sgv\bar{b}\nord\bar{c}_1$.
If $[x:\bar{a}]\bar{b}=\bar{c}_1$ the same follows from rules $\binbop{x}{\_}{\_}_{\nord,i}$, $i=1,2$ , and law \ref{sub.norm.order.weak}.
In both cases by rules $\binbop{x}{\_}{\_}_{\nord,i}$, $i=1,2$, $\bar{\Gamma}\sgv\bar{a}\nord\bar{c}$ and $\bar{\Gamma},x:\bar{a}\sgv\bar{b}\nord\bar{c}$.
A similar argument can be made for rule $\binbop{x}{\_}{\_}_{\nord,2}$.
For rule $\binbop{x}{\_}{\_}_{\nord,3}$ we have $\bar{c}=[x:\bar{c}_1]\bar{c}_2$, $\bar{\Gamma}\sgv\bar{a}\nord\bar{c}_1$ and $\bar{\Gamma},x:\bar{a}\sgv\bar{b}\wnord\bar{c}_2$.
The proposition follows from rules $\binbop{x}{\_}{\_}_{\nord,i}$, $i=1,2$.
Similar for $\binbop{x}{\_}{\_}_{\nord,4}$. 
This argument can be extended to the other constructors in an obvious way. 
\end{proof}
\noindent
\begin{law}[Preservation of norm order]
\label{sub.norm.order.pres}
For any $\bar{\Gamma}$, $x$, $\bar{a}$, $\bar{b}$, $\bar{c}$, $\bar{d}$: 
If $\bar{\Gamma}\sgv\bar{a}\wnord\bar{b}$ and $\bar{\Gamma},x:\bar{b}\sgv\bar{c}\nord\bar{d}$ then $\bar{\Gamma},x:\bar{a}\sgv\bar{c}\nord\bar{d}$.
As a consequence under the same conditions also 
$\bar{\Gamma}\sgv\bar{c}\gsub{x}{\bar{a}}\nord\bar{d}\gsub{x}{\bar{a}}$.
\end{law}
\begin{proof}
The proposition is obvious if $\bar{\Gamma}\sgv\bar{a}=\bar{b}$.
Otherwise we show the following more general proposition: Let $\bar{\Gamma}_a=(\bar{\Gamma}_1,x:\bar{a},\bar{\Gamma}_2)$ and $\bar{\Gamma}_b=(\bar{\Gamma}_1,x:\bar{b},\bar{\Gamma}_2)$ for some $\bar{\Gamma}_1$, $\bar{\Gamma}_2$.
If $\bar{\Gamma}_1\sgv\bar{a}\nord\bar{b}$ and $\bar{\Gamma}_a\sgv\bar{c}\nord\bar{d}$ then $\bar{\Gamma}_b\sgv\bar{c}\nord\bar{d}$.
The proof is by induction on $\bar{\Gamma}_a\sgv\bar{c}\nord\bar{d}$:
For rule \ostart\ we have $\bar{d}=\any{n}$ and obviously $S_{\bar{\Gamma}_a}(\bar{c})\leq S_{\bar{\Gamma}_b}(\bar{c})<n$.
For rule \ovar\ we have $\bar{c}=y$ for some $y$ and $\bar{\Gamma}_b\sgv\bar{\Gamma}_b(y)\wnord\bar{d}$. The case $x=y$ is obvious, otherwise the property follows from the inductive hypothesis.
For rule $\binbop{x}{\_}{\_}_{\nord,1}$ we have $d=[y:\bar{d}_1]\bar{d}_2$ for some $y$ $\bar{d}_1$, $\bar{d}_2$ where $\bar{\Gamma}_a\sgv\bar{c}\wnord\bar{d}_1$.
If $\bar{c}=\bar{d}_1$ then obviously $\bar{\Gamma}_b\sgv\bar{c}\wnord\bar{d}_1$ otherwise by inductive hypothesis $\bar{\Gamma}_b\sgv\bar{c}\nord\bar{d}_1$.
In both cases by $\binbop{x}{\_}{\_}_{\nord,1}$ $\bar{\Gamma}_b\sgv\bar{c}\nord\bar{d}$. Similar for $\binbop{x}{\_}{\_}_{\nord,i}$, $i=2,3,4$ and the other rules for constructors..

\noindent
The consequence essentially follows from an inductive argument using the rule \ovar.  
\end{proof}
\noindent
\begin{law}[Transitivity of norm order]
\label{sub.norm.order.trans}
For any $\bar{\Gamma}$, $\bar{a}$, $\bar{b}$, $\bar{c}$:
If $\bar{\Gamma}\sgv\bar{a}\nord\bar{b}$ and $\bar{\Gamma}\sgv\bar{b}\nord\bar{c}$ then $\bar{\Gamma}\sgv\bar{a}\nord\bar{c}$. 
\end{law}
\begin{proof}
Induction on $\bar{\Gamma}\sgv\bar{a}\nord\bar{b}$.
\begin{itemize}
\item\ostart: 
We have $\bar{b}=\any{n}$, $S_{\bar{\Gamma}}(\bar{a})<n$, and $\bar{\Gamma}\sgv\any{n}\wnord\bar{c}$ for some $n$ which by Law \ref{sub.norm.order.size} implies $n\leq S_{\bar{\Gamma}}(\bar{c})$.
From $S_{\bar{\Gamma}}(\bar{a})<S_{\bar{\Gamma}}(\bar{c})$ by Law \ref{sub.norm.order.intro} $\bar{\Gamma}\sgv\bar{a}\nord\bar{c}$.
\item\ovar: Follows directly from the inductive hypothesis.
\item$\binbop{x}{\_}{\_}_{\nord,1}$: We have $\bar{b}=\binbop{x}{\bar{b}_1}{\bar{b}_2}$ for some $x$, $\bar{b}_1$, $\bar{b}_2$ where $\bar{\Gamma}\sgv\bar{b}_1$ and $\bar{\Gamma}\sgv\bar{a}\nord\bar{b}_1$.
From $\bar{\Gamma}\sgv\binbop{x}{\bar{b}_1}{\bar{b}_2}\nord\bar{c}$ by Law \ref{sub.norm.order.body} $\bar{\Gamma}\sgv\bar{b}_1\nord\bar{c}$.
By inductive hypothesis $\bar{\Gamma}\sgv\bar{a}\nord\bar{c}$.
\item$\binbop{x}{\_}{\_}_{\nord,2}$: We have $\bar{b}=\binbop{x}{\bar{b}_1}{\bar{b}_2}$ for some $x$, $\bar{b}_1$, $\bar{b}_2$ where $\bar{\Gamma}\sgv\bar{b}_1$ and $\bar{\Gamma}\sgv\bar{a}\nord\bar{b}_2$.
From $\bar{\Gamma}\sgv\binbop{x}{\bar{b}_1}{\bar{b}_2}\nord\bar{c}$ by Law \ref{sub.norm.order.body} $\bar{\Gamma},x:\bar{b}_1\sgv\bar{b}_2\nord\bar{c}$.
By Law \ref{sub.norm.order.weak} and by inductive hypothesis $\bar{\Gamma},x:\bar{b}_1\sgv\bar{a}\nord\bar{b}_2$, $\bar{\Gamma},x:\bar{b}_1\sgv\bar{a}\nord\bar{c}$, and $\bar{\Gamma}\sgv\bar{a}\nord\bar{c}$.
\item$\binbop{x}{\_}{\_}_{\nord,3}$: We have  $\bar{a}=\binbop{x}{\bar{a}_1}{\bar{a}_2}$ and $\bar{b}=\binbop{x}{\bar{b}_1}{\bar{b}_2}$ for some $x$, $\bar{a}_1$, $\bar{a}_2$, $\bar{b}_1$, $\bar{b}_2$ 
where $\bar{\Gamma}\sgv\bar{a}_1\nord\bar{b}_1$, $\bar{\Gamma},x:\bar{a}_1\sgv\bar{a}_2\wnord\bar{b}_2$, and $\bar{\Gamma}\sgv\binbop{x}{\bar{b}_1}{\bar{b}_2}\nord\bar{c}$. 

Regarding the latter statement by definition of norm order (Table \ref{sub.nord.rules}) there are five cases.
\begin{itemize}
\item\ostart: 
This case is similar to the case \ostart\ above.
%
\item\oabsl: $\bar{c}=\binbop{x}{\bar{c}_1}{\bar{c}_2}$ for some $\bar{c}_1$, $\bar{c}_2$ where $\bar{\Gamma}\sgv\bar{b}\wnord\bar{c}_1$.
The inductive hypothesis obviously implies $\bar{\Gamma}\sgv\bar{a}\nord\bar{c}_1$ hence by rule \oabsl\ $\bar{\Gamma}\sgv\bar{a}\nord\bar{c}$.
\item\oabs: $\bar{c}=\binbop{x}{\bar{c}_1}{\bar{c}_2}$ for some $\bar{c}_1$, $\bar{c}_2$ where $\bar{\Gamma}\sgv\bar{b}_1\nord\bar{c}_1$, $\bar{\Gamma},x:\bar{b}_1\sgv\bar{b}_2\wnord\bar{c}_2$.
By a case distinction and law \ref{sub.norm.order.pres} obviously $\bar{\Gamma},x:\bar{a}_1\sgv\bar{b}_2\wnord\bar{c}_2$ hence the inductive hypothesis obviously implies $\bar{\Gamma},x:\bar{a}_1\sgv\bar{a}_2\wnord\bar{c}_2$.  
Furthermore by inductive hypothesis $\bar{\Gamma}\sgv\bar{a}_1\nord\bar{c}_1$ and hence by rule \oabs\ $\bar{\Gamma}\sgv\bar{a}\nord\bar{c}$.
\end{itemize}
The cases \oabsr\ and \oabsp\ are argued similarly to cases \oabsl\ and \oabs.
\item$\binbop{x}{\_}{\_}_{\nord,4}$: Similar to $\binbop{x}{\_}{\_}_{\nord,3}$.
\item Similar for the other rules. \qedhere
\end{itemize}
\end{proof}
\noindent
\begin{law}[Decomposition of weak norm order]
\label{sub.norm.order.decomp}
For any $\bar{\Gamma}$, $\bar{a}$, $\bar{b}$, $n$:
If $\bar{\Gamma}\sgv\binbop{x}{\bar{a}}{\bar{b}}\wnord\binbop{x}{\bar{c}}{\bar{d}}$ then 
$\bar{\Gamma},x:\bar{a}\sgv\bar{b}\wnord\bar{d}$.
Similar for the other constructors.
\end{law}
\begin{proof}
It is obviously sufficient to show that $\bar{\Gamma}\sgv\binbop{x}{\bar{a}}{\bar{b}}\nord\binbop{x}{\bar{c}}{\bar{d}}$ implies
$\bar{\Gamma},x:\bar{a}\sgv\bar{b}\nord\bar{d}$.
The proof is by by induction on $\bar{\Gamma}\sgv\binbop{x}{\bar{a}}{\bar{b}}\nord\binbop{x}{\bar{c}}{\bar{d}}$:
\ostart\ and \ovar\ are trivial.
For case $\binbop{x}{\_}{\_}_{\nord,1}$ we have $\bar{\Gamma}\sgv\binbop{x}{\bar{a}}{\bar{b}}\wnord\bar{d}$. By Law \ref{sub.norm.order.body} this implies $\bar{\Gamma},x:\bar{a}\sgv\bar{b}\nord\bar{d}$.
Similar for $\binbop{x}{\_}{\_}_{\nord,2}$.
$\binbop{x}{\_}{\_}_{\nord,i}$, $i=3,4$ directly establish the proposition. 
This argument can be extended to the other constructors in an obvious way. 
\end{proof}
\noindent
\begin{law}[Generalized inclusion includes weak norm order]
\label{sub.norm.order.basic}
For any $\bar{\Gamma}$, $\bar{a}$, $\bar{b}$:
$\bar{\Gamma}\sgv\bar{a}\ginc\bar{b}$ implies $\bar{\Gamma}\sgv\bar{a}\wnord\bar{b}$.
\end{law}
\begin{proof}
For any $\bar{\Gamma}$, $\bar{a}$, $\bar{b}$, $n$ we show three properties by simultaneous induction on typing, inclusion, and $\any{}$-inclusion: 
$\bar{\Gamma}\sgv\bar{a}:\bar{b}$ implies $\bar{\Gamma}\sgv\bar{a}\wnord\bar{b}$ ($i$), 
$\bar{\Gamma}\sgv\bar{a}\leq\bar{b}$ implies $\bar{\Gamma}\sgv\bar{a}\wnord\bar{b}$ ($ii$), and
$\bar{\Gamma}\sgv\bar{a}<\any{n}$ implies $\bar{\Gamma}\sgv\bar{a}\nord\any{n}$ ($iii$).
Note that due to law \ref{sub.nrm.fund} we can assume all auxiliary expressions in rule antecedents to be norms.

Part $i$: The proposition is trivial for \appl, obvious for \ax\ and \conv, and for \mystart\ it follows follows from \ovar.
For \weak\ the proposition follows from the inductive hypothesis and from Law \ref{sub.norm.order.weak},
for \sinc\ it follows from the inductive hypotheses of Parts $i$ and $ii$ and from Law \ref{sub.norm.order.trans}, and
for \absu\ it follows from the inductive hypothesis of Part $i$, Law \ref{sub.incl.valid}, and from $\binbop{x}{\_}{\_}_{\nord,4}$.
Similar for the other rules.

Part $ii$: For \srefl the proposition is obvious, for \sembed\ it follows from the inductive hypothesis of Part $iii$, and for
\sabs\ it follows from the inductive hypothesis, Law \ref{sub.incl.valid}, and $\binbop{x}{\_}{\_}_{\nord,3}$.
Similar for the other rules.

Part $iii$: For \sstart\ the proposition follows from \ostart\ and the definition of $\any{}$-size.
For rule \sstyp\ we have $\bar{\Gamma}\sgv\bar{a}:\bar{c}$ and $\bar{\Gamma}\sgv\bar{c}<\any{n}$ for some $\bar{c}$.
By inductive hypotheses of Parts $i$ and $iii$ we have $\bar{\Gamma}\sgv\bar{a}\wnord\bar{c}$ and $\bar{\Gamma}\sgv\bar{c}\nord\any{n}$
and hence by law \ref{sub.norm.order.trans} $\bar{\Gamma}\sgv\bar{a}\nord\any{n}$.
For rule \ssabsu\ we have $\bar{a}=[x:\bar{a}_1]\bar{a}_2$ for some $\bar{a}_1$, $\bar{a}_2$ where $\bar{\Gamma}\sgv\bar{a}_1<\any{n}$ and $\bar{\Gamma},x:\bar{a}_1\sgv\bar{a}_2<\any{n}$.
By inductive hypothesis of Part $iii$ $\bar{\Gamma}\sgv\bar{a}_1\nord\any{n}$ and $\bar{\Gamma},x:\bar{a}_1\sgv\bar{a}_2\nord\any{n}$. 
By a very simple inspection of the norm order rules (Table \ref{sub.nord.rules}) this implies $S_{\bar{\Gamma}}(\bar{a}_1)<n$ and $S_{\bar{\Gamma},x:\bar{a}_1}(\bar{a}_2)<n$ and
therefore obviously $S_{\bar{\Gamma}}(\bar{a})<n$ and by law \ref{sub.norm.order.intro} $\bar{\Gamma}\sgv\bar{a}\nord\any{n}$.
Similar for the other rules.
\end{proof}
\noindent
We now turn our attention to (regular) expressions and will eventually show a substitution property (Law \ref{sub.norm.order.sub}) where we relate the norming of a substitution and the substitution on a norming \wrt\ weak norm order. Before that we need to show two auxiliary properties about norming.
\begin{law}[Removal of elimination list]
\label{sub.norm.theta}
For all $\Lambda$, $\Theta$,  $a$:
if $\Lambda,\Theta\sngv a$ and $\Lambda\sngv a$ then $\nrm{\Lambda,\Theta}{a}=\binbop{x}{\bar{b}_1}{\bar{b}_2}$ for some $\bar{b}_1$, $\bar{b}_2$ implies $\nrm{\Lambda}{a}=\binbop{x}{\bar{b}_1}{\bar{b}_2'}$ for some $\bar{b}_2'$.
\end{law}
\begin{proof}
By induction on $\nrm{\Lambda,\Theta}{a}$ we show a more general property.
For all $\Lambda$, $\Theta_1$, $\Theta_2$, and $a$ let $\Theta=(\Theta_1, \Theta_2)$:
If $\Lambda,\Theta\sngv a$ and $\Lambda,\Theta_1\sngv a$ then 
$\nrm{\Lambda,\Theta}{a}[\Theta_1]=\binbop{x}{\bar{b}_1}{\bar{b}_2}$ for some $\bar{b}_1$, $\bar{b}_2$ implies $\nrm{\Lambda,\Theta_1}{a}[\Theta_1]=\binbop{x}{\bar{b}_1}{\bar{b}_2'}$ for some $\bar{b}_2'$.
\begin{itemize}
\item$a=\any{n}$ for some $n$: Trivial since
$\nrm{\Lambda,\Theta}{\any{n}}=\any{n}=\nrm{\Lambda,\Theta_1}{\any{n}}$.
\item$a=x$ for some $x$: 
The property follows from the inductive hypothesis.
\item$a=[x:a_1]a_2$ for some $x$, $a_1$, $a_2$: There are three cases:
\begin{itemize}
\item$\Theta_1=\Theta_2=()$: Trivial.
\item$\Theta_1=()$, $\Theta_2=(c,\Theta_2')$ for some $c$, $\Theta_2'$.
We have $\nrm{\Lambda,\Theta}{a}=[x:\nrm{\Lambda}{a_1}]\nrm{(\Lambda,x\smydef c),\Theta_2'}{a_2}$
and $\nrm{\Lambda}{a}=[x:\nrm{\Lambda}{a_1}]\nrm{\Lambda}{a_2}$ which directly implies the property.
\item$\Theta_1=(c,\Theta_1')$ for some $c$, $\Theta_1'$.
We have $\nrm{\Lambda,\Theta}{a}=[x:\nrm{\Lambda}{a_1}]\nrm{(\Lambda,x\smydef c),(\Theta_1',\Theta_2)}{a_2}$ 
and $\nrm{\Lambda,\Theta_1}{a}=[x:\nrm{\Lambda}{a_1}]\nrm{(\Lambda,x\smydef c),\Theta_1'}{a_2}$.
Assume $\nrm{\Lambda,\Theta}{a}[\Theta_1]=\nrm{(\Lambda,x\smydef c),(\Theta_1',\Theta_2)}{a_2}[\Theta_1']=[x:\bar{b}_1]\bar{b}_2$ for some $\bar{b}_1$, $\bar{b}_2$.
By inductive hypothesis $\nrm{\Lambda,\Theta_1}{a}[\Theta_1]=\nrm{(\Lambda,x\smydef c),\Theta_1'}{a_2}[\Theta_1']=[x:\bar{b}_1]\bar{b}_2'$ for some $\bar{b}_2'$.
\end{itemize}
\item$a=(a_1\:a_2)$ for some $a_1$, $a_2$: 
Since $\Lambda,\Theta\sngv a$ we know that $\nrm{\Lambda,(a_2,\Theta)}{a_1}=[x:\bar{b}]\nrm{\Lambda,\Theta}{a}$ for some $x$, $\bar{b}$.
Obviously $\nrm{\Lambda,\Theta}{a}[\Theta_1]=\nrm{\Lambda,(a_2,\Theta)}{a_1}[(a_2,\Theta_1)]$.
Assume $\nrm{\Lambda,\Theta}{a}[\Theta_1]=\nrm{\Lambda,(a_2,\Theta)}{a_1}[(a_2,\Theta_1)]=[x:\bar{b}_1]\bar{b}_2$ for some $\bar{b}_1$, $\bar{b}_2$.
By inductive hypothesis $\nrm{\Lambda,\Theta_1}{a}[\Theta_1]=\nrm{\Lambda,(a_2,\Theta_1)}{a_1}[(a_2,\Theta_1)]=[x:\bar{b}_1]\bar{b}_2'$ for some $\bar{b}_2'$.
\item The other cases can be argued in a similar style.
\qedhere
\end{itemize}
\end{proof}
\noindent
Based on this result we show a property somewhat similar to law \ref{sub.elim.prop}. 
In this property we use the notation $(a\,\Theta)$ recursively defined by $(a\,())=a$, $(a\,(b,\Theta))=((a\,b)\,\Theta)$, $(a\,(i,\Theta))=(a.i\,\Theta)$, and $(a\,(2_b,\Theta))=(a.2\,\Theta)$.
Note that $\Lambda\sngv(a\,\Theta)$ obviously implies $\Lambda,\Theta\sngv a$ but not vice versa, \eg\ $\any{0}\sngv\any{0}$ but not $\sngv(\any{0}\,\any{0})$.
\begin{law}[Elimination and generalized inclusion (2)]
\label{sub.norm.order.ext}
For all $\Lambda$, $\Theta$, $a$, $b$: 
If $\nrm{}{\Lambda}\gv\nrm{\Lambda}{b}\ginc\nrm{\Lambda}{a}$, $\Lambda\sngv(b\,\Theta)$, and
$\Lambda\sngv(a\,\Theta)$ then $\nrm{}{\Lambda}\gv\nrm{\Lambda,\Theta}{b}\ginc\nrm{\Lambda,\Theta}{a}$.
\end{law}
\begin{proof}
Induction on $\Theta$. $\Theta=()$ is obvious.
Let $\Theta=(c,\Theta')$ for some $c$, hence obviously $\Lambda\sngv((b\,c)\,\Theta')$ and $\Lambda\sngv((a\,c)\,\Theta')$
and by inductive hypothesis $\nrm{}{\Lambda}\gv\nrm{\Lambda,\Theta'}{(b\,c)}\ginc\nrm{\Lambda,\Theta'}{(a\,c)}$.
Since $\Lambda,\Theta'\sngv(b\,c)$ we know that $\nrm{\Lambda,\Theta}{b}=\binbop{x}{\bar{b}_1}{\nrm{\Lambda,\Theta'}{(b\,c)}}$ for some $x$, $\bar{b}_1$.
Similarly since $\Lambda,\Theta'\sngv(a\,c)$ we know that $\nrm{\Lambda,\Theta}{a}=\binbop{x}{\bar{a}_1}{\nrm{\Lambda,\Theta'}{(a\,c)}}$ for some $\bar{a}_1$.
By law \ref{sub.norm.theta} we have $\nrm{\Lambda}{b}=\binbop{x}{\bar{b}_1}{\bar{b}_2}$ for some $\bar{b}_2$ and similar for $\nrm{\Lambda}{a}$.
Hence since $\nrm{}{\Lambda}\gv\nrm{\Lambda}{b}\ginc\nrm{\Lambda}{a}$ by law \ref{sub.flex.prop}($iii$) $\bar{a}_1=\bar{b}_1$.
Hence obviously $\nrm{}{\Lambda}\gv\nrm{\Lambda,\Theta}{b}=\binbop{x}{\bar{b}_1}{\nrm{\Lambda,\Theta'}{(b\,c)}}\ginc\binbop{x}{\bar{a}_1}{\nrm{\Lambda,\Theta'}{(a\,c)}}=\nrm{\Lambda,\Theta}{a}$.
\end{proof}
\noindent
%
%
%
%
\noindent
We now come to the main substitution property involving norming and weak norm order. 
\begin{law}[Substitution property of norm order]
\label{sub.norm.order.sub}
For all $\Gamma$, $x$, $a$, $b$, $c$ where $\Gamma\sngv a$, $\Gamma\sngv b$, $\Gamma,x:a\sngv c$, and $\Gamma\sngv c\gsub{x}{b}$: 
If $\nrm{}{\Gamma}\gv\nrm{\Gamma}{b}\!\ginc\nrm{\Gamma}{a}$
then $\nrm{}{\Gamma}\gv\nrm{\Gamma}{c\gsub{x}{b}}\wnord\nrm{\Gamma,x:a}{c}\gsub{x}{\nrm{\Gamma}{b}}$.
\end{law}
\begin{proof}
Note that under the given assumption by Law \ref{sub.nrm.sub} with $\Lambda_1=\Gamma$, $\Lambda_2=()$, $\Theta=()$ we obtain $\nrm{\Gamma}{c\gsub{x}{b}}=\nrm{\Gamma,x\smydef b}{c}$.
For any $a$, $b$ we define $\Gamma_a=(\Gamma_1,x:a,\Gamma_2)$, $\Gamma_b=(\Gamma_1,\Gamma_2\gsub{x}{b})$ for some $\Gamma_1$, $\Gamma_2$, $x$.
For any $c$, $\Theta$ we show the more general property that $\Gamma_1\sngv a$, $\Gamma_1\sngv b$, $\nrm{}{\Gamma_1}\gv\nrm{\Gamma_1}{b}\ginc\nrm{\Gamma_1}{a}$, $\Gamma_a\sngv (c\,\Theta)$, $\Gamma_b\sngv (c\,\Theta)$ 
imply $\nrm{}{\Gamma_b}\gv\nrm{\Gamma_b,\Theta}{c}\wnord\nrm{\Gamma_a,\Theta}{c}\gsub{x}{\nrm{\Gamma_1}{b}}$.
The proof is by induction on $\nrm{\Gamma_b,\Theta}{c}$.
\begin{itemize}
\item $c=\any{n}$ for some $n$. Obvious
\item$c=x$ and therefore $\nrm{\Lambda_b,\Theta}{x}=\nrm{\Lambda_b,\Theta}{b}$.
If $\Theta=()$ then $\nrm{\Lambda_a}{x}\gsub{x}{\nrm{\Lambda_1}{b}}=\nrm{\Lambda_1}{b}=\nrm{\Lambda_b,\Theta}{b}$.
If $\Theta\neq()$ then $\nrm{\Lambda_a,\Theta}{x}\gsub{x}{\nrm{\Lambda_1}{b}}=\nrm{\Lambda_a,\Theta}{a}\gsub{x}{\nrm{\Lambda_1}{b}}=\nrm{\Lambda_a,\Theta}{a}$.
and by law \ref{sub.norm.order.ext} $\nrm{}{\Lambda_b}\gv\nrm{\Lambda_b,\Theta}{b}\ginc\nrm{\Lambda_a,\Theta}{a}$.
Therefore by law \ref{sub.norm.order.basic} $\nrm{}{\Lambda_b}\gv\nrm{\Lambda_b,\Theta}{b}\wnord\nrm{\Lambda_a,\Theta}{a}$.
\item $c=y$ for some $y\neq x$.
The two norms are either equal or the proposition follows from the inductive hypothesis. 
\item $c=[y:c_1]c_2$ for some $y\neq x$, $c_1$, $c_2$.
If $\Theta=()$ by definition of norming and substitution $\nrm{\Lambda_b}{c}=[y:\bar{c}_{1b}]\bar{c}_{2b}$
and $\nrm{\Lambda_a}{c}\gsub{x}{\bar{c}_{1b}}=[y:\bar{c}_{1a}\gsub{x}{\bar{c}_{1b}}](\bar{c}_{2b}\gsub{x}{\bar{c}_{1b}})$ where $\bar{c}_{1a}=\nrm{\Lambda_a}{c_1}$, $\bar{c}_{1b}=\nrm{\Lambda_b}{c_1}$, and $\bar{c}_{2b}=\nrm{\Lambda_b,y:c_1}{c_2}$.
By inductive hypothesis  $\nrm{}{\Lambda_b}\gv\bar{c}_{1b}\wnord\bar{c}_{1a}\gsub{x}{\nrm{\Lambda_1}{b}}$
and $\nrm{}{\Lambda_b},y:\bar{c}_{1b}\gv\bar{c}_{2b}\wnord\bar{c}_{2a}\gsub{x}{\nrm{\Lambda_1}{b}}$ where  $\bar{c}_{2a}=\nrm{\Lambda_a,y:c_1}{c_2}$
Hence by \ref{sub.nord.abs}
$\nrm{}{\Lambda_b}\gv\nrm{\Lambda_b}{c}=[y:\nrm{\Lambda_b}{c_1}]\bar{c}_{2b}
\wnord[y:\bar{c}_{1a}\gsub{x}{\nrm{\Lambda_1}{b}}]\bar{c}_{2a}\gsub{x}{\nrm{\Lambda_1}{b}}
=\nrm{\Lambda_a}{c}\gsub{x}{\nrm{\Lambda_1}{b}}$.

If $\Theta=(d,\Theta')$ for some $d$, $\Theta'$ one can argue similarly with $\bar{c}_{1a}$, $\bar{c}_{1b}$ as before and   
$\bar{c}_{2a}=\nrm{(\Lambda_a,y\smydef d),\Theta'}{c_2}$, $\bar{c}_{2b}=\nrm{(\Lambda_b,y\smydef d),\Theta'}{c_2}$.
Note that the inductive hypothesis can be applied since from $(c\,\Theta)=((c\,d)\,\Theta')$ by law \ref{sub.nrm.rd} $\Lambda_b\sngv(c_2\gsub{y}{d}\,\Theta')$
and therefore by law \ref{sub.nrm.sub} $\Lambda_b,y\smydef d\sngv(c_2\,\Theta')$, and a similar argument can be made for $\Lambda_a\sngv(c\,\Theta)$.

\item  $c=(c_1\,c_2)$ for some $c_1$, $c_2$.
By definition of norming $\nrm{\Lambda_b,\Theta}{c}=\bar{e}_b$ where $\nrm{\Lambda_b,(c_2,\Theta)}{c_1}=[y:\bar{d}_b]\bar{e}_b$ for some $y\neq x$, $\bar{d}_b$, $\bar{e}_b$.
Similarly $\nrm{\Lambda_a,\Theta}{c}\gsub{x}{\nrm{\Lambda_1}{b}}=\bar{e}_a\gsub{x}{\nrm{\Lambda_1}{b}}$ where $\nrm{\Lambda_a,(c_2,\Theta)}{c_1}=[y:\bar{d}_a]\bar{e}_a$ for some $\bar{d}_a$, $\bar{e}_a$.

By law \ref{sub.norm.theta} $\nrm{\Lambda_b}{c_1}=[y:\bar{d}_b]\bar{e}$
and $\nrm{\Lambda_a}{c_1}=[y:\bar{d}_a]\bar{e}'$ for some $\bar{e}$, $\bar{e}'$ hence  $\bar{d}_b=\bar{d}_a$.
Furthermore since $x\notin{\free}(\bar{d}_b)$ we have  $\bar{d}_b=\bar{d}_b\gsub{x}{\nrm{\Lambda_1}{b}}$.
Since $(c\,\Theta)=(c_1\,(c_2\,\Theta))$ and similar for $\Lambda_a$, by inductive hypothesis and definition of substitution $\nrm{}{\Lambda_b}\gv[y:\bar{d}_b]\bar{e}_b\wnord([y:\bar{d}_b]\bar{e}_a)\gsub{x}{\nrm{\Lambda_1}{b}}=[y:\bar{d}_b](\bar{e}_a\gsub{x}{\nrm{\Lambda_1}{b}})$
and therefore by laws \ref{sub.norm.order.decomp} and \ref{sub.norm.order.weak}  $\nrm{}{\Lambda_b}\gv\bar{e}_b\wnord\bar{e}_a\gsub{x}{\nrm{\Lambda_1}{b}}$. 
\item
Similar arguments can be made for the other operators.
\qedhere  
\end{itemize} 
\end{proof}
\subsubsection{Computable expressions}
\label{sub.compexp}
\noindent
The definition of computable expressions (see~\ref{ce}) is adapted to the adapted notions of norms and to norm order as follows:
$a\in\ce_{\Gamma}(\bar{a})$ iff 
$a\in\sn{}$, $\Gamma\rsgv a$, $\nrm{}{\Gamma}\gv\nrm{\Gamma}{a}\leq\bar{a}$, and if the following \emph{computability conditions} are satisfied: 

\noindent
\begin{itemize}[align=left]
\item[$(\alpha)$:]
For all $x$, $b$, $c$: If $a\rd\binbop{x}{b}{c}$ or $\myneg a\rd\binbop{x}{b}{c}$ then $c\in\ce_{\Gamma,x:b}(\nrm{\Gamma}{c})$ and $c\gsub{x}{d}\in\ce_{\Gamma}(\nrm{\Gamma}{c\gsub{x}{d}})$ 
for any $d\in\ce_{\Gamma}(\nrm{\Gamma}{b})$ with $\Gamma\sgv d\rtyp b$. 
\item[$(\beta)$:]
For all $b$, $c$: If $a\rd\prsumop{b}{c}$ or $\myneg a\rd\prsumop{b}{c}$ then $b\in\ce_{\Gamma}(\nrm{\Gamma}{b})$ and $c\in\ce_{\Gamma}(\nrm{\Gamma}{c})$.
\item[$(\gamma_1)$:]
For all $b$, $c$: 
$a\rd\injl{b}{c}$ implies $b\in\ce_{\Gamma}(\nrm{\Gamma}{b})$, and $a\rd\injr{b}{c}$ implies $c\in\ce_{\Gamma}(\nrm{\Gamma}{b})$.
\item[$(\gamma_2)$:]
For all $b$, $c$, $d$, $e$: If $a\rd\prdef{x}{b^c}{d}{e}$ or $\myneg a\rd\prdef{x}{b^c}{d}{e}$ then $b\in\ce_{\Gamma}(\nrm{\Gamma}{b})$ and $d\in\ce_{\Gamma}(\nrm{\Gamma}{d})$.
\item[$(\delta)$:]
For all $b_1$, $b_2$:
$a\rd\case{b_1}{b_2}$ implies both $b_1\in\ce_{\Gamma}(\nrm{\Gamma}{b_1})$ and $b_2\in\ce_{\Gamma}(\nrm{\Gamma}{b_2})$.
\end{itemize}
The condition $\alpha$ has been modified due to the fact that norms can be instantiated and that they are not compositional (see Definition \ref{sub.genNormable} and example $E.4$).
Hence we now require validity rather than normability of computable expressions. 
In condition $\alpha$  note that from $\Gamma\rsgv a$ by Law \ref{sub.val.nrm} we obtain both $\Gamma\sngv a$ and $\sngv\Gamma$.
Similarly Law \ref{sub.rd.type} implies $\Gamma\rsgv\binbop{x}{b}{c}$ and then from  $\Gamma\sgv d\rtyp b$ by Laws \ref{type.xtrct} \ref{sub.type.sub} \ref{sub.val.nrm} $\Gamma\sngv b$ and $\Gamma\sngv c\gsub{x}{d}$.
In order not to clutter the following argument about properties of computability we will not explicitly mention the use of these laws.  
The adaptations of the conditions $\gamma$ and $\delta$ are technical, in case of $\gamma$ we need to split the condition since injections and protected definitions have different norms.
The condition $\beta$ has been taken over without changes.

We show that the recursion in the above definition eventually terminates which is a core argument of this proof of strong normalisation.
\begin{law}[The definition of computable expressions terminates]%
\label{sub.comp.well}
For all $\Gamma$ and $\bar{a}$, if $\:$ $\Gamma\sgv\bar{a}$ then the definition of $\ce_{\Gamma}(\bar{a})$ terminates.
\end{law}
\begin{proof}
Proof by norm-induction on $\bar{a}$ (\ref{sub.norm.induction}). 
Consider some $a\in\sn{}$ where $\Gamma\rsgv a$ and $\nrm{}{\Gamma}\gv\nrm{\Gamma}{a}\leq\bar{a}$.
We have to show that the definition terminates for all the computability conditions for $a$.
Let $\bar{\Gamma}=\nrm{}{\Gamma}$ and $\bar{a}'=\nrm{\Gamma}{a}$.
By Law \ref{sub.norm.order.basic} $\bar{\Gamma}\sgv\bar{a}'\lesssim\bar{a}$.
\begin{itemize}
\item[$(\alpha)$:] 
Assume  $a\rd\binbop{x}{b}{c}$ or $\myneg a\rd\binbop{x}{b}{c}$ for some $x$, $b$, $c$. 
By Law~\ref{sub.nrm.rd} $\nrm{\Gamma}{a}=\nrm{\Gamma}{[x:b]c}=[x:\nrm{\Gamma}{b}]\nrm{\Gamma,x:b}{c}$.
Let $\bar{b}=\nrm{\Gamma}{b}$ and $\bar{c}=\nrm{\Gamma,x:b}{c}$.
Obviously $\bar{\Gamma},x:\bar{b}\sgv\bar{c}\nord\bar{a}'$ and therefore by by Laws \ref{sub.norm.order.trans}
and \ref{sub.norm.order.weak} $\bar{\Gamma},x:\bar{b}\sgv\bar{c}\nord\bar{a}$.
By inductive hypothesis (with $\Gamma_1=\Gamma$, $\Gamma_2=x:b$) we know that the definition of $\ce_{\Gamma,x:b}(\bar{c})$ terminates. 

Since obviously $\bar{\Gamma},x:\bar{b}\sgv\bar{c}\wnord\bar{c}$ by \oabsr\ $\bar{\Gamma},x:\bar{b}\sgv\bar{c}\nord\bar{a}'$ and therefore by law \ref{sub.norm.order.trans}
obviously $\bar{\Gamma},x:\bar{b}\sgv\bar{c}\nord\bar{a}$.
Assume some $d\in \ce_{\Gamma}(\bar{b})$ where $\Gamma\sgv d\rtyp b$: 
Since $\bar{\Gamma}\sgv\nrm{\Gamma}{d}\leq\bar{b}$ by Law \ref{sub.norm.order.basic} $\bar{\Gamma}\sgv\nrm{\Gamma}{d}\lesssim\bar{b}$.
Hence by law \ref{sub.norm.order.pres} $\bar{\Gamma}\sgv\bar{c}\gsub{x}{\bar{d}}\nord\bar{a}\gsub{x}{\bar{d}}=\bar{a}$
and by law \ref{sub.norm.order.sub}
$\bar{\Gamma}\sgv\nrm{\Gamma}{c\gsub{x}{d}}\wnord\bar{c}\gsub{x}{\bar{d}}$.

Hence by Laws \ref{sub.norm.order.trans}
and \ref{sub.norm.order.weak} $\bar{\Gamma},x:\bar{b}\sgv\nrm{\Gamma}{c\gsub{x}{d}}\nord\bar{a}$.
By inductive hypothesis (with $\Gamma_1=\Gamma$, $\Gamma_2=()$) the definition of 
$\ce_{\Gamma}(\nrm{\Gamma}{c\gsub{x}{d}})$ terminates.
\item[$(\beta)$:]
Assume $a\rd\prsumop{b}{c}$ or $\myneg a\rd\prsumop{b}{c}$ for some $b$, $c$.
By Law~\ref{sub.nrm.rd} $\nrm{\Gamma}{a}=\nrm{\Gamma}{\prsumop{b}{c}}=\prsumop{\nrm{\Gamma}{b}}{\nrm{\Gamma}{c}}$.
Let $\bar{b}=\nrm{\Gamma}{b}$ and $\bar{c}=\nrm{\Gamma}{c}$.
Obviously $\bar{\Gamma}\sgv\bar{b}\nord\bar{a}'$ and $\bar{\Gamma}\sgv\bar{c}\nord\bar{a}'$ and therefore by Law \ref{sub.norm.order.trans} $\bar{\Gamma}\sgv\bar{b}\nord\bar{a}$
and $\bar{\Gamma}\sgv\bar{c}\nord\bar{a}$.
By inductive hypothesis (with $\Gamma_1=\Gamma$, $\Gamma_2=()$) we know that the definitions of $\ce_{\Gamma}(\bar{b})$ and $\ce_{\Gamma}(\bar{c})$ terminate. 
\item[$\ldots$:]
The other cases can be argued in a similar style.
\end{itemize}
Since it terminates for all the computability conditions, the definition of $\ce_{\Gamma}(\bar{a})$ terminates.
\end{proof}
\noindent
%
\noindent
The proofs of Laws~\ref{ce.basic}(Basic properties of computable expressions) and~\ref{ce.mon.neg}(Computable expressions are closed against negation) can be essentially kept as they are. Some parts need to be restructured and adapted to account for the adapted definition of validity and of computable expressions and to account for the adapted principle of norm induction. However the basic logical arguments and the structure of the deductive steps remain unaltered.
Due to the adapted definition of computable expression the Law~\ref{ce.mon.appl} has to be adapted:
\begin{law}[Closure of computable expressions against application]
\label{sub.ce.mon.appl}
For all $\Gamma$, $a$, $b$:
$\Gamma\rsgv(a\,b)$, $a\in\ce_{\Gamma}(\nrm{\Gamma}{a})$, and $b\in\ce_{\Gamma}(\nrm{\Gamma}{b})$ imply 
$(a\,b)\in\ce_{\Gamma}(\nrm{\Gamma}{(a\,b)})$.
\end{law}
\begin{proof}
For all $\Gamma$, $a$, $b$ assume $\Gamma\rsgv(a\,b)$, $a\in\ce_{\Gamma}(\nrm{\Gamma}{a})$, and $b\in\ce_{\Gamma}(\nrm{\Gamma}{b})$.
By Law \ref{sub.val.nrm} $\Gamma\sngv(a\,b)$ which implies $\nrm{\Gamma}{a}=\binbop{x}{\bar{a}_1}{\bar{a}_2}$ for some $\bar{a}_1$, $\bar{a}_2$ where $\nrm{}{\Gamma}\gv\nrm{\Gamma}{b}\leq\bar{a}_1$.

Let $a\rd\binbop{x}{a_1}{a_2}$ for some $x$, $a_1$, $a_2$. 
From $\Gamma\rsgv(a\,b)$ by Law \ref{sub.rd.type} $\Gamma\rsgv([x:a_1]a_2\,b)$.
Hence by Law \ref{sub.type.decomp}($v$,$iv$) $\Gamma\sgv b\rtyp a_1$ and $\Gamma\sgv\binbop{x}{a_1}{a_2}\rtyp [x:a_1]b_2$ for some $b_2$.
Since by Laws \ref{valid.type} \ref{type.xtrct} $\Gamma\rsgv a_1$ and $\Gamma,x:a_1\rsgv a_2$, by Laws \ref{sub.val.nrm} \ref{sub.nrm.rd} $\nrm{\Gamma}{a_1}=\bar{a}_1$ and $\nrm{\Gamma,x:a_1}{a_2}=\bar{a}_2$.
Since $b\in\ce_{\Gamma}(\nrm{\Gamma}{b})$, by definition of computable expressions  $b\in\ce_{\Gamma}(\bar{a}_1)$.
Since $a\in\ce_{\Gamma}(\nrm{\Gamma}{a})$ and $\Gamma\sgv b\rtyp a_1$ we have $a_2\gsub{x}{b}\in\ce_{\Gamma}(\nrm{\Gamma}{a_2\gsub{x}{b}})$ which implies $a_2\gsub{x}{b}\in\sn{}$. 
Therefore by Law~\ref{sn.cond} $(a\,b)\in\sn{}$.

It remains to show the computability conditions for $(a\,b)$. 
\begin{itemize}
\item[$(\alpha)$:]
Let $(a\,b)\rd\binbop{y}{a_3}{a_4}$ or $\myneg(a\,b)\rd\binbop{y}{a_3}{a_4}$  for some $y$, $a_3$ and $a_4$.
By basic properties of reduction $a\rd\binbop'{z}{a_1}{a_2}$, $b\rd b'$, and $a_2\gsub{z}{b'}\rd\binbop{y}{a_3}{a_4}$ for some $z$, $a_1$, $a_2$, and $b'$.
Hence as argued above $\nrm{\Gamma}{a_1}=\bar{a}_1$ and $\nrm{\Gamma,x:a_1}{a_2}=\bar{a}_2$. 
By Law~\ref{sub.nrm.rd} 
$\nrm{\Gamma}{(a\,b)}=\nrm{\Gamma}{[y:a_3]a_4}$. 
Furthermore, since $b\rd b'$ by law \ref{ce.basic} and since $\nrm{}{\Gamma}\gv\nrm{\Gamma}{b}\!\ginc\bar{a}_1$ we know that $b'\in\ce_{\Gamma}(\bar{a}_1)$.
As argued we have $\Gamma\sgv b\rtyp a_1$ and hence by Law \ref{sub.rd.type} $\Gamma\sgv b'\rtyp a_1$. We now argue as follows:
\begin{eqnarray*}
&&a\in\ce_{\Gamma}(\nrm{\Gamma}{\binbop'{z}{a_1}{a_2}})\\
&\Rightarrow&\text{($a$ is computable, $b'\in\ce_{\Gamma}(\bar{a}_1)$, and $\Gamma\sgv b'\rtyp a_1$)}\\
&&a_2\gsub{z}{b'}\in\ce_{\Gamma}(\nrm{\Gamma}{a_2\gsub{x}{b'}})\\
&\Rightarrow&\text{(Law~\ref{ce.basic} since $a_2\gsub{z}{b'}\rd\binbop{y}{a_3}{a_4}$)}\\
&&[y:a_3]a_4\in\ce_{\Gamma}(\nrm{\Gamma}{a_2\gsub{x}{b'}})\\
&\Rightarrow&\text{(by Law \ref{sub.nrm.rd} $\nrm{\Gamma}{a_2\gsub{x}{b'}}=\nrm{\Gamma}{\binbop{y}{a_3}{a_4}}$)}\\
&&[y:a_3]a_4\in\ce_{\Gamma}(\nrm{\Gamma}{\binbop{y}{a_3}{a_4}})
\end{eqnarray*}
The computability condition of $\binbop{y}{a_3}{a_4}$ ensures that $a_4\gsub{y}{d}\in\ce_{\Gamma}(\nrm{\Gamma}{a_4\gsub{y}{d}})$
for any $d\in\ce_{\Gamma}(\nrm{\Gamma}{a_3})$ where $\Gamma\sgv d\rtyp a_3$.
\end{itemize}
\item[$\ldots$:]
The other cases can be argued in a similar style.
\end{proof}
\noindent
Due to the adapted definition of computable expressions the statements of Law~\ref{ce.mon} have to be slightly adapted:
\begin{law}[Closure properties of computable expressions]
\label{sub.ce.mon}
For all $\Gamma$, $x$, $a$, $b$, $c$, $d$:
\begin{itemize} 
\item[$i$:]
$\Gamma\sngv\prdef{x}{a^b}{c}{d}$, $a\in\ce_{\Gamma}(\nrm{\gamma}{a})$, $c\in\ce_{\Gamma}(\nrm{\gamma}{b})$, and $d\in\ce_{\Gamma,x:b}(\nrm{\Gamma}{d})$ 
implies $\prdef{x}{a^b}{c}{d}\in\ce_{\Gamma}(\nrm{\Gamma}{\prdef{x}{a^b}{c}{d}})$.
\item[$ii$:]
$a\in\ce_{\Gamma}(\nrm{\Gamma}{a})$ and $b\in\ce_{\Gamma}(\nrm{\Gamma}{b})$ implies $\prsumop{a}{b}\in\ce_{\Gamma}(\nrm{\Gamma}{\prsumop{a}{b}})$, 
$\injl{a}{b}\in\ce_{\Gamma}(\nrm{\Gamma}{\injl{a}{b}})$, and $\injr{a}{b}\in\ce_{\Gamma}(\nrm{\Gamma}{\injr{a}{b}})$.
\item[$iii$:]
$\Gamma\sngv\pleft{a}$ and $a\in\ce_{\Gamma}(\nrm{\Gamma}{a})$ implies
$\pleft{a}\in\ce_{\Gamma}(\nrm{\Gamma}{\pleft{a}})$.
\item[$iv$:]
$\Gamma\sngv\pright{a}$ and $a\in\ce_{\Gamma}(\nrm{\Gamma}{a})$ implies
$\pright{a}\in\ce_{\Gamma}(\nrm{\Gamma}{\pright{a}})$. 
\item[$v$:]
For all $\bar{b}_1$, $\bar{b}_2$, $\bar{c}$: $a\in\ce_{\Gamma}([x:\bar{b}_1]\bar{c})$ and $b\in\ce_{\Gamma}([x:\bar{b}_2]\bar{c})$ implies $\case{a}{b}\in\ce_{\Gamma}([x:[\bar{b}_1,\bar{b}_2]]\bar{c})$. 
\end{itemize}
\end{law}
\noindent
\begin{proof}
$\;$
\begin{itemize}
\item[$i$:]
Due to the adapted definition of computable expressions, the proof of Law~\ref{ce.mon}($i$) has to be adapted.

Since $\Gamma\sngv\prdef{x}{a^b}{c}{d}$ by definition of normable expressions we know that 
$\nrm{\Gamma}{\prdef{x}{a^b}{c}{d}}=\prdef{x}{\nrm{\Gamma}{a}^{\snrm{\Gamma}{b}}}{\nrm{\Gamma}{c}}{\nrm{\Gamma,x:b}{d}}$. 
Let $\bar{a}=\nrm{\Gamma}{a}$, $\bar{c}=\nrm{\Gamma}{c}$, $\bar{b}=\nrm{\Gamma}{b}$, and  $\bar{d}=\nrm{\Gamma,x:b}{d}$.
Assume $a\in\ce_{\Gamma}(\bar{a})$, $c\in\ce_{\Gamma}(\bar{c})$, and $d\in\ce_{\Gamma,x:b}(\bar{d})$. 
We have to show $\prdef{x}{a^b}{c}{d}\in\ce_{\Gamma}(\prdef{x}{\bar{a}^{\bar{b}}}{\bar{c}}{\bar{d}})$.

Since by definition of computable expressions $a,c,d\in\sn{}$, by Law~\ref{sn.basic}($iv$), adapted for type tags, we have $\prdef{x}{a^b}{c}{d}\in\sn{}$.
It remains to show the computability conditions:
By Law~\ref{rd.decomp}($ii$), adapted for type tags, $\prdef{x}{a^b}{c}{d}\rd e$ and $\myneg\prdef{x}{a^b}{c}{d}\rd e$ each imply $e=\prdef{x}{a'^{b'}}{c'}{d'}$ for some $a'$, $b'$, $c'$, and $d'$ where $a\rd a'$, $b\rd b'$, $c\rd c'$, and $d\rd d'$.
Note that type tags are not reduced.
Therefore, all computability conditions except $\delta_2$ are trivially satisfied.
For the the condition $\gamma_2$, by Law~\ref{ce.basic}($v$) we know that $a'\in\ce_{\Gamma}(\bar{a})$ and $c'\in\ce_{\Gamma}(\bar{c})$
\item[$ii$:]
Only the statements involving injections have been changed. The proof is straightforward, and similar to $i$, based on the norming of injections.
\item[$iii$:]
We know that either $\nrm{\Gamma,1}{a}=[\bar{b},\bar{c}]$ or $\nrm{\Gamma,1}{a}=[x\sdef\bar{b}]\bar{c}$ for some $\bar{b}$, $\bar{c}$.
The proof then proceeds as in Law~\ref{ce.mon}($iii$) with adaptations due to the adapted definition of computable expressions.
\item[$iv$:]
We know that either $\nrm{\Gamma,2}{a}=[\bar{b},\bar{c}]$ or $\nrm{\Gamma,2_{\pleft{a}}}{a}=[x\sdef\bar{b}]\bar{c}$ for some $\bar{b}$, $\bar{c}$ and  
Due to the adapted definition of computable expressions, the proof of Law~\ref{ce.mon}($iv$) has to be adapted.
\item[$v$:]
Due to the adapted definition of computable expressions, the proof of Law~\ref{ce.mon}($v$) has to be adapted.
\end{itemize}
\end{proof}
\noindent
Due to the adapted definition of computable expressions the statement of Law~\ref{ce.abs} has to be slightly adapted.
\begin{law}[Abstraction closure for computable expressions]
\label{sub.ce.abs}
For all $\Gamma$, $x$, $a$, and $b$ where $\Gamma\rsgv\binbop{x}{a}{b}$:
If $a\in\ce_{\Gamma}(\nrm{\Gamma}{a})$, $b\in\ce_{\Gamma,x:a}(\nrm{\Gamma,x:a}{b})$, and for all $c$ with $c\in\ce_{\Gamma}(\nrm{\Gamma}{a})$ and $\Gamma\sgv c\rtyp a$
we have $b\gsub{x}{c}\in\ce_{\Gamma}(\nrm{\Gamma}{b\gsub{x}{c}})$ 
then $\binbop{x}{a}{b}\in\ce_{\Gamma}(\nrm{\Gamma}{\binbop{x}{a}{b}})$.
\end{law} 
\begin{proof}
Due to the adapted definition of computable expressions the proof of Law~\ref{ce.abs} has to be slightly adapted.
\end{proof}
\subsubsection{Computability law}%
\noindent
Due to the adapted definition of computable expressions, Law \ref{norm.match.sub}(Norm matching substitution) has to be adapted.
\begin{definition}[Type-matching substitution]
\label{type.match.sub}
A substitution $\sigma_{X,B}$ where $X=(x_1,\ldots,x_n)$ and $B=(b_1,\ldots,b_n)$ is \emph{type matching \wrt\ } $\Gamma$ iff $\Gamma=(\Gamma_0,x_1:a_1,\Gamma_1\ldots x_n:a_n,\Gamma_n)$,
for some $\Gamma_0$ and $a_i$, $\Gamma_i$ and $\Gamma\sgv\sigma_{X,B}(b_i)\rtyp \sigma_{X,B}(a_i)$ where $1\leq i\leq n$.
\end{definition}
\noindent
\begin{law}[Validity and type-matching substitutions]
\label{sub.match.sub}
For all $\Gamma$, $a$: If $\Gamma\rsgv a$ then for any type-matching substitution function $\sigma_{X,B}$ we have $\sigma_{X,B}(\Gamma)\rsgv\sigma_{X,B}(a)$.
\end{law}
\begin{proof}
We show the following more general property:
For all $\Gamma$, $a$, $b$, $n$: 
\begin{itemize}
\item[$i$:] 
If $\Gamma\sgv a\rsinc\any{n}$ then $\sigma_{X,B}(\Gamma)\sgv\sigma_{X,B}(a)\rsinc\any{n}$. 
\item[$ii$:]  
If $\Gamma\sgv a\rleq b$ then $\sigma_{X,B}(\Gamma)\sgv\sigma_{X,B}(a)\rleq\sigma_{X,B}(b)$. 
\item[$iii$:] 
If $\Gamma\sgv a\rtyp b$ then $\sigma_{X,B}(\Gamma)\sgv\sigma_{X,B}(a)\rtyp\sigma_{X,B}(b)$. 
\end{itemize}
Al properties are shown simultaneously by induction on the length $m$ of $\sigma_{X,B}$ using Law \ref{sub.type.sub}.
While the case $m=0$ is obvious assume that $n>0$.
Let $\sigma_{X,B}$ be a type-matching substitution with $X=(X',x_m)$, $X'=(x_1,\ldots,x_{m-1})$, $B=(B',b_m)$, and $B'=(b_1,\ldots,b_{m-1})$ hence
$\sigma_{X,B}=\sigma_{X',B'}\gsub{x_m}{b_m}$.
For Part $i$ assume $\Gamma\sgv a\rsinc\any{n}$. By inductive hypothesis $\sigma_{X',B'}(\Gamma)\sgv\sigma_{X',B'}(a)\rsinc\any{n}$.
The interesting case is $x_m\in\dom(\Gamma)$ hence $\Gamma=(\Gamma_1,x_m:c,\Gamma_2)$ for some $\Gamma_1$, $\Gamma_2$, and $c$.
Obviously $\sigma_{X,B}(\Gamma)=(\sigma_{X',B'}(\Gamma_1),\sigma_{X,B}(\Gamma_2))$.
Since $\sigma_{X',B'}$ is type matching \wrt\ $\Gamma$ we have $\sigma_{X',B'}(\Gamma)\sgv b_m\rtyp\sigma_{X',B'}(c)$
and therefore obviously $\sigma_{X',B'}(\Gamma_1)\sgv b_m\rtyp\sigma_{X',B'}(c)$.
Hence by Law \ref{sub.type.sub} $\sigma_{X,B}(\Gamma)\sgv\sigma_{X,B}(a)\rsinc\any{n}$.
Similar arguemtns can be made for parts $ii$ and $iii$.
\end{proof}
\noindent
Similarly, Law~\ref{nrm.cesub}(Normability implies computability of all norm-matching substitutions to computable expressions) has to be adapted.
\begin{law}[Normability implies computability of all substitutions to computable expressions]
\label{sub.nrm.cesub}
For all $\Gamma$ and $a$: If $\Gamma\rsgv a$ then for any type-matching substitution function $\sigma_{X,B}$ 
with $\sigma_{X,B}(x_i)\in\ce_{\sigma_{X,B}(\Gamma)}(\nrm{\Gamma}{\sigma_{X,B}(b_i)})$, for $1\leq i\leq n$, we have $\sigma_{X,B}(a)\in\ce_{\sigma_{X,B}(\Gamma)}(\nrm{\Gamma}{\sigma_{X,B}(a)})$.
\end{law} 
\begin{proof}
For all $\Gamma$ and $a$ with $\Gamma\rsgv a$: 
The proof is by induction on $a$. We write $\sigma$ for $\sigma_{X,B}$.
\begin{itemize}
\item
$a=\any{n}$: Obvious.
\item
$a=x:$ 
We have $\Gamma\rsgv x$ and $\bar{a}=\nrm{\Gamma}{x}=x$.
There are two cases:
\begin{itemize}
\item
$x=x_i$, for some $i$.
Obviously 
$\sigma(x)=\sigma(b_i)$ and
therefore, since $\sigma(b_i)\in\ce_{\sigma(\Gamma)}(\nrm{\Gamma}{\sigma(b_i)})$ we know that $\sigma(x)\in\ce_{\sigma(\Gamma)}(\nrm{\Gamma}{\sigma(x)})$.
\item
If $x\neq x_i$, for all $i$. Hence $\sigma(x)=x$ and it is easy to see that $x\in\ce_{\sigma(\Gamma)}(x)$.
\end{itemize}
\item
$a=\binbop{x}{b}{c}$:
From $\Gamma\rsgv\binbop{x}{b}{c}$ by definition of validity $\Gamma\rsgv b$ and $\Gamma,x:b\rsgv c$.
Applying the inductive hypothesis with $\sigma$ we obtain $\sigma(b)\in\ce_{\sigma(\Gamma)}(\nrm{\Gamma}{\sigma(b)})$ 
and $\sigma(c)\in\ce_{\sigma(\Gamma,x:b)}(\nrm{\Gamma,x:b}{\sigma(c)})$. 
Consider a $d$ where $d\in\ce_{\Gamma}(\nrm{\Gamma}{\sigma(b)})$ and $\Gamma\sgv d\rtyp b$. 
If we define $X'=(x_1,\ldots,x_n,x)$, $B'=(b_1,\ldots,b_n,d)$, and $\sigma'=\sigma\gsub{x}{d}$
then obviously $\sigma'$ is type matching and substitutes to computable expressions.
Applying the inductive hypothesis with $\sigma'$ we know that $\sigma'(c)=\sigma(c)\gsub{x}{d}\in\ce_{\sigma'(\Gamma)}(\nrm{\Gamma}{\sigma(c)\gsub{x}{d}})$.
Hence by Law~\ref{sub.ce.abs} $\binbop{x}{\sigma(b)}{\sigma(c)}\in\ce_{\sigma'(\Gamma)}(\nrm{\Gamma}{\binbop{x}{\sigma(b)}{\sigma(c)}})$ 
which by definition of substitution and since obviously $\sigma'(\Gamma)=\sigma(\Gamma)$ is equivalent to $\sigma(\binbop{x}{b}{c})\in\ce_{\sigma(\Gamma)}(\nrm{\Gamma}{\sigma(\binbop{x}{b}{c})})$.
\item
$a=\prdef{x}{a_1^{a_2}}{a_3}{a_4}$:
From $\Gamma\sngv\prdef{x}{a_1^{a_2}}{a_3}{a_4}$ by definition of validity we obtain $\Gamma\rsgv a_1,a_2,a_3$, $\Gamma,x:a_2\rsgv a_4$. 
Let $\bar{a}_1=\nrm{\Gamma}{a_1}$, $\bar{a}_3=\nrm{\Gamma}{a_3}$, $\bar{a}_4=\nrm{\Gamma,x:a_2}{a_4}$, and $\bar{a}_2=\nrm{\Gamma}{a_2}$.
By inductive hypothesis with $\sigma$ we know that $\sigma_{X,B}(a_1)\in\ce_{\sigma(\Gamma)}(\bar{\sigma}(\bar{a}_1))$, $\sigma(a_3)\in\ce_{\sigma(\Gamma)}(\bar{\sigma}(\bar{a}_3))$, and
$\sigma_{X,B}(a_4)\in\ce_{\sigma(\Gamma,x:a_2)}(\bar{\sigma}(\bar{a}_4))=\ce_{(\sigma(\Gamma),x:\sigma(a_2))}(\bar{\sigma}(\bar{a}_4))$.
By Law~\ref{sub.ce.mon}($i$) we then obtain  
$\prdef{x}{\sigma(a_1)^{\sigma(a_2)}}{\sigma(a_3)}{\sigma(a_4)}\in
\ce_{\sigma(\Gamma)}(\nrm{\Gamma}{\prdef{x}{\sigma(a_1)^{\sigma(a_2)}}{\sigma(a_3)}{\sigma(a_4)}})$.
By definition of substitution this is equivalent to $\sigma(\prdef{x}{a_1^{a_2}}{a_3}{a_4})\in\ce_{\sigma(\Gamma)}(\bar{\sigma}(\prdef{x}{\bar{a}_1^{\bar{a}_2}}{\bar{a}_3}{\bar{a}_4}))$.
\item
$a=\binop{a_1}{\ldots,a_n}$:
We have $\Gamma\rsgv a$ and $\Gamma\rsgv a_i$.
By inductive hypothesis $\sigma(a_i)\in\ce_{\sigma(\Gamma)}(\nrm{\Gamma}{\bar{\sigma}(a_i)})$. 
By Law \ref{sub.match.sub} $\sigma(\Gamma)\rsgv\sigma(a)$.
Hence by Laws~\ref{ce.mon.neg}, \ref{sub.ce.mon.appl}, and the various cases of Law~\ref{sub.ce.mon} we obtain $\binop{\sigma(a_1)}{\ldots,\sigma(a_n)}\in\ce_{\sigma(\Gamma)}(\nrm{\Gamma}{\binop{\sigma(a_1)}{\ldots,\sigma(a_n)}})$.
By definition of substitution this is equivalent to $\sigma(\binop{a_1}{\ldots,a_n})\in\ce_{\sigma(\Gamma)}(\sigma(\nrm{\Gamma}{a}))$.
\qedhere
\end{itemize}
\end{proof}
This obviously implies strong normalization for expression with restricted validity.
\begin{law}[Strong normalization of expressions with restricted validity]%
\label{sn.res.valid}
For all $\Gamma$ and $a$: $\Gamma\rsgv a$ implies $a\in\sn{}$.
\end{law} 
\begin{proof}
Assume $\Gamma\rsgv a$.
By Law~\ref{sub.val.nrm} this implies $\Gamma\sngv a$. 
By Law~\ref{nrm.ce} this implies $a\in\ce_{\Gamma}(\nrm{\Gamma}{a})$. 
By Law~\ref{ce.basic}($i$) this implies $a\in\sn{}$.
\end{proof}
%
\subsection{Decidability of the typing relation}%
The approach outlined in Law \ref{decide.dtyp} must be adapted so as to always use the normal form of minimal types (Definition \ref{sub.min.type}).
\subsection{Normal forms and consistency}
Note that all of the following references to existing laws always implicitly assume a variant for such laws with restricted validity. 
The set of \emph{valid normal forms} (Definition~\ref{vnf}) is adapted by replacing $\prim$ by $\any{n}$.
Law \ref{vnf.basic} and its proof is preserved.
Properties \ref{nf.basic}, \ref{nf.regular}, and \ref{abs.decomp} and their proofs are preserved.
Law \ref{primdec}($\any{}$-declaration property) has to be adapted by replacing $\prim$ by $\any{n}$ and by adapting its proof due to the 
adapted type decomposition properties (Law \ref{sub.type.decomp}).
\begin{law}[$\any{}$-declaration property]%
\label{sub.primdec}
For all $x$, $n$, $a$, and $b$: $x:\any{n}\sgv a\rtyp b$ implies that $b\neqv x$.
\end{law}
\begin{proof}
Due to Law~\ref{sub.rd.type} (subject reduction) and \ref{type.rd} we may assume $a,b\in\nf$.
Assume that $b\eqv x$. Since $b$ is normal this means $b=x$.

The various cases of Law~\ref{sub.type.decomp} imply that $b\neq x$ in case $a$ is an abstraction, a sum, a product, a protected definition, a case distinction, or an injection.
For example, if $a=\binbop{y}{a_1}{a_2}$ for some $y$, $a_1$, $a_2$, then by Law \ref{sub.type.decomp}($iii$)
$y:a_1\sgv a_2\rtyp b_2$ and $\sgv[y:a_1]b_2\rginc x$ for some $b_2$.
By Law \ref{sub.incl.por} this implies  $\sgv[y:a_1]b_2\rincl x$
By definition $\lambda$-inclusion this implies $\sgv[y:a_1']b_2'\rleq b'$ for some  $a_1'$, $b_2'$ and $b'$ where $b'\eqv x$. 
By Law \ref{sub.leq.prop}($iii$) this implies that $b'$ is either an abstraction or a primitive constant which is obviously not possible. 

If $a\in\de$, by definition of $\de$, since $x:\any{n}\rsgv a$, $a$ can only be an variable $x$ or a negated variable $\myneg x$.
In the first case, from $x:\any{n}\sgv x\rtyp x$, by Laws \ref{sub.type.decomp}($i$) and \ref{sub.incl.por} we know that $x:\any{n}\sgv\any{n}\rincl x$ which with an argument analogous to the above case leads to a contradiction.
The second case  can be reduced to the first case using Law \ref{sub.type.decomp}($xii$). \qedhere
\end{proof}
\noindent
The consistency result~\ref{cons} is then adapted by replacing $\prim$ by $\any{n}$.
\listoftables
\bibliography{deductica}
\bibliographystyle{plain}
\printnomenclature
\printindex
\end{document}